\documentclass[a4paper,12pt,twoside,openright]{these}
\relax

\usepackage[square,comma]{natbib}
\bibpunct{(}{)}{,}{a}{}{;}
\usepackage[francais]{babel}
\usepackage{xspace}

\usepackage[french]{minitoc}

\usepackage[T1]{fontenc}
\usepackage{dimensions}

\usepackage{lscape}

\usepackage{fancyhdr}
\renewcommand{\chaptermark}[1]{\markboth{\sf\it #1}{}}

\usepackage[intlimits]{amsmath}
\usepackage{amsfonts}
\usepackage{amssymb}

\usepackage{graphics}
\usepackage{epsfig}
\usepackage{color}
\usepackage{ae,aecompl}

 \usepackage{multirow}
 \usepackage{amsmath}
 \usepackage[mathscr]{eucal}
 \usepackage{latexsym}
 \usepackage{amsbsy}
 \usepackage{float}

 \def\gsim{\;\raise0.3ex\hbox{$>$\kern-0.75em\raise-1.1ex\hbox{$\sim$}}\;}
 \def\lsim{\;\raise0.3ex\hbox{$<$\kern-0.75em\raise-1.1ex\hbox{$\sim$}}\;}
 
\newcommand{\I}{\mathrm{i}}
\renewcommand{\Re}{\mathrm{Re}}
\renewcommand{\Im}{\mathrm{Im}}

\newcommand{\bkp}{\!\!\!}
\newcommand{\bkl}{\!\!\!\!\!}
 \newcommand{\V}[1]{\boldsymbol{#1}}      
 \newcommand{\Var}{\operatorname{Var}}    
\newcommand{\bydef}{\stackrel{\mathrm{def}}{=}}
\newcommand{\abs}[1]{\vert #1\vert}
\newcommand{\Abs}[1]{\left\vert #1\right\vert} 
\newcommand{\dyn}{\operatorname{dyn}}
\newcommand{\Nph}{N_\mathrm{ph}}         

\newcommand{\OTF}{\operatorname{OTF}}

\begin{document}

\dominitoc

\pagestyle{empty}

\def\tv#1{\vrule height #1pt depth 5pt width 0pt}

\topmargin=0in
\thispagestyle{empty}

\begin{flushleft}
\end{flushleft}

\vspace{-1cm}

\begin{equation*}
\begin{array}{|c|}
\hline
\tv{15}
\mbox{
\large
UNIVERSITE PIERRE ET MARIE CURIE }\\
\mbox{
\large
PARIS 6 } \\
\hline
\end{array}
\end{equation*}

\vskip .5cm

\centerline{TH\`ESE}

\vskip 1.cm \centerline{pr{\'e}sent{\'e}e}

\centerline{pour obtenir le grade de} \vskip .6cm

\centerline{\Large \bf Docteur de l'Universit{\'e} Paris VI} \vskip .6cm

\centerline{\bf Spécialité : \textit{Astrophysique et Instrumentations
Associées}}

\vskip 0.5cm

\centerline{par} \vskip .5cm

\centerline{\Large \sc Sylvestre Lacour} \vskip 2.cm

\centerline{\Huge \bfseries
Imagerie des étoiles évoluées } \vskip .3cm \centerline{\Huge\bfseries
par interférométrie } \vskip 1.3cm
\centerline{\Huge\bfseries Réarrangement de pupille} \vskip .3cm
\vskip 3.0cm \noindent

Soutenue le 22 Janvier 2007 devant la Commission d'examen~:

\vskip 1cm

\begin{center}
\begin{tabular}{cp{.1cm}lp{.4cm}l}
 M.& & Bruno {\sc Sicardy} && Pr\'esident \\
 M.  & & John {\sc Monnier}  && Rapporteur \\
 M.  & & Christoffel {\sc Waelkens} && Rapporteur  \\
 M.  & & Denis {\sc Mourard} && Examinateur \\
 M.  & & Jean-Philippe {\sc Berger} && Examinateur \\
 M.  & & Denis {\sc Gillet} && Examinateur \\
 M.  & & Guy {\sc Perrin} && Directeur \\
 M.  & & Gérard {\sc Rousset} && Co-directeur \\
\end{tabular}
\end{center}
\topmargin=2cm

\newpage
\cleardoublepage



\pagenumbering{Roman}

\pagestyle{fancyplain}

\chapter*{Remerciements}
\markboth{Remerciements}{Remerciements}
\label{merci}
Lire les remerciements est souvent le premier geste effectué
lorsqu'une thèse vient à nous tomber entre les mains. C'est pourquoi
il est important de soigner ce chapitre. Même s'il ne conditionne pas
une hypothétique lecture du reste de la thèse, c'est sûrement ce
passage que mes amis, mes parents, et ma copine, liront en premier.

Il faut donc le (les) soigner. C'est pourquoi, je tiens, avant tout,
à remercier mes parents de m'avoir mis au monde, à mes amis de me
supporter, et à ma copine d'être à mes côtés. En fait, un remerciement
tout spécial à Laurence, sans qui le travail de relecture de Guy lui
aurait certainement fait frôler une crise de folie qui aurait pu
m'être fatale.

Parmi mes amis, certains sont aussi mes collègues, et ont ainsi
participé aux joies et aux malheurs de la thèse en astrophysique.  On
n'aime pas mettre les gens dans des cases, mais il faut remarquer que
c'est tout de même pratique pour les remerciements. C'est pour cela
que je tiens à remercier, dans un premiers temps, tout ceux qui ont
participé à la réussite des missions d'observations sur IOTA. Dans le
désordre: Anne {\sc Poncelet}, Serge {\sc Meimon}, Julien {\sc
  Woillez}, Xavier {\sc Haubois}
et Peter {\sc Schuller}. Ensuite, ceux avec qui j'ai pu profiter des joies
des écoles d'étés californiennes: Aglae {\sc Kellerer}, Myriam {\sc Benisty},
Antoine {\sc Mérand}, Guillaume {\sc Montagnier}, Laurent {\sc Pueyo}
et ``Woody''
{\sc Woodruff} (j'en oublie certainement). Enfin, il y a ceux qui m'ont tenu
compagnie sur le campus de Meudon, Evelyne {\sc Alecian}, Etienne {\sc
  Pariat} et
les autres. Parmi les personnes que j'ai la chance de pouvoir
considérer comme des amis, il me reste à remercier Guy {\sc Perrin}, mon
directeur de thèse, qui a su être un parfait encadrant tout en
acceptant la contradiction.

Dans le cadre des remerciements de cette thèse, je tiens à citer les
personnes qui m'ont aidé dans la réalisation instrumentale. Sans eux,
il n'y aurait au sous sol du LAM aucun instrument à réarrangement de
pupille, attendant aujourd'hui avec impatience l'arrivée salvatrice de Takayuki
{\sc Kotani}. Parmi mes compagnons ``vis et écrous'', il y a d'abord eu
l'équipe du GEPI,  Sébastien {\sc Croce}, Julien {\sc Gaudemard} et Thierry
{\sc Melse}. Ils m'ont apporté leur aide dès les premiers essais,
et cela en dépit d'inégales
réussites.  Ensuite vient l'entreprise ThreeBond, dont mon cousin
Raphaël {\sc Lamy}, digne représentant, a  pu utiliser une partie des
ressources  pour me fournir plusieurs séries de colles adaptées. Un
tournant instrumental a eu lieu lorsque Frédéric {\sc Chapron} a été recruté
au LESIA, et que j'ai décidé de lui confier l'étude du système. Bien
m'en a pris, car, en plus de me faire économiser du temps, cela a
permis d'aboutir à un instrument fonctionnel. La réalisation a ensuite
été confiée aux mains expertes du GEPI, et l'assemblage à l'atelier du
LESIA (un grand merci en particulier à Vartan {\sc Arslanyan} et Claude
{\sc Collin}).

Enfin, merci à mon Jury de thèse, et surtout aux deux rapporteurs,
John {\sc Monnier} et Christoffel {\sc Waelkens}, qui ont accepté de
s'attaquer à mon manuscrit en pleine période de noël.

Un dernier mot pour souhaiter bonne chance à Anne {\sc Poncelet} et
Xavier {\sc Haubois}, tout deux disciples de Guy, qui auront bientôt la joie et le plaisir de soutenir leur thèse.

\chapter*{Résumé}
\markboth{Résumé}{Résumé}
La turbulence atmosphérique est la principale limitation à la haute
résolution angulaire pratiquée en astronomie. Elle se traduit par des
variations de la phase du champ rayonné par l'astre observé. En
interférométrie, ce problème a été résolu par l'utilisation de fibres
optiques monomodes qui filtrent le front d'onde de manière à rendre
le rayonnement parfaitement cohérent. Cette technique, appliquée sur
l'interféromètre IOTA, nous a permis de mesurer avec une grande
précision les fréquences spatiales de sept étoiles évoluées.  A partir
d'une technique de déconvolution en aveugle, nous avons imagé la
surface de ces sept objets. Bételgeuse et $\mu$ Cep, deux
supergéantes rouges, ainsi que R leo, Mira et $\chi$ Cyg, trois étoiles
variables de type Mira, mais également l'étoile symbiotique CH Cyg ont montré
des structures très diverses, avec des photosphères d'apparence
fortement dissymétriques. Seule la géante rouge Arcturus n'a pas
présenté ces caractéristiques. Nous avons, notamment, pu estimer la
masse de $\chi$ Cyg à 0,88 $\pm$ 0,04 M$_\odot$ à partir de la trajectoire
balistique de la haute atmosphère de l'étoile.

Au vu de ces résultats, nous proposons d'utiliser les techniques de
filtrage spatial interférométrique pour corriger l'effet de la
turbulence au sein de la pupille d'un télescope. Cette technique, le
réarrangement de pupille, a requis le développement d'un algorithme
spécifique de réduction des données. Nous montrons qu'il permet de
reconstruire des images affranchies de l'influence de la turbulence
atmosphérique et limitées uniquement par le bruit de photon dans le
domaine visible. Nos simulations montrent qu'un tel système peut
fournir des images à la limite de diffraction des grands télescopes
avec des dynamiques d'au moins le million. Cette technique est
en cours de validation expérimentale par la construction d'un
démonstrateur en laboratoire.

\chapter*{Abstract}
\markboth{Abstract}{Abstract}
Atmospheric turbulence is an important limit to high angular
resolution in astronomy. Interferometry resolved this issue by
filtering the incoming light with single-mode fibers. Thanks to this
technique, we obtained with the IOTA interferometer very precise
measurements of the spatial frequencies of seven evolved stars. From
these measurements, we performed a blind deconvolution to restore an
image of the surface of the stars. Six of the them, Bételgeuse, $\mu$
Cep, R leo, Mira, $\chi$ Cyg and CH Cyg, feature very asymmetrical
brightness distributions. On the other hand, the Arcturus data are
extremely well fitted with a simple limb-darkened photospheric
disc. From the observations of $\chi$ Cyg, we show that the star is
surrounded by a molecular shell undergoing a ballistic motion. By
combining our dataset with spectroscopic measurements, we inferred a
mass of the star of 0.88 $\pm$ 0.04 M$_\odot$.

We propose to use the same technique of spatial filtering with
single-mode fibers to correct for the effect of turbulence in the
pupil of a telescope. Because the pupil is redundant, this technique
does require a remapping of the pupil. We developed a dedicated
algorithm to show that it was possible to reconstruct images at the
diffraction limit of the telescope free of any speckle noise. Our
simulations show that a high dynamic range (over $10^6$) could be
obtained in the visible on an 8 meter telescope. A lab experiment is
under construction to validate the concept of this new instrument.

%

\tableofcontents
\newpage
\cleardoublepage

\listoftables
\newpage

\listoffigures
\newpage
\cleardoublepage
\pagenumbering{arabic}

\chapter*{Introduction}
\markboth{Introduction}{Introduction} \addcontentsline{toc}{part}{{\large\sf\bfseries
Introduction}} \label{chap:intro}
\subsubsection{Contexte scientifique}

Les étoiles naissent, rayonnent et meurent. Le soleil, notamment, est
né il y a un peu plus de 4 milliards d'années, et débutera sa fin de
vie dans à peu près autant d'années. Dés lors, le coeur de l'étoile se
contracte, et les couches supérieures de l'étoile se dilatent. Le
diamètre de l'étoile devient si grand qu'il englobe les planètes les
plus proches, dont la Terre.  Cependant, l'étoile n'est plus le corps
compact qu'il était avant. La masse volumique du gaz à sa surface est
d'environ $10^{-3}$ kg/m$^3$, soit environ 1000 fois plus faible que la
densité de l'atmosphère terrestre. Ceci est lié à un champ
gravitationnel lui aussi très ténu.

Par ailleurs, il est fréquent d'observer dans ces étoiles évoluées la
présence d'instabilités qui génèrent des modifications considérables
de la surface stellaire. Les étoiles de type Mira, par exemple,
possèdent une photosphère dont le rayon peut varier, en quelques mois,
de la distance Terre-Soleil (1 ua) à la distance Terre-Mars (1,5
ua). Le cas des supergèantes rouges est encore plus dramatique, avec,
dans le cas de Mu Cep par exemple, une photosphère dont la surface
atteindrait la planète Saturne. Il est prédit, sur Bételgeuse
notamment \citep{2003csss...12.1024F}, l'existence de cellules de
convection de tailles comparables au rayon de l'étoile.

La surface des étoiles évoluées est cependant extrêmement difficile à
observer, car, même si ces étoiles sont imposantes comparées au
Soleil, leurs dimensions angulaires restent très faibles, de l'ordre
de 10 milli-secondes d'angle. C'est pourquoi il est utile de faire
appel à l'interféromètrie, seule technique d'observation permettant
d'obtenir la résolution spatiale nécessaire pour résoudre la
photosphère.

Cependant, observer une étoile par interférométrie fournit une
information différente de celle que nous fournirait l'imagerie
classique. Au lieu de mesurer la distribution spatiale d'intensité,
l'interférométrie nous donne des mesures de l'objet observé à des
fréquences spatiales déterminées. Ceci, considéré par certains
astronomes comme une faiblesse de l'interférométrie par rapport à
l'imagerie, est en fait un avantage dans un certain nombre de
situations. Par exemple, si l'on est capable de modéliser la structure
de l'étoile, on peut alors mesurer la position de structures très
faiblement brillantes. Nous verrons au cours de cette thèse qu'il est
notamment possible de mesurer avec une grande précision l'emplacement
de la couche moléculaire présente dans la haute atmosphère des étoiles
évoluées.

\subsubsection{De l'observation à l'instrumentation}

A l'instar de la spectroscopie, l'interférométrie apporte une
information complémentaire à l'imagerie. Un exemple est celui des
étoiles à rotation rapide, dont l'élongation peut être mesurée alors
même que l'étoile n'est pas résolue par l'interféromètre \citep{2006A&A...453.1059K}. De tels résultats sont possible grâce aux techniques
récentes de correction de l'influence de la turbulence atmosphérique
utilisées en interférométrie.

Le filtrage monomode par fibre optique est une de ces techniques
novatrices qui ont bouleversé la discipline. Elle permet de convertir
les perturbations atmosphériques en de simples variations d'amplitude
du flux lumineux \citep{1995SPIE.2476..120P}. Ces variations du flux
peuvent ensuite être mesurées, et leur influence retranchée aux
mesures effectuées.

Cette technique de filtrage de la turbulence n'est pas utilisée en
imagerie classique, où l'on préfère une correction active, en temps
réel, par le biais d'une optique adaptative. Cependant, il existe des
cas où, lorsque l'on souhaite, par exemple, observer à de courtes
longueurs d'onde, les limitations technologiques de l'optique
adaptative ne permettent pas de corriger correctement la
turbulence. Le filtrage par fibres optiques monomodes est alors une
voie instrumentale intéressante à explorer.

Comparé à l'interférométrie longue base, un tel système serait loin d'avoir la 
même résolution. Celle-ci serait limitée par la taille du télescope. 
Cependant, parce que l'on peut utiliser toute la pupille, ce système 
présenterait un certain nombre d'avantages~: 
\begin{itemize}
\item Il n'y aurait pas besoin de lignes à retard, et par conséquent pas non plus de suiveur de franges dans le cas de sources faibles
\item Toutes les fréquences spatiales présentes dans la pupille 
pourraient être mesurées en une seule fois.
\end{itemize}

Comparé à l'imagerie directe, ce système permettrait d'obtenir des
images complètement affranchies de l'influence de la
turbulence. Cependant, parce qu'il utilise un filtrage passif, un tel
système présenterait également un certain nombre d'inconvénients~:
\begin{itemize}
\item Le temps d'intégration de chaque pose serait limité par le temps de 
cohérence de l'atmosphère.
\item Le champ observable serait limité par le champ des fibres monomodes.
\end{itemize}

\subsubsection{Plan du manuscrit}

La première partie de ce mémoire de thèse est centrée sur la
problématique astrophysique des étoiles évoluées. Elle se focalise sur
la reconstruction d'images de la surface stellaire par
interférométrie. Deux techniques seront utilisées~: la première
consiste à effectuer une reconstruction d'image en aveugle, et la
deuxième, en une reconstruction paramétrique.

La deuxième partie de ce manuscrit est instrumentale et concerne
l'élaboration d'un instrument interférométrique utilisant le filtrage
spatial par fibre optique dans la pupille d'un télescope. Un premier
chapitre présente le concept de l'instrument et le deuxième permet de
fournir les bases permettant l'optimisation des paramètres de
conception de celui-ci. Nous terminerons cette thèse avec un chapitre
décrivant les réalisations expérimentales effectuées dans ce cadre.

\part*{\flushright Première partie \\ \vspace{1cm} I. Les étoiles évoluées}
\addtocontents{toc} {\protect\vskip 0.8cm \protect\begin{center}
{\protect\Large\protect\sf I. Les étoiles évoluées}
\protect\end{center} \protect\vskip -0.3cm 
}

\chapter{Introduction aux étoiles évoluées}
\begin{center}
\end{center}
\minitoc \label{sec:etoile} \vskip1cm
\newpage

\section{Les étoiles évoluées}

\subsection{La formation des étoiles géantes}

Une étoile est une boule de gaz maintenant un équilibre (stable ou
instable) entre force gravitationnelle et pression interne. Cette
dernière peut être radiative, thermique ou encore due à un gaz de
particules relativistes. Au sein de la séquence principale, la force
de pression est maintenue par la fusion de la matière contenue dans le
c\oe ur de l'étoile. C'est le cas de tout astre ayant une masse
supérieure à 0,08 M$_\odot$ (nécessaire pour permettre la fusion de
l'hydrogène) et inférieure à 100 M$_\odot$ (limite supérieure à partir de
laquelle la pression est tellement grande qu'elle engendre
l'instabilité de l'étoile). Parce qu'il s'agit justement de résister à
la pression gravitationnelle, les étoiles consomment de l'énergie. Le
``carburant'' étant nécessairement une ressource finie, il est logique
que les étoiles aient une certaine durée de vie. Elles ont été créées à
un moment précis, et s'éteindront lorsque leurs sources d'énergie
s'épuiseront. Il s'en déduit la notion d'évolution stellaire.

L'étoile naît d'un nuage de gaz auto-gravitant qui se contracte sous
l'effet de la gravitation. L'énergie potentielle gravitationnelle est
alors libérée sous forme de chaleur permettant, à terme, la fusion de
l'hydrogène (dès que la température interne atteint $\approx 10^7$ K)
qui est le premier combustible nucléaire (après le deutérium) à être
utilisé par les étoiles. S'ensuit la possible fusion de l'Hélium, du
Carbone, du Néon, de l'Oxygène, du Silicium, et plus généralement de
toutes les espèces d'indices atomiques inférieurs à celui du Fer.

En conséquence, l'évolution stellaire peut être résumée comme
suit. Durant une première phase d'évolution se forme une boule
homogène constituée principalement d'hydrogène. La transformation de
cet élément (composant A) en un composant B (Hélium), produit de
l'énergie nucléaire. Quand l'élément A est épuisé au c\oe ur de
l'étoile, la source d'énergie s'arrête, et le gradient de pression
disparaît. Le c\oe ur est alors composé de l'élément B, entouré d'une
enveloppe constituée principalement de l'élément A. Sous l'effet de sa
propre gravité, le c\oe ur se contracte et sa température
augmente. Parallèlement, l'enveloppe se dilate, et se refroidit
(Théorème du viriel). Il se peut alors que l'augmentation de la
température à la surface du c\oe ur permette la fusion en couche de A
en B, augmentant ainsi la masse du c\oe ur et sa vitesse
d'effondrement. Si la température du c\oe ur atteint un niveau
suffisant, l'élément B peut fusionner, créant une nouvelle source
d'énergie. La contraction du c\oe ur s'arrête alors, et B est
transformé en C. L'étoile se compose des éléments A, B et C répartis
respectivement dans différentes couches allant de la surface au c\oe
ur de l'étoile . Ce phénomène se répète pour des éléments D, E, F, de
masses atomiques toujours plus grandes (Carbone, Néon, Oxygène, ...),
avec toujours une stratification de l'étoile et une augmentation de la
taille de l'enveloppe stellaire.

\subsection{L'évaporation des étoiles}

\begin{figure} 
\centering 
\resizebox{\hsize}{!}{\includegraphics{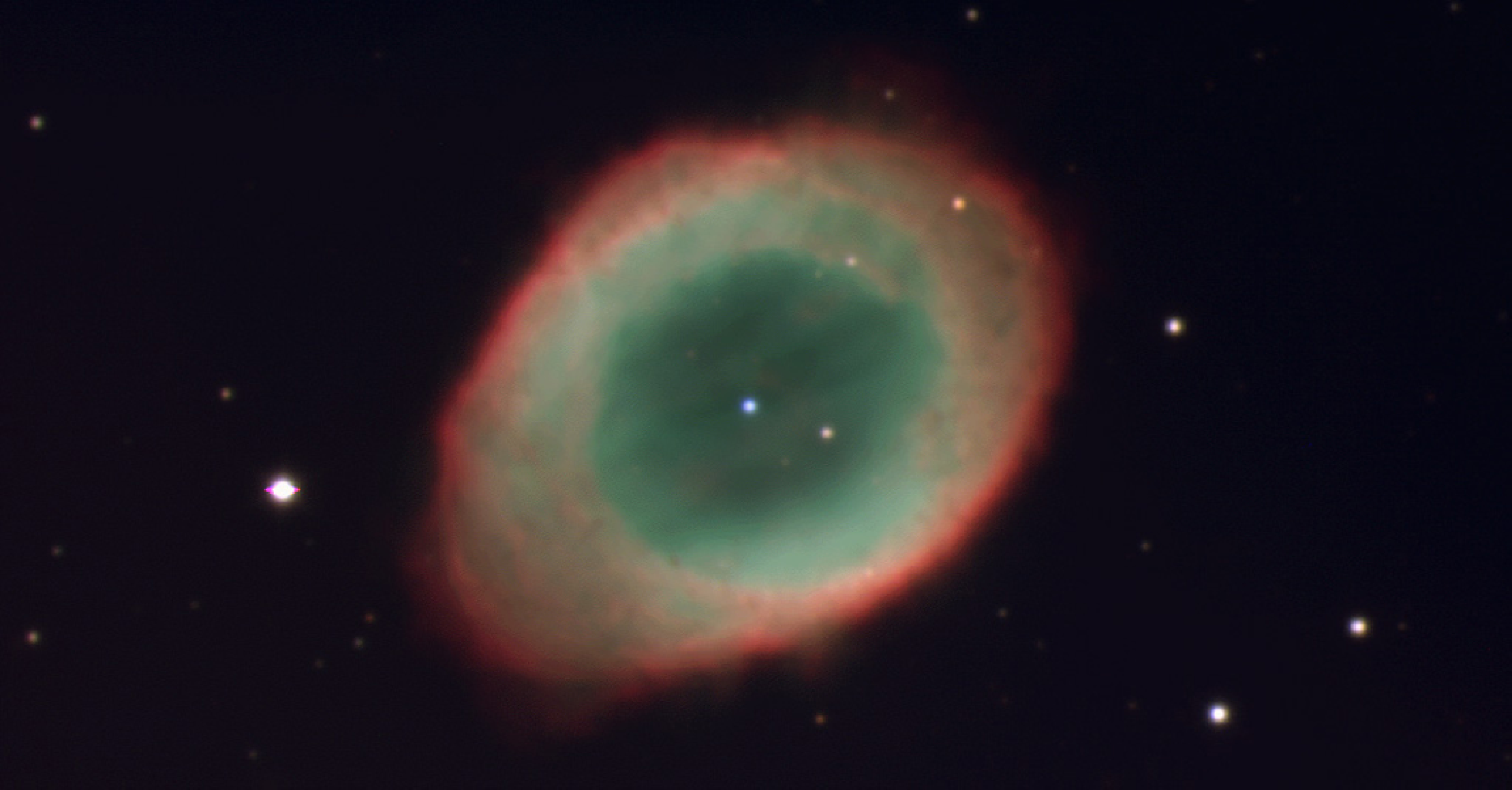}}
   \caption[Nébuleuse de la Lyre (M57)]{Nébuleuse de la Lyre (M57),
   une nébuleuse planétaire qui était autrefois une géante rouge. Le
   coeur et l'enveloppe de poussière sont devenus respectivement une
   naine blanche et un anneau de poussières entourant
   celle-ci. Crédit: W. M. Keck Observatory } \label{fig:M57}
\end{figure}

L'évolution d'une galaxie dépend principalement de la matière qui la
compose. Fruit de la nucléosynthèse primordiale, la matière originelle
est principalement composée d'Hydrogène, de Deutérium et de
Lithium. \`A partir de cette matière, des étoiles se forment, au c\oe
ur desquelles des éléments plus lourd se créent. Ces éléments sont
ensuite relâchés dans le milieu interstellaire, et participent à
la formation d'une nouvelle population d'étoiles. On note, par
exemple, la coexistence dans notre galaxie de deux populations
(Population I et II) qui se caractérisent par une proportion de métaux
(tout élément de masse atomique supérieure à l'Helium) différente. Ce
phénomène, appelé l'astration, dépend en grande partie de la capacité
des étoiles à rejeter la matière modifiée en leur c\oe ur dans le
milieu interstellaire. La maîtrise de ce processus est fondamentale à
la compréhension de l'évolution galactique. Le rejet de matières
s'opère de deux manières:
\begin{itemize}
\item Novae et Supernovae: il s'agit d'un phénomène relativiste qui se 
caractérise par une éjection massive et rapide des couches
superficielles de l'étoile. Ce taux de rejet s'obtient facilement par
l'observation de ces phénomènes brefs mais intenses.
\item Vents stellaires: les étoiles évoluées se caractérisent par une 
atmosphère extrêmement étendue et un dégagement énergétique
important. Sous cette forte pression radiative, la partie supérieure
de l'étoile peut être soufflée, relâchant ainsi cette matière dans le
milieu interstellaire (figure~\ref{fig:M57}).
\end{itemize}

Les Supernovae ont longtemps été considérées comme la source
principale de rejet d'éléments lourds dans le milieu
interstellaire. Néanmoins, cette idée a été remise en cause par la
faible quantité de supernovae observées.  Plus particulièrement, on
s'est aperçu que la plupart des étoiles éjectent 80\% de leur masse
avant leur mort, évitant ainsi de se retrouver avec une masse finale
supérieure à la masse critique de Chandrasekar (1.44 M$_\odot$), ce
qui leur épargne une fin cataclysmique. Une inconnue subsiste dans ce
scénario: comment s'opère cette perte de masse?  Sous l'effet de la
pression de radiation, les grains peuvent permettre au gaz d'acquérir
une vitesse supérieure à la vitesse d'échappement (5 à 10
km/s). Pourtant, la condensation du gaz sous forme de poussières
nécessite une température bien inférieure à celle présente dans les
couches supérieures de l'atmosphère.  Il doit, par conséquent, exister
un premier mécanisme de propulsion transportant la matière
suffisamment loin pour qu'il y ait condensation. Il semblerait que les
instabilités de l'étoile soient liées à ce phénomène. De larges
amplitudes de pulsation semblent, en particulier, être corrélés avec
une perte de masse importante \citep{1979ApJ...227..220W}. Les
instabilités régiraient ainsi la suite de l'évolution stellaire.

\newpage
\section{Les instabilités et les modes d'oscillation}

On peut distinguer trois principaux types d'instabilités. Elles se
distinguent par des échelles de temps et de tailles différentes. Il
s'agit des instabilités thermique, dynamique et convective.

\subsection{L'instabilité thermique}

La première source d'instabilité correspond à un emballement thermique.
Selon la loin des gaz parfaits:
\begin{equation}
P=\rho k T
\end{equation}
Ce qui peut aussi se traduire par:
\begin{equation}
\frac{d \, P}{P} = \frac{d \, \rho}{\rho} + \frac{d \, T}{T}
\end{equation}
Or, si l'on considère l'étoile à l'équilibre hydrostatique, on peut établir:
\begin{equation}
\frac{d\, P}{P} = 4/3 \frac{d \, \rho}{\rho}
\end{equation}
Soit
\begin{equation}
1/3 \frac{d \, \rho}{\rho} = \frac{d\, T}{T}
\end{equation}
Par conséquent, si le c\oe ur se contracte, $\rho$ augmente ainsi que la
température. Cet accroissement de température permet de lutter contre
la contraction, permettant, ainsi, un retour à l'état d'origine. De
même, si le c\oe ur se dilate, la température diminue de manière à
stopper l'expansion.

Cet équilibre est parfois rompu. C'est le cas lors du processus de
fusion au sein d'une couche. A une distance $r$ du centre de l'étoile et
à l'épaisseur $l$, l'équilibre hydrostatique
s'écrit:
\begin{equation}
\frac{d\,P}{P} = 4l/r \frac{d\,\rho}{\rho}
\end{equation}
Ainsi :
\begin{equation}
(4l/r - 1) \frac{d \, \rho}{\rho} = \frac{d \,  T}{T}
\end{equation}
Si $4l/r-1$ est négatif, l'expansion de la couche se traduit par une
augmentation de température et un emballement de la réaction de
fusion. Ceci se traduit concrètement par une pulsation de l'étoile
évoluée, expliquant des variations photométriques à des périodes de
plusieurs centaines d'années.

\subsection{L'instabilité dynamique}
\label{sec:insta_dyn}

\begin{table}
\caption{Les instabilités dynamiques}
\label{tb:periode_pulsations}
\centering
\begin{tabular}{lllll}
\hline
\hline
Type & Période & Population & Type spectral & Radial (R) ou \\
 & & & & non-radial (NR) \\
\hline
Miras & 100-700 jours & I, II & M, N, R, S & R  \\
RV Tauri & 20-150 jours & II & G, K & R \\
Cepheids & 1-50 jours & I & F6-K2 & R \\
RR Lyrae & 1.5-24 heures & II & A2-F2 & R \\
Le soleil & 5-10 min & I & G2 & NR \\
Naines blanches & 100-1000s & I, II & O, B2, A0 & NR \\
\hline
\end{tabular}
\begin{list}{}{}
\item Extrait de \citet{2001Theo_Astro}.
\end{list}
\end{table}

Elle correspond à un déplacement de matière à l'intérieur de
l'étoile. Supposons une contraction adiabatique d'une couche de
l'étoile.
\begin{equation}
	P_{\rm adia} V^\gamma=cst,
\end{equation}
La pression interne à cette couche va augmenter. Pour qu'il y ait
stabilité, il faut que la pression exercée sur cette couche augmente,
de manière à ramener le système à son état antérieur. La pression
exercée par les couches supérieures sous l'effet du champ
gravitationnel s'écrit: 
\begin{equation}
	P_{\rm grav}=\int_m^M \frac{Gm}{4 \pi r^4}  \, dm
\end{equation}
En considérant une dilatation infinitésimale de la couche ; $l=l+\varepsilon$:
\begin{equation}
	P_{\rm adia}'=P_{\rm adia}(1+3 \gamma \varepsilon)
\end{equation}
\begin{equation}
	P_{\rm grav}'=P_{\rm grav}(1+4 \varepsilon)
\end{equation}
Ainsi, pour qu'il y ait stabilité, il faut que $P_{\rm grav}' > P_{\rm
adia}'$ lorsque $\varepsilon > 0$. Soit:
\begin{equation}
	\gamma > 4/3
\end{equation}

Dans le cas d'un gaz monoatomique, $\gamma = 5/3$, ce qui confirme la
stabilité générale de l'étoile. Il existe cependant des exceptions. La
plus notable est celle d'un gaz partiellement ionisé. Lorsqu'il se
contracte, les électrons ont tendance à se recombiner et le gaz à
perdre des particules. Pour un simple gaz monoatomique ionisé entre 18
et 82\%, on peut démontrer
\citep{2001Stel_stru} que $\gamma < 4/3$.

Ce phénomène, observé à de multiples stades de l'évolution des
étoiles, est couramment appelé le ``$\kappa$ mécanisme''
\citep{1995ARA&A..33...75G,1996ARA&A..34..551G}. Une approche physique
en permet une meilleure compréhension.  Si l'on chauffe une couche de
gaz, elle se dilate, devient plus transparente, et peut ainsi émettre
son surplus d'énergie sous forme radiative.  Cependant, un gaz
partiellement ionisé, lorsqu'il se dilate, devient plus opaque du fait
de la recombinaison des électrons. Ceci augmente le réchauffement dû à
l'énergie irradiée par le centre de l'étoile, renforçant le phénomène
de dilatation. L'énergie nécessaire à l'existence de larges pulsations
peut ainsi être fournie. Ces couches de transitions entre gaz ionisé
et gaz non-ionisé sont localisées au sein de l'étoile. Il existe
principalement deux couches: celle correspondant à l'ionisation de
l'hydrogène ($1-1.5 \times 10^4$ K), et celle correspondant à l'Hélium
(He->He$^+$; $4 \times 10^4$ K). La position exacte de ces couches
dépend de la température effective de l'étoile. Pour une étoile chaude
($T_{\rm eff} = 7500$ K), ces couches sont très près de la surface, ce
qui empêche le $\kappa$ mécanisme d'entraîner suffisamment de masse
pour produire de larges oscillations. {\it A contrario}, dans les
étoiles ayant une température de surface plus faible, d'importantes
oscillations apparaissent. L'harmonique de résonance est alors
sélectionnée en fonction de la position de la couche
d'ionisation. C'est, notamment, ce qui permet aux Miras d'osciller sur
le mode fondamental (cf Table~\ref{tb:periode_pulsations}).

\subsection{Le phénomène de convection}
\label{sec:insta_conv}

Le phénomène de convection, bien que simple en son principe, s'avère
difficile à formaliser mathématiquement. Le concept peut être compris à partir d'un
déplacement vertical d'un petit élément de matière. Cet élément va
alors s'équilibrer en pression avec son nouvel environnement et se
dilater. Cette dilatation se traduit par une modification de sa
densité. Si elle devient plus faible que le milieu, la poussée
d'Archimède va l'entraîner encore plus vers le haut, produisant le
phénomène d'instabilité. Les paramètres moteurs de cette instabilité
sont le gradient de pression et le gradient de température. Un critère
d'existence de telles instabilités est celui de Schwarzschild:
\begin{equation}
	\frac{d\,\ln{T}}{d\,\ln{P}} > \frac{(\gamma -1)}{\gamma} \, .
\end{equation}
Dans l'atmosphère des étoiles, la convection apparaît dans les zones
faiblement ionisées où l'opacité génère un important gradient de
température. Cette zone d'ionisation est très étendue dans
l'atmosphère relativement froide des étoiles évoluées. Le soleil,
quant à lui, présente aussi cette zone de convection, mais sur une
épaisseur beaucoup plus faible.

\subsection{Le soleil}

\begin{figure}[h!] 
\centering 
\resizebox{\hsize}{!}{\includegraphics{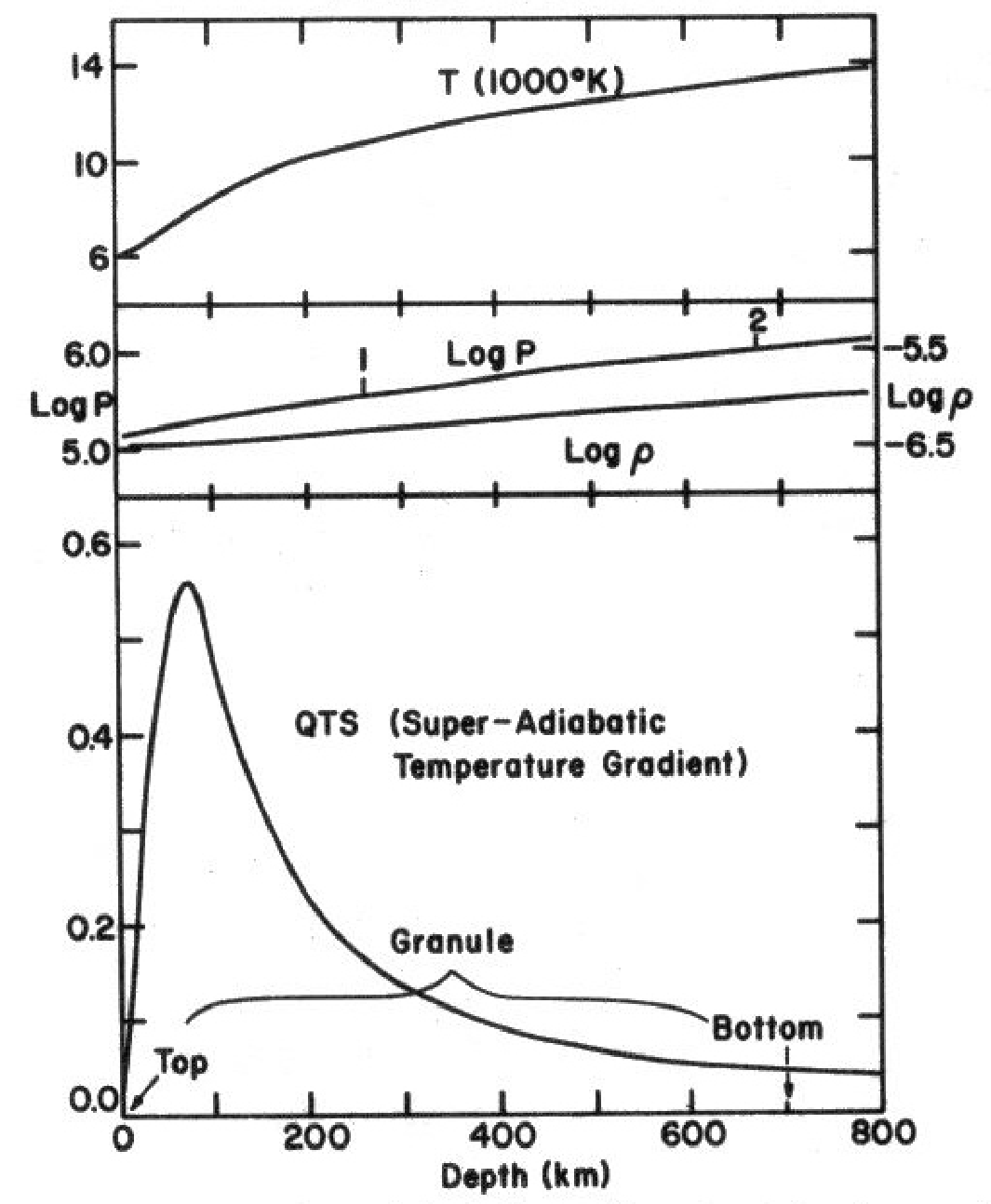}
\includegraphics{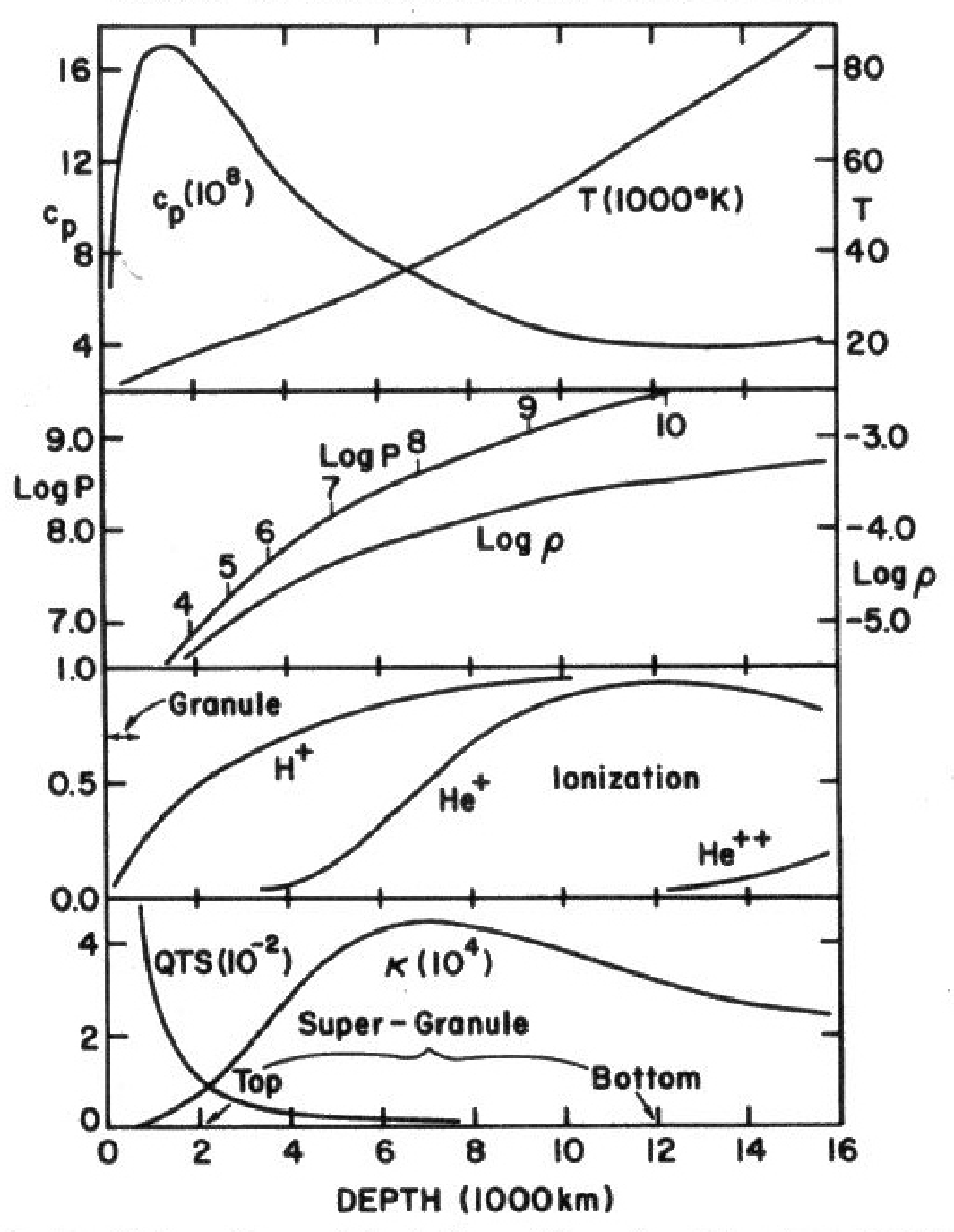}}
   \caption{ Les paramètres physiques du soleil \citep{1975ApJ...195..137S} }
   \label{fig:Schwar} 
\end{figure}

Bien que le soleil ne soit pas une étoile évoluée, cette étoile
présente néanmoins un certain nombre d'instabilités. Celles-ci sont
observés sous la forme de ``granules'' et de ``super-granules''.  Ces
observations peuvent ainsi fournir matière à réflexion pour étudier le
fonctionnement des instabilités. \citet{1975ApJ...195..137S} a
accompli ce travail dans le but de prédire les instabilités à la
surface des supergéantes. La figure~\ref{fig:Schwar} reproduit deux
graphiques de son article qui représentent les différentes valeurs
physiques à l'intérieur de notre soleil. On observe les deux
paramètres qui caractérisent deux types d'instabilités
(paragraphes~\ref{sec:insta_dyn} et~\ref{sec:insta_conv}):
\begin{itemize}
\item Instabilité dynamique: l'opacité ($\kappa$)
\item Convection: le gradient de température $QTS = d\,\ln{T}/d\,\ln{P} - (\gamma-1)/\gamma$
\end{itemize}
Par un simple raisonnement géométrique, Schwarzchild associe ensuite
l'existence des granules à la zone convective, et les super-granules à
la zone d'ionisation de l'Hydrogène et de l'Hélium. Il en conclut que
ces types d'instabilités doivent exister dans les étoiles évoluées, mais à
une échelle beaucoup plus grande.

\newpage
\section{Ma thèse dans ce contexte}

\begin{figure}[h]
\centering 
\includegraphics[width=9cm]{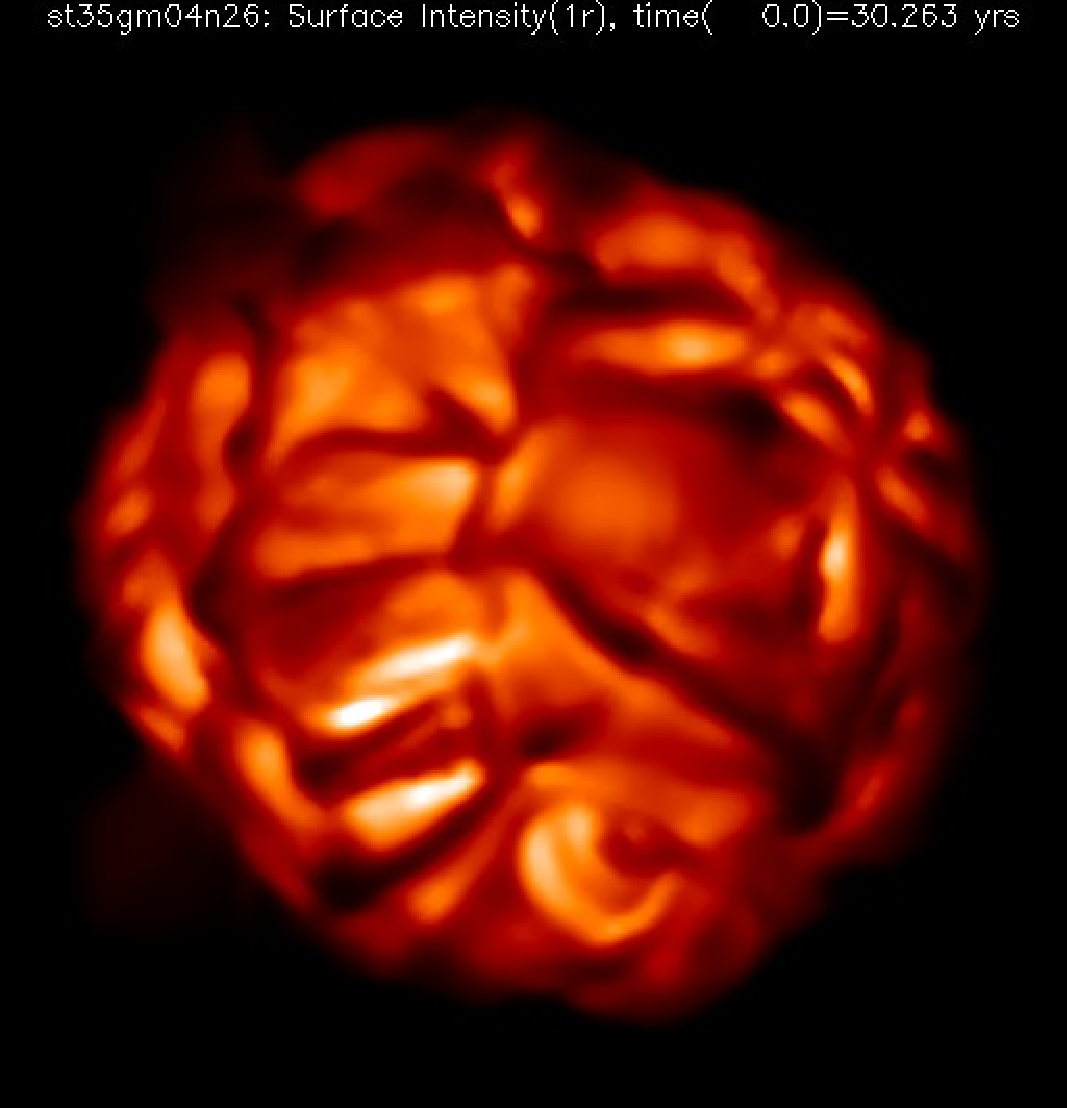}
   \caption[\citet{2003csss...12.1024F}]{ Simulation de l'effet de
   convection à la surface de super géantes
   \citep{2003csss...12.1024F} } \label{fig:Freytag}
\end{figure}

Deux questions importantes justifient ce travail de thèse:
\begin{enumerate}
\item Quelles sont les sources d'instabilité dans les étoiles évoluées? 
\item Comment la matière se trouve-t'elle éjectée de l'atmosphère?
\end{enumerate}
Ces questions, intrinsèquement liées, ont de profondes répercussions
sur l'évolution stellaire et galactique. Cependant, parce que ces
phénomènes concernent la surface stellaire et son environnement
proche, la plupart des observations restent indirectes. Comme dans le
cas du soleil, nous nous attendons à pouvoir confirmer l'existence
d'instabilités dynamiques et convectives.

Concernant les phénomènes de convection, nous disposons, d'un côté,
des mesures de variations photométriques
\citep[récemment,][]{2006MNRAS.Kiss}, et de l'autre, des simulations 
d'atmosphères convectives (Figure~\ref{fig:Freytag}). Les instabilités
dynamiques se traduisent, elles, par des pulsations de grandes
amplitudes. Elles sont mieux connues car observées par interférométrie
\citep{2004A&A...426..279P}, et elles disposent de sérieuses bases théoriques
\citep[par exemple][]{1996MNRAS.278...11F}.

Cependant, l'accélération initiale de la matière à la surface de
l'étoile est un mécanisme peu connu, notamment, parce qu'il nécessite
la connaissance de l'instabilité de l'étoile. De nombreux modèles
existent néanmoins
\citep{1988ApJ...329..299B,1996A&A...307..481B,2006A&A...456.1001C}
mais sont confrontés à très peu d'observations. En l'absence de
résolution angulaire suffisante, les observations spectroscopiques ont
cependant permis de contraindre un certain nombre de propriétés à
partir des raies d'absorptions les plus énergétiques
\citep{1979ApJ...234..548H,1982ApJ...252..697H}.

L'objectif de ce travail de thèse consiste à utiliser la haute
résolution angulaire, et notament l'interférométrie, pour tenter de
répondre à ces problématiques. On s'apperçoit alors que, face à la
complexité des phénomènes, mesurer quelques fréquences spatiales par
interférométrie ne suffit plus. C'est pourquoi il est nécessaire
d'imager entièrement la surface stellaire pour pouvoir dissocier les
différents mécanismes entrant en jeu.

\chapter{Imagerie interférométrique}
\begin{center}
\end{center}
\minitoc \label{sec:image} \vskip1cm
\clearpage

\section{Les données interférométriques}

\subsection{Le principe général}

Un objet astrophysique se caractérise par sa fonction continue de
distribution spatiale de brillance, allant de $-\infty$ à
$+\infty$. L'imagerie consiste à retrouver cette distribution. On peut
l'obtenir {\it via} un instrument imageur - télescope ou lunette - qui
permet de projeter une distribution d'intensité similaire sur un
détecteur. Néanmoins, lorsque l'on utilise un tel système, on modifie
cette fonction, en 1) la limitant en champ, 2) la limitant en
fréquence spatiale et 3) la convoluant par la réponse impulsionnelle
de l'instrument. L'image obtenue à travers un télescope est, par
conséquent, différente de la fonction de distribution de l'objet. Par
exemple, la théorie de la diffraction limite la résolution d'un
télescope à un facteur de la taille de celui-ci. Cette propriété est
une importante limitation technologique à l'obtention d'images à haute
résolution.

Pour s'en affranchir, on peut mesurer directement la valeur complexe
du champ électromagnétique provenant de l'objet observé. Le théorème
de Van Cittert-Zernike nous donne une relation directe entre la
distribution en flux d'un objet astrophysique $I(\alpha,\beta)$, et la
cohérence du champ en deux points distincts à la surface de la Terre:
\begin{equation}
  V(u,v)=TF(I(\alpha,\beta))
\end{equation}
où TF est la Transformée de Fourier normalisée, $(u,v)$ la base formée
par les deux points de mesures (en multiples de $\lambda$), et
$(\alpha,\beta)$ la coordonnée angulaire du flux observé (en radians).

Ainsi, la connaissance de la cohérence spatiale de la lumière nous
permet de reconstruire une image de l'objet, avec la transformée de
Fourier comme relation mathématique de passage.  En relevant la
cohérence du flux lumineux, on détermine une valeur correspondant à
une fréquence spatiale de l'objet. Le plan de ces fréquences spatiales
s'appelle le plan $u$-$v$. La résolution maximale de l'image
reconstruite sera limitée par la taille de la zone de mesure et, la
qualité de reconstruction par le nombre de mesures. Cependant, les
mesures ne peuvent couvrir de manière continue l'ensemble des
fréquences spatiales. Un algorithme de déconvolution devra être
utilisé pour déterminer celles manquantes. Un terme de régularisation
permettra de choisir la solution qui nous semble la plus appropriée
(voir paragraphe~\ref{sec:algo_rec_image}).

\subsection{IOTA (Infrared Optical Telescope Array)}
\label{sec:IOTA_r}

\begin{figure}[h!]
  \centering \includegraphics[width=15cm]{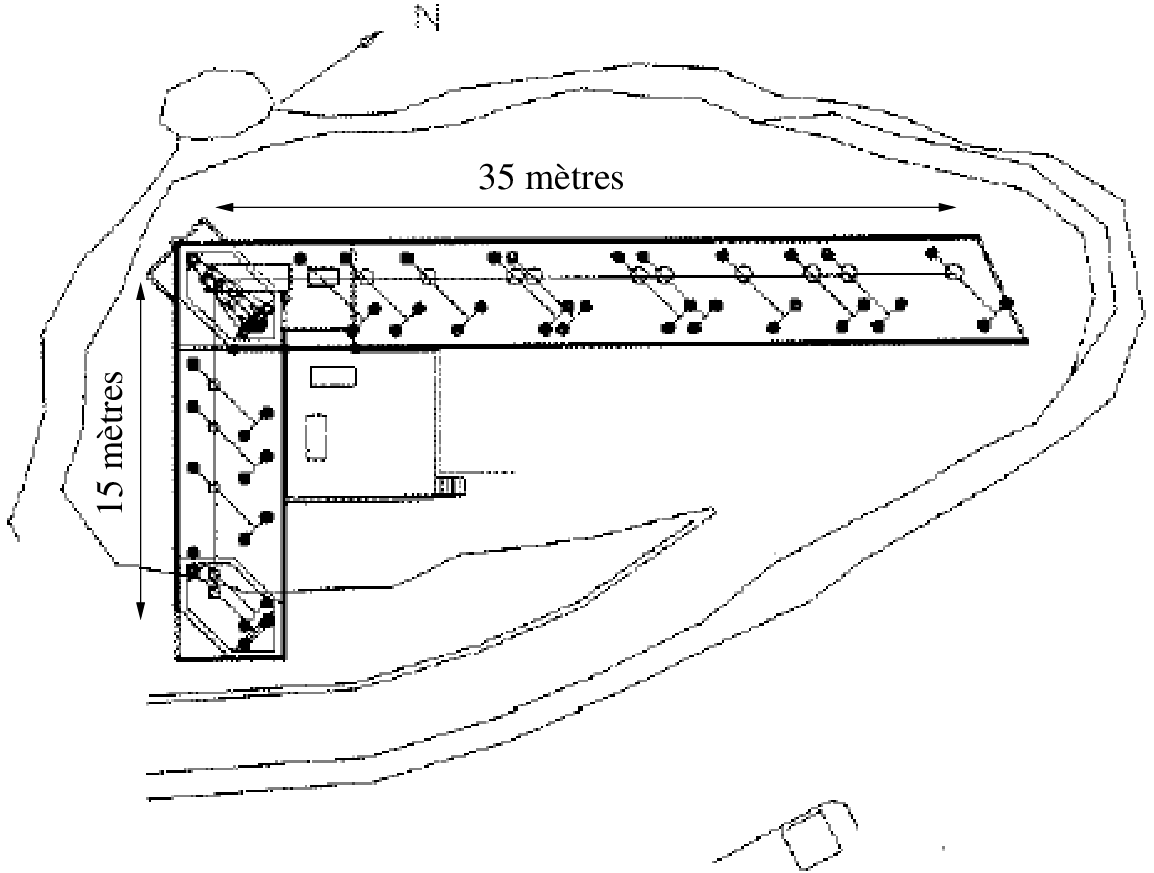}
  \includegraphics[width=5cm]{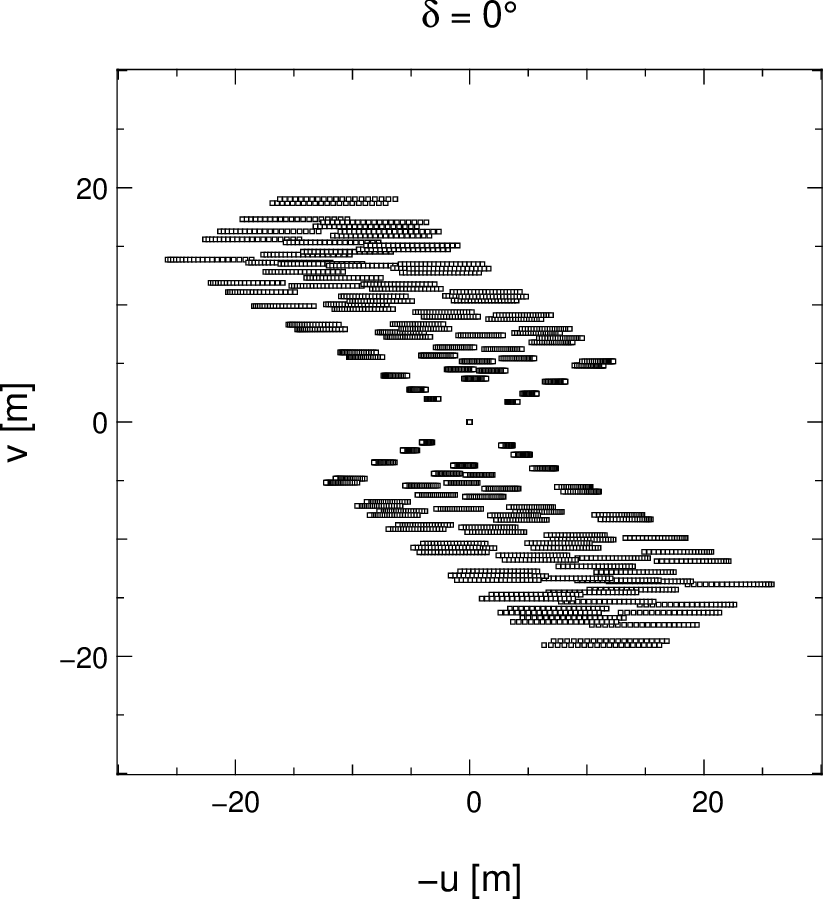}
  \includegraphics[width=5cm]{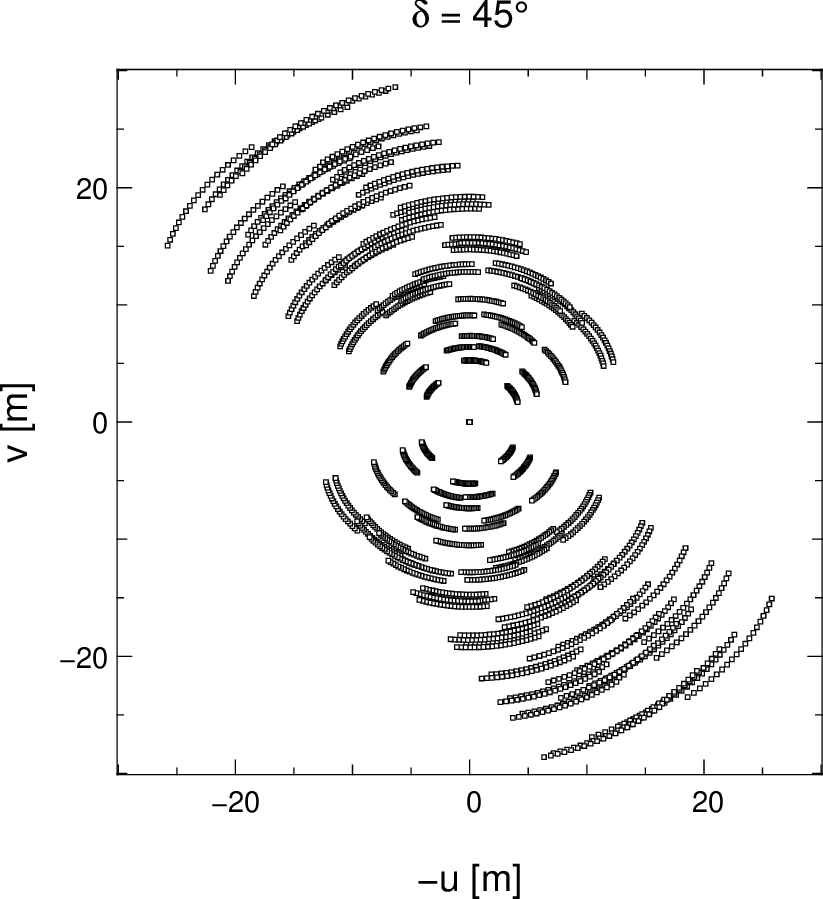}
  \includegraphics[width=5cm]{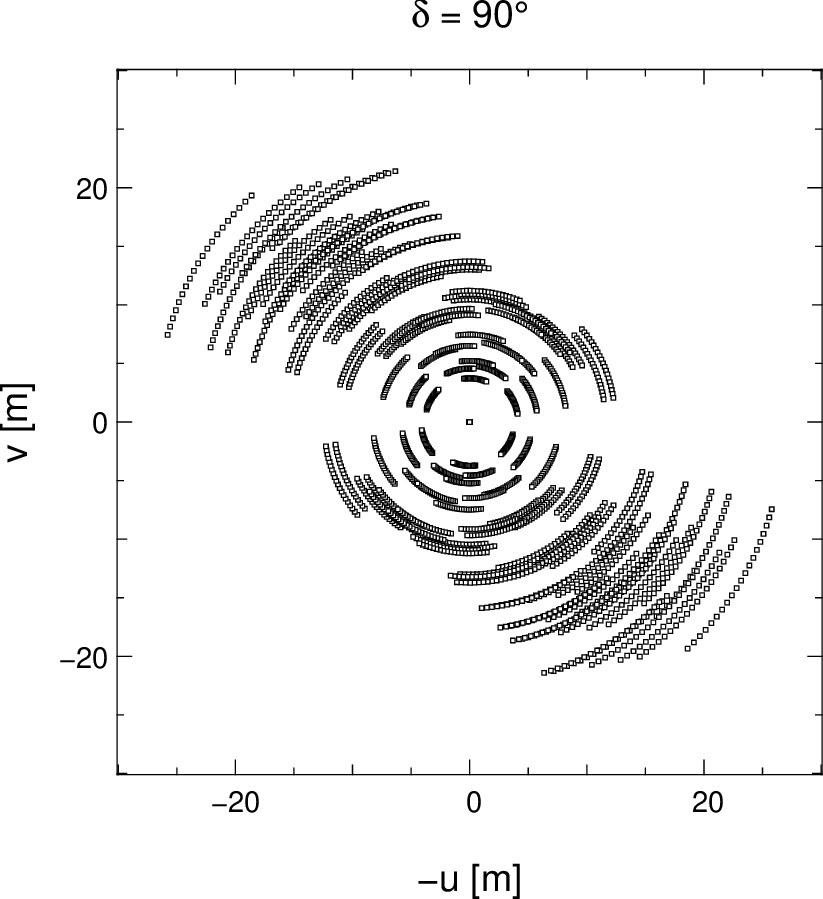}
  \caption[Couverture $u$-$v$ accessible par IOTA]{ La figure du
  dessus représente le sommet de la montagne sur lequel est installé
  IOTA. Les télescopes sont situés sur des tripodes. Il existe 17
  stations utilisables, mais toutes ne sont pas accessibles aux trois
  télescopes. Le télescope B, par exemple, est limité aux 5 stations
  du bras Sud-Est. Le domaine du plan $u$-$v$ accessible est contraint
  par la géométrie de l'interféromètre et par la position de l'étoile
  dans le ciel. Les fréquences spatiales sont représentées par les
  trois figures du dessous, ceci pour des étoiles observées pendant
  deux heures durant leur passage au méridien. La forme ``en sablier''
  entraîne une résolution non uniforme que l'on retrouvera lors de la
  reconstruction d'images. {\it Nota bene} : parce qu'il est de
  tradition de représenter le ciel avec l'Ouest à droite, les
  coordonnées correspondent, dans l'ensemble de cette thèse, aux
  référentiel ($-u$;$v$) ($u$ correspond aux coordonnées Est et $v$
  aux coordonnées Nord).  } \label{fig:IOTA}
\end{figure}

L'Infrared Optical Telescope Array (IOTA) est un interféromètre doté
de trois télescopes de 45 cm\citep{2003SPIE.4838...45T} fonctionnant
dans le proche infra-rouge. Ils sont situés sur le Mont Hopkins en
Arizona. IOTA est géré par un consortium américain rassemblant,
notamment, le {\em Smithsonian Astrophysical Observatory} et l'{\em
University of Massachusetts}. Les trois télescopes peuvent être
déplacés sur des rails, mais leur disposition reste déterminée par
l'emplacement de deux rails à 90 degrés l'un de l'autre. Le rail N-E
est d'une longueur de 35 mètres et dispose de stations tous les
multiples de 5 ou de 7 mètres. Le rail S-O dispose de 5 stations à 5,
7, 10, 14 et 15 mètres. L'ensemble des fréquences spatiales pouvant
être mesurées correspond aux combinaisons possibles projetées sur le
ciel. Cette projection dépend de la déclinaison de l'objet observé. La
longueur de base s'inscrit entre un minimum de 5 mètres et un maximum
de 38 mètres (figure~\ref{fig:IOTA}). Néanmoins, l'ensemble des
fréquences spatiales correspondant à une telle couverture n'est pas
forcément mesurable. De plus, le déplacement des télescopes nécessite
au mieux une demi-journée de travail, ce qui ne permet pas d'obtenir
plus de configurations que de nuits d'observations.

La recombinaison est de type co-axiale et est obtenue via un composant
en optique intégrée fabriqué au Laboratoire d'Astrophysique de
l'Observatoire de Grenoble. La modulation, temporelle, est effectuée
par le déplacement de deux miroirs plans mus par des piezos.  La
longueur d'onde de fonctionnement de l'optique intégrée correspond à
la bande H ($\lambda_0 = 1,65\,\mu$m et $\Delta \lambda =
0,35\,\mu$m). Certains tests ont également été conduits par d'autres
équipes permettant d'obtenir des résultats en bande K ($\lambda_0 =
2,20\,\mu$m). L'ensemble de nos données ont été acquises en bande H.

\subsection{IONIC (Integrated Optics Near-infrared Interferometric
Camera)}

\begin{figure}
  \centering
  \includegraphics[width=6cm]{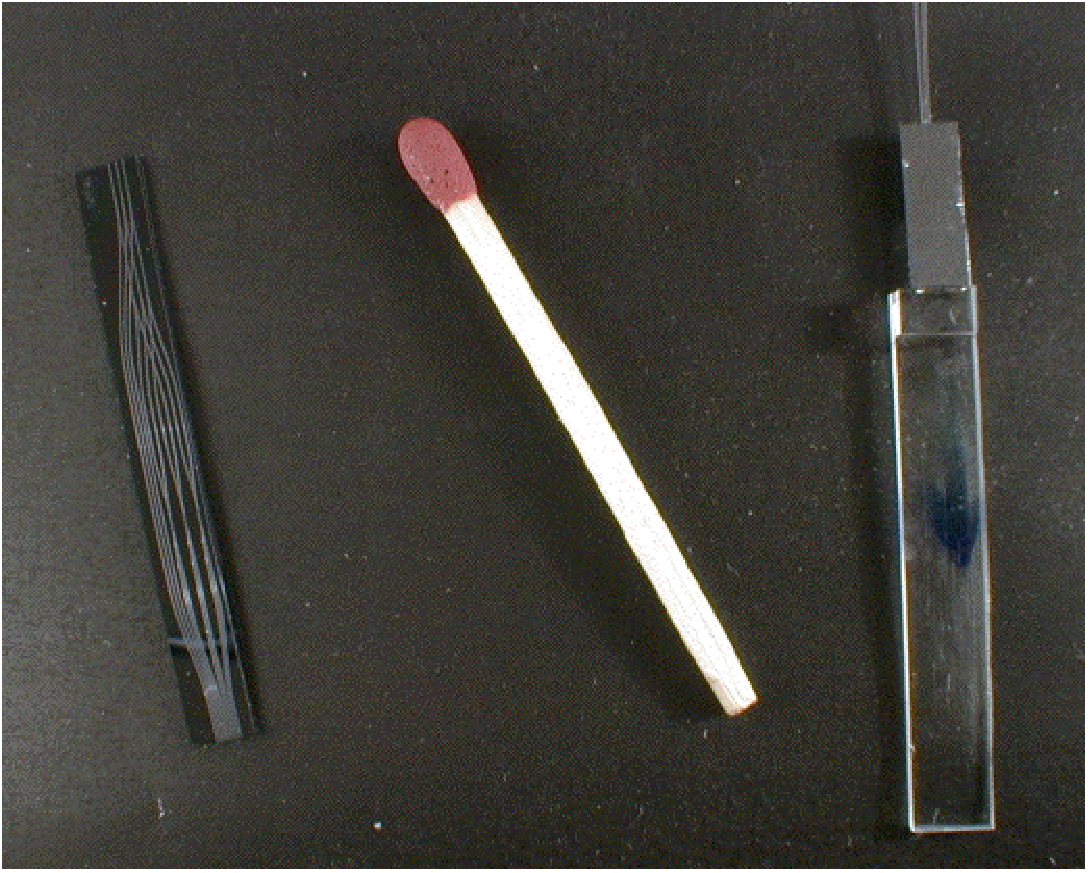} 
\vspace{.6cm}\\
  \includegraphics[width=12cm]{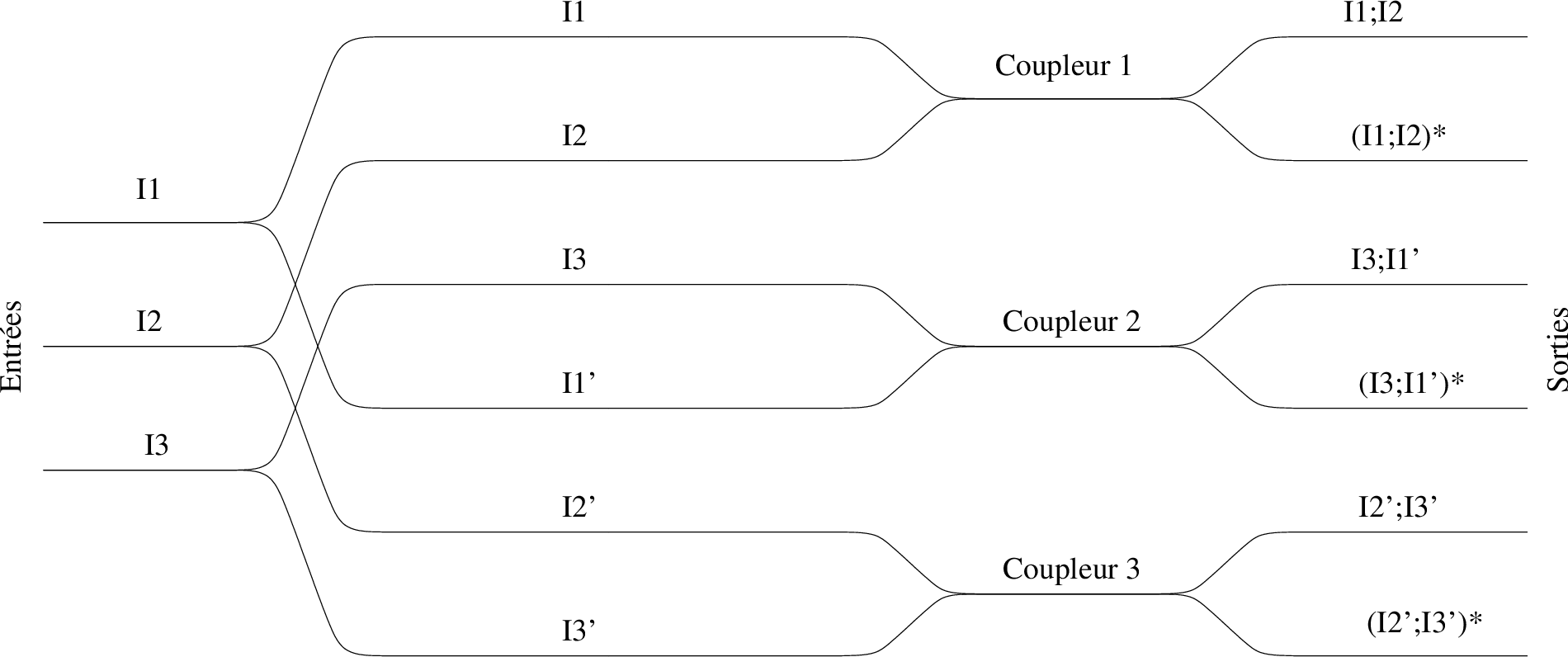} \caption[Composant en
  optique intégré fabriqué par le LAOG]{ La photo est celle d'un
  composant en optique intégré fabriqué par le LAOG ({\it
  http://www-laog.obs.ujf-grenoble.fr/activites/hra/ionic/}).  Elle
  donne une idée de la taille de ce type d'élément de
  recombinaison. La figure du bas représente le schéma optique de
  IONIC utilisé sur IOTA.  } \label{fig:schema_io}
\end{figure}

Afin de mesurer la cohérence spatiale des faisceaux provenant des
télescopes, la lumière est injectée dans des fibres optiques
monomodes. Celles-ci permettent un filtrage du rayonnement, de façon
à exclure les perturbations du front d'onde.  Les trois fibres sont
ensuite alignées sur un ``V-groove'' pour injecter la lumière dans un
recombinateur plan en optique intégrée \citep[IONIC
;][]{2003SPIE.4838.1099B}. Le circuit optique sépare ensuite le flux
provenant de chaque télescope pour le recombiner par paires via
trois coupleurs intégrés (figure~\ref{fig:schema_io}). L'utilisation
de ce type de coupleur permet d'obtenir deux sorties interférométriques
par coupleur, chacune déphasée de $\pi$ par rapport à l'autre. Les
variations de l'intensité de couplage dans les fibres peuvent ainsi
être prises en compte par l'intermédiaire de la matrice de transfert
de l'optique intégrée. Celle-ci peut être obtenue, par exemple, en
injectant de la lumière séquentiellement dans les différentes voies.

\begin{figure}
  \centering
  \resizebox{\hsize}{!}{\includegraphics{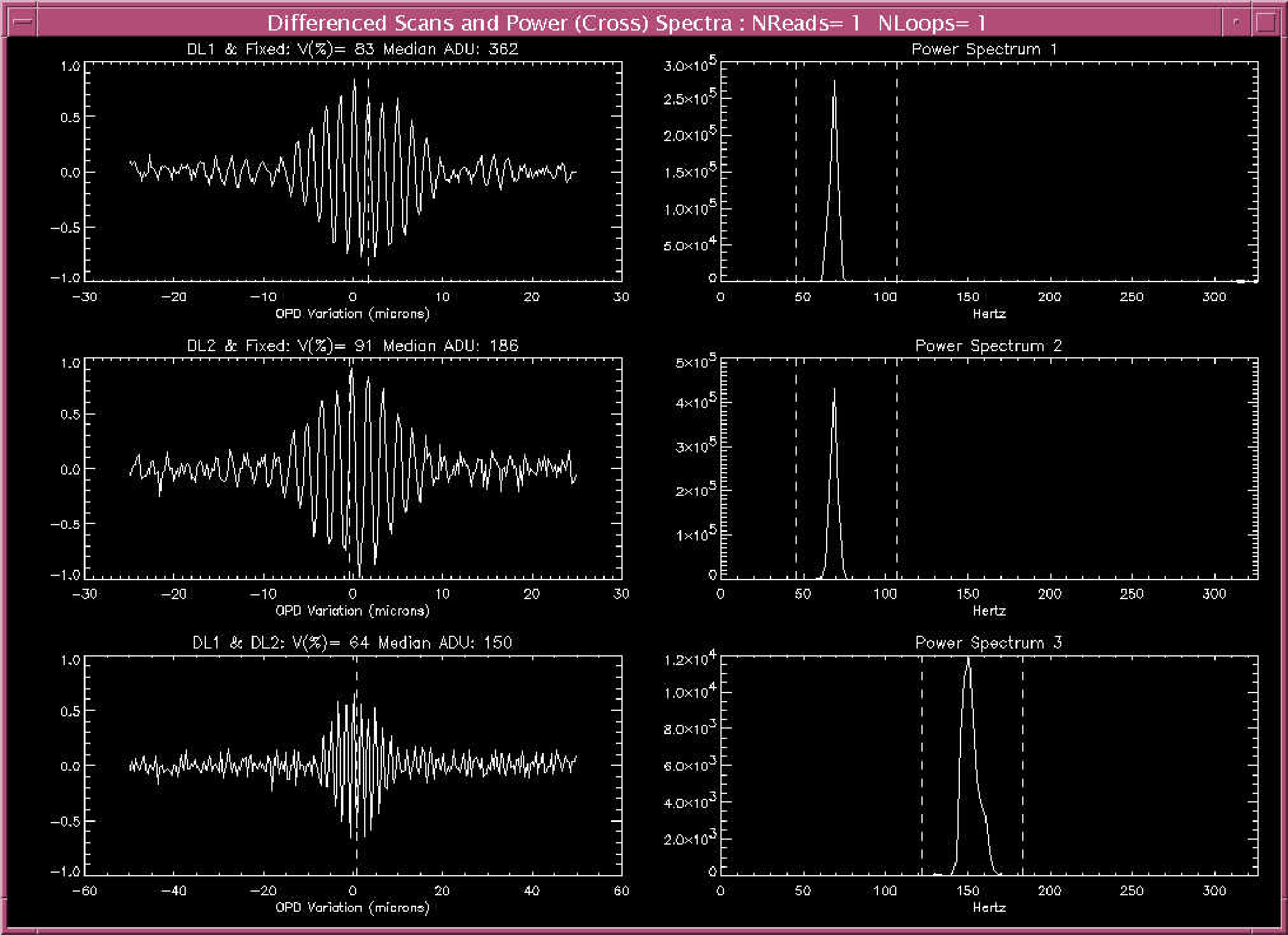}}
  \caption[Fenêtre de contrôle du suiveur de franges]{ Fenêtre de
  contrôle du suiveur de franges. On peut voir les franges
  correspondant aux trois bases ainsi que leur spectre de
  puissance. On peut constater qu'une base est modulée à une fréquence
  double par rapport aux deux autres. Ceci est dû à l'effet cumulé de
  la modulation des deux piezos. } \label{fig:Controle}
\end{figure}

Une correction rapide des variations photométriques consiste à
soustraire les deux voies déphasées issues de chaque coupleur. Ceci
peut être simplement explicité dans l'hypothèse ou l'on néglige
l'influence des facteurs de transmission des coupleurs.  Ainsi, pour
un rayonnement parfaitement cohérent, les deux sorties
interférométriques fournissent :
\begin{eqnarray}
S_{12}(t)&=&I_1(t)+I_2(t)+ \sqrt{|I_1(t)| |I_2(t)|}  \cos(\omega t) \\
S_{12}^\star(t)&=&I_1(t)+I_2(t)+ \sqrt{|I_1(t)| |I_2(t)|}  \cos(\omega t +\pi) \,.
\end{eqnarray}
Si l'on applique la soustraction, on obtient :
\begin{eqnarray}
S_{12}(t)-S_{12}^\star(t)=2\sqrt{|I_1(t)| |I_2(t)|}  \cos(\omega t) \,.
\end{eqnarray}
Soit des oscillations à la moyenne indépendante des fluctuations
photométriques. Il s'agit cependant du cas particulier où la
transmission de chacun des coupleurs est de 50\% pour chacune des
voies. Un traitement plus fin, post observation, sera nécessaire pour
tenir compte de la matrice de transfert et des variations de
l'amplitude de modulation. Cette simple soustraction permet, tout de
même, une première correction qui est utilisée par le
suiveur de frange. La figure~\ref{fig:Controle} est une copie de
l'écran de celui-ci. On peut voir les 3 figures
d'interférences corrigées des variations photométriques.

\subsection{Le traitement de données}
\label{sec:john_s}

\begin{figure}
  \centering \resizebox{\hsize}{!}{
  \includegraphics{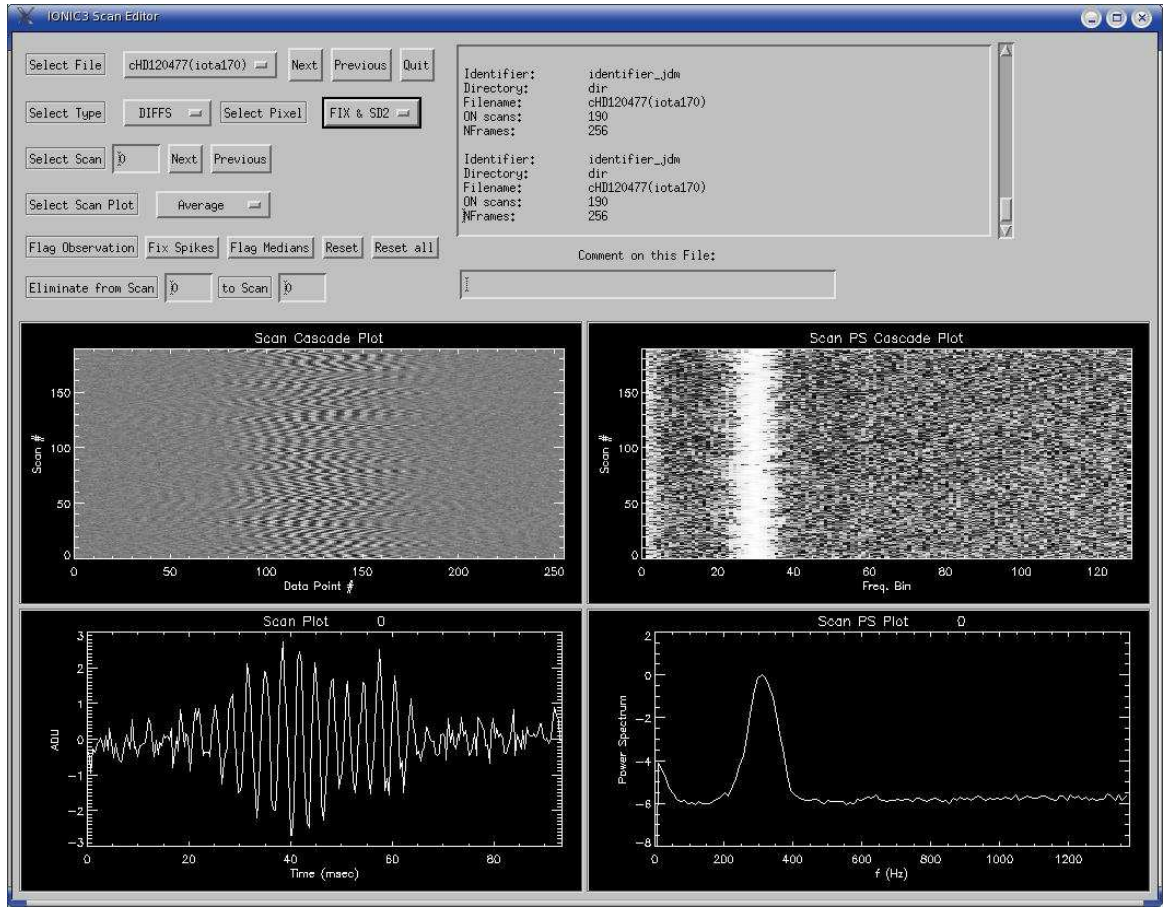}} \caption[Apparence
  graphique du logiciel ``matrix2.script'']{ Apparence
  graphique du logiciel ``matrix2.script'' conçut par J. Monnier en
  langage IDL. Cette interface permet de sélectionner les jeux de
  franges utiles, et d'éliminer ceux qui semblent corrompus. En bas à
  droite figure la densité spectrale de puissance (en unités
  logarithmiques) d'où sont extraites les visibilités non étalonnées.}
  \label{fig:johhn_algo}
\end{figure}

Une séquence d'acquisitions typique consiste en l'enregistrement d'une
ou deux séries de 200 paquets de franges obtenus en 4 minutes
environ. Chaque séquence est ensuite suivie de l'enregistrement du
fond et du flux obtenu avec chaque télescope, indépendamment des
autres. Ceci permet l'étalonnage par l'obtention de la matrice de
transfert du recombinateur. Le temps total -- y compris le pointage --
d'une séquence d'acquisitions est ainsi d'environ 10 à 20 minutes.
Les observations de la source sont entrelacées avec une séquence
d'acquisitions identique obtenue sur des étalons (aussi appelé
calibrateurs). Ceux-ci sont choisis pour être spatialement proches de
l'objet d'intérêt scientifique ($\lessapprox 10$ degrés), de
brillances similaires, de types spectraux proches, et surtout ayant des
visibilités connues avec une grande précision (source ponctuelle ou de
diamètre précisément connu). La réunion de ces conditions permet une
mesure précise de la fonction de transfert de l'interféromètre.

La réduction des données interférométriques a été réalisée sous IDL par un
logiciel conçu par John Monnier de l'{\em University of Michigan}
\citep[voir par exemple ;][]{2004ApJ...602L..57M}. Ce logiciel se
compose principalement de deux parties. La première permet de calculer
les densités spectrales de puissance, et la seconde de calibrer les
visibilités au carré. La figure~\ref{fig:johhn_algo} est une copie de
l'écran correspondant à la sélection des franges. Sur cette figure est
présenté un ensemble de 200 acquisitions que l'on peut ainsi
individuellement sélectionner ou éliminer en fonction de leur
qualité. Ceci permet ainsi d'éviter de réduire les données qui peuvent être
corrompues par un seeing trop important ou un problème
instrumental, comme l'oscillation de la ligne à retard. A partir du pic
frange que l'on peut observer dans la densité spectrale de puissance,
le logiciel fournit une visibilité au carré $V^2$, qui sera ensuite
normalisée par la visibilité des étalons.

\subsection{Les clôtures de phase}
\label{sec:ima_clo}

\begin{figure}[h]
  \centering \includegraphics[width=11cm]{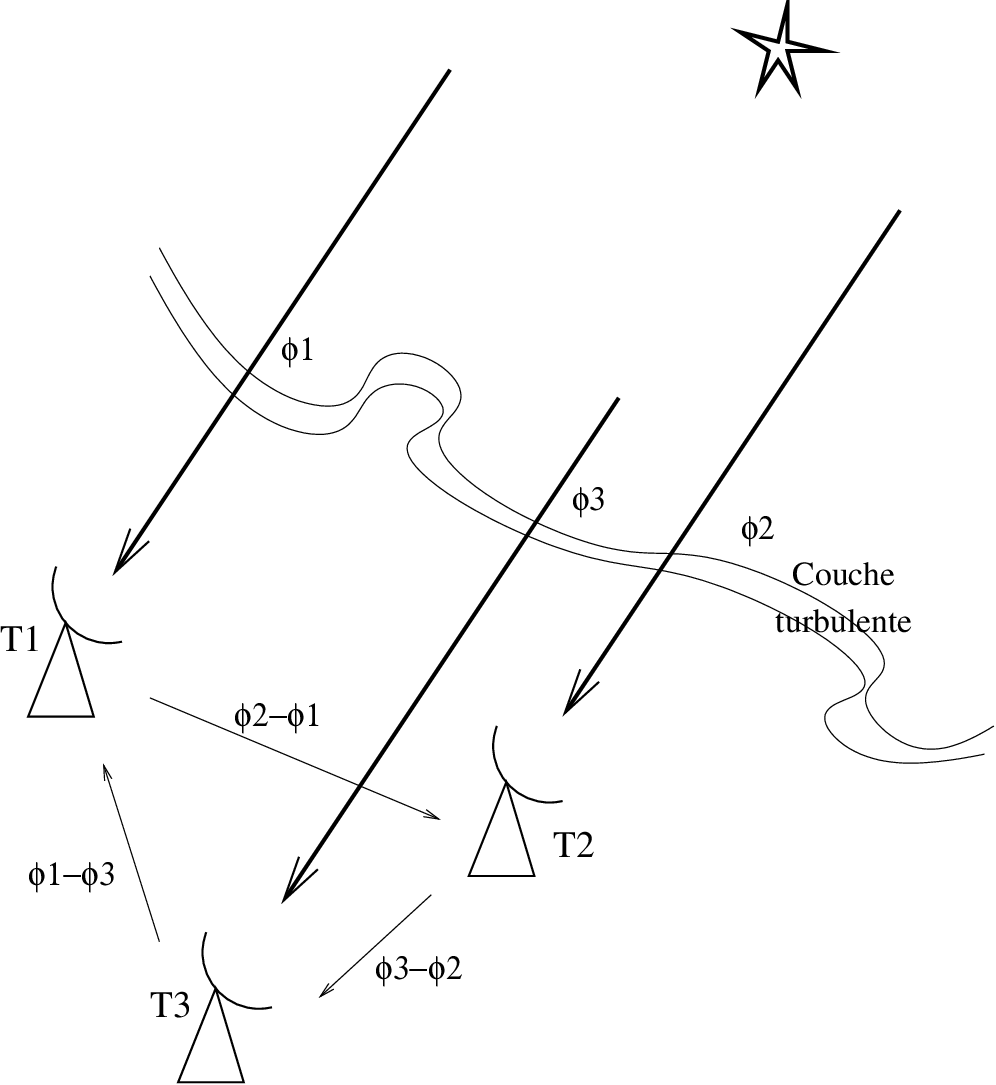}
  \caption[Illustration de l'utilité de la mesure de clôture de
  phase]{ Schéma d'un interféromètre à trois télescopes illustrant
  l'utilité de la mesure de clôture de phase. Chaque télescope reçoit
  le rayonnement astrophysique affecté par un piston atmosphérique
  ($\phi_1$, $\phi_2$ et $\phi_3$). Celui-ci est introduit par les
  variations d'indices des couches turbulentes de l'atmosphère. La
  clôture de phase permet de s'affranchir de ces termes en sommant les
  phases obtenues sur les bases formant le triangle T1-T2-T3.}
  \label{fig:schema_clo}
\end{figure}

Parallèlement à l'extraction des visibilités au carré, le logiciel de
réduction mesure également une information sur la phase des
visibilités. Il s'agit des clôtures de phase (CP).

La méthode de clôture de phase a été développée initialement par les
radiointerférométristes \citep{1958MNRAS.118..276J} pour s'affranchir
d'un terme de déphasage sur des données obtenues par interférométrie
hétérodyne. L'idée peut être comprise par le biais de la
figure~\ref{fig:schema_clo}. Dans ce schéma, le rayonnement issu de
la source astrophysique est déphasé d'un terme $\phi_i$ lorsqu'il
arrive sur chacun des télescopes. Ce terme peut être dû à un piston
atmosphérique ou à un problème de calibration astrométrique. La phase
mesurée entre les télescopes 1-2, 1-3 et 2-3 est la somme de la phase
de l'objet astrophysique (respectivement $\Phi_{12}$, $\Phi_{12}$ et
$\Phi_{12}$) et de la différence des pistons
atmosphériques. Nous mesurons de cette manière une phase
$\Phi^\ddagger_{ij}$ telle que:
\begin{equation}
\Phi^\ddagger_{ij}=\Phi_{ij}+\phi_j-\phi_i \,.
\label{eq:cloture_s}
\end{equation}
La clôture de phase est la somme des trois phases obtenues par ces
trois télescopes :
\begin{equation}
\Phi^{CP}_{ijk}=\Phi^\ddagger_{ij}+\Phi^\ddagger_{jk}+\Phi^\ddagger_{ki}\,.
\end{equation}
Ainsi, en appliquant l'équation~\ref{eq:cloture_s}, on peut montrer que
$\Phi^{CP}_{ijk}$ est indépendant des termes de piston atmosphérique,
et qu'il vaut :
\begin{equation}
\Phi^{CP}_{ijk}=\Phi_{ij}+\Phi_{jk}+\Phi_{ki}\,.
\end{equation}

En pratique, les clôtures de phase sont obtenues par l'intermédiaire
du bispectre. Nous aurons l'occasion de revoir ce point
en section~\ref{sec:clot}. Le concept mathématique, et notamment
l'influence du bruit de photon, est étudié plus en détails dans la
thèse de \citet{Thiebaut1994:PhD}.

Les clôtures de phase permettent ainsi d'obtenir une information
partielle sur la phase de l'objet astrophysique. L'avantage de cet
estimateur est sa grande robustesse aux perturbations
astrophysiques. Nous verrons que nos données de clôture de phase
sont souvent obtenues avec des précisions de quelques degrés. Ces
clôtures de phase ont deux propriétés fondamentales importantes
\begin{itemize}
\item Elles sont invariantes par rapport à la position de l'objet dans
le ciel. Ainsi, il n'est pas possible d'obtenir des valeurs
astrométriques absolues mais uniquement relatives. Nous verrons par la
suite que les reconstructions ne sont pas nécessairement centrées par
rapport à la position {\it zéro}. Ceci résulte tout simplement de
l'absence de contrainte sur la position. En pratique, nous utiliserons
un terme régulateur visant à centrer l'image
(section~\ref{sec:algo_rec_image})
\item Elles sont égales à zéro dans le cas d'un objet centro-symétrique. Plus
exactement, elles sont soit nulles, soit égales à $\pi$ lorsque la
visibilité est négative. Ceci permet de révéler clairement le passage
d'un lobe à un autre lorsque le signe de la visibilité change. Cela permet
aussi de mettre en relief la présence d'asymétries sur l'objet observé
lorsque les clôtures de phase sont différentes de $0$ ou de $\pi$. Nous
verrons que ces deux propiétés permettent une première interprétation
rapide des données interférométriques.
\end{itemize}

\clearpage
\section{Les observations}

\begin{table}[h]
\caption{Missions d'observations}
\label{tb:missions}
\centering
\begin{tabular}{llr}
\hline
\hline
Période & Objet & Observateurs \\
\hline
19 -- 28 Octobre 2004 &  -- {\it Mauvais temps} -- & GP, SL, SM et JW\\
 21 -- 31 Mai 2005 & $\chi$ Cyg, CH Cyg & SL et SM\\
 4 -- 16 Octobre 2005 & $\chi$ Cyg, Mira, Bételgeuse & GP, SL et XH\\
 28 Mars -- 6 Avril 2006 & $\chi$ Cyg, R Leo & SL, XH et PS \\
 10 -- 15 Mai 2006 & $\chi$ Cyg, R Leo, Arcturus, $\mu$ Cep & SL, AP
 et PS \\
\hline
\end{tabular}
\begin{list}{}
{\vspace{.2cm}}
\item[-] Technicien responsable de l'interféromètre,
 Marc Lacasse était généralement présent lors des missions d'observations. De
 plus, m'ont accompagné lors de ces missions : GP; Guy Perrin, SM; Serge
 Meimon, JW; Julien Woilez, XH; Xavier Haubois, PS; Peter Schuller et
 AP; Anne Poncelet.
\end{list}
\end{table}

\subsection{L'objectif}

L'objectif premier de ces missions d'observations est de parvenir à
imager par interférométrie la surface stellaire d'étoiles évoluées.
L'obtention de l'information de phase -- via les clôtures de phase --
est nécessaire pour la reconstruction d'une image complexe. Cette
information est devenue accessible sur un interféromètre fibré lors de
l'ouverture de l'observatoire IOTA à la communauté
astrophysique. Pourtant, la plupart des publications actuelles portent
uniquement sur une analyse paramétrique des données. Concernant IOTA,
nous pouvons, par exemple, citer les résultats de
\citet{2006ApJ...647..444M} sur les étoiles Herbig Ae/Be, où les
auteurs ont ajusté des modèles de disques d'accrétion sur les clôtures
de phase. Nous pouvons aussi prendre pour exemple les travaux récents
de \citet{2006astro.ph..7156R} qui ont mesuré des clôtures de phase
sur 56 étoiles de la branche asymptotique des géantes. Dans le cadre de
leur analyse paramétrique, seules quelques mesures par
étoiles ont été obtenues.

Notre approche s'avère plus difficile car l'imagerie nécessite une
connaissance exhaustive du plan des fréquences spatiales, ce qui rend
nos travaux novateurs. Alors que \citet{2006astro.ph..7156R} se sont
plutôt focalisés sur l'observation d'une grande quantité d'objets, nous
avons choisi de limiter le nombre d'étoiles observées pour étendre
autant que possible la couverture du plan $u$-$v$. Cette approche a
cependant l'inconvénient de nécessiter des changements fréquents de la
configuration de l'interféromètre, avec tous les problèmes techniques
que le déplacement des télescopes induit.

\subsection{Les missions d'observations}

\begin{table}
\caption{Liste des étoiles évoluées étudiées}
\label{tb:star}
\centering
\begin{tabular}{lrrccc}
\hline
\hline
 Objet & \multicolumn{1}{c}{$\alpha$} & \multicolumn{1}{c}{$\delta$} & Type spectral & Type & Période(s)  \\
\hline
Arcturus & 14 15 39.67 & +19 10 56.7 & K1.5III & Géante &\\
CH Cyg & 19 24 33.07 & +50 14 29.1 & M7IIIv & Symbiotique & 100-155$^{\mathrm{a}}$\\
$\chi$ Cyg & 19 50 33.92 & +32 54 50.6 & MS & Mira & 408 \\
R Leo & 09 47 33.49 & +11 25 43.6 & M8IIIe & Mira & 312 \\
Mira & 02 19 20.79 & -02 58 39.5 & M2-M7 IIIe & Mira & 332 \\
Bételgeuse & 05 55 10.31 & +07 24 25.4 & M2Iab & Supergéante& 388, 2050$^{\mathrm{b}}$\\
$\mu$ Cep & 21 43 30.46 & +58 46 48.2 & M2Ia & Supergéante & 860, 4400$^{\mathrm{b}}$\\
\hline
\end{tabular}
\begin{list}{}{}
\item[$^{\mathrm{a}}$] \citet{1992JAVSO..21...23K}
\item[$^{\mathrm{b}}$] \citet{2006MNRAS.Kiss}
\end{list}
\end{table}

C'est pourquoi nous avons choisi d'effectuer des missions
d'observations plutôt longues, avec des durées variables entre 10 et
20 jours. Il y a eu 5 missions d'observations au Mont Hopkins qui se
sont déroulées d'Octobre 2004 à Mai 2006
(tableau~\ref{tb:missions}). \`A noter la première et la dernière
mission : la première s'est traduite par quinze jours de mauvais temps
et la dernière par l'utilisation d'un tout nouveau mode de
fonctionnement qui a permis de disperser les franges et d'obtenir une
information spectro-interférométrique. Le tableau~\ref{tb:star}
répertorie les objets observés au cours de ces différentes
missions. Toutes sont des étoiles évoluées de type Mira, Géante ou
Supergéante rouge.

\subsection{Les données}

\begin{table}
\caption{Configurations des télescopes}
\label{tb:Config}
\centering
\begin{tabular}{lccccc}
\hline
\hline
Date & Configuration$^{\mathrm{a}}$ & \multicolumn{4}{c}{Objets observés} \\
\hline
24 Mai 2005&	A35-B15-C0&	&CH Cyg	\\
25 Mai 2005&	A15-B15-C0&	$\chi$ Cyg	\\
27 Mai 2005&	A15-B15-C5&	$\chi$ Cyg	\\
31 Mai 2005&	A25-B15-C10&	$\chi$ Cyg&CH Cyg	\\
1 Juin 2005&	A35-B15-C10&	$\chi$ Cyg	\\
\hline
5 Octobre 2005&	A5-B5-C0&	$\chi$ Cyg&	\\
6 Octobre 2005&	''&	$\chi$ Cyg&	Mira	\\
7 Octobre 2005&	''&	$\chi$ Cyg&	Mira&	Bételgeuse	\\
8 Octobre 2005&	A5-B15-C0&	$\chi$ Cyg&	Mira&	Bételgeuse	\\
9 Octobre 2005&	''&	$\chi$ Cyg&	\\
10 Octobre 2005&	A15-B15-C0&	$\chi$ Cyg&&	Bételgeuse	\\
11 Octobre 2005&	''&	$\chi$ Cyg&	Mira&	Bételgeuse	\\
12 Octobre 2005&	A25-B15-C0&	$\chi$ Cyg&	Mira&	Bételgeuse	\\
13 Octobre 2005&	''&	$\chi$ Cyg&	Mira	\\
15 Octobre 2005&	A30-B15-C15&&	Mira	\\
16 Octobre 2005&	''&&&	Bételgeuse	\\
\hline
29 Mars 2006&	A15-B15-C0&	$\chi$ Cyg&	R Leo 	\\
31 Mars 2006&   "&	$\chi$ Cyg	\\
2 Avril 2006&	A5-B5-C0&	$\chi$ Cyg&	R Leo	\\
3 Avril 2006&	A5-B10-C0&&	R Leo	\\
4 Avril 2006&	A25-B10-C0&&	R Leo	\\
7 Avril 2006&	A30-B15-C0&	$\chi$ Cyg&	R Leo	\\
\hline
11 Mai 2006&	A15-B5-C10&	$\chi$ Cyg&	R Leo& Arcturus&	$\mu$ Cep	\\
12 Mai 2006&	A15-B5-C0&	$\chi$ Cyg&	R Leo& Arcturus&	$\mu$ Cep	\\
13 Mai 2006&	A15-B15-C0&	$\chi$ Cyg&	R Leo& Arcturus	\\
14 Mai 2006&	A30-B15-C0&	$\chi$ Cyg&	R Leo& Arcturus&	$\mu$ Cep	\\
15 Mai 2006&	A35-B15-C21&	$\chi$ Cyg&	R Leo& &$\mu$ Cep	\\
16 Mai 2006&	A35-B15-C25&	$\chi$ Cyg&	R Leo& Arcturus&	$\mu$ Cep	\\
\hline
\end{tabular}
\begin{list}{}{}
\item[$^{\mathrm{a}}$] La notation utilisé pour désigner les configurations 
utilise la localisation des télescopes A, B et C, sur
respectivement les rails NE, SE et NE.
\end{list}
\end{table}

\begin{table}[h]
\caption{Liste des étalons utilisés}
\label{tb:calib_t}
\centering
\begin{tabular}{lcrr}
\hline
\hline
\'Etalon & Type Spectral  & Diamètre & Objet \\
\hline
\hline
HD\,176670 &    K2.5\,III   & 2,33 $\pm$  0,026 & $\chi$ Cyg\\
HD\,177808   &    M0\,III      & 2,32  $\pm$  0,030 & $\chi$ Cyg\\
HD\,180450 &    M0\,III     & 2,77  $\pm$ 0,032 & $\chi$ Cyg\\
HD\,186619 &    M0\,IIIab   & 2,19  $\pm$ 0,025 & $\chi$ Cyg\\
HD\,188149 &    K4\,III     & 1,49  $\pm$ 0,020 & $\chi$ Cyg\\
HD\,197989 &    K0\,III     & 4,44  $\pm$ 0,048 & $\chi$ Cyg\\
HD\,8512 &  K0\,IIIb  &    2,69 $\pm$   0,030 & Mira \\
HD\,16212 & M0\,III  &   3,02 $\pm$   0,032 & Mira\\
HD\,36167  & K5\,III  &     3,56 $\pm$  0,057 & Bételgeuse \\
HD\,48433  & K0.5\,III  &     2,07 $\pm$  0,027 & Bételgeuse \\
HD\,82381 & K2.5\,IIIb  & 2,09  $\pm$ 0,026 & R Leo \\
HD\,87837 & K3.5\,IIIb  &  3,22 $\pm$  0,049 & R Leo \\
HD\,120477  & K5.5\,III  &   4,46 $\pm$  0,050 & Arcturus\\
HD\,125560  & K3\,III  &   1,91 $\pm$  0,021 & Arcturus\\
HD\,129972   & G8.5III  &  1,54  $\pm$ 0,020 & Arcturus\\
HD\,176670 &    K2.5\,III   & 2,33 $\pm$  0,026 & $\chi$ Cyg\\
HD\,177808   &    M0 III      & 2,32  $\pm$  0,030 & $\chi$ Cyg\\
HD\,180450 &    M0\,III     & 2,77  $\pm$ 0,032 & $\chi$ Cyg\\
HD\,186619 &    M0\,IIIab   & 2,19  $\pm$ 0,025 & $\chi$ Cyg\\
HD\,188149 &    K4\,III     & 1,49  $\pm$ 0,020 & $\chi$ Cyg\\
HD\,197989 &    K0\,III     & 4,44  $\pm$ 0,048 & $\chi$ Cyg\\
HD\,198149  &  K0\,IV &  2,68 $\pm$  0,029 & $\mu$ Cep \\
\hline
\end{tabular}
\begin{list}{}{}
\item Les étalons sont issus des
catalogues de \citet{2002A&A...393..183B} et de \citet{2006A&A...447..783M}
\end{list}
\end{table}

\begin{figure}
\centering
\resizebox{\hsize}{!}{
\includegraphics{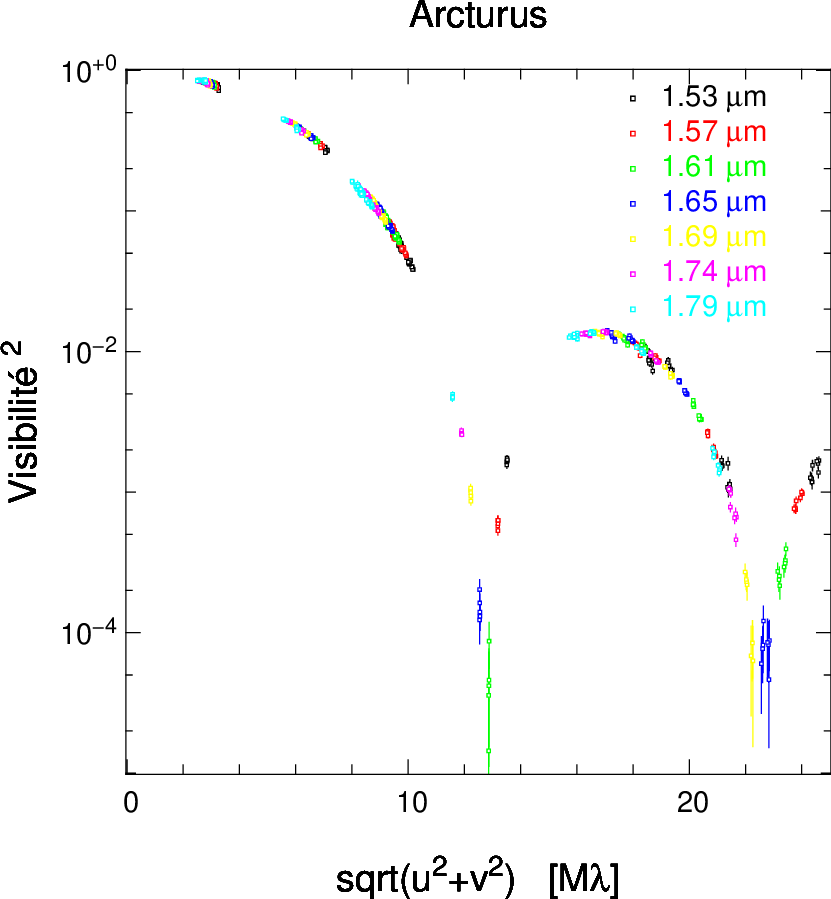}
\includegraphics{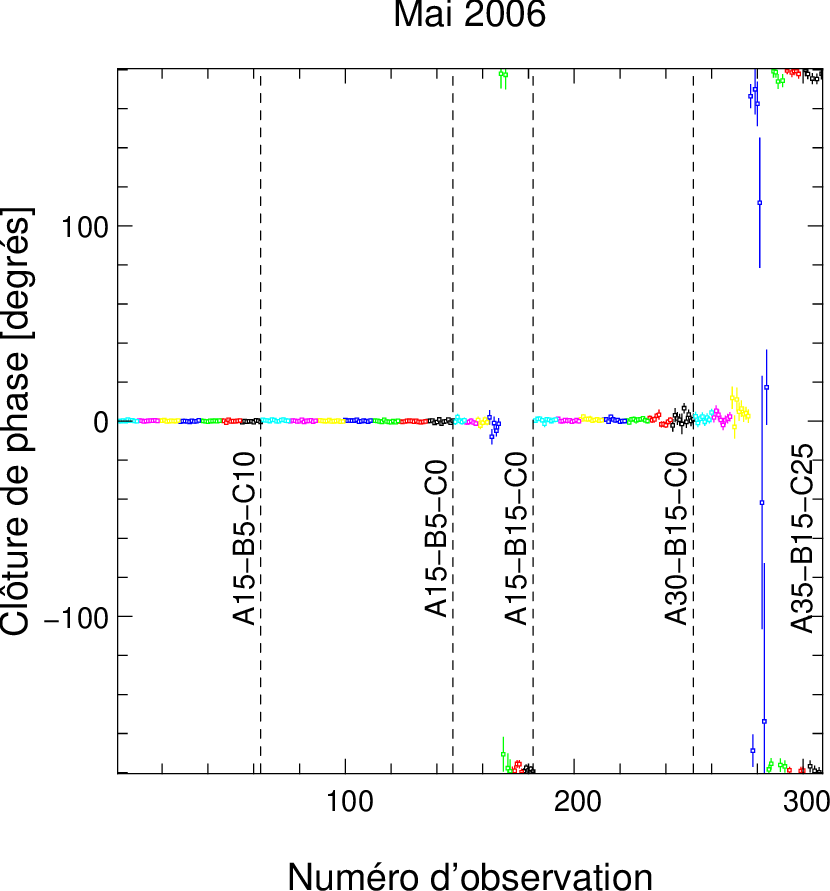}
}
{\vspace{.3cm}}
\resizebox{\hsize}{!}{
\includegraphics{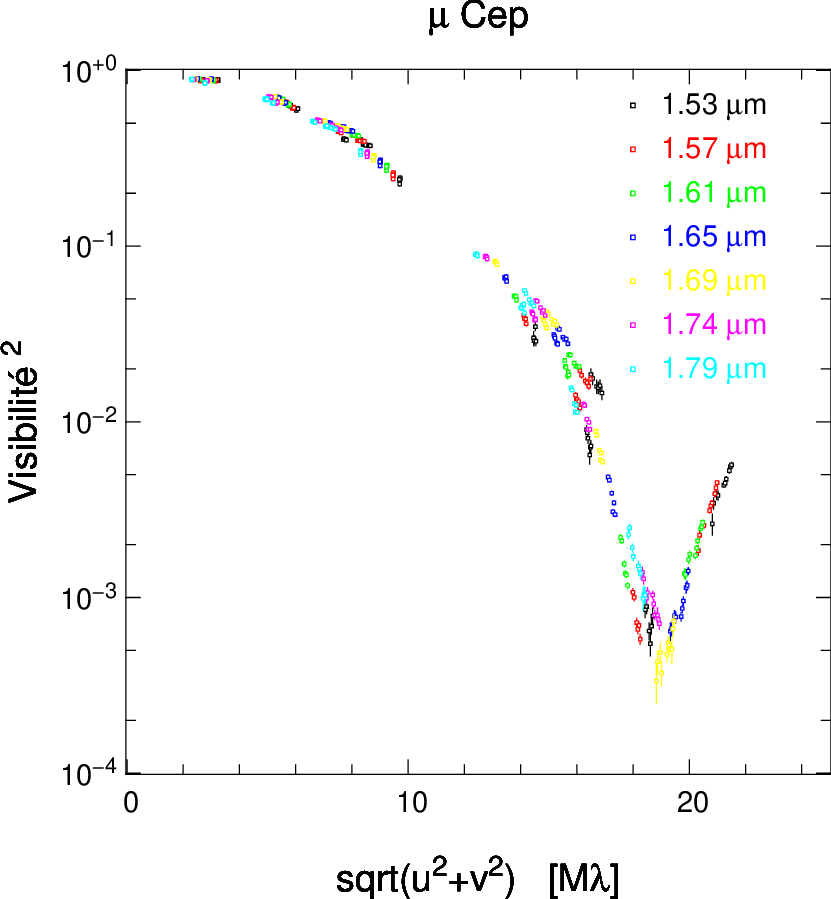}
\includegraphics{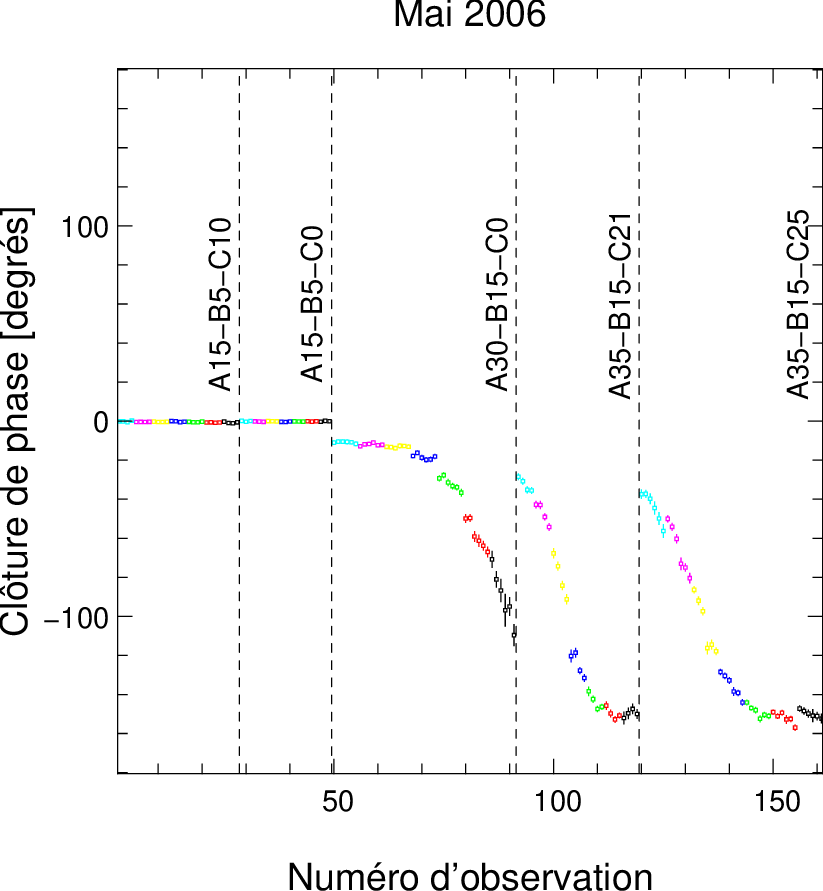}
} \caption[Données d'Arcturus et de $\mu$ Cep obtenues au cours de la
   mission de mai 2006]{ V$^2$ et CP obtenues au cours de la dernière mission
   d'observation. Les données d'Arcturus (V$^2$ et CP) sont parmis les
   plus belles données acquises. Ces mesures démontrent ainsi
   clairement la qualité de mesure que peut fournir un interféromètre
   fibré. Les clôtures de phase passent nettement de 0 à 180 degrés
   lorsque le zéro est franchi, et des visibilités très faibles
   peuvent être mesurées grâce à la technique de
   ``boot-strapping''. Les données sur $\mu$ Cep sont tout aussi
   intéressante. L'asymétrie peut être détectée à la fois sur les
   visibilités et les clôtures de phase. Cette asymétrie se voit
   nettement sur les clôtures de phase, qui passent progressivement de
   5 à 165 degrés. Ce passage étant dû à la base A35-B15, nous l'avons
   retrouvé au cours de trois nuits différentes, ce qui prouve la
   stabilité de nos mesures.} \label{fig:Donnees_6}
\end{figure}

Les données ont été réduites par le logiciel présenté en
section~\ref{sec:john_s}. Elles sont acquises lorsque deux voies, au
moins, peuvent détecter les franges. Cette condition permet d'en
déduire la présence de franges dans la troisième fenêtre
d'intégration, même si elles ne sont pas instantanément
détectables. De cette manière (cette technique est appelé
``boot-strapping'' en anglais) nous avons pu mesurer des CP et V$^2$
même pour des amplitudes de franges noyées dans le bruit de photon et
de detecteur. La figure~\ref{fig:Donnees_6}, par exemple, présente des
mesures de V$^2$ inférieures à $10^{-4}$. La liste des configurations
utilisées est présentée tableau~\ref{tb:Config}.

La liste des étoiles servant à l'étalonnage est présentée
dans le tableau~\ref{tb:calib_t}. Celles-ci ont été choisies à partir des
catalogues de \citet{2002A&A...393..183B} et de
\citet{2006A&A...447..783M} sur des critères de proximité spatiale et
de magnitude. Sur l'ensemble des étalons, nous avons constaté des
erreurs sur les visibilités au carré de l'ordre de 2\%. Ces erreurs
peuvent être dues à des incertitudes dans le calcul de leur diamètres
ou à des variations de la fonction de transfert instrumentale sur des
périodes de moins de 30 minutes. Pour tenir compte de ces
incertitudes, nous avons ajouté quadratiquement une erreur sur les
visibilités normalisées de 2\%.

Les clôtures de phase se sont révélées extrêmement fiables avec des
variations entre les différents étalons au cours de la nuit de
l'ordre du degré. Ceci est dû à la qualité et à la miniaturisation du
composant d'optique intégrée, ainsi très faiblement affecté par
d'éventuelles variations de température (le laboratoire d'optique
n'étant pas une salle isolée, nous y avons constaté des changements de
température de plusieurs degrés). Une autre source d'erreurs sur les
clôtures réside dans la différence entre le type spectral des
étalons (G8 à M0) et les étoiles évoluées que nous avons
étudiées. En pratique, des tests ont été effectués et ont révélé des
variations très faibles ($\approx 0,5$ degré) entre des étoiles chaudes (B8)
et froides (M3) \citep{Rag_CP_Cloture}. 

\begin{figure}
\centering
\resizebox{\hsize}{!}{
\includegraphics{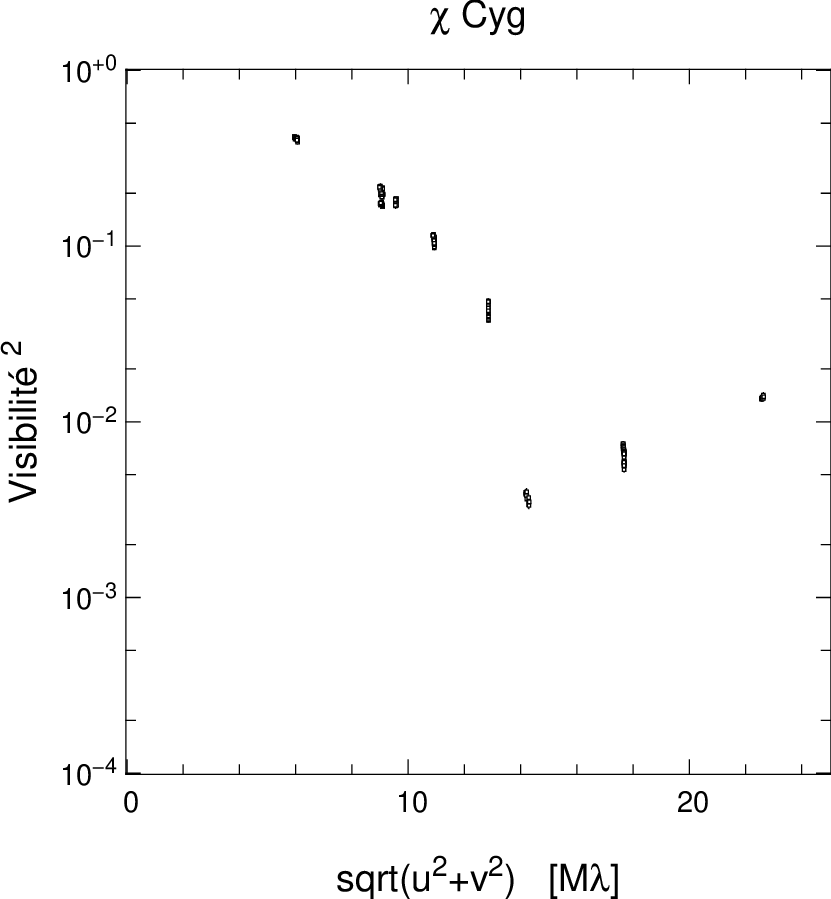}
\includegraphics{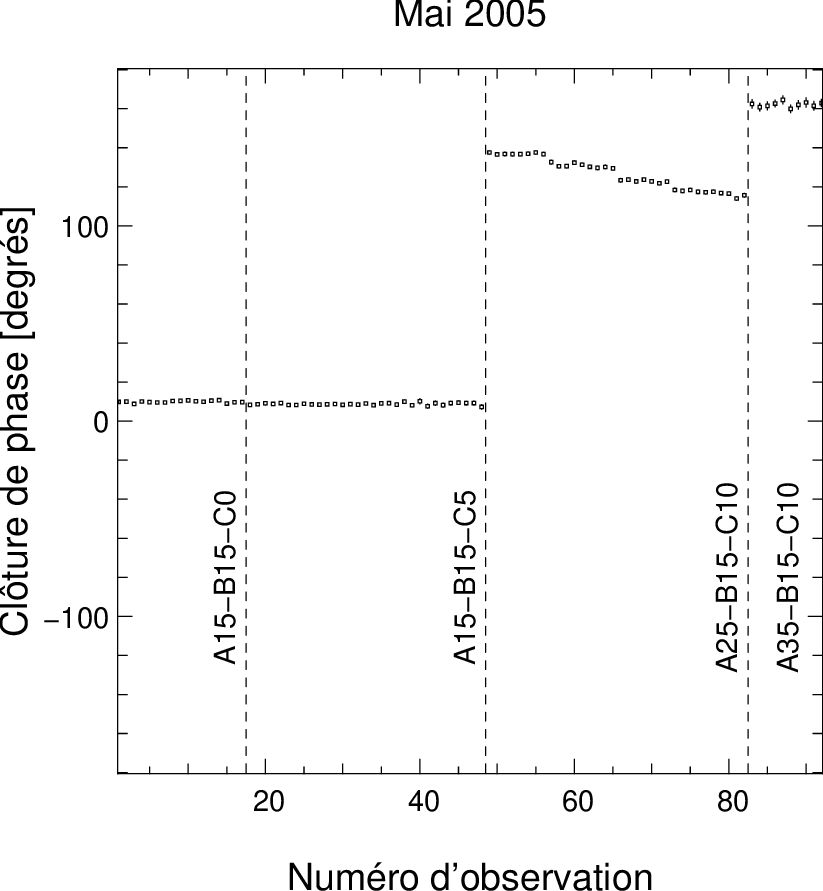}
}
{\vspace{.3cm}}
\resizebox{\hsize}{!}{
\includegraphics{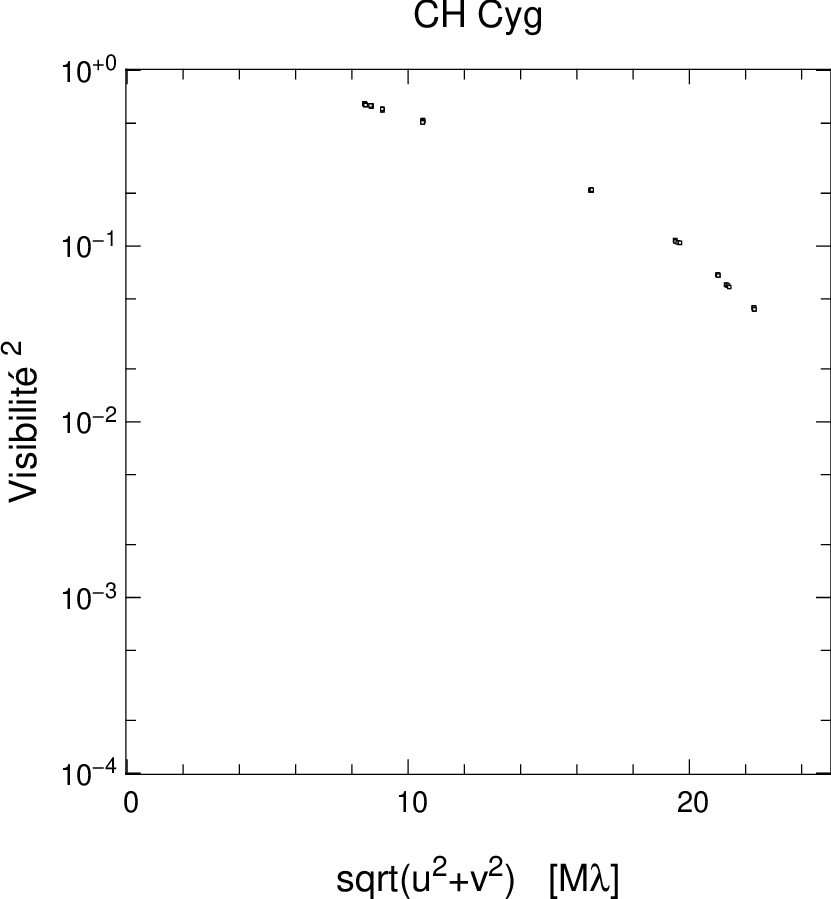}
\includegraphics{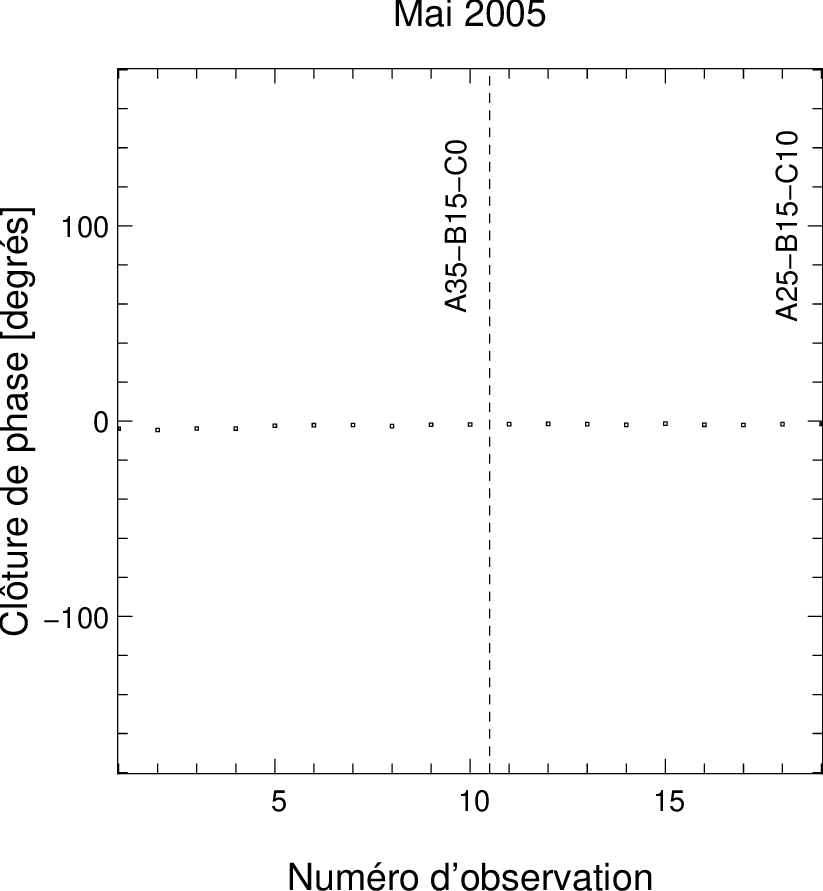}
} \caption[Données de $\chi$ Cyg et de CH Cyg obtenues au cours de la
   mission de mai 2005]{ V$^2$ et CP obtenues au cours de la première mission
   d'observation. Quatre configurations différentes ont été utilisées sur
   $\chi$ Cyg, et deux sur CH Cyg. Nous pouvons distinguer clairement le
   zéro du premier lobe de $\chi$ Cyg, ce qui n'est pas le cas pour CH
   Cyg, qui a un diamètre angulaire plus grand. Les clôtures de phase
   indiquent de fortes asymétries sur la brillance de la surface
   stellaire de $\chi$ Cyg.}  \label{fig:Donnees_1}
\end{figure}

\begin{figure}
\centering
\resizebox{\hsize}{!}{
\includegraphics{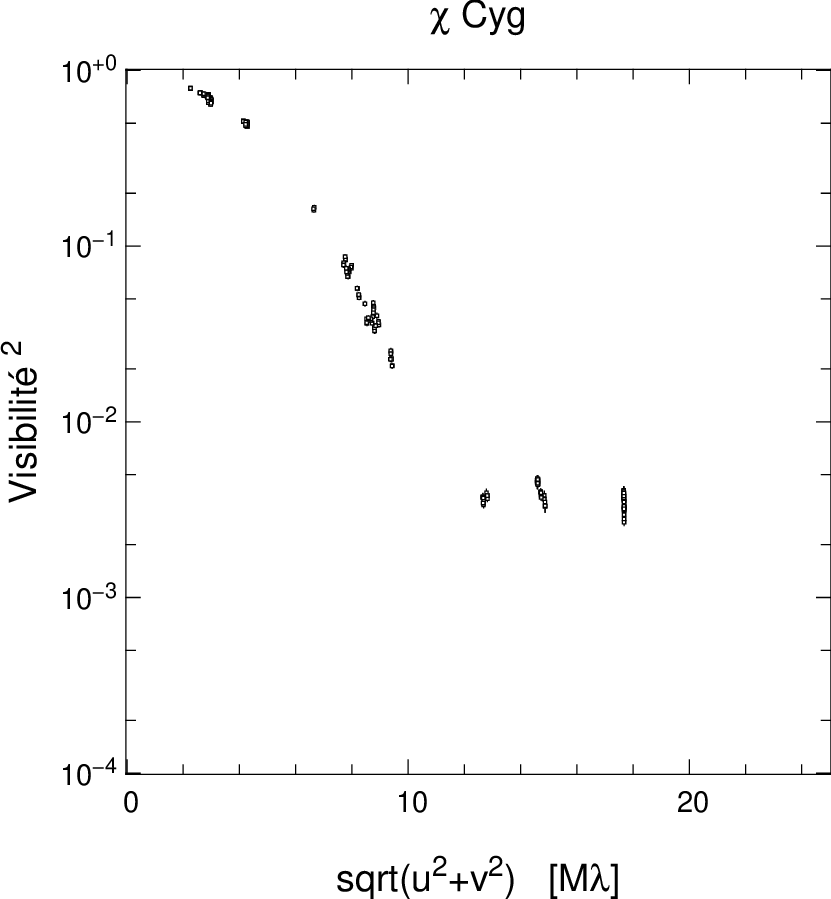}
\includegraphics{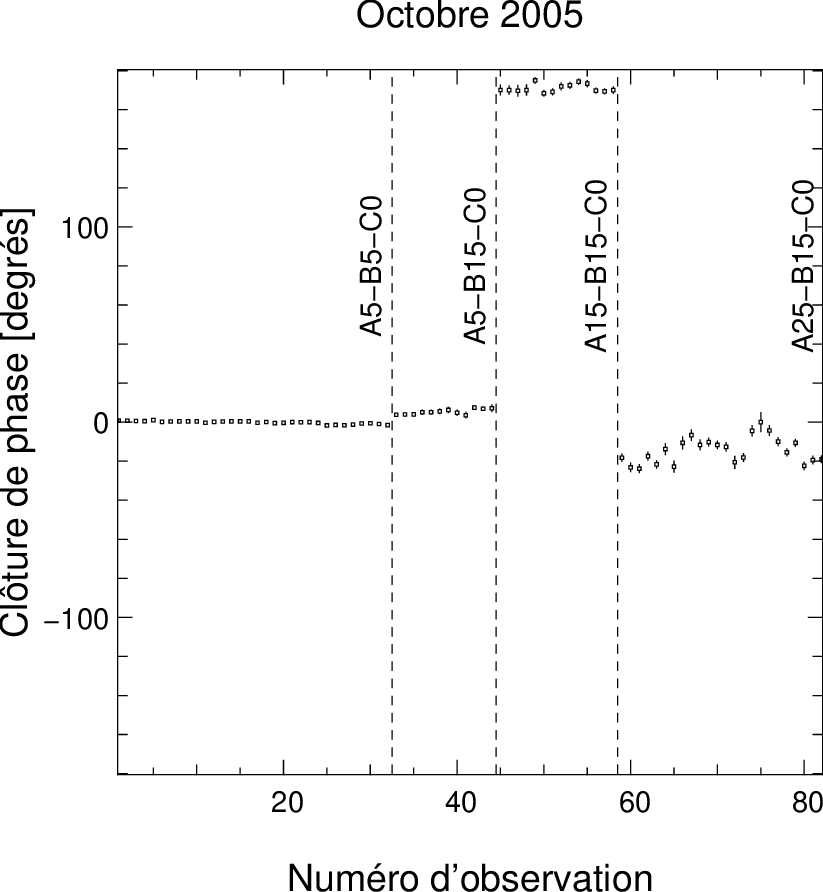}
}
{\vspace{.3cm}}
\resizebox{\hsize}{!}{
\includegraphics{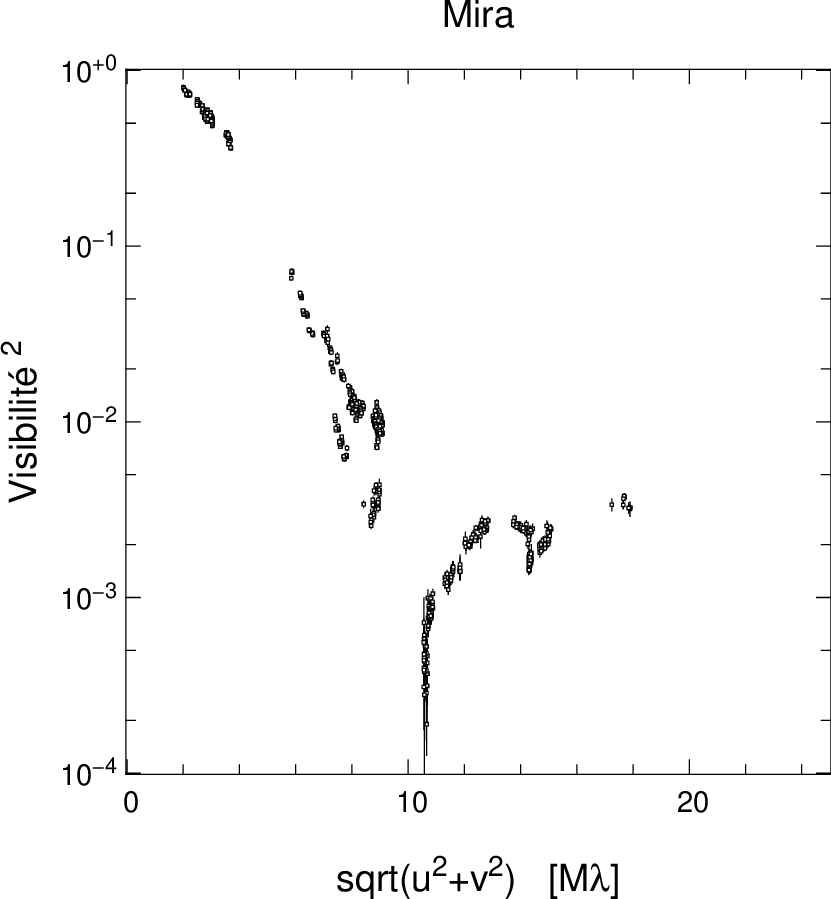}
\includegraphics{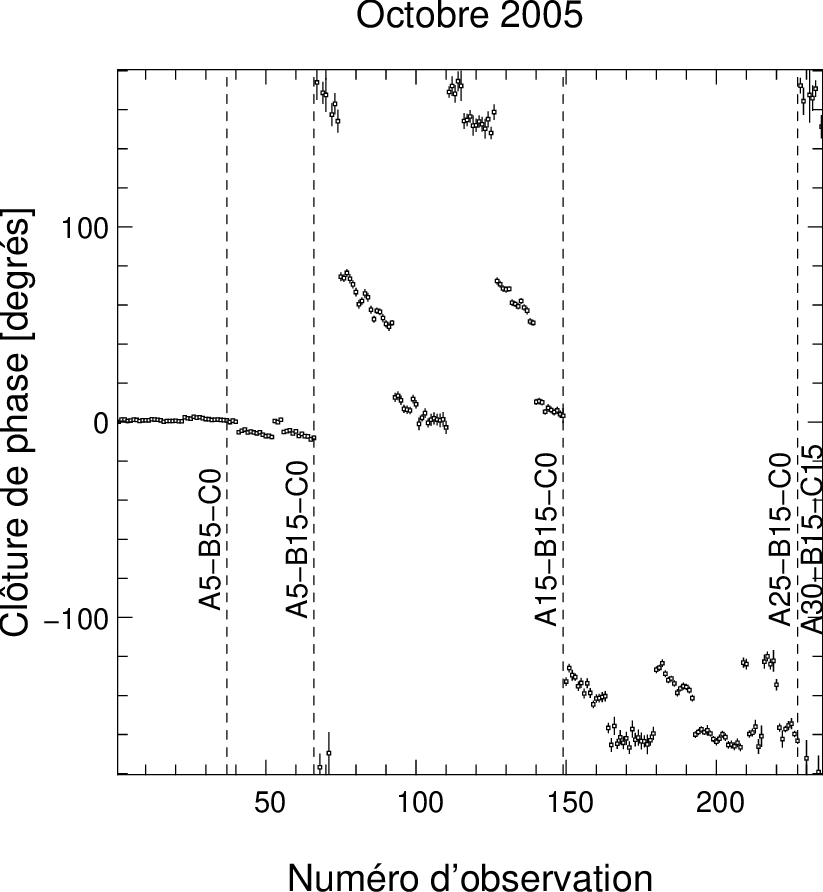}
} \caption[Données de $\chi$ Cyg et de Mira obtenues au cours de la
   mission d'octobre 2005]{ V$^2$ et CP obtenues au cours de la seconde mission
   d'observation. $\chi$ Cyg, toujours asymétrique, présente un zéro
   à une fréquence spatiale plus petite, caractéristique d'un diamètre
   plus grand. Les données de Mira témoignent d'une asymétrie
   encore plus grande, qui peut être vue à la fois sur les clôtures de
   phase et sur les visibilités. En effet, la dispersion observée aux
   fréquences proches de 8 M$\lambda$ n'est pas due à un bruit, mais
   bien à une variation de la visibilité en fonction de l'angle de
   projection des bases.}  \label{fig:Donnees_2}
\end{figure}

\begin{figure}
\centering
\resizebox{\hsize}{!}{
\includegraphics{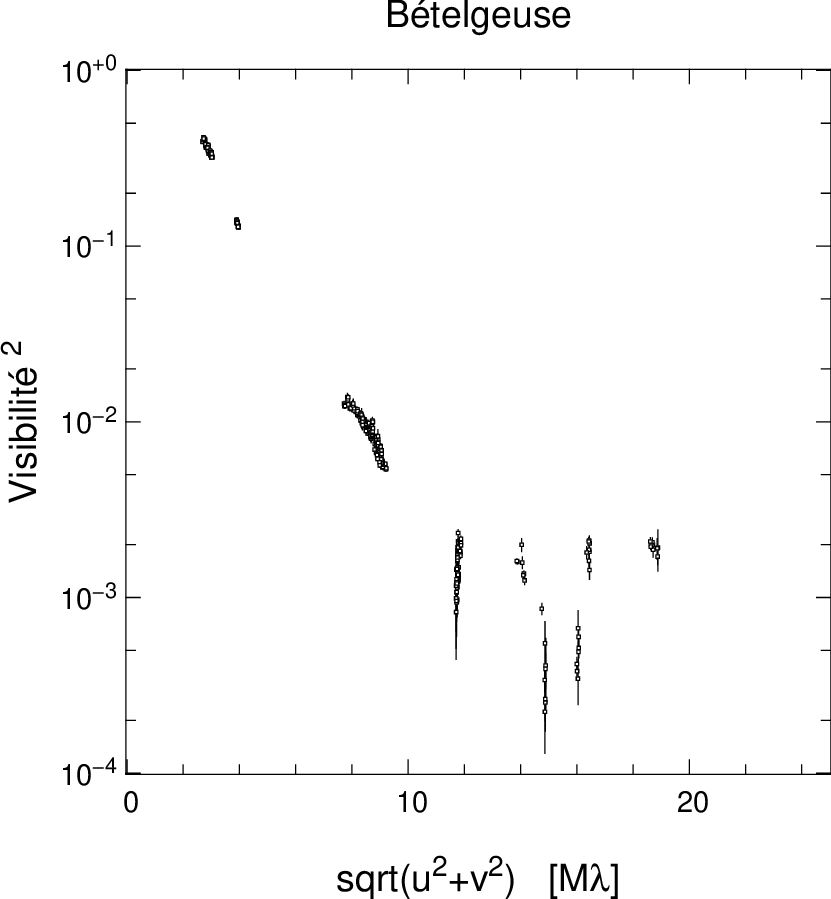} 
\includegraphics{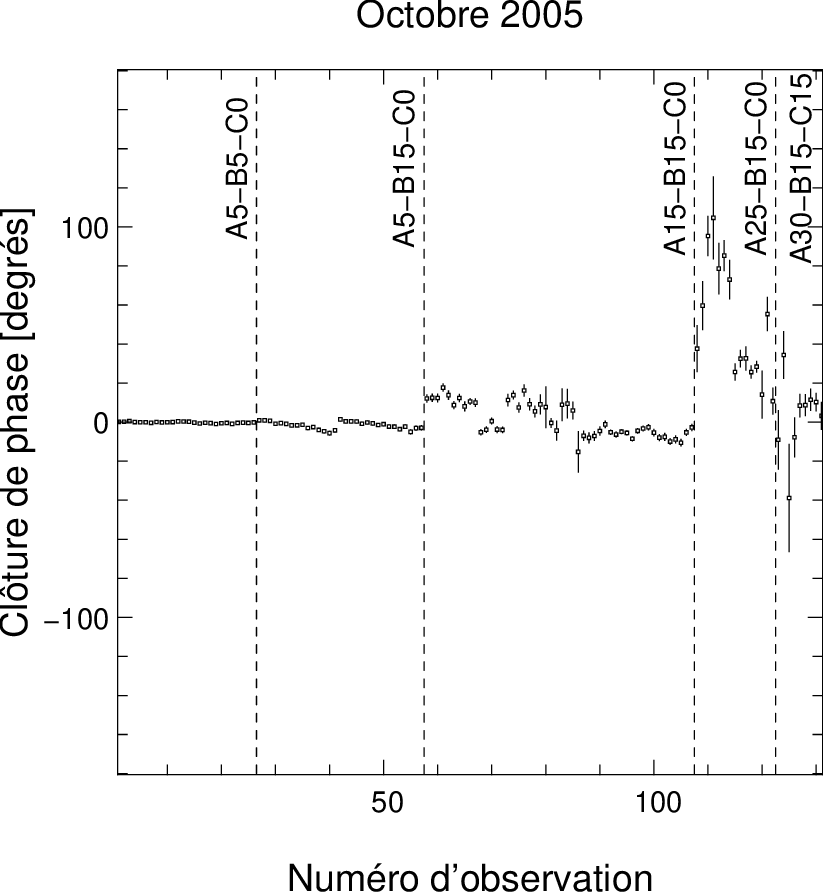}
} \caption[Données de Bételgeuse obtenues au cours de la
   mission d'octobre 2005]{ Observation de Bételgeuse au cours de la mission
   d'observation d'Octobre 2005. Nous avons ici des mesures sur les
   quatre premiers lobes de la courbe de visibilité. Le premier zéro
   se situe pour une fréquence spatiale d'environ 5 M$\lambda$, le
   second à 10 M$\lambda$, et le troisième autour de 15
   M$\lambda$. Ceci est possible grâce au large diamètre angulaire de
   Bételgeuse, qui est de 43,3 mas.  Les clôtures de phase indiquent
   des asymétries même si elles sont plus faibles que celles observées
   sur $\chi$ Cyg. Par exemple, la base A5-B15-C0, correspondant au
   second lobe, indique des clôtures de quelques degrés, à la
   différence de $\chi$ Cyg (figure~\ref{fig:Donnees_1}) qui indique
   des valeurs de l'ordre de 120 degrés. L'absence de déphasage de 180
   sur les CP est fortuit et est uniquement dû au hasard du choix des
   configurations.}  \label{fig:Donnees_3}
\end{figure}

\begin{figure}
\centering
\resizebox{\hsize}{!}{
\includegraphics{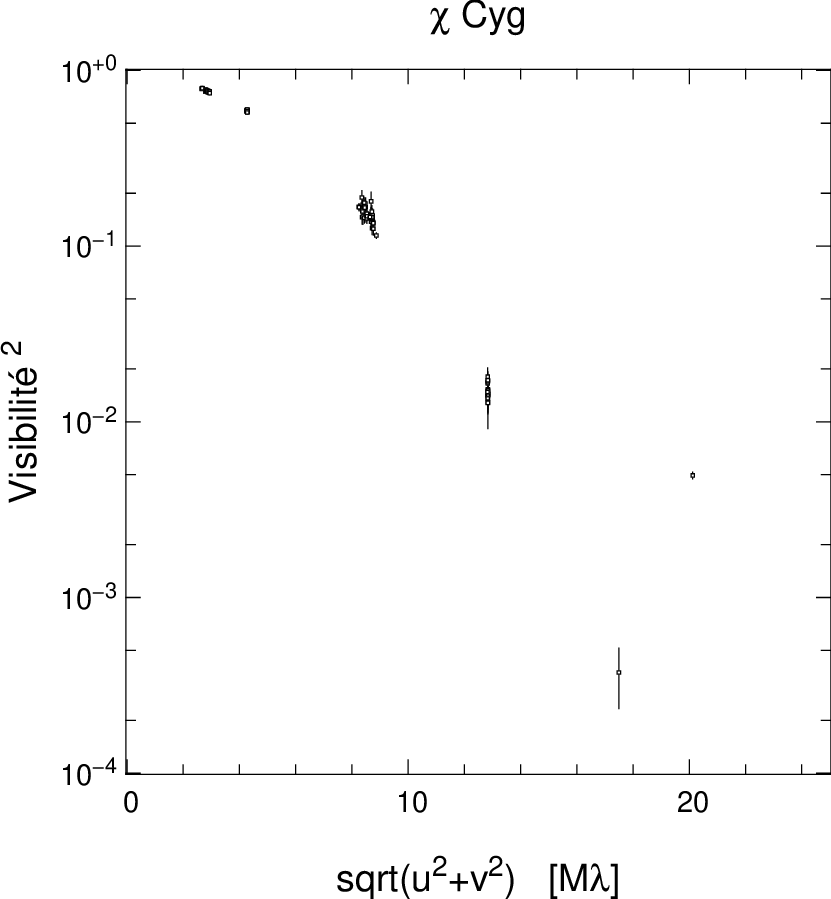}
\includegraphics{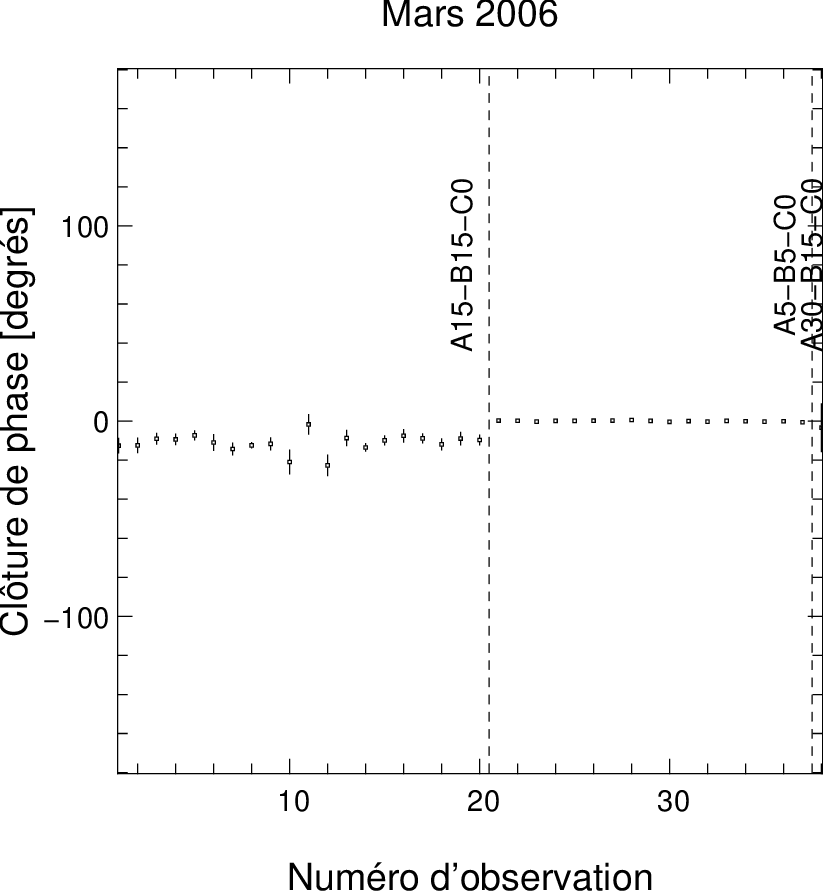}
}
{\vspace{.3cm}}
\resizebox{\hsize}{!}{
\includegraphics{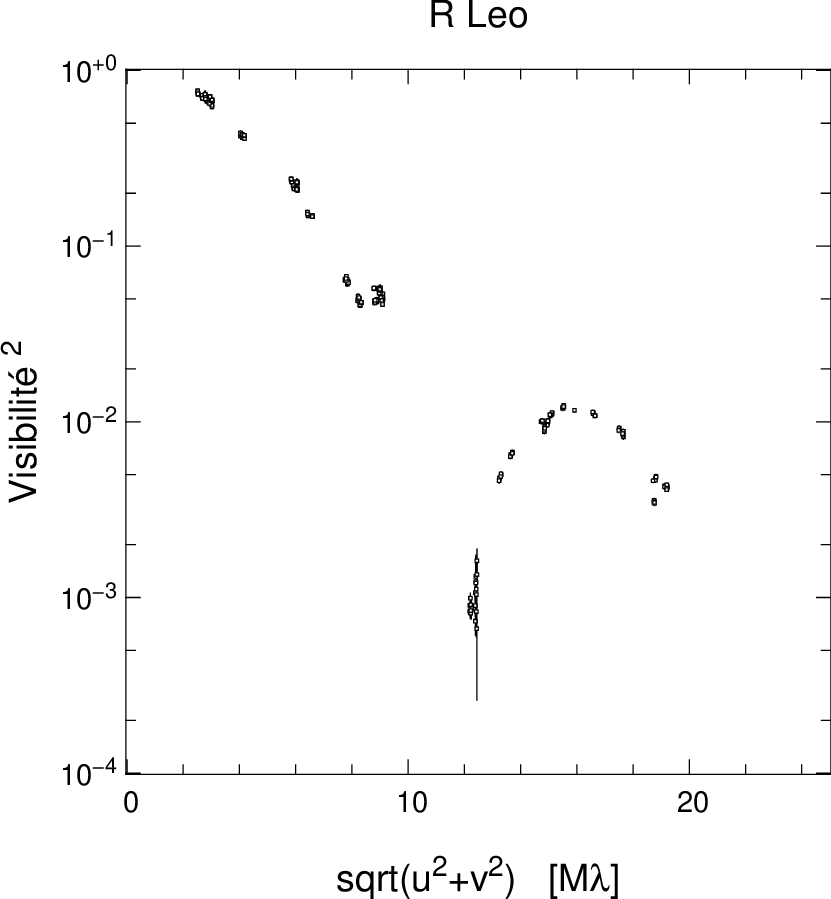}
\includegraphics{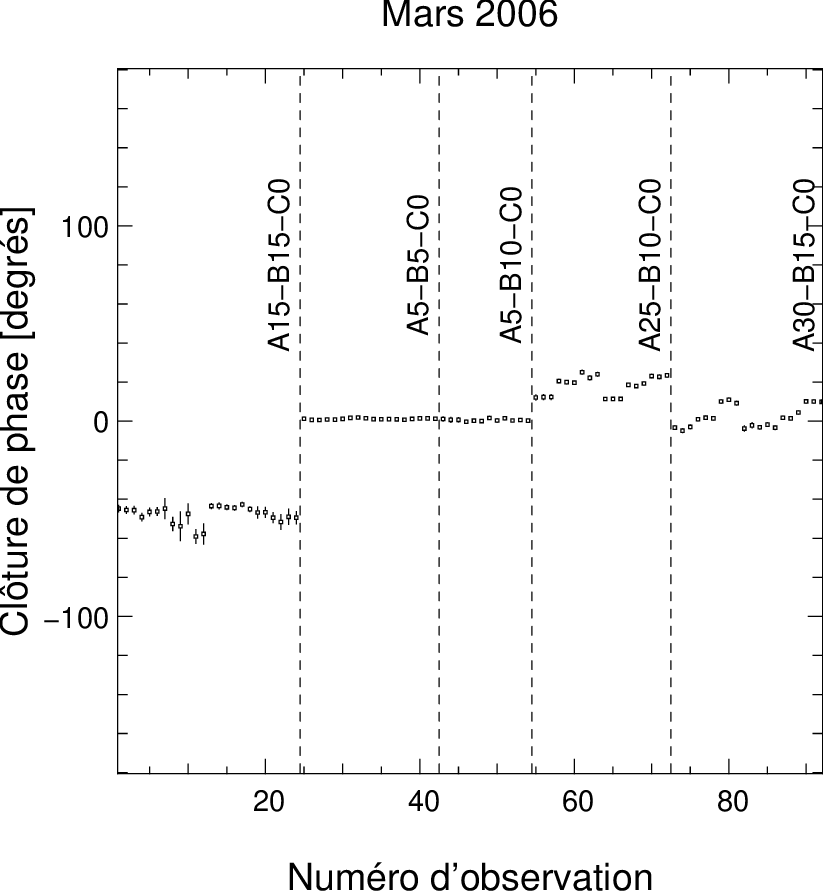}
} \caption[Données de $\chi$ Cyg et de R Leo obtenues au cours de la
   mission de mars 2006]{ V$^2$ et CP obtenues au cours de la troisième mission
   d'observation. Les données de $\chi$ Cyg sont très spartiates,
   conséquence du passage au méridien tardif de l'étoile, et donc de
   la difficulté d'observation. Le point de visibilité très faible à
   15 M$\lambda$ pourrait indiquer une petite taille angulaire de
   l'étoile (comparé aux données de Mai 2005). Nous verrons
   figure~\ref{fig:Data_Mar06} que cette valeur est la conséquence de
   l'asymétrie.  Les données sur R Leo sont de très bonne qualité,
   avec une asymétrie importante pouvant être vue à la fois sur les
   visibilités et sur les clôtures de phase. } \label{fig:Donnees_4}
\end{figure}

\begin{figure}
\centering
\resizebox{\hsize}{!}{
\includegraphics{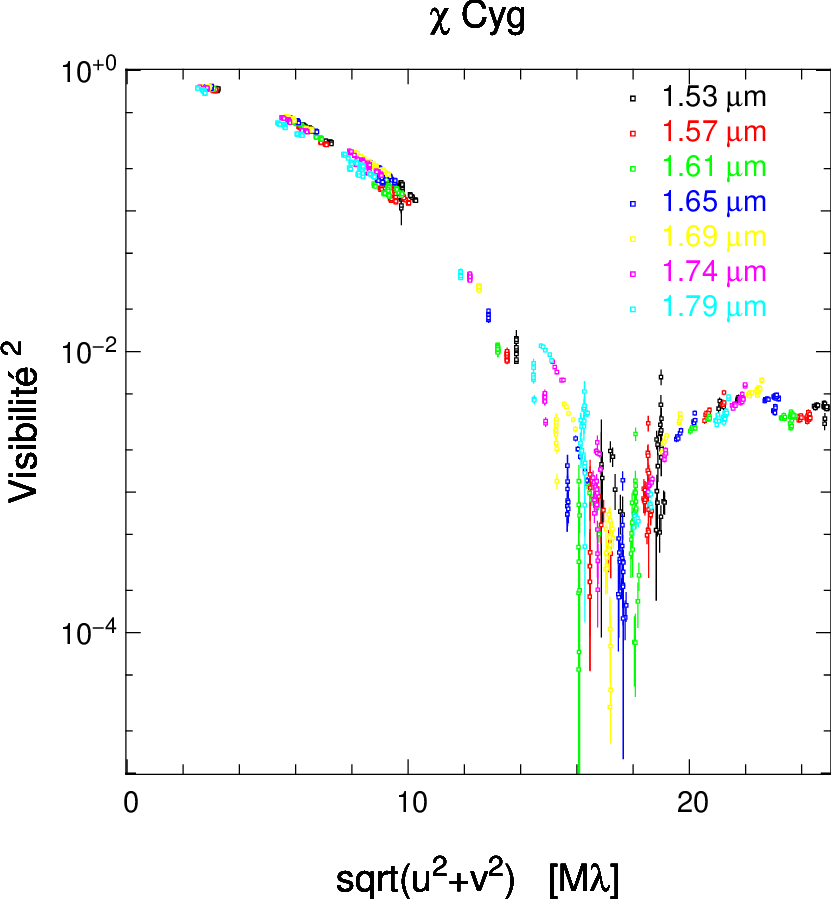}
\includegraphics{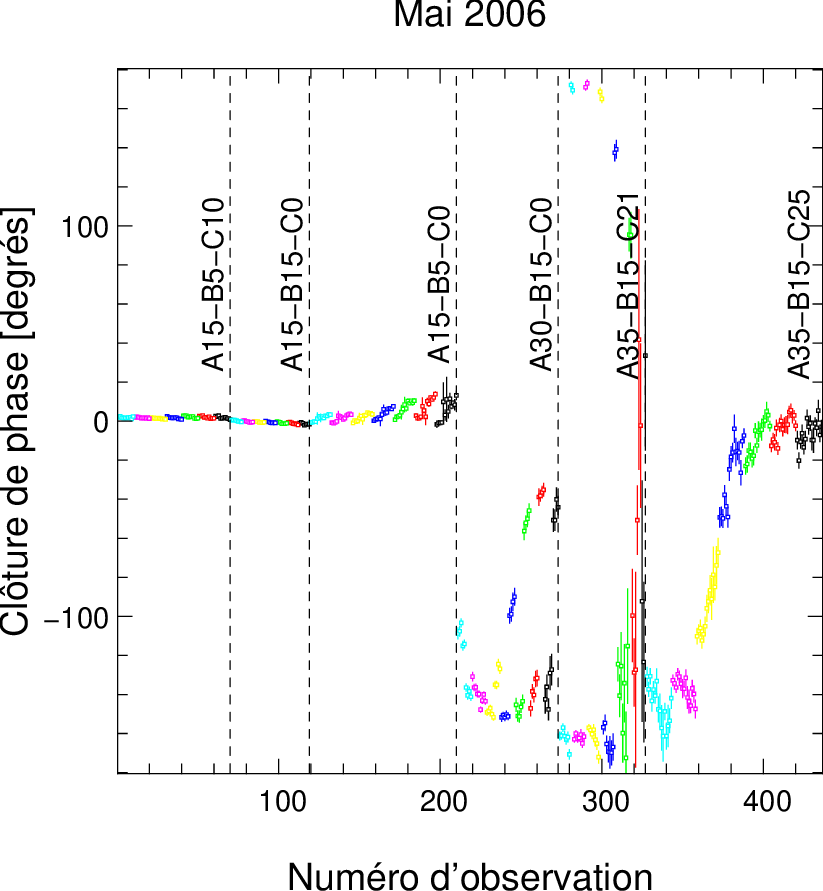}
}
{\vspace{.3cm}}
\resizebox{\hsize}{!}{
\includegraphics{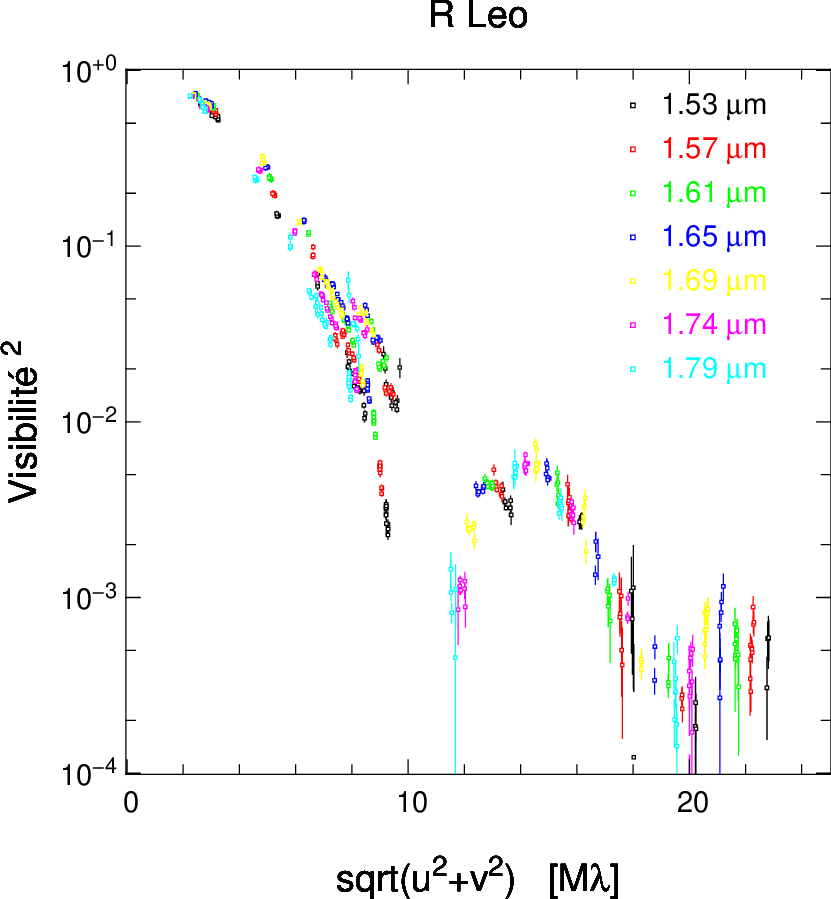}
\includegraphics{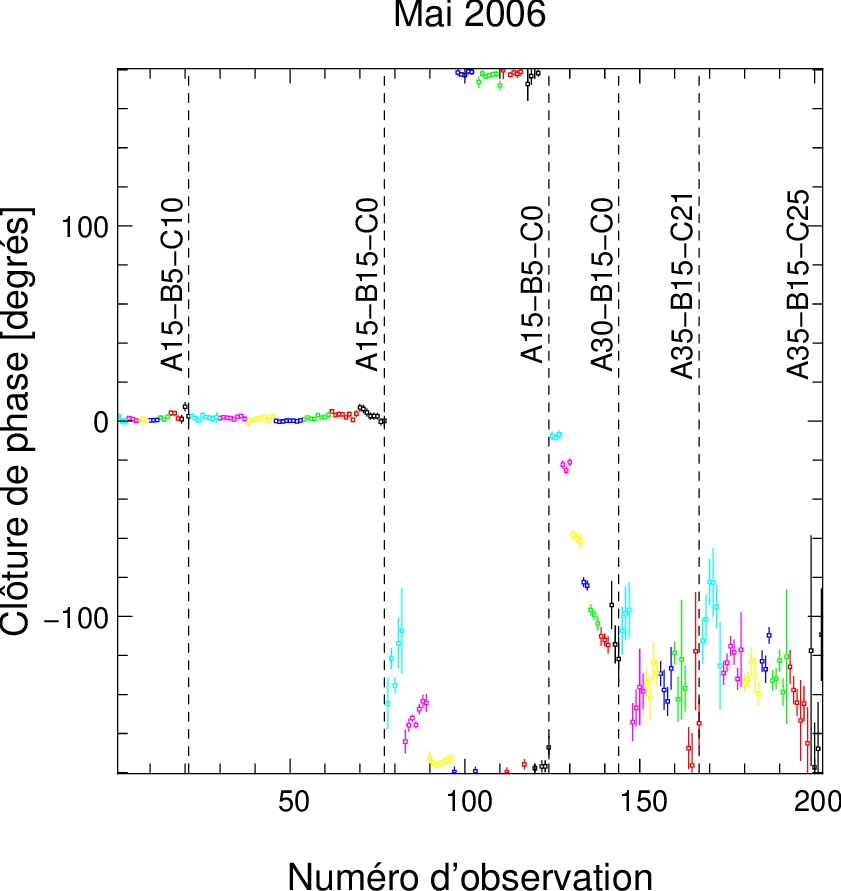}
} \caption[Données de $\chi$ Cyg et de R Leo obtenues au cours de la
   mission de mai 2006]{ V$^2$ et CP obtenues au cours de la dernière mission
   d'observation. Lors de celle-ci, nous avons positionné un prisme
   entre le détecteur et le recombinateur IONIC. Ceci nous a
   permis de disperser les franges sur sept canaux, et ainsi
   d'obtenir des mesures interférométriques simultanément à sept longueurs
   d'ondes différentes. Chaque longueur d'onde est ici représenté
   par une couleur. On peut noter un passage à zéro des visibilités de
   $\chi$ Cyg ``confu'', et le faible troisième lobe de R
   Leo. L'information spectrale sur les clôtures de phase est, elle
   aussi, très intéressante, car son attitude permet de contraindre
   spectralement l'asymétrie (voir section~\ref{sec:dispertion}).}
   \label{fig:Donnees_5}
\end{figure}

Les
figures~\ref{fig:Donnees_1},~\ref{fig:Donnees_2},~\ref{fig:Donnees_3},~\ref{fig:Donnees_4},~\ref{fig:Donnees_5}
et~\ref{fig:Donnees_6} représentent les visibilités au carré et les
clôtures de phase calibrées.  L'ordonnée des visibilités est en
logarithme, de façon à mettre en valeur l'information apportée par les
faibles visibilités. Les clôtures de phase sont comprises entre -180
et 180 degrés. La figure~\ref{fig:Donnees_1} correspond à la mission
de Mai 2005; les figures~\ref{fig:Donnees_2} et~\ref{fig:Donnees_3} à
celle d'Octobre 2005; la figure~\ref{fig:Donnees_4} à celle de Mars
2006 et enfin, les figures~\ref{fig:Donnees_5} et~\ref{fig:Donnees_6}
à celle de Mai 2006. Une première analyse de ces données est présentée
dans les légendes, de façon à mettre en relief les points notables de chacune
des observations.

Au vu de ces données, on peut remarquer, premièrement, une progression
en terme de qualité et d'efficacité. Ceci est dû à l'expérience de
l'équipe d'observation, mais aussi aux développements effectués par
l'équipe technique responsable de l'interféromètre. On peut, par
exemple, comparer la première mission d'observation (11 nuits) à la
dernière (6 nuits). Lors de la première mission, nous n'avons pu
observer qu'avec un maximum de 4 configurations sur deux objets. Lors
de la dernière mission, nous avons déplacé les télescopes chaque jour
et avons observé trois à quatre objets chaque nuit. Les conditions
météologiques favorables ont également été à l'origine de ces
résultats. La qualité du seeing nous a permit, notamment, d'obtenir
des données de grande précision, comme nous pouvons le voir sur
Arcturus (nous avons là des mesures de visibilités au carrés
inférieures à $10^{-4}$ figure~\ref{fig:Donnees_5}). A la lumière de
la qualité des données obtenues lors de cette dernière mission
d'observation, la fermeture récente de l'interféromètre IOTA est
d'autant plus regrettable.

\clearpage

\section{La reconstruction d'images interférométriques}
\label{sec:algo_rec_image}

L'objectif de ce chapitre est de fournir une vision d'ensemble des
problématiques sous-jacentes à la reconstruction d'image.  C'est
pourquoi, notamment, nous n'aborderons pas les problèmes de
non-convexité dues aux clôtures de phase. Une approche plus détaillée
peut être trouvée dans la thèse de \citet{Meimon:PhD}. 

\subsection{Le principe du problème inverse}

Résoudre un problème inverse se traduit de manière générale par
la recherche des valeurs $\V{x}$ à partir des $\V{y}$ vérifiant :
\begin{equation}
\V{y}=\V{m}(\V{x})+\V{b} \,,
\end{equation}
où $\V{y}$ sont les données, $\V{m}$ le modèle (qui peut être connu ou
inconnu), $\V{x}$ les paramètres recherchés, et $\V{b}$ les bruits
et/ou erreurs de modélisation. Le but est, alors, de trouver les {\em
meilleurs} paramètres compte tenu des données $\V{y}$ et du modèle
$\V{m}$.

Dans le cas de la reconstruction d'images interférométriques, le
problème inverse s'écrit de la manière suivante :
\begin{equation}
V(u,v)=TF(I(\alpha,\beta))+b(u,v) \,.
\label{eq:im_direct}
\end{equation}
Ceci, dans le cas où l'on possède une information sur les visibilités
complexes. Dans le cas précis des données obtenues à partir de
l'interféromètre IOTA, le problème est légèrement différent puisque
nous avons uniquement accès aux visibilités carrés et aux clôtures de
phase. Le principe est, cependant, similaire.

Pour effectuer l'inversion, reste à définir ce que sont les {\em
meilleurs} paramètres. Une première approche consiste à
effectuer un simple ajustement des données au modèle en cherchant le
maximum de vraisemblance $\mathrm{Pr}(\V{y}|\V{m}(\V{x}))$.

\subsection{Le maximum de vraisemblance}

Trouver le maximum de vraisemblance consiste à obtenir les valeurs
$\V{x}^{\rm MV}$ telles que la probabilité des mesures, étant donné le modèle,
est maximale :
\begin{equation}
\V{x}^{\rm MV} = \arg \max_{\V{x}} \mathrm{Pr}(\V{y}|\V{m}(\V{x}))
\end{equation}
Si l'on suppose les erreurs Gaussiennes et indépendantes, on peut
établir que trouver le maximum de vraisemblance équivaut à trouver le
minimum du $\chi^2$ pondéré :
\begin{eqnarray}
\V{x}^{\rm MV}  &= &  \arg \min_{\V{x}} \left[ -\log
\mathrm{Pr}(\V{y}|\V{m}(\V{x})) \right] \\
&= & \arg \min_{\V{x}} [\V{y}-\V{m}(\V{x})]^T\cdot {\rm
Cov}(\V{y})^{-1}\cdot [\V{y}-\V{m}(\V{x})]\\ &= & \arg \min_{\V{x}}
\sum_{i=1}^{n}\frac{(y_i-m_i(\V{x}))^2}{{\rm Var}(y_i)}\\
&=& \arg \min_{\V{x}} \chi^2 \,.
\end{eqnarray}

Pour en revenir à notre problème d'imagerie tel qu'établi
équation~\ref{eq:im_direct}, le $\chi^2$ s'écrit alors :
\begin{equation}
\label{eq:chi2}
\chi^2=\sum_{u,v}\frac{(V(u,v)-TF(I(\alpha,\beta)))^2}{{\rm Var}(V(u,v))} \,.
\end{equation}
Nous pouvons, cependant, montrer que la solution de $\chi^2$ minimum peut
être obtenue par
\begin{equation}
I'(\alpha,\beta)=TF^{-1}(V(u,v)) \,.
\label{eq:dirty_map}
\end{equation}
Il s'agit alors de la solution dite ``Dirty map'' qui n'est pas
nécessairement la {\em meilleure} du point de vue de l'analyse
scientifique. Plus exactement, il existe de multiples solutions pour
une valeur minimum du $\chi^2$. La solution donnée par
l'équation~\ref{eq:dirty_map} correspond au choix d'une image ayant
zéro comme valeur pour les composantes fréquentielles inconnues. Il y
a peu de chance que cette solution soit proche de la réalité.

\subsection{Le maximum \textit{a posteriori}}

\begin{figure}
\centering
\resizebox{\hsize}{!}{
\includegraphics{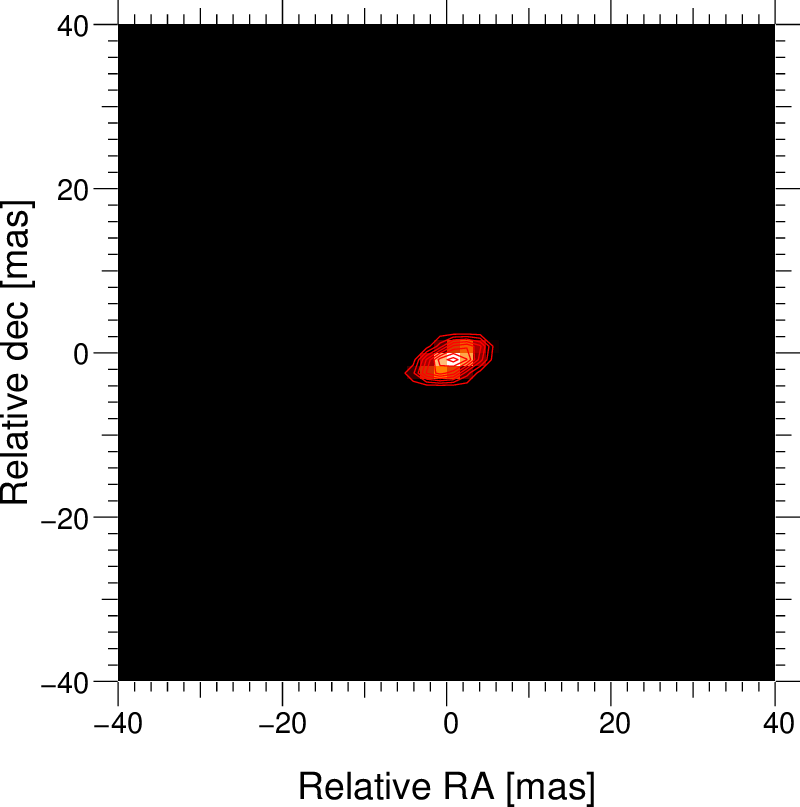}
\includegraphics{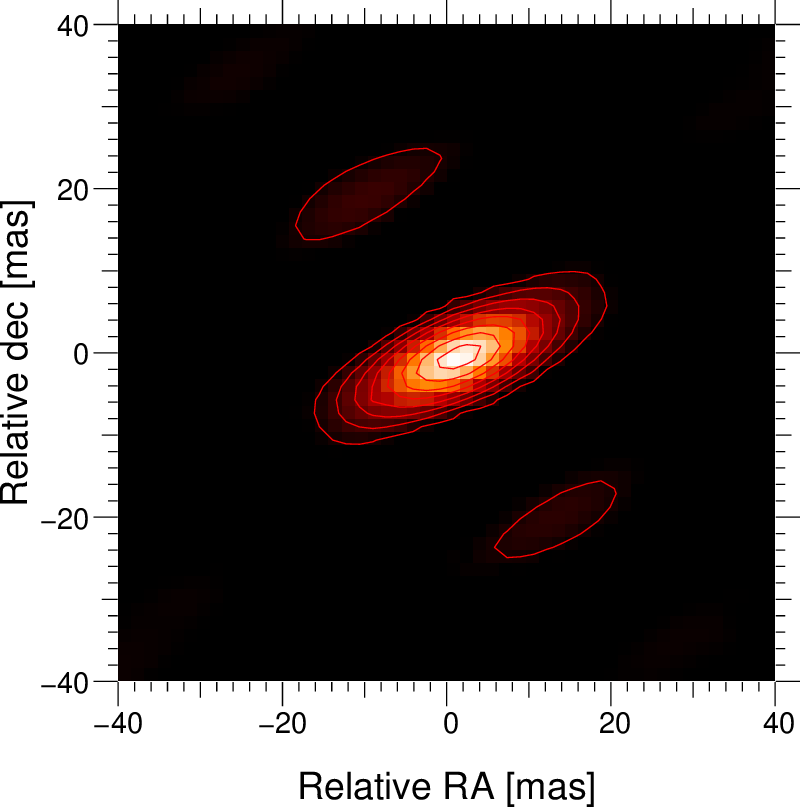}
\includegraphics{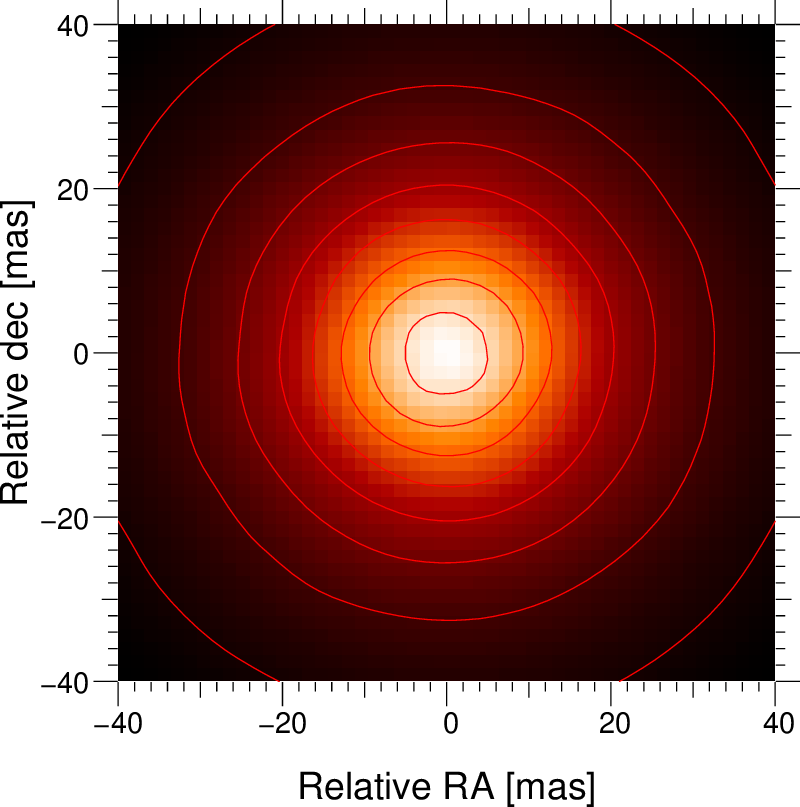}
} \caption[Reconstruction en aveugle d'une source ponctuelle simulée à
   partir de la couverture fréquentielle d'Arcturus]{ Reconstruction
   en aveugle d'une source ponctuelle simulée à partir de la
   couverture fréquentielle d'Arcturus. Nous pouvons voir l'effet de
   l'hyperparamètre de valeur: $10^6$, $10^8$ et $10^{12}$ (de gauche
   à droite). Sur l'image de droite, on retrouve la Lorentzienne qui
   correspond à la fonction de rappel : l'hyperparamètre est trop
   élevé. A l'opposé, l'image de gauche montre une résolution
   supérieure à celle de l'interféromètre (qui est de 9 $\times$ 22
   mas) : l'hyperparamètre est trop faible. Enfin, sur l'image
   centrale, on obtient une image proche de celle correspondant à la
   réponse impulsionnelle d'un télescope de la taille d'IOTA:
   l'hyperparamètre est satisfaisant. Ce choix n'est cependant pas
   forcément le même quel que soit l'objet, car il dépend aussi de la
   couverture fréquentielle et de la qualité des données.  }
   \label{fig:hyper}
\end{figure}

Le principe de la reconstruction en aveugle consiste ainsi à combler
un manque d'information - dans notre cas, les fréquences spatiales
inconnues - par un terme régulateur. On parle d'approche Bayésienne.
Lors d'un ajustement classique, on cherche les valeurs des paramètres
qui maximisent la vraisemblance
$\mathrm{Pr}(\V{y}|\V{m}(\V{x}))$. Dans le cadre du maximum {\it a
posteriori}, nous cherchons les $\V{x}^{\rm MAP}$ qui maximisent la
probabilité $\mathrm{Pr}(\V{m}(\V{x})|\V{y})$ :
\begin{equation}
\V{x}^{\rm MAP} = \arg \max_{\V{x}} \mathrm{Pr}(\V{m}(\V{x})|\V{y})
=  \arg \max_{\V{x}} \mathrm{Pr}(\V{x}|\V{y}) \,.
\end{equation}
Or, d'après la règle de Bayes :
\begin{equation}
 \mathrm{Pr}(\V{x}|\V{y}) = \frac{\mathrm{Pr}(\V{y}|\V{x}) \cdot
 \mathrm{Pr}(\V{x}) }{\mathrm{Pr}(\V{y}) } \, ,
\end{equation}
nous pouvons en déduire :
\begin{eqnarray}
\V{x}^{\rm MAP}  &= &  \arg \max_{\V{x}} \frac{\mathrm{Pr}(\V{y}|\V{x}) \cdot
 \mathrm{Pr}(\V{x}) }{\mathrm{Pr}(\V{y}) } \\ 
&= &  \arg \min_{\V{x}} \left[ -\log \left(
\frac{\mathrm{Pr}(\V{y}|\V{x}) \cdot
 \mathrm{Pr}(\V{x}) }{\mathrm{Pr}(\V{y}) }
 \right) \right]\,.\\
 &= &  \arg \min_{\V{x}} \left[ -\log
\mathrm{Pr}(\V{y}|\V{x})  -\log
\mathrm{Pr}(\V{x})\right]\,.
\end{eqnarray}
Ceci met ainsi en exergue deux termes à minimiser. Le premier, $\log
\mathrm{Pr}(\V{y}|\V{x})$ correspond, dans le cas d'un bruit
Gaussien, au $\chi^2$ que nous avons vu équation~(\ref{eq:chi2}). Le
second, $\log \mathrm{Pr}(\V{x})$, est un terme d'{\it a priori} sur
l'objet. C'est ainsi que l'on peut contraindre les fréquences
inconnues, en contraignant, par exemple, l'image à être positive, ou
plus ou moins lisse.

Sur ce modèle, le logiciel de reconstruction en aveugle que nous avons
utilisé va chercher l'image $I(\alpha,\beta)$ qui minimise la
fonction:
\begin{equation}
\chi^2 + \lambda f(I(\alpha,\beta)) \, , \quad{\rm tel\ que}\
I(\alpha,\beta) \geq 0 \,. 
\end{equation}
$\lambda$ est un hyperparamètre permettant de régler le poids de la
 fonction de régularisation $f$ par rapport au $\chi^2$.

L'algorithme de reconstruction a été développé à l'observatoire de
Lyon, sous la direction d'\'Eric Thi\'ebaut. Celui-ci utilise la fonction
de régularisation:
\begin{equation}
f(I(\alpha,\beta))=\frac{1}{1+(\alpha^2+\beta^2)/w^2}
\label{eq:t_regul}
\end{equation}
avec $w$ un paramètre à choisir en fonction de la nature de l'objet
observé.  Le choix de cette fonction s'est fait selon trois critères :
1) la fonction de régularisation doit centrer l'image, la position
n'étant pas contrainte par les clôtures de phase, 2) cette fonction
doit avoir un effet de lissage sur les données et 3) elle doit être de
codage facile pour minimiser le temps de calcul.  Ce terme de
régularisation joue le rôle d'un ``ressort'', c'est-à-dire que si la
contrainte des données devient faible, l'image aura tendance à se
rapprocher d'une Lorentzienne.  Dans le cas de la métaphore du
ressort, l'hyperparamètre correspond à la constante de raideur, et
devra donc être réglé par l'utilisateur. La Figure~\ref{fig:hyper}
montre l'influence de ce paramètre sur la reconstruction d'image. Si
le paramètre est trop grand, l'image est fortement contrainte par la
fonction de régularisation. Si le paramètre est trop faible, l'image
obtenue est sur-résolue par rapport à la résolution de
l'interféromètre, ce qui risque de faire apparaître des
artefacts. L'hyperparamètre a donc été choisi pour une valeur
intermédiaire, soit $3\times 10^7$. La largeur $w$ de la Lorentzienne
a été choisie pour une valeur avoisinant deux fois le rayon de la
photosphère.

\clearpage
\section{L'imagerie par reconstruction en aveugle}

Pour chaque étoile, nous présentons dans cette section la couverture
du plan $u$-$v$ et l'image obtenue par déconvolution en
aveugle. Nous nous contenterons ici de bréves descriptions des objets
astrophysiques ainsi que de rapides interprétations
phénoménologiques. Une analyse astrophysique plus poussée est ensuite
nécessaire, mais n'a été effectuée que pour un certain nombre de ces
objets. Nous verrons cela au Chapitre~\ref{sec:param}.

\subsection{Arcturus} \label{sec:actu_aveugle}

   \begin{figure}[h] \centering
   \includegraphics[width=6.2cm]{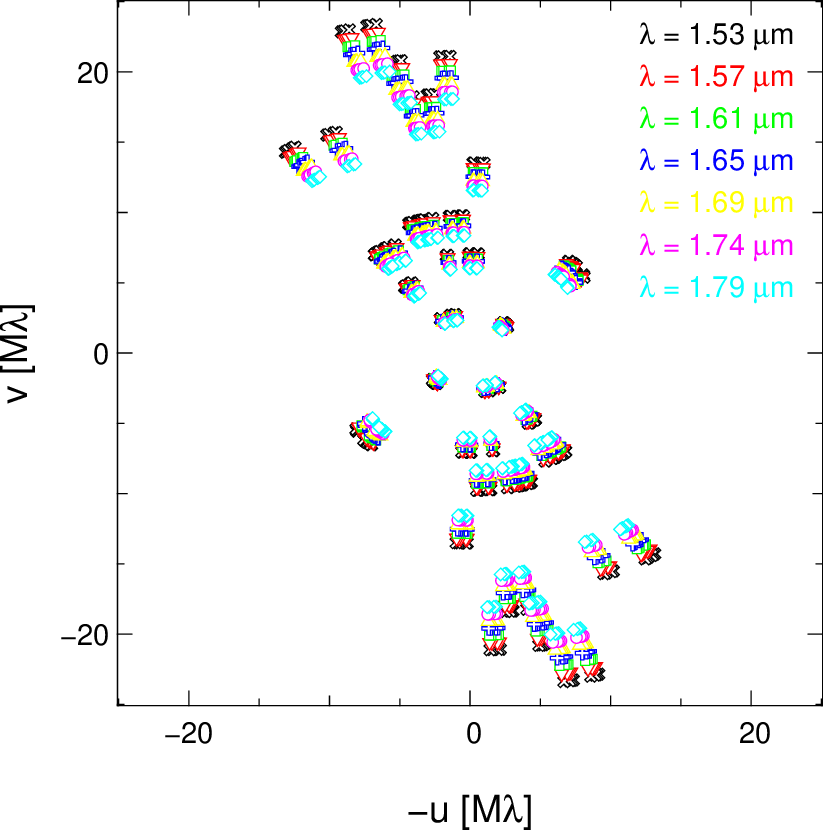}
   \includegraphics[width=7cm]{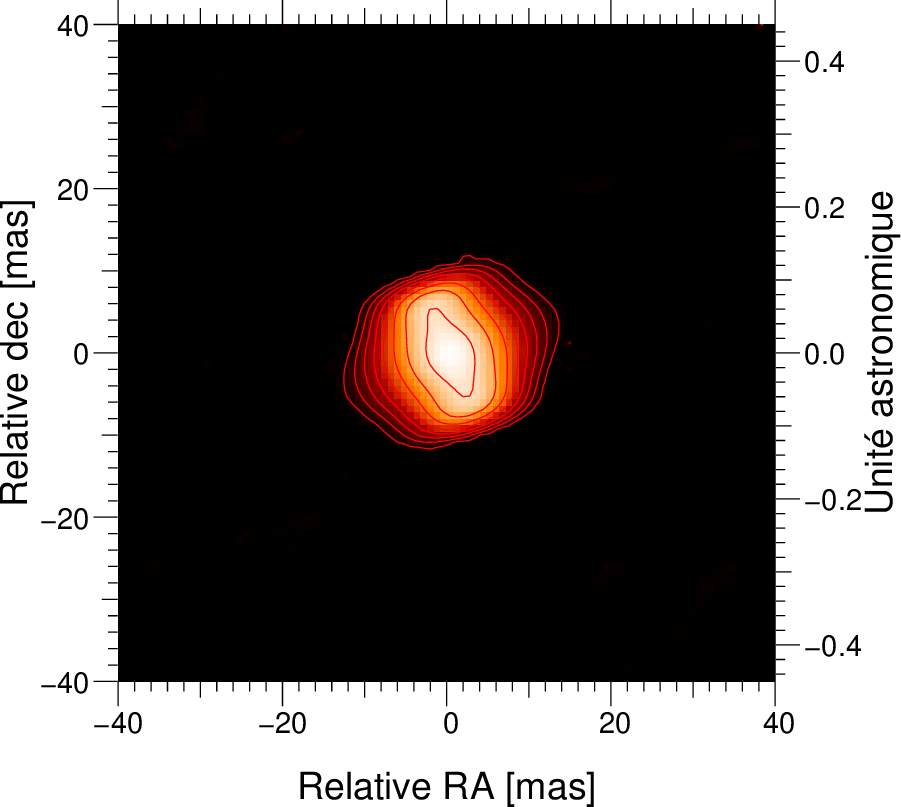}
   \caption[Plan $u$-$v$ et reconstruction en aveugle d'Arcturus]{
   Partie gauche: Plan $u$-$v$ couvert par une séquence d'observation
   de 5 jours. La couverture des fréquences spatiales est plutôt
   homogène, malgré une direction (N-NE) privilégiée.  L'utilisation
   du mode dispersé permet de bénéficier de l'influence de la longueur
   d'onde sur la fréquence spatiale mesurée. Partie droite: Image
   reconstruite par le logiciel Mira d'\'Eric Thiébaut en utilisant
   l'ensemble des longueurs d'onde et en supposant l'objet
   parfaitement achromatique. Sur l'image de droite, les courbes
   rouges correspondent aux 8 niveaux de flux allant linéairement de
   0,1 à 0,9, l'image ayant une brillance maximale normalisée à 1.
   Sur cette reconstruction, on résout clairement la géante rouge,
   sans y voir la présence d'un compagnon ou d'une asymétrie
   flagrante. } \label{fig:Arcturus} \end{figure}

Les géantes de type K sont souvent utilisées comme sources de
références pour l'étude des étoiles évoluées. Elles sont un bon
compromis entre luminosité et complexité. En effet, leur atmosphère
très compacte (du moins par rapport à l'ensemble des autres étoiles
évoluées) induit de faibles pulsations et un phénomène de convection
limité. En conséquence, on s'attend à une atmosphère simple sans
présence de couches moléculaires à grande distance de la
photospère. Arcturus (K1.5-2III, alpha Bootis) est l'une de ces
étoiles les plus connues. Elle a déjà été observée dans le visible par
l'interféromètre Mark III \citep{1996A&A...312..160Q}, ce qui a
confirmé les modèles existants, notamment, la présence d'un
assombrissement centre-bord important \citep{1977A&A....61..809M}. Des
observations plus récentes en infrarouge proche
\citep{2005A&A...435..289V} suggèrent, cependant, l'existence d'un
compagnon à cette étoile. Cette hypothèse est une parmis deux, l'autre
étant que l'atmosphère en K soit mal comprise. Si la présence d'un
compagnon à Arcturus était confirmé, cela aurait d'importantes
conséquences, cet objet étant considéré comme une référence
photométrique et spectrale
\citep[notamment pour ISO; ][]{2003A&A...400..709D}.

Les données correspondent à 5 nuits d'observation du 10 au 15 Mai
2006. Lors de ces nuits, nous avons pu utiliser le mode dispersé
d'IOTA. Pour effectuer la reconstruction d'image, nous avons supposé
l'objet achromatique, de manière à bénéficier de la couverture
fréquentielle fournie par les différentes longueurs d'ondes. Nous
avons, par la suite, validé cette hypothèse par des reconstructions
n'utilisant qu'un seul canal spectral. Notre conclusion a été que,
dans le cadre du bruit de reconstruction, l'achromaticité de l'objet
était une hypothèse convenable.  L'image obtenue
(figure~\ref{fig:Arcturus}) correspond à un disque circulaire doté
d'un assombrissement semblant plus prononcé sur l'axe horizontal. Ce
résultat est a priori surprenant. En effet, nous verrons
section~\ref{sc:Actu_test} que l'objet est parfaitement circulaire.
Cet effet est donc dû à l'algorithme de reconstruction. Une partie de
l'explication se trouve dans la géométrie de l'interféromètre
IOTA. Comme nous l'avons vu section~\ref{sec:IOTA_r}, où encore dans
la couverture $u$-$v$ de la figure~\ref{fig:Arcturus}, les bases
disponibles permettent une meilleure résolution sur l'axe N-S que sur
l'axe E-O. Cela se traduit par un déficit en informations sur les
hautes fréquences de cet axe. Ce manque est alors comblé par la
fonction de régularisation (équation~\ref{eq:t_regul}), ce qui crée
une forme plus piquée au centre, aux contours plus lisses. Il est ainsi
intéressant de noter que l'analyse classique selon laquelle l'image
est un produit de convolution entre l'objet et une réponse
impulsionnelle n'est plus vérifiée.

A partir de la variance du fond de l'image~\ref{fig:Arcturus}, nous
avons déduit une dynamique de reconstruction de 250. Le flux maximal
observé sur un pixel du fond est, lui, de 1,5\% de la brillance
maximale. Cette reconstruction d'image nous permet ainsi d'écarter la
présence d'un compagnon ayant ce niveau de flux. Nous verrons
section~\ref{sec:compa_arctu} qu'une analyse paramétrique des clôtures
de phase permet d'écarter définitevement la possibilité d'un compagnon
ayant un flux même bien inférieur.

   \begin{figure} \centering
   \includegraphics[width=5.4cm]{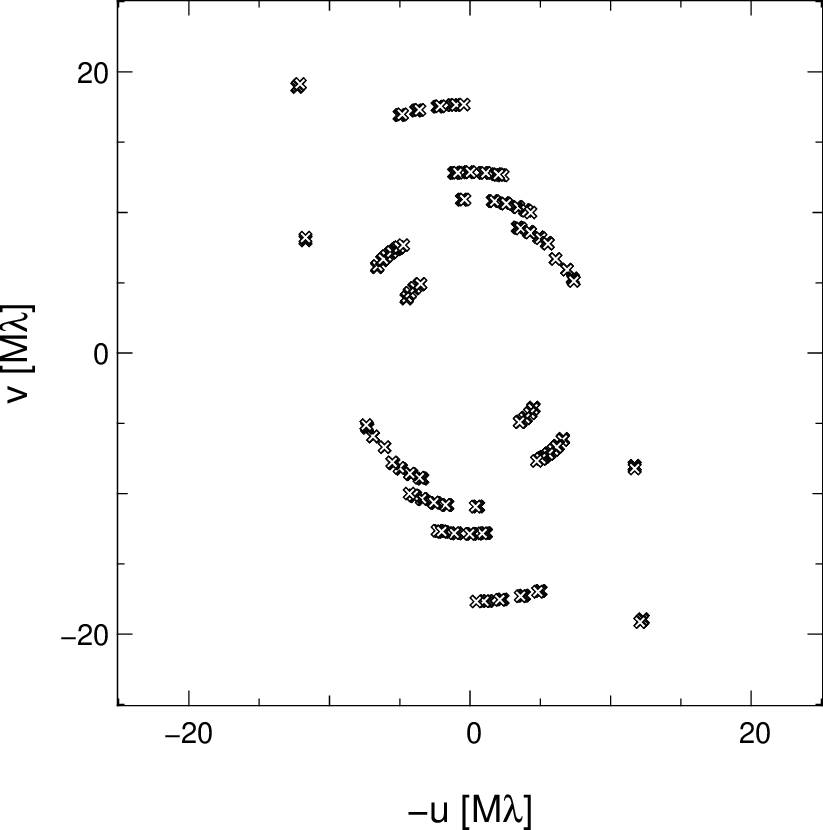}
   \includegraphics[width=6.1cm]{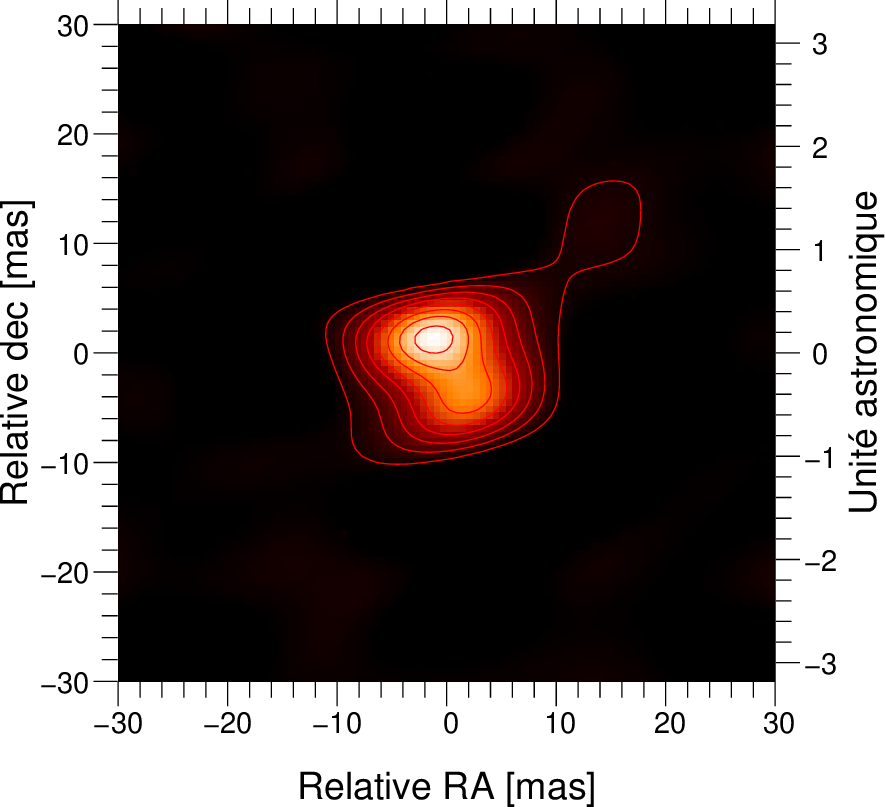}
   \includegraphics[width=5.4cm]{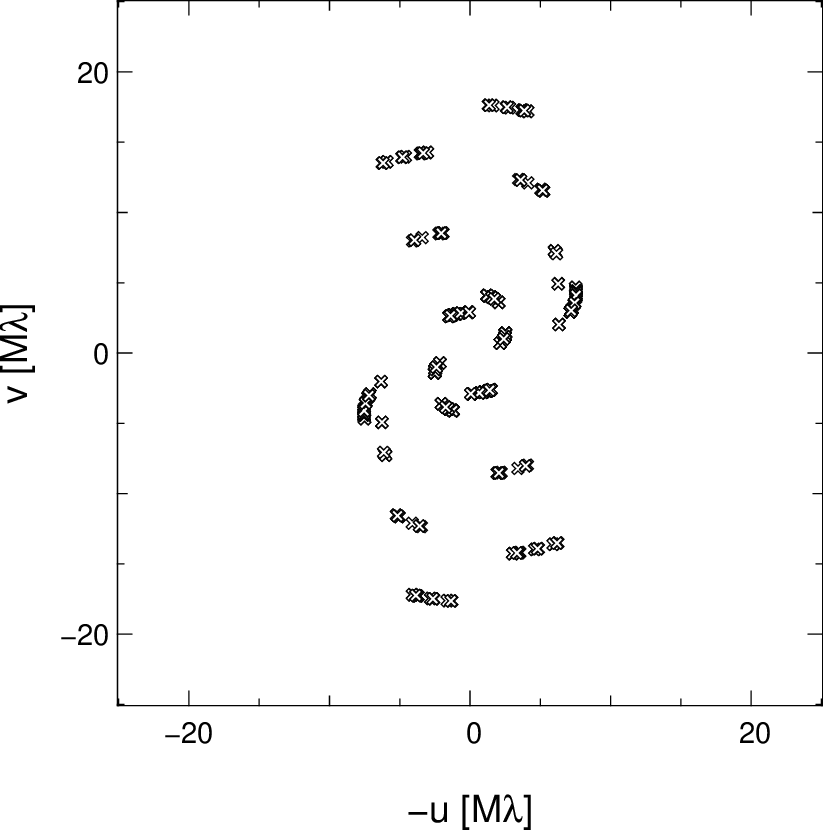}
   \includegraphics[width=6.1cm]{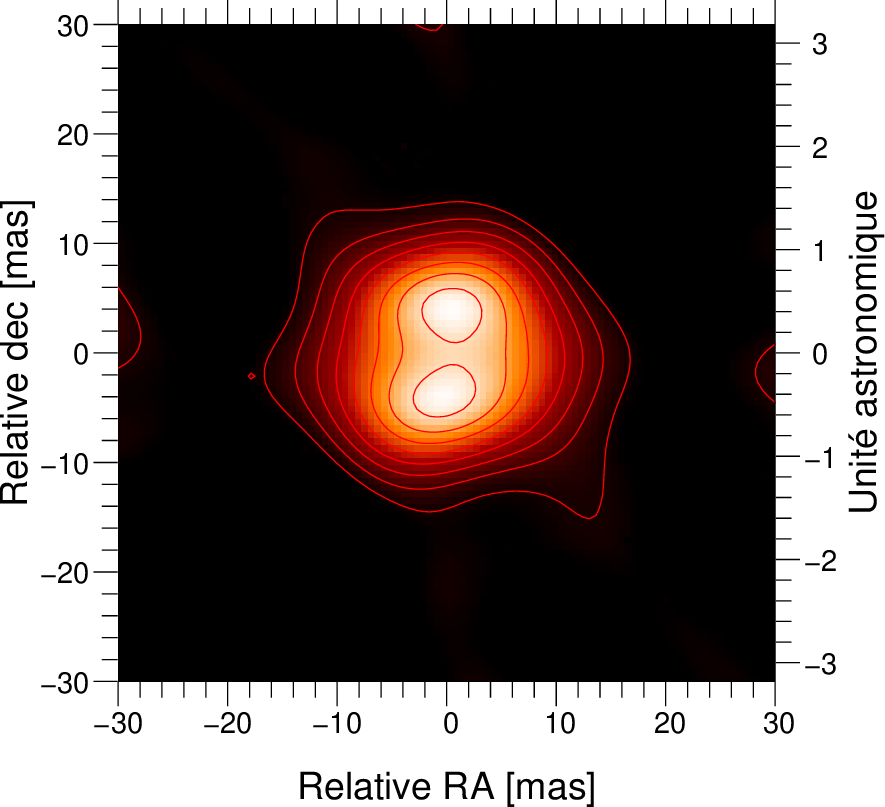}
   \includegraphics[width=5.4cm]{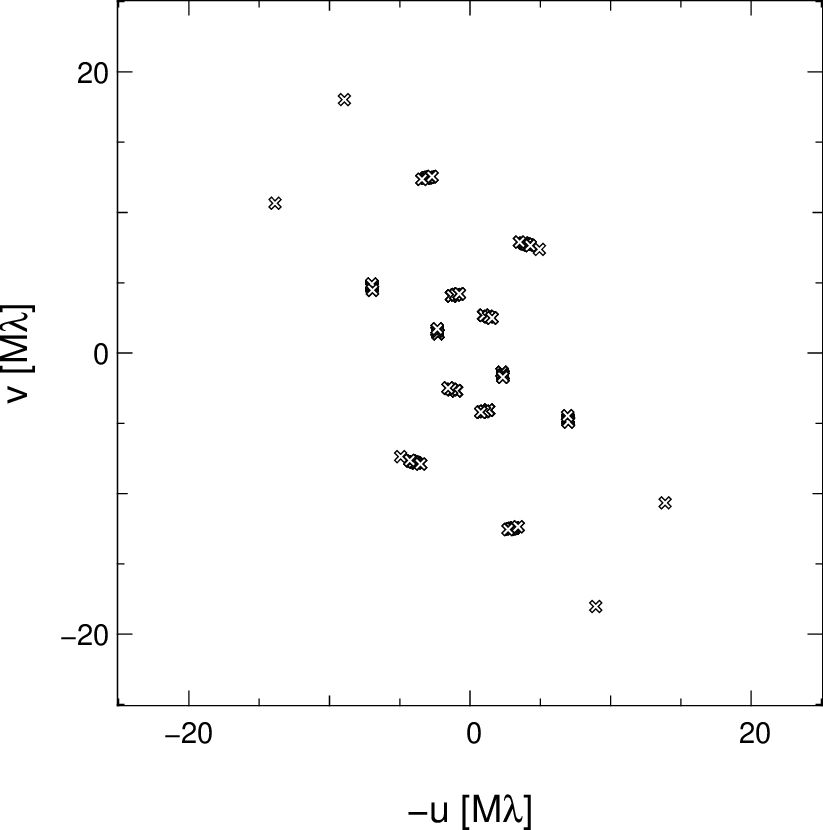}
   \includegraphics[width=6.1cm]{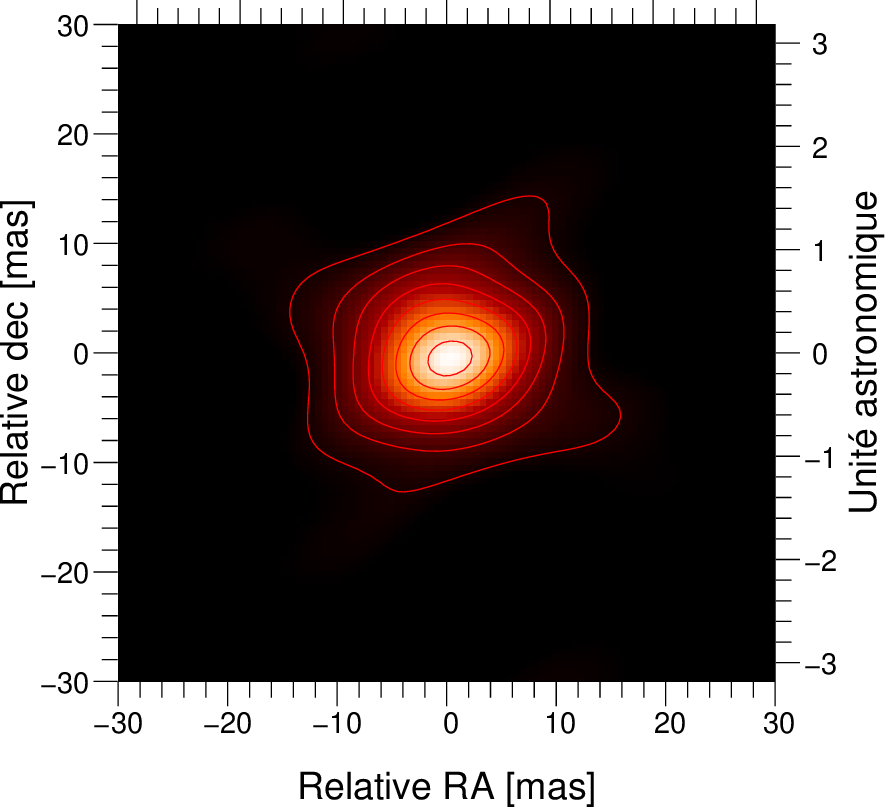}
   \includegraphics[width=5.4cm]{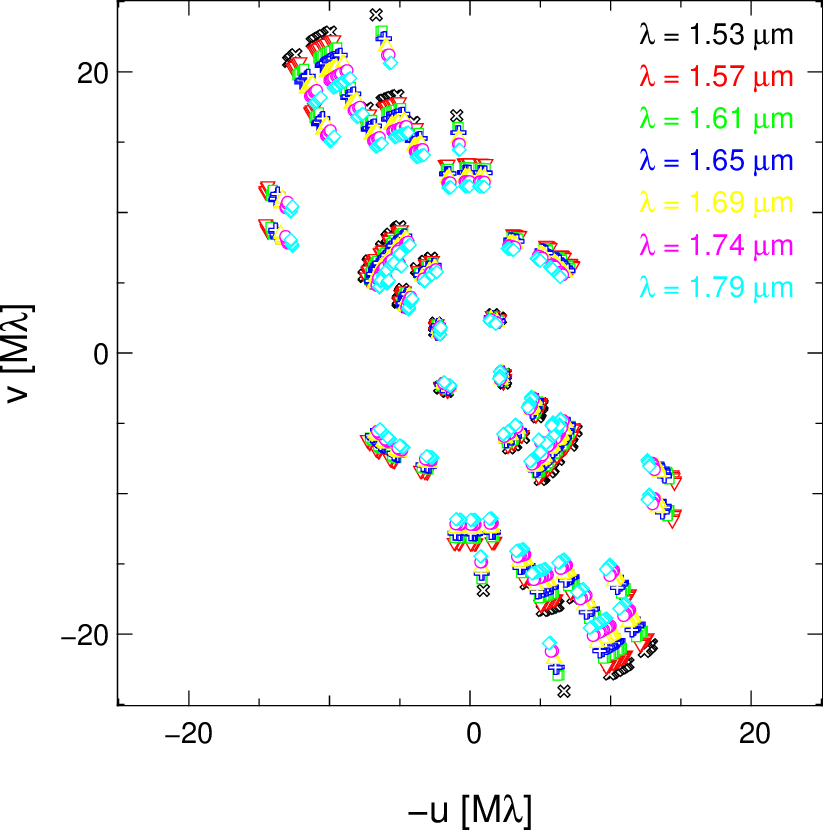}
   \includegraphics[width=6.1cm]{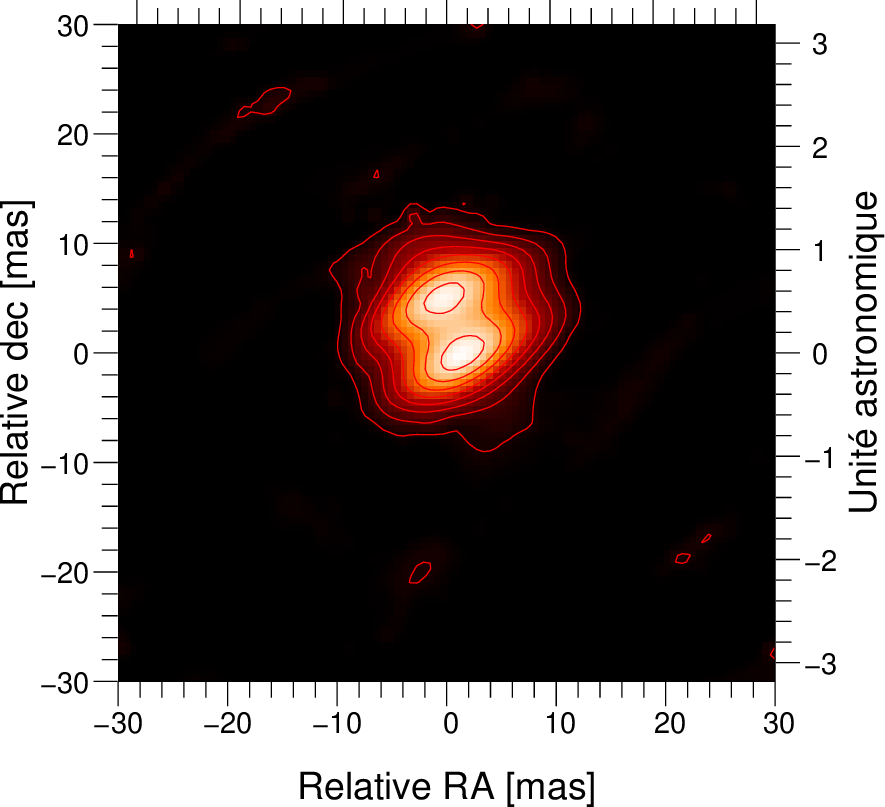}
   \caption[Plans $u$-$v$ et reconstructions en aveugle de $\chi$ Cyg]{
   Couvertures fréquentielles (gauche) et images de $\chi$ Cyg
   (droite) obtenues à quatre phases stellaires différentes (de haut
   en bas : $\phi = 0,91$, 0,24, 0,67, et 0,76). Ces images montrent
   une nette variation à la fois en taille et en morphologie.  }
   \label{fig:Chi_Cyg_rec} \end{figure}

\subsection{$\chi$ Cygni}

$\chi$ Cygni est une étoile de la branche asymptotique des géantes. De
type Mira, elle présente de fortes variations photométriques, avec une
période de 408 jours. Comme on peut nettement l'observer sur la
figure~\ref{fig:lum_cyg}, sa luminosité peut passer, en 200 jours, d'une
magnitude 3.5 à une magnitude 14 dans le visible, soit 1500 fois plus faible. Par
rapport aux autres étoiles de la classe des Mira, elle a la
particularité d'être de type S, c'est-à-dire d'avoir une atmosphère très
carbonnée, avec un ratio oxygène sur carbone d'environ 1. Ceci est la
conséquence d'un important ``dredge-up'' au cours duquel le phénomène
de convection a pu se produire suffisamment profondément dans l'étoile
pour que les métaux produits par la fusion au c\oe ur de l'étoile
puissent revenir à la surface. Pour autant, tout l'oxygène n'est pas
sous forme de monoxyde de carbone comme en témoigne la présence d'eau
ou même de masers SiO \citep{1999A&AS..139..461A}. Ceci s'explique, en
partie, par le type spectral qui passe périodiquement de S à M, au
maximum de luminosité.

Nous avons eu la chance de pouvoir observer $\chi$ Cyg au cours des 4
missions d'observation, ceci nous a permis d'obtenir une image de
l'étoile à différentes phases. Les images obtenues à partir du
logiciel de reconstruction montrent d'ailleurs clairement des
différences de morphologies aux différentes époques
(figures~\ref{fig:Chi_Cyg_rec}). Les phases d'observations sont,
chronologiquement, 0,91, 0,24, 0,67 et 0.76. Les rayons observés
varient de 1 à 1,5 unité astronomique. Les asymétries sont aussi
profondément modifiées, comme nous avons déjà pu le supposer à partir
d'une analyse empirique des clôtures de phase
figures~\ref{fig:Donnees_1},~\ref{fig:Donnees_2},~\ref{fig:Donnees_4}
et~\ref{fig:Donnees_5}. L'évolution de ces asymétries est, par ailleurs,
nettement incompatible avec un simple effet de rotation de l'astre
\citep[la période de rotation d'une telle étoile a été évaluée à 16 ans
par][]{2003A&A...397..943B}. Cette étoile est étudiée plus en détails
dans le chapitre~\ref{sec:temporelle}.

\subsection{R Leo}

   \begin{figure}[h!]
   \centering
   \includegraphics[width=6.2cm]{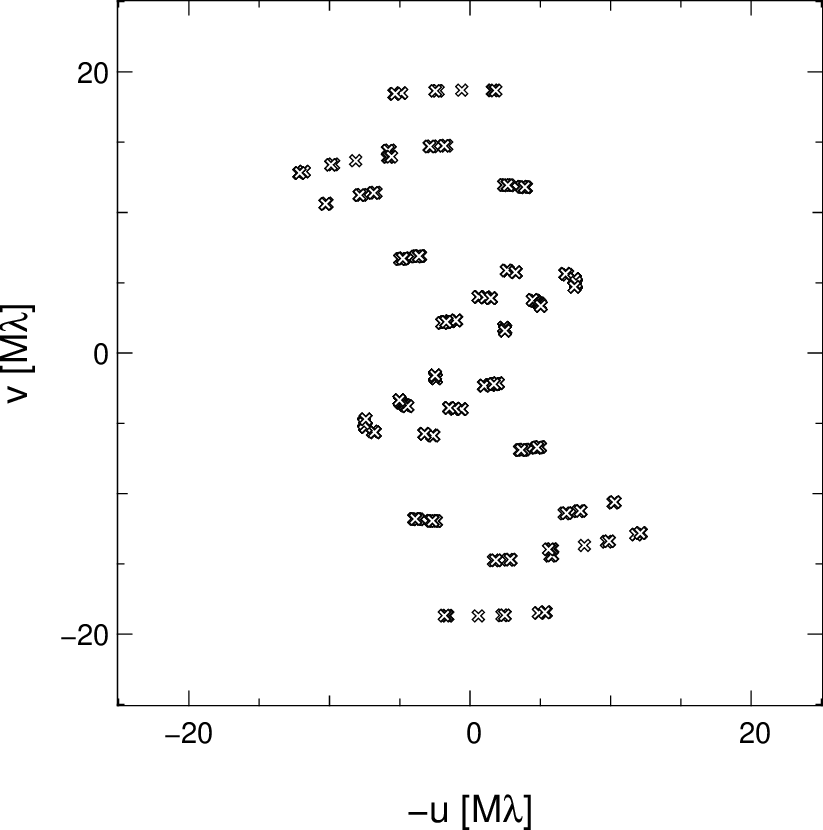}
   \includegraphics[width=7cm]{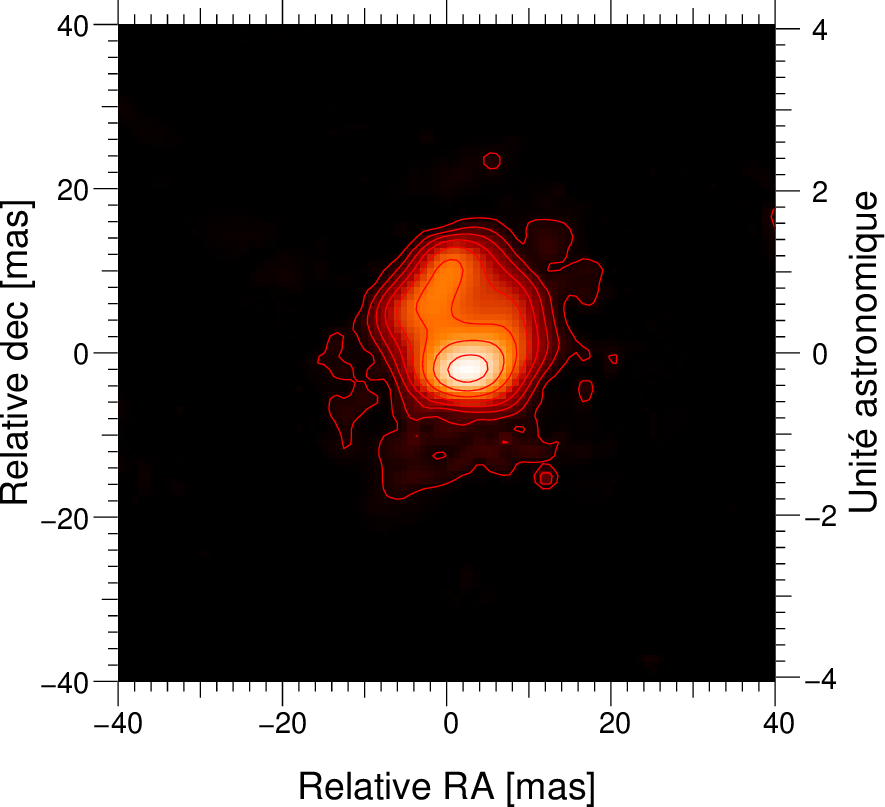}
   \includegraphics[width=6.2cm]{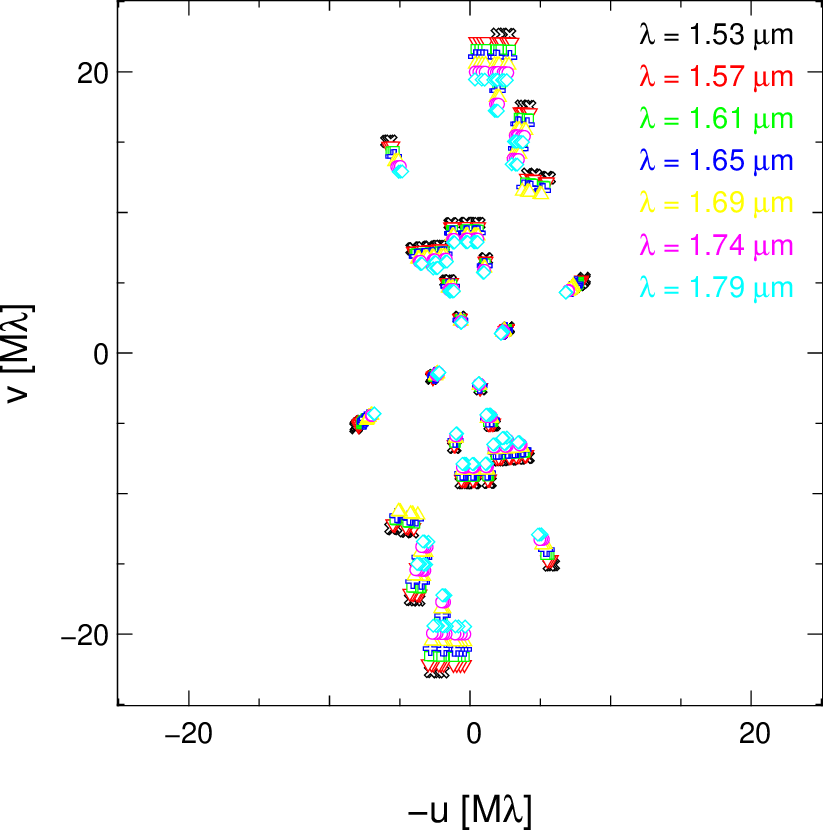}
   \includegraphics[width=7cm]{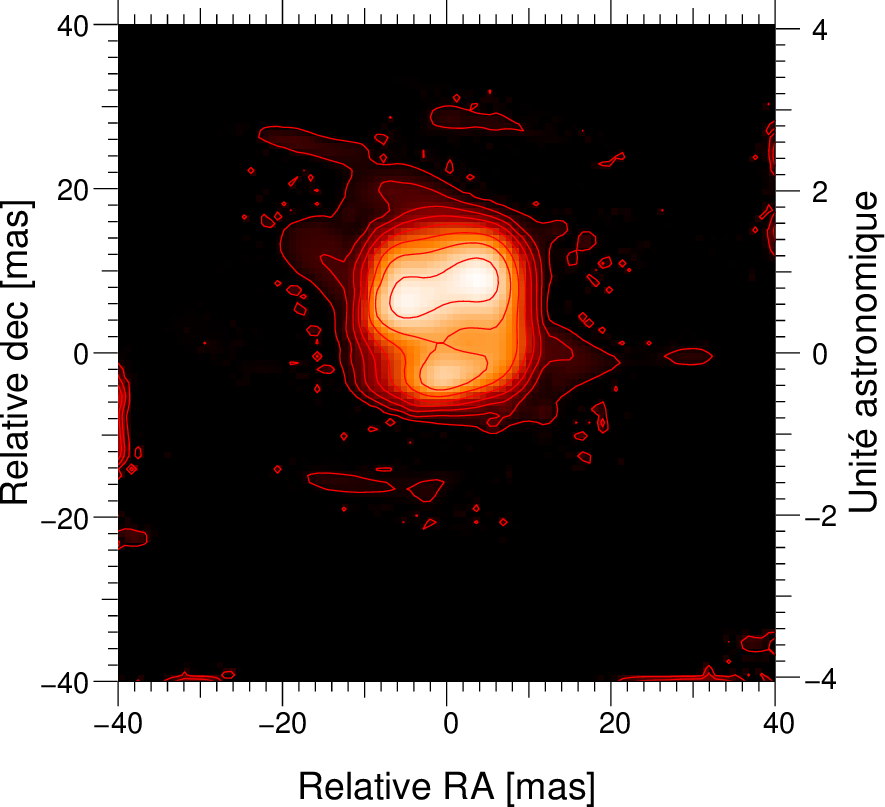}
      \caption[Plans $u$-$v$ et reconstructions en aveugle de R Leo]{ 
Plan $u$-$v$ et images de R Leo obtenues au cours des observations de
mars ($\phi = 0,92$, en haut) et mai 2005 ($\phi = 0,05$, en bas).}
         \label{fig:Rleo_rec}
   \end{figure}

R Leo est une étoile Mira des plus brillantes. Elle a été découverte
par J. A. Koch en 1782, après Mira (omicron Ceti), $\chi$ Cyg et R
Hydrae. Avec une période de 312 jours, cette étoile présente des
variations photométriques de plus de 5 magnitudes dans le visible
(figure~\ref{fig:lum_Rleo}).  Sa taille angulaire et sa luminosité en
ont fait un cas d'école dans l'étude à haute résolution angulaire des
Mira. A l'instar de $\chi$ Cyg, de multiples techniques d'observations
ont été utilisées, comme l'occultation lunaire
\citep{1991A&A...249..397D}, l'interférométrie des tavelures
\citep{1977ApJ...218L..75L,2001A&A...376..518H} le masquage de pupille
\citep{2004MNRAS.349..303J} ou encore l'interférométrie longue base
\citep[][ont observé avec les interféromètres COAST,
IOTA/FLUOR, ISI,
VLTI/VINCI]{1998MNRAS.297..462B,1999A&A...345..221P,2003ApJ...589..976W,2004A&A...426..279P,2005A&A...431.1019F}.
Ces observations ont confirmé une taille angulaire variable en
fonction de la longueur d'onde et la présence d'une couche moléculaire
proche de l'étoile. Cependant, aucune asymétrie n'a encore été
observée sur cette étoile, considérée jusqu'a maintenant comme étant à
symétrie circulaire.

Nous avons observé cette étoile lors de deux missions successives en
mars et mai 2006. Pendant ce laps de temps très court (1 mois et
demi), la morphologie de cet objet a varié considérablement, à la fois
en terme de taille et de structure
(figure~\ref{fig:Rleo_rec}). Cependant, une ressemblance a pu être
détectée dans l'importante dissymétrie que constitue les deux taches
au Nord et au Sud. Cette asymétrie est d'ailleurs clairement visible
sur les clôtures de phase (figures~\ref{fig:Donnees_6}
et~\ref{fig:Donnees_6}). L'étoile montre, de plus, un flux important
extérieur à la photosphère, que l'on ne trouvait pas sur $\chi$ Cyg et
encore moins sur Alpha Boo. La question reste ouverte sur l'origine de
ce flux. On pourrait considérer que la couche moléculaire observée par
spectro-interfemétrie \citep{2004A&A...426..279P} est la source de ce
flux.

\subsection{Mira}

   \begin{figure}[h!] \centering
   \includegraphics[width=6.2cm]{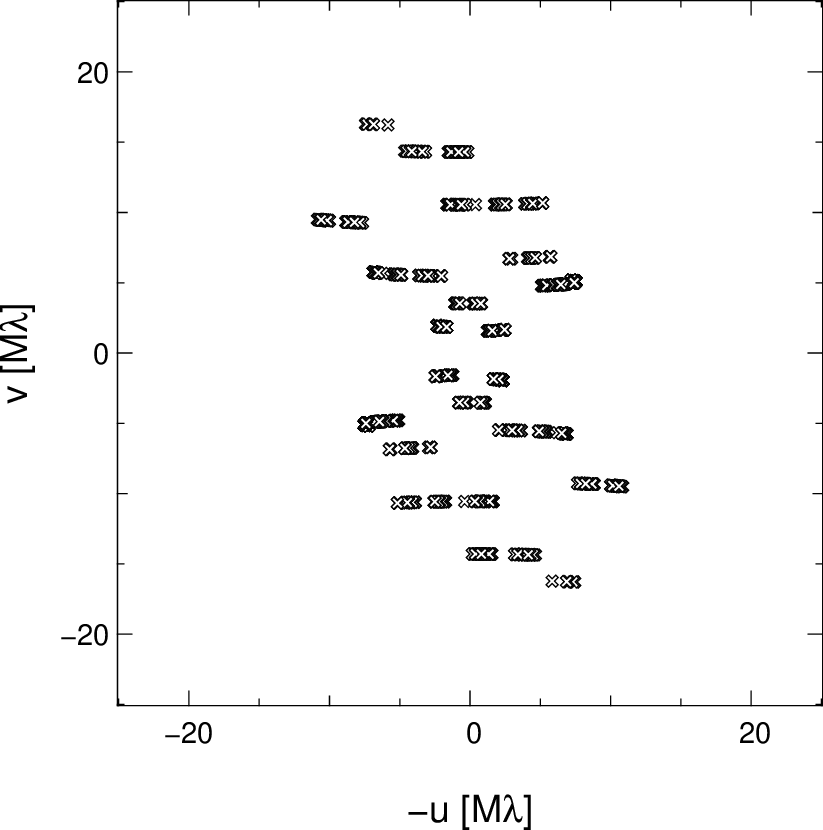}
   \includegraphics[width=7cm]{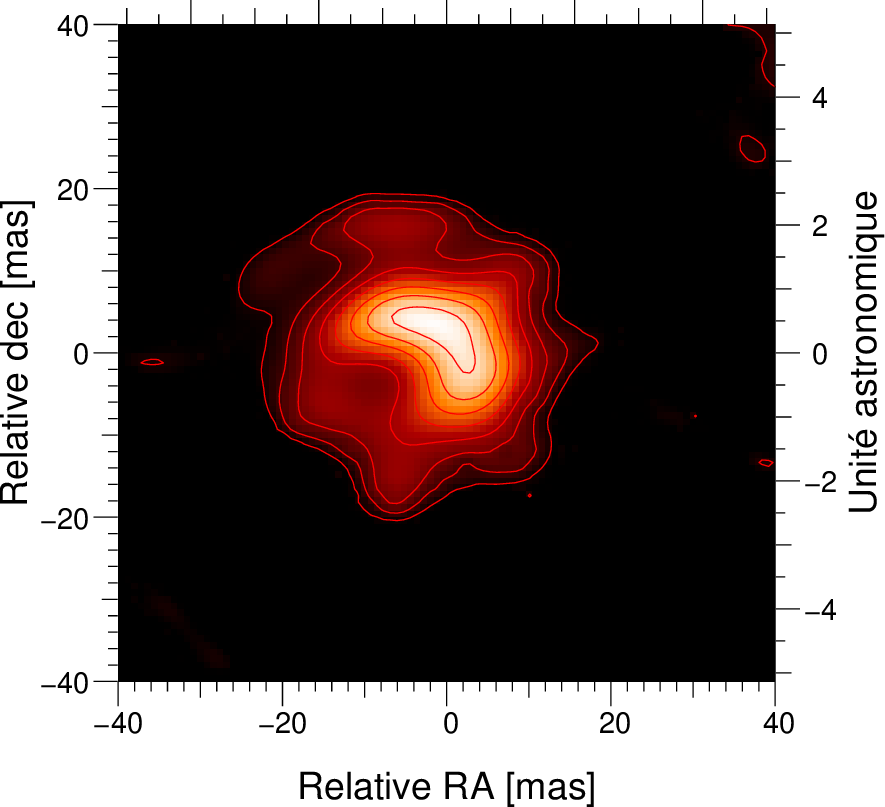}
\caption[Plan $u$-$v$ et reconstruction en aveugle de Mira]{
   Couverture fréquentielle et image de omicron Ceti obtenue via le
   logiciel de reconstruction en aveugle. La structure de l'étoile est
   profondément différente de celle des autres Miras, avec, notamment,
   la présence de deux composantes, comme si une atmosphère étendue de
   gaz chaud entourait une photosphère fortement asymétrique.  }
   \label{fig:Mira_rec}
\end{figure}

   \begin{figure}[h!]  \centering \resizebox{\hsize}{!}{
   \includegraphics{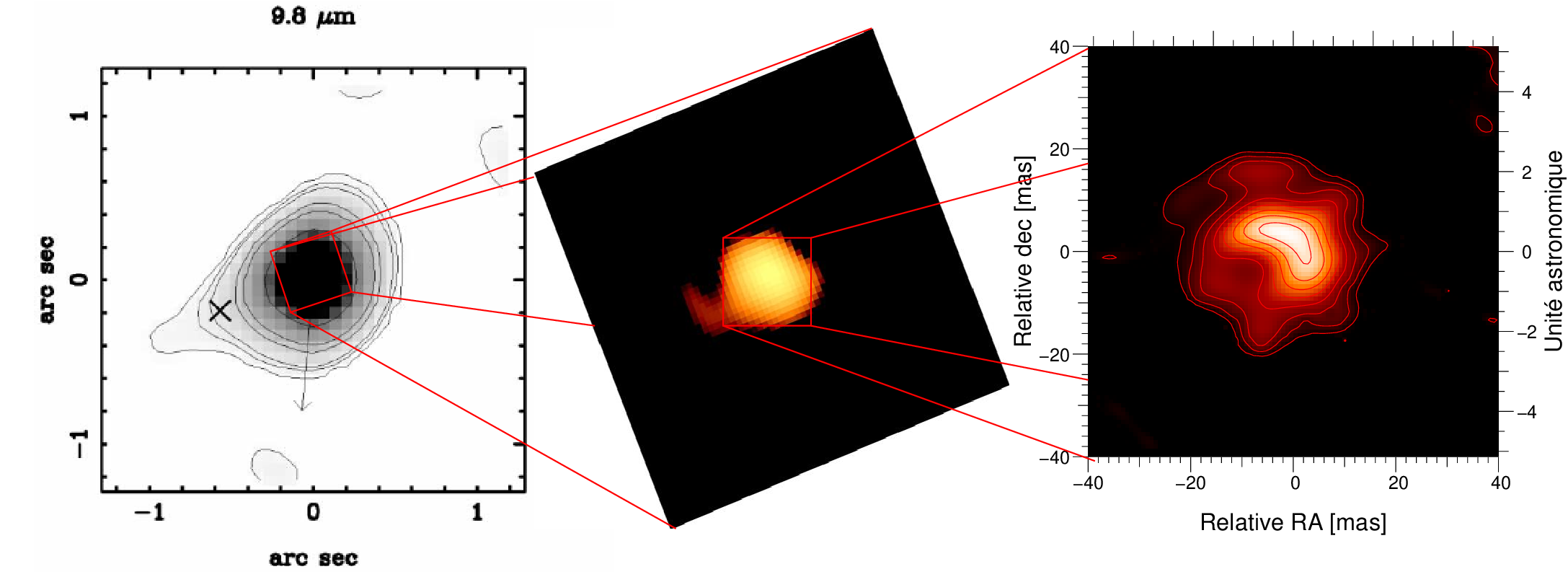}} \caption[Comparaison des
   images du sytème Mira en infrarouge moyen, UV, et infrarouge
   proche.]{ A gauche : image en infrarouge moyen du système Mira AB
   obtenue à partir du télescope IRTF \citep{2001ApJ...556L..47M}. La
   croix indique la position de Mira B telle qu'observée par Hubble en
   1997. Au centre, Mira A observée par Hubble dans l'ultraviolet
   (Communiqué de presse STScIPRC97-26, 1997). Cette image montre une
   extension de l'atmosphère en direction du compagnon. A droite,
   image telle que reconstruite à partir de nos données en bande H.}
   \label{fig:IRTF} \end{figure}

En 1596, David Fabricus a utilisé omicron Ceti comme référence de la
position de Mercure. Il a ainsi découvert la première étoile variable
de longue période. En 1642, omicron ceti a reçu son nom commun, Mira,
soit ``la merveilleuse''. Elle a, par la suite, donné son nom à
l'ensemble des étoiles ayant le même type de pulsation, les variables
Mira.

\begin{figure} \centering
   \includegraphics[width=11.2cm]{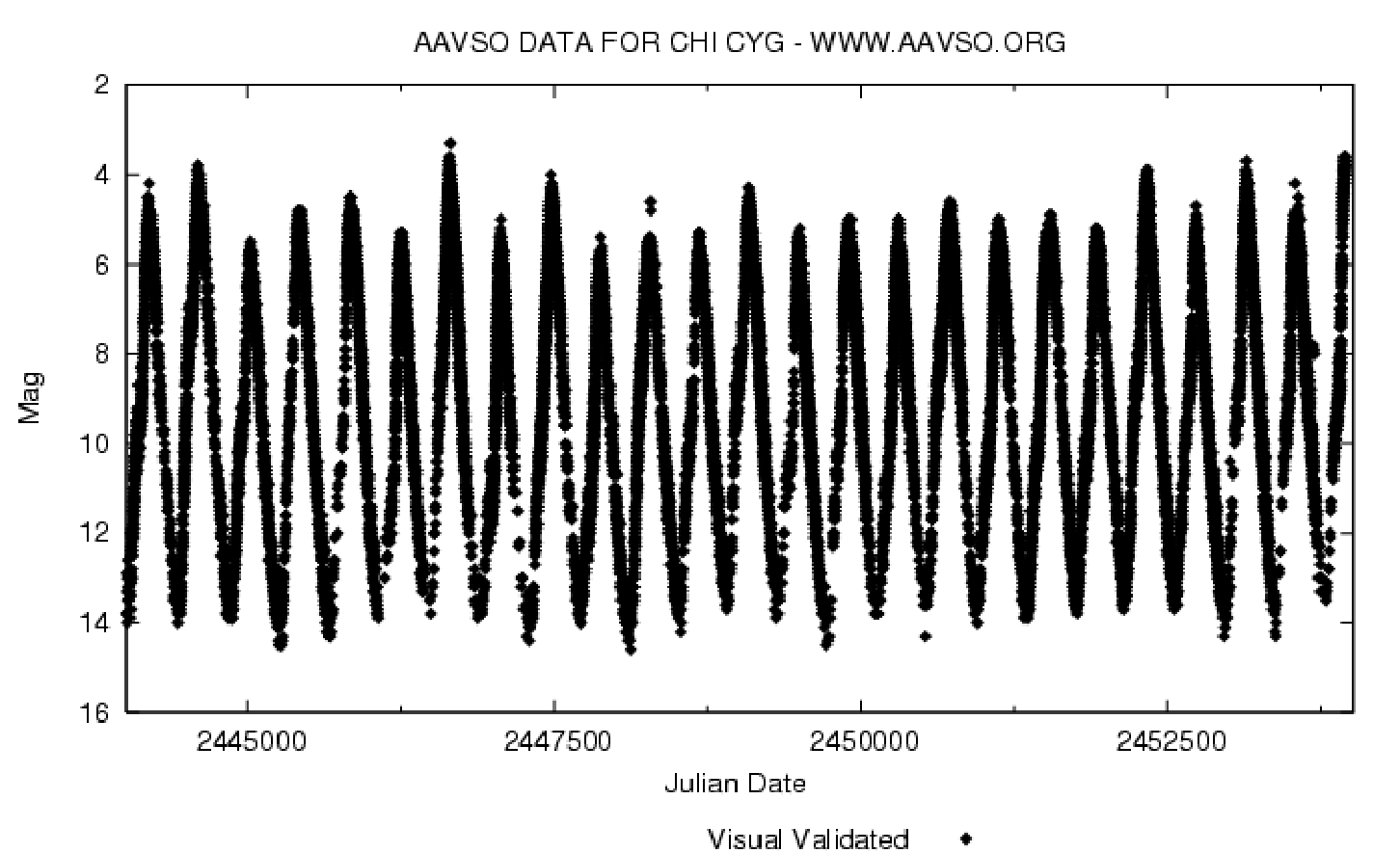}
   \includegraphics[width=11.2cm]{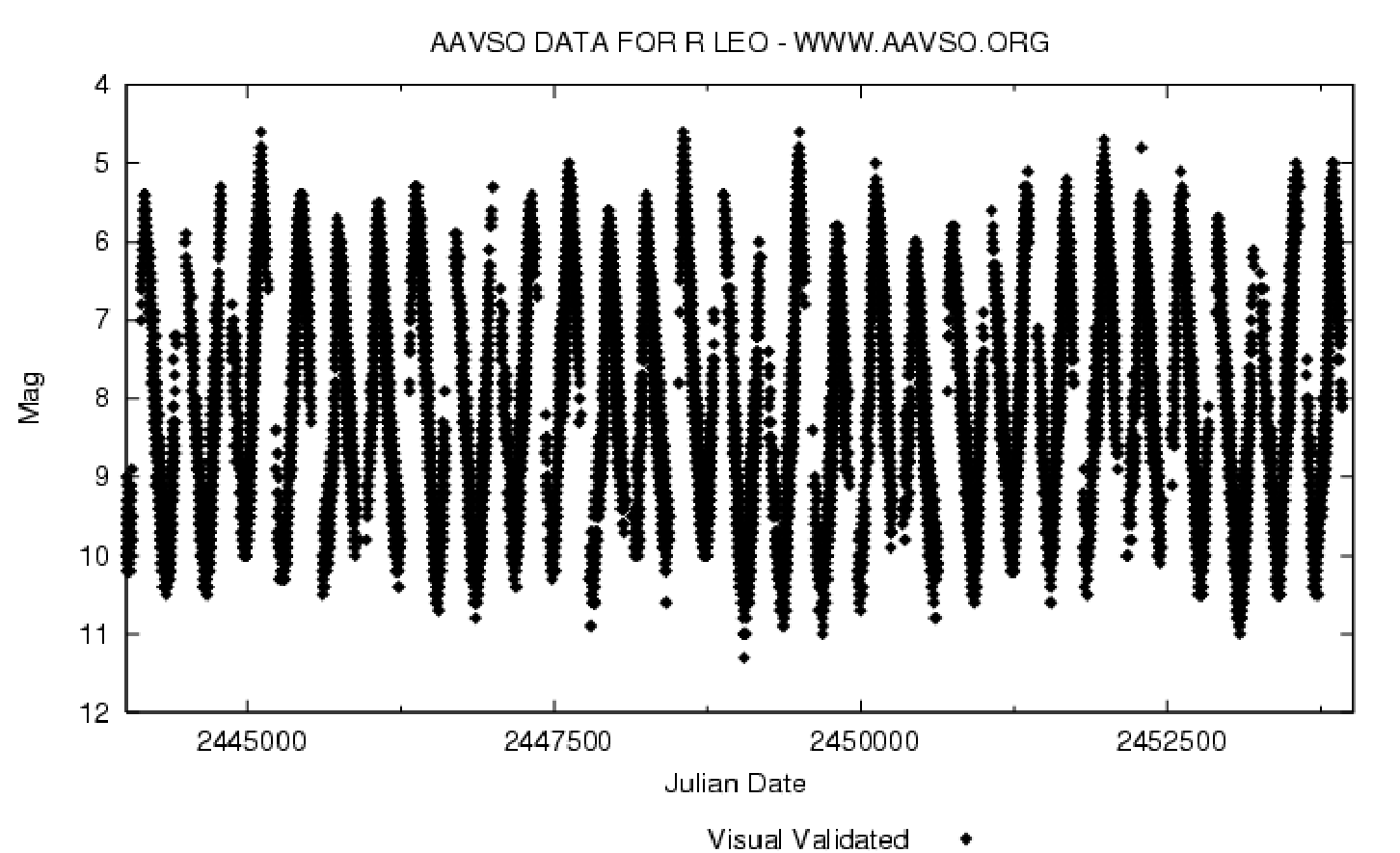}
   \includegraphics[width=11.2cm]{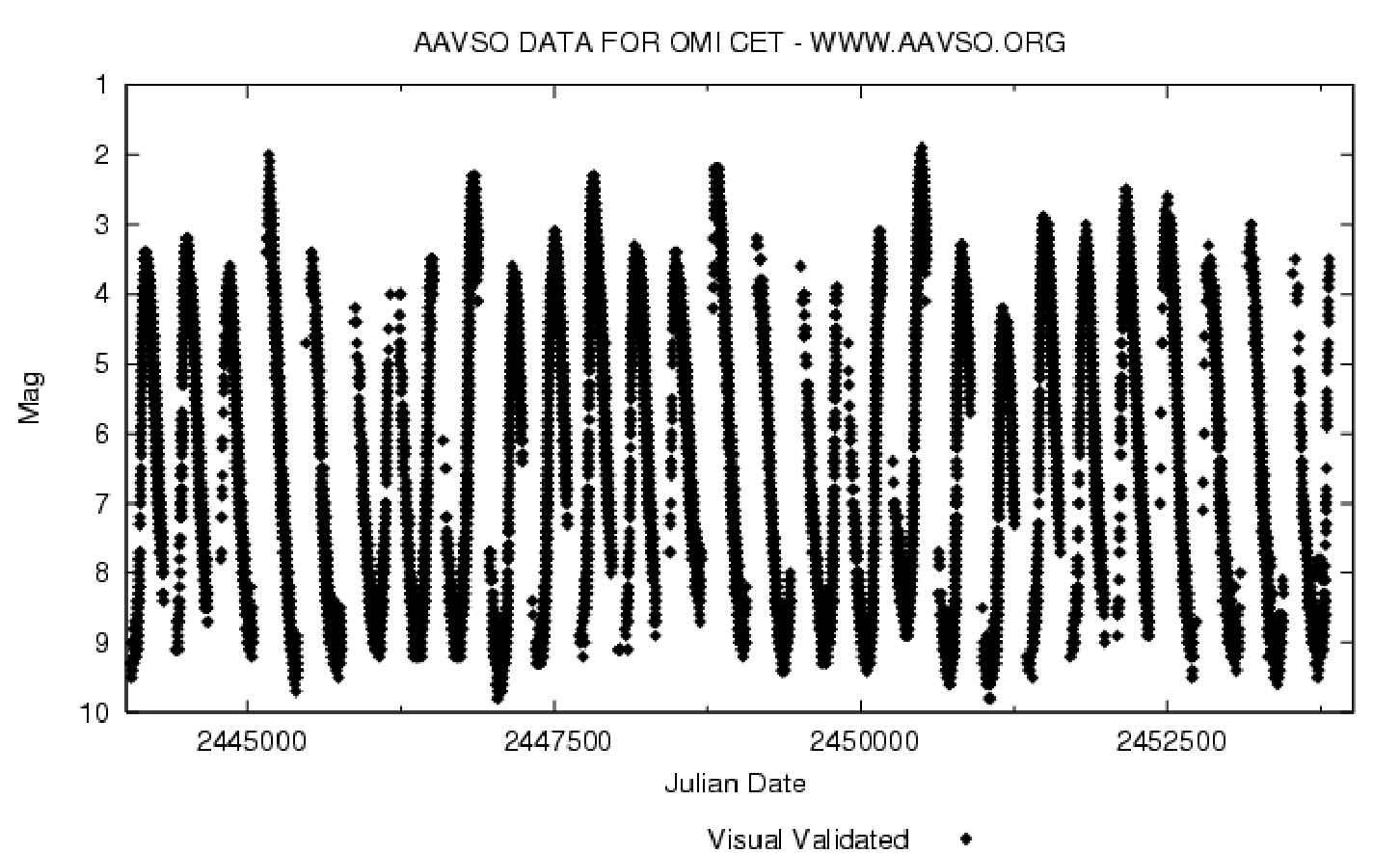} \caption[Courbes de
   lumière correspondant aux étoiles Miras : $\chi$ Cyg, R Leo, et
   Mira (o Ceti)]{Courbes de
   lumière correspondant aux étoiles Miras : $\chi$ Cyg, R Leo, et
   Mira (o Ceti). Les pulsation de ces étoiles variables sont
   clairement visibles avec des périodes respectives de 408, 312 et
   332 jours. Les modulations à plus grandes périodes sont
   interprétées comme des variations de l'opacité des couches
   moléculaires. Ces courbes sont extraites du site de l'American
   Association of Variable Star Observers.}  
   \label{fig:lum_Mira} \label{fig:lum_Rleo} \label{fig:lum_cyg}
   \end{figure}

Mira se caractérise par les types spectraux M2-M7 III, et une période
de pulsation de 332 jours (figure~\ref{fig:lum_Mira}). Elle a
récemment été sur le devant de la scène scientifique suite à
l'imagerie par le satellite Hubble d'un compagnon proche
\citep{1997ApJ...482L.175K}. \`A une distance du soleil d'environ 130
parsecs, Mira AB est, ainsi, le complexe symbiotique le plus proche que l'on
connaisse. Les étoiles sont séparées d'environ 600 mas, orbitant avec
une période d'environ 500 ans \citep{2002ApJS..139..249P}. L'étoile
secondaire serait une naine blanche
\citep{1954ApJS....1...39J,2005ApJ...623L.137K} accrétant l'atmosphère
de l'étoile principale Mira A. De multiples observations dans
l'ultraviolet (Hubble et FUSE) ont permis de calculer un taux
d'accrétion de $2,5 \times 10^{-12}$ M$_\odot$ par an
\citep{2006ApJ...649..410W}. Ce taux est très faible comparé à la
perte de masse de l'étoile principale, estimée à environ $10^{-7}$
M$_\odot$/ans \citep{1988ApJ...332..299B}.

Nous avons observé Mira A durant 5 nuits au cours du mois d'octobre
2005. Sa déclinaison étant proche de 0 degré, la synthèse d'ouverture
due à la rotation de la Terre nous a permis d'obtenir une couverture
plus homogène des fréquences spatiales. On note une structure de
l'atmosphère profondément différente des trois étoiles précédemment
observées. Deux composantes sont clairement visibles sur l'image
reconstruite figure~\ref{fig:Mira_rec}.

 Une première, d'environ 70\% du flux, est centrale et allongée dans
la direction Nord-Est. Cette composante a la propriété d'être
perpendiculaire à la direction du compagnon, et d'être dotée d'une
structure en forme de virgule qui n'est pas sans rappeler les spirales
observées dans les systèmes binaires comportant une étoile de type
Wolf-Rayet
\citep{1999Natur.398..487T}. Même si la physique de notre objet est
sensiblement différente, cette dissymétrie pourrait être expliquée par
la présence d'un troisième compagnon. Cependant, parce que nous ne
disposons pas d'observations à différentes époques, nous ne pouvons
pas exclure que l'asymétrie observée ne soit pas simplement la
conséquence de fortes variations d'opacité présentes dans les couches
supérieures de l'atmosphère, où même d'un phénomène de marée introduit
par la présence de Mira B. Pourtant un faisceau de présomption vient
conforter l'idée d'un système triple :
\begin{enumerate}
\item L'hypothèse d'un troisième compagnon proche n'est pas
nouvelle. \citet{1980A&AS...39...83B}, puis
\citet{1993ApJ...402..311K} ont étudié l'évolution du système Mira AB
par interférométrie des tavelures, et ils ont constaté la présence de
perturbations dans l'orbite de Mira B avec une période de 10-14 ans.
\item Une seconde indication vient de la présence d'un flux bipolaire 
observé dans le domaine radio à partir des vitesses radiales observées à
partir des raies de CO et de KI \citep{2000A&A...362..255J}. Ce flux
de matière, de faible vélocité, serait généré par la pression de
radiation et collimaté par la présence d'un disque équatorial
circumstellaire.
\item Une troisième indication porte sur la masse totale du système
Mira AB. \citet{2002ApJS..139..249P}, en utilisant des données acquises
par interférométrie des tavelures avec l'instrument PISCO du Pic du
Midi, ont appliqué la troisième loi de Kepler et ont estimé la
masse totale du système à 4,4 M$_\odot$. Si l'on considère, pour Mira B,
la masse typique d'une naine blanche de 0,6 M$_\odot$
\citep{1990ARA&A..28..103W}, cela signifie que Mira A à une masse de
3,8 M$_\odot$. Or, nos données nous donnent un diamétre angulaire de la
photosphère d'environ 25 mas, soit un rayon de 340 R$_\odot$ (en
utilisant la parallaxe hypparcos de 7,79 mas). A partir de ces
élements, si on utilise la
relation période-masse-rayon de \citet{Wood..89} :
\begin{equation}
\log (P) = -2.07 +1.94 \log\left(\frac{R}{R_\odot}\right) - 0.9 \log\left(\frac{M}{M_\odot}\right) \,,
\end{equation}
on obtient une masse aux alentours de 2,3 M$_\odot$. Selon ce schema,
il existerait donc une masse manquante, d'environ 1,5 M$_\odot$.
\item Enfin, un dernier argument repose sur la récente détection par 
Chandra d'émission en rayon X venant de l'étoile Mira A
\citep{2005ApJ...623L.137K}. Une telle émission est communément
produite par les disques d'accrétion autour des naines blanches. Cela
n'aurait donc pas été une surprise si elle avait été observée en
provenance de Mira B. Cependant, les étoiles de la branche
asymptotique de géante ne sont pas censées générer une telle émission,
qui n'a d'ailleurs jamais été observée précédemment. Une hypothèse
proposée par \citet{2005ApJ...623L.137K} serait la recombinaison de
champs magnétiques, suivit d'une éjection massive de matière. La
présence d'un troisième compagnon serait également une possibilité.
\end{enumerate}

La seconde composante de l'image correspond à une atmosphère, peut
être moléculaire, étendue sur deux fois la taille de la composante
centrale. Elle est excentrée par rapport à la première dans la
direction de l'étoile Mira B (Sud-Est). C'est pourquoi il semble que
cette matière soit sous l'effet du champ gravitationnel de Mira
B. Cependant, si l'on considère la répartition des masses entre Mira A
et B proposé par
\citet{2002ApJS..139..249P}, ainsi que la distance de 580 mas les
séparant, il semble difficile d'expliquer la présence de cette matière
par le seule effet de marée introduit par l'attraction
gravitationnelle de Mira B. Il serait alors possible que ce phénomène
résulte d'une conjonction de circonstances, dont la pulsation de
l'étoile serait un événement moteur.

Enfin, il est ainsi intéressant de remarquer que l'asymétrie générale
de l'étoile se retrouve aussi sur diverses observations, notamment en
infrarouge moyen avec le télescope IRTF, et dans l'ultraviolet (346
nm) avec le télescope spatial Hubble (figure~\ref{fig:IRTF}).

\subsection{Bételgeuse}

   \begin{figure}[h!]  \centering
   \includegraphics[width=6.2cm]{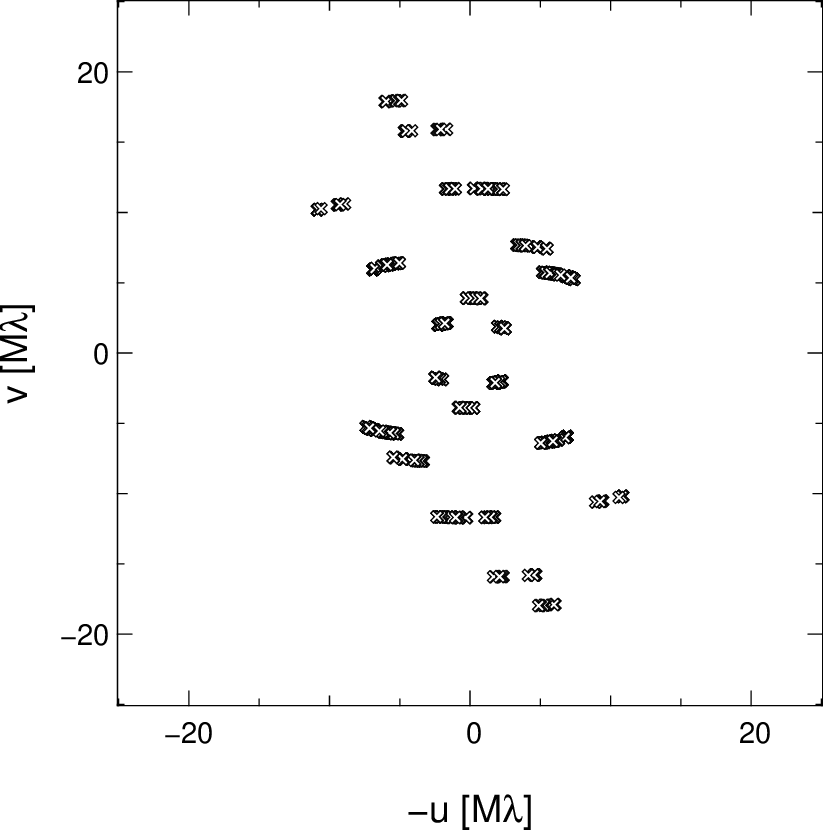}
   \includegraphics[width=7cm]{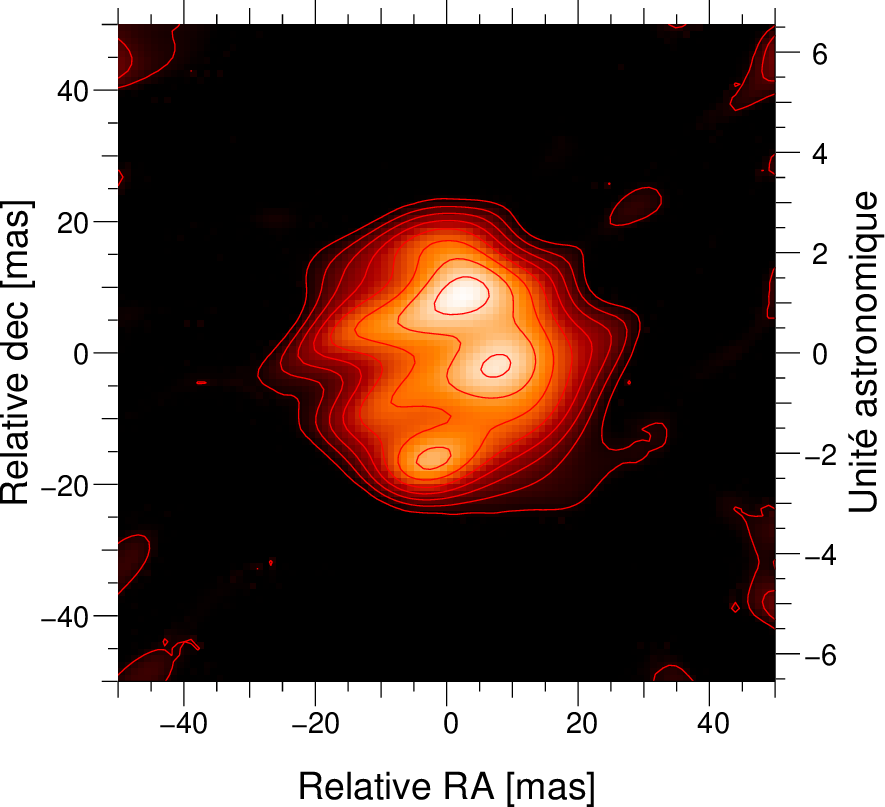}
   \caption[Plan $u$-$v$ et reconstruction en aveugle de Bételgeuse]{
   Plan $u$-$v$ et imagerie en aveugle de Bételgeuse en octobre
   2006. On peut distinguer une surface inhomogène pouvant faire
   penser aux surfaces convectives simulées par Freytag
   (figure~\ref{fig:Freytag}). } \label{fig:Betel} \end{figure}

   \begin{figure}[h!]
   \centering
   \includegraphics[width=10cm]{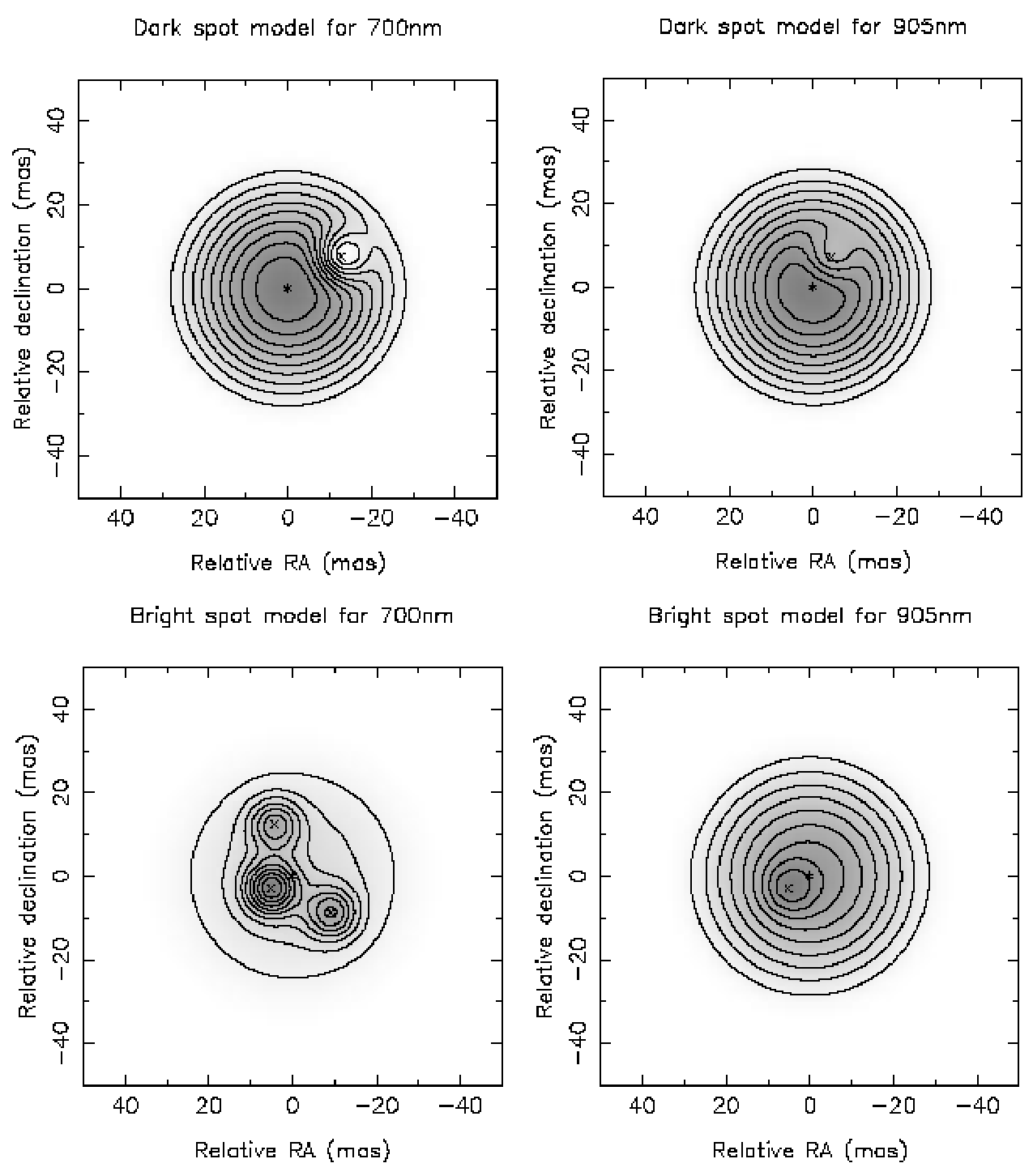}
      \caption{ Imagerie paramétrique de Bételgeuse par \citet{2000MNRAS.315..635Y}. }
         \label{fig:Betel_Y}
   \end{figure}

La constellation d'Orion est l'une des plus connues, visible de
l'hémisphère Nord comme de l'hémisphère Sud. Les deux étoiles les plus
brillantes, Rigel et Bételgeuse occupent respectivement les coins
Sud-Ouest et Nord-Est. Les premières mesures de variations de
luminosité de Bételgeuse ont été effectuées en 1836 par John
Hershel. Dans le visible, elle oscille entre des magnitudes allant de
0,2 à 1,2 (figure~\ref{fig:lum_betel}). \`A cette longueur d'onde, il
s'agit de la septième étoile la plus brillante de l'hémisphère
Nord. Cependant, comme toutes les géantes rouges, cette étoile émet
principalement dans le domaine de l'infrarouge, ce qui,
bolométriquement, en fait l'étoile la plus brillante du ciel. Ceci est
dû à sa distance, mais surtout à sa taille, d'environ 630 fois
le diamètre de notre soleil \citep[ont mesuré par interférométrie
$\theta_{\rm UD} \approx$ 43,3 mas]{2004A&A...418..675P}.

C'est en 1920 qu'a eu lieu la première mesure de diamètre
stellaire. Elle a été effectuée sur Bételgeuse par Michelson, qui a
estimé son diamètre à environ 44 mas. Parce que cela en fait l'étoile
angulairement la plus grosse du ciel -- après le soleil -- la plupart
des instruments d'interférométrie stellaire ont étudié cet
objet. Utilisant la technique d'interférométrie des tavelures,
Francois et Claude Roddier ont été les premiers à reconstruire une
image de sa surface stellaire \citep{1985ApJ...295L..21R}. Ils en ont
déduit la présence de poussières assombrissant le disque, créant une
importante asymétrie. Diverses observations de Bételgeuse s'en sont
suivies
\citep{1997MNRAS.291..819W,2000MNRAS.315..635Y}. Toutes ont confirmé
l'existence d'asymétries et, notamment, de taches, parfois sombres ou
brillantes, sur la surface. \`A titre d'exemple, la
figure~\ref{fig:Betel_Y} montre les reconstructions auxquelles 
\citet{2000MNRAS.315..635Y} ont abouti. Il s'agit dans ce cas
d'imagerie paramétrique, c'est-à-dire qu'ils ont ajusté une
photosphère tachetée à leurs données.

Contrairement à eux, la précision que nous avons sur les visibilités,
ainsi que l'étendue de la couverture fréquentielle, nous ont permis une
reconstruction en aveugle de la surface stellaire
(figure~\ref{fig:Betel}). Nous pouvons comparer ce résultat aux
travaux de \citet{2007..Xavier} qu'ils ont effectués par ajustement
paramétrique de ces mêmes données. Ils en ont déduit qu'un modèle
d'atmosphère tacheté permettait de reproduire de manière satisfaisante
les données jusqu'au troisième lobe de la fonction de
visibilité. Cependant, leur modèle ne permet plus de reproduire
précisément les données à plus hautes fréquences spatiales. Ceci est, par
conséquent, en accord général avec notre reconstruction, qui indique
une grande complexité de l'objet aux hautes fréquences spatiales. Ce
résultat corrobore l'hypothèse de
\citet{2003csss...12.1024F} concernant l'hypothèse d'une atmosphère extrêmement
convective. Les variations temporelles de ces inhomogénéités seraient
une information précieuse à recueillir dans l'avenir.

\subsection{$\mu$ Cep}

   \begin{figure}[h!] \centering
   \includegraphics[width=6.2cm]{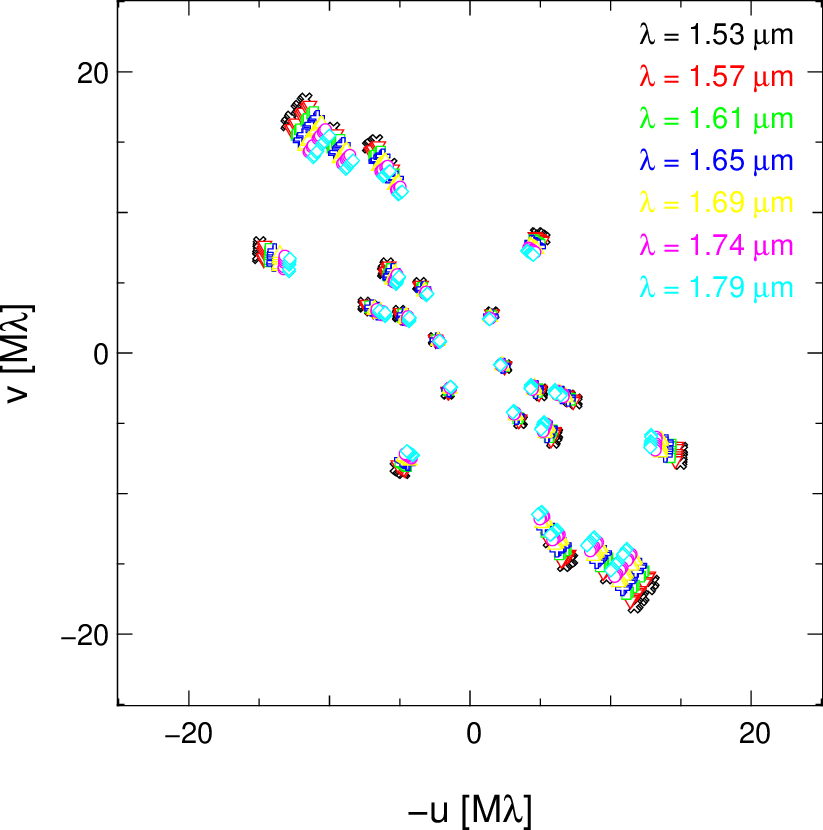}
   \includegraphics[width=7cm]{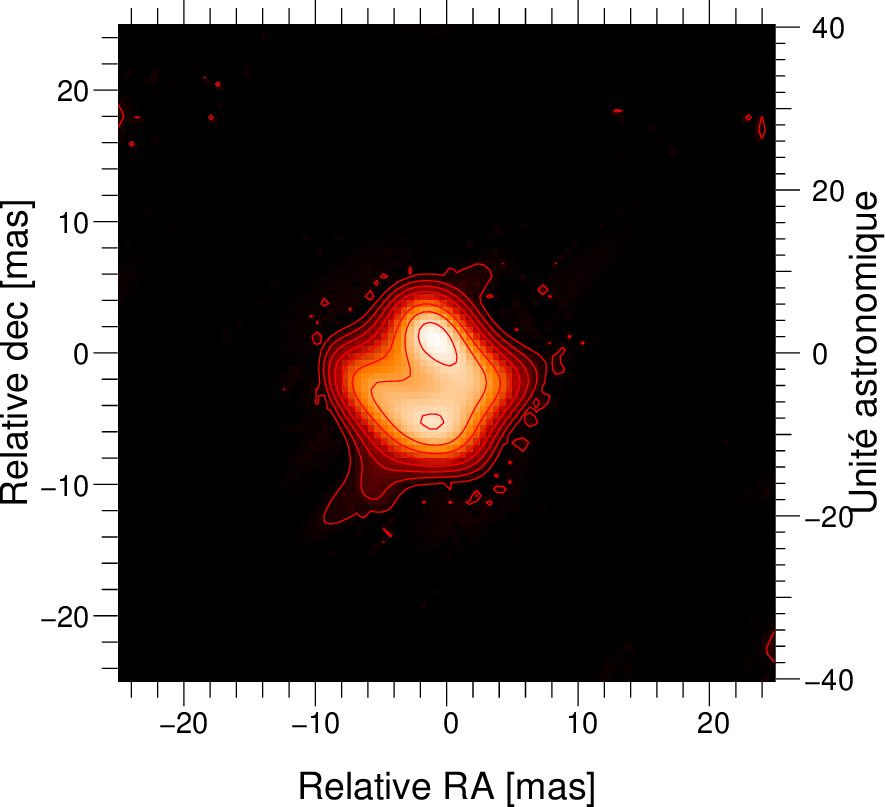} \caption{
   Plan $u$-$v$ et imagerie en aveugle de $\mu$ Cep.}
   \label{fig:mu_cep_rec} \end{figure}

La variabilité de $\mu$ Cep a été découverte en 1848 par John
Russel. Il s'agit de l'une des étoiles les plus grosses du ciel après VV
Cephei et Epsilon Aurigae. $\mu$ Cep a de nombreux points communs avec
Bételgeuse. Il s'agit d'une supergeante semi-réguliaire (de type SRc)
avec des amplitudes de variations photométriques dans le visible
d'environ 1.5 magnitude (figure~\ref{fig:lum_mucep}). Son type
spectral est M2eIa, avec des périodes de pulsation de 730 et 4400
jours \citet{2006MNRAS.Kiss}.

Nous avons pu observer cet objet courant mai 2006 en utilisant le mode
dispersé d'IOTA (figure~\ref{fig:mu_cep_rec}). Outre sa taille
remarquable, il est intéressant de noter que la reconstruction nous
donne un objet très uniforme, loin de l'image obtenue de
Bételgeuse. Ceci peut être tout simplement la conséquence d'un manque
de résolution. Cependant, on voit clairement une asymétrie dans la
structure de l'étoile.  Il est également à noter que la couche moléculaire,
telle qu'observée par \citet{2005A&A...436..317P}, n'est pas visible
dans cette image reconstruite.  Selon leurs observations, cette couche
devrait se trouver à environ un demi rayon stellaire de la
photosphère. Le flux de cette couche doit donc être inférieur à la
dynamique obtenue sur cette image ($\approx 3 \%$).  Nous verrons,
néanmoins, dans le paragraphe~\ref{sec:dispertion} que la couche
moléculaire est présente et que l'imagerie paramétrique
permet de la révéler dans nos données.

\subsection{CH Cyg}

   \begin{figure}[h!]
   \centering
   \includegraphics[width=6.2cm]{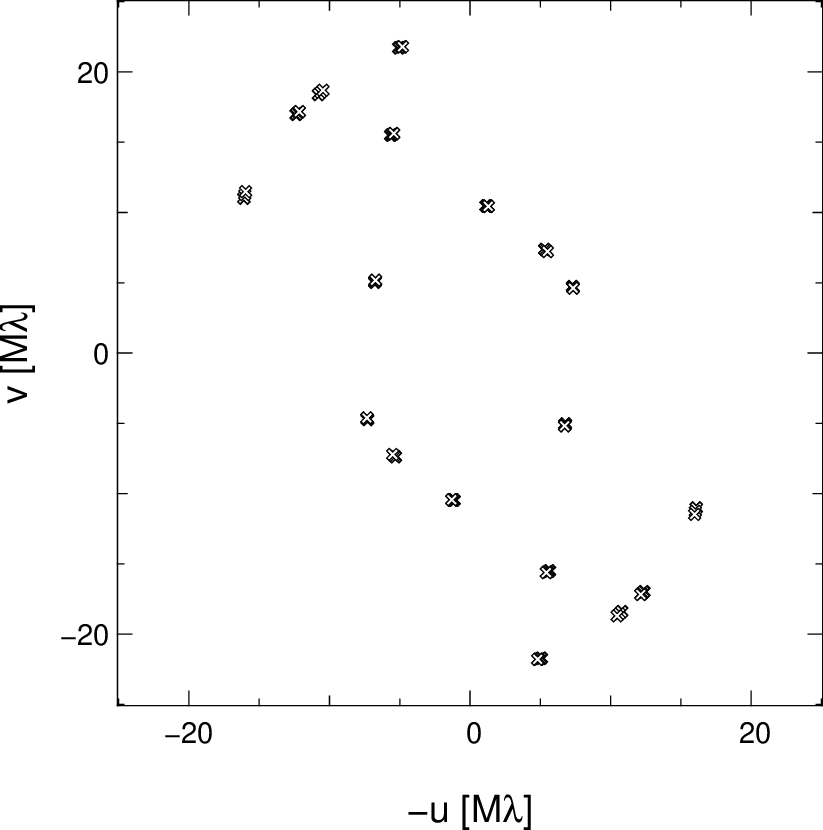}
   \includegraphics[width=7cm]{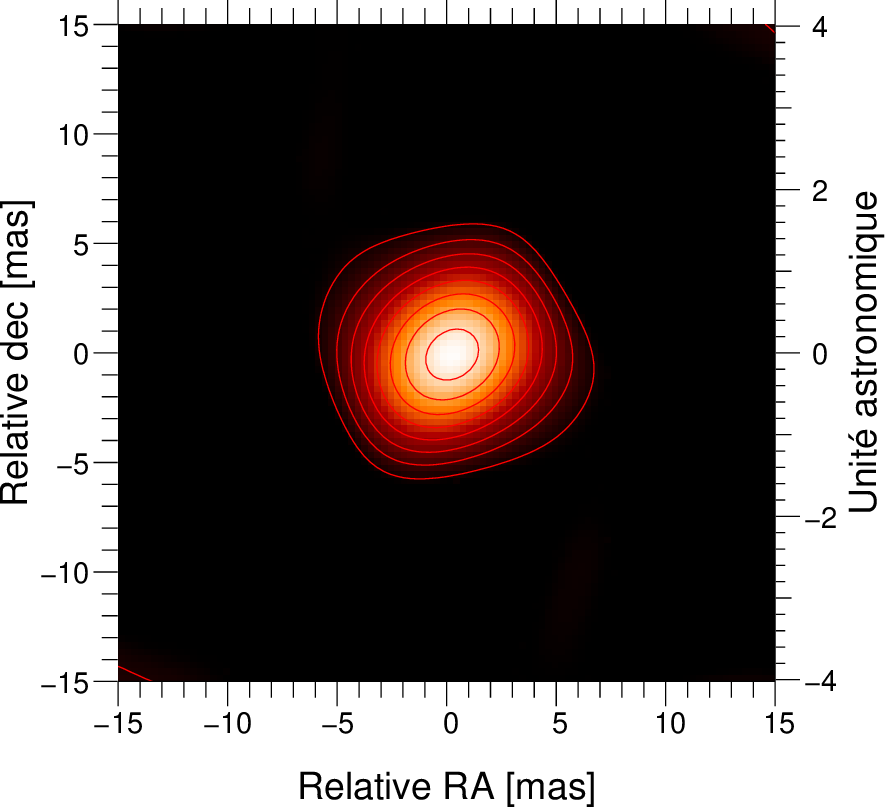}
      \caption{ Plan $u$-$v$ et imagerie en aveugle de CH Cyg.}
         \label{fig:ch_cyg}
   \end{figure}

A l'instar de Mira, CH Cyg est une étoile variable
symbiotique. L'aspect particulier de son spectre a longtemps intrigué
les chercheurs. En effet, à chaque type d'étoile correspond une lettre
(de A à M) distribuée en fonction de l'importance des raies
d'hydrogène et d'hélium. Toujours utilisée, cette classification a été
ré-arrangée afin de classer les étoiles par ordre croissant de
température (OBAFGKM). Cependant, certaines étoiles, dont CH Cyg, ne
trouvent pas de place dans cette classification. CH Cyg possède, en
effet, une température proche de celle du type M, mais avec des raies
en émission typique des étoiles de type O.

\begin{figure} \centering
 \includegraphics[width=11.2cm]{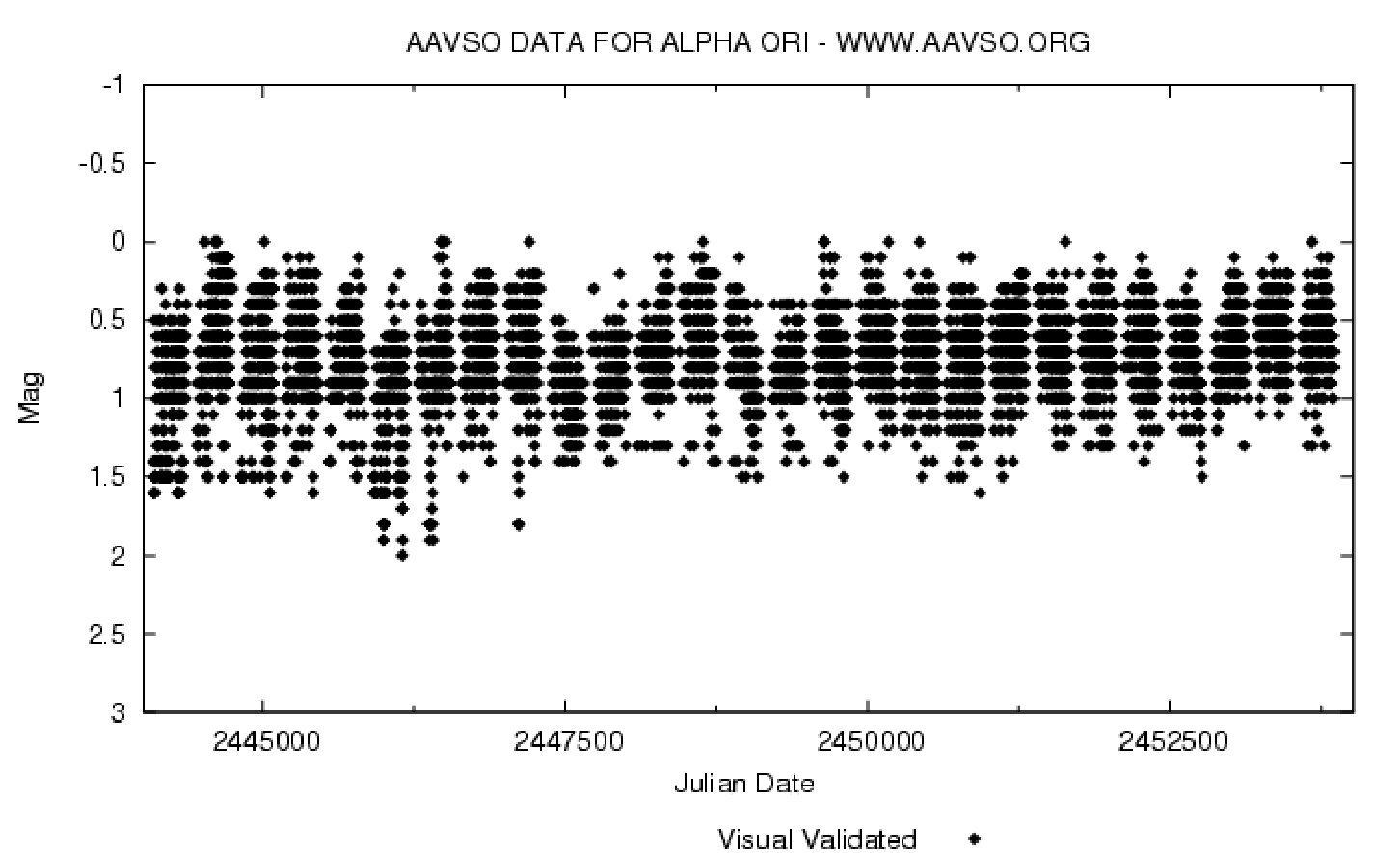}
 \includegraphics[width=11.2cm]{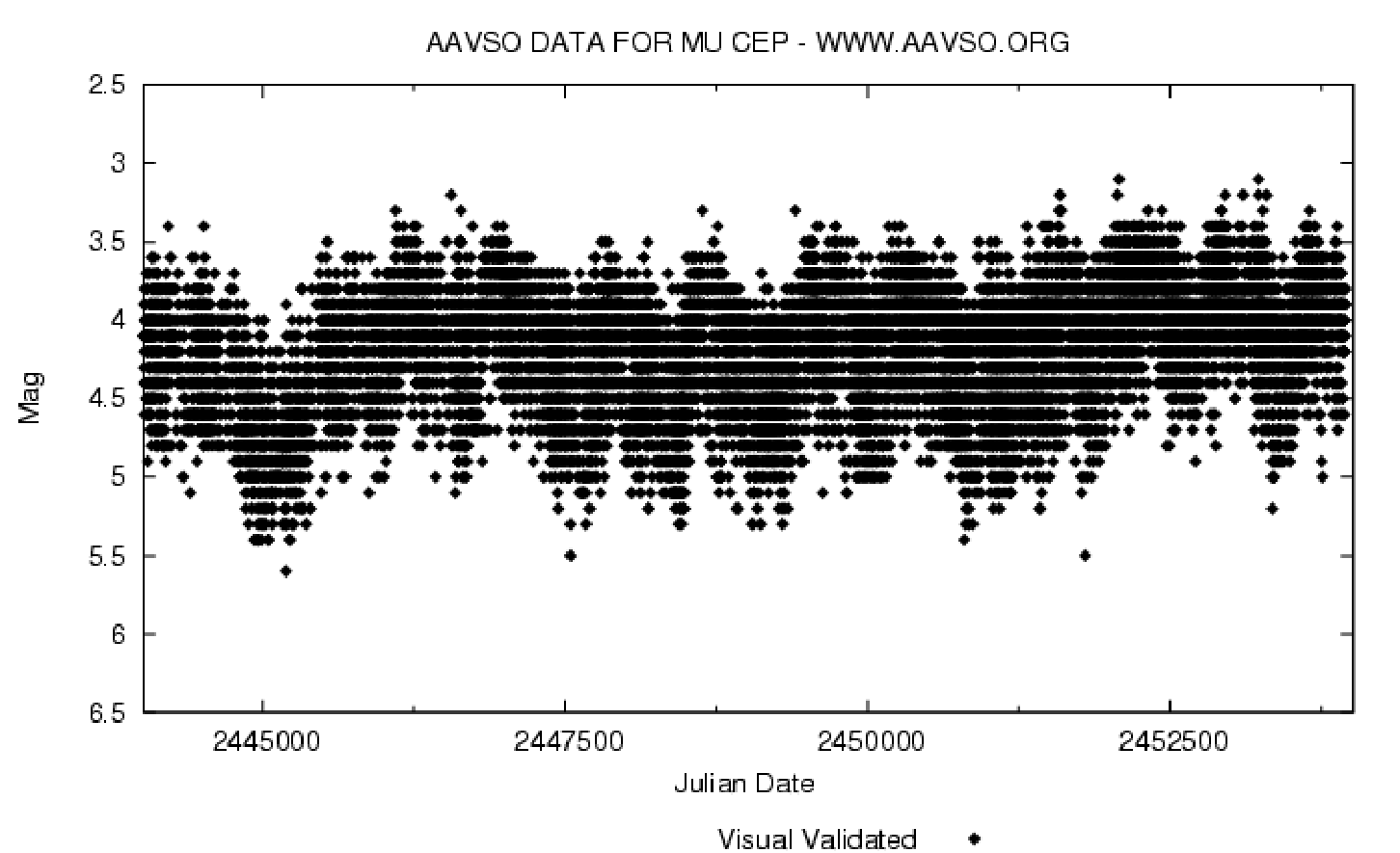}
 \includegraphics[width=11.2cm]{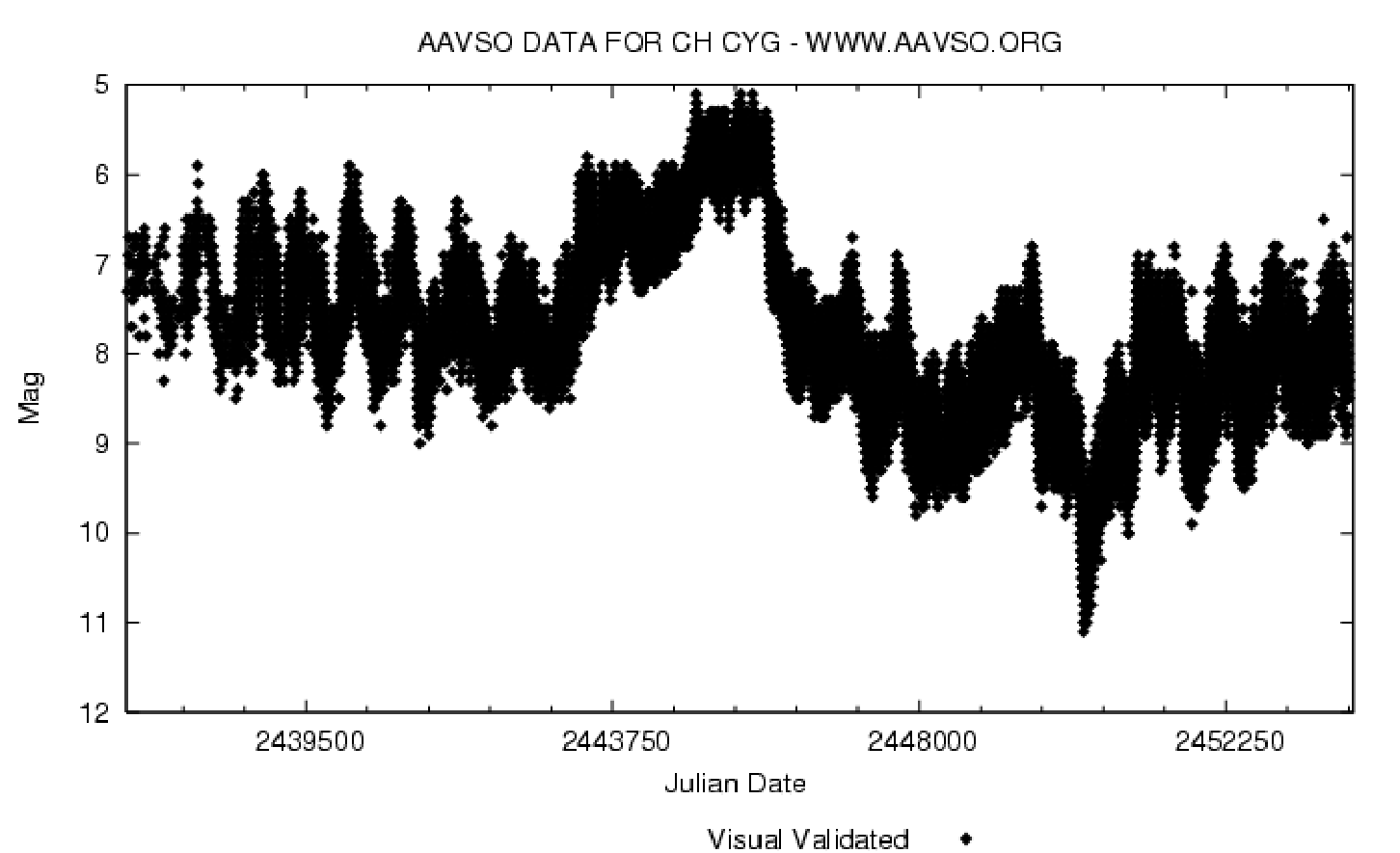} \caption[Courbes de
 lumière de Bételgeuse, $\mu$ Cep, et CH Cyg]{Courbes de
 lumière de Bételgeuse, $\mu$ Cep, et CH Cyg, obtenues à partir du
 site de l'American Association of Variable Star Observers. La courbe
 de CH Cyg est particulièrement intéressante car elle présente à la
 fois des oscillations régulières ainsi qu'une période
 particulièrement brillante interprétée comme un sursaut dans
 l'accrétion du compagnon.} \label{fig:lum_ch} \label{fig:lum_mucep}
 \label{fig:lum_betel} \end{figure}

Nous avons retenu l'idée selon laquelle CH Cyg est constituée de deux
étoiles. La première est une géante rouge et la deuxième une naine
blanche accrétant l'atmosphère de la première
\citep{1990A&A...235..219M}. Cette hypothèse est, par ailleurs,
corroborée par la courbe de lumière de l'étoile qui présente à la fois
des oscillations périodiques et d'importantes éruptions (comme celle
observée en 1986 ; cf figure~\ref{fig:lum_ch}). L'existence d'un compagnon
accrétant de la matière est confirmée par la présence de jets
observables dans le domaine de longueur d'onde radio
\citep{2001MNRAS.326..781C} et X
\citep{2004ApJ...613L..61G}. Actuellement, les astronomes s'accordent
sur la présence d'un système triple \citep{1993AJ....105.1074H}. Il
serait composé d'une naine blanche entourée d'un disque d'accrétion
orbitant avec une période de 2,07 ans autour d'une géante
semi-régulière de type M7 III. Autour de ce système en interaction
graviterait, avec une période de 14,5 ans, une troisième étoile
appartenant à la séquence principale (naine G-K).

Nous avons pu observer cette étoile durant deux nuits. La couverture
du domaine fréquentiel (figure~\ref{fig:ch_cyg}) est par
conséquent très diluée. Du fait de la petit taille angulaire de
l'étoile, et cela même en utilisant la base maximale de
l'interféromètre IOTA (38 mètres), nous n'avons pas pu atteindre le
deuxième lobe de la courbe de visibilité. Nous avons, cependant,
constaté des clôtures de phase de l'ordre de la dizaine de degrés,
démontrant la présence d'asymétries. La résolution limitée ainsi que
la faible quantité de données ne nous ont pas permis de caractériser
cette asymétrie.

\chapter{\'Etudes paramétriques}
\begin{center}
\end{center}
\minitoc \label{sec:param} \vskip1cm
\clearpage

\section{Modéliser les étoiles évoluées}

\subsection{Introduction}

Dans ce chapitre, nous présentons une technique que nous retrouvons
souvent interférométrie optique. Il s'agit d'une modélisation
paramétrique de l'étoile. Cela consiste à définir l'objet observé par
un certain nombre de paramètres (comme la taille, la brillance...) qui
conditionnent la structure de l'image. Ces paramètres sont ensuite
ajustés aux données interférométriques pour permettre de reconstruire
une image à partir des valeurs obtenues. Cette technique n'est pas
fondamentalement différente de la reconstruction d'image en aveugle
utilisée au
chapitre~\ref{sec:image}. Le principal changement réside dans le fait
que les paramètres de l'image, qui étaient précédemment aussi nombreux
que les pixels, sont maintenant réduit à un nombre plus faible
correspondant aux différentes structures possibles de l'image. Le
faible nombre de paramètres libres permet ainsi de suffisamment
contrainte l'image, ce qui permet de ne pas utiliser un {\it a priori}
contenu dans un terme de régularisation. L'{\it a priori} se trouve en
fait dans le choix du modèle.

\subsection{Le modèle géométrique}
\label{sec:model_geo}

   \begin{figure}[h]
   \centering
   \includegraphics[width=6cm]{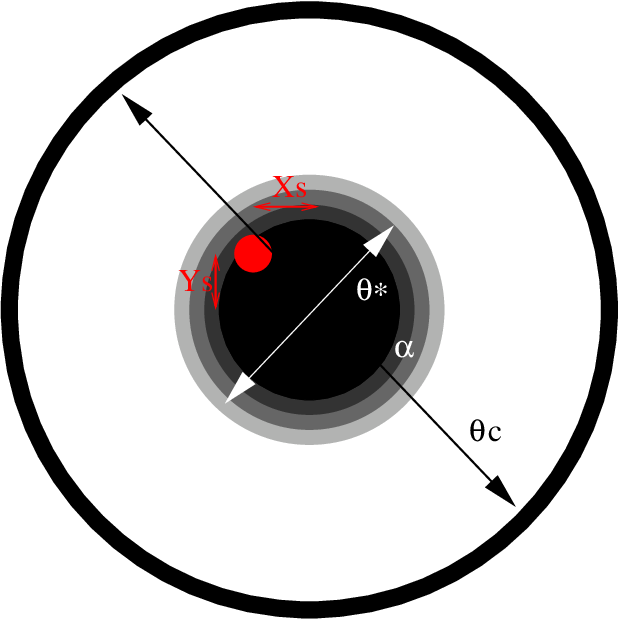}
      \caption[Modèle géométrique]{ Modèle géométrique utilisé pour l'imagerie
      paramétrique des étoiles évoluées. Sont représentés les principaux
      paramètres ($\alpha$, $\theta_\star$, $\theta_{\rm couche}$,
      $X_{\rm tache}$ et $Y_{\rm tache}$) auxquels se rajoutent la
      brillance relative de la tache et de la couche moléculaire.}
         \label{fig:model}
   \end{figure}

Le modèle va contraindre de façon stricte la reconstruction de
l'objet, son choix est en conséquence crucial. Si les paramètres
choisis ne reflètent pas la réalité, l'image obtenue n'aura pas de
sens, et pire, pourra nous induire en erreur. C'est pourquoi il est
important de sélectionner un modèle aussi proche que possible de la
physique de l'objet. Bien que la physique des étoiles évoluées soit
complexe, on peut néanmoins supposer la présence d'un certain nombre
de composantes.

Il y a, en premier lieu, la photosphère. Elle se distingue d'un simple
disque uniforme par la présence d'un assombrissement centre bord
(ACB). Celui-ci a déjà pu être observé par interférométrie sur des
étoiles ayant une température effective plus grande \citep[par exemple
sur des céphéides, cf][]{2006A&A...447..783M}. Cependant, nous ne
disposons, sur les étoiles évoluées, que de simulations d'atmosphères.
Pour les géantes \citep{1998A&A...335..637H} et les Miras
\citep{1998A&A...339..846H}, il en a été déduit que l'ACB pouvait
produire des effets très variés, mais qu'une simple loi en puissance
pouvait, cependant, parfaitement reproduire l'ensemble des cas
simulés. Cette loi s'écrit sous la forme $I(\mu)=\mu^\alpha$, où $\mu$
est le cosinus de l'angle entre la ligne de visée et la normale à la
surface de l'étoile \citep{1997A&A...327..199H}. Un tel modèle permet,
notamment, de reproduire le cas d'un disque uniforme ($\alpha=0$),
d'un disque pleinement assombri ($\alpha=1$), ou encore celui d'une
distribution d'intensité Gaussienne ($\alpha \gg 1$). La prise en
compte de cet assombrissement est fondamentale car elle influe
fortement sur la mesure du diamètre de la photosphère.

Une seconde composante provient de la présence d'une couche
moléculaire chaude, située à environ un demi rayon stellaire de la
photosphère. Suggérée par \citet{1999A&A...345..221P}, elle a été
utilisée pour la première fois pour expliquer des données
interférométriques d'étoiles Miras par \citet{2002ApJ...579..446M}. La
présence récurrente de cette couche moléculaire a ensuite pu être mise
en évidence par \citet{2004A&A...426..279P}. Sur $\chi$ Cyg notamment,
ils ont mesuré des opacités comprises entre 0,1 et 0,8, en fonction du
filtre utilisé en bande K. Parce que l'absorption de CO et H$_2$O est
bien plus faible en bande H, on s'attend à la présence de
cette couche, mais avec une opacité bien plus faible, inférieure à
0,1. Ainsi, la couche devrait être vue en émission là où l'épaisseur
géométrique est la plus grande, c'est à dire au bord de la
couche. Dans l'hypothèse d'une si faible opacité, nous avons donc
décidé de simuler la couche moléculaire par la présence d'un simple
anneau entourant l'atmosphère de l'étoile.

La troisième composante du modèle est nécessaire à l'ajustement de
clôtures de phase différentes de 0 ou 180 degrés. Il s'agit
d'introduire un terme d'asymétrie dans la brillance de la surface
stellaire. Une telle composante est souvent observée dans l'atmosphère
des étoiles évoluées \citep{2006astro.ph..7156R}. Cependant, la source
de cette asymétrie est souvent peu claire. Elle est généralement
modélisée par une ou plusieurs taches sur la surface stellaire, taches
parfois sombres ou brillantes \citep{2000MNRAS.315..635Y}. Nous avons
choisi de modéliser cette asymétrie par une composante possédant un
minimum de paramètres. Il s'agit d'une tache ponctuelle, de flux
positif ou négatif, décentrée par rapport au centre de la
photosphère. Le choix de ce modèle, simple, est justifié dans le
paragraphe~\ref{sec:clot_asym}.

La figure~\ref{fig:model} représente l'étoile et explicite les
différents paramètres que nous avons ajusté aux données. Les
variables sont : le diamètre de la photosphère ($\theta_\star$), le
coefficient d'ACB ($\alpha$), le diamètre de la couche moléculaire
($\theta_{\rm layer}$), le flux relatif de la couche moléculaire
($F_{\rm couche}/F_{\rm total}$), le flux relatif de la tache ($F_{\rm
tache}/F_{\rm total}$), et enfin la position relative de la tache
($X_{\rm tache}$ et $Y_{\rm tache}$). Les trois composantes sont par conséquent:
\begin{itemize}
\item La tache : $I(x,y)=\delta(x-X_{\rm tache})\,.\,\delta(y-Y_{\rm tache})$.
\item La couche moléculaire :   $I(r)=\delta(2r-\theta_{\rm couche})$ .
\item La photosphère : $I(r)=(1-(2r/\theta_\star)^2)^\frac{\alpha}{2}$.
\end{itemize}

A la différence d'un modèle plus physique de l'objet, comprenant, notamment,
températures et opacités \citep{2005A&A...436..317P}, ce modèle est
purement géométrique. Il permet d'être contraint uniquement par
la fonction de brillance de l'objet observé, et ne nécessite pas
d'information bolométrique ou spectrale. De plus, la transformée de
Fourier de ce modèle peut être écrite de façon analytique, permettant
des ajustements rapides et précis. L'asymétrie s'écrit de la façon
suivante:
\begin{equation}
V_{\rm tache}(u,v) =  \exp{(-2\pi \I (X_{\rm tache}\,.\, u+Y_{\rm tache}\,
  . \,v))} \, .
\end{equation}
Parce que la couche moléculaire et le disque assombri sont à symétrie
radiale, nous pouvons utiliser la fréquence spatiale radiale
$v_r=\sqrt{u^2+v^2}$ et la transformée de Hankel pour déduire :
\begin{equation}
V_{\rm couche}(v_r)=  \pi \theta_{\rm couche} \, J_0( \pi \theta_{\rm
  couche} v_r)
\end{equation}
et 
\begin{equation}
V_{\rm photosphère}(v_r) = \sum_{k \geq 0} \frac{\Gamma(\nu+1)}{\Gamma(\nu+k+1)
  k!} \left( \frac{- v_r^2}{4} \right)^k
\end{equation}
où $\nu =\frac{\alpha}{2} +1$ et $\Gamma$ la fonction gamma d'Euler.  

\subsection{Les clôtures de phase et la composante asymétrique}
\label{sec:clot_asym}

  \begin{figure}[h] \centering
   \includegraphics[width=7cm]{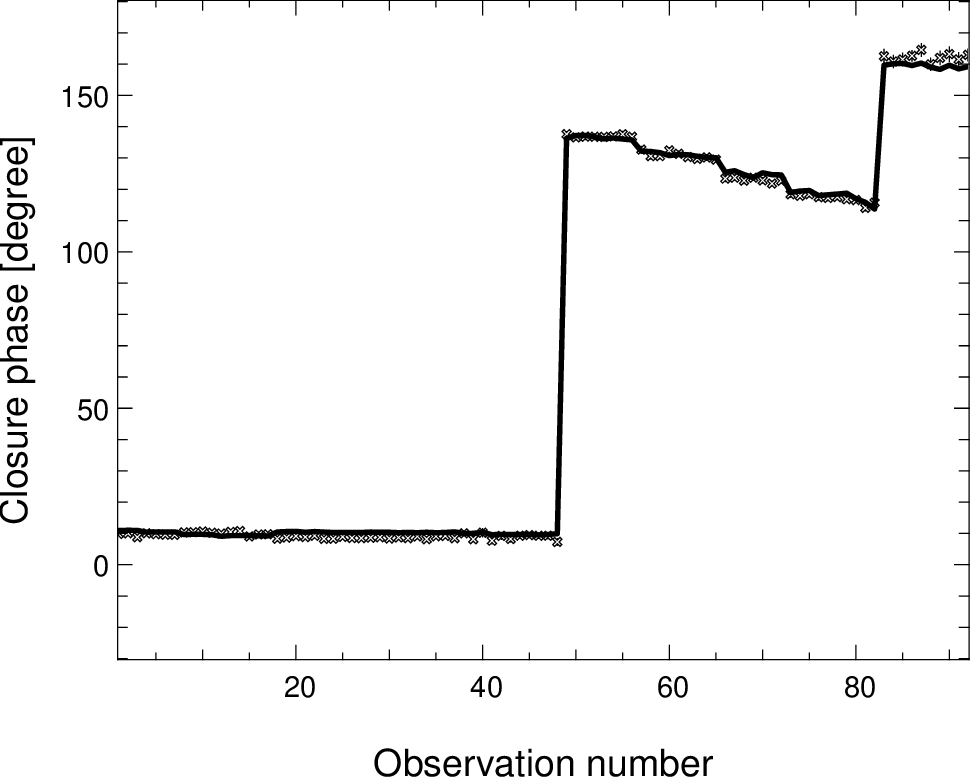}
   \includegraphics[width=7cm]{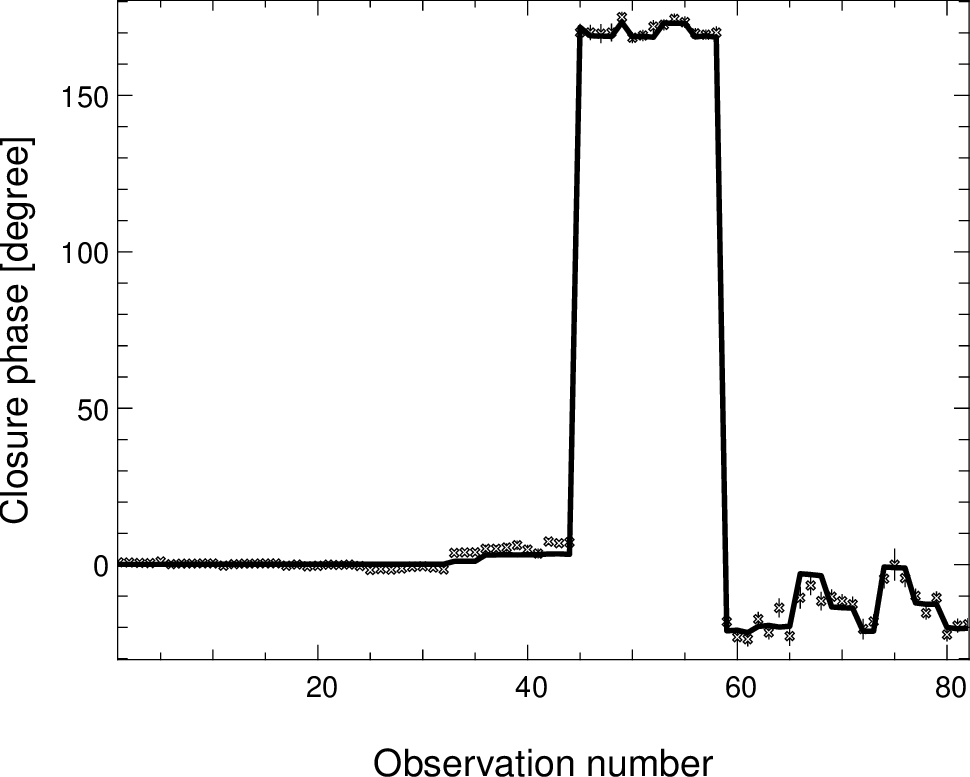}
   \includegraphics[width=7cm]{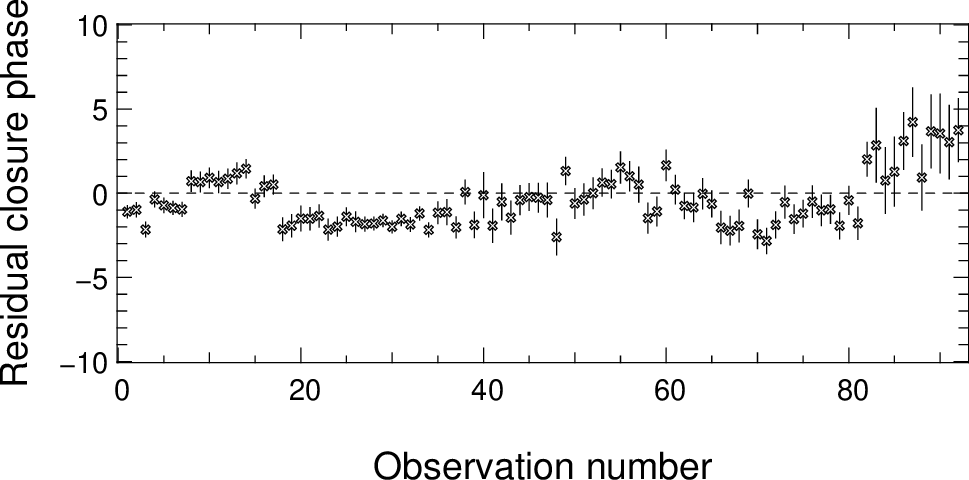}
   \includegraphics[width=7cm]{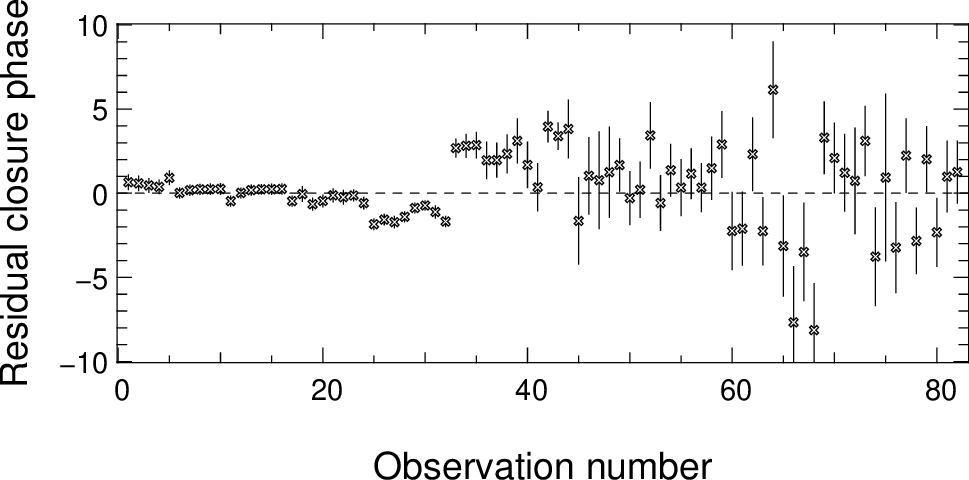}
   \caption[Ajustement d'un modèle d'asymétrie sur les données de
   $\chi$ Cyg de mai et octobre 2005]{ Clôtures de phase et résidus de
   l'ajustement d'un modèle d'asymétrie sur les données de $\chi$ Cyg
   de mai et octobre 2005. L'assymétrie est composé d'une simple tache
   décentrée par rapport au centre de la photosphère. Ce résultat a
   été obtenu par un ajustement effectué sur les clôtures de phase
   uniquement. Cette technique est décrite au
   paragraphe~\ref{sec:clot_asym}.  } \label{fig:Res_Clo} \end{figure}

\subsubsection{Dissocier la composante asymétrique de la composante symétrique}

Pour une base donnée de l'interféromètre, la mesure obtenue est une
valeur complexe, c'est-à-dire une phase et une amplitude. Cependant,
l'atmosphère introduit un déphasage des franges (le piston),
et la mesure directe de la position des franges ne nous donne pas
l'information de phase correspondant à l'objet astrophysique. Un
artifice mathématique consiste à ne pas mesurer la phase mais la
triple somme des phases correspondant aux trois lignes de bases
obtenues par les trois télescopes. Nous avons montré
en section~\ref{sec:ima_clo} que cette mesure est alors indépendante du
piston atmosphérique et ne dépend que de l'objet étudié. Les données
obtenues après réduction sont donc de deux types:
\begin{itemize}
\item l'information de module est obtenue par la mesure de l'énergie
  contenue dans les franges, ce qui correspond à une visibilité au
  carré ($V^2$).
\item l'information de phase est obtenue via la somme de trois phases,
  ce qui correspond aux clôtures de phase (CP).
\end{itemize}

L'information sur la symétrie (ou l'asymétrie) de l'étoile est
contenue dans les phases, et donc dans les clôtures de phase. En
effet, si l'objet est centro-symétrique, les visibilités complexes
sont alors réelles et les phases nulles. L'information apportée par
les clôtures est cependant difficile à interpréter. La difficulté est
renforcée par la grande précision ($\approx 1$ degré) et la faible
quantité (1 clôture pour trois lignes de base étudiées) de ces
mesures. C'est pourquoi il est intéressant d'essayer de découpler la
problématique de l'asymétrie du modèle géométrique de
l'atmosphère. Ceci peut être fait en dissociant l'image en 2
composantes distinctes.  La première composante est celle
correspondant à la partie symétrique de l'objet, dans notre cas, la
photosphère ou la couche moléculaire. Les visibilités complexes
($V_{\rm SYM}$) sont alors réelles, c'est-à-dire :
\begin{equation} 
\Im(V_{\rm SYM}) =0 \,.
\end{equation} 
La deuxième composante de l'objet est la partie asymétrique ($V_{\rm
ASYM}$). Les visibilités complexes ont alors une partie imaginaire non
nulle. C'est cette partie imaginaire qui est responsable de
l'existence de clôtures de phase différentes de 0 ou 180 degrés. Si
l'on est uniquement intéressé par l'asymétrie, il est intéressant de
chercher à ajuster cette partie imaginaire sur les clôtures de phase
sans avoir à se soucier de la composante symétrique. Ceci nécessite le
calcul des visibilités complexes par l'utilisation des valeurs mesurées
des $V^2$ :
\begin{eqnarray} 
V^{Model}&=&\Re(V^{Model})+\I\,\Im(V^{Model}) \\
 &=&\pm\sqrt{V^2-\Im(V^{Model})^2}+\I\,\Im(V^{Model}) \\
 &=&\pm\sqrt{V^2-\Im(V_{\rm ASYM})^2}+\I\,\Im(V_{\rm ASYM})
\label{eq:V_mod}
\end{eqnarray} 
La phase des visibilités résultant du modèle de l'asymétrie peut alors
être obtenue à partir de l'équation~(\ref{eq:V_mod}). Il y a, cependant,
deux remarques importantes. La première concerne les $V^2$ qui, parce
qu'il s'agit de mesures, sont sujets aux bruits. Il est donc
nécessaire de n'utiliser que les données comportant une bonne
précision sur la mesure de l'amplitude. Ceci est généralement le
cas dans nos observations avec des erreurs moyennes sur les $V^2$ de
l'ordre du pourcent. Deuxièmement, il y a une imprécision sur le signe
de la partie réelle. Notre approche a été de considérer la partie
réelle comme étant principalement due à la composante symétrique. Sur
la quasi totalité de nos objets (à l'exception notable de Mira), ceci
est effectivement le cas. Ainsi, nous avons ajusté des disques
assombris sur l'ensemble de nos données et déterminé l'emplacement des
zéro de visibilités. Nous avons ensuite utilisé l'emplacement de ces
zéros pour lever l'incertitude du signe dans
l'équation~(\ref{eq:V_mod}).

\subsubsection{Le modèle de la tache}

   \begin{figure}[h!] \centering
   \includegraphics[width=6cm]{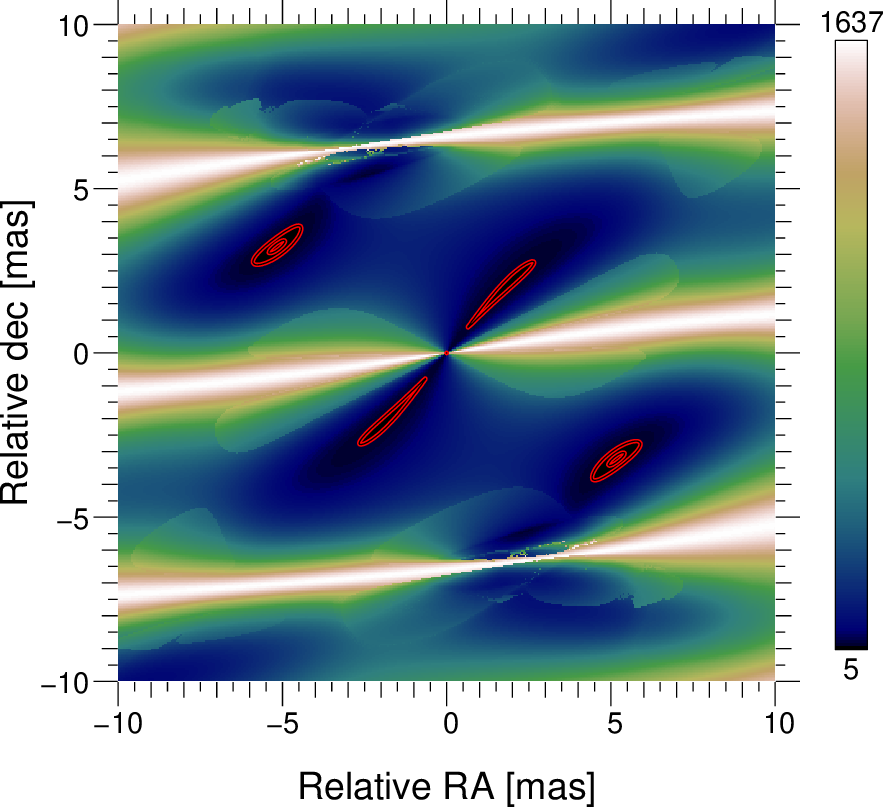}
   \includegraphics[width=6cm]{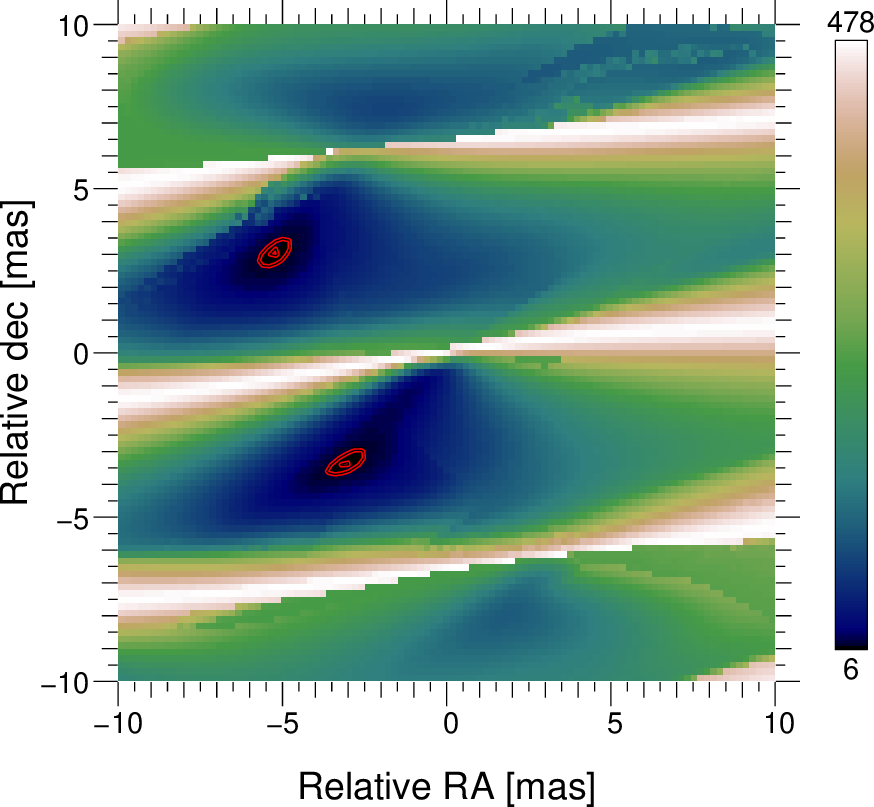}
   \includegraphics[width=6cm]{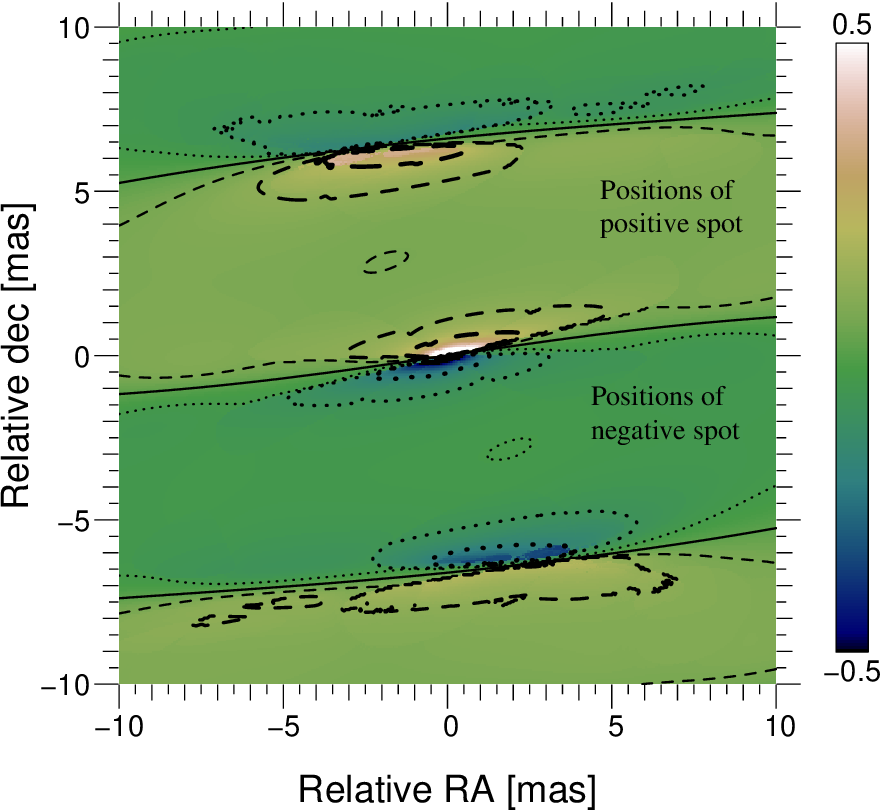}
   \includegraphics[width=6cm]{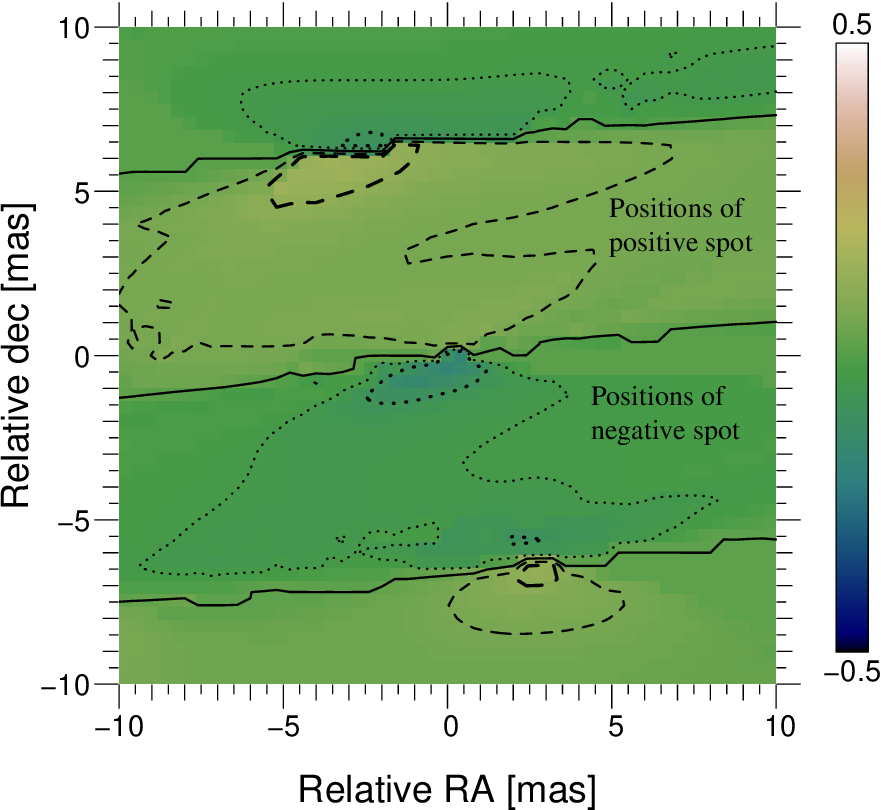}
   \caption[Cartes de $\chi^2$]{
   Figures du haut: $\chi^2$ fonction de la position de la tache par
   rapport au centre de la photosphère. Figures du bas: Amplitudes de
   la tache minimisant le $\chi^2$. La carte de gauche représente le
   $\chi^2$ issu d'un ajustement sur les clôtures de phase
   uniquement. La carte de droite représente le $\chi^2$ pour un
   ajustement sur l'ensemble des données. Les coutours rouges
   représentent les barres d'erreurs à 3 et 9 sigmas.  Le profil
   symétrique obtenu sur la carte de gauche est dû au fait que les
   clôtures de phase ne sont sensibles qu'aux asymétries et ne sont
   pas sensibles à la différence entre, par exemple, une tache
   négative sur la droite ou une tache positive sur la gauche. Cette
   incertitude disparaît lorsque l'on ajuste à la fois les clôtures de
   phase et les visibilités au carré. } \label{fig:chi2_map}
   \end{figure}

Pour en revenir à notre modèle établi en section~\ref{sec:model_geo}, la
partie imaginaire des visibilités s'écrit:
\begin{equation} 
\Im(V_{\rm ASYM}(u,v)) = F_{\rm tache}\, \sin (-2\pi (X_{\rm tache}\cdot u+Y_{\rm tache}
  \cdot v)) \,,
\end{equation} 
où $F_{\rm tache}$ est le flux de la tache relatif à la brillance
totale de l'image. Les visibilités du modèle sont alors reconstruites
à partir de la relation~(\ref{eq:V_mod}), et les clôtures sont
obtenues par la synthèse du bispectre. Cette méthode fournit des
résultats extrêmement convaincants. Surtout si l'on considère la
simplicité du modèle de l'asymétrie et la quantité de clôtures
mesurées. A titre d'exemple, la figure~\ref{fig:Res_Clo} présente le
résultat d'ajustement sur les clôtures de phase de $\chi$ Cyg observées
en mai et octobre 2005. Les résidus sur les clôtures de phase sont en
moyenne de 2 degrés, quant aux $\chi^2$ réduits, ils sont de 5,2 pour
les données de mai, et de 4,7 pour les données d'octobre. La qualité
de ces ajustements justifie l'utilisation de ce modèle simple pour
représenter l'asymétrie. Nous ne pouvons, cependant, pas exclure la
présence de plusieurs taches ou même d'une asymétrie ayant une
structure différente. En ce qui concerne les données de mai, par
exemple, nous avons pu ajuster les clôtures de phase par deux taches,
de manière à obtenir un $\chi^2$ réduit de 0,9. Cependant, le nombre
élevé de paramètres fait alors qu'ils deviennent difficiles à
contraindre.

\subsubsection{Les cartes d'asymétrie}

Les clôtures de phase ne fournissent pas un critère d'ajustement
convexe \citep{Meimon:PhD}. C'est pourquoi il est nécessaire d'établir
des cartes de $\chi^2$ pour trouver l'emplacement de l'asymétrie
vérifiant le maximum de vraisemblance. Pour chaque étoile, l'étude de
l'asymétrie commence par ce premier travail. A titre d'exemple, nous
avons reproduit figure~\ref{fig:chi2_map} les $\chi^2$ réduits obtenus
par ajustement des données acquises en mai 2005 sur $\chi$ Cyg. Pour
obtenir la carte de gauche, nous avons ajusté l'asymétrie
uniquement. La carte de droite a été obtenue par ajustement du modèle
complet défini section~\ref{sec:model_geo}. La position sur la carte
indique l'emplacement de la tache. Nous avons ici la confirmation
qu'un simple algorithme de minimisation du $\chi^2$ ne fournit pas
nécessairement la bonne solution. Avant tout ajustement, il est donc
nécessaire de bien initialiser la position de la tache. 

En second lieu, il est intéressant de noter la symétrie observée
dans la carte du $\chi^2$ obtenue par l'ajustement des clôtures de
phase uniquement. L'explication est que les clôtures de phase ne sont
sensibles qu'aux disymétries. Elles ne permettent pas de différencier
entre un flux positif d'un coté de l'étoile, ou un flux négatif (tache
sombre) de l'autre côté de l'étoile. Ce phénomène se retrouve dans
l'image du bas à droite où l'on peut voir que la carte symétrique du
$\chi^2$ devient antisymétrique pour l'amplitude de la tache. Il est
intéressant de constater que cette ambiguïté disparaît lorsque l'on
ajuste le modèle complet, comme nous pouvons le voir dans la partie de
droite de la figure~\ref{fig:chi2_map}.

\clearpage
\section{\'Etude par imagerie paramétrique d'Arcturus}
\label{sc:Actu_test}

\subsection{Une référence pour tester la qualité du  processus d'imagerie}

   \begin{figure}[h] \centering
   \includegraphics[width=7cm]{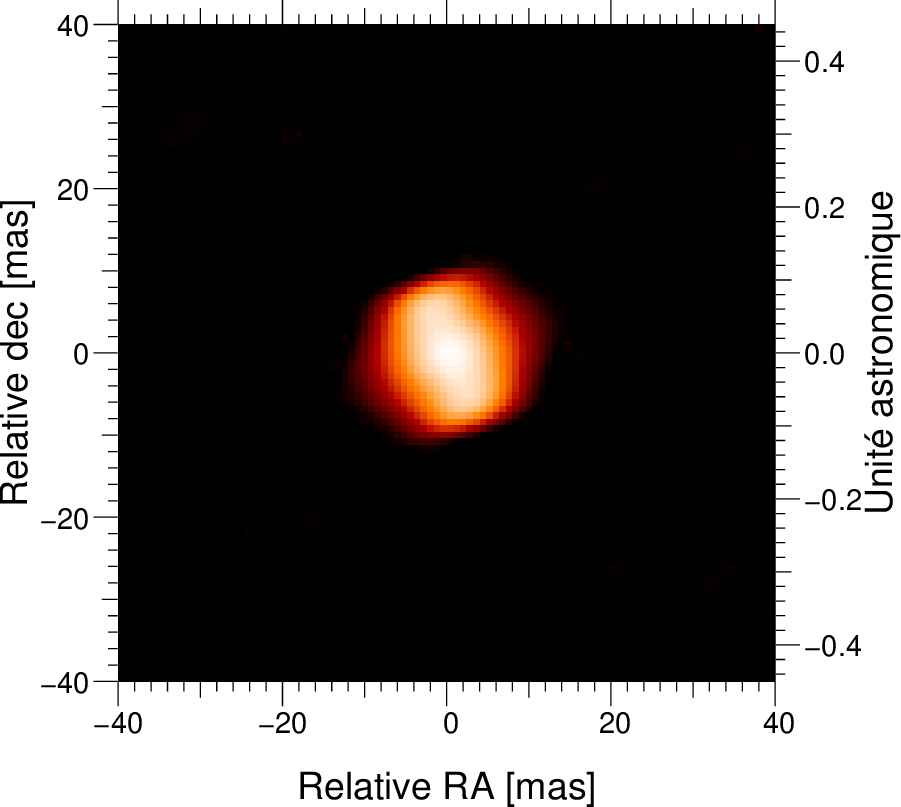}
   \includegraphics[width=7cm]{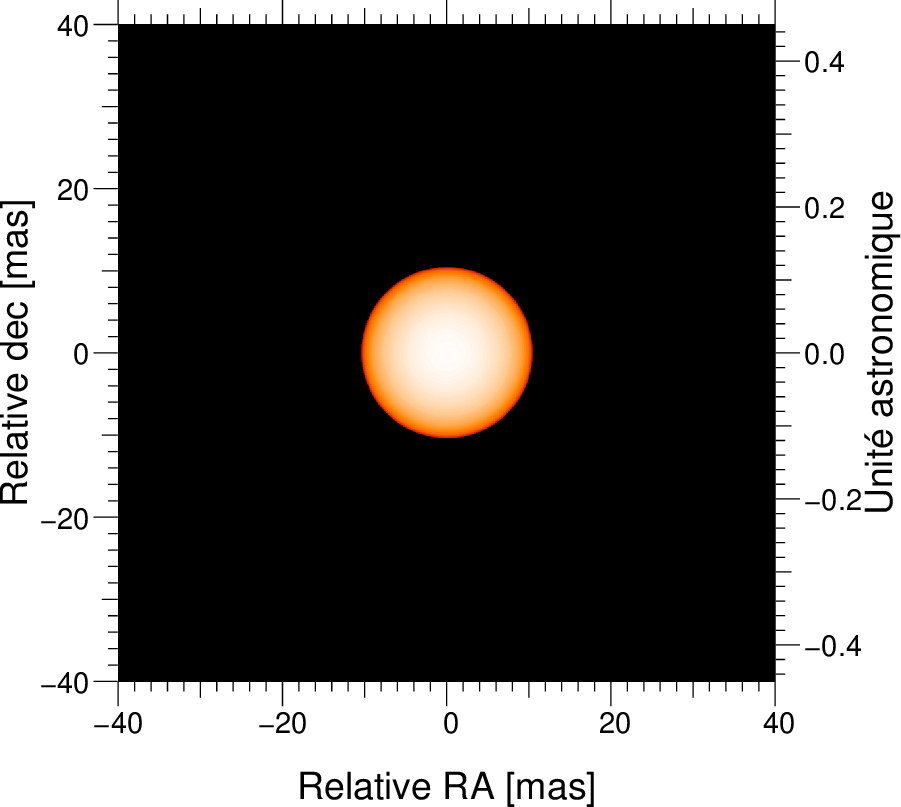}
   \caption[Comparaison entre reconstruction d'image en aveugle et
   reconstruction paramétrique de l'étoile Arcturus]{Comparaison
   entre reconstruction d'image en aveugle et reconstruction
   paramétrique de l'étoile Arcturus.  Les paramètres de l'image de
   droite sont $\alpha$ = 0,314 $\pm$ 0,003 et $\theta_\star$ = 20,91
   $\pm$ 0,01.}  \label{fig:Arctu_ims} \end{figure}

L'imagerie par reconstruction en aveugle présente de multiples
avantages. Le principal est qu'il ne contraint pas l'objet à un modèle
géométrique préétabli. Néanmoins, ce procédé ne fournit pas de taux
de confiance sur l'image obtenue. Il est nécessaire de vérifier que le
logiciel de reconstruction d'image, et, notamment, la fonction de
régularisation (équation~(\ref{eq:t_regul})), sont adaptée à nos
objets. Pour opérer cette vérification, nous avons utilisé la géante
rouge Arcturus. L'intérêt de cette étoile est sa simplicité
géométrique. En effet, les visibilités et les clôtures de phase sont
parfaitement bien modélisées par un simple disque assombri. Nous avons
utilisé une version simplifiée du modèle présenté
paragraphe~\ref{sec:model_geo} ne comportant que la composante
correspondant à la photosphère :
\begin{equation}
I(r)=\left(1-(2r/\theta_\star)^2\right)^{\frac{\alpha}{2}} \,,
\label{eq:loi_puis}
\end{equation}
où $\theta_\star$ est le diamètre de l'étoile, et $\alpha$ le
paramètre d'assombrissement \citep{1997A&A...327..199H}. L'ajustement
de ces deux paramètres nous a permis d'aboutir aux valeurs $\alpha$ =
0,314 $\pm$ 0,003 et $\theta_\star$ = 20,91 $\pm$ 0,01, pour un
$\chi^2$ réduit de 3,2.  Il est remarquable qu'un modèle aussi simple
permette un si bon ajustement de l'ensemble des données, et cela
malgré l'hypothèse d'achromaticité de l'assombrissement (voir les
résidus du modèle présentés figure~\ref{fig:Arctu_fit}). \`A partir
des paramètres $\alpha$ et $\theta_\star$, nous avons obtenu une image
de l'objet. Il est intéressant de comparer cette image avec celle
obtenue en utilisant le logiciel de reconstruction en aveugle. De
cette comparaison (figure~\ref{fig:Arctu_ims}), nous avons tiré les
conclusions suivantes:
\begin{itemize}
\item Dans la direction de résolution maximale (Nord-est), les deux
  images sont similaires, avec un assombrissement à peu près
  identique.
\item Dans l'axe à faible résolution, les deux images diffèrent
  largement. Ceci est la conséquence de l'influence du terme de
  régularisation qui domine dans l'axe où la résolution n'est pas
  suffisante.
\end{itemize}
Cependant, la structure générale des images reconstruites est
similaire. Ceci renforce la crédibilité des asymétries observées sur
les images obtenues à partir des autres données. Pourtant, nous ne
pouvons déterminer avec précision assombrissement ou taille angulaire
à partir des images reconstruites précédemment. Pour ce faire, il est
plus pertinent d'ajuster un modèle aux données.

\subsection{L'assombrissement centre-bord d'Arcturus}

   \begin{figure}[h] \centering
   \includegraphics[width=7cm]{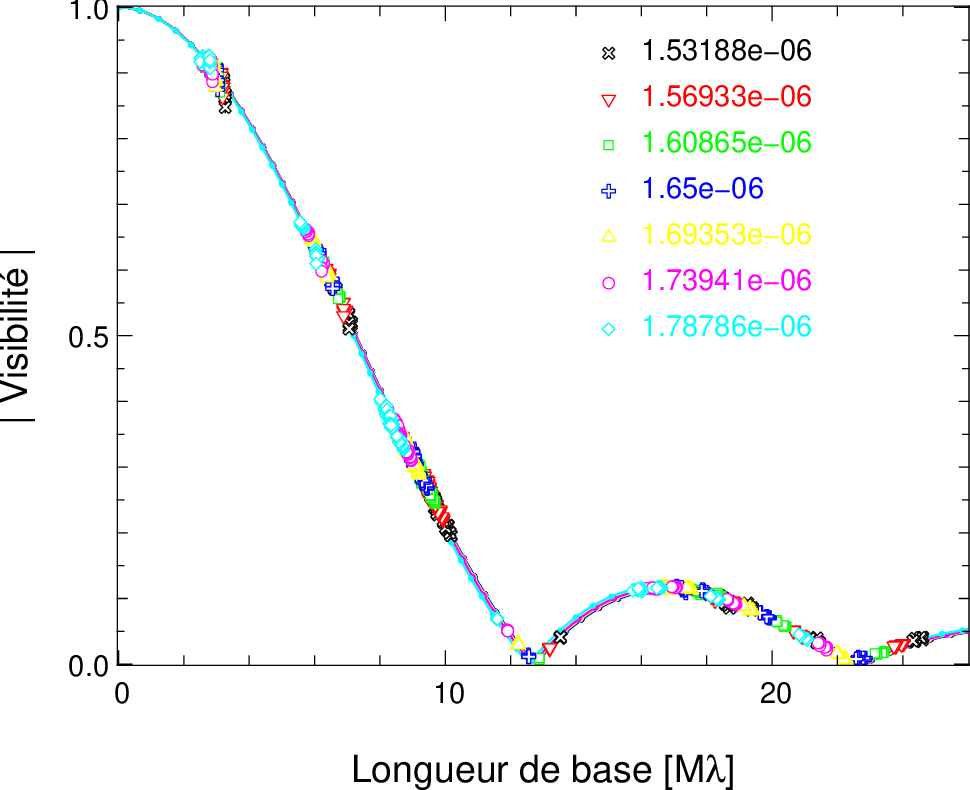}
   \includegraphics[width=7cm]{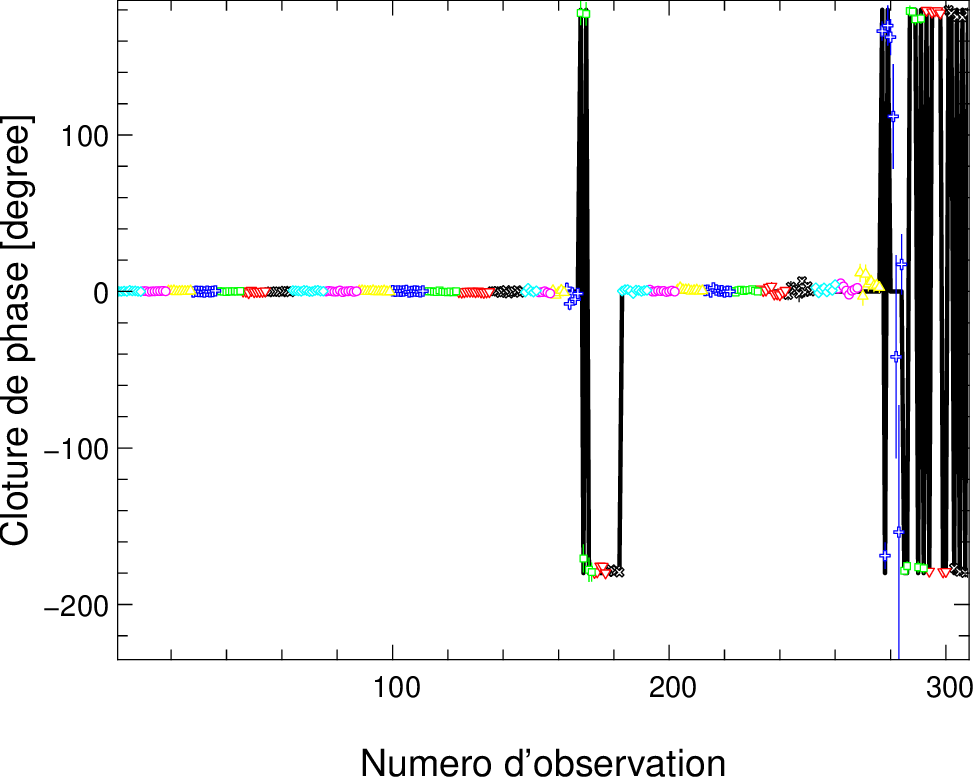}
   \includegraphics[width=7cm]{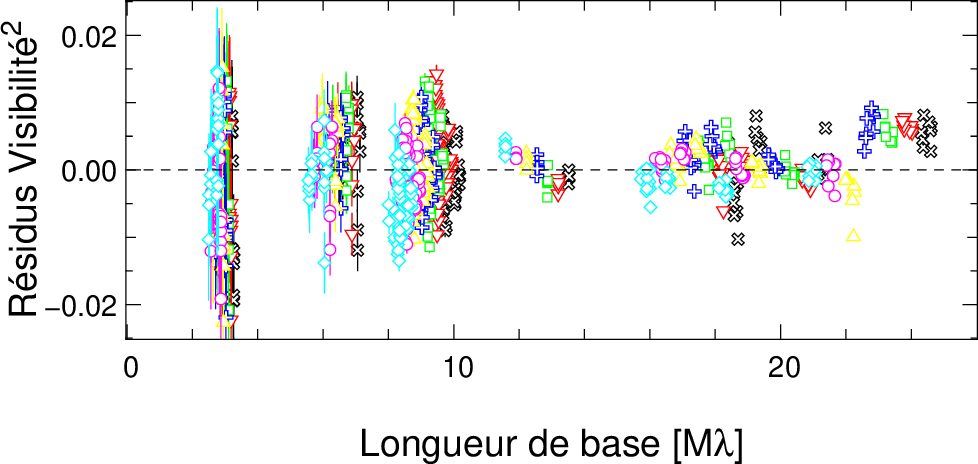}
   \includegraphics[width=7cm]{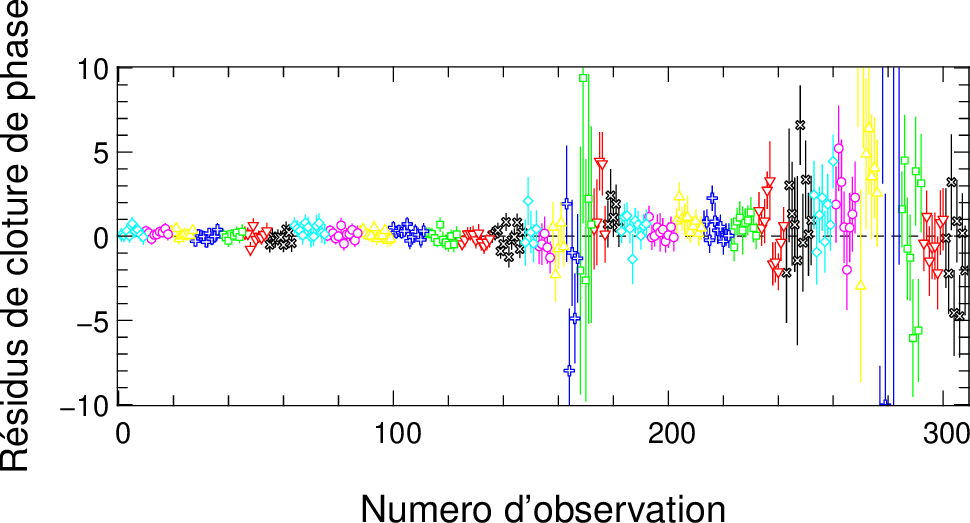}
   \caption[Visibilités et Clôtures de phase d'Arcturus]{Visibilités
   et Clôtures de phase d'Arcturus observées en mai 2006. Le modèle
   ajusté est un simple disque assombri par une loi en puissance
   (équation~(\ref{eq:loi_puis})). Le diamètre de la photosphère est
   supposé achromatique, contrairement à l'assombrissement centre
   bord. Les valeurs utilisées sont reproduites
   tableau~\ref{tb:Res_fit_Arctu}. } \label{fig:Arctu_fit}
   \end{figure}

\begin{table}
\caption{Résultat de l'ajustement du
  modèle sur les données d'Arcturus}
\label{tb:Res_fit_Arctu}
\centering
\begin{tabular}{ccc}
\hline
\hline
  $\lambda$ & ACB [$\alpha$] & $\theta_\star$ \\
\hline
1,53 $\mu$m & 0,333 $\pm$ 0,004 & 20,92 $\pm$  0,01 mas\\ 
1,57 $\mu$m & 0,327 $\pm$ 0,004 &  '' \\
1,61 $\mu$m & 0,335 $\pm$ 0,005 &  '' \\
1,65 $\mu$m & 0,341 $\pm$ 0,005 &  '' \\
1,69 $\mu$m & 0,321 $\pm$ 0,004 &  '' \\
1,74 $\mu$m & 0,301 $\pm$ 0,004 &  '' \\
1,79 $\mu$m & 0,246 $\pm$ 0,005 &  '' \\
\hline
$\chi^2$    &  \multicolumn{2}{c}{3150} \\
Degrés de liberté & \multicolumn{2}{c}{1207} \\
$\chi^2$ Réduit   &  \multicolumn{2}{c}{2,6} \\
\hline
\end{tabular}
\end{table}

L'analyse paramétrique des données ne sert pas qu'à confirmer la
validité des reconstructions par déconvolution en aveugle. Si l'objet
est suffisamment connu, l'idéal est d'ajuster un ou plusieurs modèles
d'atmosphères stellaire pour pouvoir mesurer les paramètres physiques
de l'étoile, comme la température effective ou encore l'accélération
gravitationnelle à la surface. Ce travail a été effectué, sur Arcturus
notamment, par \citet{Verhoelst:PhD} au cours de sa thèse. Bien que ce
travail de modélisation n'ait pas été produit pendant ma thèse, nous
avons pu utiliser le modèle pour estimer la dépendance du facteur
d'assombrissement $\alpha$ en fonction de la longueur
d'onde. L'ajustement et les résidus obtenus sont présentés
figure~\ref{fig:Arctu_fit}. Le tableau~\ref{tb:Res_fit_Arctu} reprend
ces résultats. Nous constatons des variations significatives de
l'assombrissement. L'interprétation astrophysique est cependant
difficile. Une voie à explorer serait de comparer ces valeurs avec les
modèles d'atmosphères présentés par
\citet{2000A&A...363.1081C}. Cependant, le modèle d'atmosphère étant
différent, cela nécessiterait certainement de re-effectuer un
ajustement à partir du modèle de \citet{2000A&A...363.1081C}. Nous
estimons que la voie à privilégier serait l'ajustement d'un modèle
complet et spécifique, comme a pu le faire \citet{Verhoelst:PhD}. Ceci
serait d'autant plus intéressant que nous pourions le contraindre à
partir de l'information spectrale. Gageons que dans un avenir proche,
nous puissions nous engager dans cette voie.

\subsection{La présence d'un compagnon}
\label{sec:compa_arctu}

\begin{figure}
  \centering
  \resizebox{\hsize}{!}{\includegraphics{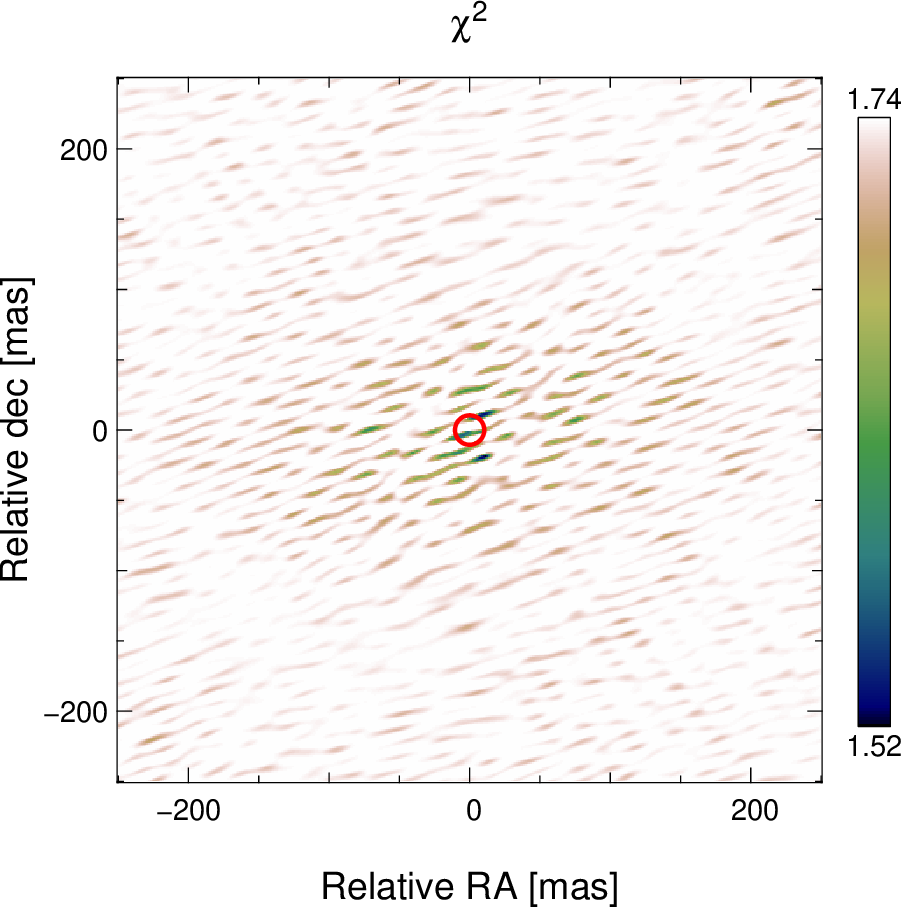}
  \includegraphics{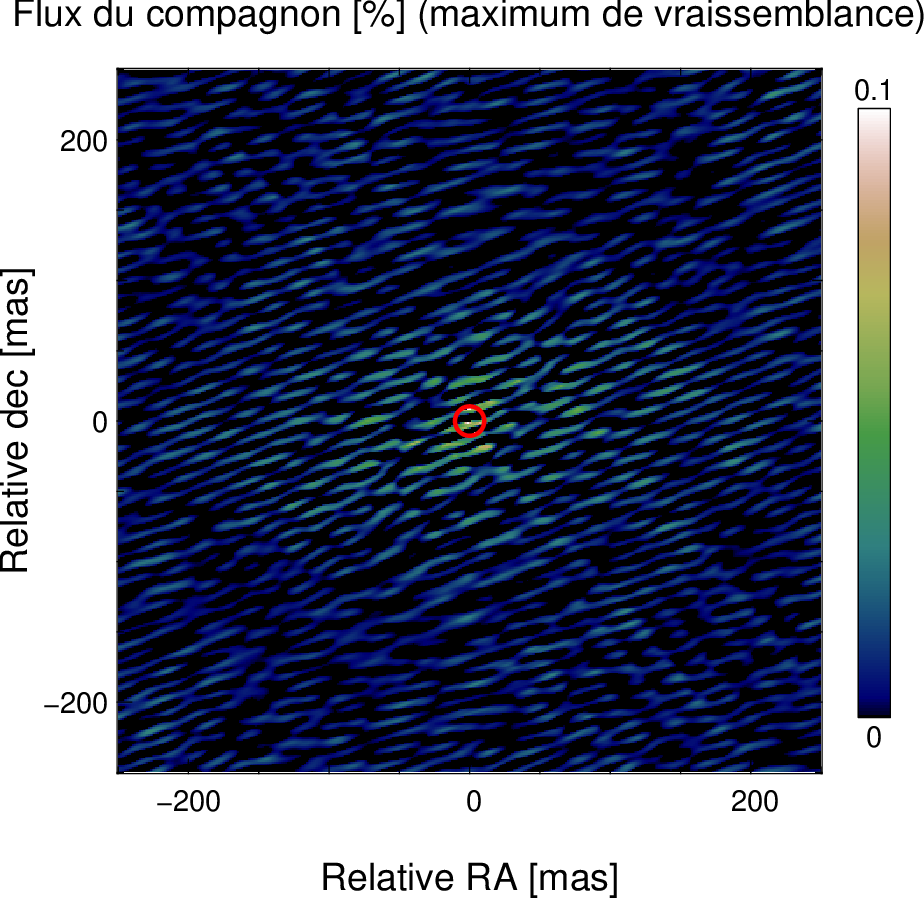}}{\vspace{.3cm}}
  \includegraphics[width=11.3cm]{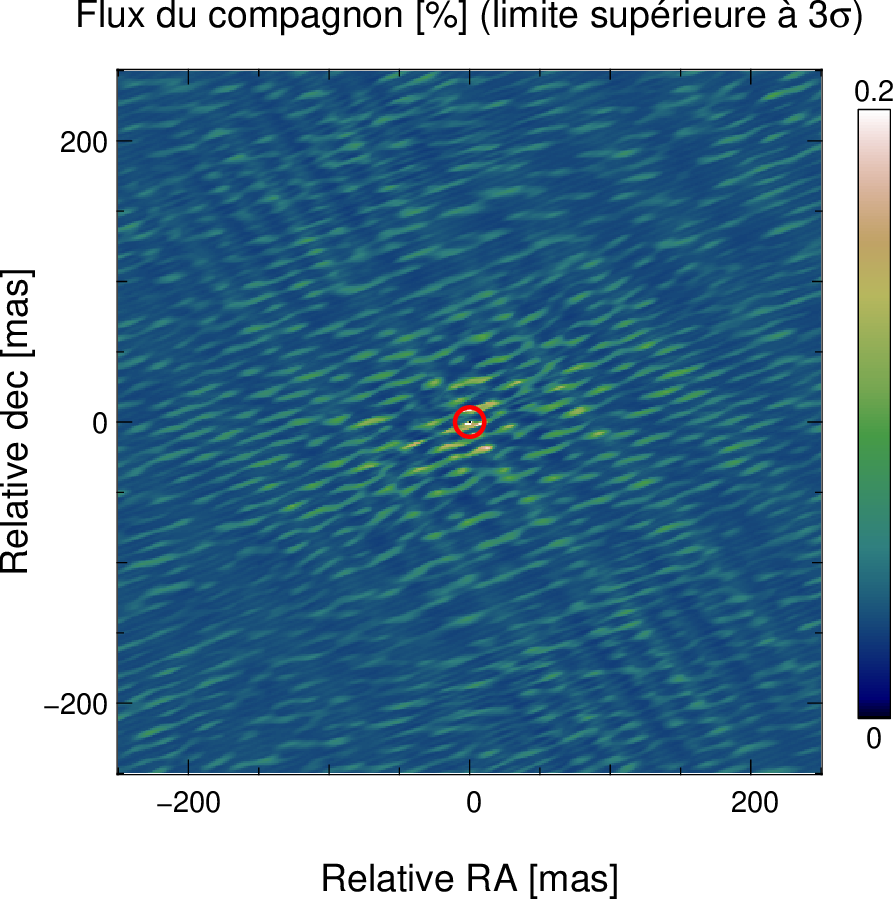}
  \caption[Recherche d'un compagnon d'Arcturus]{
  En haut à gauche est présenté le $\chi^2$ réduit, obtenu par
  ajustement de la brillance d'un compagnon positionné aux coordonnées
  définies par la figure. Le $\chi^2$ réduit maximum présent sur
  l'image est de 1,74, et est obtenu pour un compagnon inexistant
  (brillance nulle).  La figure de droite représente la brillance
  du compagnon correspondant au maximum de vraisemblance. Ce flux,
  relatif à la brillance totale de l'étoile, reste faible, avec
  des valeurs maximum autour de 0,1\% pour des positions proches de
  la photosphère (représentée par le cercle rouge). La figure du bas
  représente la limite supérieure à 3$\sigma$ du flux de l'hypothétique
  compagnon.}  \label{fig:Arctu_comp}
\end{figure}

Comme nous avons déjà pu le mentionner section~\ref{sec:actu_aveugle},
Arcturus est souvent considérée comme une étoile utile pour la
calibration spectrale \citep[ISO;][]{2003A&A...400..709D} et spatiale
\citep[masquage de pupille;][]{2000ApJ...534..907T}. Il est donc
important d'établir ou d'exclure la possibilité d'un système binaire.

Pour cela, nous avons utilisé l'approche présentée
section~\ref{sec:clot_asym}, de manière à n'ajuster que l'éventuelle
composante asymétrique. Ce choix a été dicté par plusieurs raisons :
\begin{enumerate}
\item Ce sont les clôtures de phase qui vont majoritairement contraindre
l'asymétrie, et donc le compagnon. 
\item  Les clôtures de phase ne sont que peu affectées par d'éventuels
problèmes de calibration, et fournissent des valeurs sûres et
précises.
\item Cette technique nous permet d'éviter le risque de biais introduit par un
choix de modèle d'atmosphère stellaire qui n'est pas forcément exactement
équivalent à la réalité.
\end{enumerate}

Les résultats sont présentés figure~\ref{fig:Arctu_comp}. En haut à
gauche est présenté le $\chi^2$ réduit obtenu par ajustement de la
brillance d'un compagnon positionné aux coordonnées définies par la
figure. Le maximum du $\chi^2$ réduit présent sur l'image est de 1,74,
et est obtenu pour un compagnon inexistant (brillance nulle). Les
variations de $\chi^2$ sont faibles, avec une valeur minimum de
1,52. La figure de droite représente la brillance du compagnon
correspondant au maximum de vraisemblance $F_{\rm comp}^{\rm
MV}(\alpha,\beta)$ (dans l'hypothèse d'un bruit à statistique
Gaussienne). Ce flux, relatif à la brillance totale de l'étoile, reste
toujours faible avec des valeurs maximum autour de 0,1\% pour des
positions proches de la photosphère (représenté par le cercle rouge).

Pour pouvoir écarter définitivement la présence d'un compagnon à
l'étoile, il a fallu décider d'un seuil de confiance. Nous avons
choisi d'établir la limite supérieure à 3$\sigma$, ce qui correspond à
une probabilité d'erreur de moins de 1\%. La limite supérieure de la
brillance du compagnon est donnée par la relation suivante :
\begin{equation}
F_{\rm comp}^{3\sigma}(\alpha,\beta) = F_{\rm comp}^{\rm MV}(\alpha,\beta) +
\sigma\left(F_{\rm comp}^{\rm MV}(\alpha,\beta)\right) \cdot
\sqrt{9-\chi^{2\dagger}(\alpha,\beta)} \,,
\end{equation}
où $\sigma\left(F_{\rm comp}^{\rm MV}(\alpha,\beta)\right)$ est la
déviation standard constatée sur le flux à la position
$(\alpha,\beta)$. La valeur utilisée pour $\chi^{2\dagger}$ est le
$\chi^2$ normalisé. La normalisation est ici un peu particulière, parce
qu'elle est effectuée de façon à ce que le maximum du $\chi^{2\dagger}$
soit égale au nombre de degrés de liberté. Il s'agit d'une mesure
conservatrice qui équivaut, pour ces données, à multiplier
les barres d'erreurs des clôtures de phase par le facteur $\sqrt{1,42}$.

Les valeurs de $F_{\rm comp}^{3\sigma}(\alpha,\beta)$ sont présentées
dans l'image du bas de la figure~\ref{fig:Arctu_comp}. Nous avons
ainsi contraint la brillance de l'hypothétique compagnon par deux
limites supérieures :
\begin{itemize}
\item $F_{\rm comp} \leq$ 0,18 \% pour un compagnon présent à une
distance comprise entre 10,5 et 100 mas du centre de l'étoile principale.
\item$F_{\rm comp} \leq$ 0,09 \% pour un compagnon présent à une
distance supérieure à 100 mas du centre de l'étoile principale.
\end{itemize}
A moins d'une variation spectrale très forte, la possibilité d'un
compagnon tel que proposé par \citet{2005A&A...435..289V} (rapport de
flux de 2\% en bande K et séparation d'environ 200 mas) est donc
clairement réfutée par nos données.

\clearpage

\section{\'Etude par imagerie paramétrique de $\chi$ Cyg}

\subsection{\'Etude temporelle de $\chi$ Cyg}
\label{sec:temporelle}

\begin{table}[h]
\caption{Valeurs des différents paramètres obtenues par ajustement du
  modèle sur les données}
\label{tb:Res_fit}
\centering
\begin{tabular}{lcccc}
\hline
\hline
  & Mai 2005  & Octobre 2005 & Mars-Avril 2006 & Mai 2006  \\
 & ($\phi = 0.91$) & ($\phi = 0.24$) & ($\phi = 0.67$) & ($\phi =
0.76$) \\
\hline
$\theta_\star$ [mas]&  $19.45 \pm 0.09$ & $26.25 \pm 0.08$& $23.97 \pm  0.80$&$21.27 \pm  0.11$\\
ACB [$\alpha$]             & $1.55 \pm 0.05$ & $1.08 \pm 0.03$& $2.540 \pm 0.396$&$2.343 \pm 0.051$ \\
$\theta_{\rm couche}$ [mas]& $32.22 \pm 0.15$ &  $40.75 \pm 0.37$& $35.48 \pm  0.40$&$27.13 \pm  0.13$\\
$F_{\rm couche}/F_{\rm total}$ [\%]& $6.5 \pm 0.2 $ & $4.7 \pm 0.2 $& $8.77 \pm  0.23$&$8.27 \pm  0.11$\\
$X_{\rm tache}$ [mas] & $-5.22 \pm 0.05$& $8.92 \pm 0.39$& $-2.22 \pm  0.42$&$-3.49 \pm  0.20$\\
$Y_{\rm tache}$ [mas] & $3.05 \pm 0.05$& $2.96 \pm 0.10$& $-4.24 \pm  0.34$&$-6.70 \pm  0.09$\\
$F_{\rm tache}/F_{\rm total}$ [\%] & $5.9 \pm 0.1 $&  $1.7 \pm 0.1 $& $3.72 \pm  0.26$&$1.71 \pm  0.04$\\
\hline
$\chi^2$ Réduit   & 6 & 10 & 2 & 27\\
\hline
\end{tabular}
\end{table}

   \begin{figure}[h!]  \centering
   \includegraphics[width=7cm]{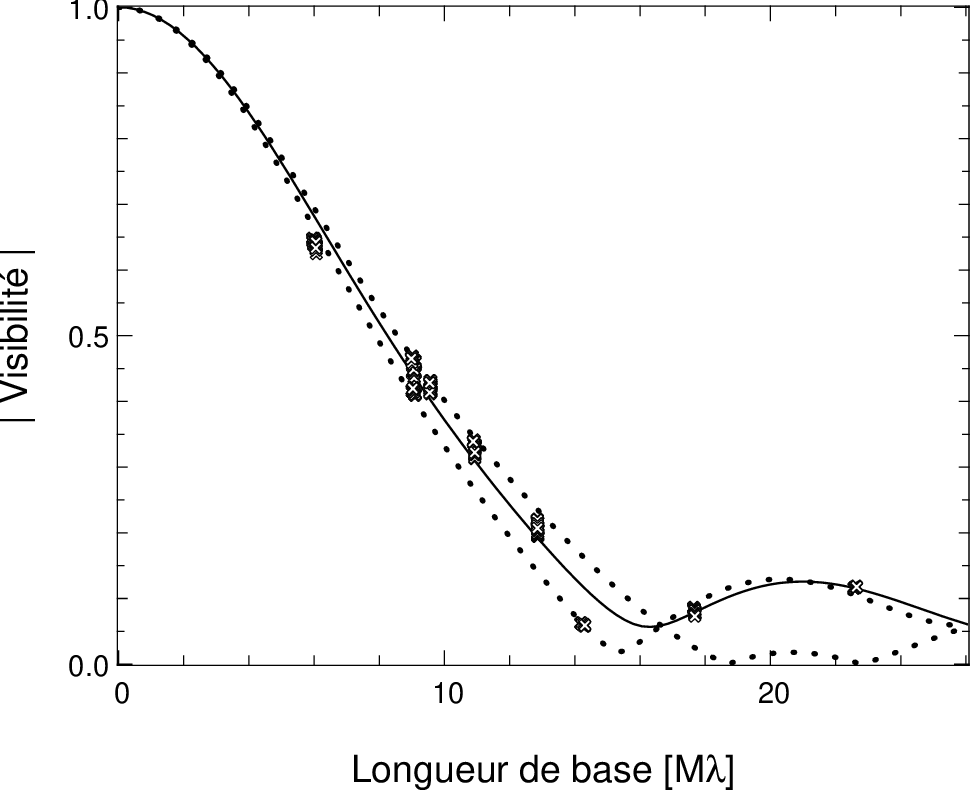}
   \includegraphics[width=7cm]{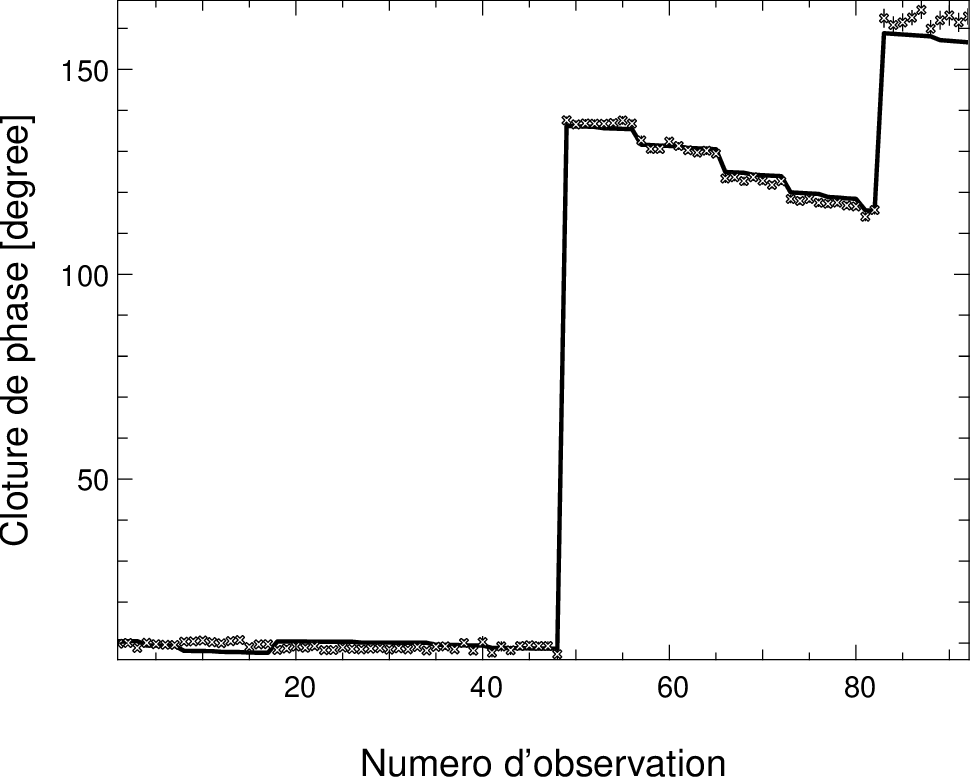}
   \includegraphics[width=7cm]{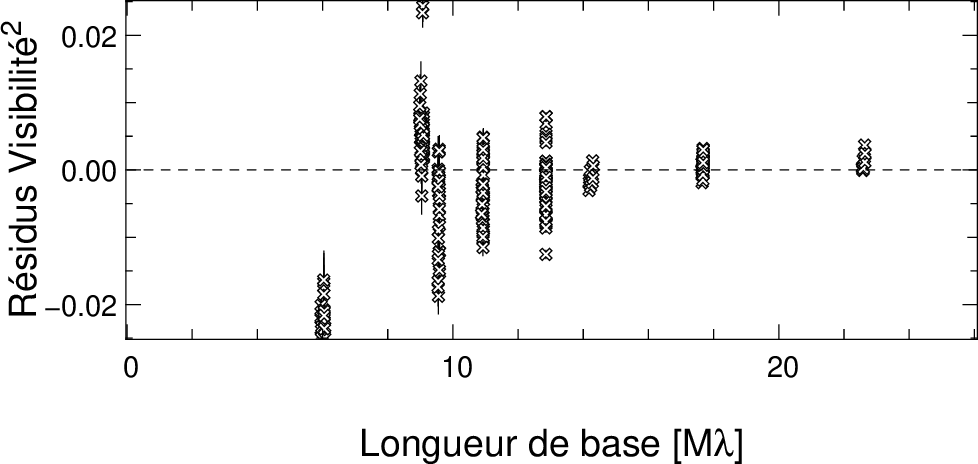}
   \includegraphics[width=7cm]{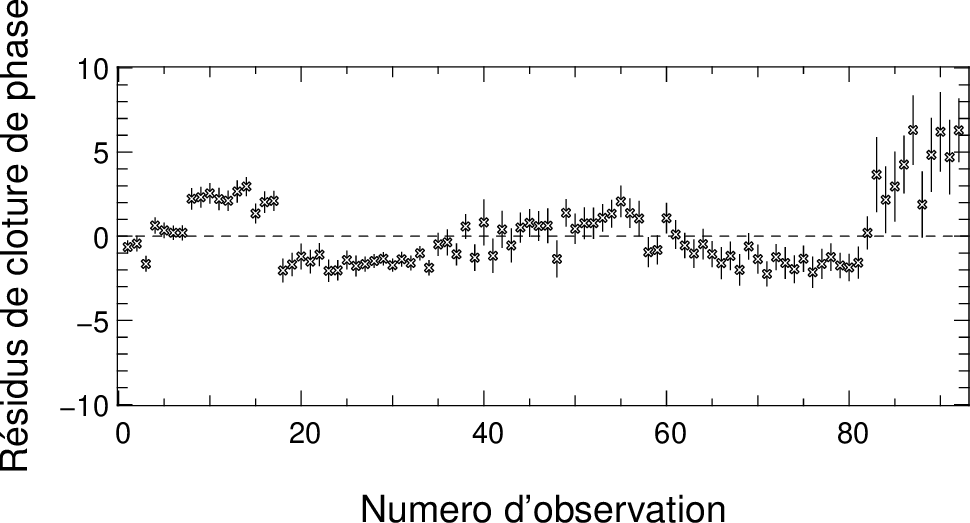}
   \caption[Ajustement des données de $\chi$ Cyg en mai 2005]{ Données
   de mai 2005. A gauche, mesure des visibilités et ajustement
   correspondant. A droite, clôtures de phase. On peut voir le modèle
   s'ajuster aux clôtures de phase, avec des résidus d'environ
   quelques degrés. Les deux courbes en pointillés représentent le
   profil radial des visibilités en direction de la tache, et à 90
   degrés. La courbe centrale correspond aux visibilités selon la
   direction de la base maximale. } \label{fig:Data_Mai05}
   \end{figure}

\clearpage

   \begin{figure}[h!]  \centering
   \includegraphics[width=7cm]{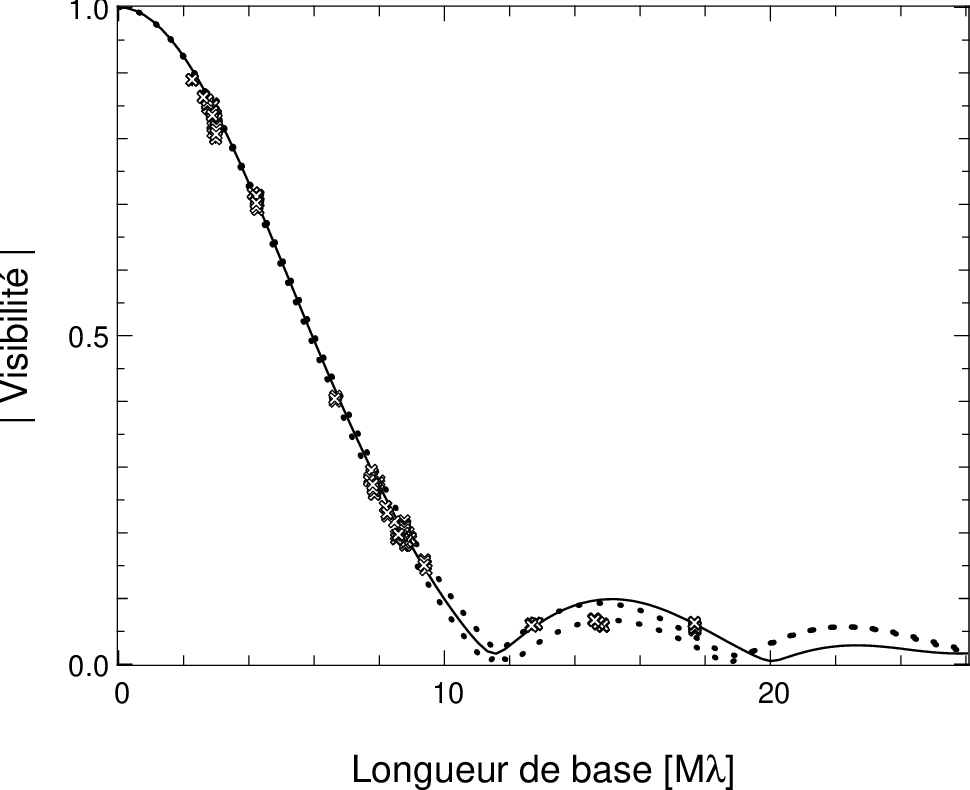}
   \includegraphics[width=7cm]{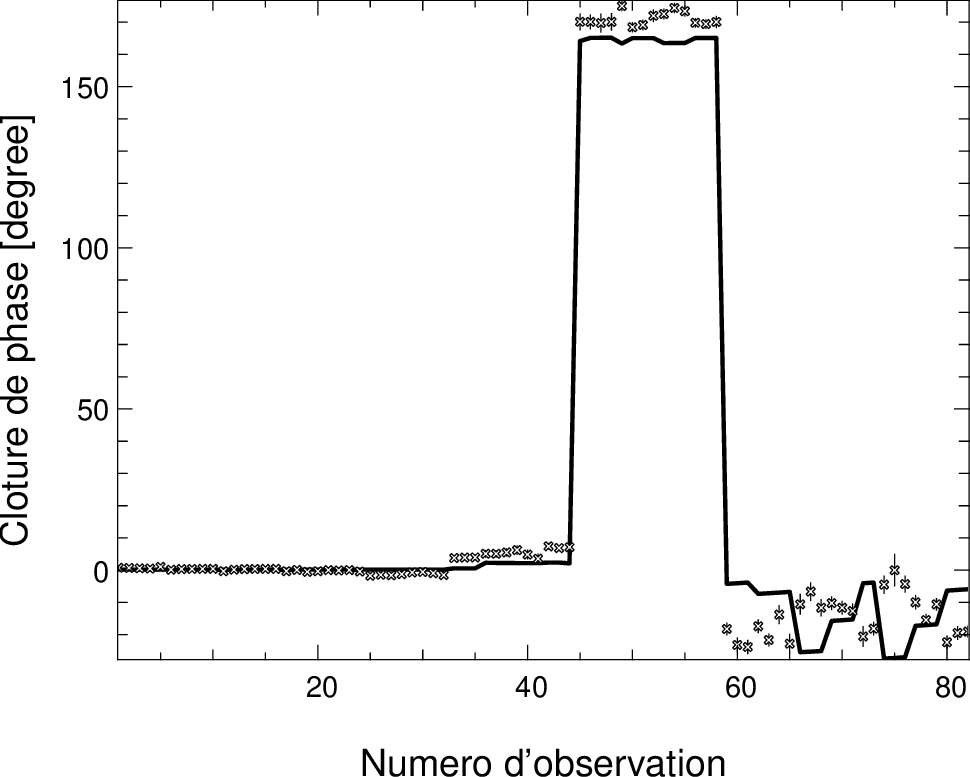}
   \includegraphics[width=7cm]{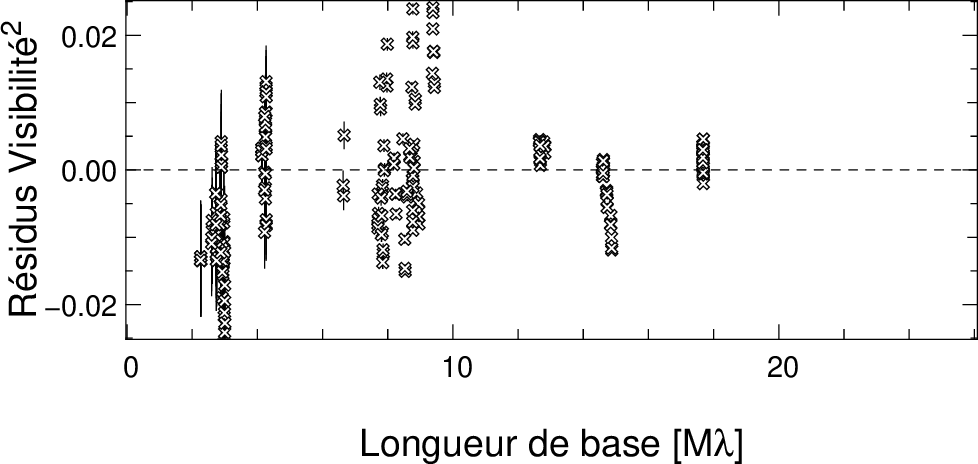}
   \includegraphics[width=7cm]{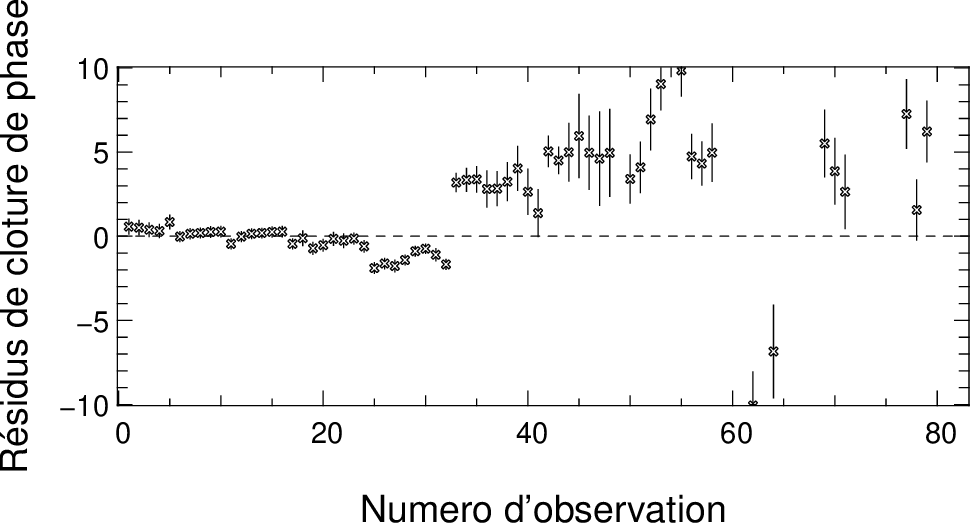}
   \caption[Ajustement des données de $\chi$ Cyg en octobre 2005]{ Idem figure~\ref{fig:Data_Mai05}, mais concernant les
   données de $\chi$ Cyg obtenues en octobre 2005. On peut noter que
   l'ajustement des clôtures de phase correspondant aux observations
   de numéros 60 à 90 est nettement moins bon que ce qui a été obtenu
   figure~\ref{fig:Res_Clo}. Ceci est dû à la contrainte des $V^2$,
   qui s'est traduite par le déplacement de la position de la tache de
   quelque milli-secondes d'angle. } \label{fig:Data_Oct05}
   \end{figure}

   \begin{figure}[h!]
   \centering
   \includegraphics[width=7cm]{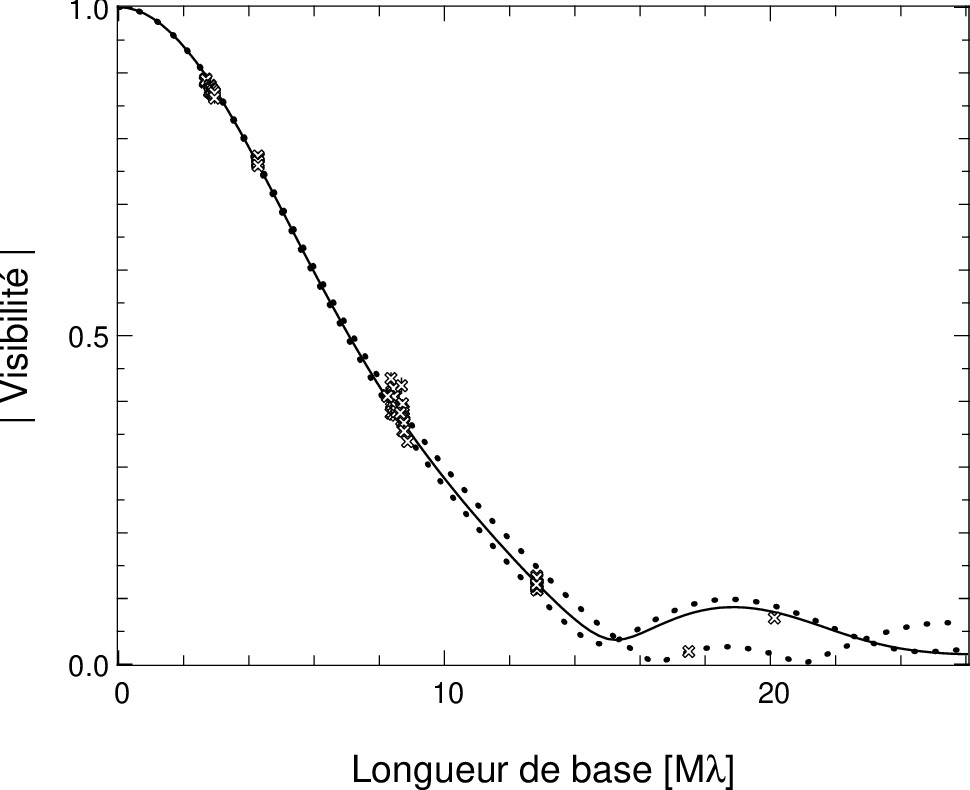}
   \includegraphics[width=7cm]{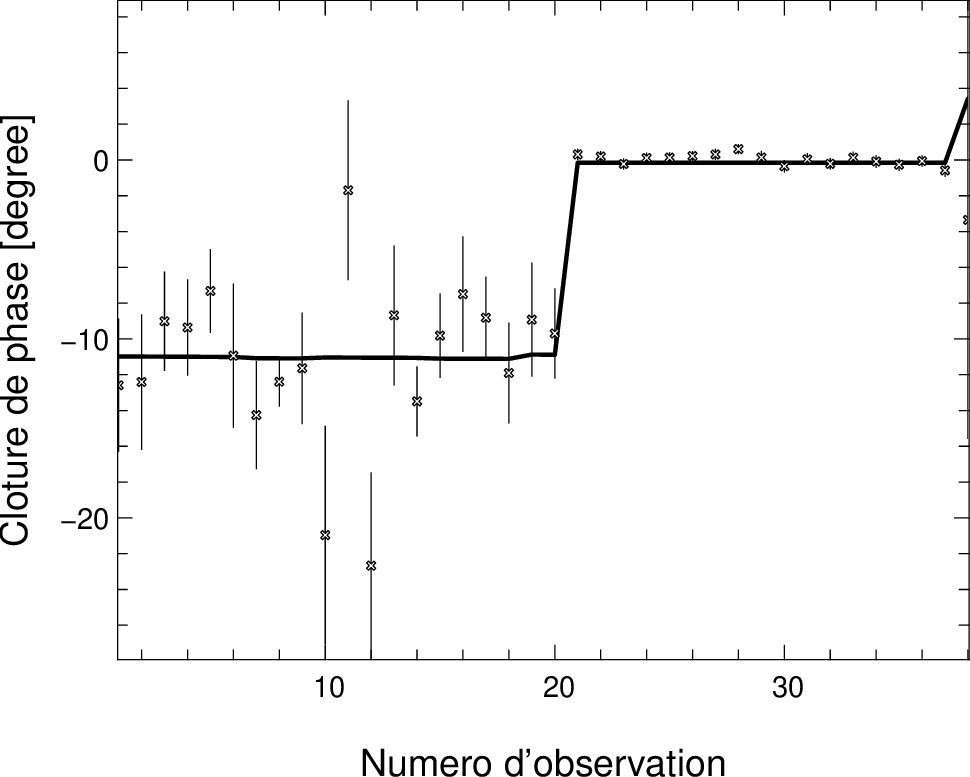}
   \includegraphics[width=7cm]{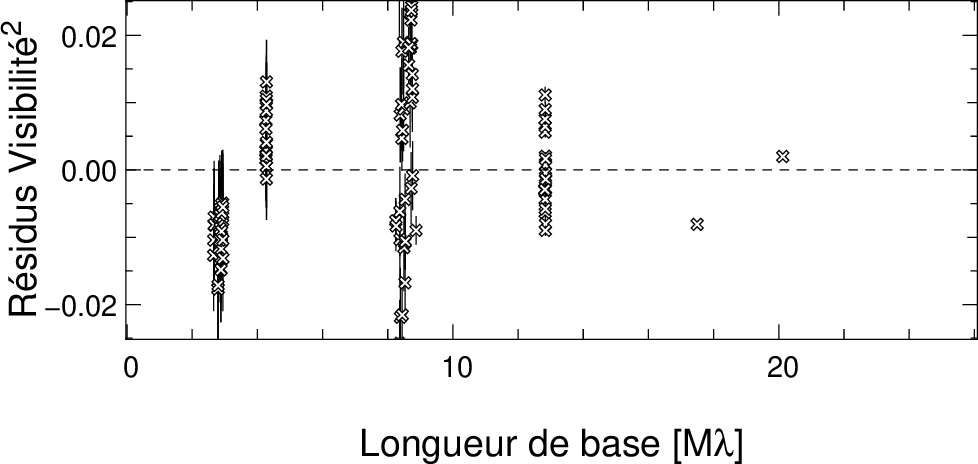}
   \includegraphics[width=7cm]{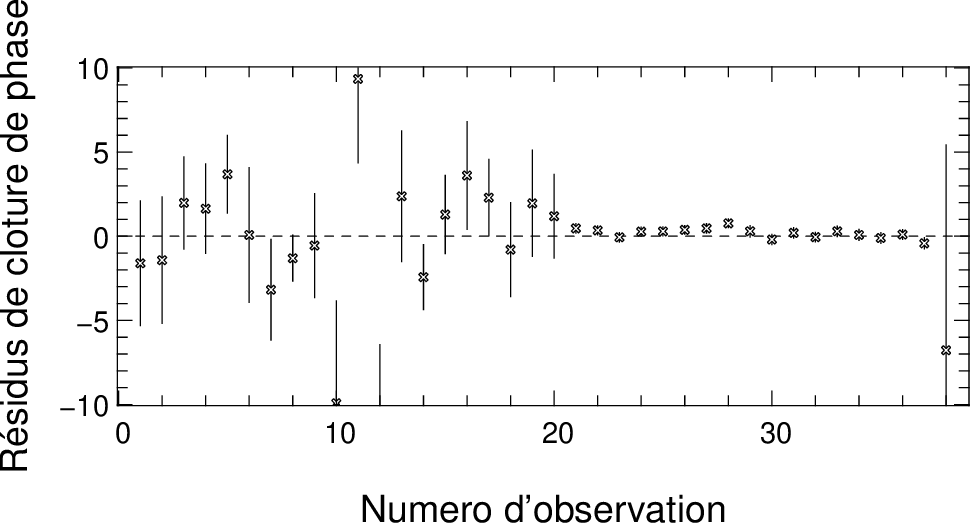}
      \caption[Ajustement des données de $\chi$ Cyg en mars 2006]{ Idem figure~\ref{fig:Data_Mai05}, mais 
      concernant les données de $\chi$ Cyg obtenues en mars 2006. }
         \label{fig:Data_Mar06}
   \end{figure}

   \begin{figure}[h!]  \centering
   \includegraphics[width=7cm]{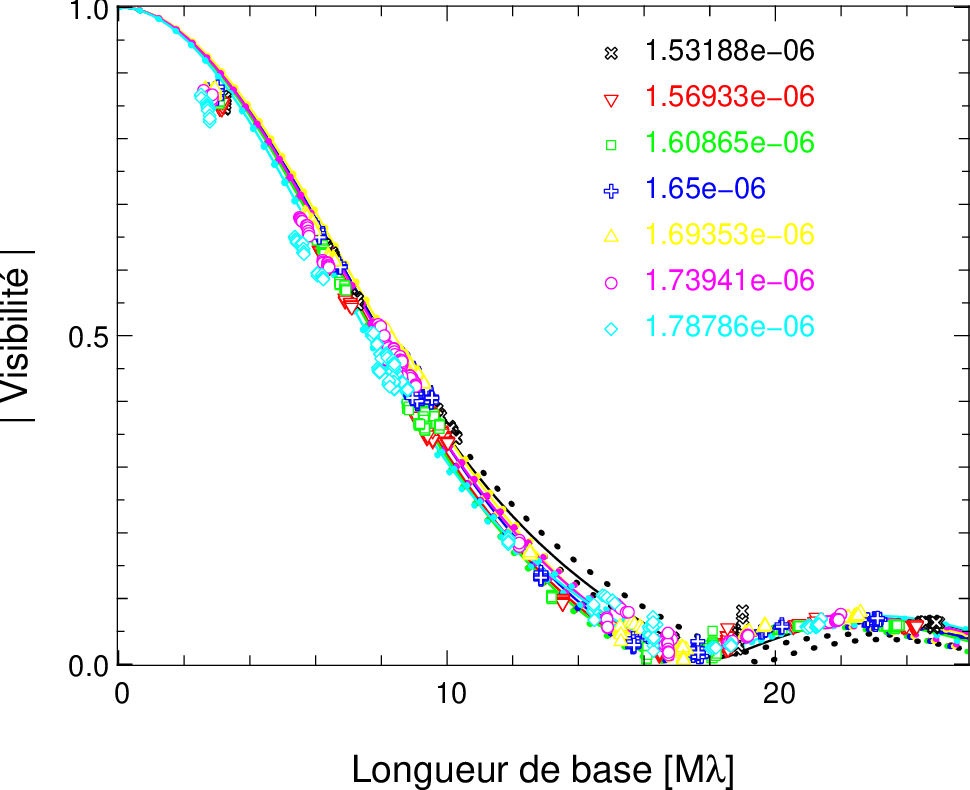}
   \includegraphics[width=7cm]{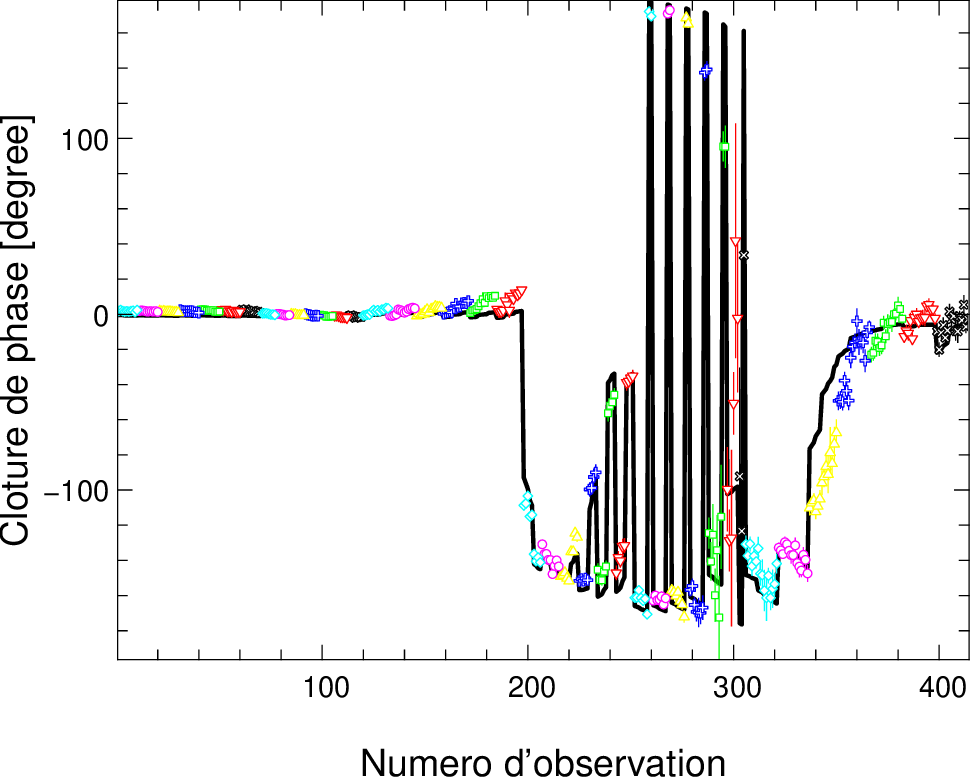}
   \includegraphics[width=7cm]{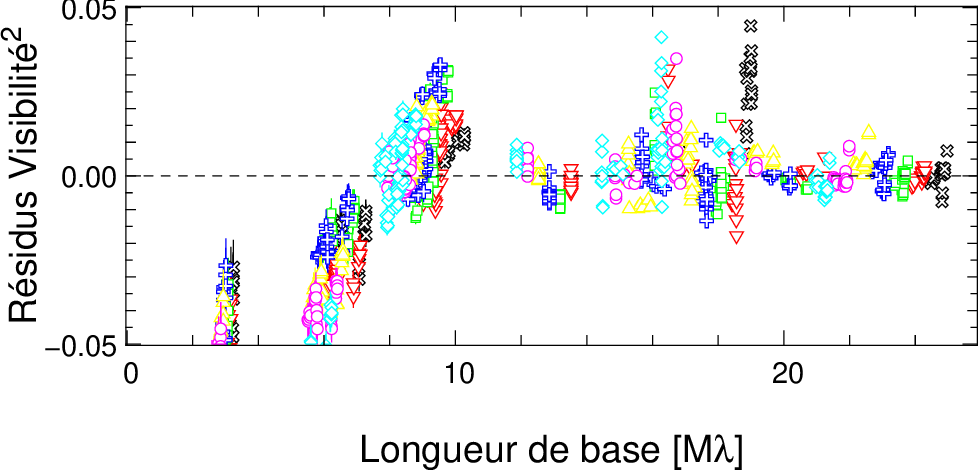}
   \includegraphics[width=7cm]{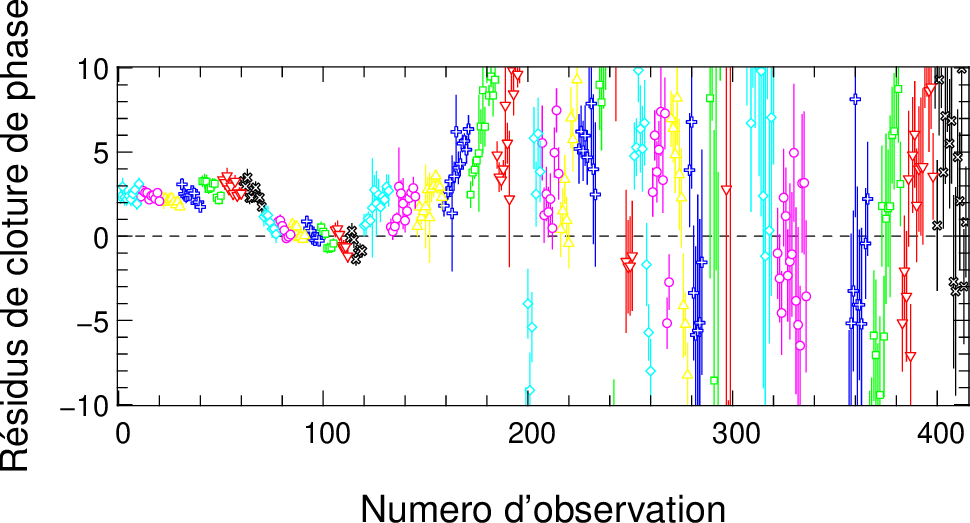}
   \caption[Ajustement des données de $\chi$ Cyg en mai 2006]{Idem figure~\ref{fig:Data_Mai05}, mais concernant les
   données de $\chi$ Cyg obtenues en mai 2006.  Les couleurs indiquent
   les différentes longueurs d'ondes obtenues grâce à l'utilisation du
   mode dispersé.}  \label{fig:Data_Mai06} \end{figure}

\begin{figure}
\centering
\includegraphics[width=7cm]{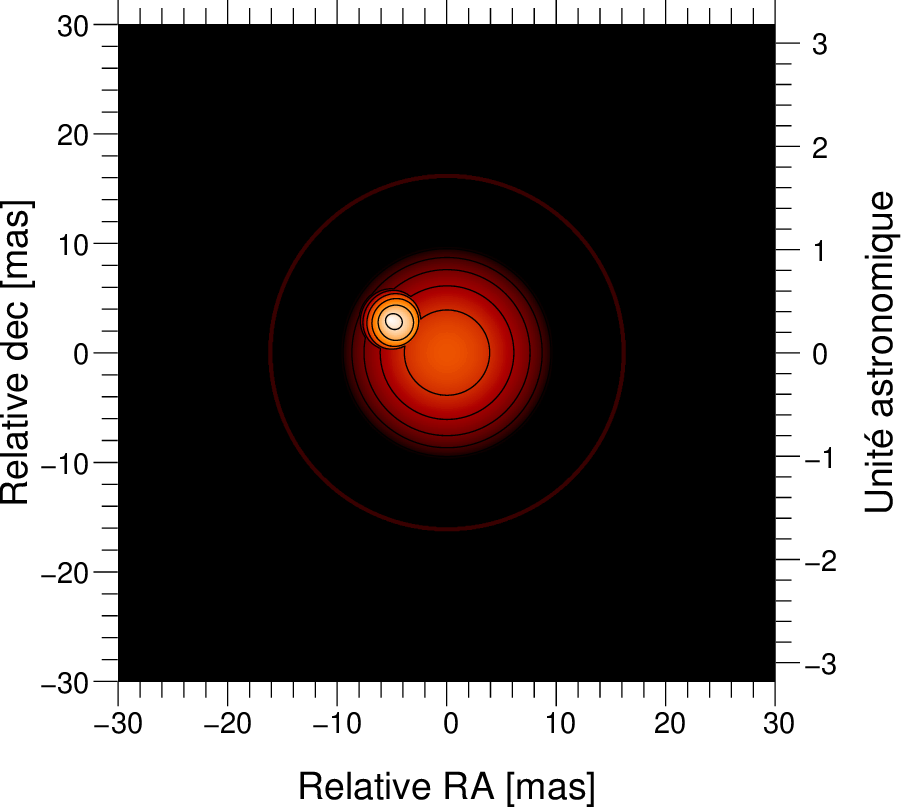}
\includegraphics[width=7cm]{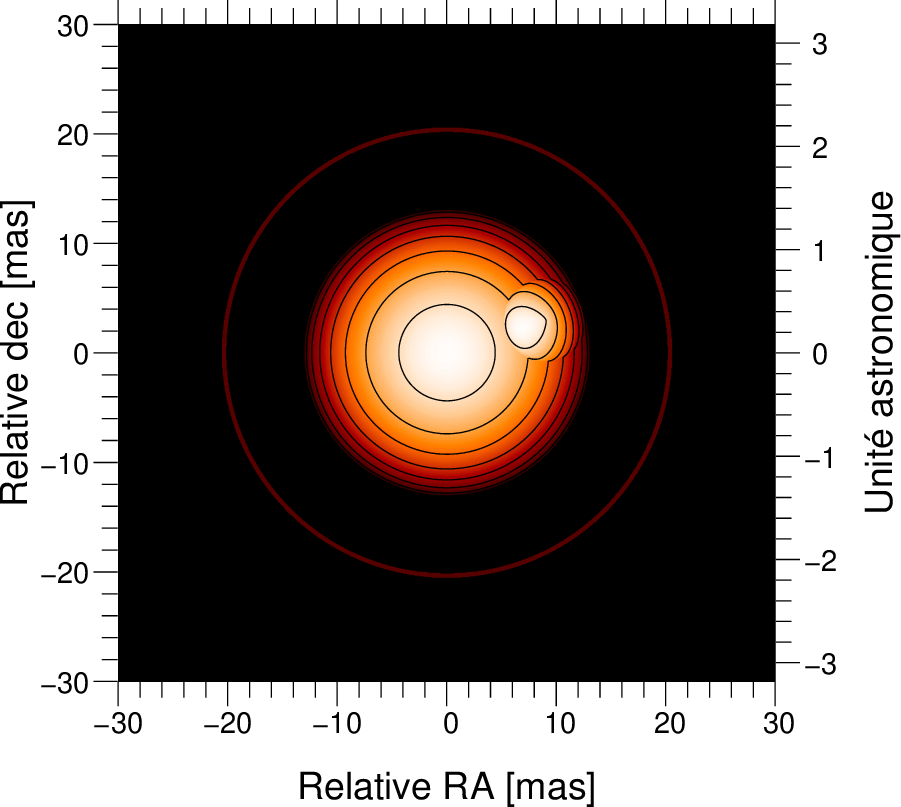}
{\vspace{.2cm}}
\includegraphics[width=7cm]{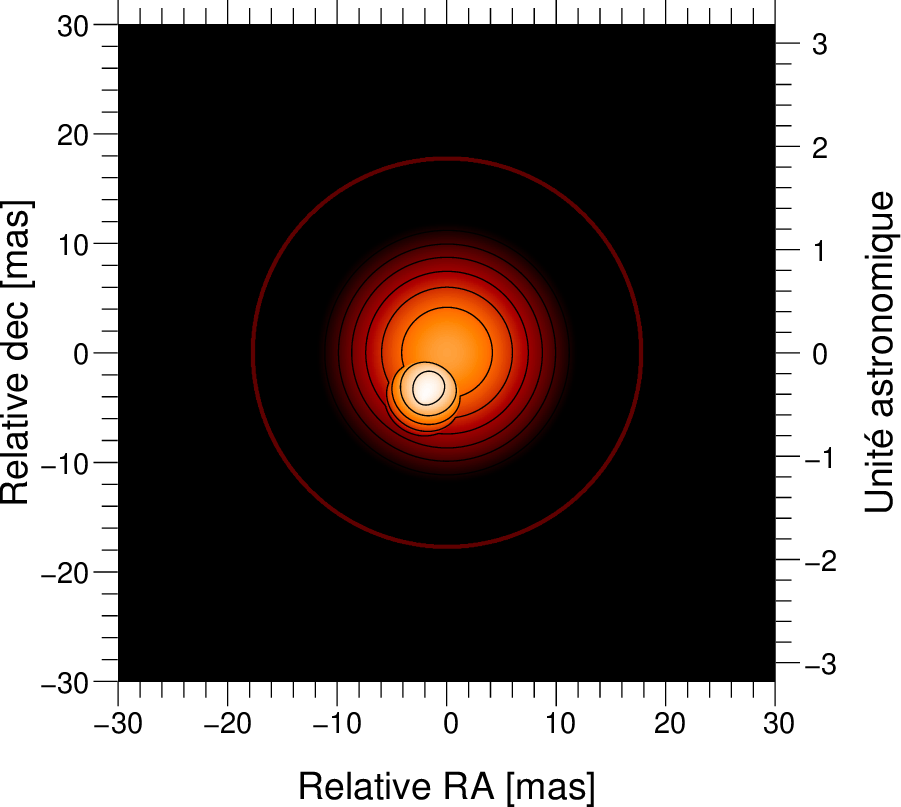}
\includegraphics[width=7cm]{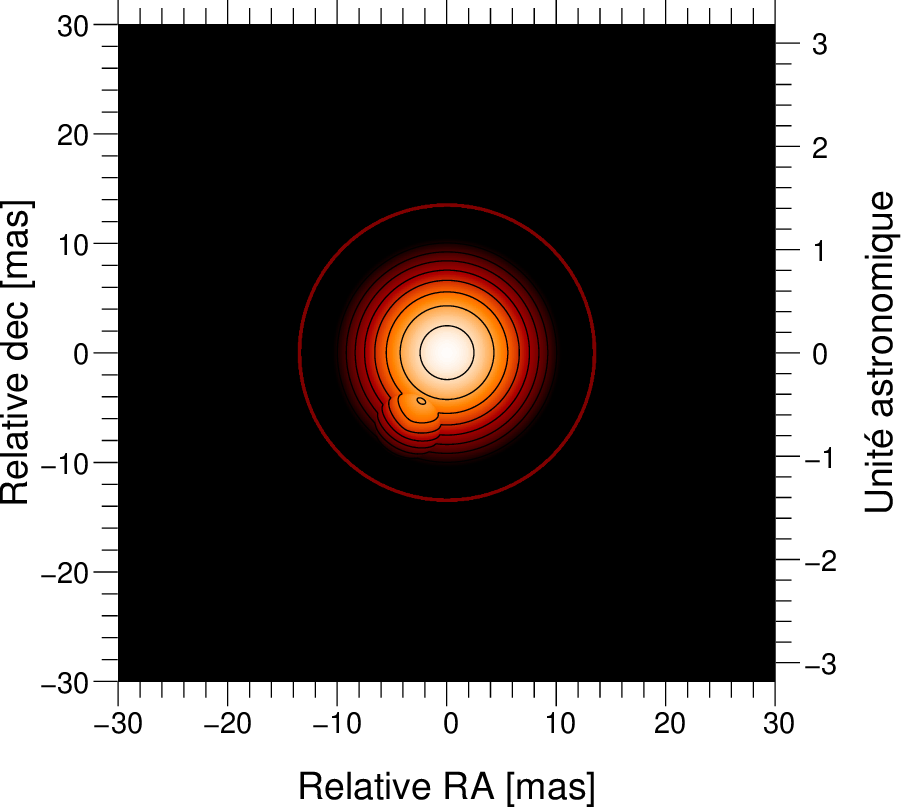}
\caption[Reconstructions par imagerie paramétrique de $\chi$ Cyg]{ 
Reconstructions d'image obtenue à partir des paramètres reportés dans
le tableau~\ref{tb:Res_fit}. En haut à gauche, $\chi$ Cyg en mai 2005,
en haut à droite, en octobre 2005, en bas à gauche en mars 2006, et
enfin en bas à droite en mai 2006.}
\label{fig:Images_param}
\end{figure}

Nous avons effectué un travail d'ajustement sur les données de $\chi$
Cyg obtenues aux mois de mai et octobre 2005, et de mars et mai
2006. Les
figures~\ref{fig:Data_Mai05},~\ref{fig:Data_Oct05},~\ref{fig:Data_Mar06}
et~\ref{fig:Data_Mai06}, montrent les ajustements obtenus. Les valeurs
optimales et les erreurs (1$\sigma$) des différents paramètres sont
reportées dans la table~\ref{tb:Res_fit}. Les $\chi^2$ réduits obtenus
sont différents de 1, ce qui signifie que les erreurs ont été
sous-estimées ou bien que le modèle n'est pas exactement fidèle à la
réalité. Pour prendre cela en compte, les barres d'erreurs des
différents paramètres ont été calculées à partir des $\chi^2$
normalisés. Cela revient à multiplier les erreurs (des $V^2$ et des
CP) par un facteur de proportionalité égal à la racine du minimum du
$\chi^2$ réduit. Il faut noter, cependant, qu'obtenir un $\chi^2$
réduit aux alentours de 5 est la preuve d'un ajustement acceptable, si
l'on considère que celui-ci a été effectué sur plusieurs centaines de
points de mesure. Le $\chi^2$ réduit de 27 obtenu sur les données de
mai 2006 est, néanmoins, troublant. Nous verrons
section~\ref{sec:cyg_couche_mol} qu'il est possible d'obtenir une
valeur bien plus faible au prix d'une complexification du modèle.

\`A partir de ces valeurs,
nous pouvons reconstruire une image dite ``paramétrique'' de l'étoile,
et ainsi observer son changement de morphologie au cours du temps
(figure~\ref{fig:Images_param}). Parce que le modèle utilisé est le
même quelle que soit la période d'observation, on peut aisément
comparer les morphologies aux différentes époques:
\begin{itemize}
\item Le diamètre de la photosphère varie. Ce n'est pas un effet de
variation d'opacité d'une quelconque couche moléculaire, mais bien
celui d'un déplacement de la limite de la photosphère. Cependant, rien ne
permet de déduire qu'il s'agit bien d'un déplacement de matière et
non pas d'une variation des propriétés du milieu.
\item La source d'asymétrie - la tache - se déplace sur la surface
stellaire bien plus vite que la simple rotation de la photosphère ne
le permettrait.
\item La position, ainsi que la brillance de la couche moléculaire,
varient au cours du temps.
\end{itemize}

\subsection{La photosphère}

\begin{figure}[h!]
  \centering
  \resizebox{\hsize}{!}{
\includegraphics{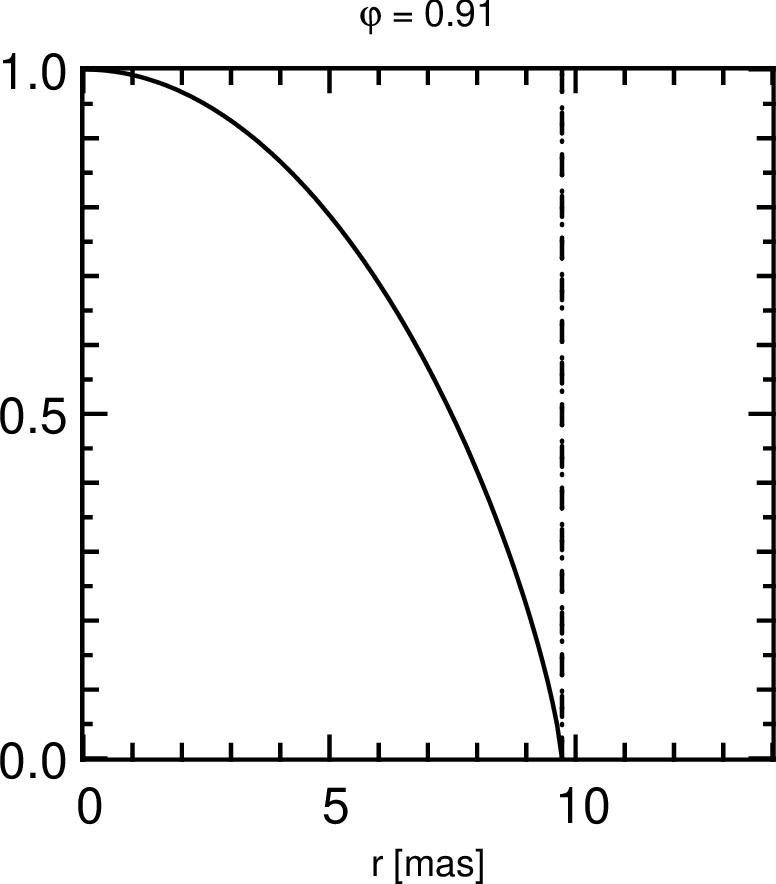}
\includegraphics{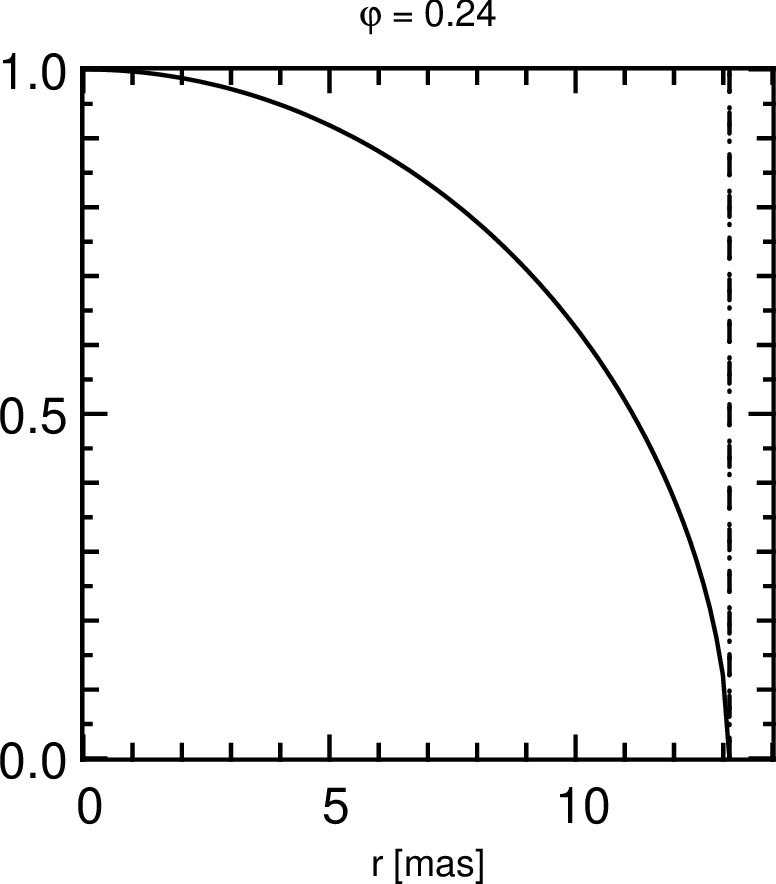}
\includegraphics{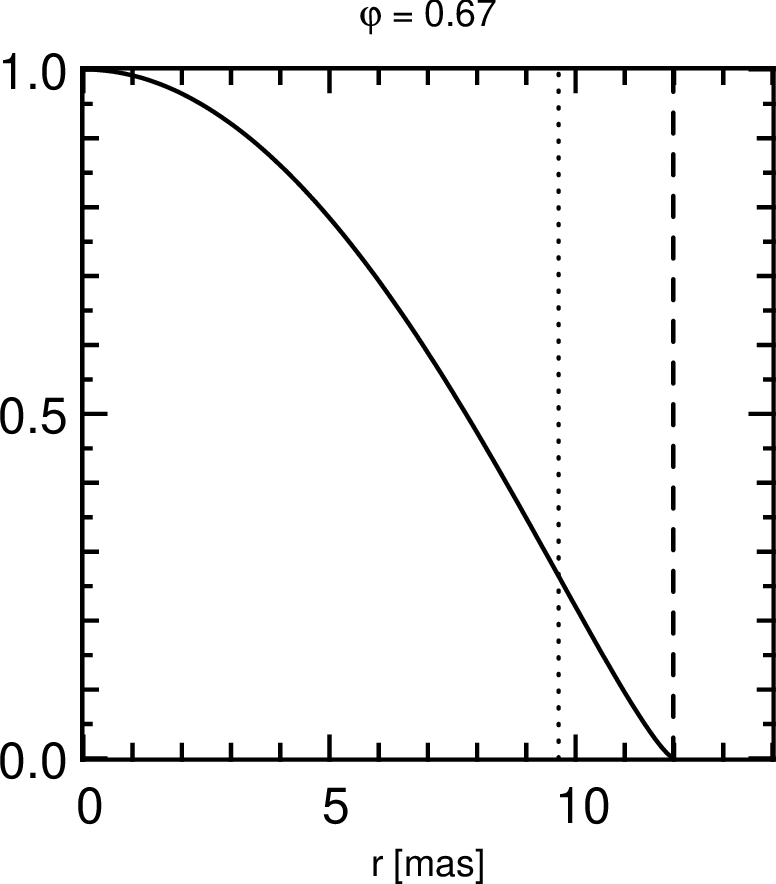}
\includegraphics{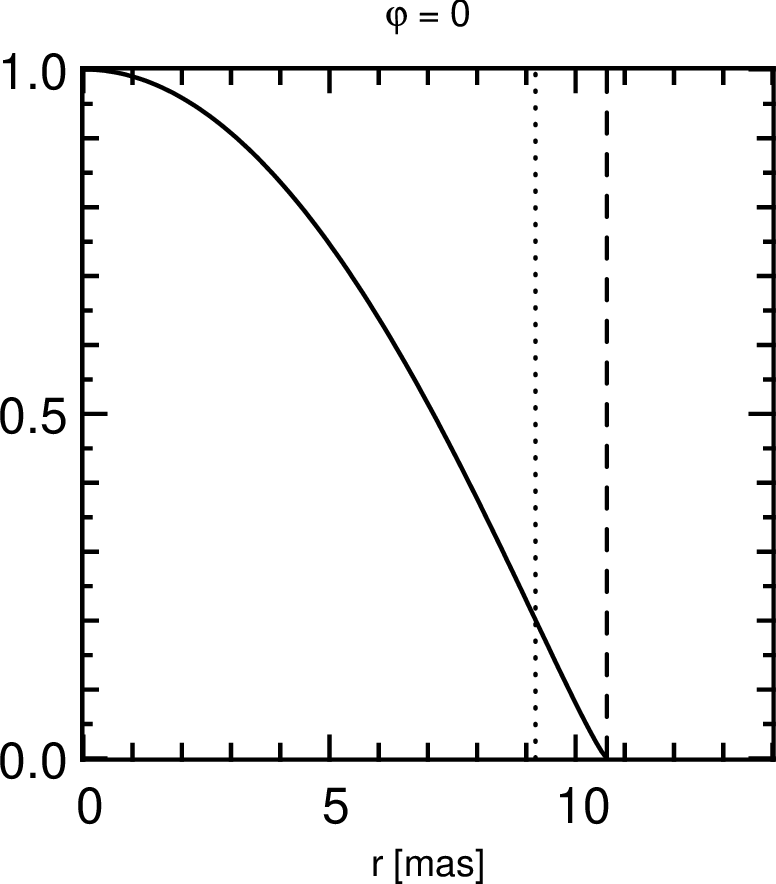}
}
  \caption[Profils d'assombrissement de $\chi$ Cyg]{Profils
  d'assombrissement de $\chi$ Cyg aux différentes phases stellaires
  ($I(r)=(1-(2r/\theta_\star)^2)^\frac{\alpha}{2}$). En pointillées les valeurs
  du diamètre de Rosseland. En tirets, les valeurs de $\theta_\star/2$.}
  \label{fig:Profils}
\end{figure}

\subsubsection{Le diamètre de Rosseland}

Les paramètres fondamentaux des étoiles variables sont la masse, la
température et le diamètre, les deux derniers termes étant variables
(la variation de la masse peut être négligée à des échelles de temps
sub-millénaires). Dans le cas d'une Mira, l'atmosphère est très
étendue, et il est donc nécessaire de s'accorder sur une définition du
diamètre. Le diamètre de Rosseland est celui généralement utilisé pour
simuler l'évolution de ces étoiles. Cette valeur, théorique, est fixée
dans les modèles, par la couche pour laquelle l'opacité, intégrée sur
$\lambda$, atteint 1. Pratiquement, une telle mesure est
impossible. Cependant, les modèles d'atmosphère montrent que, pour une
observation dans une bande proche du continu (c'est presque le cas en
bande H), une bonne approximation du diamètre de Rosseland peut être
obtenue par la mesure du point d'inflexion de l'assombrissement
centre-bord. Ceci est un argument supplémentaire justifiant
l'utilisation d'un modèle où l'assombrissement est pris en
compte. Dans le cas d'un assombrissement en loi de puissance
$I(r)=(1-(2r/\theta_\star)^2)^{\alpha/2}$, le calcul de la dérivée
nous permet d'obtenir :
\begin{eqnarray}
 \theta_{\rm Ross}&=&\theta_\star \quad {\rm si}  \quad  \alpha \leq 2
 \nonumber \\
 &=&\theta_\star/\sqrt{\alpha-1}  \quad  {\rm si} \quad  \alpha > 2 \,
 .
\label{eq:ross}
\end{eqnarray}

\subsubsection{La photosphère et l'assombrissement centre bord}

Dans le cadre de nos séries d'observations de $\chi$ Cyg, nous pouvons
en déduire la variation du diamètre de la photosphère en fonction du
temps. Nous avons utilisé les diamètres de Rosseland, corrigés de
l'assombrissement par les relations~(\ref{eq:ross}).
La figure~\ref{fig:Temp_evol} représente la variation de la taille
angulaire de $\chi$ Cygni en fonction de la phase. Sur nos données nous
avons pu ajuster une sinusoïde d'amplitude 8 mas, et de moyenne 22,4
mas. D'après la parallaxe obtenue par {\it Hypparcos} (ESA, 1997) de
9,43 $\pm$ 1,36 mas, ces variations équivalent à un rayon de l'étoile
variant de 1 au à 1,5 au. De tels rayons stellaires sont comparables
aux distances Terre-Soleil et Mars-Soleil.

L'assombrissement centre-bord varie, lui aussi, fortement en fonction
de la phase, avec des valeurs comprises entre $\alpha= 1$ et 2,5. On
constate que l'assombrissement est plus marqué lors de la contraction
de l'étoile que lors de son expansion. Ceci est en accord avec les
prédictions de \citet{1987A&A...186..200S}, ainsi qu'avec les
simulations d'assombrissement plus récentes
\citep{2002MNRAS.336.1377J}.  Il n'est d'ailleurs pas surprenant de
constater que ces valeurs sont nettement supérieures aux mesures
d'assombrissement présentes dans la littérature pour d'autres types
d'étoiles. \`A titre d'exemple,
\citet{2006A&A...453..155M} ont mesuré $\alpha=$ 0,16 sur Polaris, et
nous même n'avons obtenu {\em que} 0,30 sur Arcturus.

A la lumière du fort assombrissement obtenu, il est normal de
mesurer des tailles angulaires supérieures à ce qui a déjà été mesuré.
De manière générale, la mesure de la taille angulaire de l'objet dépend
fortement du modèle de photosphère utilisé. Par exemple,
\citet{2000MNRAS.318..381Y} ont ajusté une Gaussienne à leurs mesures
interférométriques (COAST) de $\chi$ Cyg et ont obtenu une largeur à
mi-hauteur de $13,9\pm0,8$ mas à la phase 0,83. Un tel diamètre est 44\%
inférieur à notre mesure, ce qui s'explique clairement par le choix
d'un modèle gaussien \citep{1998A&A...339..846H}. Plus récemment, 
\citet{2004A&A...426..279P} ont obtenu $21,10 \pm 0,02$ mas à $\phi=
0,24$ et $16,12 \pm 0,12$ à $\phi = 0,76$. Ces mesures sont environ
25\% inférieures aux nôtres, mais cette différence s'explique
également par le choix du modèle de la photosphère qui, dans leur cas,
était un disque uniforme, assombri par l'opacité de la couche
moléculaire.  Pour vérifier cette explication, nous avons à nouveau
fait des ajustements à $\alpha=0$. Les valeurs que nous avons obtenues
sont alors proches de celles de \citet{2004A&A...426..279P} et de
\citet{2000MNRAS.315..635Y}. Concrètement, cela signifie qu'un modèle
de photosphère assombri par le seul effet d'une couche moléculaire
risque de sous-estimer l'assombrissement, et en conséquence de
sous-estimer le diamètre angulaire de la photosphère.

\subsubsection{La température effective}

\begin{table}[h!]
\caption{Estimations de Flux bolométrique \citep{2000MNRAS.319..728W}}
\label{tb:flux}
\centering
\begin{tabular}{lrrrr}
\hline
\hline
$\phi$ & \multicolumn{1}{c}{0,91} & \multicolumn{1}{c}{0,24} & \multicolumn{1}{c}{0,67} & \multicolumn{1}{c}{0,76} \\
JD    & \multicolumn{1}{c}{2453518} & \multicolumn{1}{c}{2453653} & \multicolumn{1}{c}{2453826} & \multicolumn{1}{c}{2453867} \\
\hline
J (mag) & $0,00\pm0,15$  &   $-0.46\pm0,15$ & $0,15\pm0,15$ & $0,07\pm0,15$ \\
H (mag) & $-1,00\pm0,15$ &   $-1.65\pm0,15$ & $-1,05\pm0,15$ &  $-1,01\pm0,15$   \\
K (mag) & $-1,65\pm0,15$ &   $-2.24\pm0,15$ & $-1,73\pm0,15$  &  $-1,65\pm0,15$  \\
L (mag) & $-2,50\pm0,15$ &   $-2.84\pm0,15$ & $-2,51\pm0,15$  & $-2,48\pm0,15$  \\
$F_{\rm Bol}$ (10$^{-13}$W\,cm$^{-2}$)  
 & $6,83\pm0,38$ & $10,15\pm0,57$ & $6,80\pm0,35$ & $6,74\pm0,36$\\
\hline
\end{tabular}
\end{table}

\begin{table}[h!]
\caption{Paramètres Physiques de $\chi$ Cyg}
\label{tb:physics}
\centering
\begin{tabular}{lcccc}
\hline
\hline
$\phi$ & \multicolumn{1}{c}{0,91} & \multicolumn{1}{c}{0,24} & \multicolumn{1}{c}{0,67} & \multicolumn{1}{c}{0,76}  \\
\hline
$\theta_\star$ mas &$19.45 \pm 0.09 $ & $ 26.25 \pm 0.08 $ & $ 23.97 \pm 0.80 $ & $21.27 \pm 0.11 $\\
$R_\star^{\mathrm{b}}$&$222 \pm 31 R\ \odot $ & $ 299 \pm 42 R\ \odot$
& $ 273 \pm 39 \ R\odot$ & $243 \pm 34 \ R\odot$ \\
$T_\star$  &$2717 \pm 44 $ K & $ 2578 \pm 40 $ K& $ 2441 \pm 72 $ K& $2585 \pm 41$ K \\
 \hline
$R_{\rm couche}/R_\star$ & $1.66 \pm 0.02$& $1.55 \pm 0.02 $ & $ 1.84 \pm 0.21 $ & $ 1.48 \pm 0.03 $  \\
${T_{\rm couche}^{ex}}^\mathrm{c}$ & $2400\pm 200$ K& $3200\pm 200$ K& $ 2650\pm 200$ K& $ 2550\pm 200$ K \\
${T_{\rm couche}^{eff}}^\mathrm{d}$ & $1824 \pm 29$ K&$1795 \pm 28 $ K& $ 1723 \pm 110 $ K& $ 1994 \pm 41 $ K \\
${\tau_{\rm couche}}^{\mathrm{e}}$ & $0.043 \pm 0.002$ &$0.032 \pm 0.002  $ & $ 0.050 \pm 0.012 $ & $ 0.061 \pm 0.003 $   \\
\hline
\end{tabular}
\begin{list}{}{}
\item[$^{\mathrm{a}}$] Correction de l'assombrissement effectué par
le biais de l'équation~(\ref{eq:ross}).
\item[$^{\mathrm{b}}$] En utilisant {\it Hipparcos} distance de $106
  \pm 15$ pc.
\item[$^{\mathrm{c}}$] Température d'excitation obtenue à partir du
graphique figure~\ref{fig:Hinkle}.
\item[$^{\mathrm{d}}$] Température effective d'après
l'équation~(\ref{eq:gray}) (atmosphère grise).
\item[$^{\mathrm{e}}$] A partir de l'équation~(\ref{eq:tau}).
\end{list}
\end{table}

\begin{figure}
\centering
\includegraphics[width=11cm]{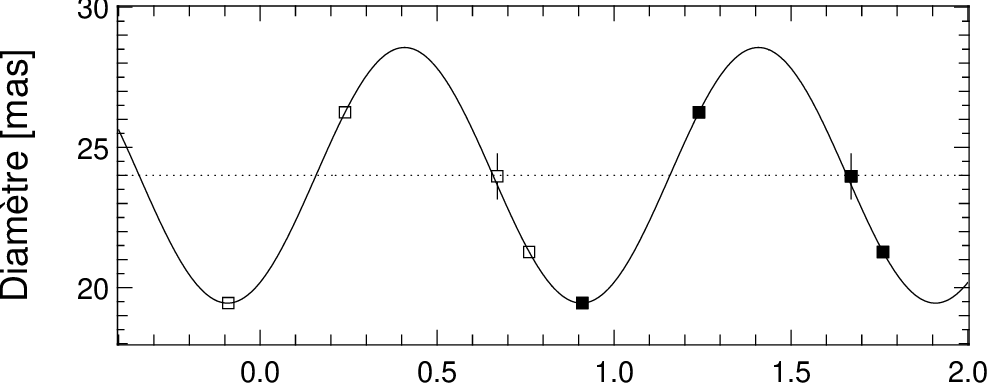}
\includegraphics[width=11cm]{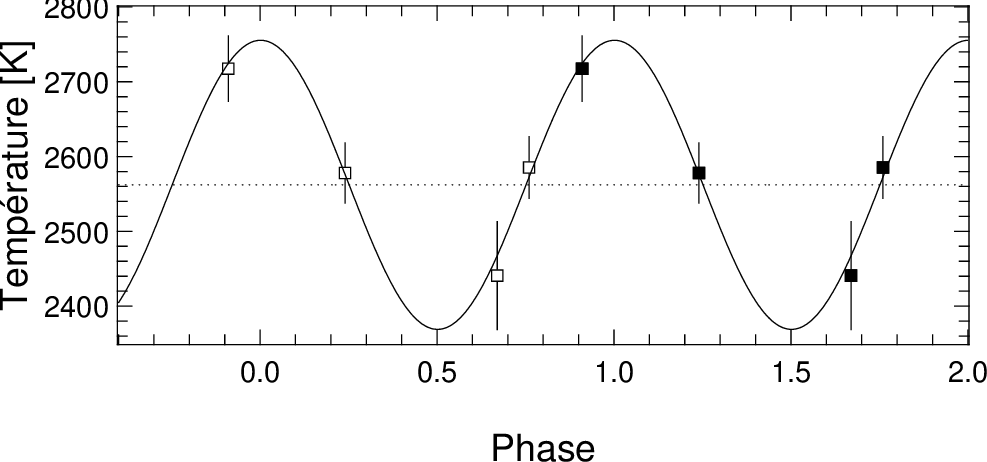}
\includegraphics[width=7cm]{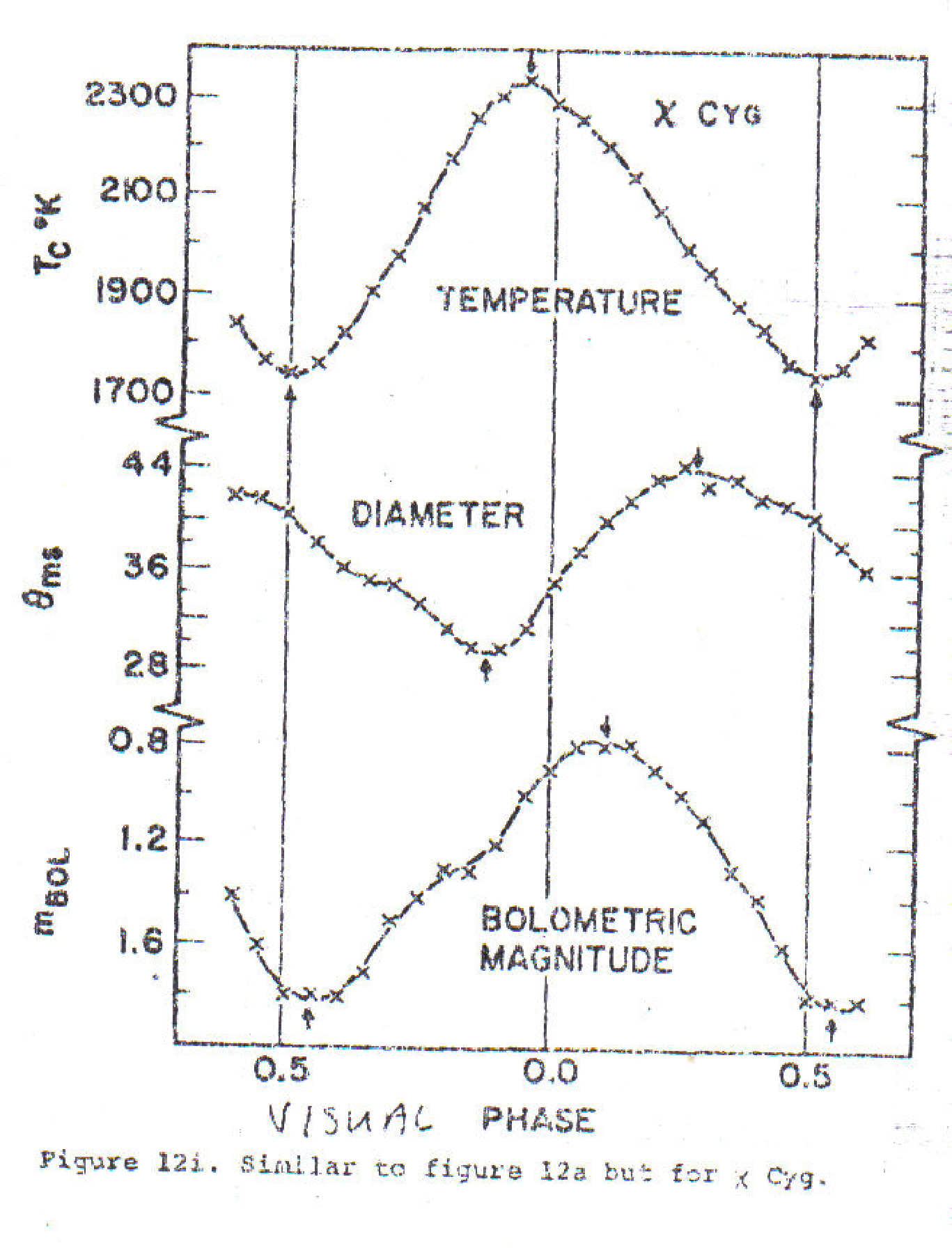}
\caption[Evolution temporelle des paramètres physiques de $\chi$
  Cyg]{ Evolution temporelle des paramètres physiques de $\chi$
  Cyg. Pour permettre une meilleure vision de la périodicité, nous
  avons reproduit ici les mesures sur deux cycles. En haut est
  représenté la variation du diamètre de la photosphère, et en bas la
  température effective. Note: Les diamètres angulaires utilisés sont
  les valeurs $\theta_\star$ représentés
  figure~\ref{fig:Profils}.  Note2 : Pour comparaison, nous avons ici
  scanné un graphique de la thèse de \citet{1973.Strecker}. Les
  températures obtenues ont été dérivées de relevés photométriques à
  3,5 microns.}
\label{fig:Temp_evol}
\end{figure}

Cette différence, considérable, sur le diamètre de l'étoile, a
également des conséquences sur la mesure de la température de la
photosphère. Celle-ci s'obtient via le flux bolométrique par la
relation:
\begin{equation}
\sigma \,.\, T_\star^4 = \frac{4}{\theta_\star^2} \,.\, F_{\rm Bol} \,,
\end{equation}
avec $\sigma$ la constante de Stefan-Boltzmann. Cette relation
s'écrit aussi, par l'utilisation d'unités adaptées, de la façon suivante
\citep{1998A&A...331..619P}:
\begin{equation}
T_\star = 7400 \left(\frac{F_{\rm Bol}}{10^{-13}\,{\rm
W\,cm^{-2}}}\right)^\frac{1}{4} \left(\frac{1\,{\rm
mas}}{\theta_\star}\right)^\frac12 \ {\rm K}\,.
\label{eq:tp}
\end{equation}

Le flux bolométrique
dépend de la phase stellaire et nécessite une mesure de flux dans
l'ensemble des différentes bandes spectrales. La phase a été obtenue
en utilisant le catalogue de mesures photométriques de l'AAVSO. De
plus, nous avons utilisé l'article de \citet{2000MNRAS.319..728W} qui
référence, pour un grand nombre d'étoiles Mira, la magnitude dans la
bande J, H, K, et L au cours du temps. Nous avons
ensuite ajusté un corps noir sur ces données, et calculé le flux
bolométrique par intégration de celui-ci.

La figure~\ref{fig:Temp_evol} représente la variation de taille
angulaire de $\chi$ Cygni en fonction du temps. Les températures,
elles aussi, suivent une courbe sinusoïdale, avec une température
maximale au diamètre minimum. Le maximum de température est déphasé
d'environ 0,5 par rapport au maximum du flux bolométrique. Cela
signifie que, même si les variations de température et les variations
du diamètre contribuent toutes deux aux variations du flux
bolométrique, celles-ci sont cependant dominées par la variation du
diamètre de la photosphére.

\subsection{La couche moléculaire }
\label{sec:cyg_couche_mol}

\subsubsection{Observations déjà existantes}

\begin{figure}[h]
  \centering
  \resizebox{\hsize}{!}{\includegraphics{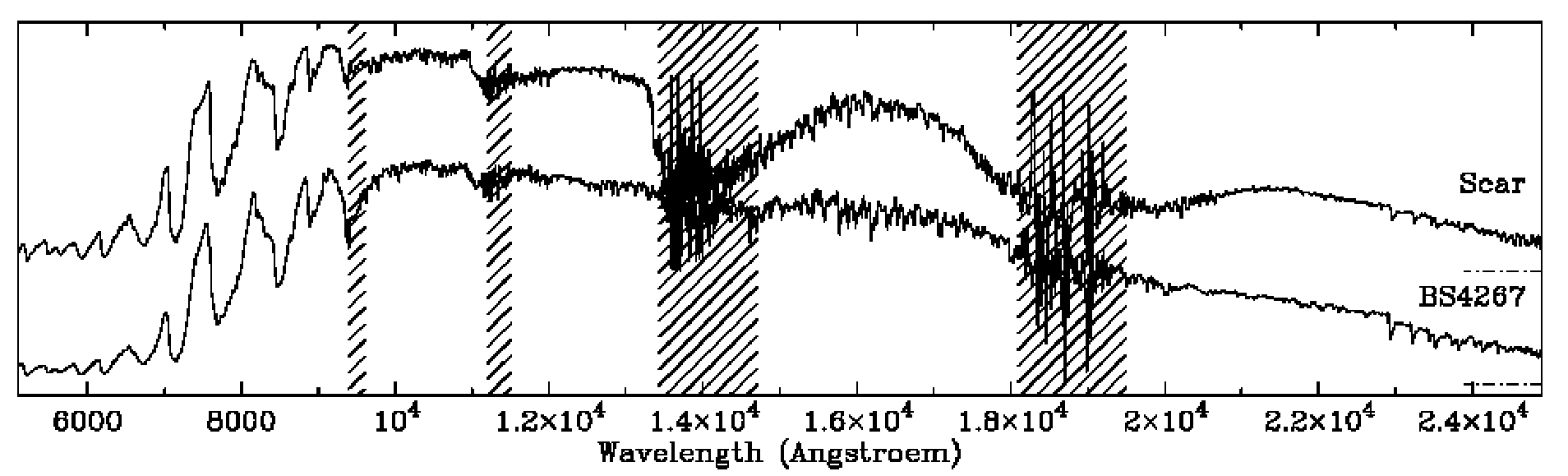}}
  \caption[Observations spectroscopiques de la géante M5 BS4267 et de
  la Mira S\,Car par \citet{2000A&AS..146..217L}]{ Observations spectroscopiques de la géante M5 BS4267 et de
  la Mira S\,Car par \citet{2000A&AS..146..217L}. En bande H, on peut
  constater l'importante absorption de l'eau (1,5 et 1,8 $\mu$m) et, dans
  une moindre mesure, du monoxyde de carbone (1,6 $\mu$m). Les bandes
  grisées marquent les zones d'absorption tellurique. }
  \label{fig:Lancon}
\end{figure}

\begin{figure}
\centering 
  \resizebox{\hsize}{!}{
\includegraphics{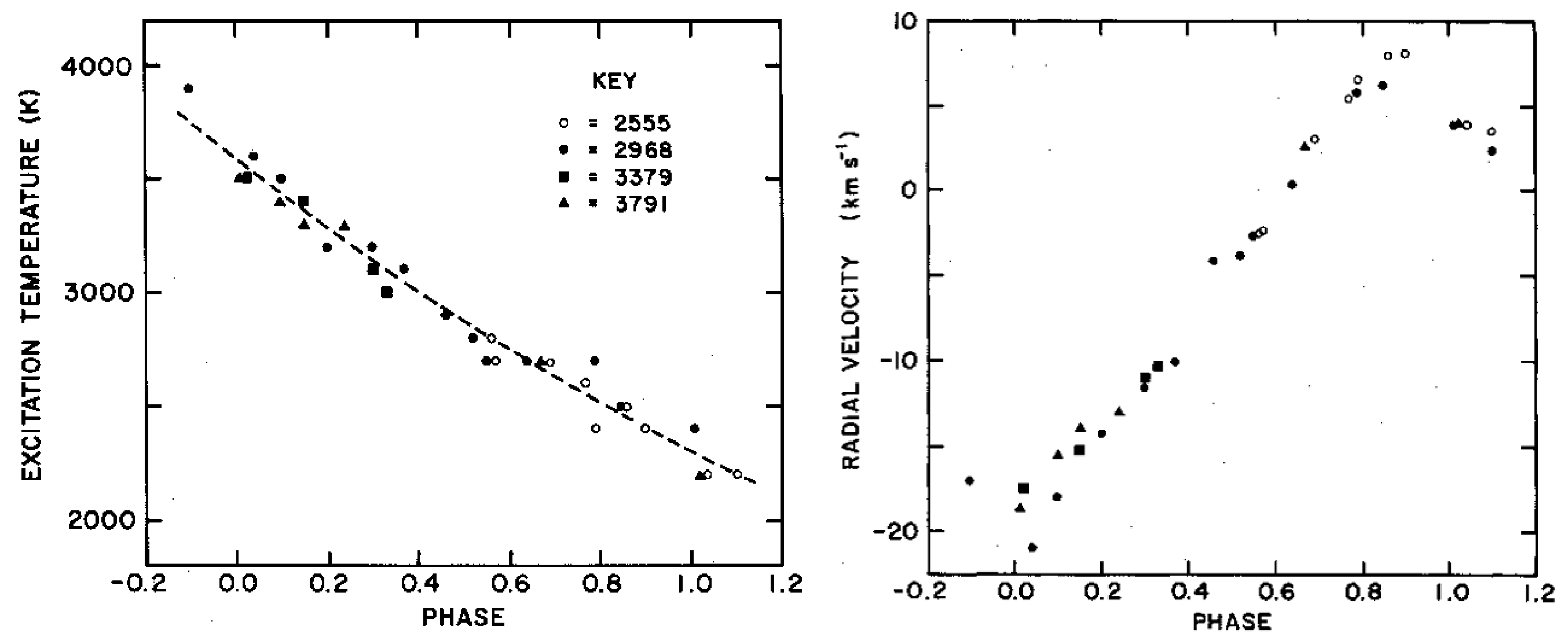}}
   \caption[Température d'excitation et vélocité radiale des raies
   d'absorptions \citep{1982ApJ...252..697H}]{ A gauche, température
   d'excitation de la couche moléculaire. A droite, vélocité radiale
   des raies d'absorptions. Ces données ont été obtenues par
   \citet{1982ApJ...252..697H} à partir des raies en absorption du CO
   ($\Delta v =3$) que nous considérerons comme traçant le déplacement
   du gaz moléculaire chaud dans l'atmosphère de la Mira $\chi$ Cyg.}
   \label{fig:Hinkle}
\end{figure}

$\chi$ Cygni est connue pour son importante émission en H$\alpha$ que
l'on présente communément comme la conséquence du passage récurrent
d'une onde de choc supersonique dans la haute atmosphère. L'absorption
de molécules comme H$_2$O et CO peut aussi être mise en évidence (voir
la Mira S\,Car figure~\ref{fig:Lancon}). La majorité de ces molécules
se situe dans l'atmosphère froide circumstellaire. Cependant, comme on
peut le voir sur nos données, une partie de ces molécules est présente
sous une forme chaude, très proche de l'étoile. Ce résultat, confirmé
par spectro interférométrie, est cependant troublant. Comment une
telle couche peut exister alors que le champ gravitationnel est
relativement intense, et alors même que la température empêche la
formation de poussières et donc d'une poussée radiative?  La réponse
se trouve certainement dans une conjonction de facteurs. Le premier
serait l'influence d'un ``effet réfrigérateur'' présent dans
l'atmosphère des Miras \citep{2000ARA&A..38..573W}. Cet effet apparaît
dans les zones où la pression diminuerait brutalement, conséquence de
la divergence entre la trajectoire du choc et celle de la matière
retombant sous l'effet de l'accélération gravitationnelle de
l'étoile. A cette endroit, la température pourrait descendre sous les
1700 K, température à partir de laquelle pourrait se former certaines
espèces de poussières à base d'alumine (Al$_2$O$_3$). Dernièrement,
\citet{2007..Perrin} ont d'ailleurs confirmé la présence d'alumine dans
l'atmosphère de Bételgeuse, apportant la preuve de sa formation à une
distance proche de la photosphère ($\approx$ 1,35 R$_\star$).

Dans le cas de l'atmosphère pulsante des étoiles Mira, la
caractérisation de l'influence de l'onde de choc sur la couche
moléculaire reste, cependant, de la spéculation. Une partie de
l'information peut être trouvée grâce aux données
spectroscopiques. Afin d'observer le déplacement de la photosphère de
$\chi$ Cyg, \citet{1982ApJ...252..697H} ont mesuré la vitesse radiale
du monoxyde de carbone fortement ionisé ($\Delta v= 3$). Ils n'ont pas
trouvé la photosphère, mais ont mesuré un gaz chaud se formant
périodiquement à la phase de pulsation nulle, en chute libre jusqu'à
la phase stellaire 0,8, et se dissociant ensuite pour réapparaître au
cycle suivant. A partir des différents niveaux d'excitation
rotationels, ils ont aussi pu en déduire la température d'excitation
de ce gaz. Ils ont alors mesuré une très haute température ($\approx$
3500 K) à la phase nulle, correspondant à la création de la couche. La
température décroît ensuite de manière exponentielle, pour converger
vers une température proche de 2000 K. Nous avons reproduit ces
résultats figure~\ref{fig:Hinkle}. Ceci concorde avec un scénario où
la couche moléculaire se forme dans la zone de post-choc, et retombe
selon une trajectoire balistique sur la surface stellaire. L'étape
suivante consiste à vérifier la compatibilité de cette hypothèse avec
nos données.

\subsubsection{La trajectoire de la couche moléculaire}

\begin{figure}[t!] \centering
   \includegraphics[width=14cm]{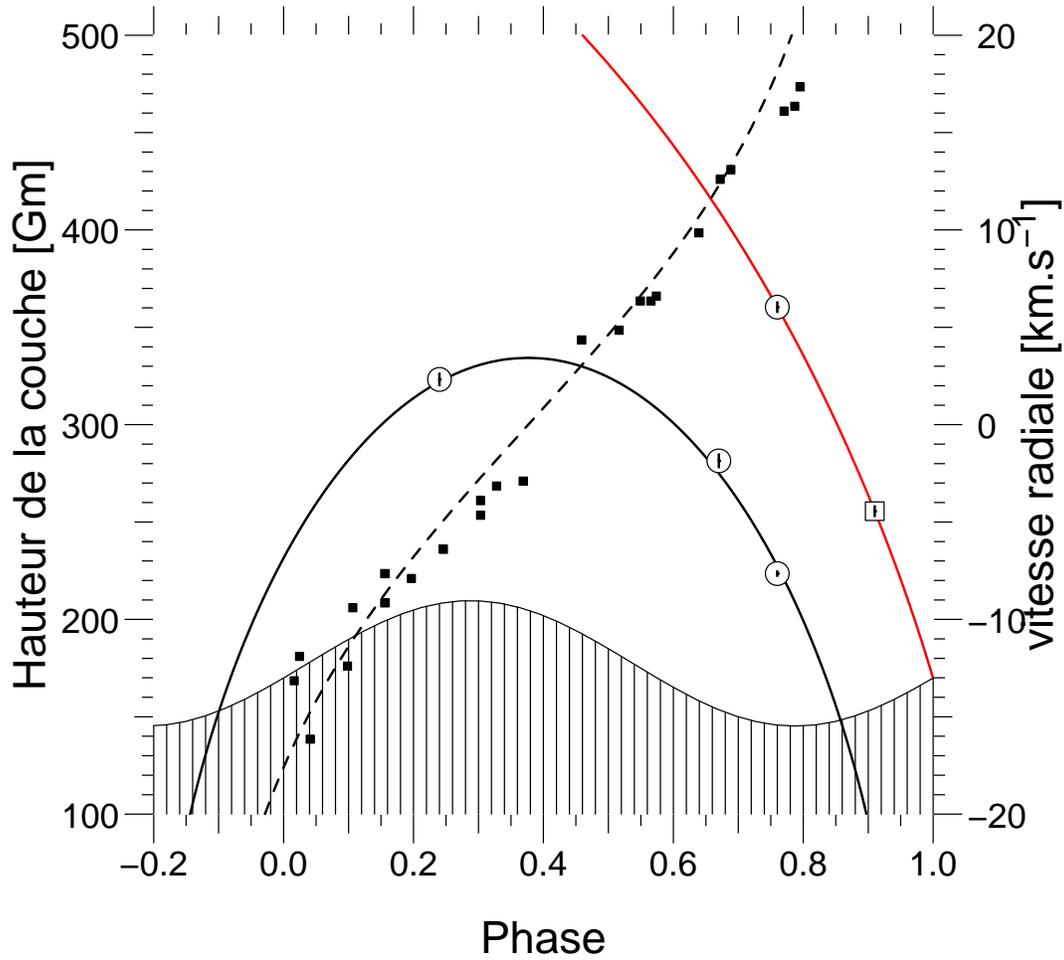}
   \caption[Modèle dynamique de l'atmosphère de $\chi$ Cyg]{ Modèle
   dynamique de l'atmosphère de $\chi$ Cyg. En traits continus sont
   représentés les trajectoires balistiques des deux couches
   moléculaires. En pointillés noir est représenté la vitesse de la
   première couche moléculaire. Les données représentées sont les
   vitesses \citep[issues de][les carrés noir]{1982ApJ...252..697H},
   et les mesures de positions de la couche moléculaire (cercles). Le
   carré correspond à la mesure de mai 2005, décalé d'une phase. Les
   erreurs sur la position des couches sont représentées par le trait
   vertical à l'intérieur des marqueurs. La zone hachurée en bas de la
   figure représente la photosphère. } \label{fig:Couche_bal}
   \end{figure}

La première étape consiste à transposer les vitesses radiales mesurées par
\citet{1982ApJ...252..697H} dans le référentiel de l'étoile. Dans leur
article, ils estiment la vitesse de l'étoile dans le référentiel
héliocentrique à 7,5 km.s$^{-1}$. Nous avons cependant utilisé une
valeur plus récente de 9,6 km.s$^{-1}$ obtenue par
\citet{1990ApJ...358..251W}. Nous avons ensuite reproduit les vitesses
de la couche moléculaire, comprises entre les phases 0 et 0,8, dans la
figure~\ref{fig:Couche_bal}. Parallèlement, nous avons reporté la
position de la couche à partir des données acquises aux phases 0,24,
0,67 et 0,76 (tableau~\ref{tb:Res_fit}), converties en mètres par le
biais de la parallaxe de l'étoile (9,43 mas). Nous avons enfin ajusté
une trajectoire balistique à l'ensemble des données, c'est à dire
ajusté à la fois aux vitesses et aux positions.

Parce que la hauteur de la couche varie considérablement,
l'accélération gravitationnelle varie elle aussi au cours de la
trajectoire. Pour en tenir compte, il a été nécessaire d'inverser le
problème. Au lieu d'ajuster les positions et vitesses, nous avons
ajusté les phases. La relation donnant le temps en fonction de la
position ($h$) est la suivante:
\begin{equation}
t(h)=t_0\pm\sqrt{\frac{h_0}{2 G M_\star}}\left(h_0
\tan^{-1}\left(\sqrt{\frac{h_0-h}{h}}\right)+\sqrt{h(h_0-h)}\right)
\,,
\label{eq:trajectoire}
\end{equation}
où $t_0$ le temps et $h_0$ la hauteur à la position maximale de la
trajectoire, $G$ la constante de gravitation universelle, et $M_\star$
la masse de l'étoile. De même, le temps peut être obtenu en fonction
de la vitesse en remplaçant dans l'équation~(\ref{eq:trajectoire}) la
position $h$ par la vitesse $v$ selon la relation :
\begin{equation}
h=\frac{2GM_\star h_0}{2GM_\star + h_0 v^2}\,.
\end{equation}

Le résultat de cet ajustement est tracé
figure~\ref{fig:Couche_bal}. On peut noter une adéquation entre
les mesures de vitesse radiale effectuées par
\citet{1982ApJ...252..697H} et nos mesures de position
par interférométrie. La hauteur maximale est de 334 $\pm$ 3 Gm
($\theta_{\rm couche} \approx 42$ mas) à une phase $\phi =$ 0,376 $\pm$
0,006. La masse de l'étoile ainsi obtenue est de 0,$88 \pm 0$,04
M$_\odot$.

Cette masse peut alors servir à confirmer le mode de pulsation de
l'étoile. Le mode (fondamental ou premier partiel), a longtemps été un
sujet de polémique entre théoriciens et observateurs
\citep{1998A&A...333..647B,1999ApJ...514L..35Y}. Il s'avère désormais
que les mesures de diamètres faites, notamment, par
\citet{1996AJ....112.2147V} ont été biaisées par l'existence de la
couche moléculaire, ce qui conduisait à surestimer les diamètres
mesurés \citep{2004A&A...426..279P}. Il est désormais admis que ces
étoiles pulsent selon leur mode fondamental. Parce que nous avons pour
$\chi$ Cyg, les trois paramètres fondamentaux : Masse, Rayon et
Période de pulsation (en jours), nous pouvons comparer nos résultats
avec la relation établie par \citet{Wood..89} :
\begin{equation}
\log (P) = -2.07 +1.94 \log\left(\frac{R}{R_\odot}\right) - 0.9 \log\left(\frac{M}{M_\odot}\right) \,.
\end{equation}
Celle-ci est obtenue à partir d'une modélisation de la structure des
Mira en présence de pulsation non-adiabatiques. \citet{Wood..89}
estime cette relation valable pour 0,$6 \lessapprox M_\star/M_\odot
\lessapprox 1$,5, et faiblement dépendante à la métallicité.
Nous avons de cette manière estimé le rayon d'une étoile de masse
0,$88 \pm 0$,04 $M_\odot$ pulsant sur son mode fondamental avec une
période de 408 jours. Le diamètre de la photosphère alors obtenu est
de $243 \pm 12$\,$R_\odot$, en négligeant l'erreur sur la parallaxe
(9,43 mas).  Ce résultat est clairement en accord avec nos mesures de
la photosphère, ce qui valide à la fois notre calcul de la masse de
l'étoile et l'hypothèse de pulsation selon le mode fondamental.

\subsubsection{L'existence d'une deuxième couche moléculaire}

\begin{figure}[t!]
\centering
\includegraphics[width=12cm]{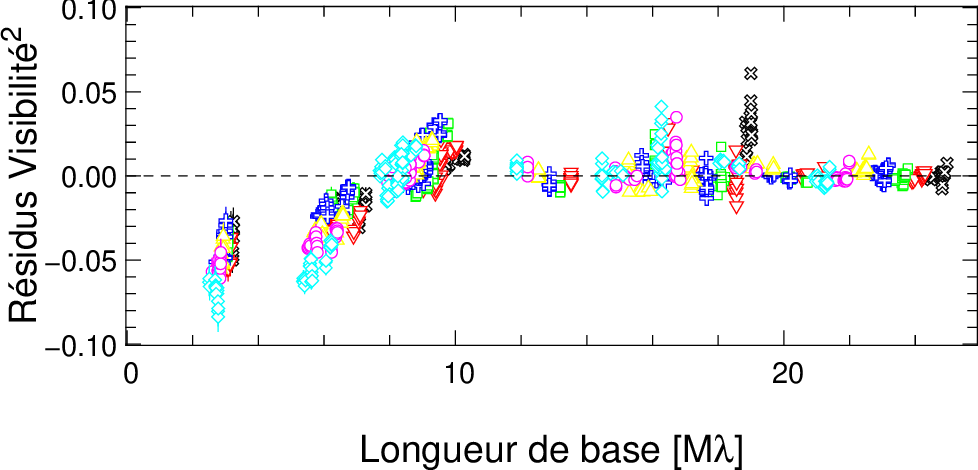}
\includegraphics[width=12cm]{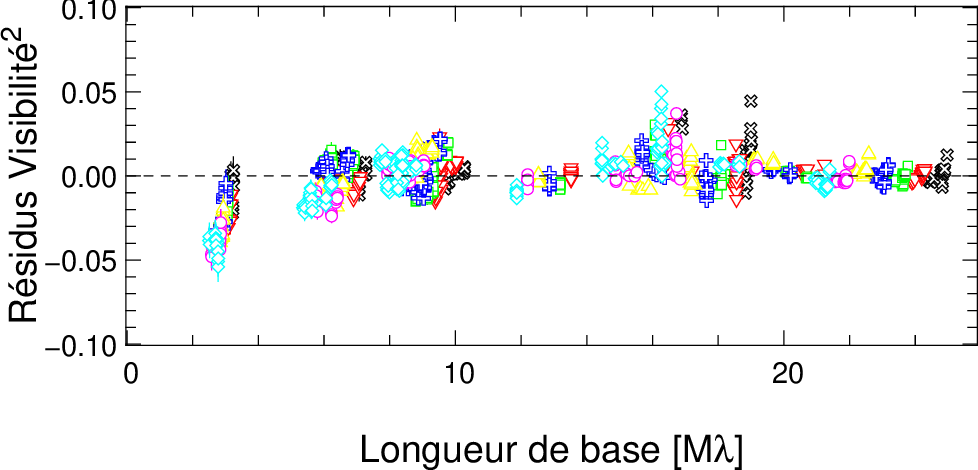}
\caption[Résidus d'ajustements des données de $\chi$ Cyg acquisent en
mai 2006]{Résidus d'ajustements des données de $\chi$ Cyg acquisent en
mai 2006. En haut, résidus pour un ajustement du modèle simple :
photosphère, couche moléculaire, et tache. En bas, résidus pour
l'ajustement d'un modèle contenant une couche moléculaire
supplémentaire. On peut remarquer que les résidus ne sont guère
modifié aux basses fréquences. On noter, cependant, une claire
amélioration de l'ajustement des basses fréquences. Le $\chi^2$ réduit
passe ainsi de 17 à 13.}
\label{fig:2iem_couche}
\end{figure}

La trajectoire de la couche moléculaire présentée
figure~\ref{fig:Couche_bal} est incompatible avec la mesure de sa
position à partir des données de mai 2005 ($\phi=0$,91). Ceci n'est
pas si surprenant sachant qu'à une telle phase, les mesures de
CO indiquent une couche soit détruite, soit au stade de
formation. Ce qui est finalement le plus surprenant, c'est que cette
couche existe. Une façon d'expliquer cette mesure est d'introduire la
présence d'une seconde couche, plus froide, et donc moins détectable à
partir des raies d'absorption $\Delta v =3$. La présence de multiples
couches moléculaires n'est d'ailleurs pas nouvelle car déjà observée,
notamment, à partir du télescope ISO \citep{1999A&A...348L..55Y}.

Pour conforter cette idée, nous avons étudié à nouveau les données
acquises en mai 2006 ($\phi=0$,76; voir
figure~\ref{fig:Data_Mai06}). Sur celles-ci, il s'avère que l'on
dénote un mauvais ajustement des basses fréquences, en partie
responsable du mauvais $\chi^2$ réduit obtenu ($\chi^2=17$). Nous
avons donc repris ces données et ajouté à notre modèle une deuxième
couche, située à une distance supérieure de la première. Les résidus
sont affichés figure~\ref{fig:2iem_couche}. L'ajustement est ainsi
nettement meilleur aux basses fréquences, avec un $\chi^2$ réduit de
13. La position de cette couche a servi à ajuster une trajectoire
balistique contrainte par la masse de l'étoile et la position observée
à $\phi=0$,91. (reportée figure~\ref{fig:Couche_bal}). Il est à noter
que cette position a été obtenue un an avant celle à $\phi=0$,76, ce
qui suppose une parfaite reproduction du phénomène entre une phase et
une autre. Cette reproductibilité est vérifiée par les données
spectroscopiques de \citet{1982ApJ...252..697H} sur la première
couche, mais il se peut que ce ne soit pas le cas pour la deuxième. La
trajectoire représentée par la courbe rouge
figure~\ref{fig:Couche_bal} n'est donc qu'une hypothèse. Elle a,
cependant, le mérite de bien mettre en évidence la présence de
plusieures couches chutant simultanément sur la surface stellaire.

\subsubsection{Température et opacité optique}

La cohérence des observations issues de différents instruments
 conforte l'hypothèse selon laquelle la couche moléculaire observée
 est bien celle détectée par \citet{1982ApJ...252..697H}. Selon ce
 scénario, la pulsation de l'étoile produit des ondes de chocs se
 répercutant dans la partie supérieure de l'atmosphère. Au contact du
 gaz retombant sur l'étoile, une zone de choc se forme et l'hydrogène
 s'ionise. Dans la zone de post-choc, les conditions de pression
 seraient alors propices à la formation de molécules qui retomberaient
 sous la forme de couche sur l'étoile \citep{2006A&A...456.1001C}. Si
 tel est bien le cas, l'opacité de cette couche devrait augmenter au
 fur et à mesure que celle-ci se contracte.  Cependant, pour obtenir
 l'opacité à partir de nos données, il est nécessaire de connaître la
 température de la couche. Une première valeur peut être obtenue
 directement à partir des mesures de la température d'excitation de la
 couche (figure~\ref{fig:Hinkle}). Néanmoins, celle-ci peut être
 différente de la température effective qui nous intéresse. Ceci
 d'autant plus que, si c'est bien dans la zone de post-choc que
 celle-ci se forme, il y a peu de chance qu'il y ait équilibre
 thermodynamique. Enfin, un dernier argument permettant d'exclure
 cette température est le fait que cette couche moléculaire est
 détectée en absorption. Cela suppose une température effective
 inférieure à celle de la photosphère, ce qui n'est clairement pas le
 cas aux phases inférieures à 0,5. Nous avons néanmoins, pour
 information, reporté ces valeurs dans le tableau~\ref{tb:physics}.
 C'est pourquoi nous avons estimé qu'il était préférable de calculer
 la tempèrature effective à partir de l'approximation d'une atmosphère
 grise \citep[équation (6a) dans][]{1997ApJ...476..327R}:
\begin{equation}
  T_{\rm
  couche}^4=T_\star^4 \left(
  1 - \sqrt{1-(\theta_\star/\theta_{\rm couche})^2} \right) \,.
\label{eq:gray}
\end{equation}
Cette hypothèse est peu réaliste si l'on considère les pulsations de
la photosphère et donc la propagation de multiples chocs dans les
couches supérieures. Nous pouvons, cependant, en tirer un ordre de
grandeur de l'épaisseur optique par la relation :
\begin{equation}
\frac{F_{\rm couche}}{F_\star}=\frac{B(\lambda,T_{\rm couche})}{B(\lambda,T_\star)}\,.\,\frac{\theta_{\rm couche}^2}{\theta_\star^2}\,.\,\frac{1-\exp{(-\tau_{\rm couche}})}{\exp{(-\tau_{\rm couche})}}\,.
\end{equation}
Soit, avec $B(\lambda,T)$ est la fonction de Planck:
\begin{equation}
\tau_{\rm couche} = \ln{\left(1+\frac{F_{\rm couche}}{F_\star} \,.\, \frac{
    B(\lambda,T_\star) \, \theta_\star^2}{ B(\lambda,T_{\rm
    couche}) \, \theta_{\rm couche}^2} \right)} \,.
\label{eq:tau}
\end{equation}
Les valeurs sont consignées dans le tableau~\ref{tb:physics}. Nous
pouvons noter que l'opacité croît bien entre les phases $\phi = 0$,24,
0,67 et 0,76. Ceci correspond bien à une contraction de la
couche. Nous avons cherché à vérifier si ces variations étaient bien
proportionnelles à $\theta_{\rm couche}^{-2}$ mais n'avons pu
l'établir. Cela signifie que la contraction de la couche s'accompagne
aussi de modifications de ses propriétes optiques et donc
très certainement de sa constitution moléculaire.

Enfin, un dernier point d'intérêt notable sont les faibles épaisseurs
optiques mesurées. Ces valeurs sont significativement inférieures à ce
qui a été observé par \citet{2004A&A...426..279P}, ce qui est en
accord avec l'hypothèse de faible opacité des couches moléculaires en
bande H. Ceci justifie la simplification de la représentation
géométrique de la couche moléculaire par un anneau brillant (le modèle
a été explicité section~\ref{sec:model_geo}).

\subsection{L'asymétrie de l'étoile }

\begin{figure}
  \centering
  \resizebox{\hsize}{!}{
\includegraphics{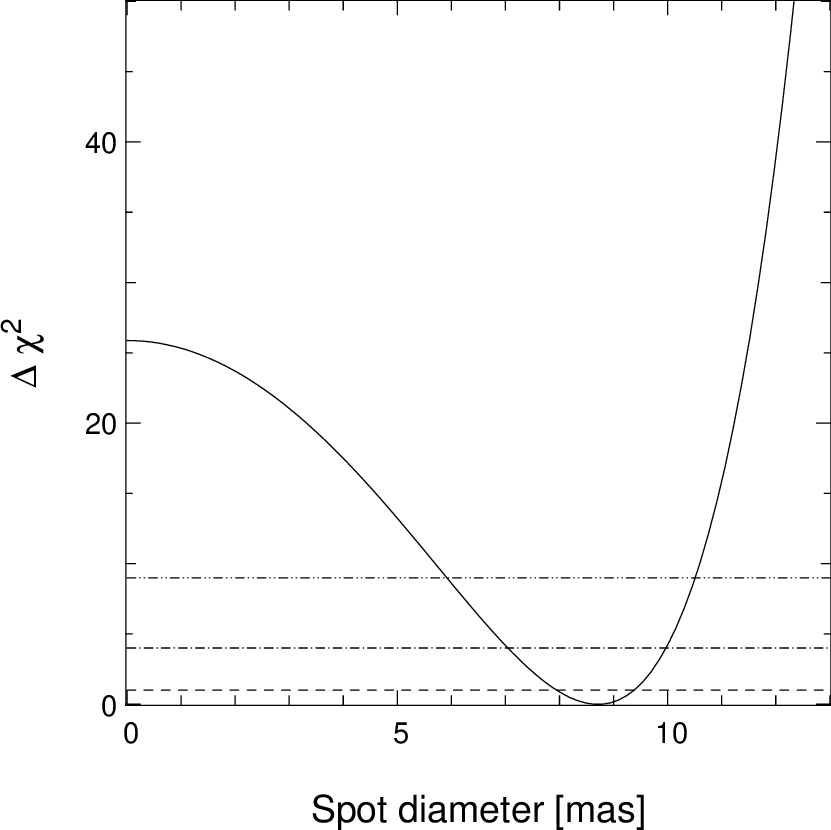}
\includegraphics{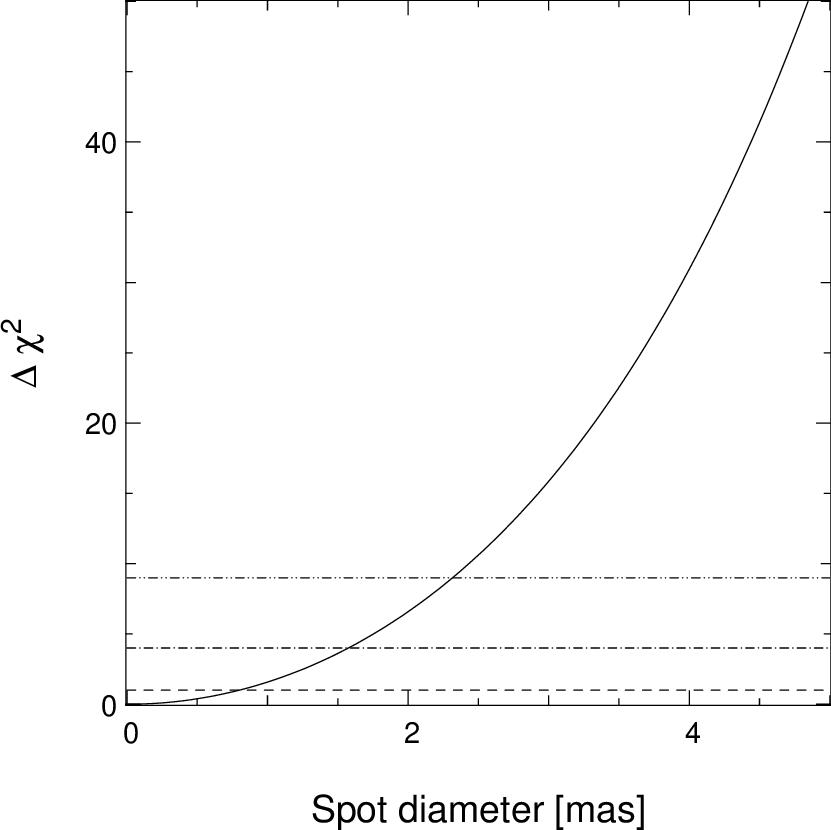}
} \caption[$\Delta \chi^2$ en fonction du diamètre de la tache]{ $\Delta \chi^2$ en fonction du diamètre de la tache. Il
  est obtenu à partir du $\chi^2$ non-réduit par la relation
  :$\Delta\chi^2=\chi^2-\min(\chi^2)$. A gauche il s'agit de
  l'ajustement d'une tache de diamètre variable ajusté sur les données
  de $\chi$ Cyg de mai 2005. A droite, même chose pour les données
  d'octobre 2005. Sur les données de mai 2005, la tache semble être
  extrêmement localisée à la surface de l'étoile. Les traits
  horizontaux représentent les limites à 1, 2 et 3$\sigma$}
  \label{fig:Spot_diam}
\end{figure}

\begin{table}
\caption{Géométrie de l'asymétrie}
\label{tb:asym_f}
\centering
\begin{tabular}{lcccc}
\hline
\hline
$\phi$ & \multicolumn{2}{c}{0.91} & \multicolumn{2}{c}{0.24} \\
\hline
\multicolumn{5}{c}{Tache ponctuelle} \\
\hline
$X_{\rm tache}$ [mas]& \multicolumn{2}{c}{$-5.18\pm0.09$} & \multicolumn{2}{c}{$12.1\pm1.21$} \\
$Y_{\rm tache}$ [mas] & \multicolumn{2}{c}{$3.23\pm0.07$} & \multicolumn{2}{c}{$-1.45\pm0.26$} \\
$F_{\rm tache}$ & \multicolumn{2}{c}{$6.1\pm0.1\%$} & \multicolumn{2}{c}{$5.0\pm0.7\%$} \\
$\chi^2$ & \multicolumn{2}{c}{5.2} & \multicolumn{2}{c}{4.7} \\
\hline
\multicolumn{5}{c}{Tache résolue} \\
\hline
$X_{\rm tache}$ [mas] & \multicolumn{2}{c}{$-5.18\pm0.09$} & \multicolumn{2}{c}{$10.37\pm0.29$} \\
$Y_{\rm tache}$ [mas] & \multicolumn{2}{c}{$3.23\pm0.07$} & \multicolumn{2}{c}{$-2.51\pm0.20$} \\
$F_{\rm tache}$ & \multicolumn{2}{c}{$6.1\pm0.1\%$} & \multicolumn{2}{c}{$5.0\pm0.6\%$} \\
$\theta_{\rm tache}^{\mathrm{a}}$ & \multicolumn{2}{c}{$\leq 0.73$} & \multicolumn{2}{c}{$8.71\pm0.65$} \\
$\chi^2$ & \multicolumn{2}{c}{5.2} & \multicolumn{2}{c}{4.0} \\
\hline
\multicolumn{5}{c}{Deux taches ponctuelles} \\
\hline
$X_{\rm tache}$ [mas] & $-5.56\pm0.11$ & $-2.86\pm0.18$ & $12.46\pm0.75$ & $14.2\pm3.57$\\
$Y_{\rm tache}$ [mas] & $3.04\pm0.07$ & $7.27\pm0.22$ & $-2.07\pm0.67$ & $-9.05\pm1.71$\\
$F_{\rm tache}$ & $ 7.7\pm0.1 \%$ & $3.5 \pm 0.7\%$ & $2.2\pm0.5 \%$ & $1.1\pm0.7\%$ \\
$\chi^2$ & \multicolumn{2}{c}{0.9} & \multicolumn{2}{c}{3.4} \\
\hline
\end{tabular}
\end{table}

Nous avons vu que la pulsation de l'étoile est importante. Comparée à
celle-ci, les petites variations photométriques à la surface de
l'étoile peuvent paraître négligeables. Cependant, ces variations de
quelques pourcents de la luminosité totale de l'étoile, ramenées à une
faible section de l'étoile, pourraient être le signe de phénomènes
extrêmement violents au sein de la photosphère. En fait, personne ne
sait vraiment quel est le phénomène à l'origine de ces
asymétries. Nous verrons d'ailleurs section~\ref{sec:disp_asym} quelques
explications possible, que nous testerons à partir des données
interférométriques.

Un certain nombre de résultats peuvent néanmoins être tirés de nos
données obtenues en bande large. Plus exactement, nous pouvons tenter
de répondre aux questions d'ordre géométriques, notamment : 1) Quelle
est la vitesse d'évolution de l'asymétrie? 2) L'asymétrie est-elle
due à une seule tache ou à plusieurs, formant une structure complexe?
3) Quelle est la taille, ou du moins, l'ordre de grandeur de ces taches.

A la première interrogation un début de réponse peut être apporté à
partir de nos données sur $\chi$ Cyg. La figure~\ref{fig:Images_param}
apporte un bon élément de réponse. Ces reconstructions permettent,
d'ailleurs, une interprétation plus aisée que celle des reconstructions
en aveugle effectuées chapitre~\ref{sec:image}. On note qu'entre mai
2005, octobre 2005, et mars/mai 2006, la tache s'est considérablement
déplacée. Il nous est donc impossible de savoir s'il s'agit de la même
tache qui se serait déplacée, ou s'il s'agit de différentes taches
apparaissant et disparaissant. Le temps d'évolution est donc inférieur
aux périodes séparant ces missions, soit environ 6 mois. Nous notons,
cependant, que la tache s'est faiblement déplacée entre mars et mai
2006. Il semblerait donc que le temps séparant ces deux missions (1
mois et demi) soit à peu près celui régissant l'évolution de ces
taches.

Pour tenter de répondre aux deux autres interrogations, nous avons ajusté à
nos données des modèles plus complexes. Le premier modèle, que nous
avons utilisé jusqu'ici, est celui d'une tache ponctuelle, non-résolue
par l'interféromètre. Le second modèle consiste en cette même tache,
mais avec un diamètre ajustable. Le dernier modèle utilisé consiste en
deux taches. Pour se limiter à l'étude de l'asymétrie, nous avons
utilisé l'algorithme présenté section~\ref{sec:clot_asym}, de façon à
effectuer un ajustement sur les clôtures de phase uniquement. Les
résultats, obtenus sur les données de $\chi$ Cyg de mai et octobre
2005, sont reportés tableau~\ref{tb:asym_f}. Nous notons une
amélioration du $\chi^2$ lorsque l'on augmente la compléxité du
modèle. Cependant, les valeurs deviennent alors incertaines
avec souvent plusieurs minimum ayant des $\chi^2$ proches. Il faut
noter que, par exemple, les résultats obtenus dans le cas de deux
taches ponctuelles sont trés incertains avec de nombreux minimums
possibles à 3$\sigma$. On peut dire, cependant, qu'il est probable que
l'asymétrie soit générée par plusieurs taches, malgré la qualité des
ajustements obtenus à partir d'un modèle composé d'une seule tache.

Le modèle composé d'une tache de dimention variable nous a également
permis de jeter un premier regard sur la dimention de ces taches.  Le
résultat est contrasté. Nous avons affiché figure~\ref{fig:Spot_diam}
le $\chi^2$ non-réduit auquel nous en avons retranché sa valeur
minimum. Les traits horizontaux représentent les limites à 1, 2 et
3$\sigma$. Nous pouvons en déduire que la tache semble extrêmement
localisée en mai 2005 avec une taille inférieure à 2,4 mas
(3$\sigma$), au contraire d'octobre 2005 ($\theta_{\rm tache}= 8$,$7
\pm 2$,5). L'information sur la dimention de la tache est une
information importante car elle permet d'obtenir un rapport de flux
entre la tache et la photosphère. Avec un flux relatif de 6\% et une
taille inférieure à 2,4 mas, la tache observée en mai 2005 est
extrêmement brillante, avec un rapport de brillance surfacique entre
la photosphére et la tache supérieur à 4. Une telle différence de
brillance est conséquente, et devrait permettre de contraindre bon
nombre de modèles.

\clearpage
\section{\'Etude des données spectro-interférométriques}
\label{sec:dispertion}

\begin{figure}
  \centering
  \resizebox{\hsize}{!}{\includegraphics{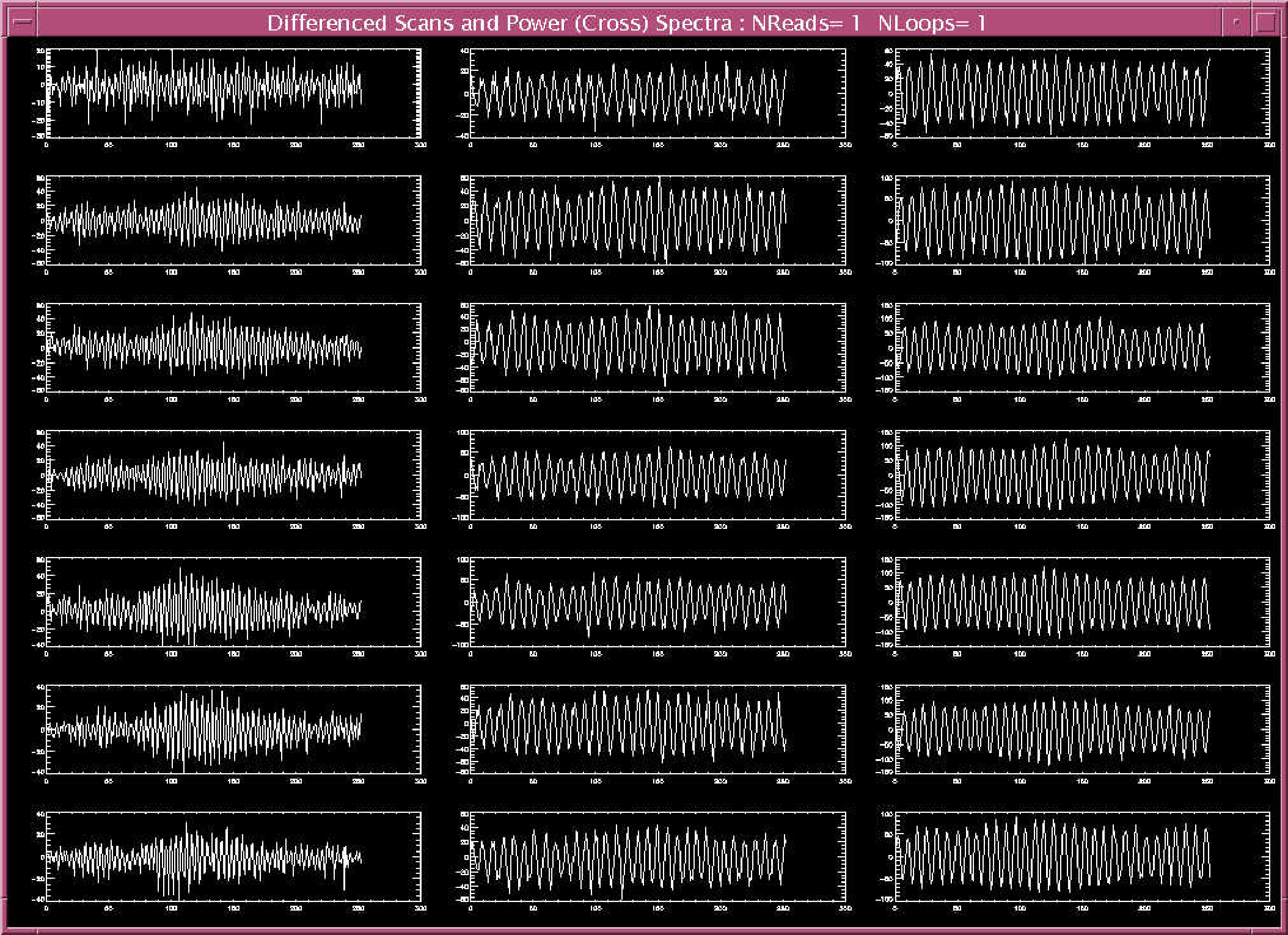}}
  \caption[Interface de contrôle du mode dispersé d'IOTA]{ Interface
  de contrôle d'IOTA. On peut voir l'ensemble des franges obtenues à
  sept longueurs d'onde différentes pour les trois lignes de
  bases. Données acquises sur le calibrateur HD\,87837.}
  \label{fig:fringes-disp}
\end{figure}

   \begin{figure}
   \centering
   \includegraphics[width=7cm]{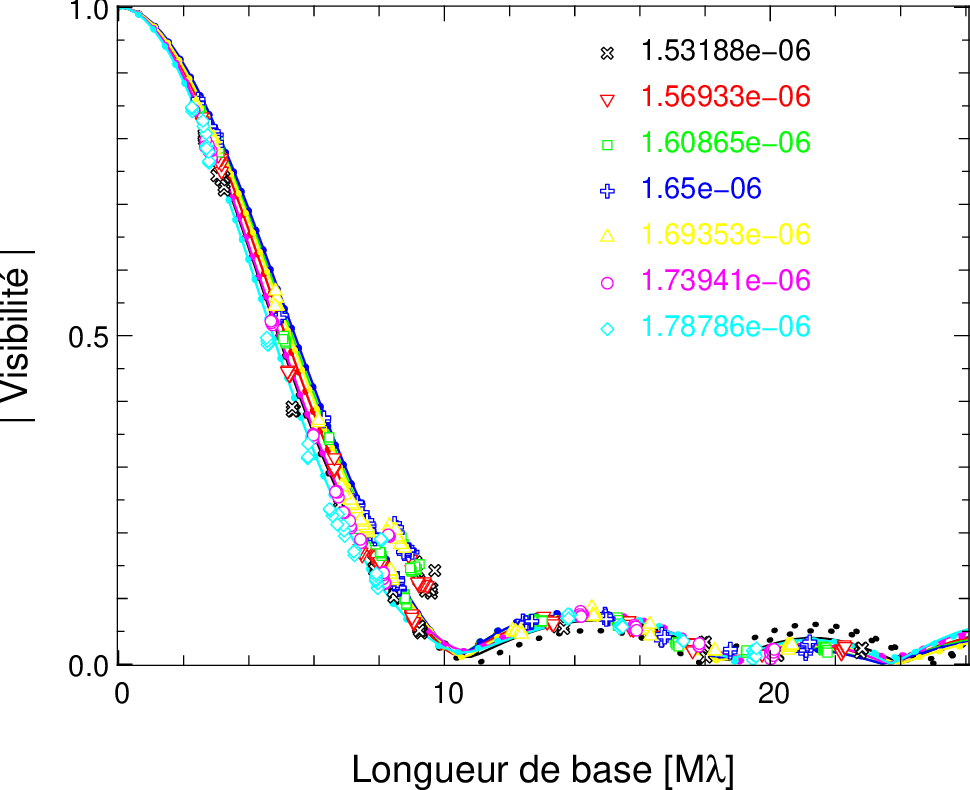}
   \includegraphics[width=7cm]{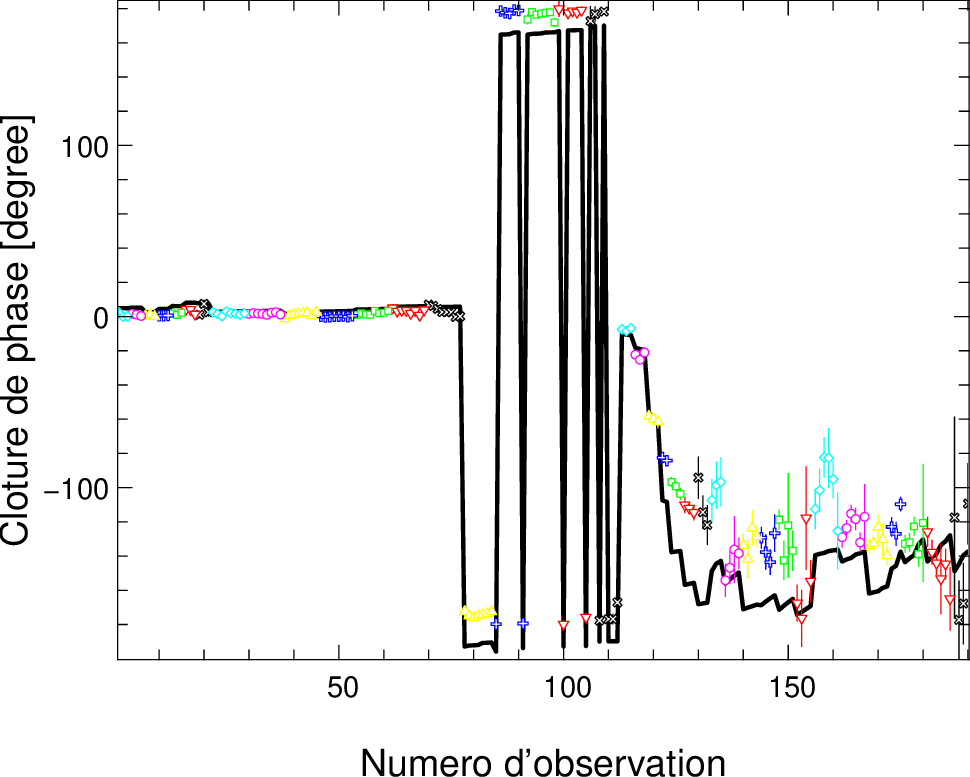}
   \includegraphics[width=7cm]{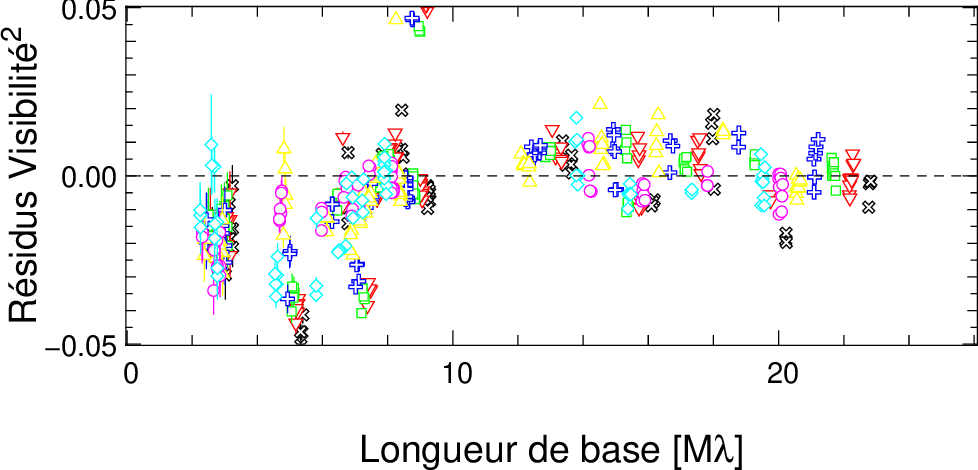}
   \includegraphics[width=7cm]{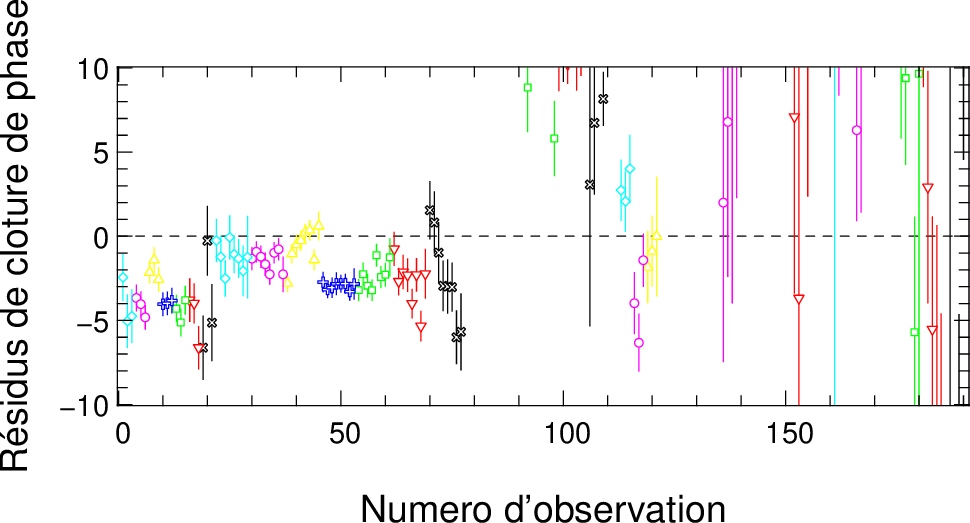}
      \caption[Ajustement des données de R Leo en mai 2006]{Observation en mode dispersé de R Leo en mai 2006
      et meilleur ajustement obtenu du modèle présenté figure~\ref{fig:model}.}
         \label{fig:Data_Rleo}
   \end{figure}

   \begin{figure}
   \centering
   \includegraphics[width=7cm]{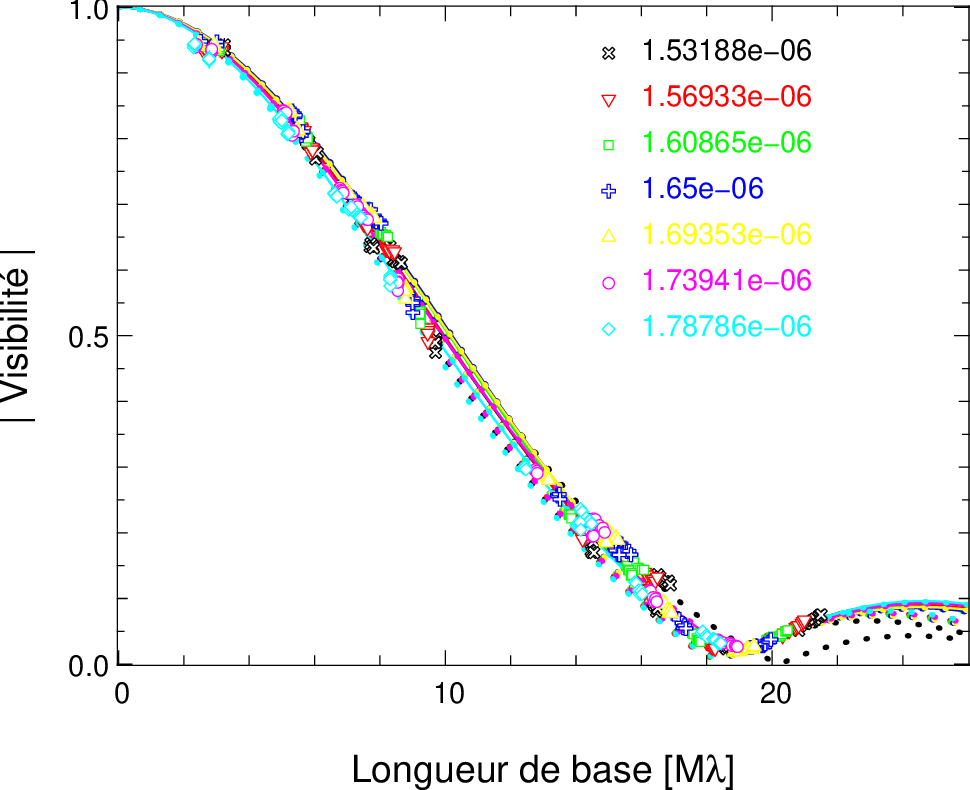}
   \includegraphics[width=7cm]{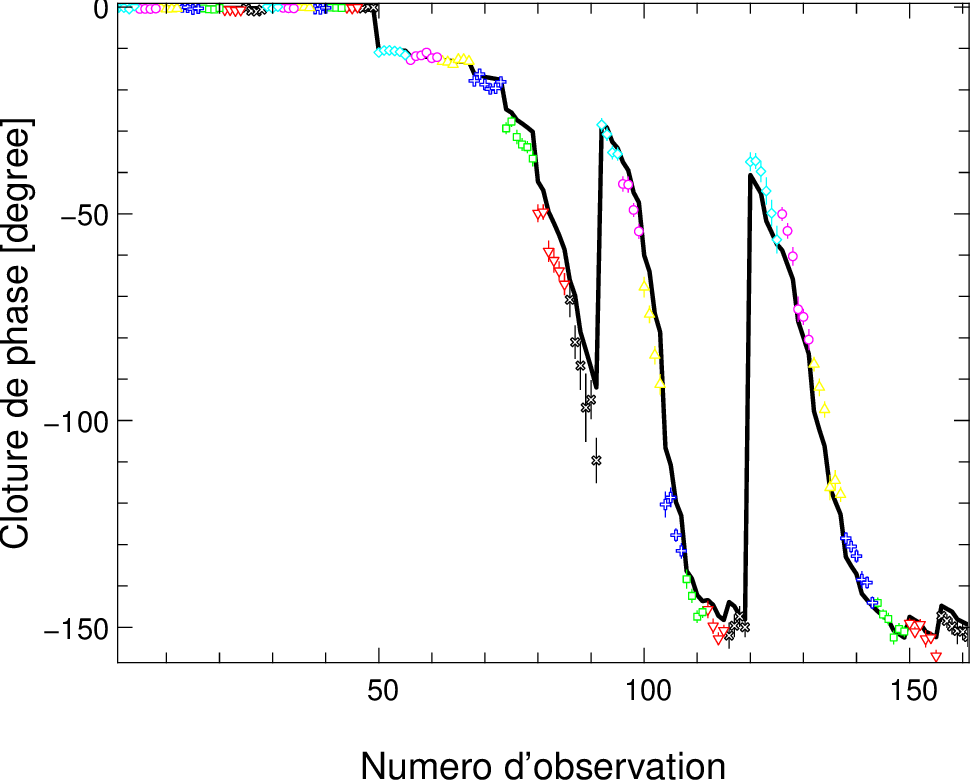}
   \includegraphics[width=7cm]{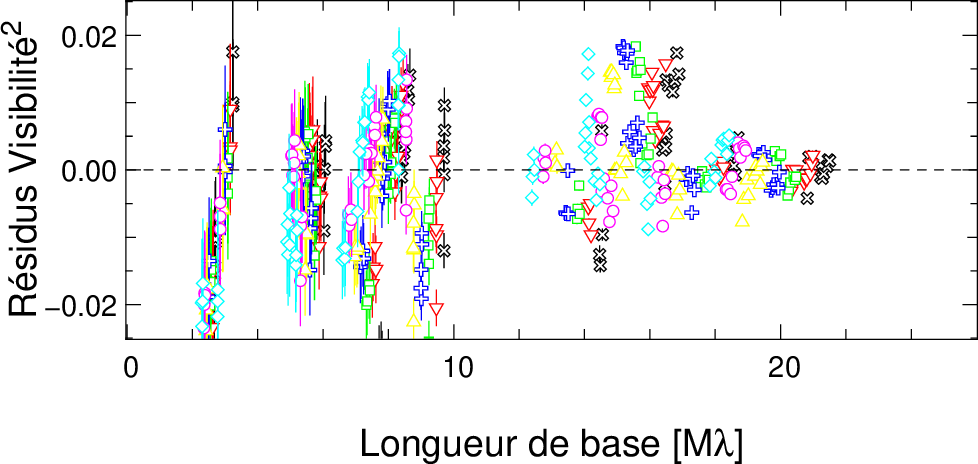}
   \includegraphics[width=7cm]{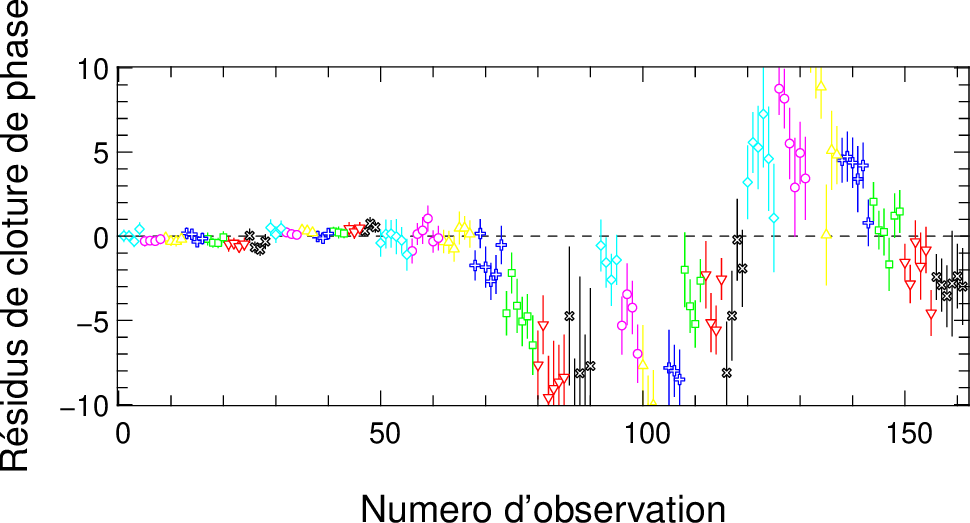}
      \caption[Ajustement des données de $\mu$ Cep en mai 2006]{Observation en mode dispersé de $\mu$ Cep en mai 2006
      et meilleur ajustement obtenu du modèle présenté figure~\ref{fig:model}.}
         \label{fig:Data_MuCep}
   \end{figure}

\subsection{Les données dispersées}

Lors de la dernière mission d'observation sur IOTA, nous avons eu la
chance de pouvoir utiliser le tout nouveau mode ``dispersé''. Ce mode
consiste à disposer un prisme devant la caméra, de façon à répartir
les signaux sur plusieurs pixels en fonction de la longueur
d'onde. Comme le montre la figure~\ref{fig:fringes-disp}, nous avons
ainsi pu récolter simultanément l'information sur sept canaux
spectraux, allant de 1,5 à 1,9 microns. Cette dispersion nous a permis
d'avoir une meilleure couverture du plan fréquentiel comme nous
l'avons vu en section~\ref{sc:Actu_test}.  Elle nous a également
permis d'en déduire une information plus riche sur la composition de
la couche moléculaire et les raisons de la présence d'asymétries.

Nous avons appliqué le modèle utilisé dans la précédente section mais
avec les trois paramètres (ACB, Flux couche, Flux tache) dépendant de
la longueur d'onde. Nous avons effectué ce travail sur trois objets:
$\chi$ Cyg, R Leo et $\mu$ Cep. Les données sont représentées
figures~\ref{fig:Data_Mai06},~\ref{fig:Data_Rleo}
et~\ref{fig:Data_MuCep}. Le tableau~\ref{tb:spectro} récapitule les
résultats obtenus. Les $\chi^2$ réduits obtenus sont respectivement de
6, 17 et 24. L'ajustement est tout à fait raisonable pour $\mu$ Cep,
ce qui n'est pas vraiment le cas pour les deux Miras. En conséquence,
il semblerait que la structure des Miras soit plus compliquée que ce
qui est accessible par notre modèle. Nous avons vu, par exemple, que
l'ajustement de $\chi$ Cyg pouvait être nettement amélioré par un
modèle avec deux couches moléculaires. Pour R Leo, les données
semblent encore plus complexes, avec certainement la présence de
multiples taches sur la photospère. Nous avons néanmoins poursuivit
l'étude sur la base du modèle simple. Ces résultats devront donc être
affinés dans le futur.

\begin{table}
\caption{Mesures spectro-interférométriques des paramétres de R Leo,
  $\chi$ Cyg, et $\mu$ Cep}
\label{tb:spectro}
\centering
\begin{tabular}{cccc}
\hline
\hline
$\lambda$ & ACB [$\alpha$] & $F_{\rm couche}/F_{\rm total}$ [\%]  & $F_{\rm tache}/F_{\rm total}$ [\%] \\
\hline
 & \multicolumn{3}{c}{R Leo ($\theta_\star =29,56 \pm  0,18$ mas)}  \\
\hline
 1,53 $\mu$m&$ 1,05 \pm 0,05$&$ 11,3 \pm  0,9$&$  1,1 \pm  0,3$ \\ 
 1,57 $\mu$m&$ 1,04 \pm 0,05$&$  7,6 \pm  0,6$&$  1,5 \pm  0,2$ \\ 
 1,61 $\mu$m&$ 1,13 \pm 0,05$&$  5,1 \pm  0,6$&$  1,6 \pm  0,2$ \\ 
 1,65 $\mu$m&$ 1,18 \pm 0,05$&$  4,2 \pm  0,6$&$  1,6 \pm  0,2$ \\ 
 1,69 $\mu$m&$ 1,18 \pm 0,06$&$  6,4 \pm  0,5$&$  1,1 \pm  0,2$ \\ 
 1,74 $\mu$m&$ 1,06 \pm 0,07$&$ 11,0 \pm  0,8$&$  2,0 \pm  0,2$ \\ 
 1,79 $\mu$m&$ 0,90 \pm 0,07$&$ 13,4 \pm  0,8$&$  1,8 \pm  0,3$ \\ 
\hline
 & \multicolumn{3}{c}{$\chi$ Cyg ($\theta_\star =20,96 \pm  0,10$ mas) } \\
\hline
 1,53 $\mu$m&$ 2,74 \pm 0,10$&$  9,7 \pm  0,5$&$  1,6 \pm  0,2$ \\ 
 1,57 $\mu$m&$ 2,06 \pm 0,05$&$  8,0 \pm  0,2$&$  1,6 \pm  0,1$ \\ 
 1,61 $\mu$m&$ 1,99 \pm 0,05$&$  8,1 \pm  0,2$&$  1,4 \pm  0,1$ \\ 
 1,65 $\mu$m&$ 2,17 \pm 0,05$&$  7,1 \pm  0,2$&$  1,7 \pm  0,1$ \\ 
 1,69 $\mu$m&$ 2,35 \pm 0,05$&$  6,7 \pm  0,2$&$  2,0 \pm  0,1$ \\ 
 1,74 $\mu$m&$ 2,36 \pm 0,05$&$  8,5 \pm  0,2$&$  1,9 \pm  0,1$ \\ 
 1,79 $\mu$m&$ 2,31 \pm 0,05$&$ 11,6 \pm  0,2$&$  1,8 \pm  0,1$ \\ 
\hline
& \multicolumn{3}{c}{$\mu$ Cep ($\theta_\star =16,85 \pm  0,15$ mas)} \\
\hline
 1,53 $\mu$m&$ 1,53 \pm 0,08$&$  3,0 \pm  0,3$&$  2,8 \pm  0,1$ \\ 
 1,57 $\mu$m&$ 1,50 \pm 0,08$&$  2,4 \pm  0,3$&$  2,4 \pm  0,1$ \\ 
 1,61 $\mu$m&$ 1,54 \pm 0,08$&$  1,4 \pm  0,3$&$  2,2 \pm  0,1$ \\ 
 1,65 $\mu$m&$ 1,62 \pm 0,08$&$  1,3 \pm  0,3$&$  2,2 \pm  0,1$ \\ 
 1,69 $\mu$m&$ 1,65 \pm 0,08$&$  2,0 \pm  0,3$&$  2,4 \pm  0,1$ \\ 
 1,74 $\mu$m&$ 1,61 \pm 0,08$&$  3,6 \pm  0,2$&$  2,7 \pm  0,1$ \\ 
 1,79 $\mu$m&$ 1,50 \pm 0,08$&$  4,7 \pm  0,3$&$  2,7 \pm  0,1$ \\ 
\hline
\end{tabular}
\end{table}

\subsection{La dépendance spectrale de la couche moléculaire}

\begin{table}
\caption{Paramètres physiques de R Leo, $\chi$ Cyg et $\mu$ Cep
  observés en mai 2006}
\label{tb:calib}
\centering
\begin{tabular}{lccc}
\hline
\hline
 & R Leo & $\chi$ Cyg & $\mu$ Cep \\
\hline
$\phi$ & 0,05 & 0,76 & ... \\
$F_{\rm Bol}$ (10$^{-13}$W\,cm$^{-2}$)  
 & $23,63\pm2,73$ & $6,74\pm0,36$ & $17,62\pm0,26$ \\
Parallaxe$^{\mathrm{a}}$ & $9,87 \pm 2,07$ mas & $9,43 \pm 1,36$ mas
& $0,62 \pm 0,52$ mas\\
\hline
$R_\star^{\mathrm{a}}$ & $322 \pm 71 R_\odot $ & $ 209 \pm 32 R_\odot$ & $ 2925 \pm 1487 R_\odot$ \\
$T_\star$ &  $3001 \pm 96 $ K & $ 2780 \pm 54 $ K& $ 3693 \pm 30 $ K \\
\hline
$R_{\rm couche}/R_\star$ & $1,35 \pm 0,02 $ & $ 1,48 \pm 0,03 $ & $ 1,53 \pm 0,04 $  \\
${T_{\rm couche}}^{\mathrm{b}}$ & $2274 \pm 73 $ K& $ 2002 \pm 33 $ K& $ 2598 \pm 21 $ K\\
\hline
$\frac{{\rm d}(F_{\rm tache}/F_{\rm total})}{{\rm d}\lambda}$ [$\times
  10^4$ m$^{-1}$] & $1,6 \pm
1,7$ & $1,9 \pm 0,9$ & $ 1,5 \pm 1,3$ \\
$T_{\rm tache}- T_{\rm \star}$ & $-500\pm 550$ K & $-590 \pm 240$ K &
$-470 \pm 370$ K \\
\hline
\end{tabular}
\begin{list}{}{}
\item[$^{\mathrm{a}}$] {\it Hipparcos} (ESA, 1997).
\end{list}
\end{table}

Pour chacune de ces étoiles, nous avons estimé le flux bolomètrique,
et ainsi obtenu la température effective de l'étoile
(relation~(\ref{eq:tp})). Nous avons obtenu des températures pour la
photosphère comprises entre 2700\,K ($\chi$ Cyg) et 3700\,K ($\mu$
Cep). A partir du rapport $R_{\rm couche}/R_\star$ nous avons ensuite
utilisé l'équation~(\ref{eq:gray}) pour en déduire la température de
la couche moléculaire. Nous avons enfin estimé l'épaisseur optique de
celle-ci à partir de la relation~(\ref{eq:tau}). Ces résultats sont
reportés tableau~\ref{tb:calib}.

\begin{figure}[h!]  \centering
   \includegraphics[width=5cm]{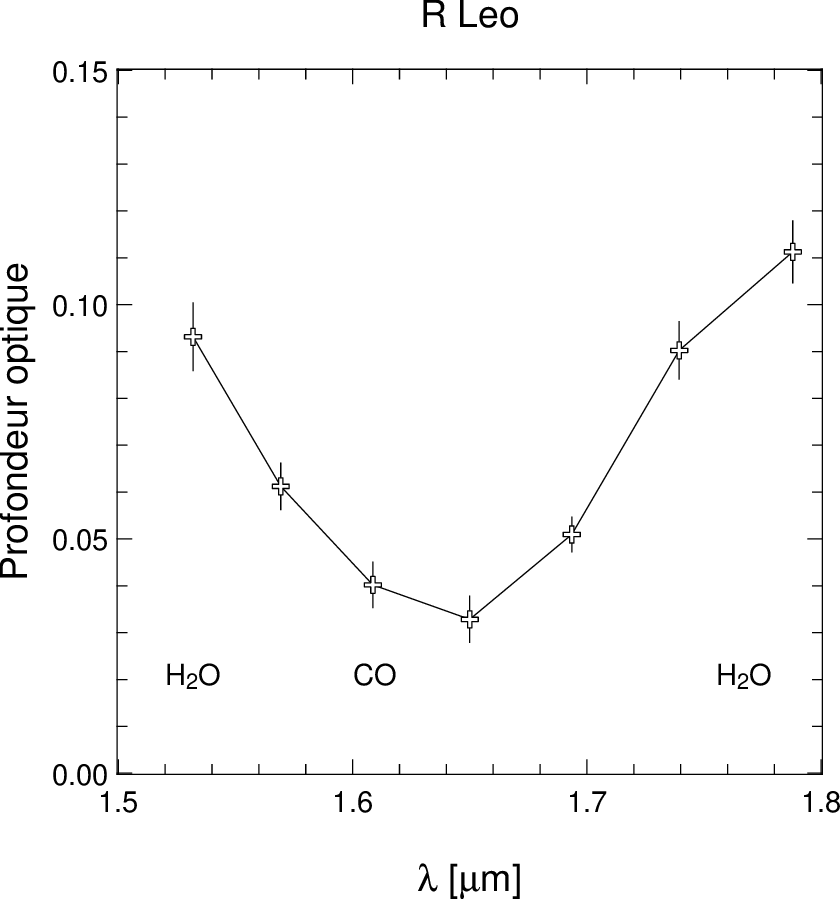}
   \includegraphics[width=5cm]{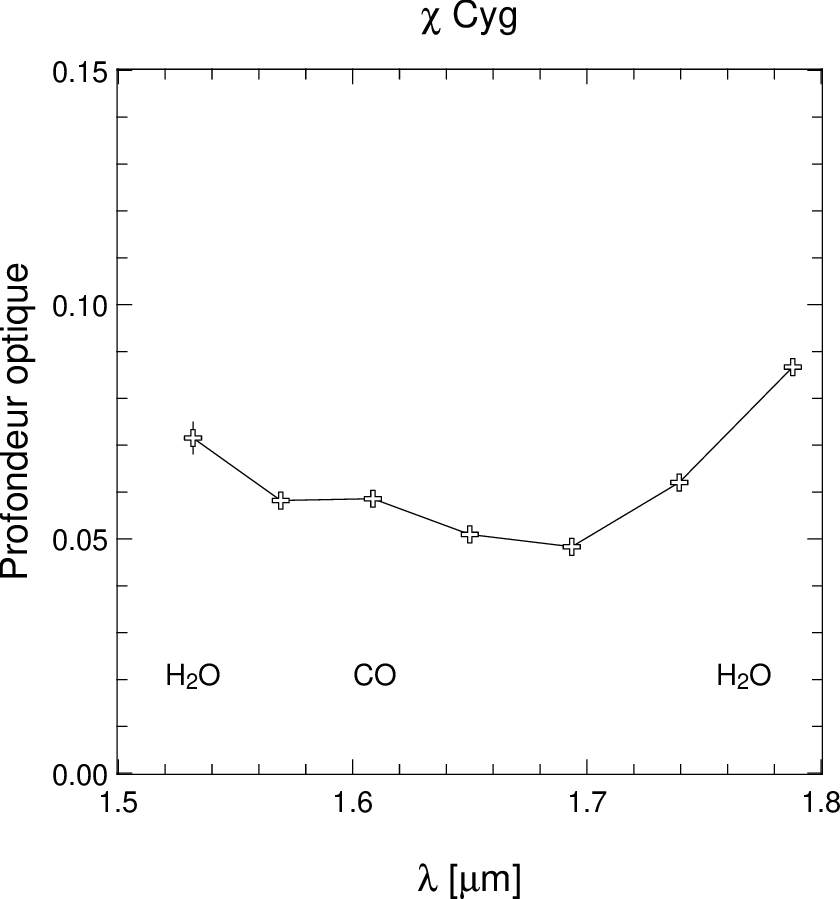}
   \includegraphics[width=5cm]{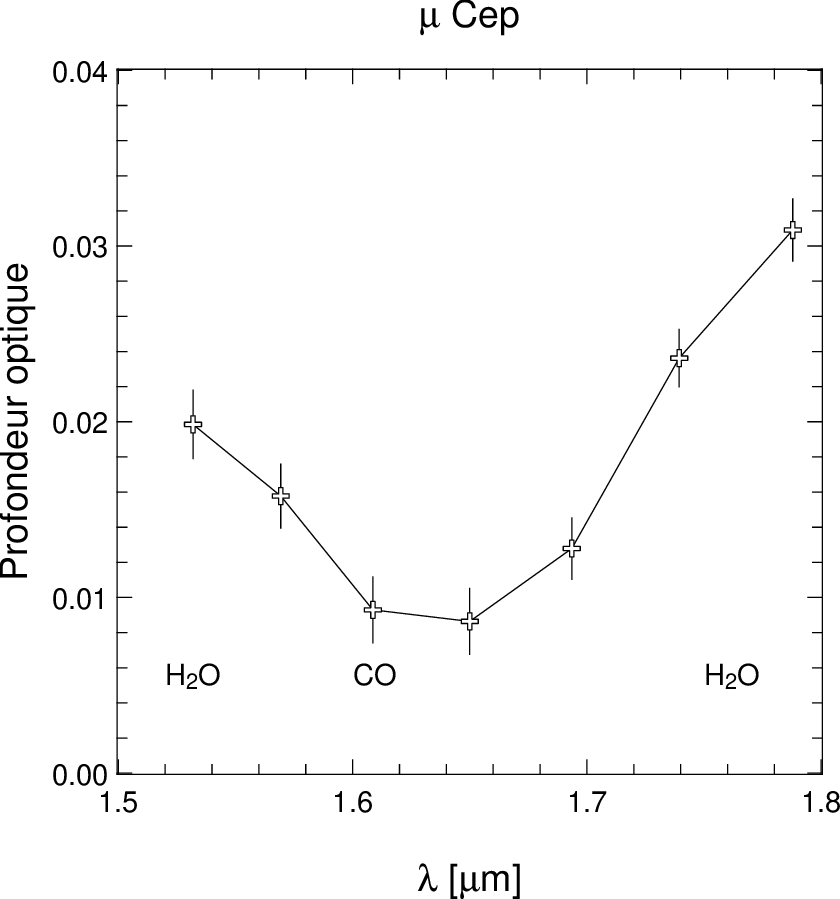}
   \includegraphics[width=5cm]{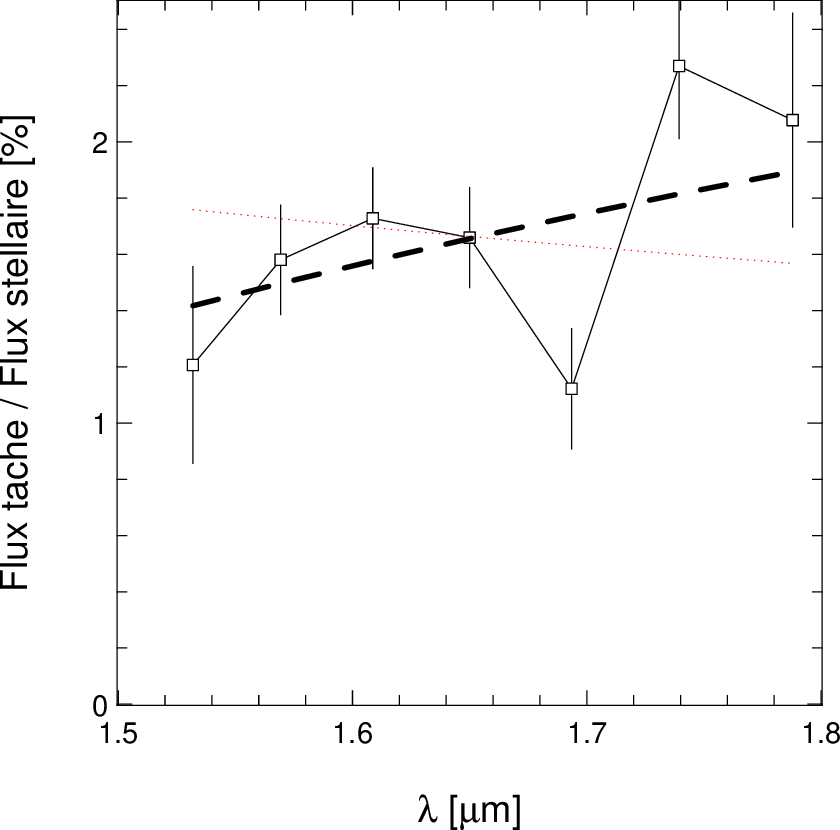}
   \includegraphics[width=5cm]{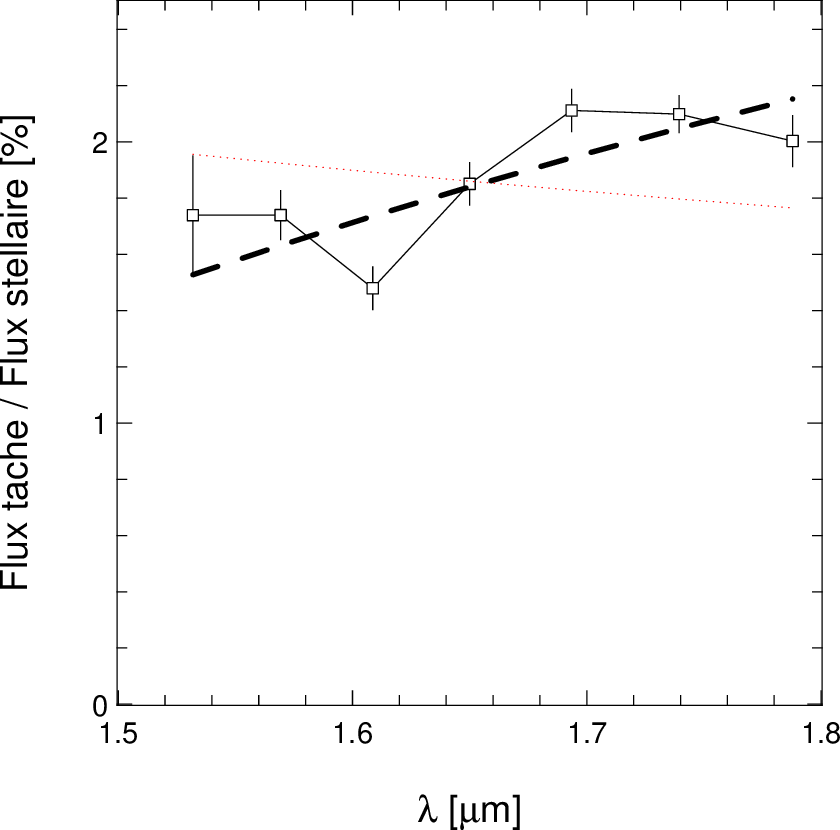}
   \includegraphics[width=5cm]{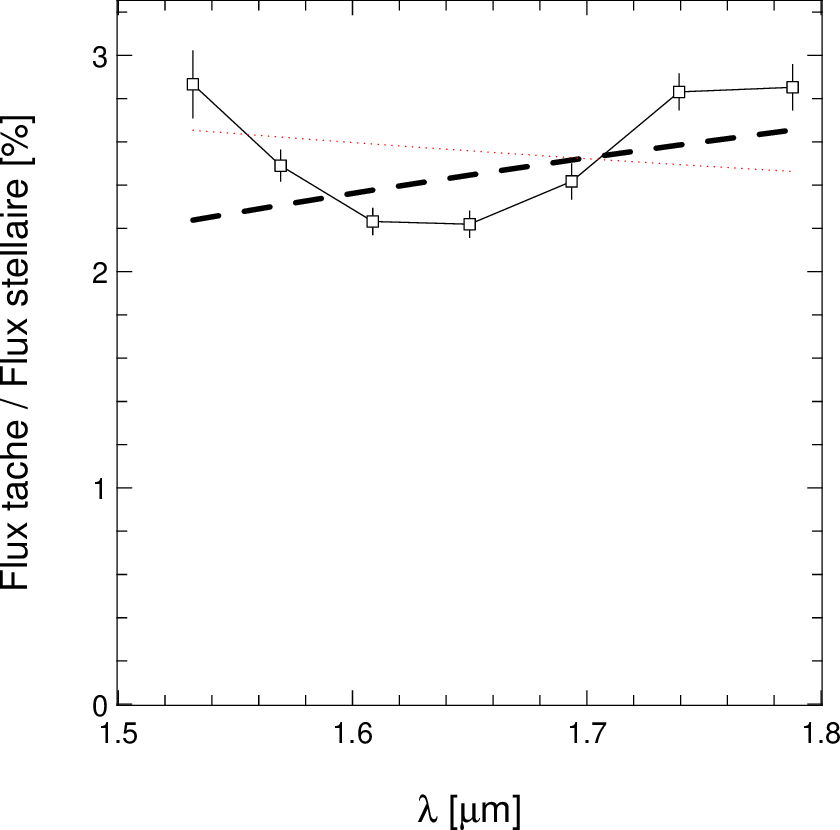}
   \includegraphics[width=5cm]{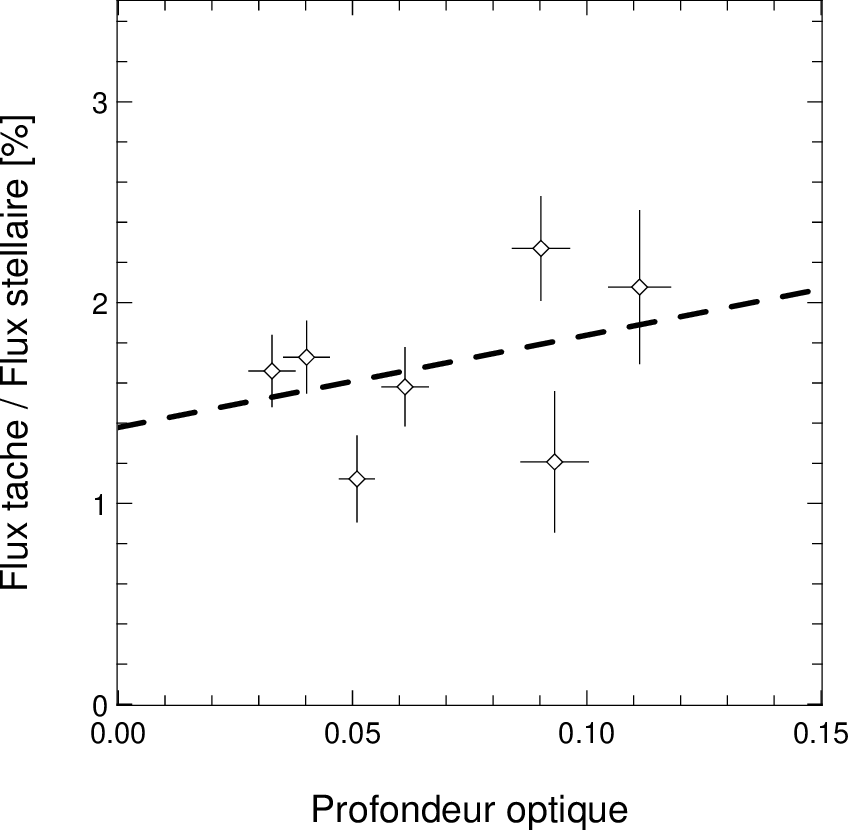}
   \includegraphics[width=5cm]{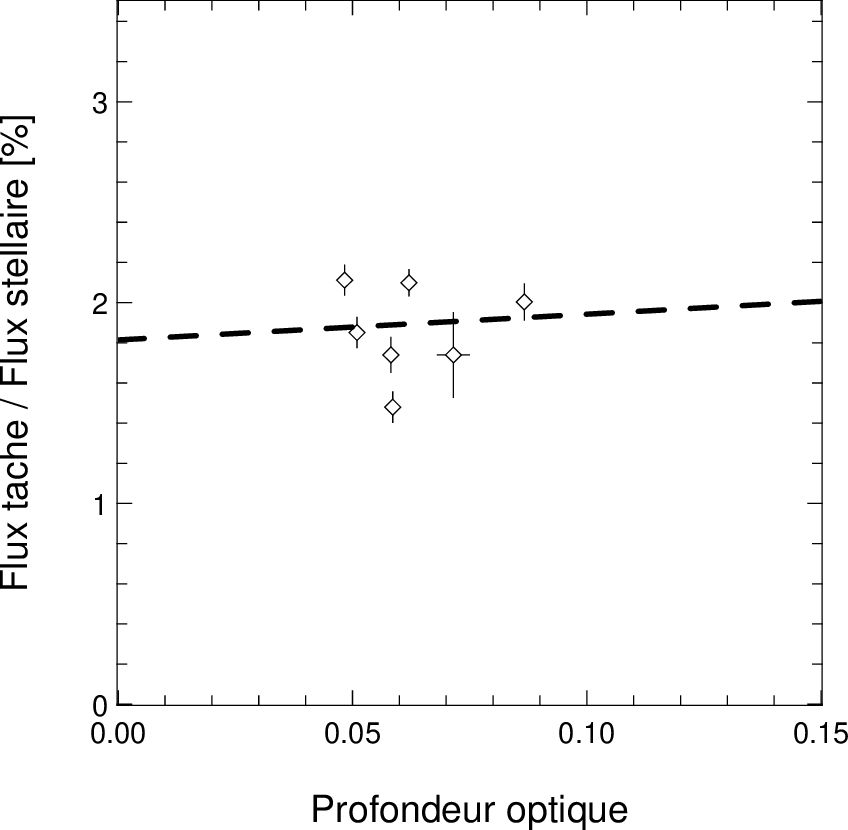}
   \includegraphics[width=5cm]{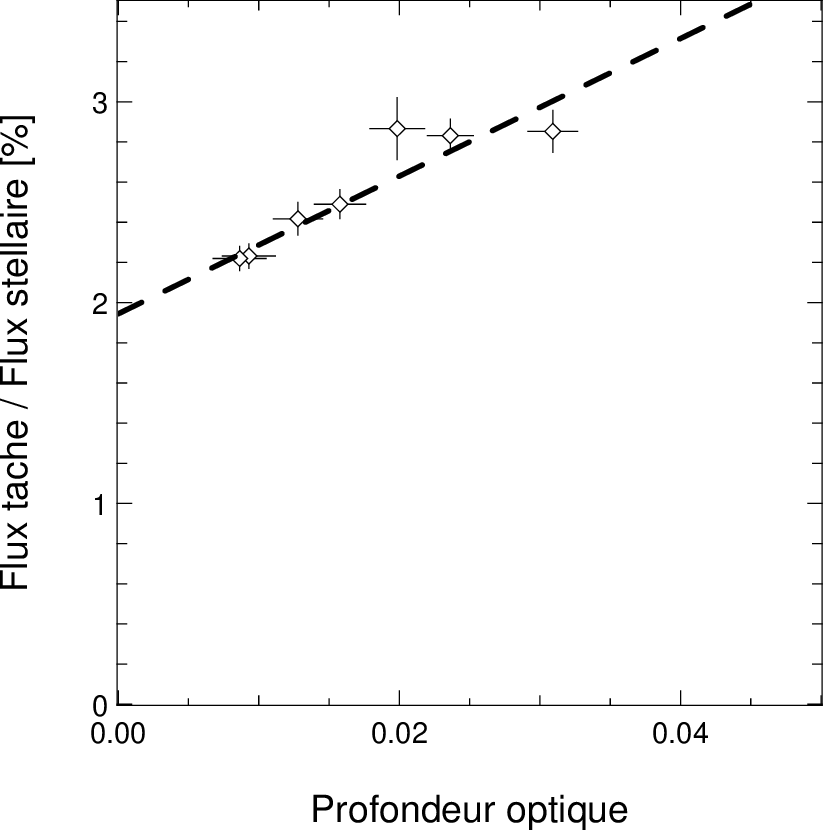}
   \caption[Représentation des résultats spectro-interférométriques
   pour R Leo, $\chi$ Cyg et $\mu$ Cep]{ Représentation des résultats
   spectro-interférométriques pour R Leo, $\chi$ Cyg et $\mu$ Cep. En
   haut, profondeur optique de la couche moléculaire. Au centre, flux
   relatif de la tache responsable de l'asymétrie. La courbe en
   pointillé rouge représente un modèle de dépendance spectrale pour
   une tache ayant une température de 4000\,K. En bas, flux de la
   tache en fonction de l'opacité de la couche. Les lignes en
   pointillés sont le résultat de l'ajustement d'une fonction affine
   sur les données.}  \label{fig:spectro} \end{figure}

Les trois graphiques du haut de la figure~\ref{fig:spectro}
représentent l'opacité optique de la couche en fonction de la longueur
d'onde. On note que l'opacité est plus importante sur les bords de la
bande H. Ceci s'observe pour les couches des trois étoiles, et
correspond aux zones d'absorption de la molécule d'eau. Une deuxiéme
molécule est également présente dans notre bande spectrale. Il s'agit du
monoxyde de carbone (CO) qui absorbe à cette longueur d'onde lorsqu'il
est dans son troisième niveau d'excitation vibrationelle. La forêt de
raies ainsi créé s'étale sur une grande partie de la bande H, avec un
maximum à 1,6 $\mu$m. On peut noter que l'opacité de la couche à cette
longueur d'onde est plus marquée sur $\chi$ Cyg que sur $\mu$ Cep ou R
leo. Parce que l'atome d'oxygène forme en priorité du
monoxyde de carbonne, nous pouvons en déduire une information sur
le rapport carbone sur oxygène dans l'atmosphère de l'étoile. Nous
avons obtenu des rapports $\tau_{1,61\, \mu {\rm m}}/\tau_{1,53\, 
\mu{\rm m}}$ de 0,43, 0,82 et 0,46 pour, respectivement, R Leo, $\chi$
Cyg et $\mu$ Cep. Ceci est une indication d'un rapport C/O deux fois
plus élevé dans l'atmosphère de $\chi$ Cyg que dans celle de R Leo où
$\mu$ Cep. Ce résultat est en accord avec le type spectral, S, de
l'étoile.

\subsection{La dépendance spectrale de l'asymétrie}
\label{sec:disp_asym}

L'amplitude de la tache en fonction de la longueur d'onde est
particulièrement intéressante. Comme nous l'avons dit, il existe de
multiples possibilités pouvant expliquer la présence d'asymétries sur
ces étoiles: une opacité variable de la couche, des inhomogénéités en
température dues à des phénomènes de convection, la présence de
pulsations non radiales, etc... Nos données spectro-interférométriques
nous offrent une information nouvelle, dont l'interprétation n'est pas
évidente. Nous avons cherché à confirmer (ou infirmer) deux
hypothèses. La première est celle d'une asymétrie due à des variations
d'opacité de la couche moléculaire. La deuxiéme est celle d'une
asymétrie due à des variations de température à la surface de
l'étoile.

\subsubsection{L'asymétrie et l'absorption moléculaire}

Dans un premier temps, nous avons cherché à relier le spectre de
l'asymétrie à l'absorption due aux molécules d'eau et de monoxyde de
carbone. Puisque l'information sur la composition de
l'atmosphère peut être trouvée par le biais de l'absorption moléculaire
dans la couche, nous avons cherché à relier l'amplitude du flux de la
tache (tableau~\ref{tb:spectro}) à l'opacité de la couche
moléculaire. Les résultats sont présentés figure~\ref{fig:spectro}. Il
est à noter que nous n'avons pas représenté le flux de la tache
par rapport au flux total, mais par rapport au
flux stellaire :
\begin{equation}
\frac{F_{\rm tache}}{F_\star}=\frac{F_{\rm tache}}{F_{\rm total}} \cdot \left({1-\frac{F_{\rm couche}}{F_{\rm total}}}\right)^{-1}\,,
\end{equation}
ceci pour s'affranchir des variations spectrales dues à l'absorption
de la couche moléculaire.

Sur R Leo et $\chi$ Cyg on ne constate pas de corrélations entre
l'absorption moléculaire et le flux de la tache. Ceci signifie que
la  présence de l'asymétrie n'est pas reliée à l'absorption
moléculaire. Ainsi, l'asymétrie ne serait pas causée par
des variations d'opacité des hautes couches de l'atmosphère.

Cependant, nous constatons une forte corrélation entre opacité et flux
de la tache sur $\mu$ Cep. Ainsi, sur $\mu$ Cep du moins, une partie
de l'asymétrie est due à la présence des molécules. Si ce résultat est
comfirmé, il signifie que les taches que nous observons sont dues à la
présence d'inhomogénéité dans l'environement proche de
l'étoile. Cependant, la seule présence de la couche moléculaire ne
peut pas être responsable de l'asymétrie, car il n'existe pas de
relation de proportionnalité entre l'amplitude de l'asymétrie et
l'épaisseur optique de la couche. En conséquence, si l'asymétrie est
bien due à des variations d'épaisseur optique, la majorité de cette
asymétrie doit être la conséquence de la présence de poussières.

\subsubsection{Estimation de la température effective de l'asymétrie}

L'hypotèse de variations photométriques à la surface de l'étoile dues
a l'absorption de la couche moléculaire est souvent avancée. Une
seconde hypothèse, vérifiable grâce aux données spectroscopiques, est
celle de la présence de variations de température dans la
photosphère. Cette possibilité, expliquée par la présence importante de
cellules de convection, est souvent envisagée pour expliquer la présence
de taches \citep[notamment sur Beltelgeuse;][]{2003csss...12.1024F}.

Une estimation peut être faite sur la différence de température
nécessaire pour produire de telles asymétries. Une hypothèse raisonnable
serait celle de la présence de cellules de convection de tailles inférieures au
rayon de l'étoile, recouvrant environ un vingtième de la surface
stellaire visible \citep{1975ApJ...195..137S}. Pour que cela puisse
apparaître sous la forme d'une tache ayant un flux de 2\%, la
température de la tache doit vérifier :
\begin{equation}
2\%=\frac{B(\lambda,T_{\rm tache})-B(\lambda,T_\star)}{10\, B(\lambda,T_\star)}\,.
\end{equation}
Pour une température de photosphère de 3000\,K,
nous pouvons en déduire que la température de la tache devrait
avoisiner les 4000\,K.

   \begin{figure} \centering
   \includegraphics[width=9cm]{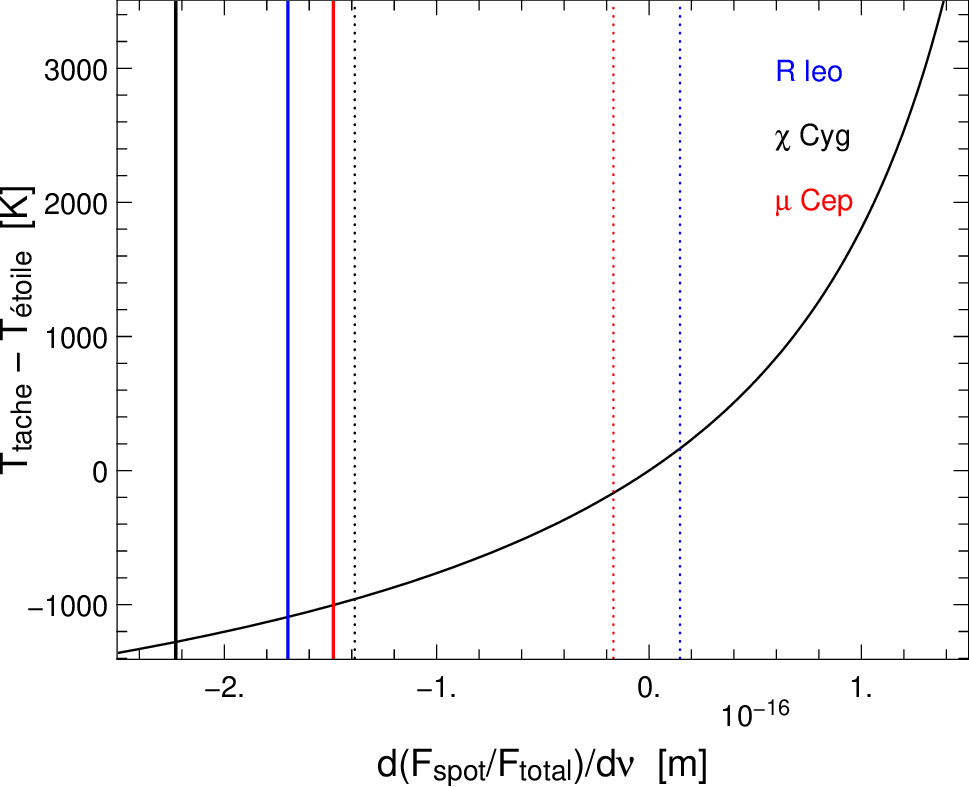} \caption[Ecarts
   de température entre la tache et la photosphère]{ Ecarts
   de température entre la tache et la photosphère en fonction de la
   dérivée de l'amplitude relative de la tache. Les lignes verticales
   correspondent aux mesures effectuées, et les lignes en pointillés
   aux barres d'erreur à 1 $\sigma$. Les hypothèses utilisées sont une
   tache de flux relatif de 2\%, et une température de photosphère de
   3000 Kelvins.}  \label{fig:delta_temp} \end{figure}

Or, nos données spectrales ne nous permettent pas une mesure directe
de la température. Cependant, une information peut être obtenue à
partir de la dépendance en fonction de la longueur d'onde du flux
relatif de la tache.  En effet, le modèle du corps noir nous donne le
flux d'énergie par unité de fréquence ($\nu=1/\lambda$) sous la forme de la
loi de Planck:
\begin{equation}
B(\nu,T)=\frac{2 h \nu^3}{c^2}\frac{1}{\exp(x)-1}
\label{eq:corps_noir_planck}
\end{equation}
où
\begin{equation}
x = \frac{h \nu}{k T}\,.
\end{equation}
Nous pouvons en déduire le gradient du flux en fonction de la fréquence :
\begin{equation}
\frac{\partial B(\nu,T)}{\partial \nu} = \frac{2 h
  \nu^2}{c^3}\left(x \frac{-\exp(x)}{(\exp(x)-1)^2}+\frac{3}{\exp(x)-1}\right)
\label{eq:der_planck}
\end{equation}
que nous souhaiterions relier au gradient du flux de la tache. Cela
peut se faire en écrivant la dérivé de la façon suivante :
\begin{equation}
\frac{\partial F_{\rm tache}/F_\star}{\partial \nu} =
\frac{F_{\rm tache}}{F_\star} \left(\frac{1}{F_{\rm tache}}
\frac{\partial F_{\rm tache} }{\partial \nu}
- \frac{1}{F_\star } \frac{\partial F_\star }{\partial \nu} \right)
\end{equation}
ou encore :
\begin{equation}
\frac{\partial F_{\rm tache}/F_\star}{\partial \nu} =
\frac{F_{\rm tache}}{F_\star} \left(\frac{1}{B(\nu,T_{\rm
    tache})} \frac{\partial B(\nu,T_{\rm tache})}{\partial \nu}
- \frac{1}{B(\nu,T_\star)} \frac{\partial B(\nu,T_\star)}{\partial \nu} \right)
\end{equation}
Or, d'après les équations~(\ref{eq:corps_noir_planck})
et~(\ref{eq:der_planck}), on peut établir que :
\begin{equation}
\frac{1}{B(\nu,T)} \frac{\partial B(\nu,T)}{\partial \nu}
= \frac{1}{\nu}\left( \frac{-x}{(1-\exp(-x)}+3\right)
\end{equation}
et ainsi relier le gradient de l'amplitude de la tache à sa
température :
\begin{equation}
\frac{\partial F_{\rm tache}/F_\star}{\partial \nu} =
\frac{F_{\rm tache}}{F_\star} \frac{h}{k}
\left(\frac{1}{T_\star \left(1-\exp(\frac{- h \nu}{k  T_\star})
\right)}- \frac{1}{T_{\rm tache} \left(1-\exp(\frac{- h \nu}{k T_{\rm
tache}}) \right)}\right)\,.
\label{eq:tem_dif}
\end{equation}
En supposant une tache ayant un flux $F_{\rm tache}/F_\star = 2\%$ et
une température $T_{\rm tache}=4000$\,K, ainsi qu'une photosphère
ayant une température $T_\star=3000$\,K, on obtient un gradient
de 6,$8 \times 10^{-17}$. Nous avons représenté ces gradients en
pointillés rouge sur les graphiques du centre de la
figure~\ref{fig:spectro}. Il est à noter que la valeur postive du
gradient par rapport au nombre d'onde équivaut à un gradient négatif
par rapport à la longueure d'onde ($\nu=c/\lambda$). Même si ces
résultats sont à la limite du bruit sur nos mesures, on peut voir que,
de manière générale, le gradient sur le flux de la tache est de signe
contraire à ce qui serait attendu, avec des valeurs égales à -1,$7 \pm
1$,$8\times 10^{-16}$, -2,$2 \pm 0$,$8
\times 10^{-16}$ et -1,$5 \pm 1$,$3 \times 10^{-16}$ pour
respectivement, R Leo, $\chi$ Cyg et $\mu$ Cep.

A la lumière de ces résultats, il semblerait que, si une température
devait être dérivée de ces résultats, elle soit plutôt inférieure à la
surface de la photosphère. Ce résultat est représenté graphiquement
par la figure~\ref{fig:delta_temp}. La courbe représente l'écart de
température entre la tache et l'étoile obtenu par la
relation~(\ref{eq:tem_dif}). Un certain nombre d'hypothèses ont été
faites pour pouvoir effectuer ce tracé. Nous avons notamment supposé
une température de photosphère de 3000\,K et une tache ayant un flux
moyen de 2\%. Sur ce graphique nous avons également représenté nos
mesures du gradient du flux de l'asymétrie, obtenue par l'ajustement
des données de la figure~\ref{fig:spectro}. Les 3 lignes verticales
repésentent les valeurs de maximum de vraisemblance obtenues R Leo
(bleu), $\chi$ Cyg (noir) et $\mu$ Cep (rouge).  Les lignes en
pointillées représentent les limites supérieures à 1$\sigma$. A la
lumière de ces résultats, il semble peu probable que les
inhomogénéités aperçues à la surface de l'étoile soient le fruit de
telles cellules de convection.

\clearpage

\section{Perspectives}

   \begin{figure} \centering
   \includegraphics[width=7.6cm]{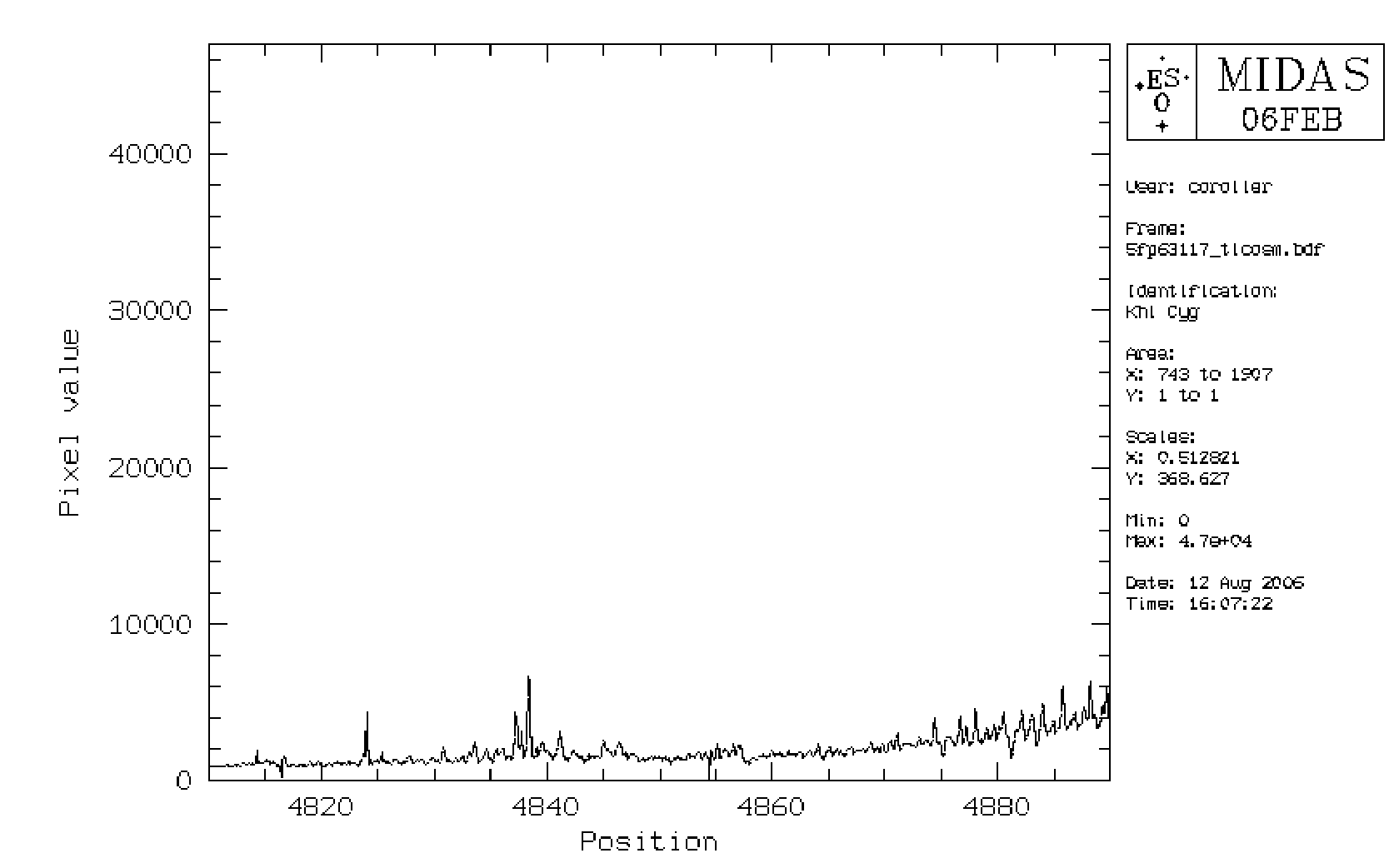}
   \includegraphics[width=7.6cm]{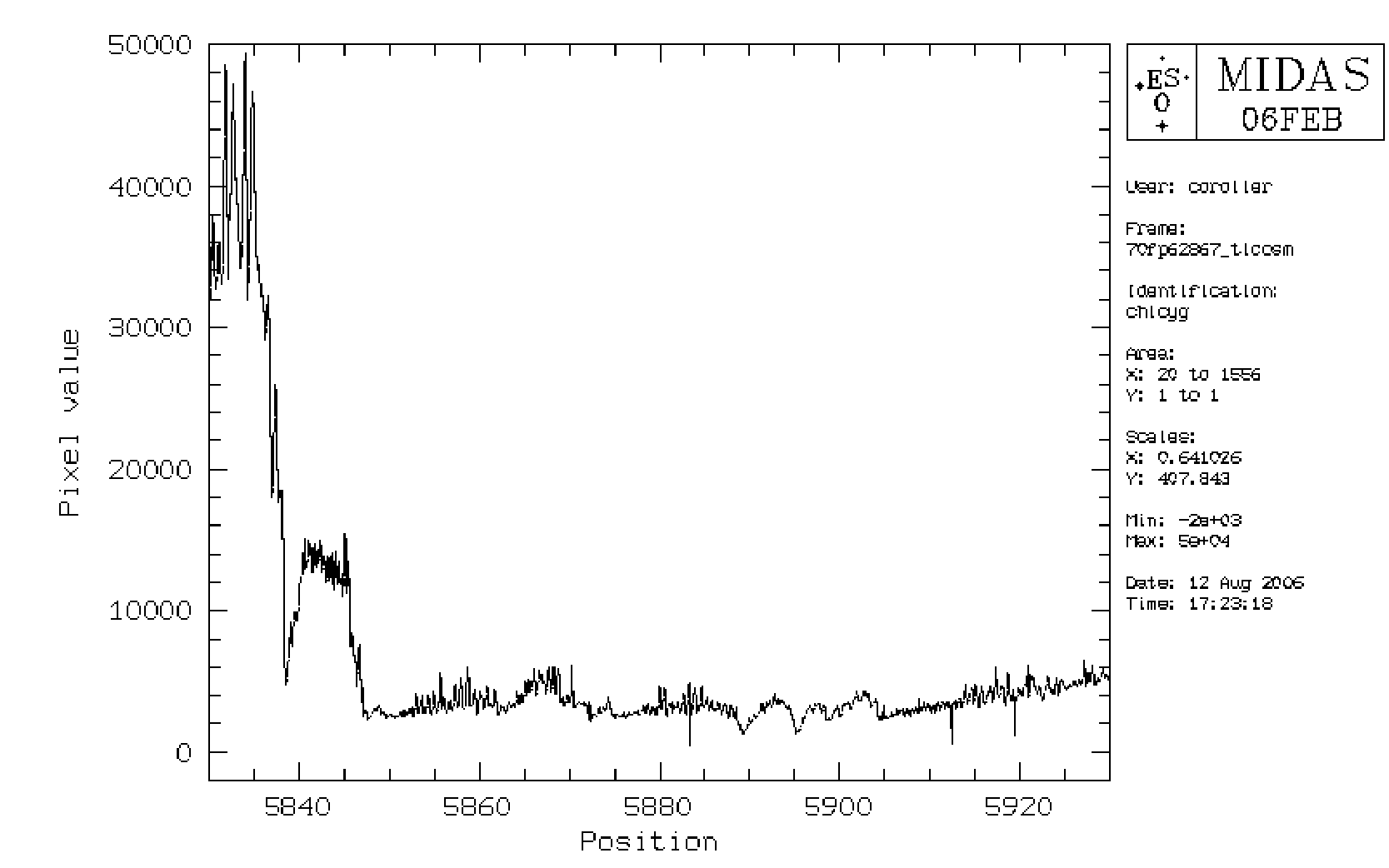}
   \includegraphics[width=7.6cm]{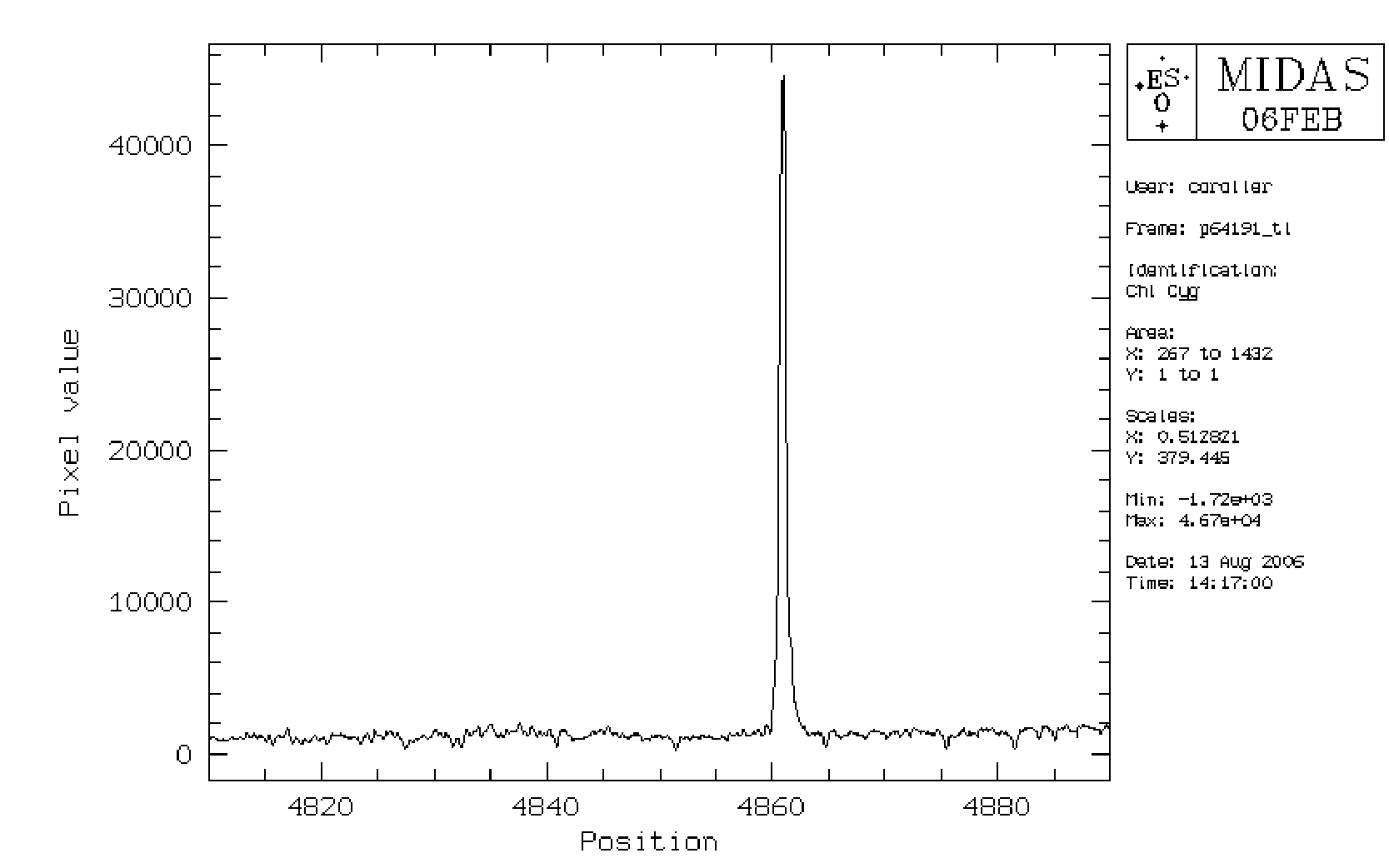}
   \includegraphics[width=7.6cm]{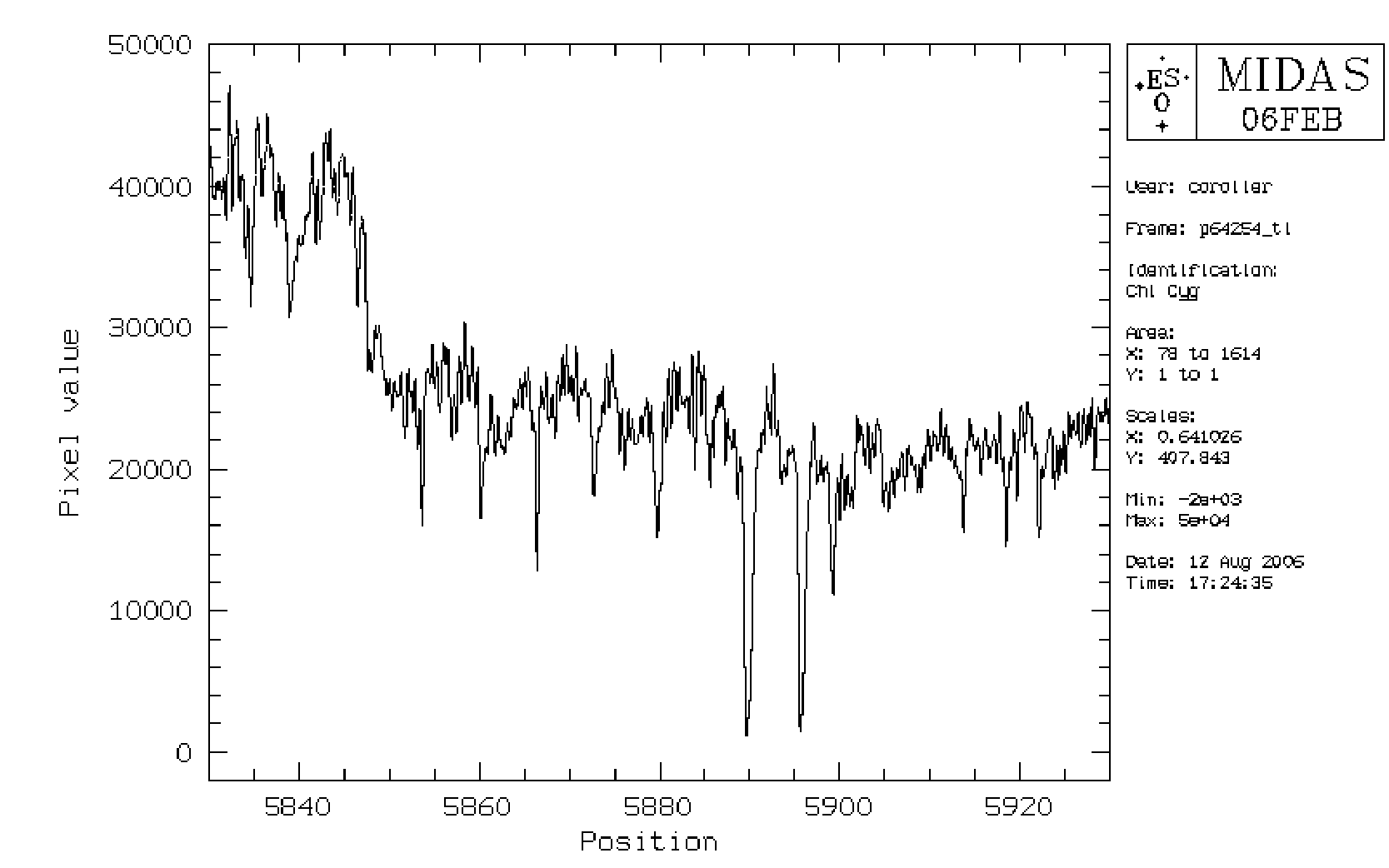}
   \caption[Observations effectuées à l'Observatoire de Haute provence
   de H$\beta$ et du doublet du Sodium sur $\chi$ Cyg]{ Observations
   effectuées à l'Observatoire de Haute provence sur le télescope de
   152 cm et le spectromètre AURELIE. \`A gauche les observations de
   H$\beta$ aux phases stellaires 0,8 et 0,96. \`A droite,
   observations du doublet du Sodium aux phases 0,74 et 0,96.}
   \label{fig:aurelie} \end{figure}

Ces travaux montrent clairement l'intérêt de conjuguer la
spectroscopie et l'interférométrie. Dans la
section~\ref{sec:temporelle}, nous avons pu mettre en évidence la
position et la dynamique de la couche moléculaire.  La
spectro-interférométrie est l'étape suivante, que nous avons effectuée
section~\ref{sec:dispertion}. Elle permet, outre d'obtenir
l'information spatiale à haute résolution angulaire, d'ajouter une
information sur la nature de l'objet (composition et température).

La spectro-interférométrie à très haute résolution spectrale est la
prochaine étape à franchir. Un tel instrument fournirait, en plus de
l'information spatiale, une information dynamique via l'effet
Doppler. Dans ce cadre, nous avons mené des études spectroscopiques
préparatoires sur $\chi$ Cyg. La Figure~\ref{fig:aurelie} montre les
observations des raies H$\beta$ (émission) et Sodium
(absorption). Nous avons pu observer de fortes variations temporelles,
à la fois en terme de vitesse radiale et en terme
d'amplitude. L'obtention d'une telle résolution sur ces raies, par le
biais de la spectro-interférométrie, nous permettrait d'obtenir une
mesure précise de leurs positions et de leurs vitesses.

Un instrument permettant de telles mesures n'est pas si futuriste
qu'il n'y paraît. AMBER, par exemple, permet déja d'effectuer des mesures
interférométriques avec une résolution de 12\,000 dans le proche
infrarouge. Il est, de plus, intéressant de noter que lorsque l'on
diminue la longueur d'onde, on augmente la résolution spatiale. \`A
une longueur d'onde de 550 nm, un télescope de 10 mètres a
potentiellement la même résolution qu'un interféromètre de 30 mètres
en bande H. En pratique, la limitation est alors la turbulence de
l'atmosphère, qui devient trés importante pour les faibles longueurs
d'onde. Les techniques interférométriques sont, cependant, bien
adaptées à la calibration de l'effet de la turbulence.

C'est pourquoi une partie conséquente de ma thèse a été consacrée à
l'élaboration d'un instrument interférométrique fonctionnant dans le
domaine de longueur d'onde visible. Parce qu'un télescope de 10 mètres
est suffisant pour résoudre les étoiles et leur environnement proche,
nous avons conçu cet instrument dans l'hypothèse d'une utilisation à
partir d'un télescope simple. La deuxième partie de ce manuscrit
traite de ce projet.


\cleardoublepage

\begin{figure}
  \centering  \resizebox{\hsize}{!}{\includegraphics{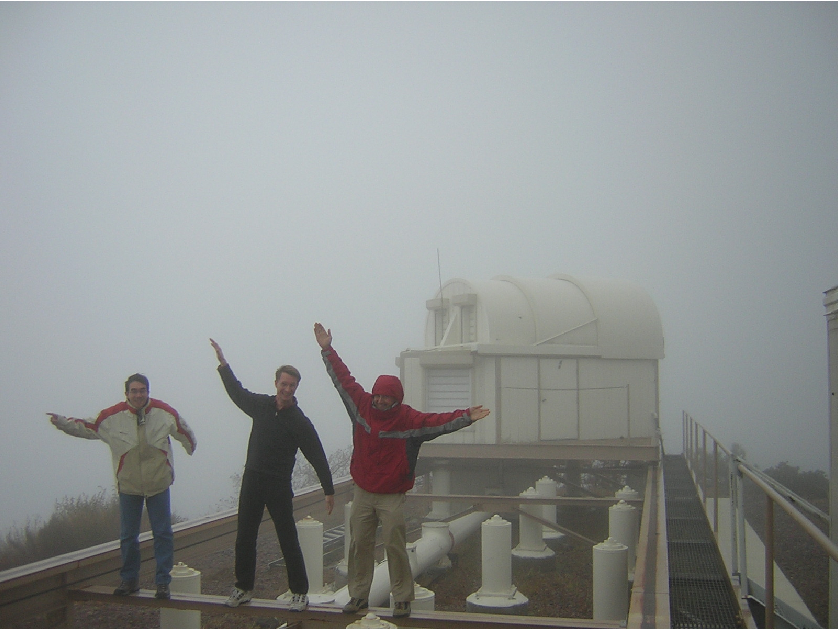}}
\end{figure}

\part*{\flushright Deuxième partie \\ \vspace{1cm} II. Le
  réarrangement de pupille}
\addtocontents{toc} {\protect\vskip 0.8cm \protect\begin{center}
{\protect\Large\protect\sf II. Le
  ré-arrangement de pupille} \protect\end{center} \protect\vskip
-0.3cm }

\chapter{Le principe}
\begin{center}
\end{center}
\minitoc \label{sec:concept} \vskip1cm

\clearpage
\section{De l'imagerie directe à l'interférométrie}

\subsection{L'imagerie directe en présence de turbulences}

   \begin{figure}[h]
   \centering
   \includegraphics[width=10cm]{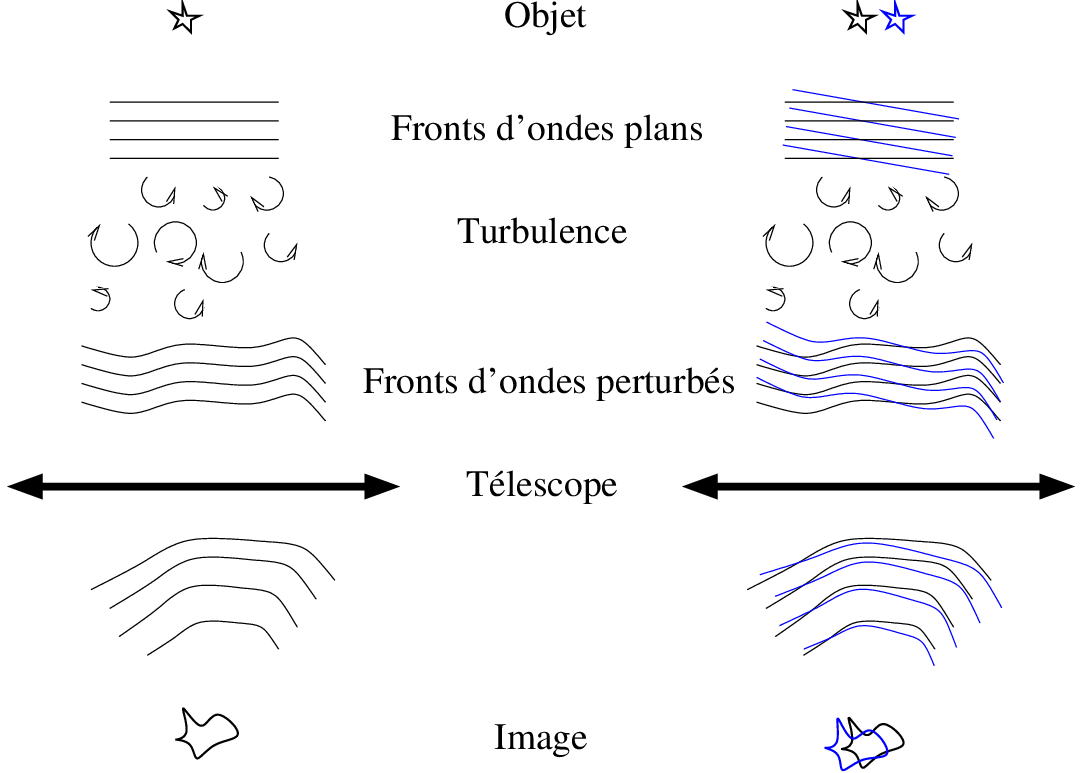}
   \caption[Schéma de l'influence de la turbulence atmosphérique sur
   la création d'une image]{Schéma de l'influence de la turbulence atmosphérique sur
   la création d'une image. Le front de l'onde électromagnétique
   provenant d'une source à l'infinie est plane en l'absence de
   perturbations. En se propageant à travers l'atmosphère terrestre, le
   rayonnement perd en cohérence et, focalisé par le télescope, est
   diffracté sous la forme d'une ``tache'' sur le détecteur. Pour une
   source multiple, l'image obtenue correspond à la convolution de
   l'objet par cette ``tache''.}
         \label{fig:image_turb}
   \end{figure}

Une propriété fondamentale de l'imagerie directe est le lien de
convolution qui existe entre l'objet observé et l'image obtenue. Cette
relation est valable pour un champ plus petit que l'angle
d'isoplanétisme~:
\begin{equation}
I(\alpha,\beta)=(O*S)(\alpha,\beta)\,.
\end{equation}
Ainsi, l'image $I(\alpha,\beta)$ est obtenue par le produit de
convolution ($*$) entre l'objet $O(\alpha,\beta)$ et la réponse
impulsionnelle de l'instrument $S(\alpha,\beta)$.

En présence de l'atmosphère terrestre, la réponse impulsionnelle d'un
télescope de taille supérieure au paramètre de Fried ($r_0$) possède
deux propriétés importantes~:
\begin{itemize}
\item Une fréquence de coupure qui limite la résolution
  angulaire de nos observations. Ainsi, comme le montre -- schématiquement -- la
  figure~\ref{fig:image_turb}, deux objets distincts peuvent
  apparaître confondus sur le détecteur. Cette limite en résolution
  est fixée soit par la taille du télescope ($\alpha=\lambda/D$), soit
  par la turbulence dans le cas d'une pose longue.
\item La réponse impulsionnelle varie au cours du temps et dépend
  fortement des paramètres atmosphériques. Il est possible d'en faire
  la moyenne au cours du temps -- comme on le fait lors d'une longue
  pose -- mais on perd alors en résolution, fixée désormais par le
  paramètre de Fried~: $\alpha=\lambda/r_o$.
\end{itemize}

La solution réside, a priori, dans la correction de la réponse
impulsionnelle de l'instrument en temps réel, via un miroir
déformable. C'est ce que l'on appelle l'optique adaptative. On
restitue ainsi une image quasiment invariable au cours du temps qui peut être
intégrée sur le détecteur. Ce mode est idéal pour un objet de faible
intensité. Cependant, cette technique se trouve vite confrontée à des
limites technologiques, par exemple lorsque l'on veut observer aux
courtes longueurs d'ondes ($<1\mu$m) ou dans le cas des futurs
télescopes de grandes tailles (ELT).

L'alternative -- qui peut être également considérée comme une approche
complémentaire -- consiste à effectuer un travail de déconvolution
post-observation.  A partir de poses courtes, la restauration de
l'image à la limite de diffraction du télescope peut être vue
comme un simple problème de déconvolution. Cette approche est utilisée
dans les techniques d'interférométrie des tavelures
\citep{1970A&A.....6...85L} et de masquage de pupille
\citep{1987Natur.328..694H}.

\subsection{L'interférométrie des tavelures et le masquage de pupille}
\label{sec:spl_mask}

\begin{figure}
  \centering \resizebox{\hsize}{!}{
  \includegraphics{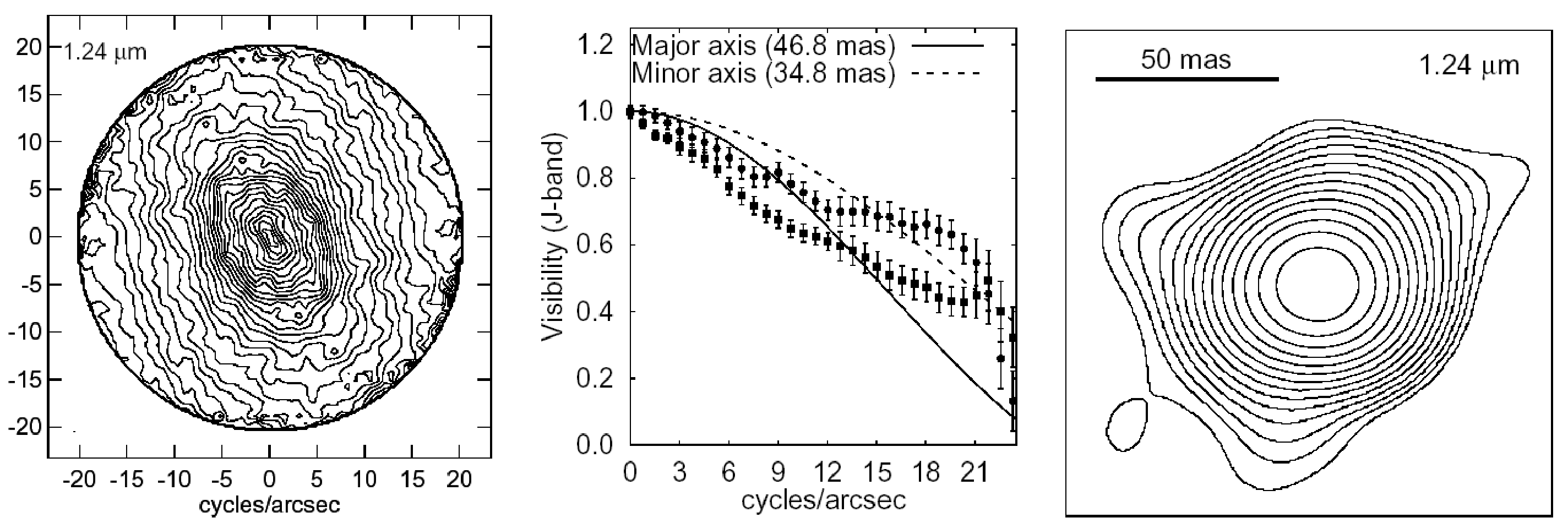} } \resizebox{\hsize}{!}{
  \includegraphics[scale=.82]{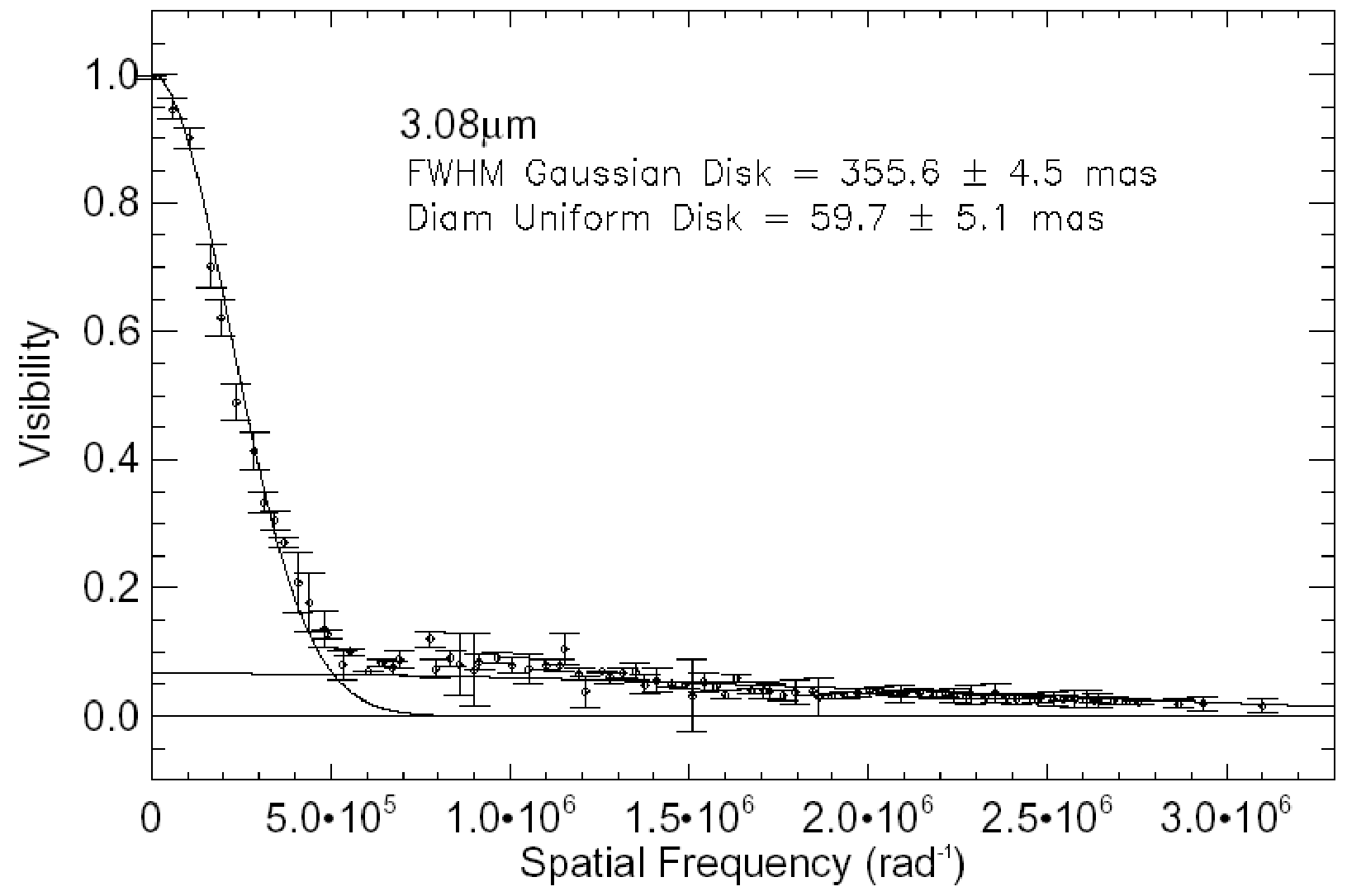}
  \includegraphics{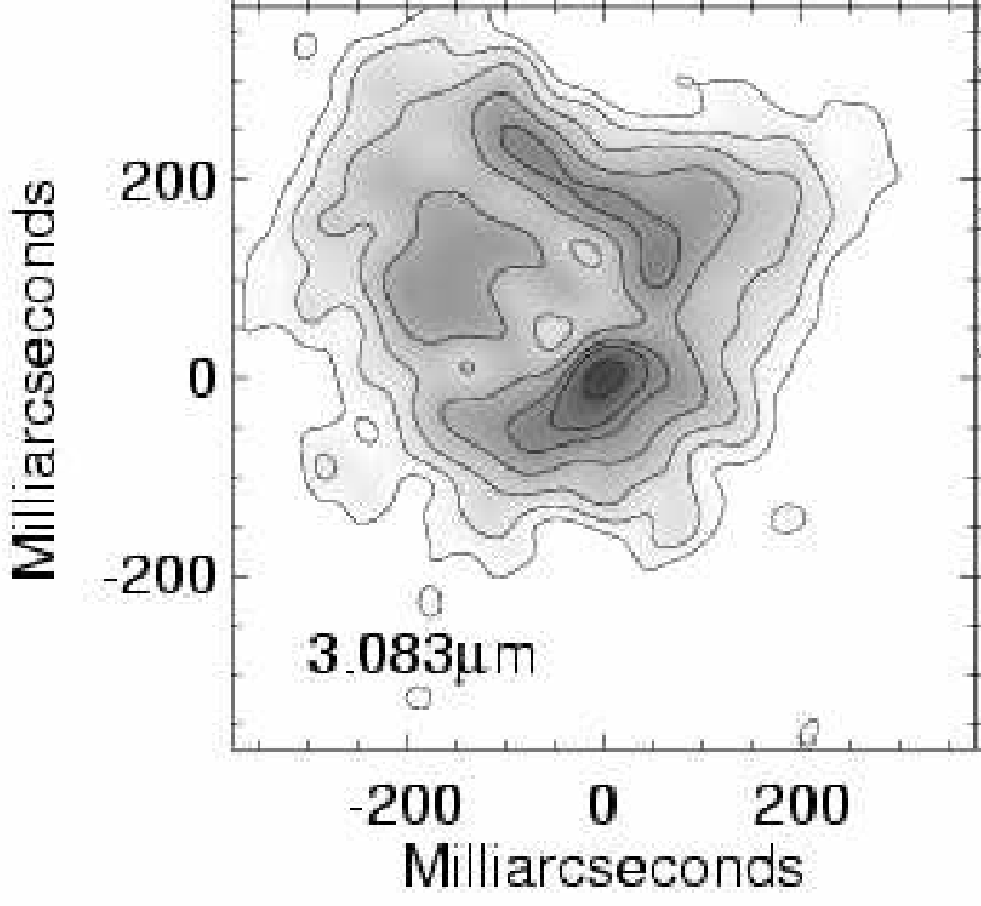} } \caption[Observations de
  B{[e]} MWC\,349A \citep{2002A&A...395..891H} et IRC+10216
  \citep{2005ApJ...624..352T}]{ Ces deux résultats observationnels,
  même s'ils sont obtenus sur des objets différents, présentent le
  type de résultats obtenus en interférométrie des tavelures et en
  masquage de pupille. En haut~: Observation de l'étoile B[e]
  MWC\,349A par \citet{2002A&A...395..891H} (interférométrie des
  tavelures) avec le télescope de 6 mètres SAO. En bas~: Observation
  de l'étoile carbonée IRC+10216 par \citet{2005ApJ...624..352T}
  (masquage de pupille) en utilisant le télescope de 10 mètres Keck
  I. A gauche sont représentés les plans des fréquences spatiales.
  Contrairement au masquage de pupille, l'interférométrie des
  tavelures donne une fonction continue correspondant à la surface
  d'autocorrélation de la pupille (figure du haut à gauche).  }
  \label{fig:speckl_mask}
\end{figure}

L'objectif des deux techniques d'interférométrie des tavelures et de
masquage de pupille est le même, à savoir, retrouver la distribution
de brillance de l'objet ($O(\alpha,\beta)$) à partir d'une image
($I(\alpha,\beta)$). Ce problème de déconvolution s'écrit dans le
domaine des fréquences spatiales~:
\begin{equation}
TF(I(\alpha,\beta))=TF(O(\alpha,\beta))\, . \, TF(S(\alpha,\beta))\,.
\end{equation}

Sous cette forme, on peut établir un lien direct avec
l'interférométrie longue base classique en notant que
$TF(O(\alpha,\beta))$ correspond à la cohérence du champ
électromagnétique dans la pupille $V(u,v)$ (cf. théorème de Zernike
Van Cittert), si celui-ci n'est pas perturbé par l'atmosphère. Par
analogie, $TF(I(\alpha,\beta))$ représente la cohérence du champ
mesuré, et, enfin, $TF(S(\alpha,\beta))$ la fonction de transfert
instrumentale.

L'interférométrie des tavelures et le masquage de pupille consistent
tous deux à mesurer la cohérence du champ électromagnétique affranchi
des turbulences. L'objectif, à savoir la mesure la plus précise
possible des visibilités $V(u,v)$, est le même que pour
l'interférométrie longue base. Des publications récentes comme celles
de \citet{2002A&A...395..891H} et \citet{2005ApJ...624..352T} en sont
l'illustration. Ils présentent les visibilités dans le domaine
fréquentiel de la même façon que nous les avons présentées dans la
première partie de cette thèse (figure~\ref{fig:speckl_mask}).

La différence entre masquage de pupille et interférométrie des
tavelures réside dans le choix de la fonction de transfert optique
$TF(S(\alpha,\beta))$. Celle-ci est égale à l'autocorrélation du champ
électrique dans la pupille. Afin d'utiliser un maximum de photons,
l'interférométrie des tavelures utilise une pupille
pleine. L'inconvénient réside dans ce que les hautes fréquences sont
extrêmement atténuées par les turbulences atmosphériques. A l'opposé, le
masquage de pupille utilise un masque non-redondant pour occulter une
partie de la pupille. En contrepartie d'une perte de flux, on obtient
alors une fonction de transfert assez stable sur l'ensemble des
fréquences disponibles, l'échelle des sous-pupilles étant fixée par
l'importance des perturbations atmosphériques.

\subsection{L'approche interférométrique}

La problématique des techniques de déconvolution consiste à retrouver
la cohérence du champ d'un objet astrophysique débarrassé de la
turbulence atmosphérique.  La précision de cette mesure caractérise la
qualité de l'image reconstruite.  Les erreurs produites par les
techniques classiques (figure~\ref{fig:speckl_mask}) donnent une idée
des progrès restant à accomplir dans le domaine. On note des erreurs
de l'ordre de 10\% sur les visibilités obtenues par interférométrie
des tavelures, et de l'ordre de 5\% sur les visibilités obtenues par
masquage de pupille.

Or, dans le cadre de nos travaux sur les étoiles évoluées, nous avons
observé des visibilités précises à quelques pourcents. Cette
différence est surprenante lorsque l'on compare la difficulté relative
entre l'interférométrie longue base et le masquage de pupille. En
effet, nos données ont été obtenues sur plusieurs jours avec des
télescopes déplacés sur plusieurs dizaines de mètres. A l'opposé, un
télescope offre la possibilité de mesurer de multiples fréquences
spatiales instantanément, sans avoir recours à un système complexe de
lignes à retard. Une précision au moins égale au \% devrait être
atteinte en appliquant les méthodes d'interférométrie longue base à la
mesure de la cohérence du champ électrique dans la pupille d'un
télescope.

\citet{1992A&A...255..462M} ont déjà suivi un tel raisonnement.  Ils
 ont appliqué la méthode de retournement de pupille. Pour cela, ils
 ont dupliqué la pupille et l'on conjuguée avec elle-même par le biais
 d'une modulation temporelle de la différence de marche optique. Ils
 ont ainsi publié un certain nombre de résultats astrophysiques
 \citep{1992A&A...255..462M,1992A&A...260..510M}. Cependant, au prix
 d'un concept compliqué, ils n'ont amélioré que faiblement la qualité
 des données comparées aux résultats obtenus avec des méthodes plus
 classiques comme l'interférométrie des tavelures. Une des raisons
 principales est qu'ils n'ont pas disposé de filtrage spatial, et que,
 seule la taille des pixels sur le detecteur venait limiter la taille
 de la zone de cohérence à faire interférer.  Il faut, cependant,
 noter que la qualité des mesures interférométriques était alors loin
 de celle obtenue actuellement. Les bouleversements technologiques
 récents du domaine que sont, notamment, l'apparition de l'optique
 guidée et du filtrage monomode, ont considérablement changé les
 perspectives. \`A la lumière de ces développements technologiques
 récents, il est intéressant de s'interroger à nouveau sur la
 problématique de la mesure interférométrique du champ électrique dans
 la pupille.

\clearpage
\section{La mesure du champ complexe dans la pupille}

Lorsque l'on conçoit un interféromètre, le choix technologique est
fondamental. Il conditionne la qualité et la sensibilité de
l'interféromètre. Il s'agit de la première étape à valider avant
l'élaboration de ce système. Deux points cruciaux sont à étudier, à
savoir, le filtrage du front d'onde et le système de recombinaison.
Ils conditionnent la précision des données obtenues par les
interféromètres longue base actuels.

\subsection{Le filtrage du front d'onde}

   \begin{figure}[h]
   \centering
   \includegraphics[width=10cm]{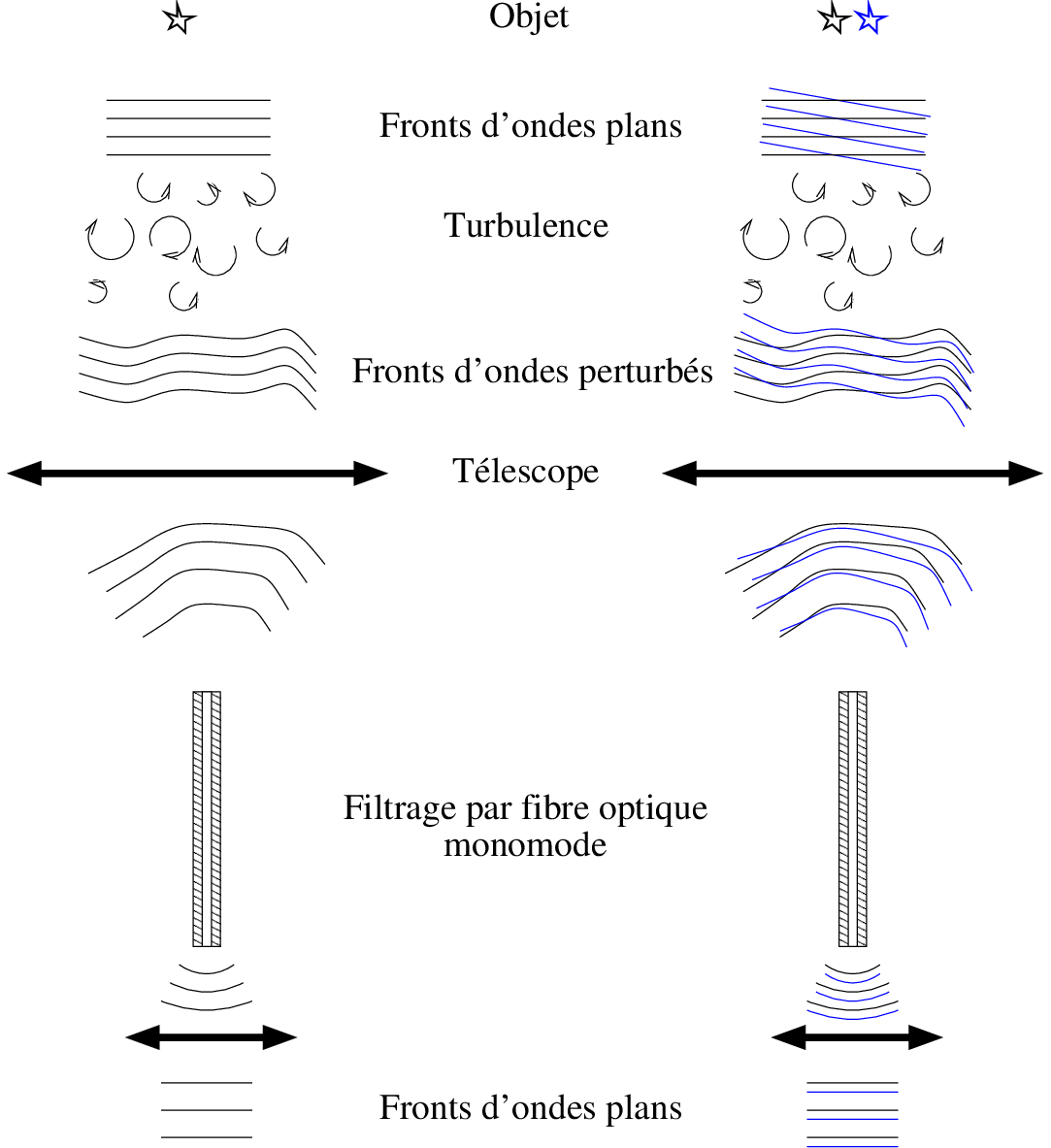}
   \caption[Schéma de l'influence du filtrage spatial sur des fronts
   d'ondes perturbés]{Schéma de l'influence du filtrage spatial sur des fronts
   d'ondes perturbés. Quel que soit la  déformation ou l'angle
   d'incidence, les fronts d'ondes à la sortie d'une fibre monomode
   sont lisses et plans. Les perturbations de phase deviennent alors
   des variations d'intensité.}
         \label{fig:turb_filtre}
   \end{figure}

   \begin{figure}[h] \centering
     \includegraphics[width=10cm]{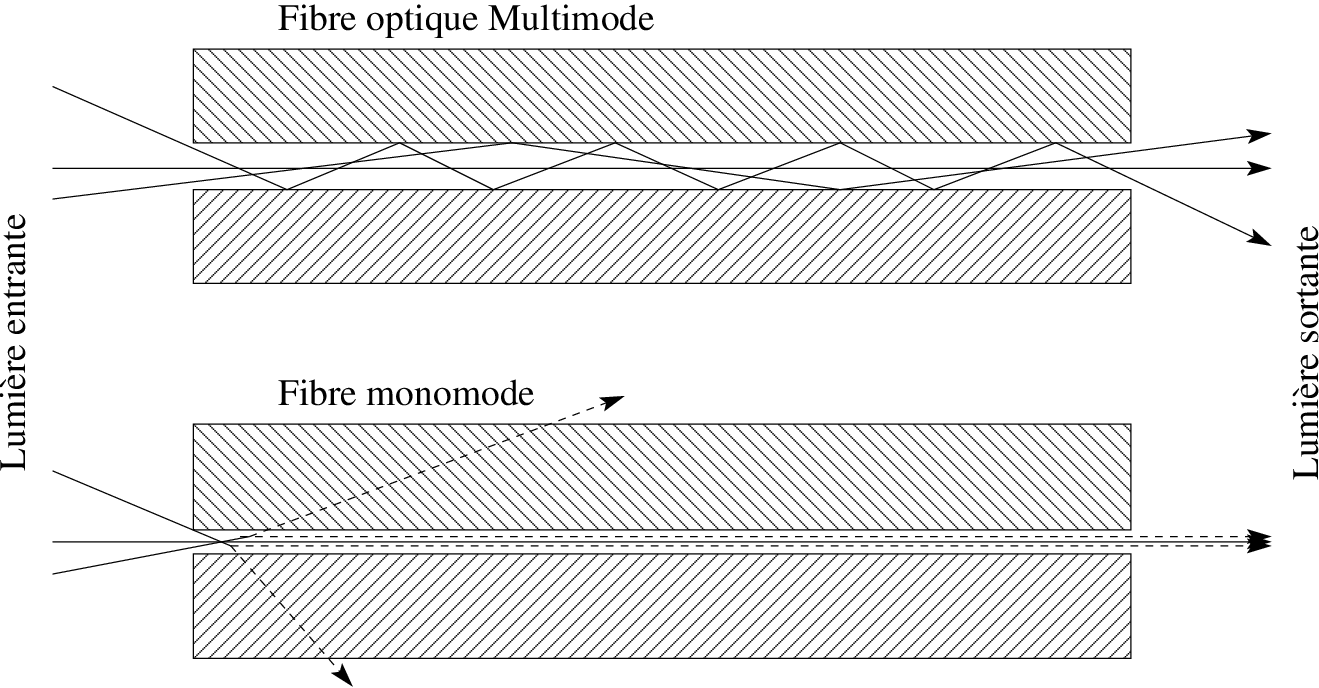}
     \caption[Schéma de l'injection dans une fibre optique monomode]{
     Dans une fibre optique monomode, la lumière, arrivant sous
     différents angles, a des trajectoires différentes dans le coeur
     de la fibre, et, en conséquence, plusieurs parcours
     optiques. Dans une fibre optique monomode, la lumière n'est plus
     réfléchie mais guidée selon le mode fondamental de la
     fibre. Lorsque la lumière incidente possède plusieurs modes, tous
     les autres modes sont rejetés.}  \label{fig:fibre_schema}
     \end{figure}

Pour mesurer précisément la cohérence du champ électrique provenant de
l'objet astrophysique, il est nécessaire de ne pas subir de perte de
cohérence provenant de la turbulence atmosphérique. Cette perte est
due aux déformations des fronts d'onde qui, en l'absence
d'atmosphère, arrivent plans (figure~\ref{fig:turb_filtre}).  Il faut,
en conséquence, trouver un moyen de corriger ces déformations. Un
miroir déformable via une optique adaptative (méthode active) peut
être une solution, tout comme un trou filtrant ou une fibre
optique monomode (méthode passive). 

De ces trois solutions, seul le filtrage par fibre optique monomode
permet d'obtenir un front d'onde parfaitement plan. Le rayonnement
filtré par une fibre monomode est ainsi parfaitement
cohérent. Cependant, la correction se traduit par une perte de
couplage qui dépend de la perturbation sur l'onde incidente. Ainsi,
une fibre monomode convertit des perturbations de phase en des
perturbations d'amplitude \citep{1997A&AS..121..379C}.

La figure~\ref{fig:turb_filtre} reflète l'avantage et l'inconvénient
du filtrage spatial passif.  L'intérêt réside dans un lissage parfait
du front d'onde~: l'onde lumineuse en sortie de fibre est parfaitement
cohérente, comme elle le serait en l'absence de turbulence
atmosphérique. L'inconvénient de ce filtrage est que, à l'échelle
d'une pupille, il nous fait perdre l'information spatiale de l'objet
observé. Dans le cas précis de la figure~\ref{fig:turb_filtre}, le
front d'onde bleu est initialement incliné par rapport au front d'onde
noir. Après filtrage monomode, les deux fronts d'onde sont parallèles,
et seul demeure un déphasage entre les deux. Ceci est une propriété
des fibres optiques monomodes que nous avons explicitée en
figure~\ref{fig:fibre_schema}. Elle provient du fait que seul le mode
fondamental est transporté par la fibre. \`A la sortie de la fibre,
deux paramètres seulement caractérisent le front d'onde~: la phase
($\phi$) et le gain ($g$). Nous utiliserons par la suite le
coefficient de transmission complexe de la fibre~:
\begin{equation}
G=g \exp(\I \phi) \,.
\label{eq:Gi}
\end{equation}
Si l'information spatiale à l'échelle d'une fibre est perdue, nous
pouvons cependant obtenir une information via la mesure de la
cohérence entre deux fibres. La mesure de la cohérence $\mu$ entre les
rayonnements issus de deux fibres de coefficients de transmission
$G_1$ et $G_2$ est~:
\begin{equation}
\mu=V\,G_1\,G_2^\star\,.
\end{equation}
où $V$ est la visibilité de l'objet à la fréquence spatiale définie
par la position relative des deux pupilles. Si l'on connaît les
facteurs de transmission des fibres ($G_1$ et $G_2$), on peut ainsi
obtenir une mesure précise de la visibilité de l'objet via $\mu$. Il
est important de préciser que la valeur de $\mu$ est une mesure
empirique de la cohérence. Il ne s'agit pas du facteur de cohérence,
puisque celui-ci n'est pas normalisé.

\subsection{Le choix du mode de recombinaison}

\subsubsection{Solutions de recombinaisons co-axiales}

   \begin{figure}[h]
   \centering 
\resizebox{\hsize}{!}
{   \includegraphics{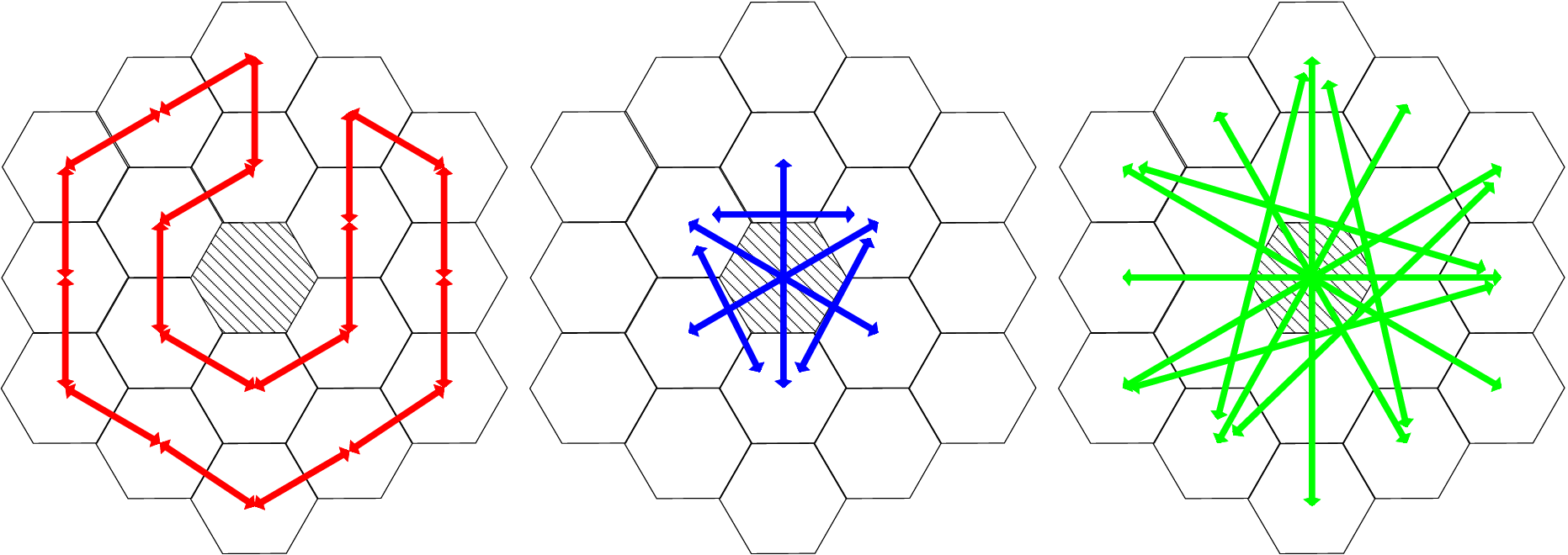} }
   \caption[Exemple de possibilité offerte dans le cas d'un système
   disposant de recombinaisons interférométriques par paires]{Exemple de possibilité offerte dans le cas d'un système
   disposant de recombinaisons interférométriques par paires. Le flux
   pourrait être distribué en trois réseaux interférométriques
   d'importance. Le rouge servirait à l'étalonnage, le bleu à la
   mesure des basses fréquences spatiales, et le vert aux hautes
   fréquences.}
         \label{fig:abcd}
   \end{figure}

   \begin{figure}
   \centering {
   \includegraphics[width=6cm]{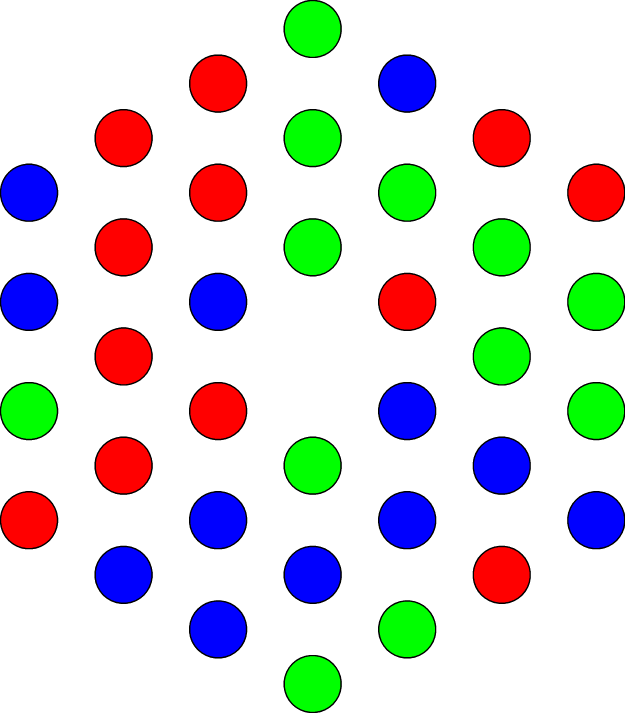} \hfill
   \includegraphics[width=6cm]{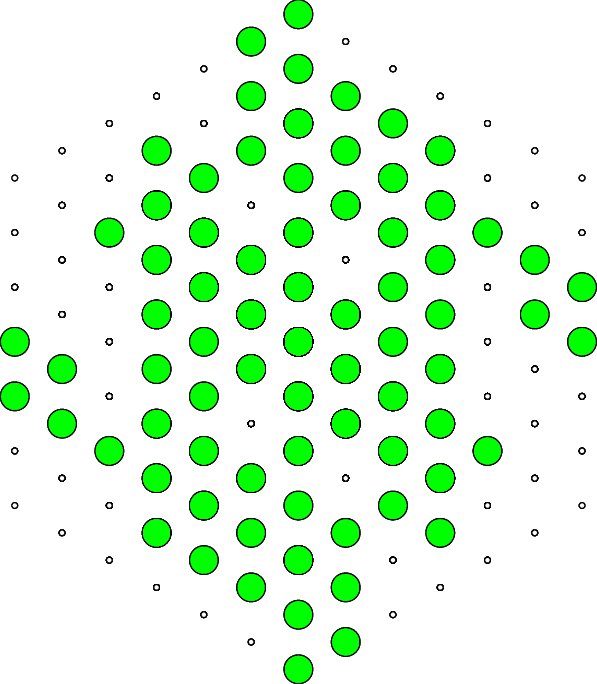}  \vspace{1cm}}{
   \includegraphics[width=6cm]{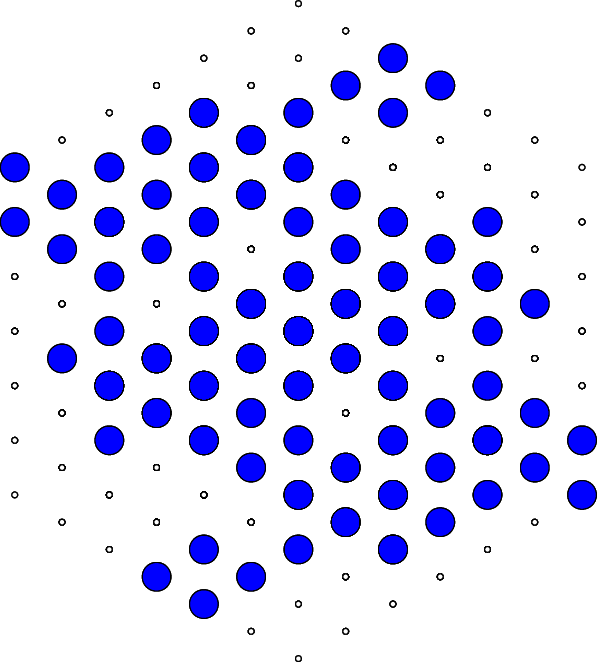} \hfill}
   \includegraphics[width=6cm]{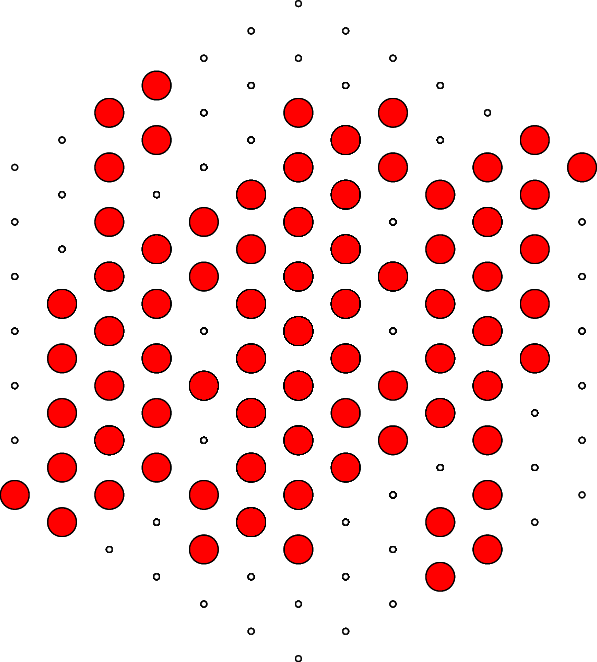} 
   \caption[Exemple de possibilité offerte dans le cas d'un système à
   recombinaison mutli-axiale]{Exemple de possibilité offerte dans le cas d'un système à
   recombinaison mutli-axiale. Ici, les 36 éléments de la pupille
   (figure du haut à gauche) sont séparés en 3 sous-groupes. Les trois
   autres figures représentent les plans $u$-$v$ obtenus à partir de
   chaque sous-groupe. Toutes les fréquences spatiales ne sont pas
   forcément accessibles, mais cette technique permet, néanmoins,
   d'obtenir un compromis entre information fréquentielle et
   sensibilité de l'instrument.}
         \label{fig:decoup}
   \end{figure}

Le filtrage spatial est une des clefs des précisions obtenues en
interférométrie. La deuxième provient de la technique de
recombinaison. Sur IOTA, nous avons vu que la recombinaison s'opérait
par une optique intégrée, associée à une modulation temporelle. Il
s'agissait d'une recombinaison co-axiale.  Elle est, d'ailleurs,
utilisée sur la plupart des interféromètres fibrés actuels ({IOTA
/IONIC}, CHARA/FLUOR, VLTI/VINCI). Devant la présision des mesures
obtenues avec l'interférométre IOTA, nous avons envisagé, dans un
premier temps, d'utiliser la même méthode de modulation. Concrètement,
la modulation temporelle présente de sérieux avantages en
interférométrie. L'une de ses caractéristiques
principales est qu'elle rend la cohérence invariable par rapport a un
piston fixe. En effet, si l'amplitude de modulation est
suffisamment grande, l'influence d'un piston statique est un simple
décalage des franges.

 Cependant, lorsque l'interféromètre est doté de multiples télescopes,
un codage temporel devient problématique. Le flux de chaque
télescope doit alors être modulé par une fréquence singulière. Chaque
fréquence doit être choisie de façon à ce que la modulation entre les
fréquences des différents télescopes ne génère pas deux fréquences
identiques. Ceci conduit à choisir des fréquences de modulation
non-redondantes. Suivant le nombre de télescopes, on peut alors
aboutir à de grandes différences entre les hautes fréquences et les
basses fréquences. Les franges observées trop rapidement seraient
alors soumises à un bruit de photon et de détecteur important, et les
franges observées trop lentement subiraient l'influence d'un bruit de
piston dynamique important.

Un moyen de contourner ce problème serait de recombiner les sous-pupilles par
paires. Ceci rend la modulation temporelle plus complexe, parce qu'elle
nécessite autant de systèmes interférométriques que de faisceaux au
carré. C'est face à cette complexité qu'un recombinateur en optique
intégrée de type ABCD devient nécessaire. Par rapport à un système
multi-axial, cette solution a l'avantage de pouvoir diviser le flux de
chaque faisceau et, suivant une configuration optimale, permet la
recombinaison des seules bases intéressantes. Une telle idée est
illustrée par la figure~\ref{fig:abcd}, dans le cadre de la
recombinaison de 18 éléments d'un télescope. Il est proposé ici de
diviser le flux de chaque segment en deux. Une partie de ce flux
alimente un premier réseau interférométrique (en rouge) qui servirait
à la mesure des pistons et des gains.  Le reste du flux sert à
extraire l'information astrophysique par la mesure des différentes
fréquences spatiales (en bleu et en vert). Ainsi, seules les
recombinaisons nécessaires sont effectuées.

\subsubsection{Solutions de recombinaisons multi-axiales}

Une autre possibilité est
celle d'une recombinaison multi-axiale.  Historiquement, les
premières franges stellaires ont été obtenues par l'interféromètre
multi-axial de Michelson, à modulation spatiale
\citep{1920ApJ....51..257M}. Ce mode de recombinaison a l'avantage de
la simplicité. Ainsi, à partir d'une
seule lentille, on peut recombiner plus d'une centaine de
faisceaux. De part sa simplicité technique, la précision sur les
visibilités mesurées devrait donc être optimale. La contrepartie de ce
système est la sensibilité qui décroît proportionnellement au nombre
de télescopes $M$. Ceci est dû au nombre de fréquences spatiales qui
augmente proportionnellement à $M^2$, alors que le flux est proportionnel
à $M$. Un compromis peut être obtenu en choisissant une recombinaison
par groupe (illustration figure~\ref{fig:decoup})

Au cours de cette thèse, nous avons principalement étudié ce type de
recombinateur. Néanmoins, un recombinateur par paire en optique
intégrée constitue un développement technologique intéressant à
expérimenter dans le futur.

\subsection{Le réarrangement de pupille}
\label{sc:concept_rea}

   \begin{figure}[h] \centering
   \includegraphics[width=9cm]{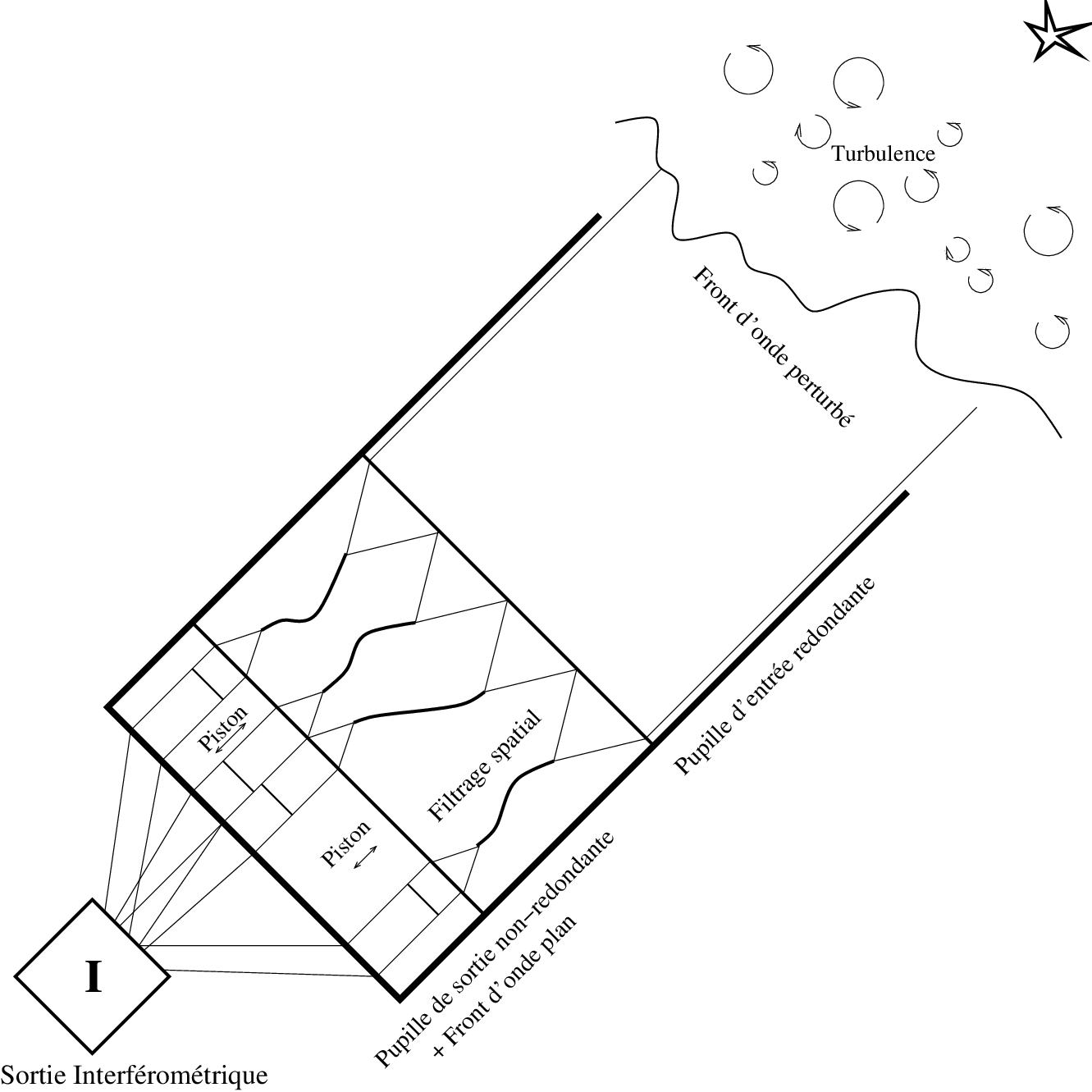} \caption[Concept
   du réarrangeur de pupille]{Concept du réarrangeur de pupille. La
   pupille d'entrée est divisée en sous-pupilles dont le flux est
   focalisé dans une fibre optique. Les fibres sont ensuite
   redistribuées selon une configuration non-redondante, puis
   collimatées pour former la pupille de sortie. Celle-ci est ensuite
   focalisée sur le détecteur.}  \label{fig:concept} \end{figure}

Dans le cas de l'imagerie classique, ce sont les pertes de cohérence
qui limitent la dynamique des images reconstruites. Cette limite est
présente en interférométrie des tavelures, mais aussi lorsque l'on
utilise une optique adaptative
\citep{2005A&A...441.1205C,2006A&A...447..397C}.  C'est la redondance
de la pupille qui en est la cause. En effet, dans une pupille, de
multiples vecteurs identiques existent, et s'additionnent pour former
la fonction de transfert optique. Or, ces vecteurs s'additionnent de
manière complexe, et lorsque des perturbations de phases existent
entre les différents vecteurs, ceux-ci s'additionnent de manière
incohérente.

Une manière de s'affranchir de ce problème est d'utiliser un masque
non-redondant. Il s'agit de la technique de masquage de pupille que
nous avons vue briévement section~\ref{sec:spl_mask}. Le masque permet
de sélectionner un certain nombre de fréquences spatiales, qui, parce
que générées par une unique paire d'ouvertures, contribuent à la
fonction de transfert instrumentale sans perte de cohérence. 

Le principal inconvénient de la technique de masquage de pupille est
qu'elle nécessite de bloquer la lumière sur une majeure partie de la
pupille. La sensibilité d'un tel instrument est donc très faible. Une
solution consiste à d'augmenter la taille des sous-pupilles. Pour une taille
de sous pupille égale au paramètre de Fried $r_0$, la cohérence entre
2 sous-pupilles est alors d'environ 36\%, ce qui entraîne des
variations de la fonction de transfert optique difficiles à calibrer.

Le problème de la fluctuation de la cohérence, bien connu
en interférométrie longue base, a été résolu par l'utilisation de
fibres optiques. Ainsi, les perturbations de phases sont échangées contre
des fluctuations en amplitudes, plus faciles à étalonner. Le même
principe peut être appliqué à la technique de masquage de pupille. Un
tel concept a déja été proposé par \citet{1998SPIE.3350....2C}. Nous
proposons de pousser plus en avant le concept de l'utilisation de fibres
optiques, en réarrangeant la pupille, de manière à ce que la totalité
de la pupille du télescope soit utilisée.  Le concept auquel nous avons abouti est
présenté figure~\ref{fig:concept}. Le front d'onde perturbé de l'objet
astrophysique est filtré par une matrice de fibres optiques. Celles-ci
sont réarrangées selon une configuration non-redondante et le flux
sortant de chaque fibre est individuellement collimaté pour former une
nouvelle pupille. Dans la suite de ce manuscrit, lorsque l'on évoquera
la pupille d'entrée, il s'agira de la pupille du télescope, aux fronts
d'ondes perturbés par l'atmosphère. La pupille de sortie correspond à
la pupille non-redondante filtrée par les fibres optiques.

Le champ dans la pupille est ensuite focalisé sur le
détecteur. L'image obtenue est une tache de diffraction modulée par
des paquets de franges. \`A partir de l'amplitude et de la position des
franges on peut déduire une mesure de la cohérence complexe $\mu$. \`A
chaque vecteur fréquence $\mathbf{u}_k$ est associé une valeur
$\mu_k$.  En l'absence de réarrangement de la pupille, la 
cohérence est reliée à la visibilité de l'objet $V(\mathbf{u}_k)=V_k$
par la relation~:
\begin{equation}
\mu_k=V_k \sum_{(i,j)\in\mathcal{B}_{k}} G_i G_j^\star \,,
\end{equation}
où $\mathcal{B}_{k}$ est l'ensemble des paires de sous-pupilles
$(i,j)$ telles que les vecteurs position des sous-pupilles
$(\mathbf{r}_i,\mathbf{r}_j)$ vérifient:
\begin{equation}
\mathcal{B}_{k} = \Bigl\{\,(i,j) \ :\ (\mathbf{r}_i - \mathbf{r}_j)/\lambda = \mathbf{u}_k
  \,\Bigr\} \,.
\label{eq:mu_sum}
\end{equation}
et $G_i$ et $G_j$ sont les facteurs de transmission complexes tels
qu'établis par la relation~(\ref{eq:Gi}).

La somme de ces vecteurs complexes déphasés a alors un impact
important sur l'amplitude de la cohérence. Pour des pupilles à
plusieurs $r_0$ d'écart, les déphasages peuvent être supérieurs à
$\pi$. L'addition présentée dans l'équation~(\ref{eq:mu_sum}) fait
alors converger l'amplitude vers zéro, rendant la mesure très
difficile. La technique de réarrangement en une pupille non-redondante
a pour but d'éviter cet écueil. Elle contraint la pupille de sortie de
façon à ce qu'il n'existe qu'une seule paire de sous-pupilles
vérifiant $\mathcal{B}_{k}$.  La cohérence pour chaque paire de
sous-pupilles sera ainsi, sans ambiguïté, associée à une unique
fréquence spatiale observée sur le détecteur. Pour deux pupilles $i$,
et $j$, le terme de cohérence $\mu_{(i,j)}$ observé est alors
proportionnel aux facteurs de transmission complexes $G_i$ et $G_j$ et
à la cohérence $V_k$ du champ de l'objet observé~:
\begin{equation}
\mu_{(i,j)}=V_k G_i G_j^\star \,.
\label{eq:mu}
\end{equation}
La particularité du réarrangement de la pupille est qu'il ne modifie
pas les visibilités complexes de l'objet $V_k$
\citep{1992A&A...253..641T}. Elles sont ainsi égales à celles
mentionnées dans l'équation~(\ref{eq:mu_sum}) et correspondent à la transformé
de Fourier de l'objet observé à la fréquence spatiale $\mathbf{u}_k =
(\mathbf{r}_i - \mathbf{r}_j)/\lambda$ ($\mathbf{r}_i$ et
$\mathbf{r}_j$ sont les vecteurs position des sous-pupilles dans la
pupille d'entrée).

\clearpage
\section{L'estimation des visibilités complexes}

\subsection{Les estimateurs de clôture}
\label{sec:clot}
 
L'utilisation de fibres optiques permet, en restreignant l'effet de la
turbulence à deux inconnues par fibres $g$ et $\phi$
(equation~(\ref{eq:Gi})), de poser le problème de manière exacte. Pour
pouvoir retrouver les termes de visibilité de notre objet, il reste,
cependant, à mesurer ces deux termes. Une façon simple d'obtenir
l'amplitude de transmission de chaque fibre consiste à extraire une
partie du flux. On parle alors d'étalonnage
photométrique. Cependant, cette méthode ne permet pas de mesurer le
déphasage $\phi$ et, surtout, nous fait perdre une partie du flux. Une
seconde méthode consiste à utiliser, non pas les termes de cohérence
complexe $\mu_{(i,j)}$, mais les clôtures de phase et d'amplitude.

Le fonctionnement de ces termes de clôtures peut être
compris en développant l'équation~(\ref{eq:mu}) sous la forme~:
\begin{equation}
\mu_{(i,j)}=|V_{(i,j)}| g_i g_j\, \exp\left(\I (\arg(V_{(i,j)}) + \phi_i - \phi_j)\right)
\end{equation}
où $V_k=V_{(i,j)}=|V_{(i,j)}| \exp(\I\arg(V_{(i,j)}))$, et $G_i=g_i
\exp(\phi_i)$ le terme de transmission complexe de la fibre $i$.  La
clôture de phase s'obtient à partir de la phase du bispectre~:
\begin{eqnarray}
\mu^{(3)}_{(i,j,k)}&=&\mu_{(i,j)}\mu_{(j,k)}\mu_{(k,i)}\nonumber\\
&=& |g_i|^2 |g_j|^2 |g_k|^2  |V_{(i,j)}| |V_{(j,k)}| |V_{(k,i)}| \exp(\I
(\arg(V_{(i,j)})+\arg(V_{(j,k)})+\arg(V_{(k,i)})))\nonumber\\
&=& |g_i|^2 |g_j|^2 |g_k|^2  V_{(i,j)} V_{(j,k)} V_{(k,i)}  \,.
\nonumber
\\
\end{eqnarray}
La phase du bispectre ne dépend alors plus que de la somme des phases
de l'objet~:
\begin{equation}
\arg(\mu^{(3)}_{(i,j,k)})=\arg(V_{(i,j)} V_{(j,k)} V_{(k,i)})\,.
\end{equation}
 De même, il existe la clôture d'amplitude~:
\begin{eqnarray}
\mu^{(4)}_{(i,j,k,l)}&=&\frac{\mu_{(i,j)}\mu_{(k,l)}}{\mu_{(i,k)}\mu_{(j,l)}}\nonumber\\
&=&\frac{|V_{(i,j)}||V_{(k,l)}|}{|V_{(i,k)}||V_{(j,l)}|}\frac{\exp(\I(\arg(V_{(i,j)})+\arg(V_{(k,l)})-\arg(V_{(i,k)})-\arg(V_{(j,l)})))}{\exp(\I(2\phi_j-2\phi_k))}
\nonumber
\\
\end{eqnarray}
dont l'amplitude ne dépend plus que de l'amplitude du produit des isibilités:
\begin{equation}
|\mu^{(4)}_{(i,j,k,l)}|=\left|\frac{V_{(i,j)}V_{(k,l)}}{V_{(i,k)}V_{(j,l)}}\right|\,.
\end{equation}
 On peut même poursuivre ce raisonnement en imaginant un estimateur
qui soit indépendant à la fois de la phase et de l'amplitude~:
\begin{eqnarray}
\mu^{(6)}_{(i,j,k,l,m,n)}=\frac{\mu_{(i,j)}\mu_{(k,l)}\mu_{(m,k)}\mu_{(j,n)}}{\mu_{(i,k)}
  \mu_{(j,l)}\mu_{(m,j)}\mu_{(k,n)}} =\frac{V_{(i,j)}V_{(k,l)}V_{(m,k)}V_{(j,n)}}{V_{(i,k)}V_{(j,l)}V_{(m,j)}V_{(k,n)}}
\,. 
\end{eqnarray}

Ces estimateurs sont intéressants parce qu'ils peuvent êtres cumulés
sur de nombreuses acquisitions alors même que la turbulence varie
fortement. Cependant, ils ont plusieurs inconvénients. Le premier est
qu'ils ne sont pas optimaux lorsque le signal sur bruit des
acquisitions instantanées est faible. Le second inconvénient est que
ces estimateurs ne fournissent pas un critère convexe. Lorsque l'on va
chercher à retrouver les visibilités complexes de l'objet recherché,
de multiples minima pourront apparaître, posant des problèmes de
déconvolution. Enfin, c'est en terme de difficultés de calculs que se
pose le troisième inconvénient. En effet, pour optimiser le rapport
signal sur bruit, il est préférable d'utiliser toutes les clôtures
disponibles. Pour $M$ sous-pupilles, cela correspond à
$\frac{M(M-1)(M-2)}{6}$ clôtures de phase et
$\frac{M(M-1)(M-2)(M-3)}{24}$ clôtures d'amplitude. Pour 100
sous-pupilles, on devrait ainsi travailler avec environ 4 millions de
clôtures.

\subsection{Un problème bien posé}

\begin{figure}[ht]
  \centering
{  \includegraphics[scale=.33]{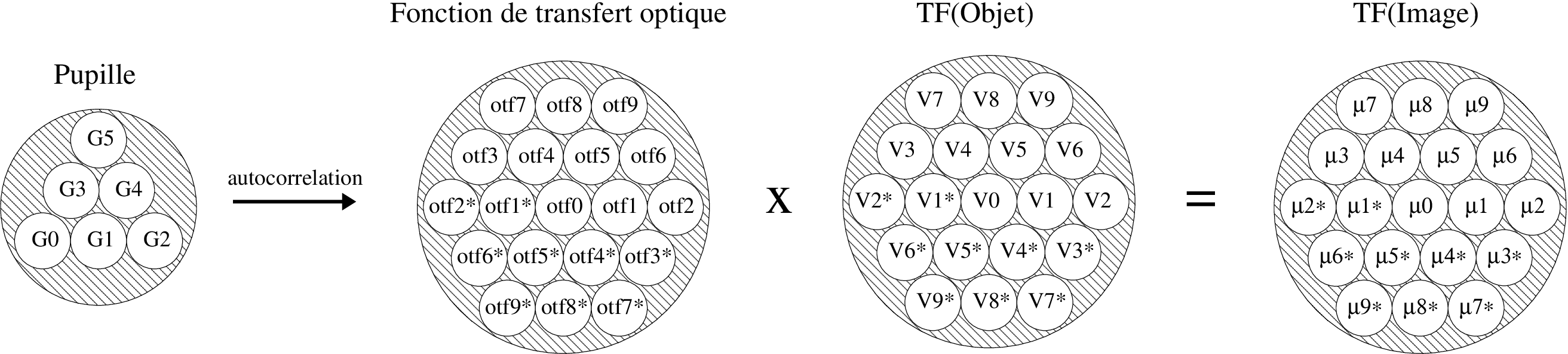}
 }\vspace{1cm}
{ \includegraphics[scale=.33]{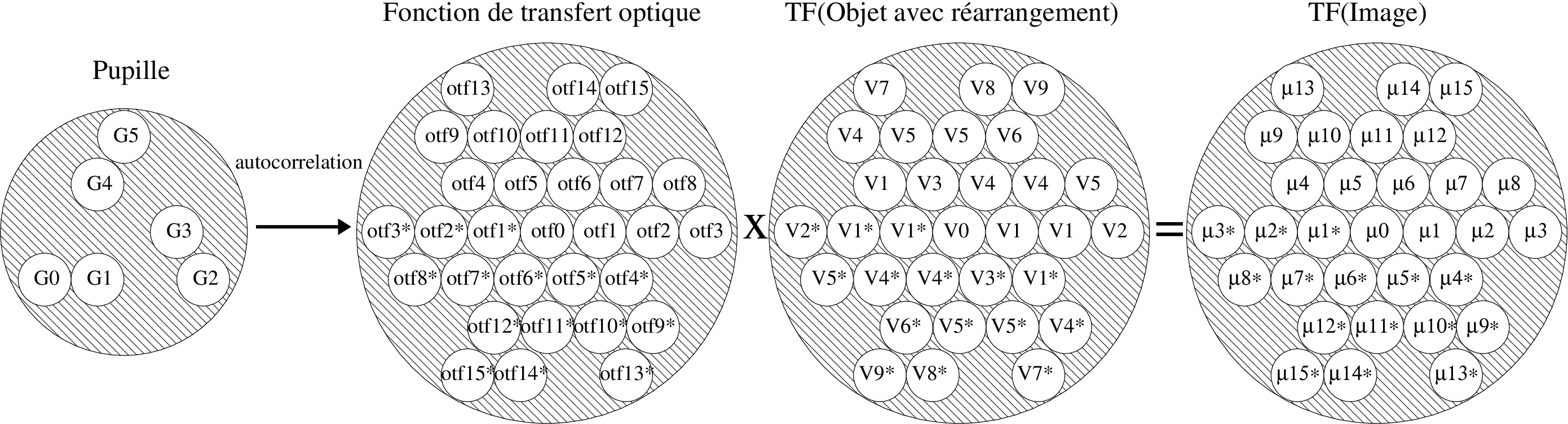} }
  \caption[Principe de l'imagerie de Fourier]{ Principe de l'imagerie
  de Fourier. La fonction de transfert optique est l'autocorrélation
  de la pupille. Celle-ci est alors multipliée par la transformée de
  Fourier de l'objet pour donner la transformée de Fourier de l'image
  observé. Dans le cas d'un système à réarrangement (figure du bas),
  le principe est le même, à la différence que la fonction de
  transfert optique est multipliée aux visibilités réarrangées.}
  \label{fig:remapped}
\end{figure}

Face aux difficultés concernant l'utilisation des clôtures, nous avons
choisi une approche différente, se voulant optimale dans le cadre d'un
bruit de statistique gaussienne. Il s'agit de considérer le problème
mathématique tel que nous l'avons établi par l'équation~(\ref{eq:mu}) et
de chercher les inconnues complexes $G_i$ et $V_k$ les plus proches au
sens du maximum de vraisemblance \citep{1985..Goodman}. Cependant,
avant d'en venir à l'algorithme de déconvolution
(section~\ref{sc:algo_eric}), il a fallu établir que le problème était
bien posé. Ceci se traduit par la démonstration de l'unicité de la
solution.

Nous avons pris l'exemple de la figure~\ref{fig:remapped}. La pupille
d'entrée du télescope est triangulaire et divisée en 6 sous-pupilles
de facteurs de transmission complexes [$G_0$,..,$G_5$]. La figure du
haut présente un filtrage sans réarrangement, et celle du bas un
réarrangement en une configuration non-redondante. Le problème de la
déconvolution d'une pupille sans réarrangement est strictement
identique à celui posé par l'interférométrie des tavelures, et fournit
un excellent moyen de comparaison.

Dans les deux cas, la fonction de transfert optique est
l'autocorrélation de la pupille. Sans réarrangement, cela correspond à
10 termes [otf$_0$,..,otf$_9$]. Avec réarrangement, la fonction de
transfert optique est composée de 16 termes
[otf$_0$,..,otf$_{15}$]. La fonction de transfert optique est ensuite
multipliée par les visibilités de l'objet [$V_0$,..,$V_9$] pour donner les
termes de cohérence complexe $\mu$. Ce sont les termes de visibilité
que l'on souhaite obtenir (sans ambiguïté) par la mesure de la cohérence
des franges sur le détecteur.

On peut ainsi établir une liste d'équations à inverser. Dans le cas de
la pupille sans réarrangement, il s'agit des équations~:
{\footnotesize
\begin{eqnarray}
\mu_0 &=&\sum_{i=0}^5 |G_i|^2 \nonumber \\
\mu_1 &=&V_1 \, ( G_0 G_1^\star +G_1 G_2^\star + G_3 G_4^\star) \nonumber\\
\mu_2 &=&V_2 \,  G_0 G_2^\star \nonumber \\
\mu_3 &=&V_3 \,  G_2 G_3^\star \nonumber \\
\mu_4 &=&V_4 \, ( G_4 G_5^\star +G_1 G_3^\star + G_2 G_4^\star)\nonumber\\
\mu_5 &=&V_5 \, ( G_0 G_3^\star +G_3 G_5^\star + G_1 G_4^\star)\nonumber\\
\mu_6&=&V_6 \, G_0 G_4^\star \nonumber \\
\mu_7&=&V_7 \, G_2 G_5^\star \nonumber \\
\mu_8&=&V_8 \, G_1 G_5^\star \nonumber \\
\mu_9&=&V_9 \, G_0 G_5^\star 
\end{eqnarray} }
Le nombre d'observables réelles est de 17, pour 9 valeurs complexes
([$\mu_1$,..,$\mu_9$]) et une valeur réelle ($\mu_0$). Le nombre
d'inconnues réelles s'élève à 18 pour les visibilités
([$V_1$,..,$V_9$]; puisque par définition $V_0=1$) et à 12 pour les
facteurs de transmission ([$G_0$,..,$G_5$]). Parce que le nombre de
mesures disponibles est inférieur au nombre d'observables, il ne peut
donc pas y avoir unicité de la solution. Ce problème peut, néanmoins,
être inversé dans le cas de l'interférométrie des tavelures, si
l'utilisateur contraint les visibilités par un a priori sur l'objet
observé. Cette technique est décrite en détail par
\citet{Thiebaut_Conan-1995-bdec}.

Lorsque l'on réarrange la pupille, le nombre d'équations augmente,
tout en conservant le même nombre d'inconnues~:
{\footnotesize
\begin{eqnarray}
\mu_0 &=&\sum_{i=0}^5 |G_i|^2 \nonumber \\ 
\mu_1 &=&V_1 \, G_0 G_1^\star \nonumber \\
\mu_2 &=&V_1 \, G_1 G_2^\star \nonumber \\
\mu_3 &=&V_2 \, G_0 G_2^\star \nonumber \\
\mu_4 &=&V_1 \, G_3 G_4^\star \nonumber \\
\mu_5 &=&V_3 \, G_2 G_3^\star \nonumber \\
\mu_6 &=&V_4 \, G_4 G_5^\star \nonumber \\
\mu_7 &=&V_4 \, G_1 G_3^\star \nonumber \\
\mu_8 &=&V_5 \, G_0 G_3^\star \nonumber \\
\mu_9 &=&V_4 \, G_2 G_4^\star \nonumber \\
\mu_{10}&=&V_5 \, G_3 G_5^\star \nonumber \\
\mu_{11}&=&V_5 \, G_1 G_4^\star \nonumber \\
\mu_{12}&=&V_6 \, G_0 G_4^\star \nonumber \\
\mu_{13}&=&V_7 \, G_2 G_5^\star \nonumber \\
\mu_{14}&=&V_8 \, G_1 G_5^\star \nonumber \\
\mu_{15}&=&V_9 \, G_0 G_5^\star \,.
\label{eq:non-red}
\end{eqnarray} }
Le nombre d'observables réelles est ici de 31 ([$\mu_1$,..,$\mu_{15}$]
et $\mu_0$), pour toujours 30 inconnues ([$V_1$,..,$V_9$] et
[$G_0$,..,$G_5$]). Il est donc possible que le problème soit
inversible. Pour le prouver, nous avons appliqué un logarithme à ces
équations. Le logarithme de nombres complexes fournit deux séries
d'équations, l'une sur les phases et l'autre sur les logarithmes des
amplitudes, selon le principe~:
\begin{eqnarray}
\ln(G_i)&=& \ln(g_i) + \I \phi_i  \nonumber\\
\ln(V_k)&=& \ln(|V_k|) + \I \arg(V_k)  \nonumber\\
\ln(\mu_i)&=& \ln(|\mu_i|) + \I \arg(\mu_i)  
\end{eqnarray}
Mise à part la condition de normalisation $\mu_0$, le système
d'équation~(\ref{eq:non-red}) se trouve ainsi linéarisé. Nous avons
écrit, sous forme matricielle, le système d'équations régissant les
amplitudes~: { \footnotesize
  \begin{equation}
    \left(\!
    \begin{array}{l}
      \ln(|\mu_1|) \\
      \ln(|\mu_2|) \\
      \ln(|\mu_3|) \\
      \ln(|\mu_4|) \\
      \ln(|\mu_5|) \\
      \ln(|\mu_6|) \\
      \ln(|\mu_7|) \\
      \ln(|\mu_8|) \\
      \ln(|\mu_9|) \\
      \ln(|\mu_{10}|) \\
      \ln(|\mu_{11}|) \\
      \ln(|\mu_{12}|) \\
      \ln(|\mu_{13}|) \\
      \ln(|\mu_{14}|) \\
      \ln(|\mu_{15}|)
    \end{array}
    \!\right) = \left(
    \begin{array}{rrrrrrrrrrrrrrrr}
      1&\bkp1&\bkp0&\bkp0&\bkp0&\bkp0&\bkp1&\bkp0&\bkp0&\bkp0&\bkp0&\bkp0&\bkp0&\bkp0&\bkp0 \\
      0&\bkp1&\bkp1&\bkp0&\bkp0&\bkp0&\bkp1&\bkp0&\bkp0&\bkp0&\bkp0&\bkp0&\bkp0&\bkp0&\bkp0 \\
      1&\bkp0&\bkp1&\bkp0&\bkp0&\bkp0&\bkp0&\bkp1&\bkp0&\bkp0&\bkp0&\bkp0&\bkp0&\bkp0&\bkp0 \\
      0&\bkp0&\bkp0&\bkp1&\bkp1&\bkp0&\bkp1&\bkp0&\bkp0&\bkp0&\bkp0&\bkp0&\bkp0&\bkp0&\bkp0 \\
      0&\bkp0&\bkp1&\bkp1&\bkp0&\bkp0&\bkp0&\bkp0&\bkp1&\bkp0&\bkp0&\bkp0&\bkp0&\bkp0&\bkp0 \\
      0&\bkp0&\bkp0&\bkp0&\bkp1&\bkp1&\bkp0&\bkp0&\bkp0&\bkp1&\bkp0&\bkp0&\bkp0&\bkp0&\bkp0 \\
      0&\bkp1&\bkp0&\bkp1&\bkp0&\bkp0&\bkp0&\bkp0&\bkp0&\bkp1&\bkp0&\bkp0&\bkp0&\bkp0&\bkp0 \\
      1&\bkp0&\bkp0&\bkp1&\bkp0&\bkp0&\bkp0&\bkp0&\bkp0&\bkp0&\bkp1&\bkp0&\bkp0&\bkp0&\bkp0 \\
      0&\bkp0&\bkp1&\bkp0&\bkp1&\bkp0&\bkp0&\bkp0&\bkp0&\bkp1&\bkp0&\bkp0&\bkp0&\bkp0&\bkp0 \\
      0&\bkp0&\bkp0&\bkp1&\bkp0&\bkp1&\bkp0&\bkp0&\bkp0&\bkp0&\bkp1&\bkp0&\bkp0&\bkp0&\bkp0 \\
      0&\bkp1&\bkp0&\bkp0&\bkp1&\bkp0&\bkp0&\bkp0&\bkp0&\bkp0&\bkp1&\bkp0&\bkp0&\bkp0&\bkp0 \\
      1&\bkp0&\bkp0&\bkp0&\bkp1&\bkp0&\bkp0&\bkp0&\bkp0&\bkp0&\bkp0&\bkp1&\bkp0&\bkp0&\bkp0 \\
      0&\bkp0&\bkp1&\bkp0&\bkp0&\bkp1&\bkp0&\bkp0&\bkp0&\bkp0&\bkp0&\bkp0&\bkp1&\bkp0&\bkp0 \\
      0&\bkp1&\bkp0&\bkp0&\bkp0&\bkp1&\bkp0&\bkp0&\bkp0&\bkp0&\bkp0&\bkp0&\bkp0&\bkp1&\bkp0 \\
      1&\bkp0&\bkp0&\bkp0&\bkp0&\bkp1&\bkp0&\bkp0&\bkp0&\bkp0&\bkp0&\bkp0&\bkp0&\bkp0&\bkp1
    \end{array}
    \right) \cdot \left( \!
    \begin{array}{l}
      \ln(g_0) \\
      \ln(g_1) \\
      \ln(g_2) \\
      \ln(g_3) \\
      \ln(g_4) \\
      \ln(g_5) \\
      \ln(|V_1|) \\
      \ln(|V_2|) \\
      \ln(|V_3|) \\
      \ln(|V_4|) \\
      \ln(|V_5|) \\
      \ln(|V_6|) \\
      \ln(|V_7|) \\
      \ln(|V_8|) \\
      \ln(|V_9|)
    \end{array}
    \! \right) \, .
    \label{eq:matrix_A}
  \end{equation}}
Ce système d'équations n'est pas parfaitement inversible. Plus
exactement, pour un nombre de colonnes de 15, le rang de cette matrice
est de 14. L'inconnue restante est un terme de normalisation des
facteurs de transmission. On peut lever la dégénérescence en utilisant
la relation à fréquence nulle qui fixe la somme des gains~:
\begin{equation}
\mu_0 =\sum_{i=0}^5 |G_i|^2 \,. 
\end{equation} 

De même, nous avons établi le système d'équations régissant les phases~:
{ 
\footnotesize
  \begin{equation}
  \left(\!
  \begin{array}{l}
    \arg(\mu_1) \\
    \arg(\mu_2) \\
    \arg(\mu_3) \\
    \arg(\mu_4) \\
    \arg(\mu_5) \\
    \arg(\mu_6) \\
    \arg(\mu_7) \\
    \arg(\mu_8) \\
    \arg(\mu_9) \\
    \arg(\mu_{10}) \\
    \arg(\mu_{11}) \\
    \arg(\mu_{12}) \\
    \arg(\mu_{13}) \\
    \arg(\mu_{14}) \\
    \arg(\mu_{15})
  \end{array}
  \!\right) = \left(
  \begin{array}{rrrrrrrrrrrrrrr}
    1&\bkl-1&\bkp0&\bkp0&\bkp0&\bkp0&\bkp1&\bkp0&\bkp0&\bkp0&\bkp0&\bkp0&\bkp0&\bkp0&\bkp0 \\
    0&\bkp1&\bkl-1&\bkp0&\bkp0&\bkp0&\bkp1&\bkp0&\bkp0&\bkp0&\bkp0&\bkp0&\bkp0&\bkp0&\bkp0 \\
    1&\bkp0&\bkl-1&\bkp0&\bkp0&\bkp0&\bkp0&\bkp1&\bkp0&\bkp0&\bkp0&\bkp0&\bkp0&\bkp0&\bkp0 \\
    0&\bkp0&\bkp0&\bkp1&\bkl-1&\bkp0&\bkp1&\bkp0&\bkp0&\bkp0&\bkp0&\bkp0&\bkp0&\bkp0&\bkp0 \\
    0&\bkp0&\bkp1&\bkl-1&\bkp0&\bkp0&\bkp0&\bkp0&\bkp1&\bkp0&\bkp0&\bkp0&\bkp0&\bkp0&\bkp0 \\
    0&\bkp0&\bkp0&\bkp0&\bkp1&\bkl-1&\bkp0&\bkp0&\bkp0&\bkp1&\bkp0&\bkp0&\bkp0&\bkp0&\bkp0 \\
    0&\bkp1&\bkp0&\bkl-1&\bkp0&\bkp0&\bkp0&\bkp0&\bkp0&\bkp1&\bkp0&\bkp0&\bkp0&\bkp0&\bkp0 \\
    1&\bkp0&\bkp0&\bkl-1&\bkp0&\bkp0&\bkp0&\bkp0&\bkp0&\bkp0&\bkp1&\bkp0&\bkp0&\bkp0&\bkp0 \\
    0&\bkp0&\bkp1&\bkp0&\bkl-1&\bkp0&\bkp0&\bkp0&\bkp0&\bkp1&\bkp0&\bkp0&\bkp0&\bkp0&\bkp0 \\
    0&\bkp0&\bkp0&\bkp1&\bkp0&\bkl-1&\bkp0&\bkp0&\bkp0&\bkp0&\bkp1&\bkp0&\bkp0&\bkp0&\bkp0 \\
    0&\bkp1&\bkp0&\bkp0&\bkl-1&\bkp0&\bkp0&\bkp0&\bkp0&\bkp0&\bkp1&\bkp0&\bkp0&\bkp0&\bkp0 \\
    1&\bkp0&\bkp0&\bkp0&\bkl-1&\bkp0&\bkp0&\bkp0&\bkp0&\bkp0&\bkp0&\bkp1&\bkp0&\bkp0&\bkp0 \\
    0&\bkp0&\bkp1&\bkp0&\bkp0&\bkl-1&\bkp0&\bkp0&\bkp0&\bkp0&\bkp0&\bkp0&\bkp1&\bkp0&\bkp0 \\
    0&\bkp1&\bkp0&\bkp0&\bkp0&\bkl-1&\bkp0&\bkp0&\bkp0&\bkp0&\bkp0&\bkp0&\bkp0&\bkp1&\bkp0 \\
    1&\bkp0&\bkp0&\bkp0&\bkp0&\bkl-1&\bkp0&\bkp0&\bkp0&\bkp0&\bkp0&\bkp0&\bkp0&\bkp0&\bkp1
  \end{array}
  \right) \cdot \left(\!
  \begin{array}{c}
    \phi_0 \\
    \phi_1 \\
    \phi_2 \\
    \phi_3 \\
    \phi_4 \\
    \phi_5 \\
    \arg(V_1) \\
    \arg(V_2) \\
    \arg(V_3) \\
    \arg(V_4) \\
    \arg(V_5) \\
    \arg(V_6) \\
    \arg(V_7) \\
    \arg(V_8) \\
    \arg(V_9)
  \end{array}
  \!\right) \, .
  \label{eq:matrix_P}
  \end{equation}
} Le rang de cette matrice est de 12.  Trois termes manquent pour que
les $\mu$ définissent sans ambiguïté l'ensemble des phases. Le premier
correspond à une référence de phase des facteurs de
transmission. Puisque les visibilités ne sont influencées que par des
différences de marche, nous pouvons utiliser une valeur arbitraire
pour cette phase de référence ($\phi_0=0$). Le choix de cette valeur
n'a pas d'importance car il ne contraint pas les visibilités. Il
reste alors deux termes inconnus qui portent, eux, sur les
visibilités. Il s'agit d'un ``tip'' et d'un ``tilt'', qui
caractérisent de la même façon un basculement (virtuel) de la surface
d'onde ou un décentrement du centroïde de l'objet observé. Cette
indétermination peut être expliquée par le fait que le système ne peut
différencier un tip/tilt dû à la turbulence ou à l'image. Un moyen de
contourner le problème consistera, lors de la reconstruction d'image,
à fixer le barycentre de brillance de l'objet au centre de l'image.

Cependant, même si une inversion matricielle permet de retrouver les
visibilités à partir des mesures de la cohérence $\mu$, une telle
technique nécessite le calcul du logarithme des mesures. Sur des
données bruitées, ceci détériore considérablement la qualité des
observations. C'est pourquoi, nous avons développé un algorithme
spécifique permettant un ajustement des inconnues directement sur les
mesures de la cohérence $\mu_{i,j}$.

\clearpage
\section{L'algorithme de déconvolution}
\label{sc:algo_eric}

Cet algorithme utilise la propriété d'unicité de la solution pour
éviter de faire appel à un terme de régularisation. En conséquence, il
ne peut être utilisé que sur des systèmes interférométriques
redondants. Cependant, la simple redondance n'est pas une condition
suffisante. Pour vérifier la propriété d'unicité, il est nécessaire
d'établir le système d'équations tel que nous l'avons vu dans l'exemple
par les relations~(\ref{eq:matrix_A}) et~(\ref{eq:matrix_P}).

\subsection{Le maximum de vraisemblance}

L'objectif consiste à trouver les valeurs $G_i$ et $V_k$ vérifiant à la
fois l'équation~:
\begin{equation}
\mu_{i,j}=V_k\, G_i\, G_j^\star \,
\label{eq:algo_mu}
\end{equation}
et la relation de normalisation~:
\begin{equation}
\mu_0 =\sum_{i} |G_i|^2 \,. 
\label{eq:g_norm}
\end{equation}

Comme explicité dans le paragraphe~\ref{sc:concept_rea}, $G_i$ et
$G_j$ sont les transmissions complexes du champ dans les fibres $i$ et
$j$, $\mu_{i,j}$ est une mesure de la cohérence entre $i$ et $j$,
$\mu_0$ l'amplitude à fréquence nulle, et $V_k$ est la visibilité de
l'objet astrophysique à la $k^{\rm ieme}$ fréquence spatiale
$\mathbf{u}_k = (\mathbf{r}_i - \mathbf{r}_j)/\lambda$.

Le maximum de vraisemblance \citep{1985..Goodman} s'écrit alors~:
\begin{equation}
  \label{eq:chi2_b}
  \chi^2 = \sum_{(i,j)}\!\!
  w_{i,j}\,\left\vert\mu_{i,j} - V_k\,G_i\,G_j^\star \right|^2 +w_0
  \left\vert \mu_0 - \sum_i \abs{G_i}^2  \right|^2
\end{equation}
où $w_{i,j}$ et $w_0$ sont les poids statistiques, tels que~:
\begin{equation}
  \label{eq:weight}
  w_{i,j}
  = \frac{1}{\Var\bigl(\Re(\mu_{i,j})\bigr)}
  = \frac{1}{\Var\bigl(\Im(\mu_{i,j})\bigr)}
\end{equation}
et
\begin{equation}
  w_{0}
  = \frac{1}{\Var\bigl(\mu_{0}\bigr)}
  \,.
\end{equation}
L'équation~(\ref{eq:chi2_b}) peut aussi s'écrire sous la forme de
plusieurs sommes, une pour chaque fréquence spatiale:
\begin{equation}
  \label{eq:chi2_c}
  \chi^2  = \sum_k \sum_{(i,j)\in\mathcal{B}_k}\!\!
  w_{i,j}\,\left\vert\mu_{i,j} - G_i\,G_j^\star\,V_k\right\vert^2+w_0
  \left\vert \mu_0 - \sum_i \abs{G_i}^2  \right|^2
\end{equation}
avec $\mathcal{B}_k$ les paires de sous-pupilles ($i$,$j$) telles que
leurs vecteurs position vérifient $(\mathbf{r}_i -
\mathbf{r}_j)/\lambda = \mathbf{u}_k$.

Résoudre ce problème au sens du maximum vraisemblance consiste à
trouver les valeurs complexes $\V{V}$ et $\V{G}$ qui minimisent le
$\chi^2$ de l'équation~(\ref{eq:chi2_c}). Deux problèmes se posent alors
:
\begin{itemize}
\item Le signal sur bruit d'une seule pose peut être très faible. Il
  faudra alors effectuer une minimisation du $\chi^2$ sur l'ensemble des
  acquisitions. Pour une série de 10\,000 poses, et pour une centaine
  de sous-pupilles, cela correspond à une minimisation sur environ un
  million de termes simultanément.
\item Le $\chi^2$ s'écrit sous la forme d'un polynôme du sixième degré
vis-à-vis des inconnues, il y a en conséquence de fortes chances qu'il
soit non-convexe. L'application directe d'un algorithme de
minimisation peut alors mener à de mauvais résultats.
\end{itemize}
En collaboration avec Eric Thi\'ebaut de l'observatoire de Lyon, nous
avons développé un algorithme adapté, inspiré des méthodes
d'auto-calibration utilisées en radio-interférométrie
\citep{1981MNRAS.196.1067C}.

\subsection{Les visibilités}
\label{sc:algo_visib}

Dans l'hypothèse où les $\V{G}$ sont connus, l'équation~(\ref{eq:chi2_c}) est
alors quadratique. Trouver le minimum n'est plus qu'un simple
problème de moindre carré, avec une solution telle que~:
\begin{equation}
  \label{eq:1st-optim-cond-V}
  \frac{\partial\chi^2}{\partial V_k} = 0\,,\quad\forall k 
\end{equation}
où l'on définit, par linéarité, la dérivée de la quantité réelle $\chi^2$
par le  complexe $V_k$ de la façon suivante:
\begin{equation}
  \label{eq:cmplx-deriv}
  \frac{\partial\chi^2}{\partial V_k} \bydef
  \frac{\partial\chi^2}{\partial \Re\left(V_k\right)}
  + \I\,\frac{\partial\chi^2}{\partial \Im\left(V_k\right)} \,.
\end{equation}
  Alors~:
\begin{eqnarray}
  \frac{\partial\chi^2}{\partial V_k}
  &=& 2\!\sum_{(i,j)\in\mathcal{B}_k}\!
      w_{i,j}\,\left(G_i\,G_j^\star\,V_k - \mu_{i,j}\right)\,G_i^\star\,G_j
      \nonumber\\
  &=& 2\,V_k\!\sum_{(i,j)\in\mathcal{B}_k}w_{i,j}\,\abs{G_i}^2\,\abs{G_j}^2
     -2\!\sum_{(i,j)\in\mathcal{B}_k}w_{i,j}\,\mu_{i,j}\,G_i^\star\,G_j\,.
     \label{eq:chi2-pder-Vk}
\end{eqnarray}
Résoudre l'équation~(\ref{eq:1st-optim-cond-V}) à partir de l'expression
des derivées
partielles de l'équation~(\ref{eq:chi2-pder-Vk}) donne~:
\begin{equation}
\label{eq:best-obj-vis}
  V_k^\dagger =
    \frac{
      \displaystyle\sum_{(i,j)\in\mathcal{B}_k}\!
      w_{i,j}\,G_i^\star\,G_j\,\mu_{i,j}
    }{
      \displaystyle\sum_{(i,j)\in\mathcal{B}_k}\!
      w_{i,j}\,\abs{G_i}^2\,\abs{G_j}^2
    }\,.
\end{equation}
$V_k^\dagger$ est alors la valeur optimale au sens des moindres
carrés, pourvu que l'on connaisse les transmissions complexes.
Concrètement, l'équation~(\ref{eq:best-obj-vis}) permet, dans la suite
de ce paragraphe, de s'attaquer à un problème plus simple, n'ayant que
les facteurs de transmission complexes comme inconnues.

\subsection{Les facteurs de transmission complexes}
\label{sc:algo_transmi}

La deuxième partie de cet algorithme consiste à ajuster les
coefficients de transmission des fibres. Nous pouvons écrire sous une
seconde forme le $\chi^2$~:
\begin{equation}
  \label{eq:reduced-chi2}
  \chi^{2\dagger}  = \sum_k \sum_{(i,j)\in\mathcal{B}_k}\!\!
  w_{i,j}\,\left\vert\mu_{i,j} - G_i\,G_j^\star\,V_k^\dagger\right\vert^2+w_O
  \left\vert \mu_0 - \sum_i \abs{G_i}^2  \right|^2
\end{equation}
où $V_k^\dagger$ est donné par la relation~(\ref{eq:best-obj-vis}).  \`A
la valeur optimale du $\chi^{2\dagger}$ on doit avoir~:
\begin{equation}
  \label{eq:1st-optim-cond-G}
  \frac{\partial\chi^{2\dagger}}{\partial G_i} = 0\,,\quad\forall i\,.
\end{equation}
Or, sachant que $V_k=V_k^\dagger$ minimise le $\chi^2$, on peut en
déduire \citep{2006astro.ph.10458L} que le minimum global doit vérifier~:
\begin{equation}
  \label{eq:1st-optim-cond-G-alt}
  \frac{\partial\chi^{2\dagger}}{\partial G_i} =
  \left.\frac{\partial\chi^2}{\partial G_i}
  \right\vert_{\V{V}=\V{V}^\dagger}
  =  0\,,\quad\forall i\,.
\end{equation}
Concrètement, pour trouver la dérivée partielle des $\chi^{2\dagger}$,
il suffit de calculer la dérivée partielle des $\chi^2$ où les $V_k$
sont considérés comme fixes et obtenus par la
relation~(\ref{eq:best-obj-vis}). Ainsi, les dérivées partielles par
rapport aux $G_i$ sont~:
\begin{eqnarray*}
  \frac{\partial\chi^{2\dagger}}{\partial G_i} &=&
  \left.\frac{\partial\chi^2}{\partial G_i}
  \right\vert_{{V_k}={V_k}^\dagger}\\
  &=&
  -2\,\sum_k \!\sum_{j:(i,j)\in\mathcal{B}_k}\!\!\!\!
  w_{i,j}\,\left(\mu_{i,j} - G_i\,G_j^\star\,V_k^\dagger\right)\,G_j\,V_k^{\dagger\star} \\
  && - 2\,\sum_k \!\sum_{j:(j,i)\in\mathcal{B}_k}\!\!\!\! 
  w_{j,i}\,\left(\mu_{j,i}^\star -
  G_i\,G_j^\star\,V_k^{\dagger\star}\right)\,G_j\,V_k^\dagger \\ 
  && - 4\, w_0\, G_i \,\left(  \mu_0 - \sum_i |G_i|^2 \right) \\
  &=& 2\,G_i\,\left( \sum_k \abs{V_k^\dagger}^2\,\left[
      \sum_{j:(i,j)\in\mathcal{B}_k}\!\!\!\! w_{j,i}\,\abs{G_j}^2
    + \!\!\sum_{j:(j,i)\in\mathcal{B}_k}\!\!\!\! w_{i,j}\,\abs{G_j}^2
      \right] + 2\, w_0 \, \left[\sum_i \abs{G_i}^2 - \mu_0 \right] \right)
  \\
  && -2\,\sum_k \left[
      \sum_{j:(i,j)\in\mathcal{B}_k}\!\!\!\! w_{i,j}\,\mu_{i,j}\,G_j\,V_k^{\dagger\star}
    + \!\!\sum_{j:(j,i)\in\mathcal{B}_k}\!\!\!\! w_{j,i}\,\mu_{j,i}^\star\,G_j\,V_k^\dagger
      \right]\,.
\end{eqnarray*}
\`A partir de cette dernière relation, nous avons établi un algorithme
itératif permettant de déterminer le coefficient de transmission
complexe $G_i$ en supposant les autres $G_{j:j\not=i}$
connus. L'équation récurrente est la suivante~:
\begin{equation}
  G_i^{(n+1)}
  =
  \frac{\displaystyle
    \sum_k \left[
      \sum_{j:(i,j)\in\mathcal{B}_k}\!\!\!\!
      w_{i,j}\,\mu_{i,j}\,G_j^{(n)}\,{V_k^{\dagger(n)}}^\star
      + \!\!\!\sum_{j:(j,i)\in\mathcal{B}_k}\!\!\!\!
      w_{j,i}\,\mu_{j,i}^\star\,G_j^{(n)}\,V_k^{\dagger(n)}
      \right]
  }{\displaystyle
    \sum_k \Abs{V_k^{\dagger(n)}}^2\,\left[
      \sum_{j:(i,j)\in\mathcal{B}_k}\!\!\!\! w_{j,i}\,\Abs{G_j^{(n)}}^2
      + \!\!\!\sum_{j:(j,i)\in\mathcal{B}_k}\!\!\!\! w_{i,j}\,\Abs{G_j^{(n)}}^2
      \right]+ 2\, w_0 \, \left[\sum_i \Abs{G_i^{(n)}}^2 - \mu_0 \right]
  }\,.
  \label{eq:update}
\end{equation}
où~:
\begin{equation}
  \label{eq:best-obj-vis_2}
  V_k^{\dagger(n)} =
    \frac{
      \displaystyle\sum_{(i,j)\in\mathcal{B}_k}\!
      w_{i,j}\,G_i^{(n)\star}\,G_j^{(n)}\,\mu_{i,j}
    }{
      \displaystyle\sum_{(i,j)\in\mathcal{B}_k}\!
      w_{i,j}\,\abs{G_i^{(n)}}^2\,\abs{G_j^{(n)}}^2
    }\,.
\end{equation}

\subsection{Le cas d'acquisitions multiples}

Les techniques de déconvolution nécessitant une atmosphère figée, une
séquence d'observation est généralement composée de multiples
observations. Il semble illusoire d'espérer pouvoir mesurer avec
suffisamment de précision les valeurs complexes $\V{V}$ dans le cadre
d'une seule acquisition. C'est pourquoi nous avons extrapolé cet algorithme pour
prendre en compte la présence de multiples acquisitions.

Les hypothèse faites ici sont~:
\begin{itemize}
\item un instrument stable, sans rotation de la pupille, de façon à ce que
  les vecteurs base correspondant aux $\mathcal{B}_k$ ne changent
  pas
\item un objet invariant, aux visibilités $V_k$ fixes
\item des coefficients de transmission $G_{i,t}$ variables, et 
  des mesures de la cohérence elles aussi dépendantes du temps ($\mu_{i,j,t}$)
\end{itemize}
Le $\chi^2$ s'écrit alors:
\begin{equation}
  \label{eq:chi2-multi}
  \chi^2 = \sum_t \left[ \sum_k \sum_{(i,j)\in\mathcal{B}_k}\!\!
  w_{i,j,t}\,\left\vert\mu_{i,j,t} - G_{i,t}\,G_{j,t}^\star\,V_k\right\vert^2+w_{0,t}
  \left\vert \mu_{0,t} - \sum_i \abs{G_{i,t}}^2  \right|^2 \right] \,.
\end{equation}
Nous pouvons alors en déduire, comme nous l'avons fait aux
paragraphes~\ref{sc:algo_visib} et~\ref{sc:algo_transmi}, un algorithme itératif qui permet
d'obtenir les $G_{i,t}$ optimaux en supposant que les $G_{j:j\not=i,t}$
soient connus:
\begin{equation}
  G_{i,t}^{(n+1)}
  =
  \frac{\displaystyle
    \sum_k \left[
      \sum_{j:(i,j)\in\mathcal{B}_k}\!\!\!\!
      w_{i,j,t}\,\mu_{i,j,t}\,G_{j,t}^{(n)}\,{V_k^{\dagger(n)}}^\star
      + \!\!\!\sum_{j:(j,i)\in\mathcal{B}_k}\!\!\!\!
      w_{j,i,t}\,\mu_{j,i,t}^\star\,G_{j,t}^{(n)}\,V_k^{\dagger(n)}
      \right]
  }{\displaystyle
    \sum_k \Abs{V_k^{\dagger(n)}}^2\,\left[
      \sum_{j:(i,j)\in\mathcal{B}_k}\!\!\!\! w_{j,i,t}\,\Abs{G_{j,t}^{(n)}}^2
      + \!\!\!\sum_{j:(j,i)\in\mathcal{B}_k}\!\!\!\! w_{i,j,t}\,\Abs{G_{j,t}^{(n)}}^2
      \right]+ 2\, w_{0,t} \, \left[\sum_i \Abs{G_{i,t}^{(n)}}^2 - \mu_{0,t} \right]
  }\,,
  \label{eq:update-multi}
\end{equation}
où~:
\begin{equation}
  \label{eq:best-obj-vis-multi}
  V_k^{\dagger(n)} =
    \frac{
      \displaystyle \sum_t 
\sum_{(i,j)\in\mathcal{B}_k}\!
      w_{i,j,t}\,G_{i,t}^{(n)\star}\,G_{j,t}^{(n)}\,\mu_{i,j,t} 
    }{
      \displaystyle \sum_t 
 \sum_{(i,j)\in\mathcal{B}_k}\!
      w_{i,j,t}\,\abs{G_{i,t}^{(n)}}^2\,\abs{G_{j,t}^{(n)}}^2 
    }\,.
\end{equation}
Ainsi, même si les coefficients de transmission des fibres restent
peu connus, les visibilités sont obtenues par une moyenne
pondérée sur l'ensemble des acquisitions, ce qui permet une
détermination optimale.

\subsection{Résumé de l'algorithme}
\label{sec:algo_res}

Notre algorithme se compose donc des étapes suivantes~:
\begin{enumerate}
\item initialisation:  choisir des coefficients de transmision
  ($\V{G}^{(0)}$) initiaux et mettre $n$ à zéro.
\item générer les visibilités $\V{V}^{(n)}$ à partir des facteurs
  complexes de transmission $\V{G}^{(n)}$ et de
  l'équation~(\ref{eq:best-obj-vis-multi})
\item Si l'algorithme converge, arrêter; sinon poursuivre à l'étape suivante
\item calculer $\V{G}^{(n+1)}$ à partir de la relation~(\ref{eq:update-multi})
\item effectuer $n:=n+1$ et retourner à l'étape 2
\end{enumerate}

Cet algorithme itératif a le mérite d'être simple à mettre en \oe
uvre, et peu gourmand en mémoire. Il peut ainsi traiter plusieurs
dizaines de milliers d'acquisitions simultanément, ce qui est
nécessaire pour les objets de faible brillance ou pour obtenir des
images à très grande dynamique. Il a, cependant, un certain nombre
d'inconvénients dont il faut être conscient. Premièrement, il n'est
pas fait démonstration de la convexité du problème. Il est en
conséquence possible que le résultat obtenu ne soit pas le minimum
global. Seule la pratique permettra de mettre en évidence la
robustesse du procédé. Une autre source de problèmes peut venir du
caractère itératif de l'équation~(\ref{eq:update-multi}). Des
oscillations stables peuvent apparaître, empêchant la découverte de la
condition d'optimalité. Une solution serait de coupler l'algorithme à
un autre algorithme de minimisation du $\chi^2$ plus
classique. Puisque le maximum de vraisemblance est une somme de
carrés, nous pourrions utiliser un Levenberg-Marquardt
\citep{More1977LevenbergMarquardtAlgorithm} couplé avec un algorithme
de région de confiance \citep{More_Sorensen-1983-trust_region_step} de
façon à résoudre ce problème avec certitude.  En pratique, cet
algorithme n'a conduit à aucun problème de convergence.

Bien sûr, ce travail est le fruit de nombreux travaux déjà
existants. En premier lieu, l'autocalibration nous a permis de
simplifier le problème, qui ne suppose plus de trouver simultanément
les visibilités et les facteurs de transmission, mais seulement ces
derniers. Alors que \citet{1981MNRAS.196.1067C} proposaient de
l'utiliser pour le calcul des phases uniquement, nous l'avons étendu
aux calculs des phases et des
amplitudes. L'équation~(\ref{eq:update-multi}) est ainsi très proche de
l'algorithme itératif proposé par \citet{Matson1991:phase} pour
traiter le bispectre \citep[aussi amélioré par][]{Thiebaut1994:PhD}.

\subsection{Le cas de données dispersées spectralement}

\subsubsection{La dépendance en longueur d'onde du facteur de
  transmission complexe}

Lorsque l'on dispose de données dispersées spectralement, on a accès
à des mesures de cohérence $\mu_{i,j,t,\lambda}$ qui sont fonctions de la
longueur d'onde. Les facteurs de transmission complexes
$G_{i,t,\lambda}$ peuvent alors être obtenus par l'algorithme itératif
de l'équation~(\ref{eq:update-multi}) et les visibilités $V_{k,\lambda}$
par la relation~(\ref{eq:best-obj-vis-multi}).

Cependant, si on peut établir la fonction de dépendance entre
$G_{i,\lambda}$ et la longueur d'onde, on augmente considérablement la
sensibilité de l'instrument. De cette manière, on peut
utiliser l'information sur la totalité des longueurs d'onde pour
estimer les $G_{i,\lambda}$.

Cela est possible dans le cas de faibles perturbations à l'echelle
d'une sous-pupille. Il est alors nécessaire de calculer
l'expression du couplage du champ électrique dans une fibre optique
monomode. Nous expliciterons ce calcul dans le chapitre suivant. Nous
allons néanmoins nous servir de l'expression de l'amplitude complexe
couplée dans la fibre établie équation~(\ref{eq:A_qul})~:
\begin{equation}
A =
 \iint_{-\infty}^{+\infty} U_\circ(u,v)\, 2 \eta  \sqrt{ \frac{3}{\pi}}\,
\exp\left(-6  (u^2+v^2) \eta^2 \right)
  dudv\,,
\label{eq:A_norm_s1}
\end{equation}
où $\eta$ est le rapport entre l'ouverture de la lentille et de la
fibre, $u$ et $v$ les coordonnées dans le plan pupille (en unités de
diamétre de la sous-pupille) et $U_\circ(u,v)$ le champ normalisé rayonné par l'astre
observé. A partir des perturbations atmosphériques de la phase
$\phi(u,v)$ du champ
dans la pupille, on peut établir~:
\begin{eqnarray}
U_\circ(u,v) = & \displaystyle \frac{2\exp(\I\phi(u,v))}{\sqrt{\pi} }  & {\rm si}\ \sqrt{u^2+v^2} \leq
1/2 \\ & 0 & {\rm sinon}.
\end{eqnarray}
En remplaçant l'expression du champ électrique dans
l'équation~(\ref{eq:A_norm_s1}), on obtient ainsi l'amplitude complexe
de couplage dans le cas général de perturbations atmosphériques:
\begin{equation}
A =
 \iint_{\sqrt{u^2+v^2}\leq 1/2} \frac{ 4 \eta 
 \sqrt3}{ \pi } \,
\exp\left(\I\phi(u,v)-6  (u^2+v^2) \eta^2 \right)
  dudv\,,
\end{equation}
Reste à faire intervenir un terme fondamental, le piston atmosphérique
moyen sur la sous-pupille. Nous utiliserons comme définition du
piston la moyenne pondérée de la phase~:
\begin{equation}
\Psi=\frac{\displaystyle \iint_{\sqrt{u^2+v^2}\leq 1/2} \phi(u,v)\, 
\exp\left(-6(u^2+v^2) \eta^2 \right)
  dudv}{\displaystyle \iint_{\sqrt{u^2+v^2}\leq 1/2} 
\exp\left(-6(u^2+v^2) \eta^2 \right)
  dudv}
\,.
\end{equation}
De cette manière, il est possible d'écrire clairement les résidus de
phase dans la sous-pupille:
\begin{equation}
\tilde{\phi}(u,v)=\phi(u,v)-\Psi
\end{equation}
Et de sortir de l'expression de l'amplitude complexe de couplage le
déphasage introduit par le piston moyen~:
\begin{equation}
A = \exp(\I\Psi)
 \iint_{\sqrt{u^2+v^2}\leq 1/2} \frac{ 4 \eta 
 \sqrt3}{ \pi } \,
\exp\left(\I\tilde\phi(u,v)-6  (u^2+v^2) \eta^2 \right)
  dudv\,.
\label{eq:A_norm_s3}
\end{equation}

Nous allons faire ici l'approximation des faibles perturbations. Cette
approximation est possible lorsque la variance de la phase est faible.
Concrètement, cela signifie que la taille de la sous-pupille a
été choisie de façon à être proche où inférieure à $r_0$.
On a alors $\tilde\phi(u,v) < 1$. On peut ainsi développer le champ
électrique selon~:
\begin{equation}
\exp\left(\I\tilde\phi(u,v)\right)\approx
1+\I\tilde\phi(u,v)-\frac{1}{2}\tilde\phi^2(u,v)
\end{equation}
et introduire cet approximation dans l'équation~(\ref{eq:A_norm_s3})~:
\begin{eqnarray}
A &\approx& \exp(\I\Psi) 
\iint_{\sqrt{u^2+v^2}\leq 1/2} \frac{ 4 \eta 
 \sqrt3}{ \pi } \,
\exp\left(-6  (u^2+v^2) \eta^2 \right)
  dudv + \nonumber\\
&& \exp(\I\Psi)
\iint_{\sqrt{u^2+v^2}\leq 1/2} \I\tilde\phi(u,v)\ \frac{ 4 \eta 
 \sqrt3}{ \pi } \,
\exp\left(-6  (u^2+v^2) \eta^2 \right)
  dudv - \nonumber\\
&& \exp(\I\Psi)
\iint_{\sqrt{u^2+v^2}\leq 1/2} \frac{1}{2}\tilde\phi^2(u,v)\ \frac{ 4 \eta 
 \sqrt3}{ \pi } \,
\exp\left(-6  (u^2+v^2) \eta^2 \right)
  dudv   \,.
\end{eqnarray}
Or, on peut montrer que le deuxième terme, imaginaire, est nul. Ceci
peut se démontrer de la façon suivante~:
\begin{eqnarray}
\lefteqn{\iint_{\sqrt{u^2+v^2}\leq 1/2} \tilde\phi(u,v) \,
\exp\left(-6  (u^2+v^2) \eta^2 \right)  dudv} \nonumber\\
&=&\iint_{\sqrt{u^2+v^2}\leq 1/2} (\phi(u,v)-\Psi) \,
\exp\left(-6  (u^2+v^2) \eta^2 \right)  dudv \nonumber\\
&=&\iint_{\sqrt{u^2+v^2}\leq 1/2} \phi(u,v)
\exp\left(-6  (u^2+v^2) \eta^2 \right)  dudv -  \Psi
\iint_{\sqrt{u^2+v^2}\leq 1/2}  
\exp\left(-6  (u^2+v^2) \eta^2 \right)  dudv \nonumber\\
&=&0
\end{eqnarray}
Ainsi, le couplage s'écrit sous la forme d'un terme complexe
dû au piston moyen sur la pupille, et un terme réel:
\begin{equation}
A \approx\exp(\I\Psi)\frac{4\eta\sqrt3}{\pi}
\iint_{\sqrt{u^2+v^2}\leq 1/2}
(1+\frac{1}{2}\tilde\phi^2(u,v))\exp\left(-6  (u^2+v^2) \eta^2 \right)
dudv\,.
\end{equation}
Puisque l'on souhaite faire ressortir la dependance du couplage en
fonction de la longueur d'onde, il est intéressant d'écrire les
termes de déphasage sous la forme de différences de marche. On écrit de
cette manière $\tilde\phi_0=\tilde\phi\lambda$ et
$\Psi_0=\Psi\lambda$~:
\begin{eqnarray}
A &\approx&\exp(\I\Psi_0/\lambda)\underbrace{\frac{4\eta\sqrt3}{\pi}
\iint_{\sqrt{u^2+v^2}\leq 1/2}
\exp\left(-6  (u^2+v^2) \eta^2 \right)
dudv}_{\mbox{= $A_1$}}\nonumber\\
&&\exp(\I\Psi_0/\lambda)\frac1{\lambda^2}\underbrace{\frac{2\eta\sqrt3}{\pi}
\iint_{\sqrt{u^2+v^2}\leq 1/2}
\tilde\phi_0^2(u,v)\exp\left(-6  (u^2+v^2) \eta^2 \right)
dudv}_{\mbox{= $A_2$}}\,.
\end{eqnarray}

Nous avons de cette façon écrit, dans l'hypothèse de faible
perturbations de phase à l'échelle d'une sous-pupille, le couplage
complexe en fonction de 3 paramètre réels indépendants de la longueur
d'onde $\Psi_0$, $A_1$ et $A_2$~:
\begin{equation}
A \approx \exp(\I\Psi_0/\lambda) (A_1+ A_2/\lambda^2) \,.
\end{equation}
Chacun de ces paramètres varie en fonction du temps et de la
fibre. Ils dépendent donc de $t$ et de $i$. A partir de cette
expression du couplage, on peut déduire la dépendance en longueur
d'onde des facteurs de transmission complexes. Il faut cependant tenir
compte du spectre de l'objet observé. Celui-ci peut être 
 dérivé à partir des fréquences spatiales nulles
$\mu_{0,\lambda}$, intégrées sur l'ensemble des observations pour avoir
un meilleur signal sur bruit~:
\begin{equation}
\mu_{0,\lambda} = \sum_t \mu_{0,t,\lambda} \,.
\end{equation}
On montre ainsi que le facteur de transmission peut être dérivé de
l'ensemble des longueur d'onde à partir de 3 paramètres réels~:
\begin{equation}
G_{i,t,\lambda} =  \sqrt{\mu_{0,\lambda}} (K_{i,t}+M_{i,t}/\lambda^2) \exp(\I\Psi_{i,t}/\lambda)\,.
\end{equation}

\subsubsection{L'ajustement aux données}

\begin{figure}[t]
  \centering \resizebox{\hsize}{!}{
  \includegraphics{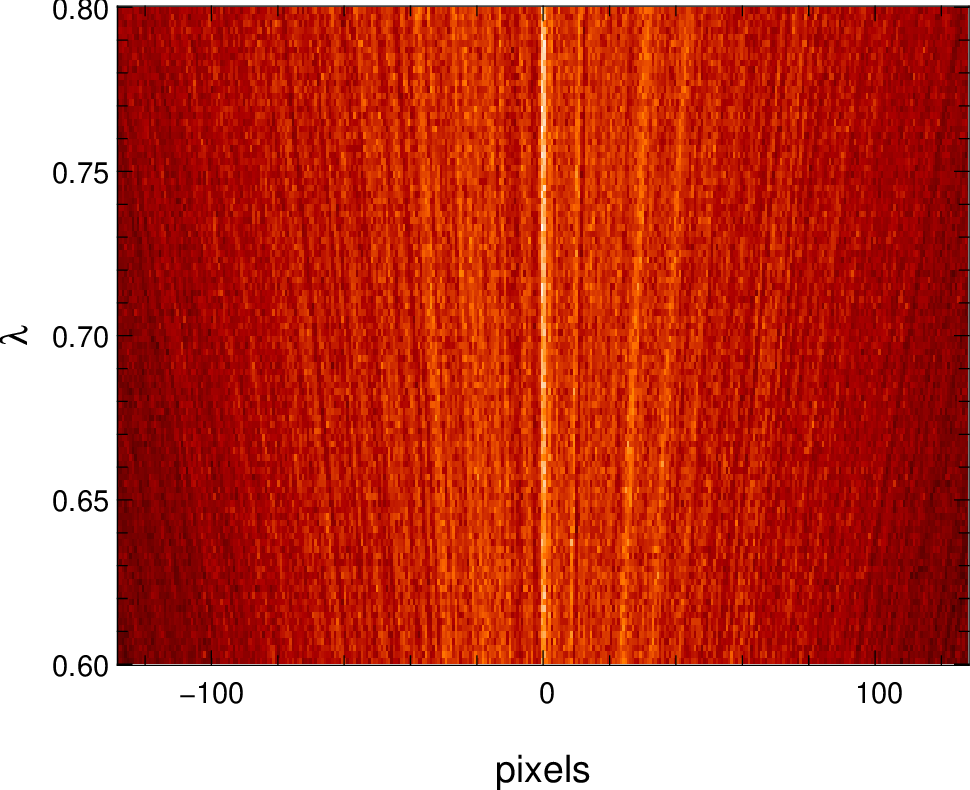}
  \includegraphics{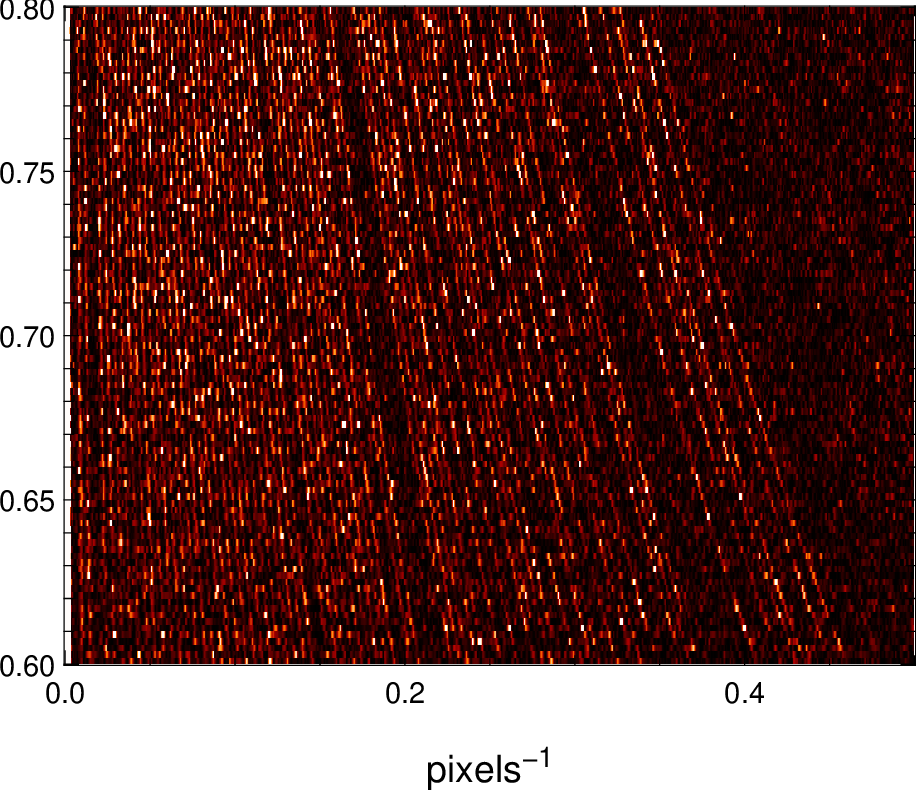} } \caption[Simulation de
  données spectro-interférométriques]{ Simulation de données
  spectro-interférométriques. Malgré le bruit de photon présent dans
  les données, on peut voir nettement les franges dans le domaine
  spatial (figure de gauche) et les pics franges dans le domaine des
  puissances spectrales (figure de droite). Les fibres sont
  positionnées selon une configuration non-redondante à une dimension,
  d'emplacements 1, 3, 6, 20, 31, 41, 54, 63, 70, 78, 90 et 96 en
  unités de longueurs arbitraires.  } \label{fig:yo}
\end{figure}

Nous avons établi une relation entre les facteurs de transmission
complexes et trois termes réels, un piston $\Psi_{i,t}$ et deux gains
$K_{i,t}$ et $M_{i,t}$. L'étape suivante consiste à obtenir ces valeurs. Une technique
simple consiste à insérer directement les transmissions complexes dans le
$\chi^2$ établi par l'équation~(\ref{eq:chi2-multi})~:
\begin{eqnarray}
  \label{eq:chi2-multi_lambda}
  \lefteqn{\chi^2 =}\nonumber \\
& {\displaystyle \sum_\lambda \sum_t  \sum_k
  \sum_{(i,j)\in\mathcal{B}_k}\!\!}  w_{i,j,t,\lambda}\left\vert\mu_{i,j,t,\lambda} -
  \mu_{0,\lambda}\left(K_{i,t}+\frac{M_{i,t}}{\lambda^2}\right)\left(K_{j,t}+\frac{M_{j,t}}{\lambda^2}\right)\exp\left(\I
  \frac{ \Psi_{i,t}-\Psi_{j,t}}{\lambda}\right)
  V_{k,\lambda}\right\vert^2 \nonumber \\
&   + \, \,\sum_\lambda \sum_t  w_{0,t,\lambda} \left\vert \mu_{0,t,\lambda} - \sum_i
   \Abs{K_{i,t}+\frac{M_{i,t}}{\lambda^2}}^2 \mu_{0,\lambda}  \right|^2  \,.
\end{eqnarray}

Cependant, il faut savoir que, au moment précis où nous passons de la
recherche de termes complexes à la recherche d'une phase et d'une
amplitude, nous perdons une grande partie de la qualité de
l'algorithme. En effet, alors que nous pouvions additionner une grande
quantité de phaseurs complexes jusqu'à obtenir un signal sur bruit
adéquat, ce n'est plus le cas si l'on fait intervenir les phases, ou
plus précisément, le piston. Il est également intéressant de noter que
l'on se trouve alors face à un problème très proche de celui d'un
chercheur de franges en interférométrie, qui se confronte à la
difficulté de trouver le piston enroulé sur les phaseurs complexes.

C'est pour ces raisons que nous avons choisi de retarder au maximum la
recherche du piston atmosphérique. Il est ainsi préférable de le
calculer à partir des valeurs issues de l'algorithme obtenu
paragraphe~\ref{sec:algo_res}.  L'équation du $\chi^2$ permet ensuite,
à partir des valeurs estimées $G_{i,t,\lambda}^{(n)}$, de calculer les
valeurs de maximum de vraisemblance du piston différentiel et des
gains, à $i$ et $t$ donnés~:
\begin{equation}
\label{eq:chi2_lambda}
\chi^2_{i,t} = \sum_\lambda w^g_{i,t,\lambda}
 \Abs{G_{i,t,\lambda}^{\flat(n)} - \sqrt{\mu_{0,\lambda}}\,\left(K_{i,t}^{(n)}+\frac{M_{i,t}^{(n)}}{\lambda^2}\right) 
  \exp\left(\I \frac{ \Psi_{i,t}^{(n)}}{\lambda} 
 \right)}^2\,,
\end{equation}
où les poids statistiques $w^g_{i,t,\lambda}$ correspondent maintenant
à l'inverse de la variance des coefficients de transmission:
\begin{equation}
  w^g_{i,t,\lambda}
  = \frac{1}{\Var\bigl(\Re(G_{i,t,\lambda})\bigr)}
  = \frac{1}{\Var\bigl(\Im(G_{i,t,\lambda})\bigr)}
  \,.
\end{equation}

Le faible nombre de paramètres (3) présents dans le $\chi^2_{i,t}$ de
l'équation~(\ref{eq:chi2_lambda}), rend la minimisation possible par un
algorithme tel qu'un Levenberg-Marquardt. Néanmoins, il faut remarquer
que la vraisemblance n'est pas un critère convexe vis-à-vis de la
phase.  Ce problème peut être résolu en partant d'une grille de
conditions initiales. Cette grille serait composée, par exemple, de
l'ensemble~:
\begin{equation}
\Psi_{\rm init} = n\frac{\lambda_0}{2}\,,\quad\forall\, n \in \mathrm{N}\quad{\rm
  tq} \quad \Abs{\Psi_{\rm init}} < \Psi_{\rm max}
\end{equation}
où $\Psi_{\rm max}$ serait une estimation du piston maximum.  Cette
grille, couplée à l'algorithme de minimisation, permet d'obtenir le
minimum global pour le piston $\Psi_{i,t}^{(n)}$ et les gains
$K_{i,t}^{(n)}$ et $M_{i,t}^{(n)}$. On peut alors utiliser
l'équation~(\ref{eq:best-obj-vis-multi}) pour obtenir les visibilités~:
\begin{equation}
  \label{eq:best-obj-vis-multi_lo}
  V_{k,\lambda}^{\dagger(n)} =
    \frac{
      \displaystyle \sum_t 
\sum_{(i,j)\in\mathcal{B}_k}\!
      w_{i,j,t,\lambda}\, \left(K_{i,t}^{(n)}+\frac{M_{i,t}^{(n)}}{\lambda^2}\right) \left(K_{j,t}^{(n)}+\frac{M_{j,t}^{(n)}}{\lambda^2}\right)   \exp\left(2\I \pi\frac{\Psi_{j,t}-\Psi_{i,t}}{\lambda} \right)\,\mu_{i,j,t,\lambda} 
    }{
      \displaystyle \sum_t 
 \sum_{(i,j)\in\mathcal{B}_k}\!
      w_{i,j,t,\lambda}\,\Abs{K_{i,t}^{(n)}+\frac{M_{i,t}^{(n)}}{\lambda^2}}^2 
\Abs{K_{j,t}^{(n)}+\frac{M_{j,t}^{(n)}}{\lambda^2}}^2 
    }\,.
\end{equation}

\subsubsection{L'algorithme modifié}

La version de l'algorithme avec dispersion spectrale se compose en conséquence
des étapes suivantes~:
\begin{enumerate}
\item initialisation:  choisir des coefficients de transmisions
  ($\V{G^{(0)}}$) initiaux et mettre ${n}$ à zéro.
\item ajuster sur les coefficients de transmission complexes les
  gains $\V{K^{(n)}}$ et $\V{M^{(n)}}$, et les pistons $\V{\Psi^{(n)}}$
  (faire attention à la non-convexité du piston)
\item générer les visibilités $\V{V^{(n)}}$ à partir des
  gains normalisés, des pistons et de l'équation~(\ref{eq:best-obj-vis-multi_lo})
\item Si l'algorithme converge, arrêter; sinon poursuivre à l'étape suivante
\item calculer $\V{G^{(n+1)}}$ à partir de la relation~(\ref{eq:update-multi})
\item effectuer $n:=n+1$ et retourner à l'étape 2
\end{enumerate}

L'algorithme présenté ici est encore en développement, mais il a le
mérite de pouvoir servir de base à une étude de performance d'un
instrument à dispersion spectrale. Le temps a malheureusement manqué
pour pouvoir simuler ses performances dans des situations
réalistes. Nous n'avons, notamment, pas pu tester la robustesse de
cette algorithme par rapport à celui sans dispersion spectrale. Comme
nous l'avons vu, le risque principal est celui de l'indétermination du
piston à un facteur de $2\pi$. Une telle erreur récurrente sur le
piston aurait pour conséquence de biaiser nos résultats. Des travaux
en cours tentent de répondre à cette question, notamment dans le cas
de signaux fortement bruités.

Un point intéressant à explorer est la statistique
temporelle du piston. On pourrait envisager un algorithme prenant en
compte la variation du piston au cours du temps, pour obtenir de
meilleures estimations. Ceci permettrait, en utilisant la
continuité temporelle des variations de phase, de diminuer le risque
de sauts intempestifs de $2\pi$.

\clearpage
\section{La dynamique de reconstruction}

\subsection{Une approximation analytique de la dynamique}
\label{sec:analy}

Ce système permet d'obtenir les visibilités de l'objet observé avec
une très grande précision. Pour estimer la dynamique d'un tel
procédé, nous avons adopté deux approches. 

La première technique est
basée sur une approximation analytique proposée par
\citet{2002.Baldwin}. Elle donne l'expression suivante de la dynamique
:
\begin{equation}
  \dyn = \sqrt{\frac{n}{(\delta V/V)^2+(\delta \phi)^2}} \, ,
\end{equation}
Dans cette expression, $n$ est le nombre total de mesures, $(\delta
V/V)$ l'erreur en amplitude et $\delta \phi$ l'erreur sur la phase.
En considérant comme seule source de bruit, un bruit de photon,
l'erreur sur les fréquences spatiales mesurées est de $\sqrt{\Nph}$ où
$\Nph$ est le nombre total de photons. Pour des interférences ayant un
facteur de cohérence de 1, et un nombre $M$ de sous-pupilles, le
rapport signal sur bruit du pic frange est alors~:
\begin{equation}
  V / \delta V = \frac{\sqrt{\Nph}}{M} \, ,
\label{eq:sb_pour_V}
\end{equation}
De plus, \citep{1985..Goodman}, dans le cadre d'un bruit blanc, lie
l'erreur sur l'amplitude à l'erreur sur la phase par la relation~:
\begin{equation}
\delta \phi \approx \frac{\delta V}{ V} \, .
\end{equation}

On peut en déduire l'approximation suivante de la dynamique de
l'image reconstruite~:
\begin {equation}
\dyn = \sqrt{\frac{M(M-1)}{2 M^2/\Nph}} \approx \sqrt{\frac{\Nph}{2}} \, .
\label{eq:dyn}
\end{equation}
Cette expression permet de déduire deux conclusions importantes pour
cet instrument~:
\begin{itemize}
\item Une dynamique infinie est théoriquement accessible pourvu que
  l'on observe la source suffisamment longtemps
\item La limitation due au bruit de photon est accentuée parce que la
  localisation spatiale des photons est perdue
\end{itemize}
Ce dernier point a pour conséquence de limiter de façon importante la
dynamique possible. En effet, obtenir une dynamique de $10^{10}$
nécessiterait $\approx 10^{20}$ photons, soit plusieurs
centaines de jours d'intégration sur un télescope de dix mètres pour
une source de magnitude 0 dans le visible.

\subsection{Les simulations de l'instrument}
\label{sec:simu_c4}

\subsubsection{Les caractéristiques de la simulation}

\begin{figure}[h]
  \centering \resizebox{\hsize}{!}{
  \includegraphics{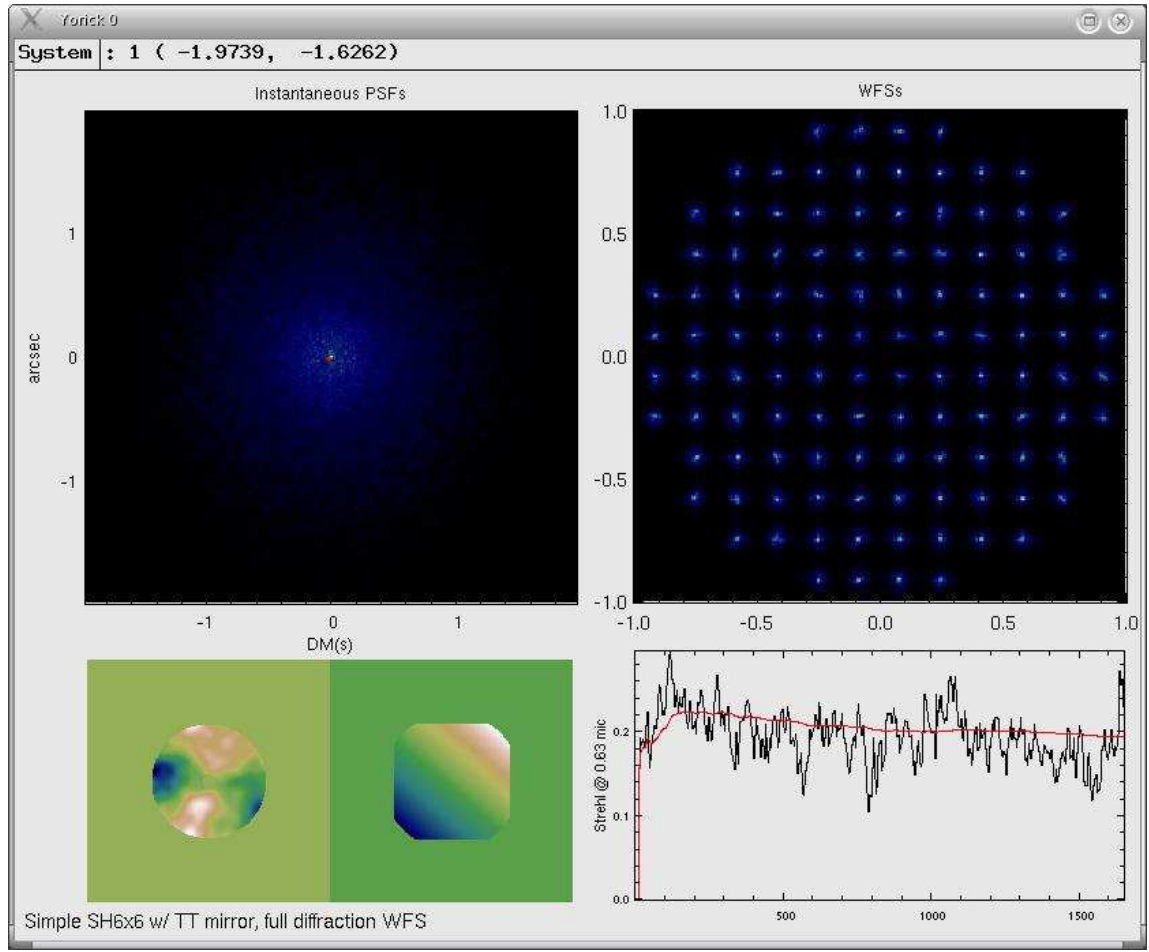} }
 \caption[Logiciel de
  simulation d'optique adaptative ``YAO'']{ Logiciel de simulation
  d'optique adaptative ``YAO''. A droite est représenté le détecteur
  du Shack-Hartmann, à gauche, la réponse impulsionnelle
  instantanée. On peut voir, en bas à gauche, la position du miroir
  déformable et du miroir tip/tilt. Le Strehl obtenu à chaque
  itération est représenté par la courbe en bas à droite.  }
  \label{fig:yao}
\end{figure}

Pour effectuer ces simulations, nous avons utilisé
``YAO''\footnote{http://www.maumae.net/yao/}, un logiciel de
simulation d'optique adaptative développé par François Rigaut. Toute
la programmation a été faite en langage
Yorick\footnote{http://yorick.sourceforge.net/}, un langage
interprété sous licence GNU à l'écriture proche du C.

``YAO'' nous a permis de simuler un télescope optique de 8 mètres,
doté d'une optique adaptative classique optimisée pour fonctionner
dans l'infrarouge. Nous avons ensuite simulé des conditions de
turbulences atmosphériques -- que nous avons choisies plutôt bonnes --
correspondant à un $r_0$ de 20 centimètres à une longueur d'onde de
630 nanomètres. Pour cela, nous avons utilisé une atmosphère composée
de 4 couches de turbulences distinctes à des altitudes de 0, 400, 6\,000
et 9\,000 mètres. La vitesse de déplacement de ces couches est une
variable importante car elle conditionne le temps de cohérence. Nous
avons utilisé des vitesses de déplacement allant de 6 à 20 m.s$^{-1}$
en fonction de la hauteur de la couche turbulente. L'optique
adaptative consiste en un système de détection de front d'onde de type
Shack-Hartmann et un miroir déformable doté de 12x12 actionneurs. La
fréquence de la boucle d'asservissement est de 500 Hz, avec un gain de
0,6 et un retard de correction de 4 ms. L'étoile guide est une étoile
de magnitude 5. La figure~\ref{fig:yao} est une copie d'écran de
l'affichage ``YAO''. On voit, à droite, le détecteur du Shack-Hartmann
et, à gauche, la réponse impulsionnelle instantanée. En bas à gauche
est représenté l'état du miroir déformable et du tip/tilt. La courbe
en bas à droite représente le Strehl obtenu après chaque itération. On
peut noter un Strehl moyen aux alentours de 0,2, ce qui reste élevé
pour une observation à 630 nm. Il faut cependant rester conscient que
cette optique adaptative est théorique et ne reflète pas certaines
autres causes de limitations, comme les erreurs de calibration ou
celles de la matrice d'inversion.

Le réarrangement de la pupille est effectué par la division de la
pupille d'entrée en 132 sous-pupilles hexagonales. Chacune de ces
sous-pupilles est filtrée par le mode fondamental d'une fibre optique
monomode. Le taux d'injection maximum dans les fibres, via une
sous-pupille hexagonale, est de 78\%. Cependant, à une longueur d'onde
de 630 nm, le taux d'injection est bien plus faible. Nos simulations
nous ont permis de mesurer un taux d'injection de $\approx 5\%$ pour
des sous-pupilles de tailles 5 $r_0$, et de $\approx 20\%$ lorsque
l'optique adaptative est activée. Les 132 sous-pupilles sont ensuite
réarrangées selon une configuration non-redondante en deux dimensions,
de manière à produire un total de 8\,646 fréquences spatiales
distinctes.

Le temps d'acquisition total a été fixé à 40 secondes. Cependant, le
temps de cohérence de l'atmosphère nécessite des acquisitions très
rapides, dont nous avons choisi de fixer la durée à 4
millisecondes. Cette période d'acquisition correspond à deux
déplacements du miroir déformable, pour permettre de prendre en compte
l'effet de ces déplacements au cours des acquisitions. Nous avons
ensuite ajouté un bruit Gaussien sur le détecteur pour prendre en
compte le bruit de photon (il s'agit d'une approximation du bruit
Poissonnien valable lorsque le nombre de photons est important). La
quantité de photons a été déterminée pour une bande passante de 60 nm
et un taux de couplage moyen dans les fibres de 5\% (en l'absence
d'optique adaptative). Aucun bruit de détecteur n'a été ajouté car le
système a été défini pour fonctionner dans le visible. Nous nous
sommes en conséquence placé dans le cas de l'utilisation d'une caméra
à comptage de photon.

L'effet de la chromaticité de la lumière a été
ignoré. Il a été pris le parti de considérer, lors de cette
simulation, que le problème du chromatisme était un problème technique,
pouvant être géré lors de la conception de l'instrument. Il existe, en
effet, plusieurs solutions techniques, dont celle consistant à
disperser les franges sur le détecteur.

\subsubsection{La fonction de transfert optique}

\begin{figure}
  \centering
  \resizebox{\hsize}{!}{
    \includegraphics{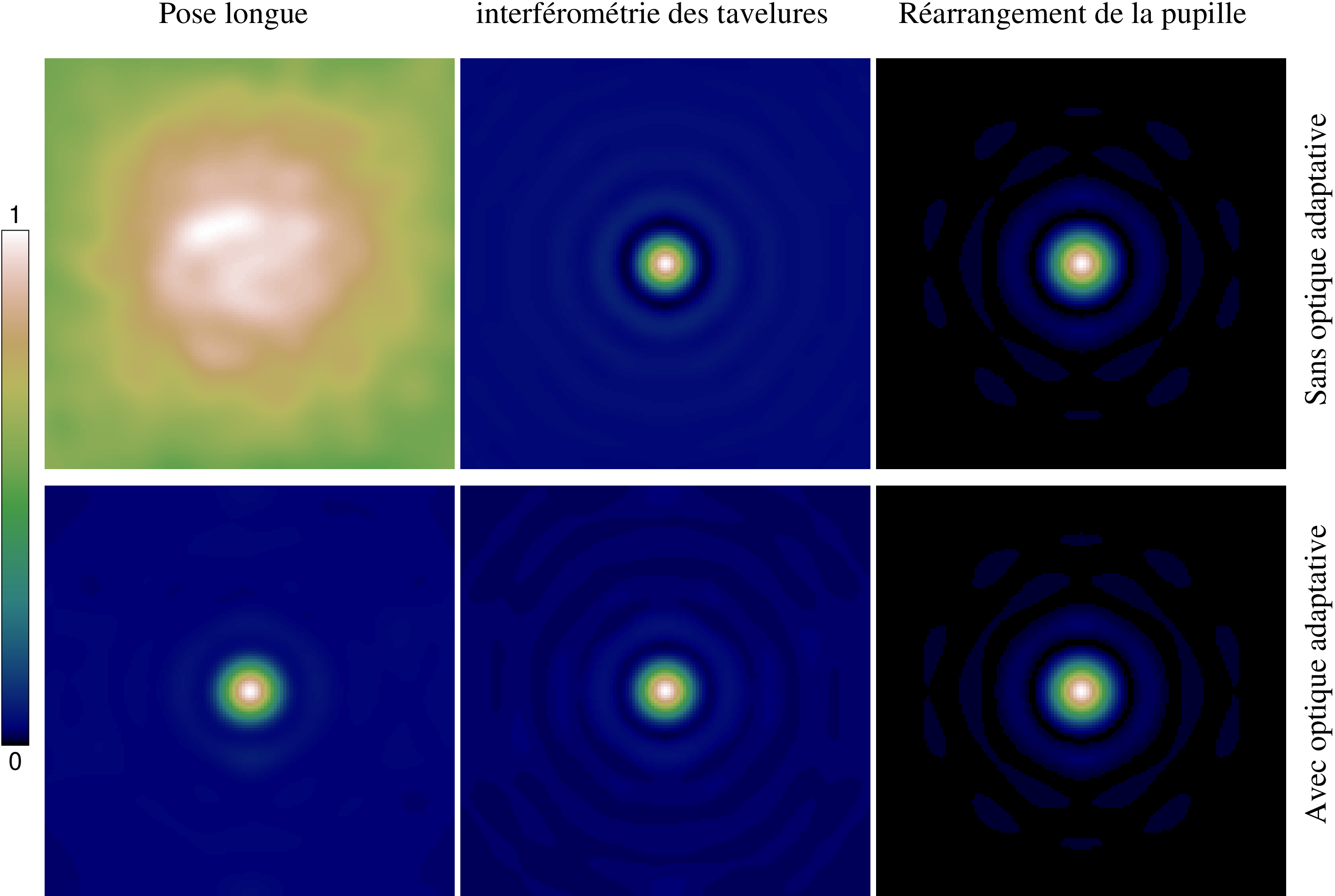}   }
  \caption[Simulations de réponses impulsionnelles]{ Images des réponses impulsionnelles obtenues à partir de
  trois techniques différentes. L'acquisition consiste en 10\,000
  poses de 4 ms sur un télescope de 8 mètres, dans le domaine visible,
  et en présence d'une atmosphère turbulente ($r_0=20$ cm). Le calcul
  de la transformé de Fourier de ces images a été obtenu par les
  équations~(\ref{eq:lp}),~(\ref{eq:sp}) et~(\ref{eq:rp}). Elles
  correspondent respectivement à une acquisition longue pose, à
  une déconvolution par interférométrie des tavelures et à un système avec
  réarrangement de pupille. Tous les cas ont été calculés avec ou sans
  la présence d'une optique adaptative. Une coupe horizontale de chacune
  de ces images est présentée figure~\ref{fig:simu2}.
}
  \label{fig:simu1}
\end{figure}

\begin{figure}[h]
  \centering
  \resizebox{\hsize}{!}{
    \includegraphics{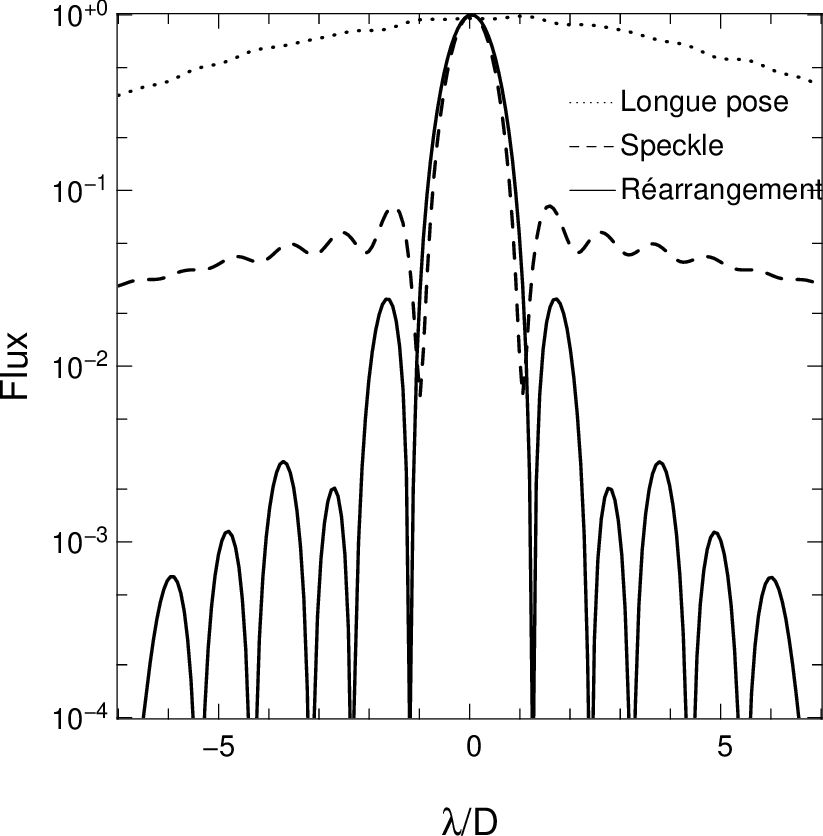} 
    \includegraphics{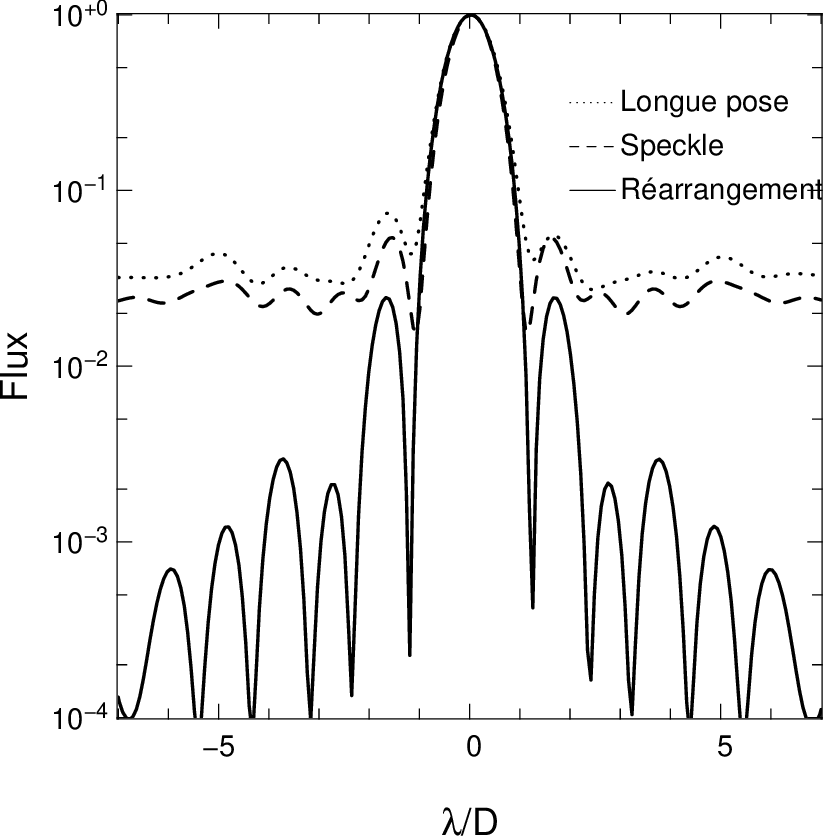}   }
  \caption{ Coupes horizontales des réponses impulsionnelles présentées
  figure~\ref{fig:simu1}}
  \label{fig:simu2}
\end{figure}

Le premier test que nous avons effectué a consisté à imager les
fonctions de transfert instrumentales, ou plutôt, les réponses
impulsionnelles. \`A titre de comparaison, nous avons étudié
trois cas. Ceux-ci ont le mérite de pouvoir être calculés simplement, à
partir des coefficients de transmission complexes de la pupille.

La première fonction de transfert optique (FTO) est celle que l'on
obtiendrait pour une longue pose. En utilisant une notation discrète du
champ dans la pupille (les $\V{G}$), la fonction de transfert s'écrit
à la fréquence $\mathbf{u_k}$:
\begin{equation}
  \OTF_k=\left\langle
  \sum_{(i,j)\in\mathcal{B}_k}\!\!
  G_i\,G_j^\star
  \right\rangle\,.
  \label{eq:lp}
\end{equation}
Cette valeur correspond à la moyenne temporelle des fonctions de
transfert instantannées. De manière similaire, on peut déduire une fonction de
transfert optique dans le cas d'un traitement post détection proche de
l'interférométrie des tavelures. Dans ce type de méthode, il est
souvent effectué la somme de la densité spectrale de puissance de l'image d'un côté, et le
bispectre de l'autre. Le biais introduit par le bruit de photon est
ensuite retranché, et l'image obtenue par déconvolution. En négligeant
l'influence de la phase, on peut simuler une réponse impulsionnelle sous
la forme suivante~:
\begin{equation}
  \OTF_k = \sqrt{\Bigl\langle\,
    \Bigl\vert\sum_{(i,j)\in\mathcal{B}_k}\!\!
    G_i \, G_j^\star\Bigr\vert^2\,
    \Bigr\rangle}
  \label{eq:sp}
\end{equation}
Enfin, la technique que nous proposons, permet de calculer l'influence
des facteurs de transmission $\V{\widetilde{G}}$ par l'algorithme
présenté section~\ref{sc:algo_eric}. La fonction de transfert optique
est alors la suivante~:
\begin{equation}
  \OTF_k = \left\langle
  \sum_{(i,j)\in\mathcal{B}_k}
  \frac{G_i \, G_j^\star}{\widetilde{G}_i \, \widetilde{G}_j^\star}
  \right\rangle
  \label{eq:rp}
\end{equation}
Il faut noter, cependant, que cette OTF est physiquement échantillonnée,
ce qui se traduit pratiquement par une limitation du champ spatial
observé.

Les résultats sont représentés figure~\ref{fig:simu1} sous la forme de
réponses impulsionnelles. Les images de la partie supérieure de la
figure correspondent aux résultats obtenus avec un front d'onde
non-corrigé, et la partie inférieure avec l'activation de l'optique
adaptative. Un premier résultat est que l'optique adaptative, même si
elle n'est pas conditionnée pour fonctionner aux longueurs d'onde du
visible, permet de gagner en résolution. Cependant, la dynamique est
très faible, de l'ordre de 20. Nous avons tracé figure~\ref{fig:simu2}
les coupes horizontales des réponses impulsionnelles. On peut noter
l'intérêt de la technique d'interférométrie des tavelures (aussi
appelée technique speckle) qui permet, même sans optique adaptative, de
restituer une image à la limite de diffraction. La dynamique maximale
est alors obtenue lorsque l'on conjugue une technique speckle avec une
optique adaptative. La dynamique reste cependant faible, inférieure à
50.

Lorsque l'on utilise l'algorithme présenté section~\ref{sc:algo_eric}
pour estimer les coefficients de transmission, on peut voir sur les
panneaux de la figure~\ref{fig:simu1} qu'une très grande dynamique
peut être obtenue. On arrive, d'ailleurs, à reconstituer une fonction
de transfert très proche de la tache d'Airy. On peut voir
figure~\ref{fig:simu2} que les anneaux d'Airy ne sont pas parfaitement
déterminés. Ceci est dû au filtrage hexagonal de la pupille par les
fibres optiques. Ce résultat montre qu'il est possible d'obtenir une
dynamique supérieure à $10^3$ à une distance angulaire de quelques
$\lambda/D$ de l'étoile centrale. Cependant, parce que la réponse
impulsionnelle est stable et connue, il est possible de l'ajuster pour
obtenir une dynamique encore plus grande. C'est ce que nous avons fait
par la suite pour reconstruire des images à très haute dynamique.

\subsubsection{Les images reconstruites}

\begin{figure}
  \centering
    \includegraphics[width=14cm]{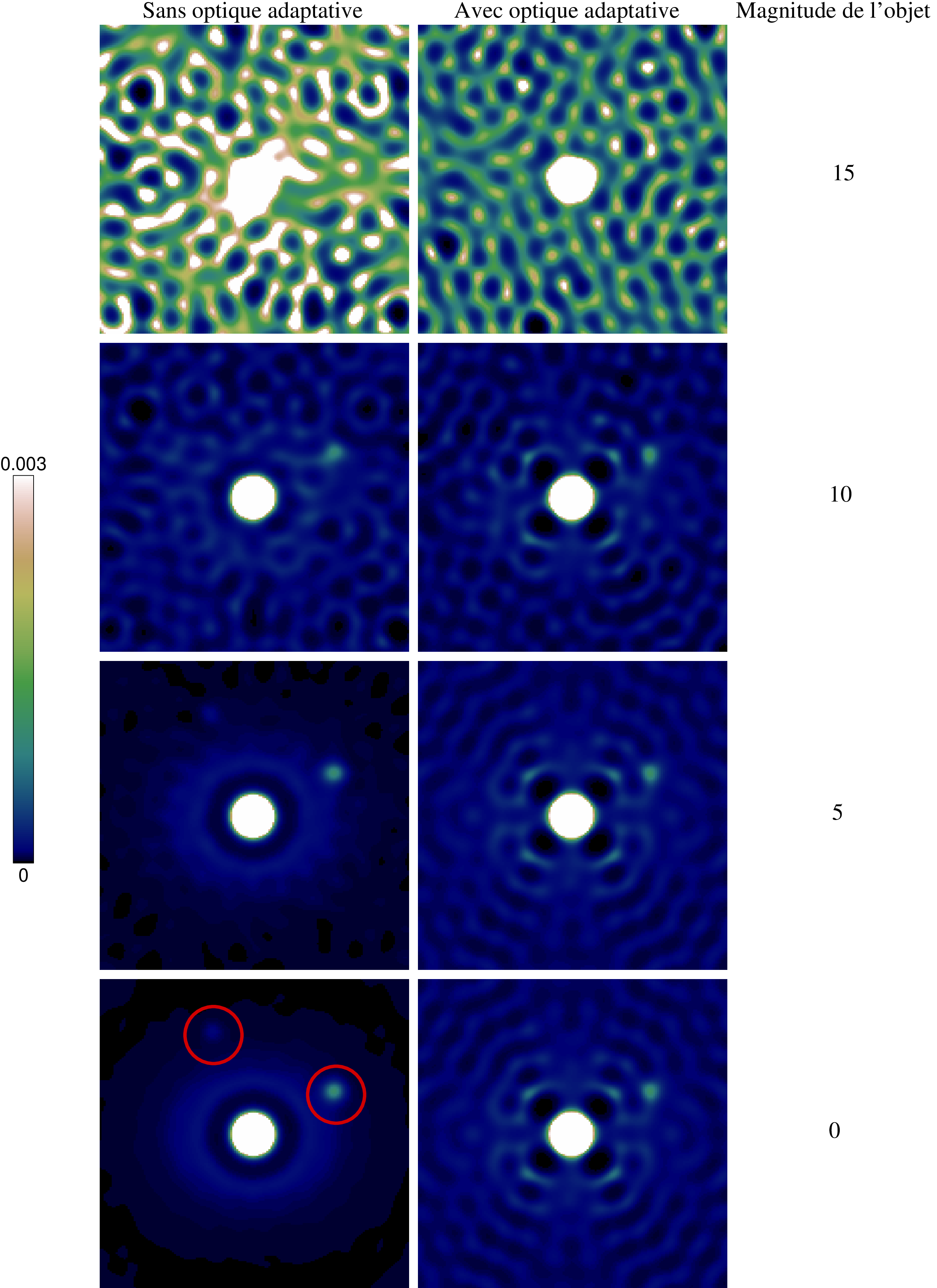}   
  \caption[Images reconstruites par la
  méthode de réarrangement de pupille]{ Ces simulations présentent des images reconstruites par la
  méthode de réarrangement de pupille. L'objet central est une étoile
  de magnitude variable (entre 0 et 15). L'environnement de l'étoile
  consiste en un disque d'accrétion de brillance $1/100$ celle de
  l'étoile, et de deux compagnons de flux respectifs un millième et
  un dix-millième. Les compagnons sont entourés en rouge sur l'image
  reconstruite en haut à gauche. }
  \label{fig:simu3}
\end{figure}

\begin{figure}
  \centering
  \resizebox{\hsize}{!}{
    \includegraphics{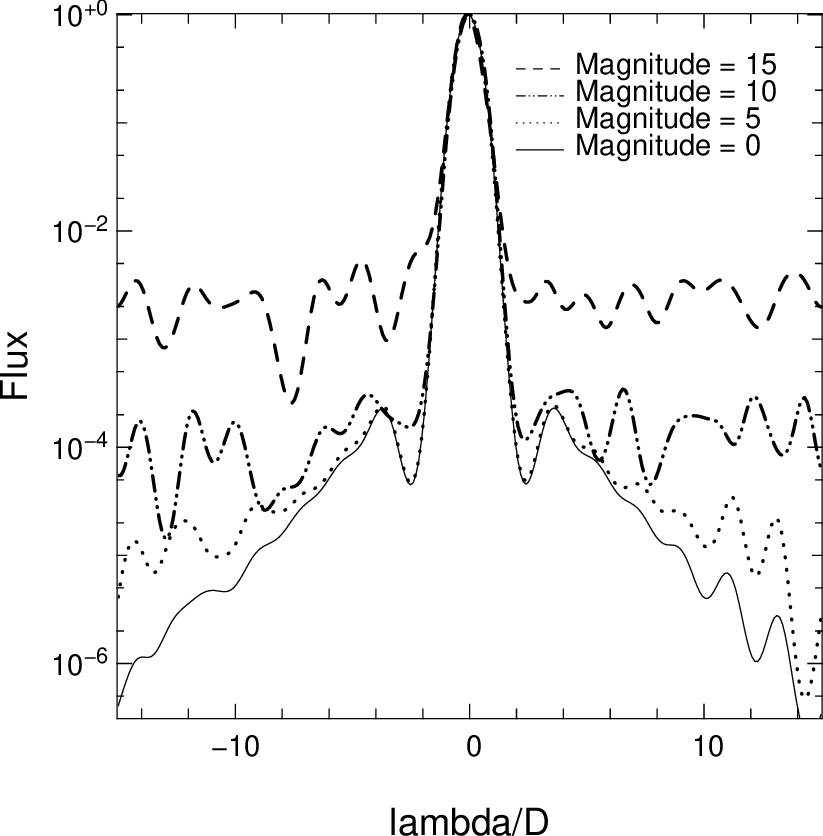} 
    \includegraphics{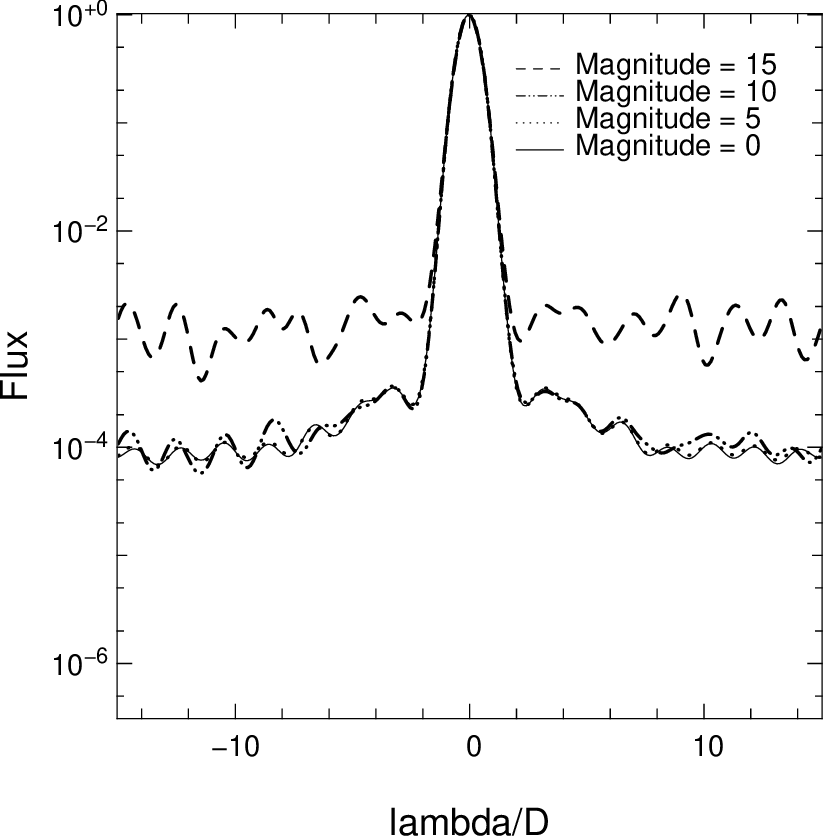}}
  \caption{ Coupes horizontales des images reconstruites présentées
  figure~\ref{fig:simu3}  }
  \label{fig:simu4}
\end{figure}

L'étape suivante consiste à reconstruire une image à partir d'un objet
complexe. Pour cela, nous avons considéré une étoile centrale de
magnitude variable (0, 5, 10, 15 mag), entourée d'un disque
circumstellaire. Le disque a un rapport de brillance avec l'étoile
centrale de $10^{-2}$, ainsi qu'une répartition décroissante
exponentielle. Nous avons aussi ajouté à ce disque deux compagnons,
de brillances $1/1\,000$ et $1/10\,000$.

Comme nous les avons précédemment calculés, les facteurs de
transmission complexes $\V{G}$ sont obtenus par l'algorithme de
l'équation~(\ref{eq:update-multi}) sur les 10\,000 acquisitions. Nous en
avons déduit les visibilités de l'objet par la
relation~(\ref{eq:best-obj-vis-multi}). Enfin, nous avons utilisé les
visibilités pour reconstruire une image. Il est intéressant de noter
que l'on a alors le choix de la réponse impulsionnelle. Ainsi, si on
multiplie les visibilités par une fonction Gaussienne, la réponse
impulsionnelle sera une Gaussienne. Si on les multiplie par une
fonction de transfert en forme de cône, on obtiendra alors une réponse
impulsionnelle en forme de tache d'Airy. Ce choix dépend en
conséquence de nos objectifs scientifiques. Dans notre étude, parce
que l'on souhaite mettre en évidence la dynamique maximale, nous avons
multiplié nos visibilités par une fonction Gaussienne. Au prix d'une
légère perte en résolution spatiale, nous avons pu  obtenir des
dynamiques nettement supérieures aux limitations dues aux anneaux
d'Airy.

Ces résultats sont présentés figures~\ref{fig:simu3}
et~\ref{fig:simu4}. Dans le cadre des données simulées sans optique
adaptative (AO), le bruit de photons est clairement la cause de la
limitation en dynamique. On peut voir que lorsque la magnitude de
l'objet augmente, la dynamique diminue linéairement. Nous avons
représenté dans le tableau~\ref{tb:dyn} la dynamique des images en
fonction de la magnitude de l'objet. La dynamique a été obtenue à
partir de la variance d'une section bruitée de l'image. Celle-ci est
comparée à la dynamique théorique telle que nous l'avons calculée
section~\ref{sec:analy}. Nous pouvons constater une dynamique
effective très proche de la dynamique théorique, calculée à partir du
bruit de photons seulement. Ainsi, il a effectivement été reconstruit
une image avec une dynamique de $10^6$. Ceci est une confirmation de
la limite instaurée par le bruit de photons et non plus par le bruit
des turbulences atmosphériques. Ceci est remarquable pour des
observations obtenues avec un télescope de 8 mètres aux longueurs
d'onde visibles.

Notre travail a également permis de mettre en lumière l'influence de
l'optique adaptative. Comme l'indique le tableau~\ref{tb:dyn},
l'utilisation de cette technique permet de gagner en dynamique lorsque
l'objet est faiblement brillant (mag > 10). Cependant, nous avons noté
que la dynamique était limitée aux alentours de $1,5 \times 10^4$. Une
analyse de nos calculs nous a permis d'aboutir à la conclusion que
l'optique adaptative limitait bien la dynamique. Ceci est dû au petit
déplacement du miroir. En effet, la boucle d'asservissement déplace le
miroir toutes les 2 ms, alors que le temps d'intégration d'une pose est
de 4 ms. Ce résultat ne remet cependant pas en cause l'utilité de
conjuguer notre système à une optique adaptative. Il montre,
néanmoins, l'intérêt d'une étude détaillée d'un tel système, où
l'optique adaptative serait configurée pour ne pas introduire de
bruit. Cela pourrait se faire, par exemple, en autorisant des
déplacements du miroir déformable uniquement entre deux poses.

\begin{table}
\caption{Dynamique des images de la figure~\ref{fig:simu3}}
\label{tb:dyn}
\centering
\begin{tabular}{c c c c c}
\hline \hline
 &\multicolumn{2}{c}{Sans AO} &\multicolumn{2}{c}{Avec
  AO} \\
Magnitude & $\sqrt{\Nph/2}^{\mathrm{a}}$ & D.R.$^{\mathrm{b}}$ &
 $\sqrt{\Nph/2}^{\mathrm{a}}$ & D.R.$^{\mathrm{b}}$ \\
\hline
0 & $1,1 \times 10^6$ & $0,9 \times 10^6$ & $2,4 \times 10^6$ & $1,8 \times 10^4$   \\
5 & $1,1 \times 10^5$ &  $1,5 \times 10^5$ & $2,4 \times 10^5$ & $1,7 \times 10^4$   \\
10 & $1,1 \times 10^4$ &  $1,3 \times 10^4$ & $2,4 \times 10^4$ & $1,6 \times 10^4$  \\
15 & $1,1 \times 10^3$ & $0,8 \times 10^3$  & $2,4 \times 10^3$ & $1,2 \times 10^3$   \\
\hline
\end{tabular}
\begin{list}{}{}
\item[$^{\mathrm{a}}$] Dynamique prédite par l'équation~(\ref{eq:dyn}).
\item[$^{\mathrm{b}}$] Dynamique mesurée à partir de l'écart type
  constaté sur le fond des images reconstruite de la figure~\ref{fig:simu3}.
\end{list}
\end{table}

\chapter{L'optimisation des paramètres de l'instrument}
\begin{center}
\end{center}
\minitoc \label{ch:param} \vskip1cm
\clearpage
\section{Introduction}

Il  existe deux types de paramètres entrant en jeu lors de la conception
de l'instrument. Un certain nombre d'entre eux sont liés à
l'environnement de l'instrument et ne peuvent être modifiés. Il s'agit
de~:
\begin{itemize}
\item $D$, le diamètre du télescope
\item $r_0$, le paramètre de Fried. Il correspond au diamètre d'une
 surface cohérente dont la variance de la phase ($\sigma_\phi$) est
 inférieure à 1 radian. Pour un système comprenant une optique
 adaptative, nous utiliserons à la place de $r_0$ le paramètre de
 Fried généralisé $\rho_0$ \citep{2000JOSAA..17..903C}.
\item Les bruits de photon et de détecteur. 
\end{itemize}
Ensuite, en fonction de ceux-ci, il faut choisir un certain nombre de
paramètres physiques optimisant les performances finales. Ces
paramètres libres sont~:
\begin{itemize}
\item $M$, le nombre de sous-pupilles et de fibres optiques
\item $d$, la taille des sous-pupilles
\item $\eta$, le rapport entre l'ouverture numérique des lentilles et
  celle des fibres
\item $\Delta \lambda/\lambda_0$ la bande passante spectrale
\item $D'/d'$, le rapport maximal de taille entre la sous-pupille de sortie et la
  distance séparant deux sous-pupilles.
\end{itemize}
Il s'agit de simuler l'influence de chacune de ces valeurs
dans le cadre des paramètres fixes. Par exemple, le diamètre du
télescope contraint le nombre de sous-pupilles par la relation~:
\begin{equation}
M d^2 \leq D \,.
\end{equation}
Or, ces paramètres influent souvent sur différents facteurs. Nous nous
focaliserons au cours de ce chapitre sur la sensibilité de
l'instrument et les sources d'erreurs sur les visibilités (voir
tableau~\ref{tb:choix} ci-dessous).

\begin{table}[h]
\caption{Influence des caractéristiques de l'instrument}
\label{tb:choix}
\centering
\begin{tabular}{lccccc}
\hline
\hline
& $d$ & $\eta$ & $\Delta \lambda/\lambda_0$ & $D'/d'$ & Paragraphe\\
\hline
Champ de la fibre & \checkmark & \checkmark & &&\ref{sc:champ_fibre} \\
Bruit de confusion & \checkmark & \checkmark & &&\ref{sc:confusion} \\
Champ de l'interféromètre & & & \checkmark &\checkmark & 
\ref{sc:Champ_interfero}  \\
Bruit de piston & & & \checkmark & \checkmark & \ref{sc:piston} \\
Sensibilité  & \checkmark & \checkmark & \checkmark & \checkmark &  \\
\hline
\end{tabular}
\end{table}

\clearpage

\section{L'injection dans les fibres monomodes}
\label{sc:inject}

\subsection{La relation entre plan pupille et plan image}

Au cours de ce chapitre, nous serons régulièrement amené à calculer
l'amplitude complexe du rayonnement émis par la source
astrophysique. Pour cela, il est parfois intéressant de l'établir dans
le plan pupille du telescope, mais aussi dans le plan du détecteur
(pour former une image) ou encore dans le plan de la fibre optique (pour
calculer le taux d'injection). La correspondance entre le champ dans
le plan pupille et celui dans le plan focal est décrite par la relation de
Fraunhoffer.

Concrètement, cette relation établit qu'un champ électrique dans le système de
coordonnées de la pupille $E_\circ(x_1,y_1)$ se diffracte de façon à produire un
champ $E_\bullet(x_2,y_2)$ à une distance $f$ telle que:
\begin{equation}
E_\bullet(x_2/f,y_2/f) \propto TF(E_\circ(x_1/\lambda,y_1/\lambda))\,,
\end{equation}
où $TF$ est l'opérateur transformée de Fourier. Sans optique, cette
relation est vérifiée dans le cas où $f$ est suffisament grand ($f \gg
\frac{\pi(x_1^2+y_1^2)}{\lambda}$). Lorsque l'on utilise une lentille
(ou un mirroir) pour faire converger la lumière, cette relation est
vérifiée au point focal.

Dans ce chapitre, nous utilisons le sigle $\circ$
pour désigner le champ dans le plan pupille, et $\bullet$ le champ dans
le plan focal. Les coordonnées respectivement utilisées sont ($u$,$v$)
en unité de longueur d'onde, et $(\alpha,\beta)$ en unité de distance
focale. Ceci permet de simplifier la relation de Fraunhoffer qui, en
faisant usage de ces coordonnées conjuguées, s'écrit~:
\begin{equation}
E_\bullet(\alpha,\beta) \propto TF(E_\circ(u,v))\,.
\label{eq:Fraun}
\end{equation}

\subsection{Eléments théoriques des fibres}

   \begin{figure}[h!]
   \centering
   \includegraphics[width=10cm]{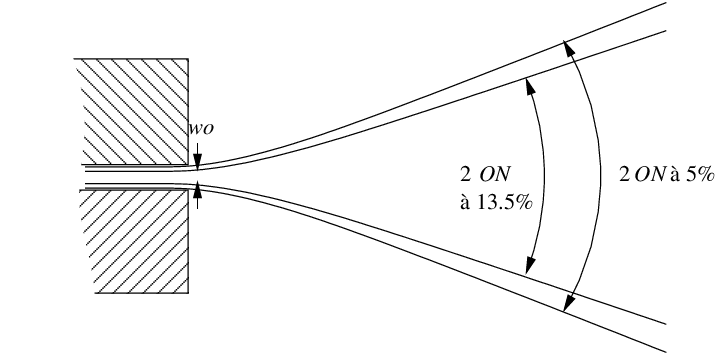}
   \caption[Flux en sortie de fibre]{ Flux en sortie de fibre. La diffraction crée un champ
   divergeant, quasi-Gaussien. Il peut être caractérisé par une largeur
   fixée par un seuil de 5\% ($e^{-3}$) ou de 13,5\% ($e^{-2}$). Dans le
   c\oe ur de la fibre, le champ est défini par le diamètre modal
   $w_0$. L'ouverture numérique correspond à un angle que nous
   définirons au cours de cette thèse par le niveau d'intensité de
   5\%.  }
         \label{fig:mesure_ON}
   \end{figure}

Certain paramètres fondamentaux caractérisent les fibre optiques
monomodes. Ils sont déterminés par la physique de la fibre et servent
de référence lors de l'achat des fibres. Il s'agit de~:
\begin{itemize}
\item L'ouverture numérique $ON$, angle dû à la diffraction de la
  lumière à la sortie de la fibre. Cette valeur est quasi
  achromatique.  Dans le cas d'une fibre à saut d'indices, elle est
  fixée par l'indice de la fibre et de son c\oe ur: $ON =
  \sqrt{n_c^2-n_g^2}$. Cette valeur est importante car elle peut être
  déterminée expérimentalement à partir du seuil de 5\% du flux maximum
  (figure~\ref{fig:mesure_ON}).
\item Le diamètre du mode fondamental $w_0$ (en anglais, MFD pour Mode
  Field Diameter) est une longueur chromatique et ne correspond pas
  nécessairement au diamètre du coeur. Alors que l'ouverture numérique
  est définie par un seuil d'intensité de 5\%, le diamètre est
  caractérisé par le seuil de 13,5\%. La relation entre ces deux
  valeurs est la suivante~:
\begin{equation}
w_0=\frac{\sqrt{6}}{\pi} \frac{ \lambda}{ON}\,.
\label{eq:w0_ON}
\end{equation}
\item La fréquence de coupure $\lambda_c$. Il s'agit de la limite
  spectrale pour que la fibre se comporte comme une fibre
  monomode. Aux longueurs d'ondes inférieures, d'autres modes
  apparaissent. L'utilisation d'une fibre monomode se fait
  généralement dans le domaine spectral $\lambda_c < \lambda < 1,3
  \lambda_c$. Aux longueurs d'ondes supérieures, la fibre devient
  sensible aux courbures et le facteur de transmission
  décroît. $\lambda_c$ est relié à l'ouverture numérique et au rayon
  du c\oe ur
  de la fibre ($a$) par la relation:
\begin{equation}
\lambda_c=\frac{2 \pi a ON}{2,405} \,.
\end{equation}
\end{itemize}

Le champ dans la fibre est à symétrie circulaire, avec une amplitude proche d'une
fonction Gaussienne. Cette approximation donne~:
\begin{equation}
E(r) \propto  \exp\left(\frac{-4r^2}{w_0^2}\right)\,.
\end{equation}
Dans un plan image fictif défini par une focale $f$, le champ en  coordonnées angulaires
donne~:
\begin{equation}
E_\bullet(\alpha,\beta) \propto \exp\left(-\frac{4(\alpha^2+\beta^2)f^2}{w_0^2}\right)\,.
\end{equation}
Nous pouvons alors utiliser la transformée de Fourier qui lie le champ
dans le plan pupille au champ du plan focal (équation~(\ref{eq:Fraun}))
pour en déduire le champ pupillaire associé au mode fondamental de la
fibre:
\begin{equation}
E_\circ(u,v) \propto
\exp\left(-\frac{\pi^2 w_0^2(u^2+v^2)}{4f^2}\right) \, .
\end{equation}
qui peut également s'écrire d'après l'équation~(\ref{eq:w0_ON})~:
\begin{equation}
E_\circ(u,v) \propto
\exp\left(-\frac{3 (u^2+v^2)\lambda^2}{2f^2\,ON^2}\right) \, 
\label{eq:champ_fibre}
\end{equation}
Il est intéressant de noter que l'amplitude du champ dans la pupille
est une Gaussienne de largeur indépendante de la longueur d'onde
(parce que $u$ et $v$ sont en unités de longueur d'onde, $u\lambda$ et
$v\lambda$ représentent des distances achromatiques).

\subsection{L'efficacité de couplage}

L'amplitude complexe couplée dans la fibre $A$ est le produit scalaire
normalisé du mode fondamental de la fibre ($E$) par le champ électrique
incident ($U$):
\begin{equation}
A =
\frac{ \iint_{-\infty}^{+\infty}
  U_\bullet(\alpha,\beta)\,.\,E_\bullet^\star(\alpha,\beta)\ d\alpha
  d\beta}{\iint_{-\infty}^{+\infty}
  |U_\bullet(\alpha,\beta)|^2 d\alpha
  d\beta\,\times\,\iint_{-\infty}^{+\infty}
  |E_\bullet(\alpha,\beta)|^2d\alpha d\beta} \,.
\end{equation}
Elle peut être calculée de la même façon par le recouvrement des champs
dans le plan pupille (théorème de Parceval-Plancherel):
\begin{equation}
A =
\frac{\iint_{-\infty}^{+\infty} U_\circ(u,v)\,.\,E^\star_\circ(u,v)\
  dudv}{\iint_{-\infty}^{+\infty}|U_\circ(u,v)|^2
  dudv\,\times\,\iint_{-\infty}^{+\infty}|E_\circ(u,v)|^2dudv} \,.
\label{eq:int_rec}
\end{equation}

L'équation du champ de la fibre dans le plan pupille a été établie en
équation~(\ref{eq:champ_fibre}). En faisant intervenir le diamètre de la
sous-pupille ($d$) et le rapport d'ouverture numérique entre la lentille
d'injection et la fibre~:
\begin{equation}
\eta=\frac{ON_{\rm lentille}}{ON_{\rm fibre}}=\frac{d}{2fON}\,,
\label{eq:eta}
\end{equation}
on peut obtenir une forme simple, normalisée~:
\begin{equation}
E_\circ(u,v) = \frac{ 2 \eta \lambda}{ d} \sqrt{ \frac{3}{\pi}} 
\exp\left(-6 \frac{ (u^2+v^2)\lambda^2}{d^2} \eta^2 \right) \, ,
\label{eq:Eo}
\end{equation}
où $u$, $v$ et $d$ sont respectivement les coordonnées dans le plan
pupille et le diamètre de la sous-pupille. L'amplitude complexe
couplée peut ainsi s'écrire pour un quelconque champ pupillaire
$U_\circ(u,v)$ normalisée~:
\begin{equation}
A =
 \iint_{-\infty}^{+\infty} U_\circ(u,v)\,\frac{ 2 \eta \lambda}{ d} \sqrt{ \frac{3}{\pi}}\,
\exp\left(-6 \frac{ (u^2+v^2)\lambda^2}{d^2} \eta^2 \right)
  dudv\,.
\label{eq:A_qul}
\end{equation}
L'efficacité de couplage correspond à l'énergie injectée dans la fibre
par rapport à l'énergie totale incidente. Elle est égale au carré du
module de l'amplitude complexe~:
\begin{equation}
\rho = |A|^2 \,.
\label{eq:eff_coup}
\end{equation}


\subsection{L'injection dans le cas d'un front d'onde plan}

   \begin{figure}[h]
   \centering \resizebox{\hsize}{!}{
\includegraphics{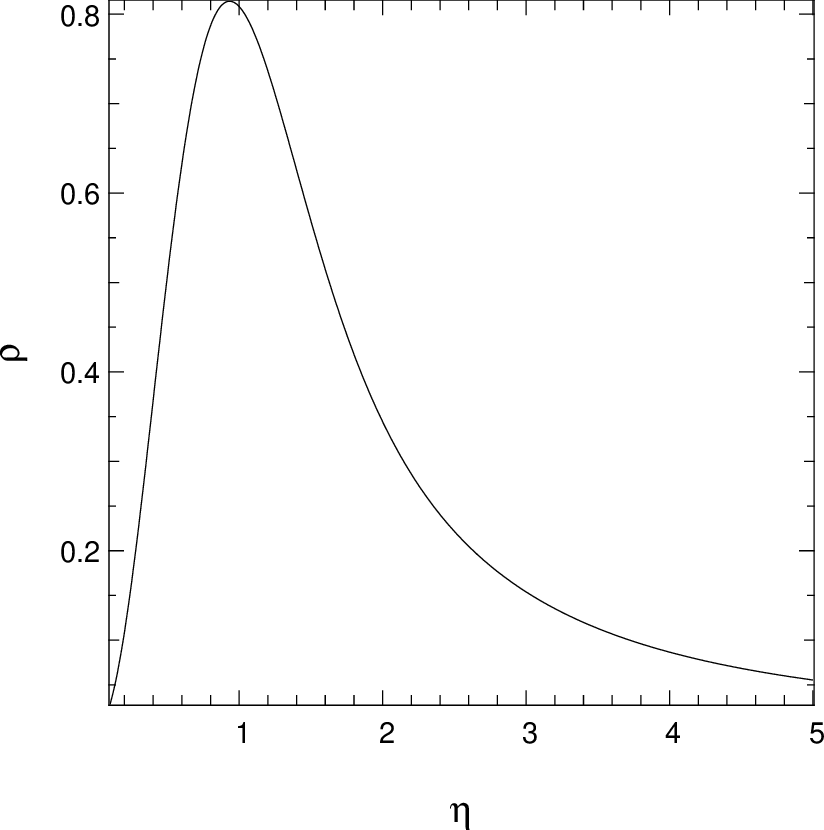}
\includegraphics{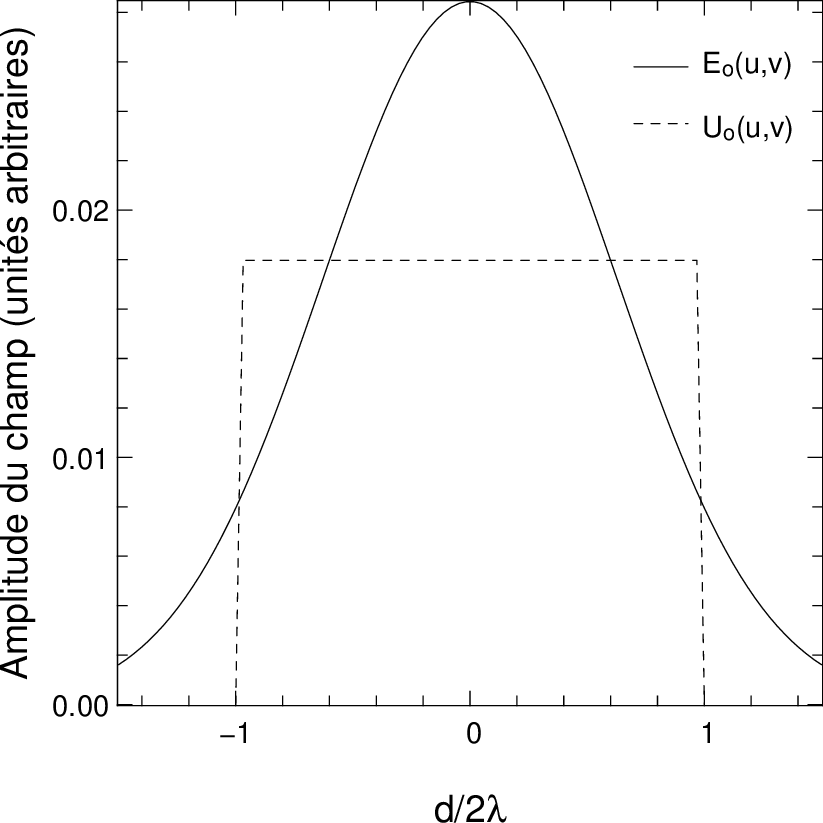}}
   \caption[Efficacité de couplage en fonction du rapport
   d'ouverture $\eta$]{ A gauche~: efficacité de couplage en fonction du rapport
   d'ouverture $\eta = ON_{\rm lentille} / ON_{\rm fibre}$. Dans le
   cas d'un front d'onde incident cohérent, l'efficacité de couplage
   est maximale pour $\eta =0,92$. A droite~: le champ électrique
   associé à la fibre et à l'onde incidente dans le plan pupille.
}
         \label{fig:champ_fibre}
   \end{figure}

Pour une pupille pleine, et en l'absence de perturbations
atmosphériques, le champ provenant de l'objet astrophysique non résolu
s'écrit de façon normalisé~:
\begin{equation}
U_\circ(u,v)= \left\{ \begin{array}{cl}
\displaystyle \frac{2\lambda}{\sqrt{\pi} d}  &\mbox{si $ \sqrt{u^2+v^2} \leq d/2\lambda $} \\
  0 &\mbox{sinon.}
       \end{array} \right.
\end{equation}
D'où l'amplitude complexe de couplage~:
\begin{equation}
A = \iint_{\sqrt{u^2+v^2} \leq d/2\lambda}  \frac{4\sqrt{3} \eta \lambda^2}{\pi
  d^2} 
\exp\left(-6 \frac{ (u^2+v^2)\lambda^2}{d^2} \eta^2 \right)       dudv \,,
\end{equation}
et~:
\begin{equation}
A = \frac{2}{\sqrt{3} \eta} \left( 1 -\exp\left(-\frac{3 \eta^2}{2}\right)\right)  \,.
\end{equation}
Par conséquent, l'efficacité de couplage s'écrit~:
\begin{equation}
\rho = \frac{4}{3  \eta^2} \left(1 -\exp\left(-\frac{3
\eta^2}{2}\right)\right)^2  \,.
\end{equation}
Nous l'avons représentée figure~\ref{fig:champ_fibre}, ainsi que les
 champs pupillaires correspondant à son maximum. Celui-ci est
obtenu pour $\eta = 0,92$ et correspond à une efficacité de couplage
de 81 \%.

\subsection{La sensibilité dans le cas d'un front d'onde perturbé}

   \begin{figure}[h]
   \centering
\resizebox{\hsize}{!}{
   \includegraphics{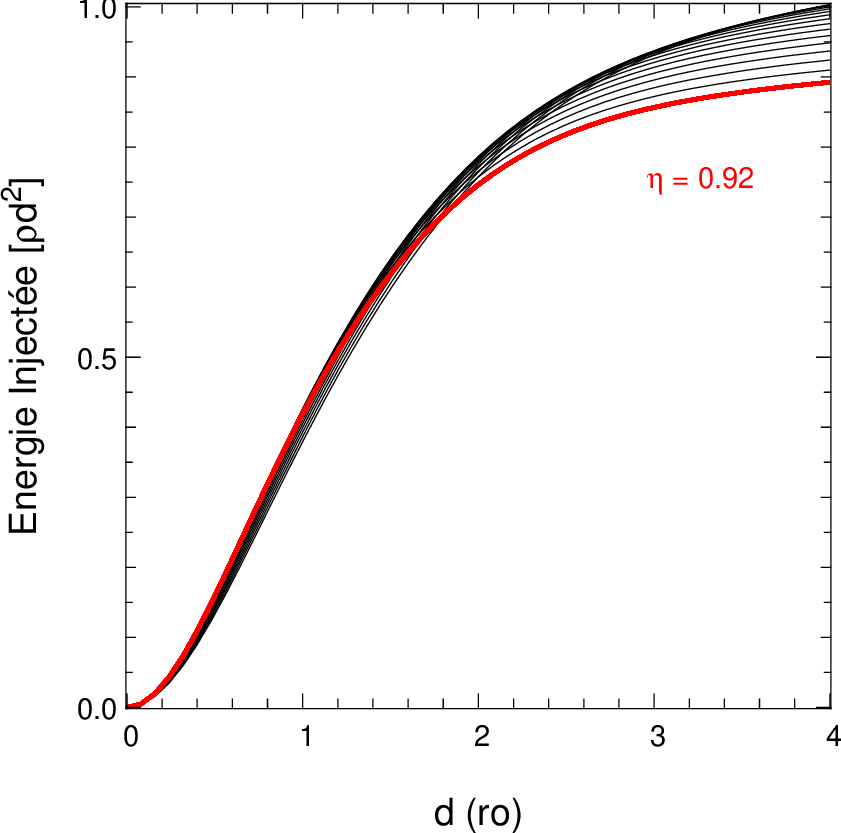}
   \includegraphics{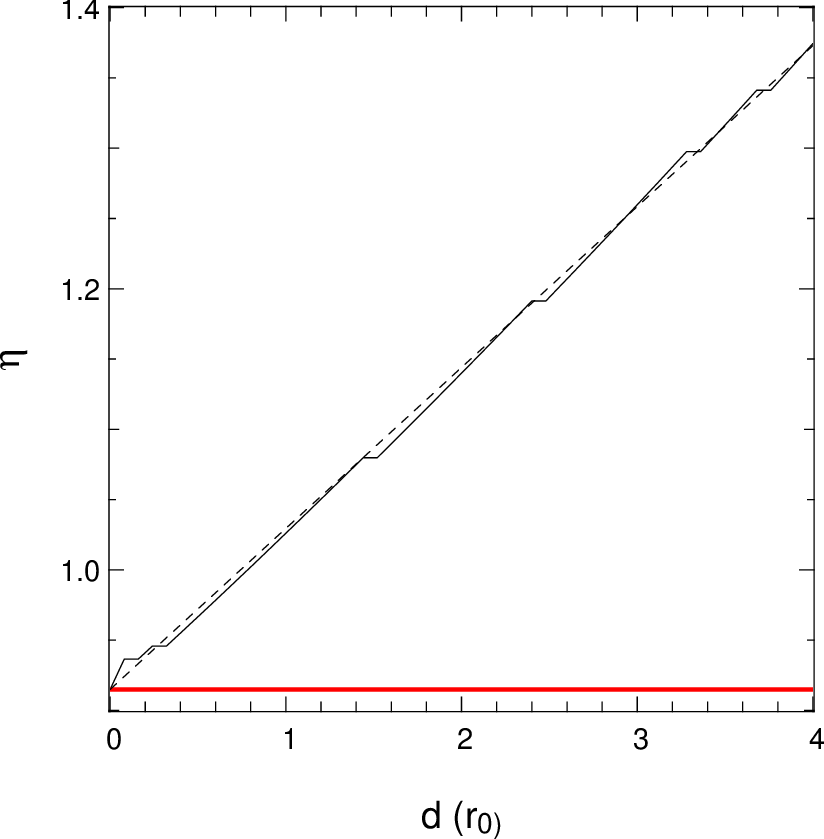}}
   \caption[\'Energie injectée dans une fibre en présence de turbulences
     atmosphériques]{ A gauche~: $\rho d^2$, une quantité proportionnelle à
     l'énergie injectée dans une fibre, en présence de turbulences
     atmosphériques. Pour une petite pupille l'énergie injectée est
     maximale pour $\eta=0,92$. Lorsque la taille de la pupille
     augmente, le $\eta$ maximisant l'énergie couplée change. Nous
     l'avons représenté sur la courbe de droite, et approximé par la
     relation~: $\eta = 0.115\,d/r_0+0.92$ (courbe en pointillés).  }
         \label{fig:inject_turb}
   \end{figure}

Lorsque le bruit de détecteur domine, la sensibilité de l'instrument
est liée au nombre de photons reçus par pixel, et non directement à celui 
collecté par l'ensemble de la pupille. Plus exactement, à nombre de photons fixe,
plus on utilise de fibres, plus la sensibilité décroît.

Ainsi, une caractérisation de la sensibilité peut être obtenue à
partir de l'énergie couplée en moyenne dans une fibre. Cette énergie
est proportionnelle à l'efficacité de couplage et à l'aire de la
sous-pupille. Dans le cas d'un front d'onde incident parfaitement
plan, la sensibilité maximale est obtenue pour $\eta= 0,92$ et un
diamètre de sous-pupille maximal. Lorsque le front d'onde est perturbé
par l'atmosphère, la sensibilité maximale n'est plus obtenue pour une
valeur unique de $\eta$. Nous avons utilisé un logiciel de simulation
de fronts d'ondes perturbés pour en déduire une moyenne de
l'efficacité de couplage en fonction du rapport $d/r_0$. La thèse
d'\citet{Assemat:PhD} décrit ce logiciel en détails.

Nos résultats montrent que l'augmentation de la taille de la pupille
n'entraîne pas obligatoirement une augmentation notable de l'énergie
injectée. \`A $r_0$ fixé, l'énergie est proportionnelle à $\rho
d^2$. Nous avons représenté cette valeur en fonction de $d$ sur le
graphique de gauche de la figure~\ref{fig:inject_turb} ($d$ est donné
en unité de $r_0$). La courbe rouge représente la valeur optimale en
l'absence de perturbations ($\eta =0,92$). Les autres courbes
correspondent à différentes valeurs de $\eta$ comprises entre 0,9 et
1,5. Un certain nombre de conclusions peuvent en être tirées~:
\begin{itemize}
\item Lorsque $d/r_0 < 2$, le choix du facteur de couplage n'est pas
  déterminant si $0,9 < \eta < 1,5$.
\item L'efficacité de couplage optimale est obtenue pour une ouverture
  numérique de la fibre plus faible. Nous avons ajusté une loi affine
  aux données de la figure~\ref{fig:inject_turb}, qui nous donne la
  valeur optimale du rapport d'ouverture en fonction de la taille de
  la sous pupille~:
\begin{equation}
\eta = \frac{ON_{\rm lentille}}{ON_{\rm fibre}} = 0.115\,d/r_0+0.92 \,.
\end{equation}
\item Lorsque $d/r_0 > 3$, l'énergie injectée n'augmente plus que
marginalement.
\end{itemize}

\textbf{A la lumière de ces résultats, une valeur optimale en terme de
sensibilité est obtenue pour une taille de sous-pupille $d = 3 r_0$ et un
rapport d'ouvertures numériques $\eta=1,25$.}

\clearpage
\section{Le champ de la fibre et le phénomène de confusion}

La résolution de l'instrument est fixée par le diamètre $D$ de la
pupille totale du télescope. Le nombre d'éléments de résolution est
déterminé par le champ observé sur le ciel. Celui-ci est déterminé par
le champ vu par chaque fibre individuelle, et s'obtient par un calcul
d'efficacité de couplage similaire à celui effectué dans la section
précédente.

\subsection{Le champ dans le cas d'un front d'onde plan}
\label{sc:champ_fibre}

\begin{figure}[h]
   \centering
\resizebox{\hsize}{!}{
   \includegraphics{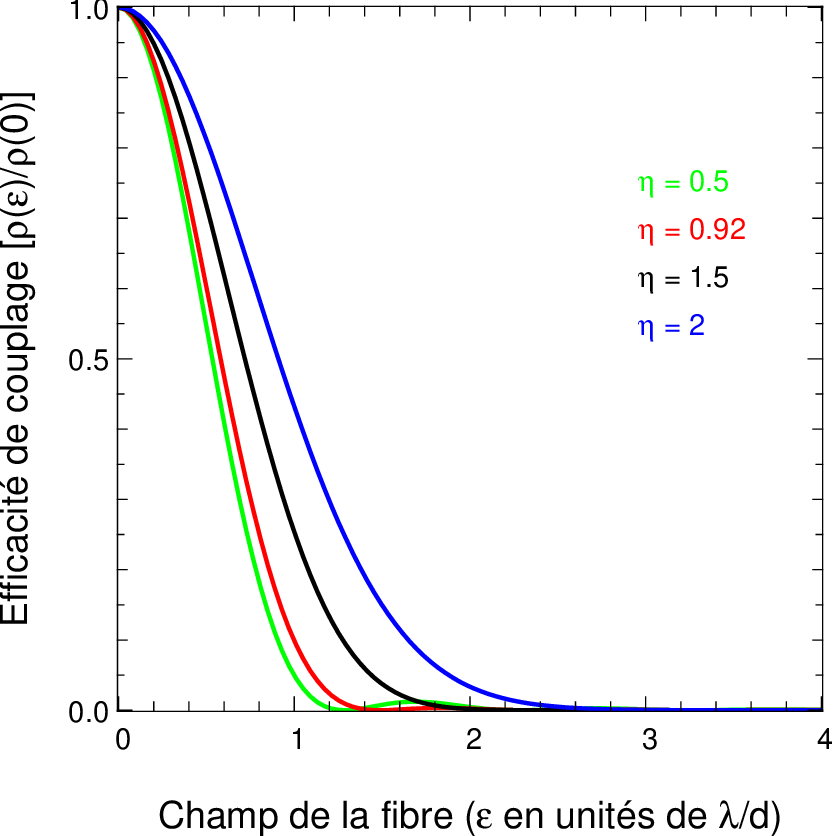}
   \includegraphics{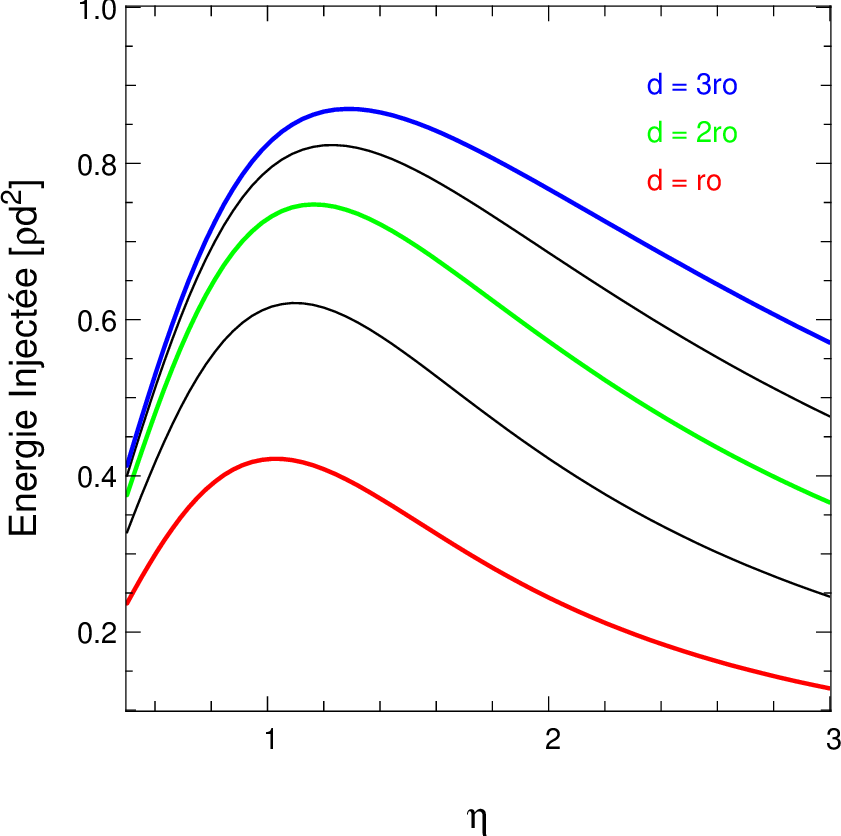}} \caption[Champ de la fibre
   en l'absence de perturbations atmosphérique. \'Energie couplée en
   fonction du paramètre $\eta$ en présence de perturbations
   atmosphérique]{ A gauche~: efficacité de couplage en fonction de la
   position de l'étoile dans le ciel, en l'absence de perturbations
   atmosphérique. Le champ injecté dans la fibre est en unités de
   $\lambda/d$, et peut ainsi être aisément comparé à l'élément de
   résolution de l'instrument qui est de $\lambda/D$. A droite~:
   l'énergie couplée en fonction du paramètre $\eta$, en présence de
   turbulences atmosphériques. Lorsque ce paramètre augmente, le champ
   augmente. Cependant, passée une valeur optimale, l'accroissement de
   $\eta$ se traduit également par une diminution du flux couplé dans
   la fibre.}  \label{fig:Champ}
\end{figure}

Lorsque l'objet observé n'est plus sur l'axe du télescope, le champ
incident est incliné par rapport au champ de la fibre. L'efficacité de
couplage pour une source hors axe définit ainsi le champ de la
fibre. Pour une source distante dans la direction de $u$ d'un angle
$\varepsilon$ de l'axe du système, on obtient l'expression suivante du
champ~:
\begin{equation}
U_\circ(u,v)= \left\{ \begin{array}{cl}
\displaystyle  \frac{2 \lambda\exp(-2 \I \pi u \varepsilon)}{\sqrt{\pi} d}  &\mbox{si $ \sqrt{u^2+v^2} \leq d/2\lambda $} \\
  0 &\mbox{sinon}
       \end{array} \right.\,,
\end{equation}
et une efficacité de couplage~:
\begin{equation}
\rho = \left| \iint_{\sqrt{u^2+v^2} \leq d/2\lambda}  \frac{4\sqrt{3} \eta \lambda^2\exp(-2 \I \pi u \varepsilon)}{\pi
  d^2} 
\exp\left(-6 \frac{ (u^2+v^2)\lambda^2}{d^2} \eta^2
  \right)       dudv \right|^2\,,
\end{equation}
que l'on peut aussi écrire, après les changements de variables
$u'=2\lambda u/d$ et $v'=2\lambda v/d$~:
\begin{equation}
\rho = \left| \iint_{\sqrt{u'^2+v'^2} \leq 1}  \frac{ \sqrt{3} \eta
\exp\left(- \I \pi u' \varepsilon d/\lambda\right)}{\pi} 
\exp\left(- 3 \frac{ (u'^2+v'^2)}{2} \eta^2
  \right)       du'dv' \right|^2\,,
\label{eq:conf}
\end{equation}

La figure~\ref{fig:Champ} représente le champ ($\varepsilon$ en unités
de $\lambda/d$) pour différentes valeurs
de $\eta$. Celui-ci est fortement limité par l'injection dans la
fibre. Par exemple, pour un télescope de diamètre $D=7d$, l'élément de
résolution est de $\lambda/7d$. Or, pour $\eta=0,92$, nous avons un
champ total à mi-hauteur de $0,6\lambda/d$. Il n'est alors composé que
de 4 éléments de résolution. C'est pourquoi il est préférable de
choisir $\eta$ aussi grand que possible.  Pour pouvoir décider d'un
compromis entre champ et flux injecté, nous avons représenté ces
valeurs dans la partie de droite de la figure~\ref{fig:Champ}.

\textbf{Accroître $\eta$ permet d'augmenter le champ, mais cela se
  fait au prix d'une perte en sensibilité. Il s'agit alors d'obtenir
  un compromis entre champ et sensibilité}.

\subsection{Le bruit de confusion}
\label{sc:confusion}

\begin{figure}[h]
   \centering
\resizebox{\hsize}{!}{
   \includegraphics{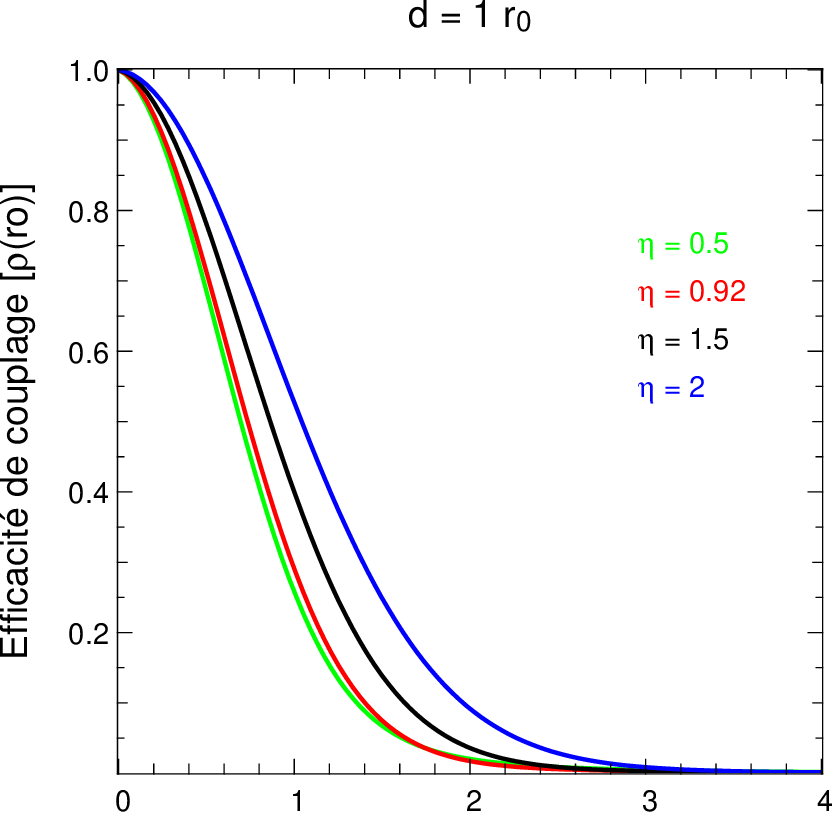}
   \includegraphics{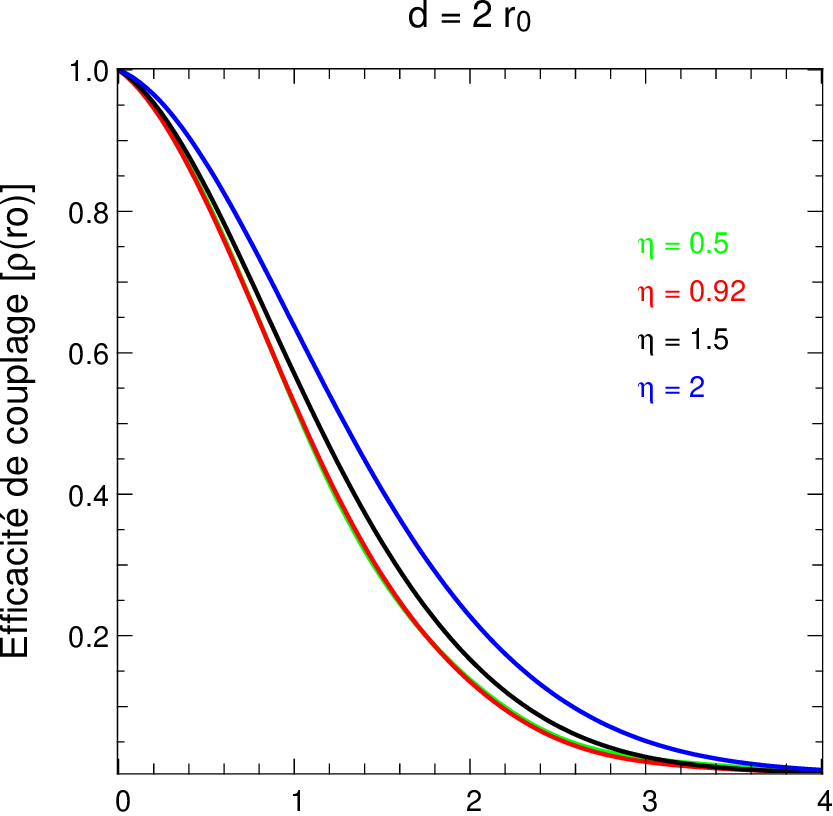}
   \includegraphics{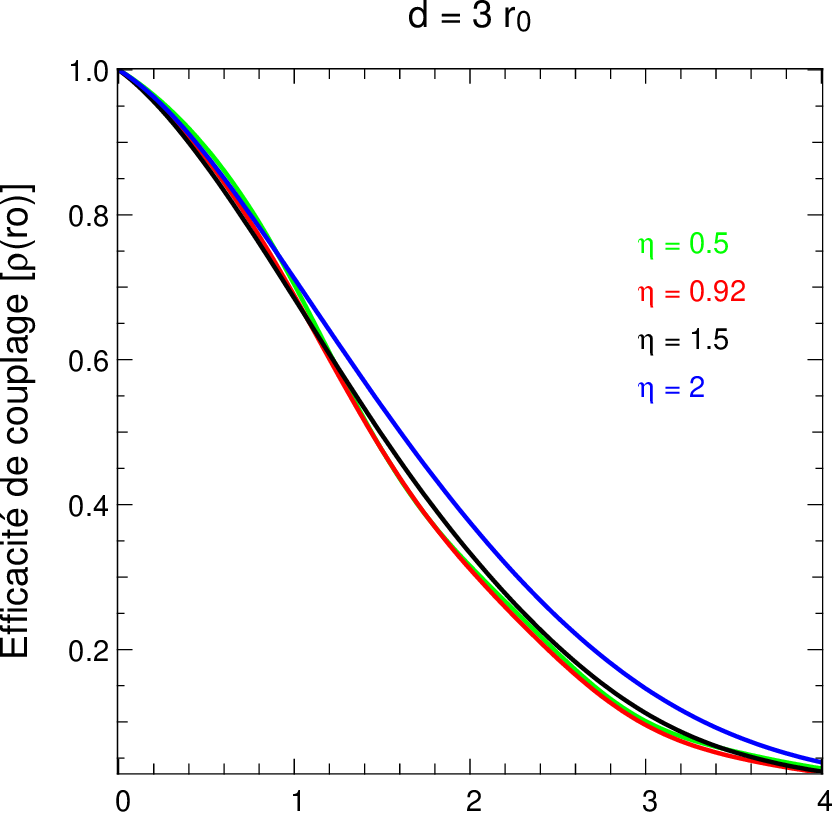}}
\resizebox{\hsize}{!}{
   \includegraphics{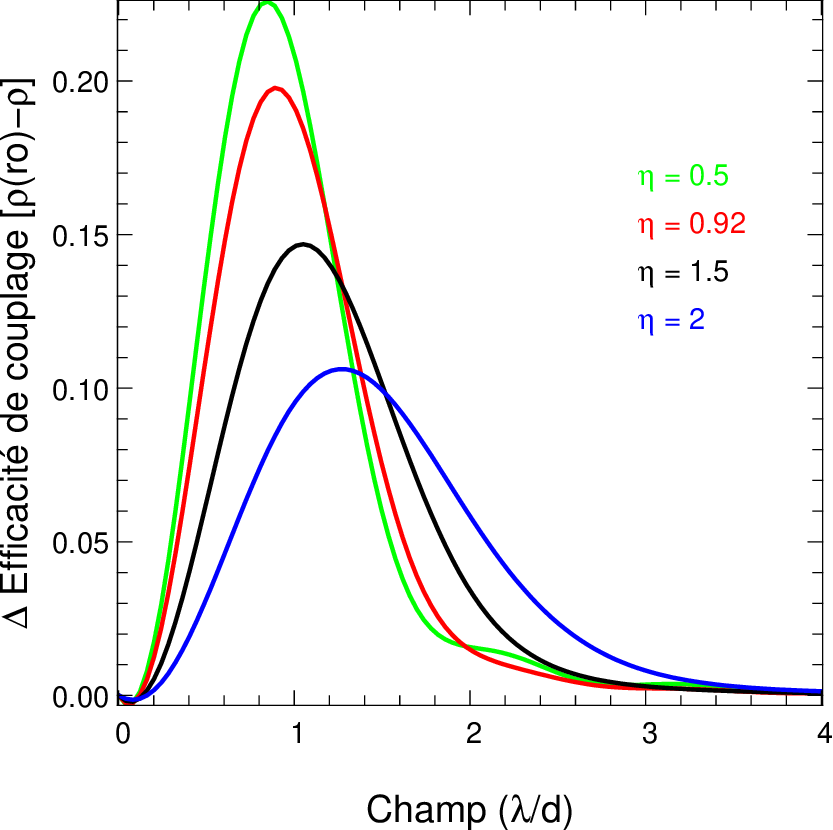}
   \includegraphics{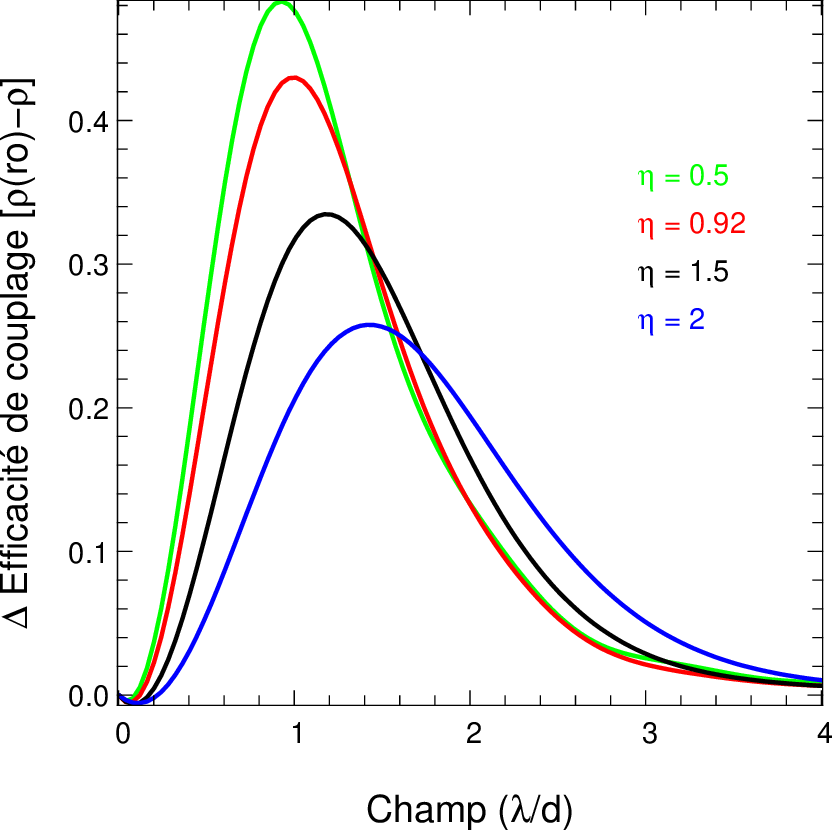}
   \includegraphics{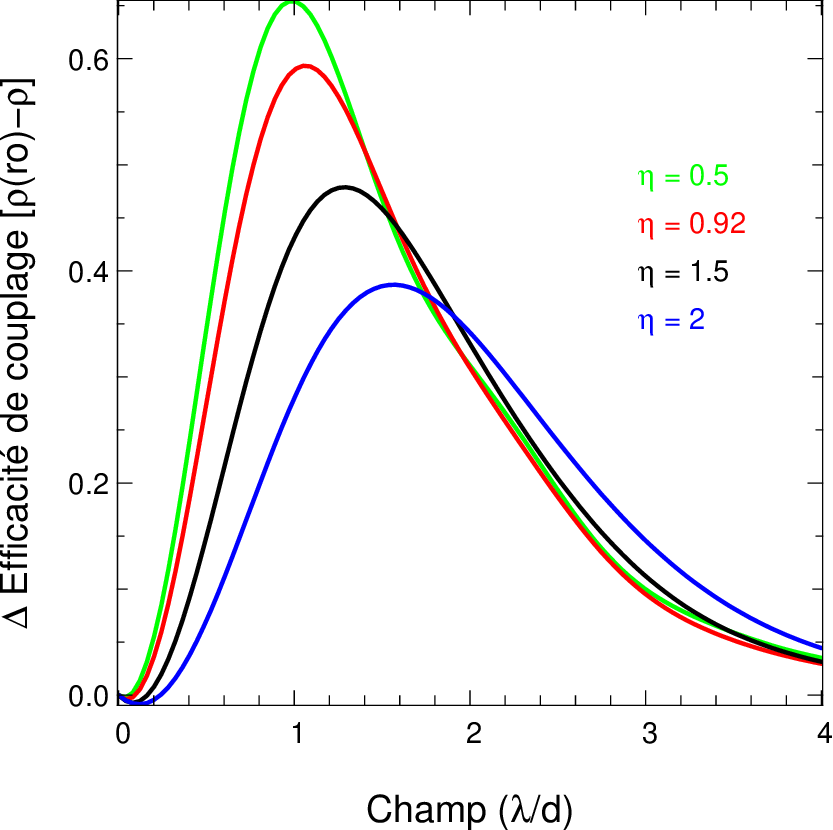}} \caption[Effet du bruit de
   confusion]{ Partie supérieure~: champ de vue moyen de l'instrument
   en présence de turbulences. Trois cas sont étudiés où $d=r_0$,
   $2r_0$ et $3r_0$.  Partie inférieure~: écart d'efficacité de
   couplage entre le champ sans (graphique de gauche de la
   figure~\ref{fig:Champ}) et en présence de turbulences. On peut
   noter une zone critique (entre 1 et 2 $\lambda/d$) où la présence
   d'un objet introduirait un bruit de confusion important.}
   \label{fig:Champs}
\end{figure}

Le bruit de confusion est également un paramètre qui doit entrer en
jeu lors de la caractérisation de l'instrument. Ce bruit correspond à
l'influence de la turbulence sur le champ de vue des fibres. Le tip et
le tilt atmosphériques, notamment, font que les fibres ne
``regardent'' pas toutes nécessairement dans la même direction. La
visibilité alors mesurée peut ainsi être faussée.  \`A ce titre,
l'observation d'un système binaire peut s'avérer problématique si le
flux injecté dans les fibres ne provient pas simultanément des deux
objets, ou si les rapports de flux injectés pour les deux objets
varient d'une fibre à l'autre \citep{2002A&A...387..366G}.

Pour mesurer cet effet, nous avons utilisé l'équation~(\ref{eq:conf}) à
laquelle nous avons ajouté la présence de perturbations
atmosphériques. Nous en avons déduit le champ de vue moyen des fibres
en présence de turbulences. Les résultats sont présentés
figure~\ref{fig:Champs} pour trois tailles de sous-pupille: $r_0$,
$2r_0$ et $3r_0$. Sans surprise, la présence de turbulences augmente le
champ, d'un facteur pouvant aller jusqu'à trois dans le cas
$d=3r_0$. La différence entre le champ d'une fibre sans et avec
turbulence est représentée dans les trois graphiques du bas de la
figure~\ref{fig:Champs}. Ceux-ci révèlent clairement un espace du champ
(entre 1 et 2 $\lambda/d$) où la présence d'un objet introduirait une
source d'erreur dans la mesure des visibilités. Cependant, il n'est
pas aisé d'en déduire un critère pour estimer l'influence de ce flux
sur les visibilités générées par l'algorithme itératif du Chapitre
3. Nous pouvons néanmoins retirer de cette étude que le bruit de
confusion diminue lorsque $\eta$ augmente, ou lorsque le rapport
$d/r_0$ diminue.

\textbf{En conclusion, et en considérant la taille d'une sous-pupille
choisie de façon à maximiser la sensibilité de l'instrument, il est
préférable de choisir un $\eta$ le plus élevé possible afin de
maximiser le champ et de minimiser le bruit de confusion. A cet égard,
augmenter $\eta$ de 10\% par rapport à sa valeur de couplage optimale
ne fait diminuer que marginalement le taux de couplage (voir
tableau~\ref{tb:recap_param}).}


\clearpage
\section{Recombinaison et bande spectrale}
\label{sec:recom_spec}

\subsection{La modulation spatiale en monochromatique}

\begin{figure}[h]
   \centering
\resizebox{\hsize}{!}{
   \includegraphics{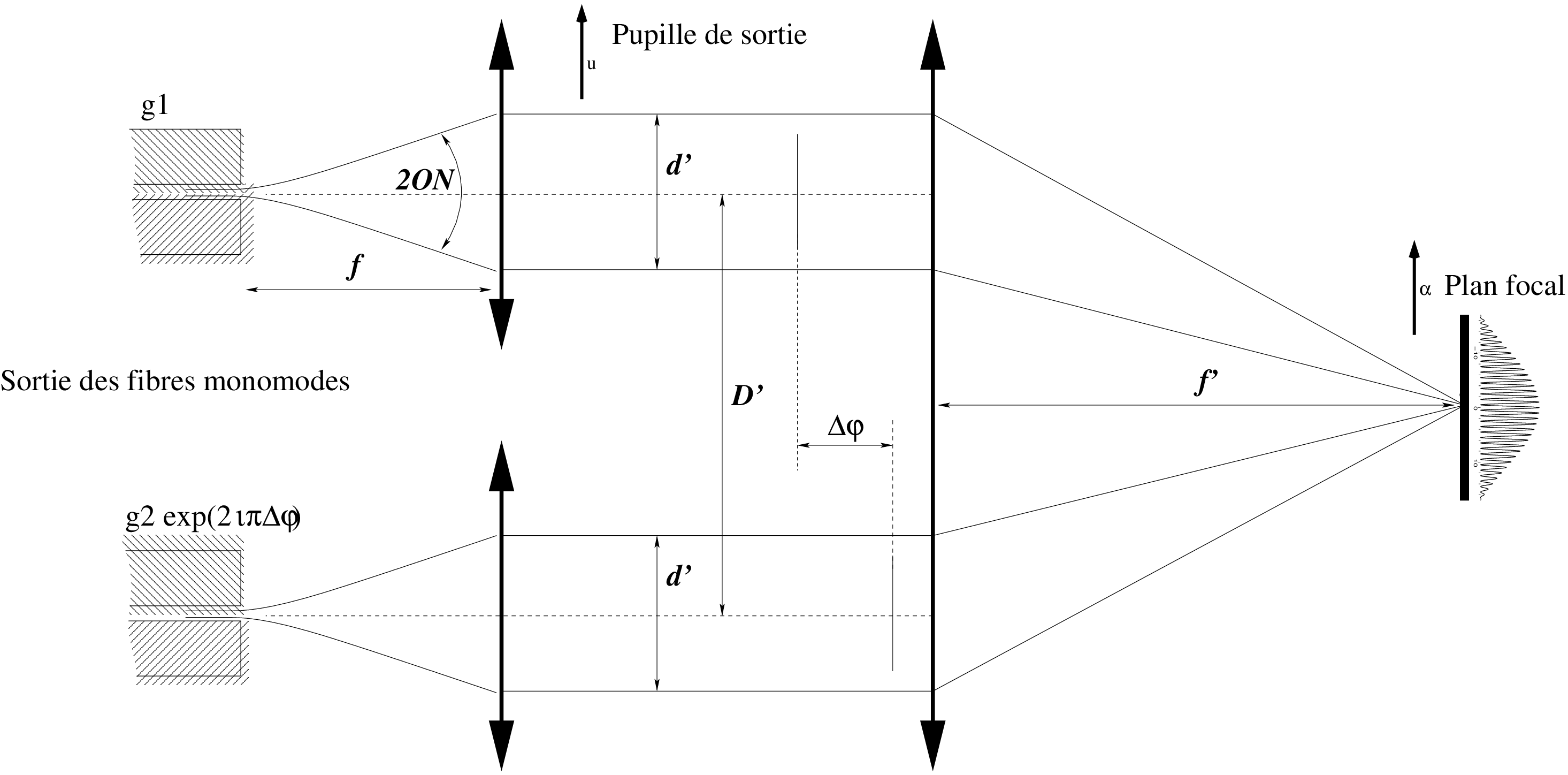}}
   \caption{Schéma d'un système de recombinaison à deux faisceaux}
   \label{fig:schema_pup_s}
\end{figure}

Le champ électrique dans la pupille de sortie est composé du champ
sortant de chacune des fibres optiques. Pour effectuer une analyse détaillée de
l'influence du mode de recombinaison, nous avons travaillé sur un
système simple de recombinaison à deux faisceaux.  Le schéma de la
figure~\ref{fig:schema_pup_s} représente un tel système. Les
paramètres physiques entrant en jeu dans la recombinaison sont~: 
\begin{itemize}
\item $d'$, le diamètre d'un faisceau de sortie défini tel que~:
  $d'=2 f\,ON$
\item $D'$, la distance séparant les deux faisceaux dans le plan
  pupille
\item $f'$ la focale de la lentille de recombinaison. Cette focale
  intervient dans le calcul de la largeur de la tache de
  diffraction. Nous nous sommes affranchis de cette valeur par
  l'utilisation des coordonnées angulaires du plan focal ($\alpha=x/f'$
  et $\beta=y/f'$)
\end{itemize} 
Interviennent également le champ transmis par chacune des fibres
($g_1$ et $g_2$) et leur déphasage ($\Delta \phi$). \`A partir de
l'équation~(\ref{eq:champ_fibre}), nous avons déduit le champ dans la
pupille de recombinaison~:
\begin{equation}
E_{1\circ}(u,v) = g_1
\exp\left(-\frac{6  (u^2+v^2)\lambda^2}{d'^2}\right) * \delta(u-D'/2\lambda)
\end{equation}
pour le premier faisceau, et pour le second~:
\begin{equation}
E_{2\circ}(u,v) = g_2
\exp\left(\I\Delta \phi-\frac{6  (u^2+v^2)\lambda^2}{d'^2}\right) * \delta(u+D'/2\lambda)\,.
\end{equation}
La transformée de Fourier (relation~(\ref{eq:Fraun})) nous donne le champ
dans le plan focal image, soit
\begin{equation}
E_{1\bullet}(\alpha,\beta) = g_1 
\exp\left(-\frac{\pi^2 d'^2(\alpha^2+\beta^2)}{6\lambda^2}\right) \exp(\I\pi
D'/\lambda\alpha )\,,
\end{equation}
et
\begin{equation}
E_{2\bullet}(\alpha,\beta) = g_2 
\exp\left(-\frac{\pi^2 d'^2
  (\alpha^2+\beta^2)}{6\lambda^2}\right) \exp(\I\Delta \phi -  \I\pi
D'/\lambda\alpha)\,,
\label{eq:E2o}
\end{equation}

L'image obtenue sur le détecteur correspond à l'énergie de la somme
des champs électriques des deux fibres. S'ils sont cohérents (source
ponctuelle), ils s'additionnent de manière complexe, s'ils ne le sont
pas, ce sont les modules au carré qui s'additionnent.  Dans la suite
de ce raisonnement, nous supposerons une source ponctuelle, cohérente
($V(u,v) =1$). Ainsi~:
\begin{equation}
I(\alpha,\beta)=\left|E_{1\bullet}(\alpha,\beta)+E_{2\bullet}(\alpha,\beta)\right|^2\,
\end{equation}
nous donne~:
\begin{equation}
I(\alpha,\beta) =  \left|g_1 \exp(\I\pi D'/\lambda\alpha) +g_2\exp(\I\Delta
\phi -\I\pi D'/\lambda\alpha) \right|^2
\exp\left(-\frac{\pi^2 d'^2
  (\alpha^2+\beta^2)}{3\lambda^2}\right) \,.
\end{equation}
Et, en développant, nous arrivons alors à l'expression suivante~:
\begin{equation}
I(\alpha,\beta) = \left(
g_1^2+g_2^2+2g_1g_2\cos(2\pi D'/\lambda\alpha-\Delta\phi) \right)
\exp\left(-\frac{\pi^2 d'^2
  (\alpha^2+\beta^2)}{3\lambda^2}\right)\,.
\label{eq:image}
\end{equation}
Deux termes importants caractérisent l'image sur le détecteur. Le
premier correspond à la modulation dans laquelle réside
l'information sur la cohérence et la phase des franges.  Le deuxième
terme correspond à l'enveloppe de l'image. Il s'agit d'une Gaussienne
qui est l'image des c\oe urs des fibres optiques projetée
sur le détecteur.  Alors que la largeur de cette Gaussienne dépend de
la taille de la sous-pupille de sortie $d'$, la fréquence de
modulation des franges dépend de l'écart entre deux sous-pupilles
$D'$. Pour une fréquence d'échantillonage suffisante ($D'>2d'$), il
est tout à fait possible, au bruit de photons et de détecteur près, de
mesurer avec précision l'amplitude et la phase des franges. Cela change
lorsque l'on considère l'aspect chromatique de la lumière.

\subsection{La modulation spatiale en polychromatique}
\label{sc:inf_spect}

\begin{figure}[h]
   \centering \includegraphics[width=11cm]{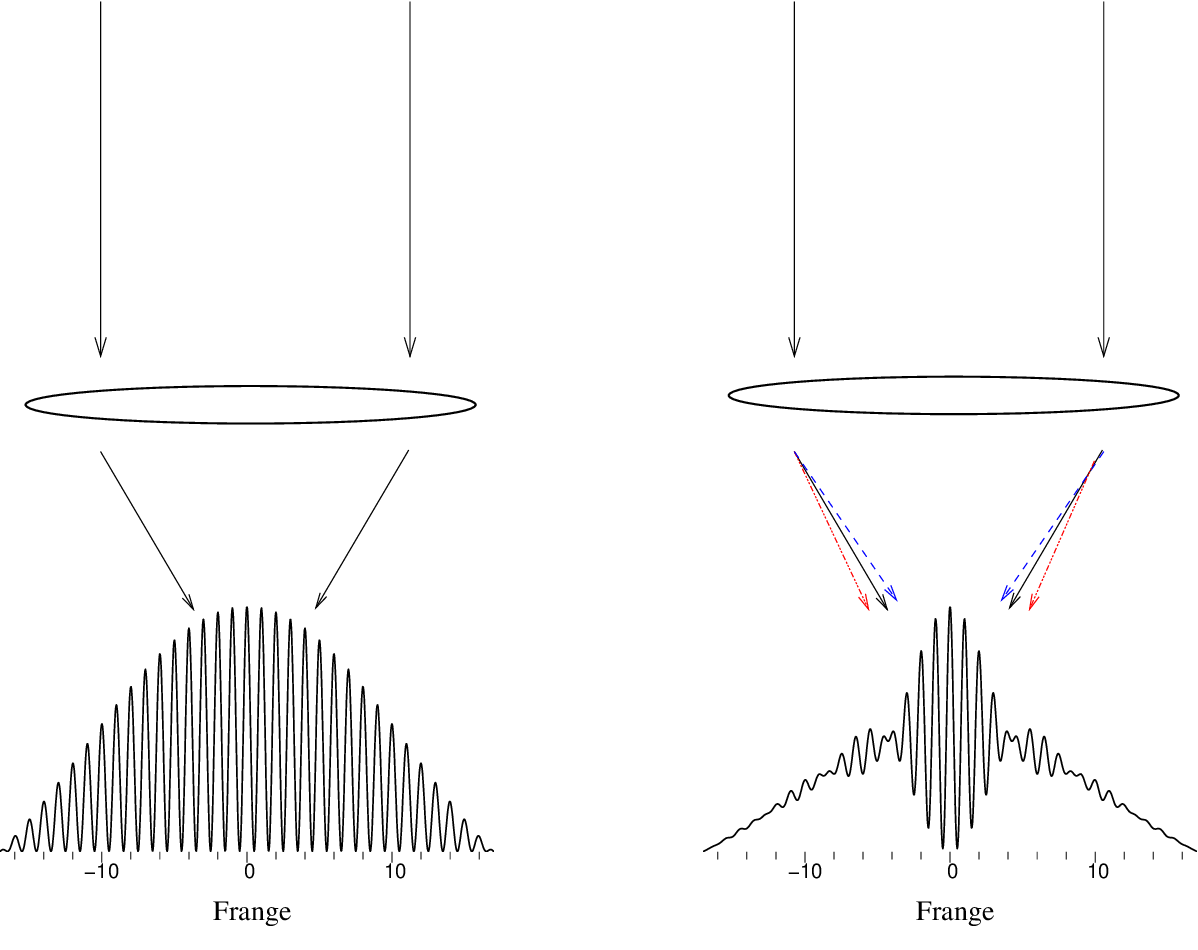}
   \caption[Influence du chromatisme sur les franges]{ Influence du
   chromatisme sur les franges. A gauche~: franges achromatiques. La
   longueur de cohérence est infinie et les franges sont présentes sur
   la totalité de l'image. A droite~: le chromatisme entraîne la
   modulation des franges par une fonction en sinus cardinal.}
   \label{fig:chrom}
\end{figure}

Si l'objet astrophysique observé est achromatique, l'image
obtenue est, quant à elle, chromatique. Elle correspond à la somme des
images aux différentes longueurs d'onde. Pour une bande spectrale
$\Delta\lambda$, on obtient par conséquent~:
\begin{equation}
I(\alpha,\beta) =
\int_{\lambda_0-\Delta\lambda/2}^{\lambda_0+\Delta\lambda/2}
I(\alpha,\beta,\lambda) d\lambda \,,
\end{equation}
où $\lambda_0$ correspond à la longueur d'onde centrale. Cependant,
avant d'entreprendre l'intégration, il est important d'étudier
l'aspect chromatique du déphasage entre les champs des deux fibres
optiques. Plus exactement, puisque le piston, mesuré en différence de
marche, est achromatique, le déphasage est proportionnel à la longueur
d'onde. Nous pouvons alors obtenir un terme de déphasage achromatique
$\Delta \phi_0$ en effectuant le changement de variable suivant~:
\begin{equation}
\Delta\phi_0=\Delta\phi\frac{\lambda}{\lambda_0}.
\end{equation}
Ainsi, le caractère chromatique de l'image observée apparaît entièrement
lorsqu'elle est écrite sous la forme~:
\begin{equation}
I(\alpha,\beta,\lambda) = S(\lambda) \left(
g_1^2+g_2^2+2g_1g_2\cos\left(\frac{ 2\pi D'\alpha -
\lambda_0\Delta\phi_0}{\lambda}\right) \right) \exp\left(-\frac{\pi^2
d'^2 (\alpha^2+\beta^2)}{3\lambda^2}\right)\,.
\label{eq:image_c_o}
\end{equation}
Par souci de simplification, nous utiliserons une écriture en fonction
du nombre d'onde $\nu=1/\lambda$~:
\begin{equation}
I(\alpha,\beta,\nu) = S(\nu) \left(
g_1^2+g_2^2+2g_1g_2\cos\left((2\pi D'\alpha -
\frac{\Delta\phi_0}{\nu_0})\nu\right) \right) \exp\left(-\frac{\pi^2
d'^2 (\alpha^2+\beta^2)\nu^2}{3}\right)\,.
\label{eq:image_c}
\end{equation}

 L'influence de la longueur d'onde porte sur trois termes distincts:
\begin{itemize}
\item L'objet ayant une certaine ``couleur'', celle-ci induit une
  variation spectrale d'intensité $S(\nu)$. Dans le cas d'une étoile
  de température 6000\,K observée aux longueurs d'ondes du visible,
  $S(\nu)$ est à peu près constant. L'hypothèse $S(\nu)=S(\nu_0)$ sera
  utilisée par la suite.
\item L'enveloppe est également modifiée par le chromatisme. La largeur
  de la Gaussienne diminue avec la longueur d'onde. Pour simplifier les
  calculs suivants, nous négligerons l'influence chromatique de
  l'enveloppe et nous l'approximerons par une valeur moyenne:
  $$\exp\left(-\frac{\pi^2 d'^2
  (\alpha^2+\beta^2)\nu_0^2}{3}\right)\,.$$
\item Le terme de modulation sera lui aussi affecté par la
  superposition de franges aux différentes fréquences. Ce phénomène
  crée une perte de cohérence spatiale moyenne (voir
  figure~\ref{fig:chrom}) qui est l'objet du travail suivant.
\end{itemize}
Sur la base de ces hypothèses,  nous pouvons ainsi intégrer l'image sur
le nombre d'onde:
\begin{eqnarray}
\lefteqn{I(\alpha,\beta)
 =} \nonumber\\
&&\displaystyle \int_{\nu_0-\Delta\nu/2}^{\nu_0+\Delta\nu/2}
S(\nu_0) \left(
g_1^2+g_2^2+2g_1g_2\cos\left((2\pi D'\alpha -
\frac{\Delta\phi_0}{\nu_0})\nu\right) \right) \exp\left(-\frac{\pi^2
d'^2 (\alpha^2+\beta^2)\nu_0^2}{3}\right) d\nu =\nonumber\\
&& S(\nu_0)\Delta\nu
\left(g_1^2+g^2+\frac{2g_1g_2}{\Delta\nu}
\int_{\nu_0-\Delta\nu/2}^{\nu_0+\Delta\nu/2}
\cos\left((2\pi D'\alpha -
\frac{\Delta\phi_0}{\nu_0})\nu\right)d\nu
\right)  \exp\left(-\frac{\pi^2
d'^2 (\alpha^2+\beta^2)\nu_0^2}{3}\right)\nonumber\\
\end{eqnarray}
Or~:
\begin{eqnarray}
\lefteqn{\int_{\nu_0-\Delta\nu/2}^{\nu_0+\Delta\nu/2}
\cos\left((2\pi D'\alpha -
\frac{\Delta\phi_0}{\nu_0})\nu\right)d\nu =}\nonumber\\
&&\Delta \nu 
{\rm \ sinc}\left((2\pi D'\alpha\nu_0 -
\Delta\phi_0)\frac{\Delta\nu}{2\nu_0}\right)
\cos\left(2\pi D'\nu_0\alpha -
\Delta\phi_0\right)
\end{eqnarray}
On peut ainsi séparer la fonction image sous la forme de trois termes~:
\begin{eqnarray}
\lefteqn{I(\alpha,\beta)
d\nu =} \nonumber\\
&& \underbrace{S(\nu_0)\Delta\nu
g_1^2  \exp\left(-\frac{\pi^2
d'^2 (\alpha^2+\beta^2)\nu_0^2}{3}\right)}_{\mbox{Contribution de la
première fibre ($=I_1(\alpha,\beta)$)}} +\nonumber\\
&& \underbrace{S(\nu_0)\Delta\nu
g_2^2 \exp\left(-\frac{\pi^2
d'^2 (\alpha^2+\beta^2)\nu_0^2}{3}\right)}_{\mbox{Contribution de la
deuxième fibre ($=I_2(\alpha,\beta)$)}} +\nonumber\\
&&2\underbrace{ S(\nu_0)\Delta\nu g_1g_2{\rm \ sinc}\left((2\pi D'\nu_0\alpha -
\Delta\phi_0)\frac{\Delta\nu}{2\nu_0}\right)
\exp\left(-\frac{\pi^2
d'^2 (\alpha^2+\beta^2)\nu_0^2}{3}\right)}_{\mbox{Amplitude de
modulation ($=A_{(1,2)}(\alpha,\beta)$)}}
\cos\left(2\pi D'\nu_0\alpha -
\Delta\phi_0\right) \nonumber\\
\label{eq:Iab_dec}
\end{eqnarray}
Les deux premiers termes correspondent au flux obtenu par chaque
 fibre indépendemment. Le troisième terme correspond au terme de
modulation généré par la cohérence de l'onde lumineuse. Ainsi~:
\begin{equation}
I(\alpha,\beta)=I_1(\alpha,\beta)+I_2(\alpha,\beta)+2A_{(1,2)}(\alpha,\beta)\cos\left((2\pi D'\alpha -
\lambda_0\Delta\phi_0)\nu_0\right)\,.
\end{equation}

\subsection{L'influence de la bande passante sur le facteur de cohérence}
\label{sec:fact_cohe}

A partir de la définition du facteur de cohérence:
\begin{equation}
\mu=\frac{\displaystyle\iint_{-\infty}^{+\infty}
A_{(1,2)}(\alpha,\beta)d\alpha d\beta}{\displaystyle\sqrt{\iint_{-\infty}^{+\infty}
I_1(\alpha,\beta)d\alpha d\beta \cdot \iint_{-\infty}^{+\infty}
I_2(\alpha,\beta)d\alpha d\beta }}\,,
\end{equation}
on peut établir~:
\begin{eqnarray}
\mu&=&\frac{\displaystyle\iint_{-\infty}^{+\infty}g_1g_2{\rm \ sinc}\left((2\pi D'\nu_0\alpha -
\Delta\phi_0)\frac{\Delta\nu}{2\nu_0}\right)
\exp\left(-\frac{\pi^2
d'^2 (\alpha^2+\beta^2)\nu_0^2}{3}\right)  d\alpha d\beta}{ \displaystyle\sqrt{
\iint_{-\infty}^{+\infty} g_1^2  \exp\left(-\frac{\pi^2
d'^2 (\alpha^2+\beta^2)\nu_0^2}{3}\right)d\alpha d\beta
\cdot
\iint_{-\infty}^{+\infty} g_2^2  \exp\left(-\frac{\pi^2
d'^2 (\alpha^2+\beta^2)\nu_0^2}{3}\right)d\alpha d\beta
}}\nonumber\\
&=&\frac{\displaystyle\iint_{-\infty}^{+\infty} {\rm \ sinc}\left((2\pi D'\nu_0\alpha -
\Delta\phi_0)\frac{\Delta\nu}{2\nu_0}\right)
\exp\left(-\frac{\pi^2
d'^2 (\alpha^2+\beta^2)\nu_0^2}{3}\right)  d\alpha d\beta}{\displaystyle
\iint_{-\infty}^{+\infty}   \exp\left(-\frac{\pi^2
d'^2 (\alpha^2+\beta^2)\nu_0^2}{3}\right)d\alpha d\beta}\nonumber\\
&=&\frac{\displaystyle\int_{-\infty}^{+\infty} {\rm \ sinc}\left((2\pi D'\nu_0\alpha -
\Delta\phi_0)\frac{\Delta\nu}{2\nu_0}\right)
\exp\left(-\frac{\pi^2
d'^2 \alpha^2\nu_0^2}{3}\right)  d\alpha}{\displaystyle
\int_{-\infty}^{+\infty}   \exp\left(-\frac{\pi^2
d'^2 \alpha^2\nu_0^2}{3}\right)d\alpha}\nonumber\\
&=&\frac{\sqrt{\pi}d'\nu_0}{\sqrt{3}}\displaystyle\int_{-\infty}^{+\infty} {\rm \ sinc}\left((2\pi D'\nu_0\alpha -
\Delta\phi_0)\frac{\Delta\nu}{2\nu_0}\right)
\exp\left(-\frac{\pi^2
d'^2 \alpha^2\nu_0^2}{3}\right)  d\alpha
\,.
\end{eqnarray}
Enfin, en effectuant le changement de variable $\gamma=d'\alpha\nu_0$
en peut encore simplifier l'écriture de $\mu$~:
\begin{equation}
\mu=\sqrt{\frac{\pi}{3}}\displaystyle\int_{-\infty}^{+\infty} {\rm \ sinc}\left((2\pi \frac{D'}{d'}\gamma -
\Delta\phi_0)\frac{\Delta\nu}{2\nu_0}\right)
\exp\left(-\frac{\pi^2
\gamma^2}{3}\right)  d\gamma
\,.
\label{eq:coherance}
\end{equation}

\begin{figure}
   \centering
\resizebox{\hsize}{!}{
   \includegraphics{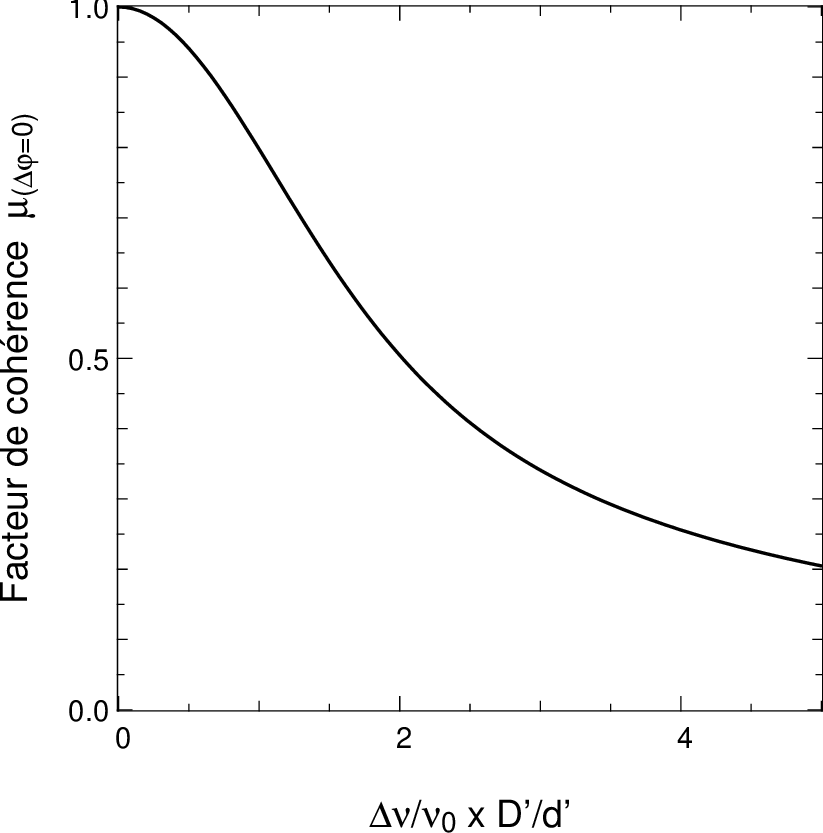}
   \includegraphics{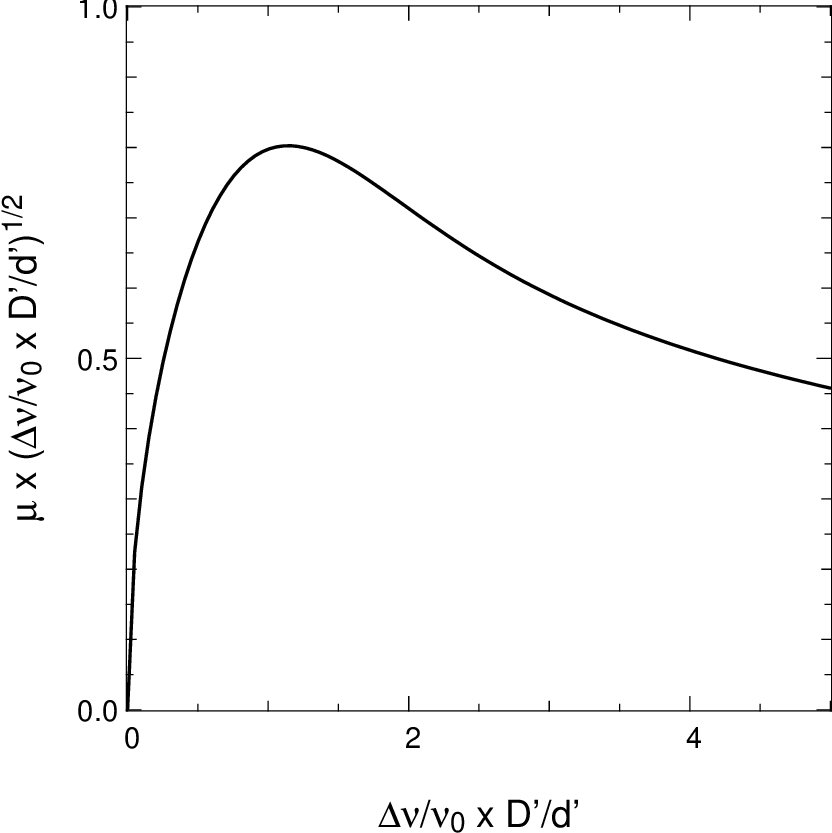}} \caption[Représentation
   du facteur de cohérence]{ A gauche~: facteur de cohérence $\mu$ en
   fonction du rapport $\frac{D'\Delta\nu}{d' \nu_0}$. Le facteur de
   cohérence est supérieur à 0,5 pour une valeur de
   $\frac{D'\Delta\nu}{d' \nu_0}$ inférieure à 2. A droite~: le
   facteur de cohérence est multiplié par un terme proportionnel à la
   racine carré de la bande passante. Pour une configuration donnée,
   ce graphique permet de choisir la bande passante optimisant le
   rapport signal sur bruit de photons.}  \label{fig:coher}
\end{figure}

Le facteur de cohérence dépend de trois termes~: la bande
spectrale $\frac{\Delta\nu}{\nu_0}$, le taux de dilution de la pupille
de sortie $\frac{D'}{d'}$ et le déphasage $\Delta\phi_0$. Les deux
premiers paramètres sont à déterminer lors de la conception de
l'instrument. Pour disposer d'une sensibilité optimale de
l'instrument, il est important d'avoir une cohérence $\mu$
maximale. Le choix du compromis entre bande passante et diamètre de
pupille de sortie peut être fait dans l'hypothèse d'un déphasage nul
($\Delta \phi_0 =0$). Le facteur de cohérence s'écrit alors~:
\begin{equation}
\mu_{(\Delta\phi_0=0)}=\sqrt{\frac{\pi}{3}}\displaystyle\int_{-\infty}^{+\infty} {\rm \
sinc}\left( \frac{\pi D'\Delta\nu}{d'\nu_0}\gamma \right)
\exp\left(-\frac{\pi^2
\gamma^2}{3}\right)  d\gamma
\,.
\end{equation}
La figure~\ref{fig:coher} représente $\mu_{(\Delta\phi_0=0)}$ en
fonction du rapport $\frac{D'\Delta\nu}{d' \nu_0}$. Nous pouvons voir
que plus cette valeur est grande, plus la perte d'efficacité
interférométrique est importante. Le problème est d'aboutir à un
compromis entre efficacité interférométrique et largeur de bande
spectrale. Le rapport signal sur bruit optimum peut être un critère de
choix. Dans le plan $u$-$v$, le signal du pic frange est proportionnel
à $\mu_{(\Delta\phi_0=0)} N_{\rm photons}$ alors que le bruit de
photons est proportionnel à $\sqrt{N_{\rm photons}}$. En considérant
que $N_{\rm photons}$ est proportionnel à la bande spectrale, le
signal sur bruit sera proportionnel à $\mu_{(\Delta\phi_0=0)}
\sqrt{\Delta\nu}$. La courbe correspondante est présentée dans la
partie droite de la figure~\ref{fig:coher}. Elle permet d'obtenir une
valeur optimale de la bande passante pour un rapport $d'/D'$ déterminé
:
\begin{equation}
\frac{\Delta\nu}{\nu_0} =  1,15 \frac{d'}{D'}
\label{eq:delta_lam2}
\end{equation}
Cependant, il faut souligner que, dans un système à plusieurs fibres,
la distance $D'$ séparant deux sous-pupilles dépend de la paire de
fibres utilisée, et est donc variable.

\textbf{Si l'on considère l'instrument dans son ensemble, on peut
  établir une relation entre la largeur de la bande passante et le
  diamètre de la pupille de sortie $\max(D'/d')$. A partir des
  configurations non-redondantes que nous avons générées
  section~\ref{sec:nr_rearr}, nous avons estimé que $<D'/d'> \approx
  1/2 \max(D'/d')$, ce qui permet de déterminer une bande passante
  optimale pour~:
\begin{equation}
\frac{\Delta\nu}{\nu_0} \approx 2 \frac{d'}{\max(D')}
\label{eq:delta_lambda}
\end{equation}
Il faut noter que la bande spectrale élargit les pics franges dans la
densité spectrale de puissance de l'image. Notamment, lorsque
$\frac{\Delta\nu}{\nu_0} > \frac{d'}{D'}$ il existe un risque
de confusion des fréquences spatiales. Ceci doit être pris en compte
lors du calcul de la configuration non-redondante (voir
section~\ref{sc:remap_chrom}).  }

\subsection{Le champ interférométrique}
\label{sc:Champ_interfero}

\begin{figure}
   \centering
\resizebox{\hsize}{!}{
   \includegraphics{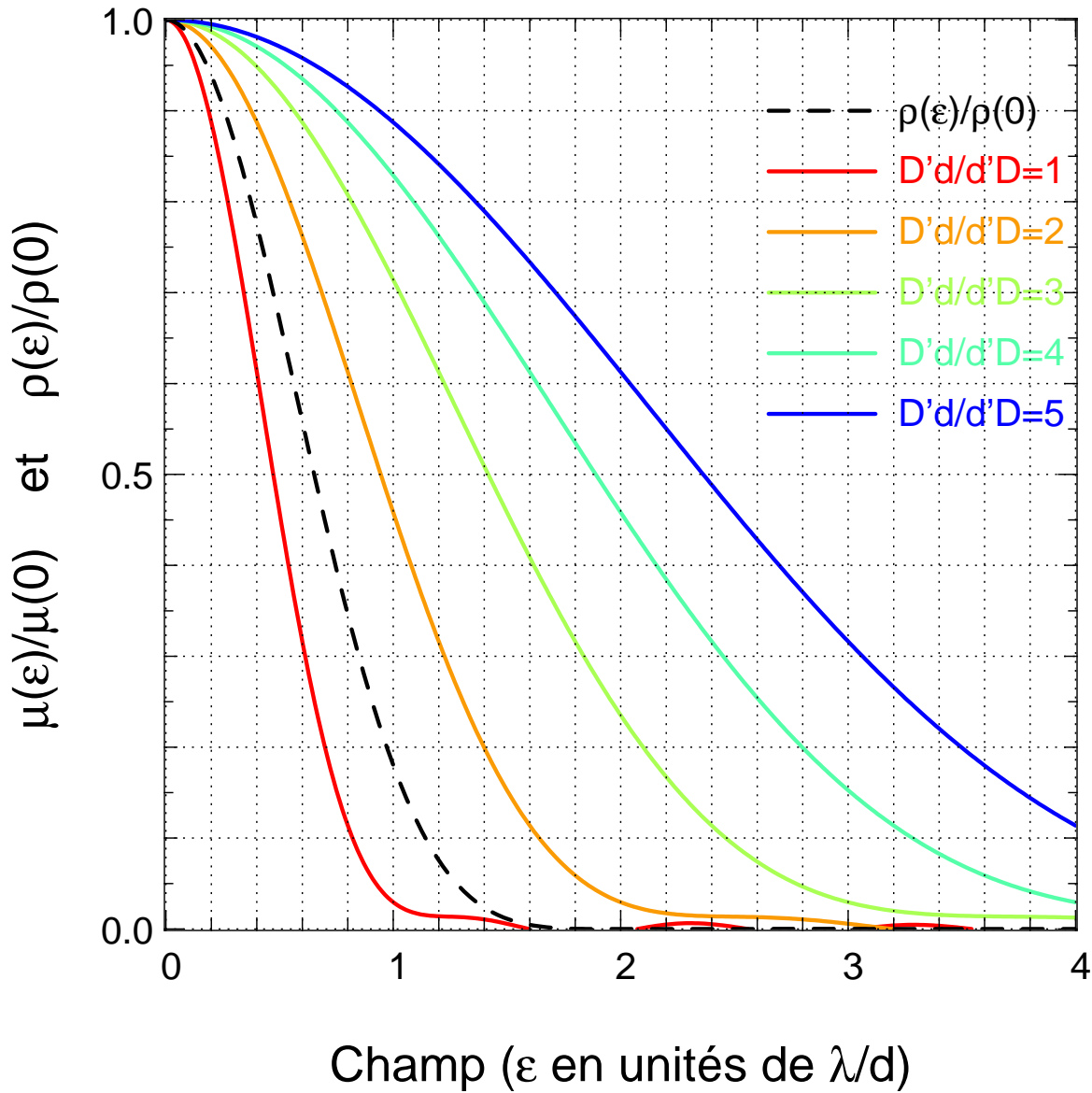}} \caption[Champ de l'interféromètre]{Les courbes de
   couleur représentent le facteur de cohérence normalisé pour
   différentes valeurs de $D'd/Dd'$. Le facteur de cohérence permet
   ainsi de définir le champ de l'interférométre. La bande passante
   utilisée pour tracer ce graphique est $\frac{\Delta\nu}{\nu_0} = 2
   \frac{d'}{\max(D')}$.  La courbe en pointillés représente le champ
   de la fibre en entrée du télescope, pour $\eta=1$,25. A partir de
   ce graphique, nous avons déduit qu'en terme d'efficacité
   interférométrique, il faut chercher à maximiser le rapport
   $D'd/Dd'$ pour l'ensemble des fibres. De plus, pour que le champ
   soit limité par la fibre, il faut que ${D'}/{d'} \geq 2 {D}/{d}$. }
   \label{fig:champ_int}
\end{figure}

Le champ interférométrique est communément délimité par la position où
un objet hors-axe ponctuel aurait un contraste moyen de ses franges de
50\%. Dans le cas d'un système binaire par exemple, un compagnon se
trouvant à la limite du champ interférométrique ne contribuera aux
franges observées qu'à hauteur de 50\% de son flux. Sa contribution
diminue lorsque le compagnon s'éloigne encore plus du champ de
l'interférométre.

Ce champ peut aussi être calculé à partir du facteur de cohérence,
tel que nous l'avons établi dans la section précédente.  Lorsque
l'objet est hors axe, les champs sont déphasés d'une valeur qui dépend
du diamètre de la pupille d'entrée $D$ et de l'angle d'inclinaison
$\varepsilon$. Le déphasage introduit s'écrit~:
\begin{equation}
\Delta\phi_0=2\pi\varepsilon D/\lambda \,.
\label{eq:phi_alpha}
\end{equation} 
Notons que, pour se rapprocher de la définition du champ représenté
dans le cas d'une fibre par la figure~\ref{fig:champ_fibre}, nous
avons réintroduit ici sun terme de longueur d'onde
$\lambda=1/\nu$. Par ailleurs, en utilisant
l'équation~(\ref{eq:coherance}) et~(\ref{eq:phi_alpha}), nous obtenons
l'écriture du facteur de cohérence en fonction de l'inclinaison
$\varepsilon$~:
\begin{equation}
\mu(\varepsilon) = \sqrt{\frac{\pi}{3}}\displaystyle\int_{-\infty}^{+\infty} {\rm \ sinc}\left(\pi \frac{D'\Delta\nu}{d'\nu_0}\left(\gamma -
\frac{Dd'}{dD'}\cdot\frac{\varepsilon d}{\lambda}\right)\right)
\exp\left(-\frac{\pi^2
\gamma^2}{3}\right)  d\gamma
\,.
\end{equation} 
Cette expression met en exergue l'importance du paramètre
${Dd'}/{dD'}$ sur le champ inteférométrique. Ce facteur joue le role
de bras de levier sur lequel il va falloir jouer lors du réarrangement
de la pupille.

Pour traduire l'effet de la perte en efficacité interférométrique due
à l'inclinaison, nous avons représenté figure~\ref{fig:champ_int} le
facteur de cohérence normalisé $\mu(\varepsilon)/\mu(0)$ en fonction
du terme $\varepsilon$, en unités de $\lambda/d$. Pour tracer ce
graphique, nous avons choisi d'utiliser une valeur conservatrice de la
bande passante, de manière à se situer dans la situation la plus
défavorable. La valeur maximale de bande passante que l'observateur
serait ammené à utiliser a été établie section~\ref{sec:fact_cohe} par
$\frac{\Delta\nu}{\nu_0} = 2 \frac{d'}{\max(D')}$. Face au champ de
l'interférométre, nous avons affiché le champ d'une fibre tel
qu'établit par la relation~(\ref{eq:conf}) dans la
section~\ref{sc:champ_fibre}, pour $\eta=1$,25. Ce champ est
représenté en pointillés sur la figure~\ref{fig:champ_int}. On voit
nettement dans cette figure l'influence cruciale du choix des rapports
${Dd'}/{dD'}$.

\textbf{Pour que le
champ de l'interféromètre soit déterminé par le champ de la fibre, il
faut que le réarrangement de toutes les paires de fibres vérifie la relation~:
\begin{equation}
\frac{D'}{d'} \geq  2 \frac{D}{d}\,.
\end{equation}
Pour une configuration donnée, le choix de la position d'une fibre dans
la pupille de sortie en fonction de sa position dans la pupille
d'entrée devra donc être effectué de façon à respecter cette
contrainte. De manière générale, pour maximiser le facteur de
cohérence d'une source hors axe, le réarrangement doit chercher à
maximiser les différentes valeurs de ${Dd'}/{dD'}$.}

\subsection{Le bruit du piston atmosphérique}
\label{sc:piston}

\begin{figure}
   \centering
\resizebox{\hsize}{!}{
   \includegraphics{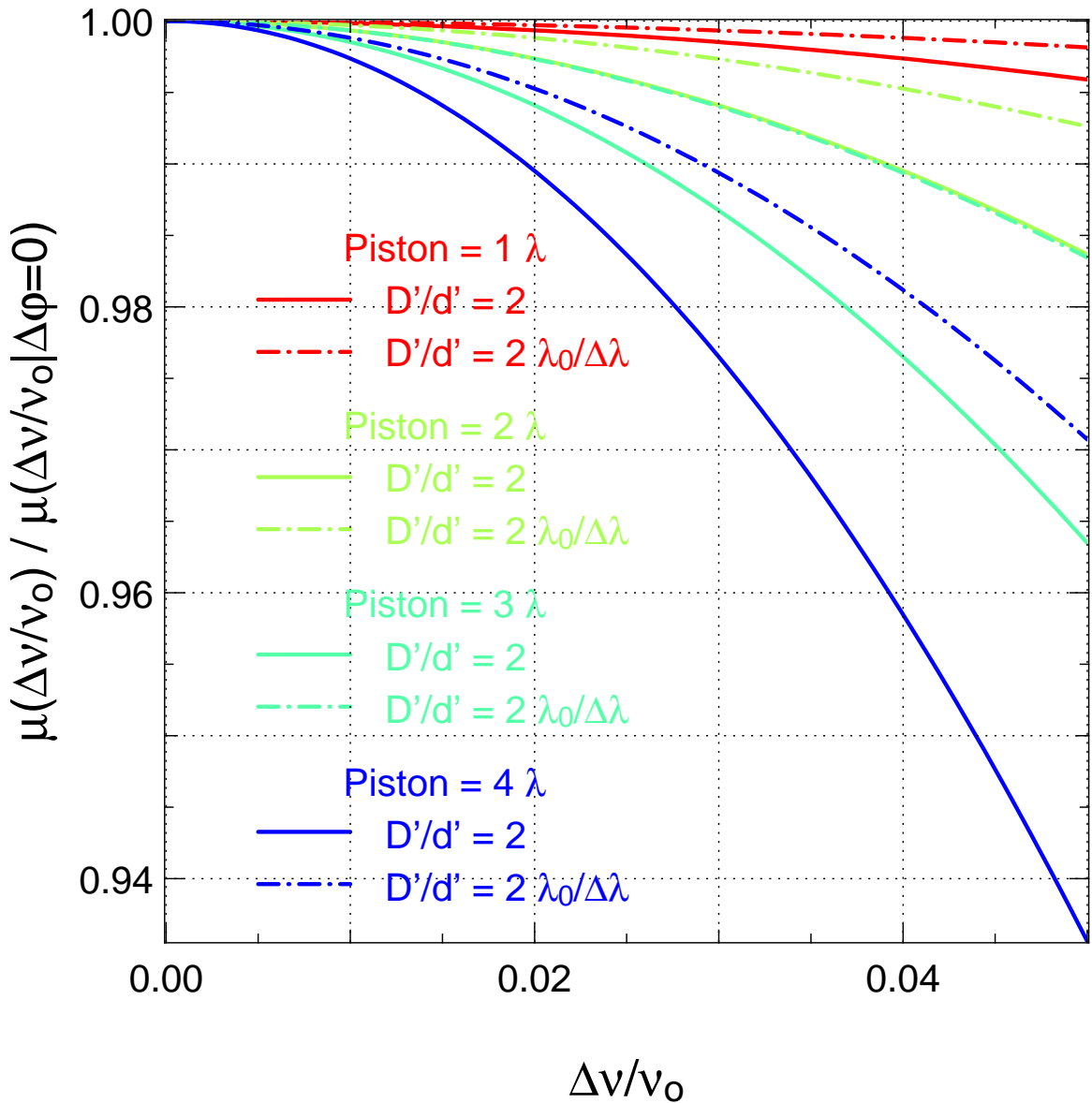}} \caption[Biais du piston
   atmosphérique]{ Effet de la bande passante et du piston
   atmosphérique sur l'efficacité interférométrique. Les deux
   situations limites envisagées sont $D'/d = 2
   \lambda_0/\Delta\lambda$ et $D'/d' =2$. Cette perte en efficacité
   interférométrique introduit un bruit sur la mesure des visibilités
   normalisées.  } \label{fig:piston}
\end{figure}

Lorsque le rayonnement de la source passe à travers l'atmosphère, il
traverse un milieu inhomogéne, d'indice variable. Les variations
de l'indice des couches atmosphériques produisent ce que l'on appelle un
piston différentiel. Concrètement, ce piston se traduit par un
déphasage entre les différentes sous-pupilles, qui produit un effet
identique a ce que l'on a vu pour une source hors axe. 

C'est pourquoi l'effet du piston atmosphérique peut être, de la même
façon, déduit du facteur de cohérence.  Puisque la différence de
marche du piston atmosphérique est achromatique, le déphasage ($\Delta
\phi_0$) est, lui, très chromatique. Par exemple, pour un piston de 3
$\mu$m le déphasage donne $\Delta\phi_0 \approx 4\pi$ en bande H et
$\Delta\phi_0 \approx 9\pi$ dans le visible. \`A partir de
l'équation~(\ref{eq:coherance}), nous avons pu déduire la perte
d'efficacité interférométrique en fonction de la bande passante
utilisée ${\Delta\nu}/{\nu_0}$~:
\begin{equation}
\mu(\frac{\Delta\nu}{\nu_0})=\sqrt{\frac{\pi}{3}}\displaystyle\int_{-\infty}^{+\infty} {\rm \ sinc}\left((2\pi \frac{D'}{d'}\gamma -
\Delta\phi_0)\frac{\Delta\nu}{2\nu_0}\right)
\exp\left(-\frac{\pi^2
\gamma^2}{3}\right)  d\gamma
\,.
\end{equation}
Interviennent aussi dans cette équation le piston $\Delta
\phi_0$ et de la séparation entre les
deux sous-pupilles $D'/d'$ 

La figure~\ref{fig:piston} présente le rapport
$\mu(\frac{\Delta\nu}{\nu_0})/\mu(\frac{\Delta\nu}{\nu_0})_{(\Delta\phi_0=0)}$ en fonction de la bande passante pour
différentes amplitudes de piston. Les courbes pleines et en pointillés
correspondent à deux cas~: $D'/d' = 2$ et $D'/d' = 2
\lambda_0/\Delta\lambda$. Ces deux situations permettent de définir
les conditions limites correspondant aux longueurs de base maximales
et minimales. Nos résultats montrent que plus la bande spectrale est
grande, plus l'influence du piston atmosphérique sur les visibilités
est importante.  Ainsi, pour un piston de 4 fois la longueur d'onde et
une bande spectrale de 0,02, la perte de visibilité sera de 1\%.
Cette perte est multipliée par quatre pour une bande spectrale deux
fois plus faible. La raison de cette dépendance par rapport à la bande
passante peut s'expliquer en terme de longueur de cohérence. Lorsque
celle-ci est plus courte, l'interférogramme est alors fortement
atténué par la figure de diffraction. Lorsque la longueur de cohérence
est plus grande, l'atténuation de la figure de diffraction est plus
faible

Il est important de noter qu'il s'agit d'un biais statistique, pouvant
peut être mitigé par plusieurs observations. Cependant, il ne se
moyenne pas à zero, et le caractère non stationnaire de la turbulence
rend l'étalonnage très difficile. C'est pourquoi il ne faut pas
s'attendre à une diminution importante de l'erreur estimée par la
figure~\ref{fig:piston}.


\clearpage
\section{Le choix du réarrangement}
\label{sec:nr_rearr}

\subsection{Les contraintes du réarrangement}

Le réarrangement de pupille est une technique de codage de
l'information. Pour un processus d'imagerie classique, l'information
sur la turbulence et l'objet astrophysique est mélangée. Pour
pouvoir distinguer les perturbations instrumentales de l'information
astrophysique, il est nécessaire de coder l'information manquante à
des fréquences différentes, qui, dans la pratique, doivent être
supérieures.

La qualité de ce codage dépend du choix la configuration de la pupille
de sortie. Pour avoir une sensibilité maximale, il faut~:
\begin{itemize}
\item utiliser un minimum de pixels, soit une plage minimale de
fréquences spatiales (lorsque l'on est limité par le bruit du
détecteur).
\item une bande spectrale la plus large possible, et par conséquent
  minimiser les rapports $D'/d'$ (équation~(\ref{eq:delta_lambda})).
\end{itemize}
Or, le théorème de Shannon fixe le nombre de pixels nécessaire à un
bon échantillonage du plan fréquentiel par~: $N_{\rm pixels} = 2
\max(D')/d'$. Ainsi, les deux conditions précédentes conduisent à
minimiser le rapport $\max(D')/d'$. Il faut déterminer la
configuration la plus compacte possible, tout en fournissant le moyen
de dissocier chacune des fréquences spatiales présentes. Pour cela, il
faut que chaque fréquence spatiale soit distincte (configuration
non-redondante) et séparée d'au moins 1 pixel$^{-1}$ des autres
fréquences présentes. Ces conditions reviennent à disposer les
sous-pupilles sur une grille de maillage de 1 pixel$^{-1}$. Cette
grille doit être rectangulaire si les pixels le sont, et carré pour
des pixels carrés. Parce que les pupilles ne peuvent se superposer, on
obtient de plus un maillage optimum lorsque 1 pixel$^{-1}$ correspond
à la taille $d'$ d'une sous-pupille. A partir de ces conditions,
obtenir la configuration non-redondante la plus compacte possible
nécessite le développement d'un algorithme spécifique.

Il faut également choisir entre une configuration non-redondante à une
dimension ou deux dimensions. Ce choix dépend de l'objectif
scientifique. Si l'on souhaite obtenir une
information spectrale de l'objet observé, il faut disperser les
franges, et pour cela disposer les fibres sur une seule
dimension. Dans le cas contraire, utiliser une configuration à
deux dimensions permet un arrangement plus compact.

\subsection{L'algorithme}

\begin{figure}[h]
   \centering
\resizebox{\hsize}{!}{
   \includegraphics{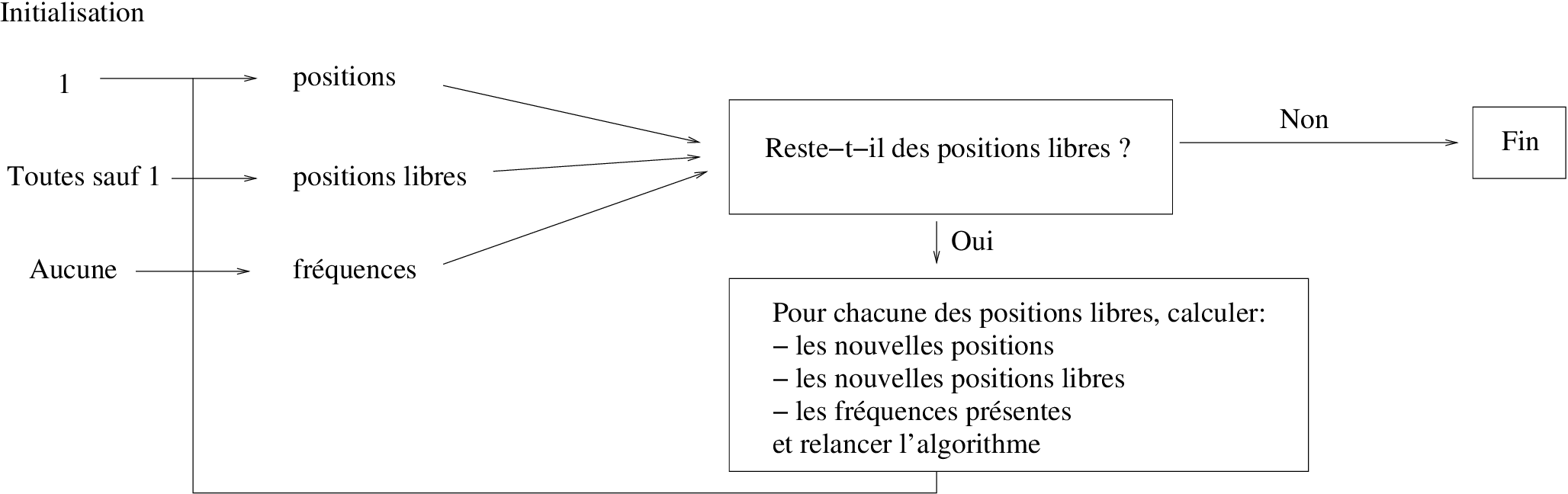}} \caption[Schéma de
   l'algorithme utilisé pour déterminer les configurations
   non-redondantes optimales]{Schéma de l'algorithme utilisé pour
   déterminer les configurations non-redondantes optimales. Elles sont
   toutes testées par l'utilisation d'une technique récursive.}
   \label{fig:algo}
\end{figure}

Si trouver une configuration non-redondante est quelconque est aisé~\citep[par
exemple il existe des solutions proposées par][]{1970..Golay}, il est
autrement plus difficile de déterminer la configuration qui minimise la
valeur $\max(D')/d'$. Par exemple, un algorithme simple donnant une
configuration non-redondante fixe la position des
fibres de la façon suivante~:
\begin{equation}
X_{(n)}=2^n d' \,.
\label{eq:algo_base}
\end{equation}
Un tel algorithme implique $\max(D')/d'=2^n$, une valeur loin d'être
optimale. Pour trouver les configurations optimales, nous avons
développé un algorithme de recherche de configurations
non-redondantes.  Le problème que nous avons tenté de résoudre est le
suivant~: pour un rapport maximal $D'/d'$ donné, quel est le nombre
maximum de sous-pupilles placées sur une maille d'unités $d'$ qui
satisfont la propriété de non-redondance?

Cet algorithme a été programmé dans le langage Yorick. Il est schématisé
figure~\ref{fig:algo}.  Dans le cadre
d'une configuration redondante à une dimension, il est le suivant~: \\
\texttt{
func Config(positions,positions\_libres,frequences,dimension,N\_but) \\
\{ \\
  N\_pos=dimsof(positions)(2); \\
  if (N\_pos >= N\_but) \\
    write,"N\_pos = "+pr1(N\_pos)+" ---
    Positions = "+pr1(positions);  \\
  for (i=positions(0)+1;i<=dimension;i++) \\
    if (positions\_libres(i))\\
      \{ \\
        frequences2=grow(frequences,i-positions); \\
        positions\_libres2=positions\_libres; \\
        positions\_libres2(frequences2*(frequences2 <=
	dimension-i)+i)=0; \\
        positions2=grow(positions,i); \\
        Config,positions2,positions\_libres2,frequences2,dimension,N\_but;\\
      \} \\
\} \\
}

Un des paramètres de la fonction est la variable
\texttt{N\_but}. Elle permet de préciser le nombre de
sous-pupilles que l'on cherche à disposer dans la grille. La fonction
se lance de la façon suivante~: \\ \texttt{ dimension=18; \\
positions=[1]; \\ positions\_libres=array(short(1),dimension); \\
N\_but=6; \\
Config,positions,positions\_libres,frequences,dimension,N\_but; \\ }
On obtient au bout de quelques secondes~: \\ \texttt{ N\_pos = 6 ---
Positions = [1,2,5,11,13,18] \\ N\_pos = 6 --- Positions =
[1,2,5,11,16,18] \\ N\_pos = 6 --- Positions = [1,2,9,12,14,18] \\
N\_pos = 6 --- Positions = [1,2,9,13,15,18] \\ N\_pos = 6 ---
Positions = [1,3,8,14,17,18] \\ N\_pos = 6 --- Positions =
[1,4,6,10,17,18] \\ N\_pos = 6 --- Positions = [1,5,7,10,17,18] \\
N\_pos = 6 --- Positions = [1,6,8,14,17,18] \\ }

Cet algorithme, récursif, a la propriété de calculer toutes les
possibilités de configurations non-redondantes. Il n'affiche un
résultat que si le nombre de fibres est supérieur à
\texttt{N\_but}. Par conséquent, le temps de réponse devient très long
lorsque le nombre de fibres dépasse $\approx 15$, mais la valeur
optimale au rapport $D'/d'$ est toujours fournie.  Dans le cadre d'une
configuration de la pupille à 2 dimensions nous nous sommes également
servis de cet algorithme. Pour cela, nous avons utilisé une maille à 2
dimensions, que nous avons déplié sous la forme d'un tableau à une
dimension. Il a ensuite été fait appel à la fonction \texttt{Config} à
l'instar d'une configuration à une dimension.

\subsection{Les configurations optimales}

\begin{figure}[h!]
   \centering
\resizebox{\hsize}{!}{
   \includegraphics{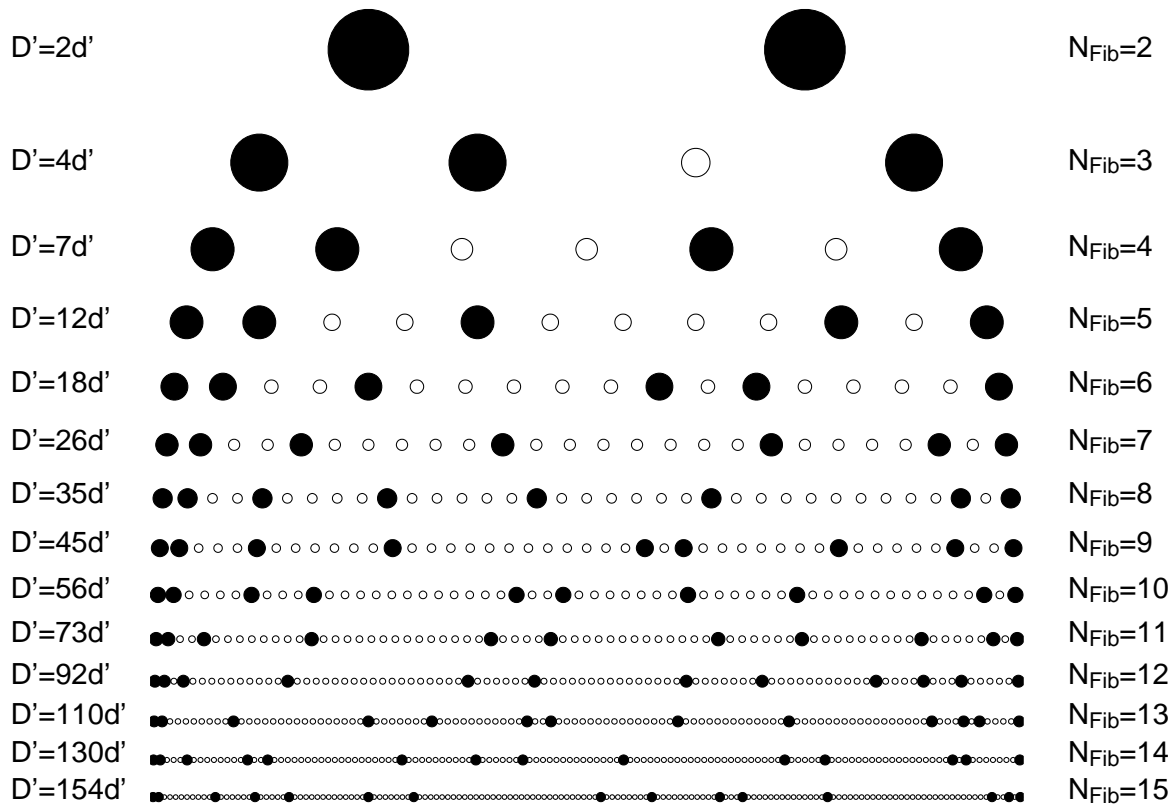}} \caption[Séries non-redondantes
   à une dimension]{ Séries non-redondantes à une dimension. Ces
   configurations sont optimales car le plus compact possible. }
   \label{fig:red1d}
\end{figure}

\begin{figure}[h!]
   \centering
   \includegraphics[width=4.7cm]{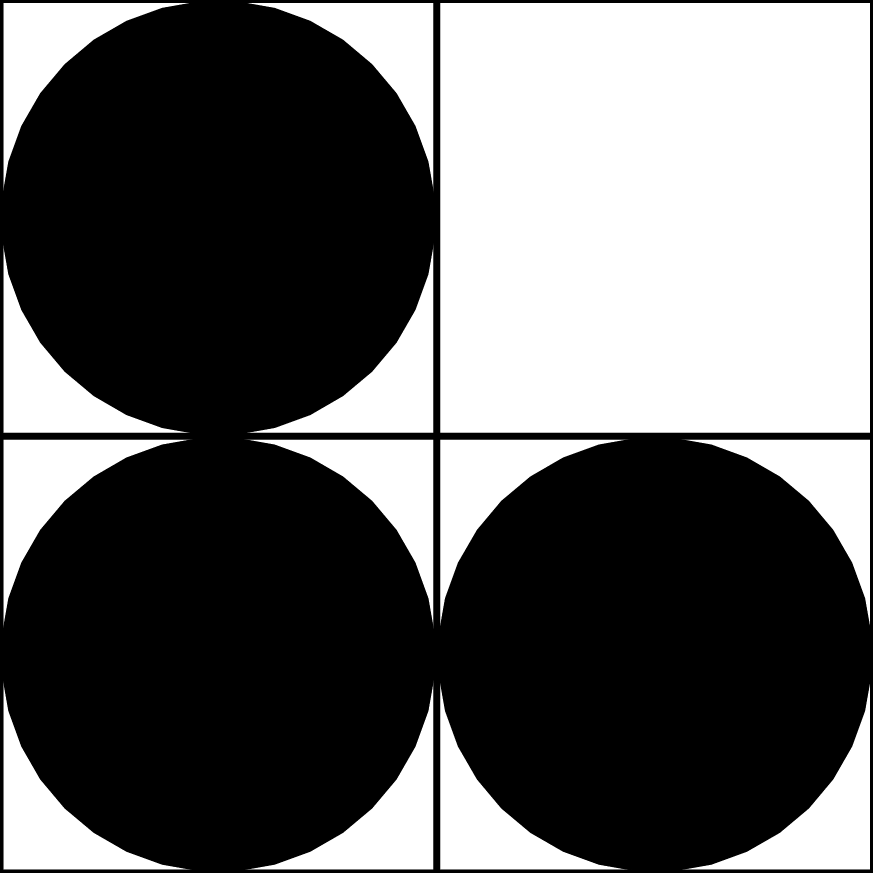}
\hfill
   \includegraphics[width=4.7cm]{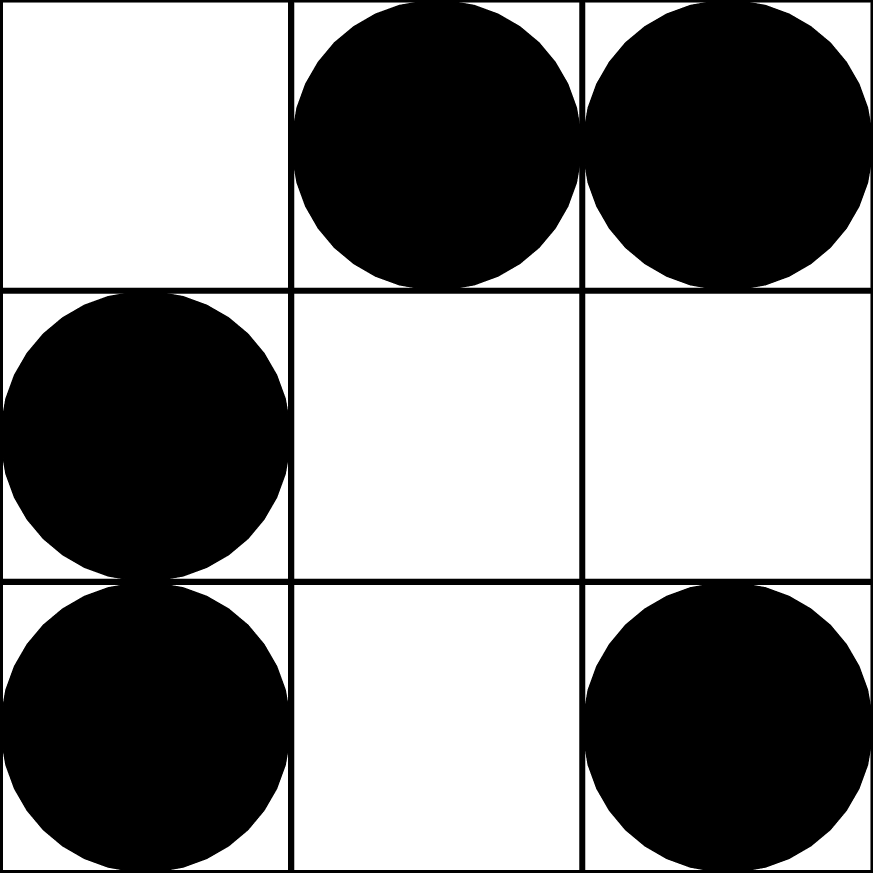}
\hfill
   \includegraphics[width=4.7cm]{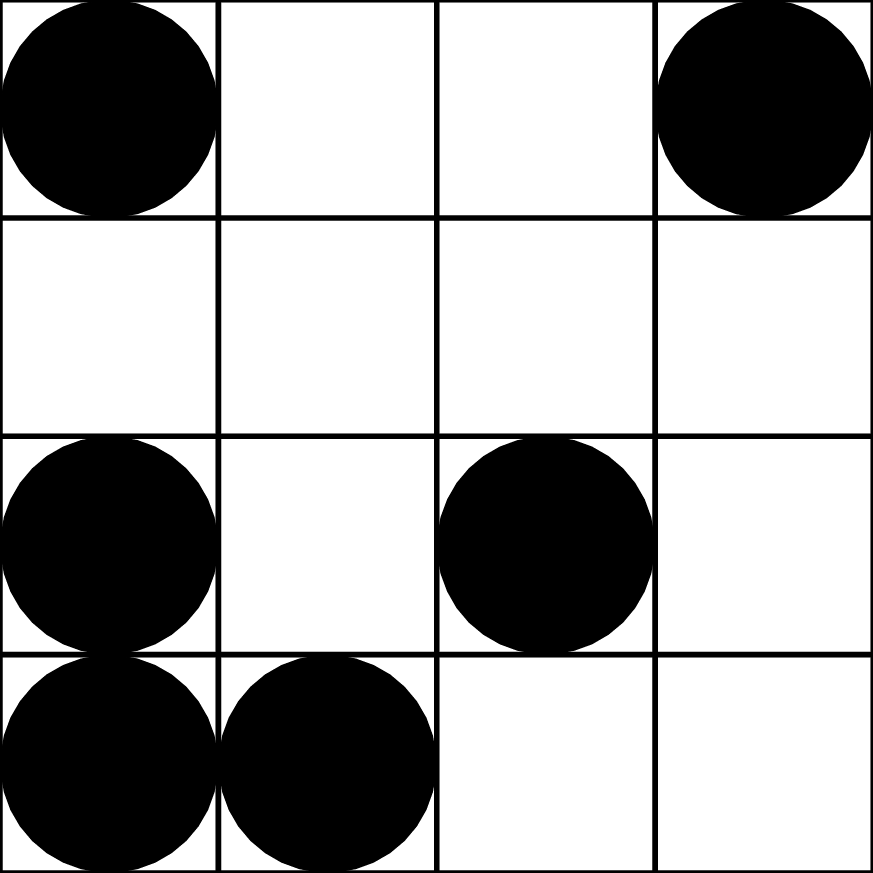} \vspace{.3cm}
\hfill
   \includegraphics[width=4.7cm]{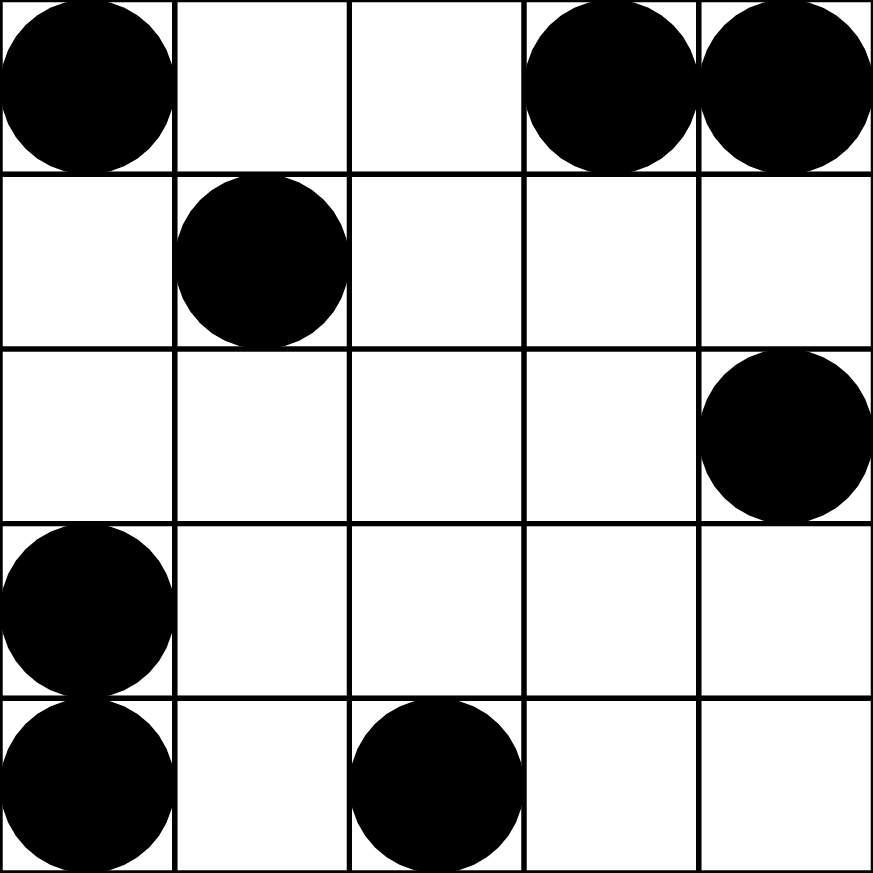}
\hfill
   \includegraphics[width=4.7cm]{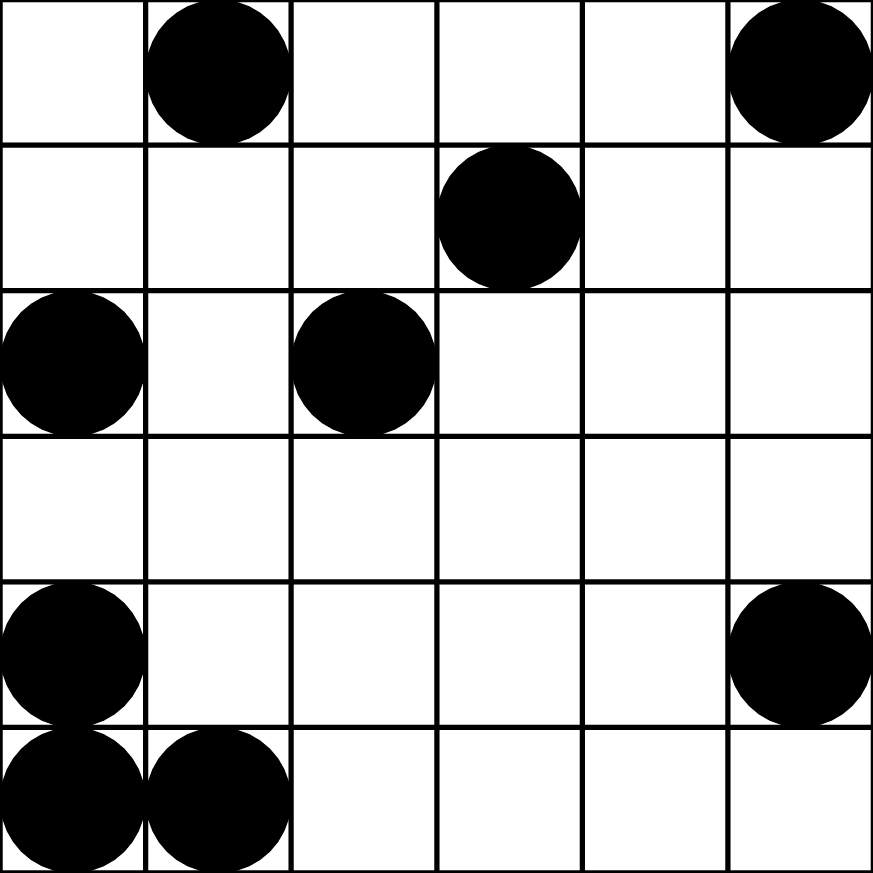}
\hfill
   \includegraphics[width=4.7cm]{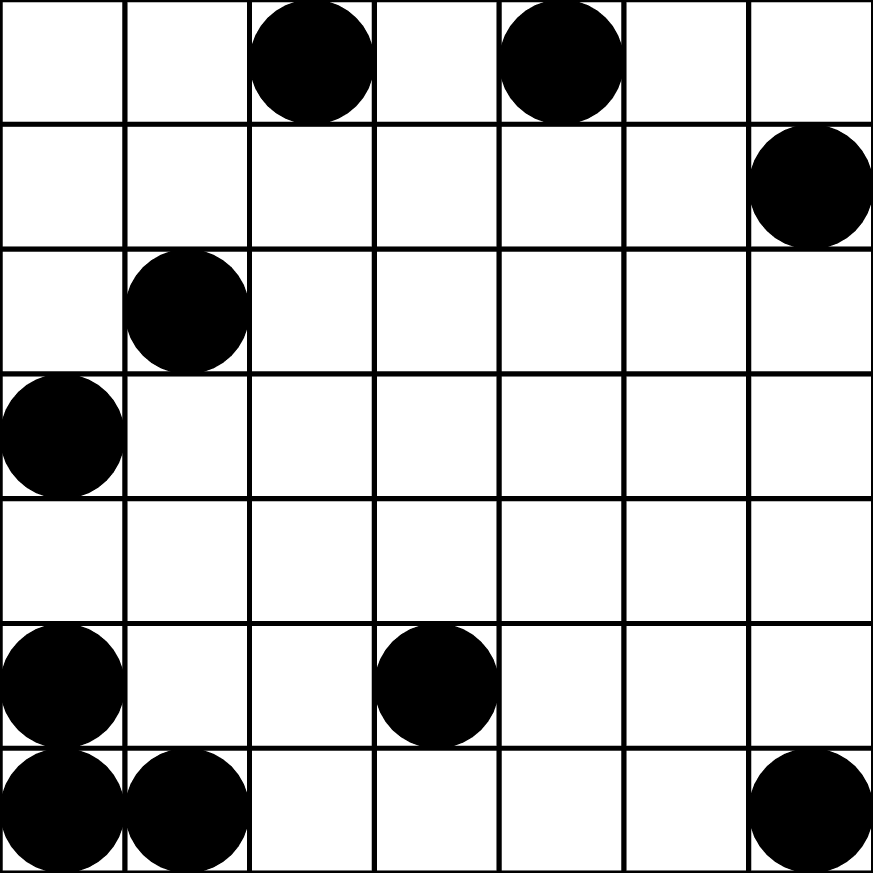} \vspace{.3cm}
\hfill
   \includegraphics[width=4.7cm]{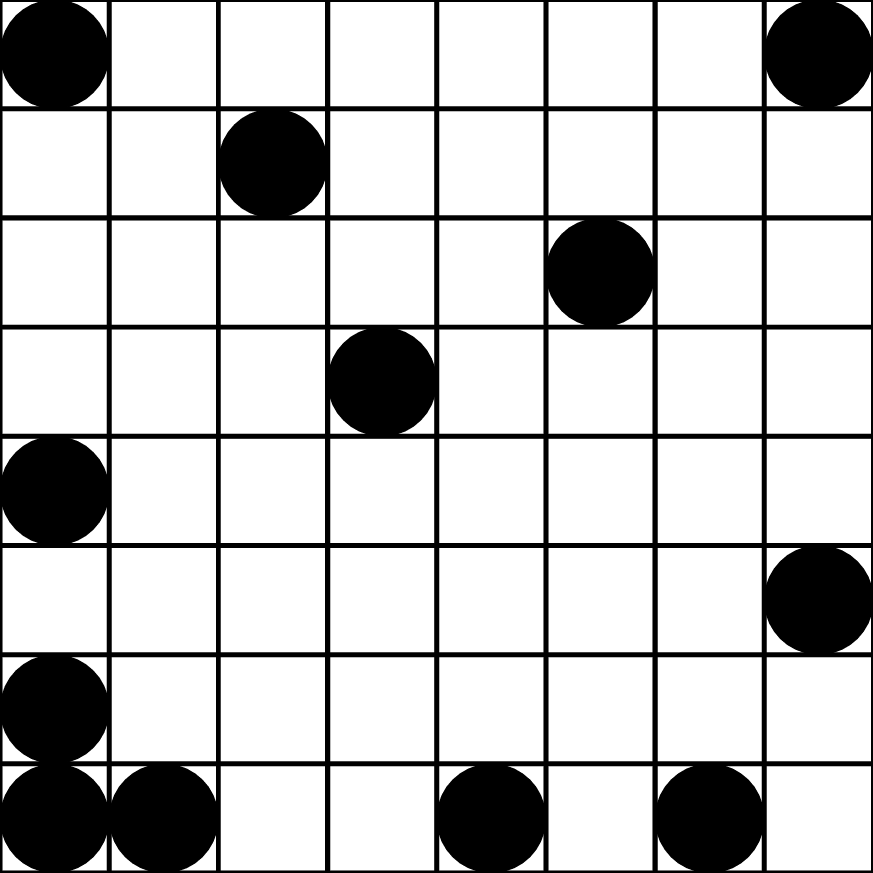}
\hfill
   \includegraphics[width=4.7cm]{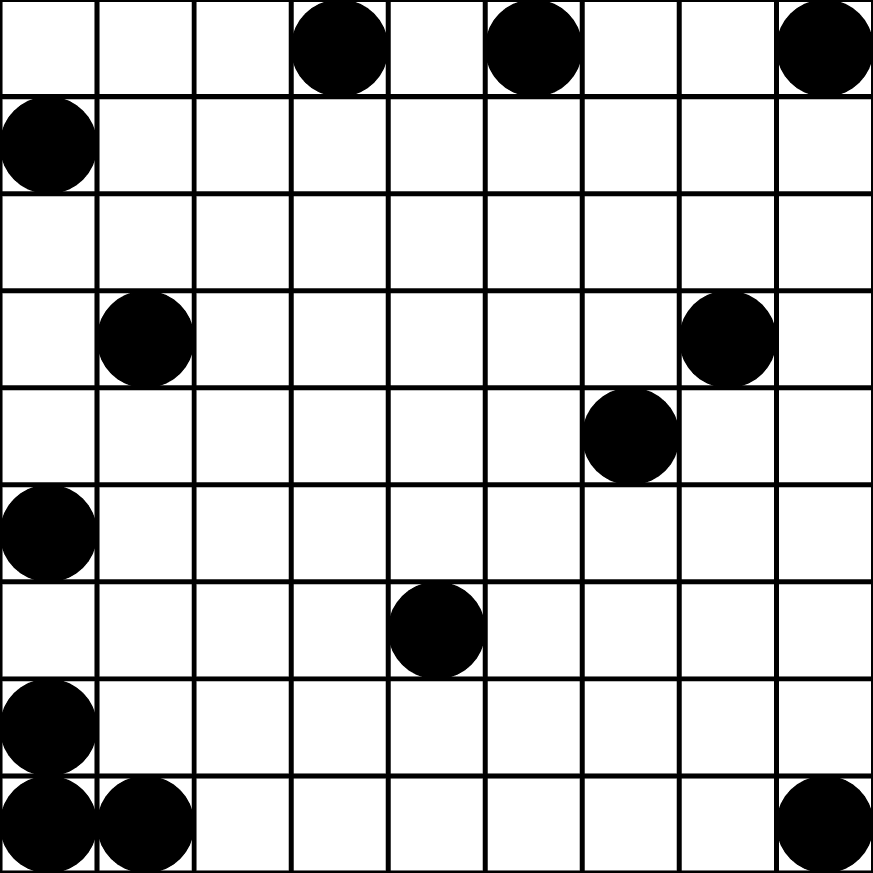}
\hfill
   \includegraphics[width=4.7cm]{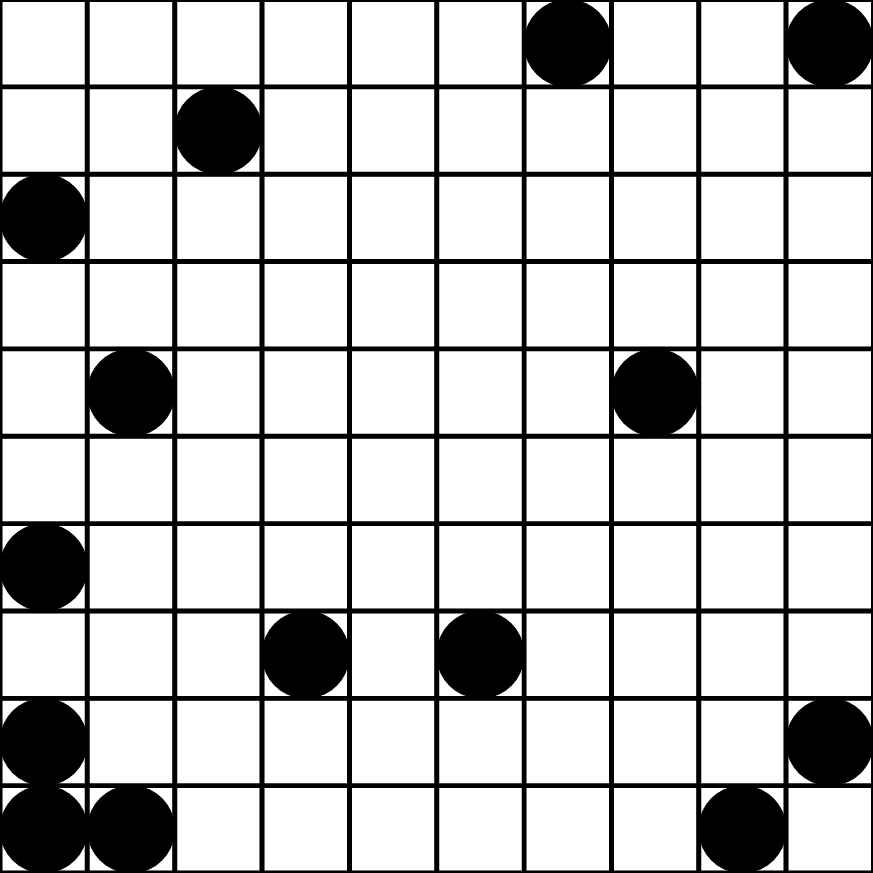} \vspace{.3cm}
\hfill
   \includegraphics[width=4.7cm]{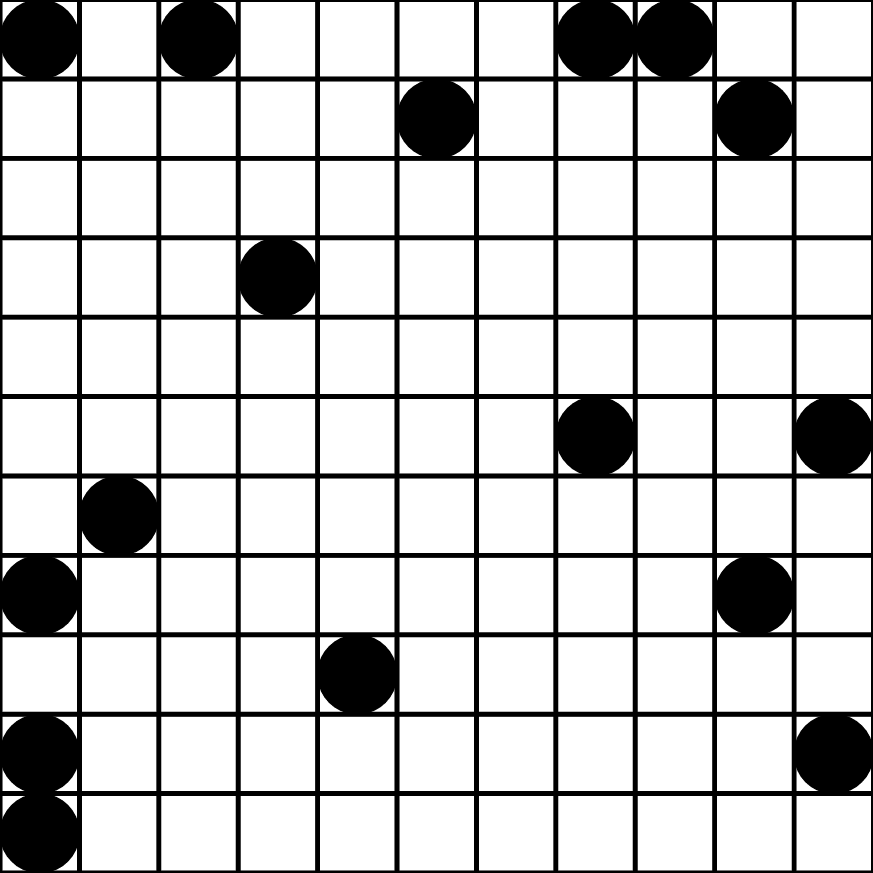}
\hfill
   \includegraphics[width=4.7cm]{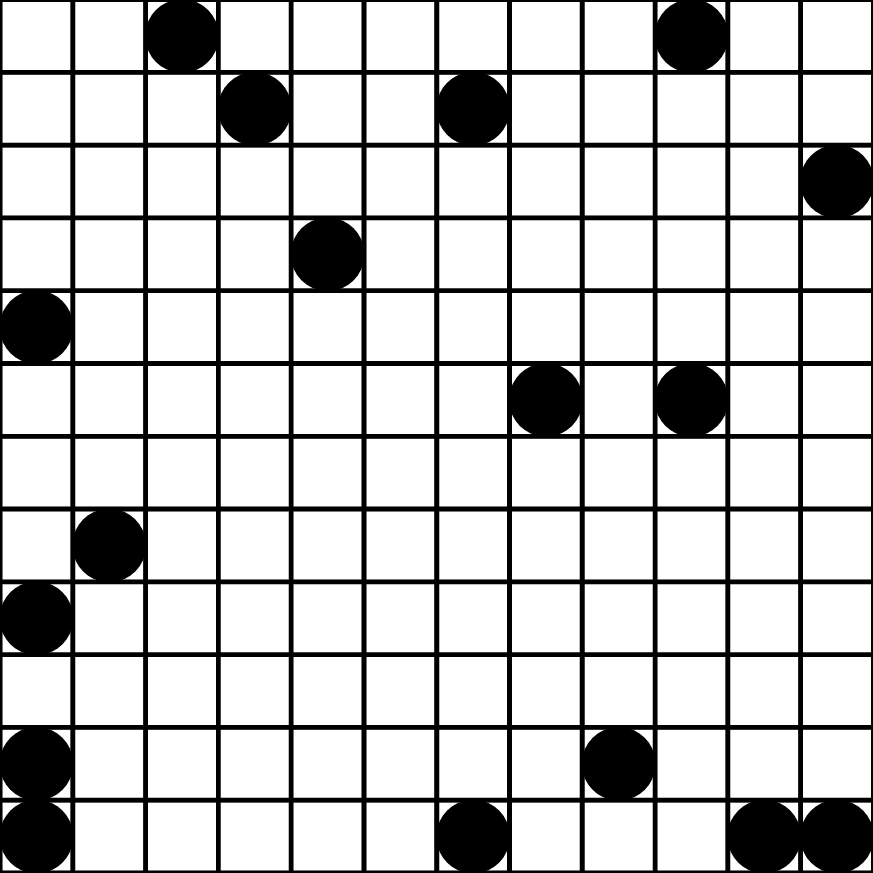}
\hfill
   \caption[Séries non-redondantes à deux dimensions]{ Séries non-redondantes à deux dimensions. Ces
   configurations sont optimales car le plus compact possible. }
   \label{fig:red2d}
\end{figure}

\begin{figure}[h!]
   \centering
   \includegraphics[width=7.cm]{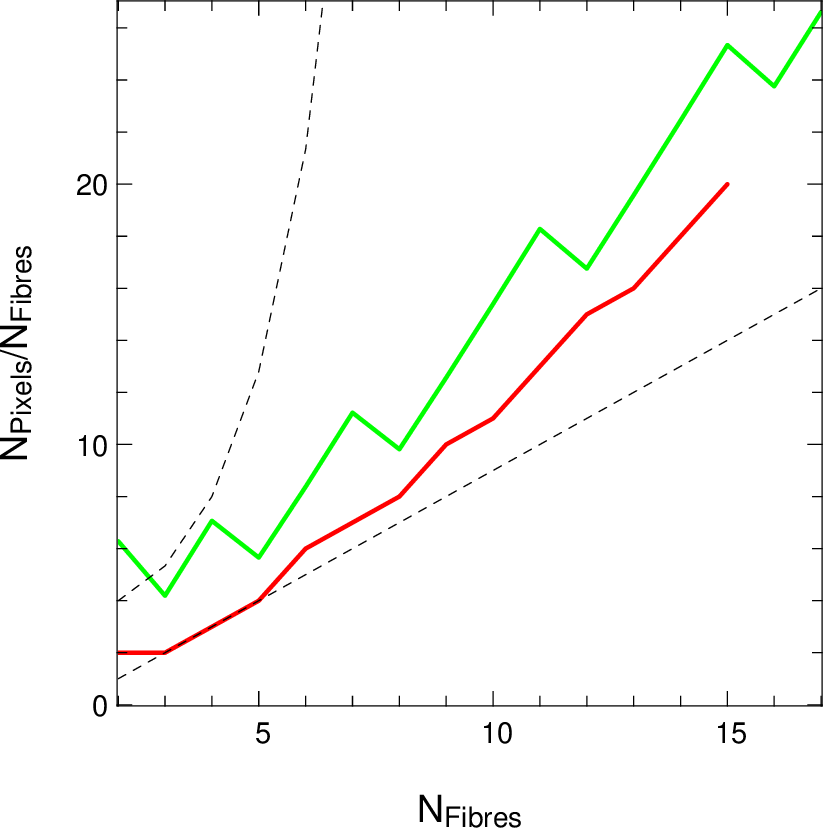}\hfill
   \includegraphics[width=7.cm]{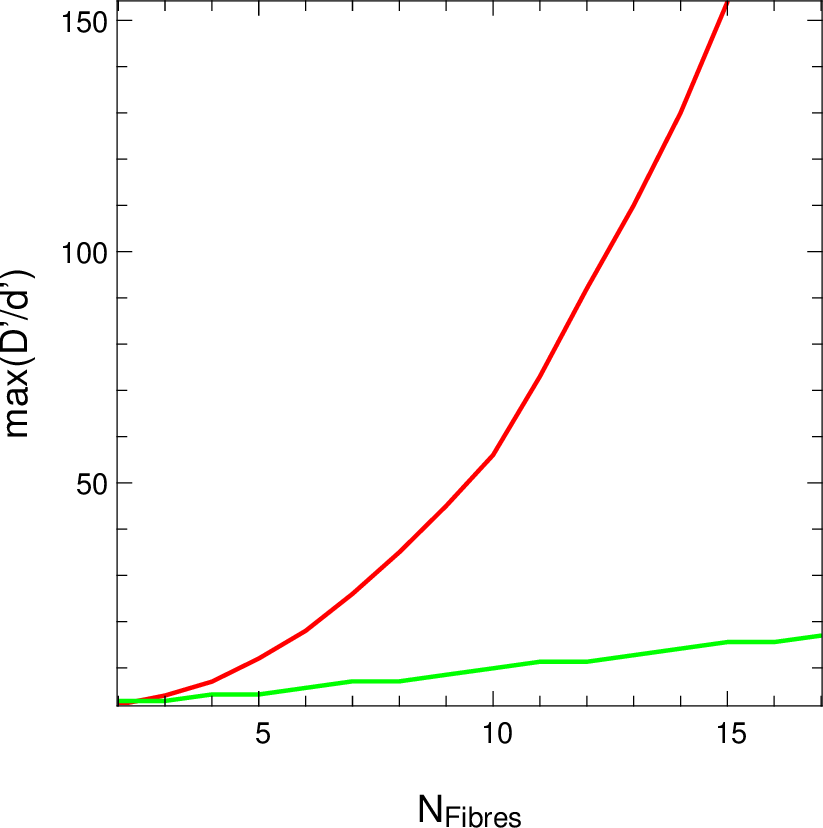}\vspace{.15cm}
   \includegraphics[width=7.cm]{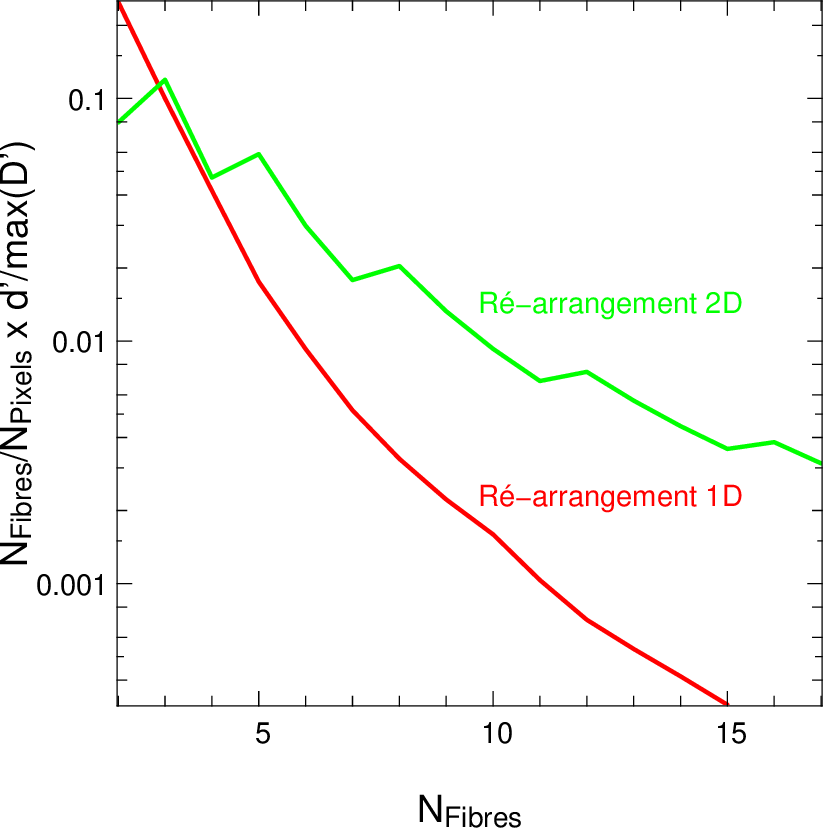}\hfill
   \includegraphics[width=7.cm]{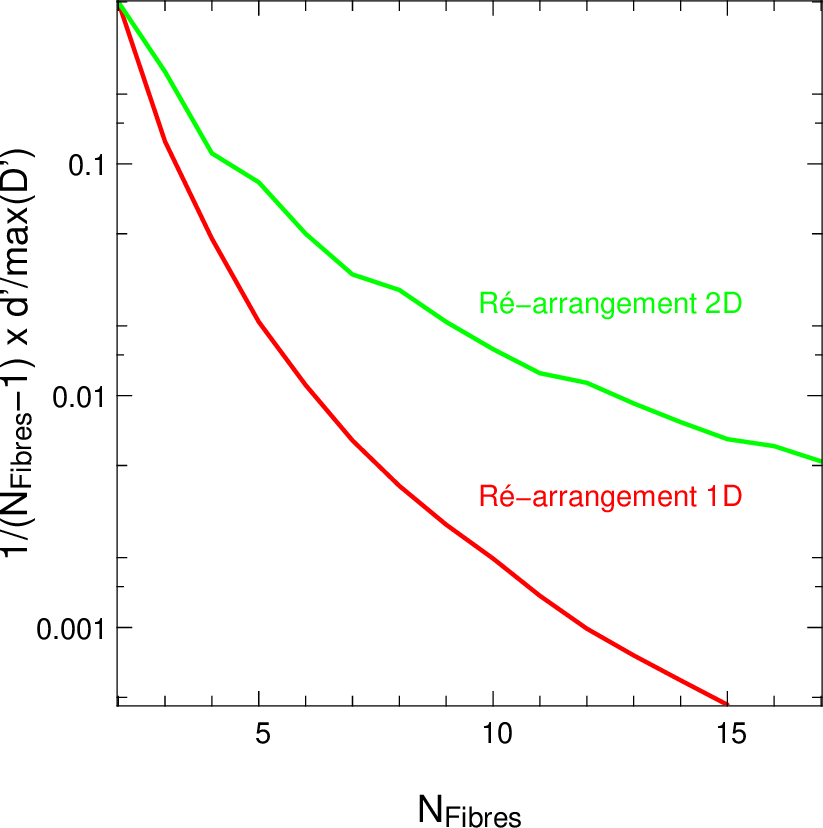}
   \caption[Effet du réarrangement sur la sensibilité de
   l'instrument]{ En haut à gauche~: Nombre de pixels par fibre en
   fonction du nombre de fibres. La courbe verte correspond à un
   réarrangement à 2 dimensions, et la rouge à un réarrangement à une
   dimension. Les courbes en pointillés représentent deux situations
   extrêmes où $N_{\rm Pixels}= N_{\rm Fibres} (N_{\rm Fibres-1})$ et
   ou $N_{\rm pixels}= 2^{N_{\rm Fibres}+1}$. En haut à droite~: Base
   maximale en fonction du nombre de fibres. Dans le cas d'un
   réarrangement à une dimension, il s'agit de la longueur du
   tableau. Dans celui à deux dimensions, il s'agit du diamètre du
   tableau. En bas~: le rapport $N_{\rm Fibres}/N_{\rm Pixels} \times
   d'/\max(D')$ est proportionnel au flux par pixel, alors que le
   rapport $1/(N_{\rm Fibres}-1) \times d'/\max(D')$ est proportionnel
   au flux par fréquence spatiale. Ces deux valeurs traduisent la
   sensibilité de l'instrument au bruit de lecture et au bruit de
   photons. } \label{fig:rearrg}
\end{figure}

Les figures~\ref{fig:red1d} et~\ref{fig:red2d} représentent les
configurations les plus compactes possibles. Plus précisément, il est
impossible d'ajouter une fibre dans aucun des
maillages. D'autres configurations non-redondantes ayant le même
nombre de fibres peuvent néanmoins exister. Les solutions obtenues
sont telles que, pour un nombre de fibres donné, le nombre de pixels
nécessaires pour respecter un bon échantillonage sera le minimum. Ce
nombre de pixels est, d'après le théorème de Shannon, de $2
\max(D')/d'$ dans le cas d'un maillage à une dimension, et de $(2
\max(D')/d')^2 \pi/4$ pour un maillage à deux dimensions. Le rapport
$\pi/4$ provient du caractère circulaire de l'image de la fibre.

Le rapport du nombre de pixels sur celui des fibres est présenté dans
la figure~\ref{fig:rearrg} (haut-gauche). La courbe verte représente
le rapport dans le cadre d'un arrangement à 2 dimensions, et la courbe
rouge dans celui à une dimension. La première courbe en pointillés
donne les valeurs de codage limites telles que $\max(D')/d'=2^{N_{\rm
Fibres}}$ (algorithme~\ref{eq:algo_base}) et $N_{\rm Pixels}=N_{\rm
Fibres} (N_{\rm Fibres}-1)$. La deuxième courbe correspond au minimum
théorique de codage où l'information sur chaque fréquence spatiale est
codée par 2 pixels. Nous nous en éloignons lorsque le nombre de fibres
augmente, la contrainte de non-redondance devenant plus forte.

Le rapport $N_{\rm Fibres}/N_{\rm Pixels}$ nous donne la répartition
du flux sur les pixels. Néanmoins, le rapport $D'/d'$ va également
conditionner le flux en limitant la bande spectrale utilisable.  Nous
avons établi section~\ref{sc:inf_spect} qu'il existe un optimum en
rapport signal sur bruit tel que~: $\Delta\lambda/\lambda_0 = 2
d'/\max(D')$.  Ainsi, pour estimer une valeur proportionnelle au flux
par pixel, nous avons représenté sur la figure~\ref{fig:rearrg} le
rapport $N_{\rm Fibres}/ N_{\rm Pixels} \times d'/\max(D')$. Cette
courbe traduit la sensibilité de l'instrument au bruit de lecture et
met clairement en évidence l'avantage d'un réarrangement en 2
dimensions. De la même façon, si l'on remplace le nombre de pixels par
celui des fréquences spatiales, on aboutit à une valeur qui traduit la
sensibilité de l'instrument au bruit de photons. Il s'agit du rapport
$1/(N_{\rm Fibres}-1) \times d'/\max(D')$ qui donne, à un facteur de
proportionnalité près, le nombre de photons dans chaque pic frange. 

De cette façon, les deux graphiques du bas de la
figure~\ref{fig:rearrg} montrent l'intérêt, en terme de sensibilité au
bruit de photons (figure de droite) et du détecteur (figure de
gauche), d'utiliser un réarrangement à deux dimensions. Cependant, un
certain nombre de pistes sont à explorer afin d'améliorer la
sensibilité~:
\begin{itemize}
\item utiliser un détecteur à comptage de photons pour s'affranchir de
  la sensibilité au bruit de lecture.
\item faire interférer les fibres par groupes. Les conditions de
  déconvolution de l'algorithme section~\ref{sc:algo_eric} doivent
  néanmoins êtres respectées, c'est à dire avoir au moins autant
  d'équations que d'inconnues.
\item dans le cas d'un réarrangement à une dimension, disperser la
  lumière sur une bande spectrale la plus large possible. Le signal
  sur bruit par pixels n'en sera pas amélioré, mais ceci augmentera la
  limite en sensibilité due au bruit de photons.
\end{itemize}

\subsection{L'effet du chromatisme sur le codage fréquentiel}
\label{sc:remap_chrom}

\begin{figure}[h]
   \centering
   \includegraphics[width=7cm]{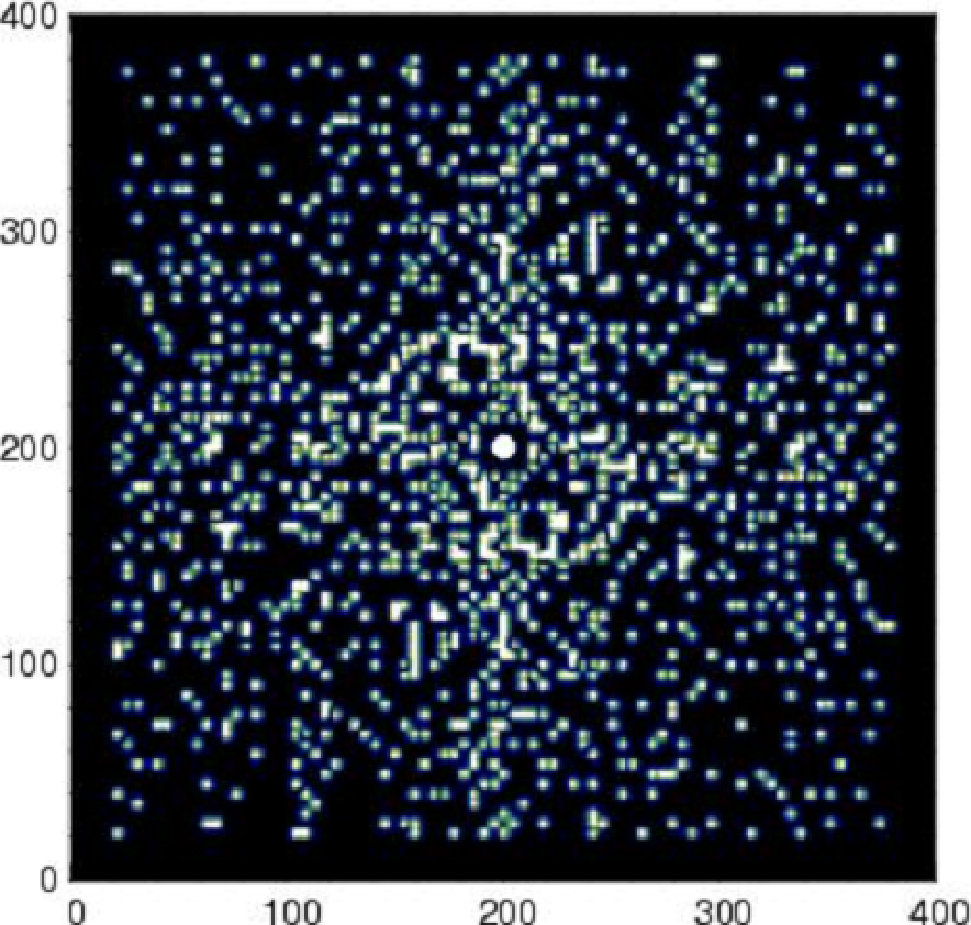}\hfill
   \includegraphics[width=7cm]{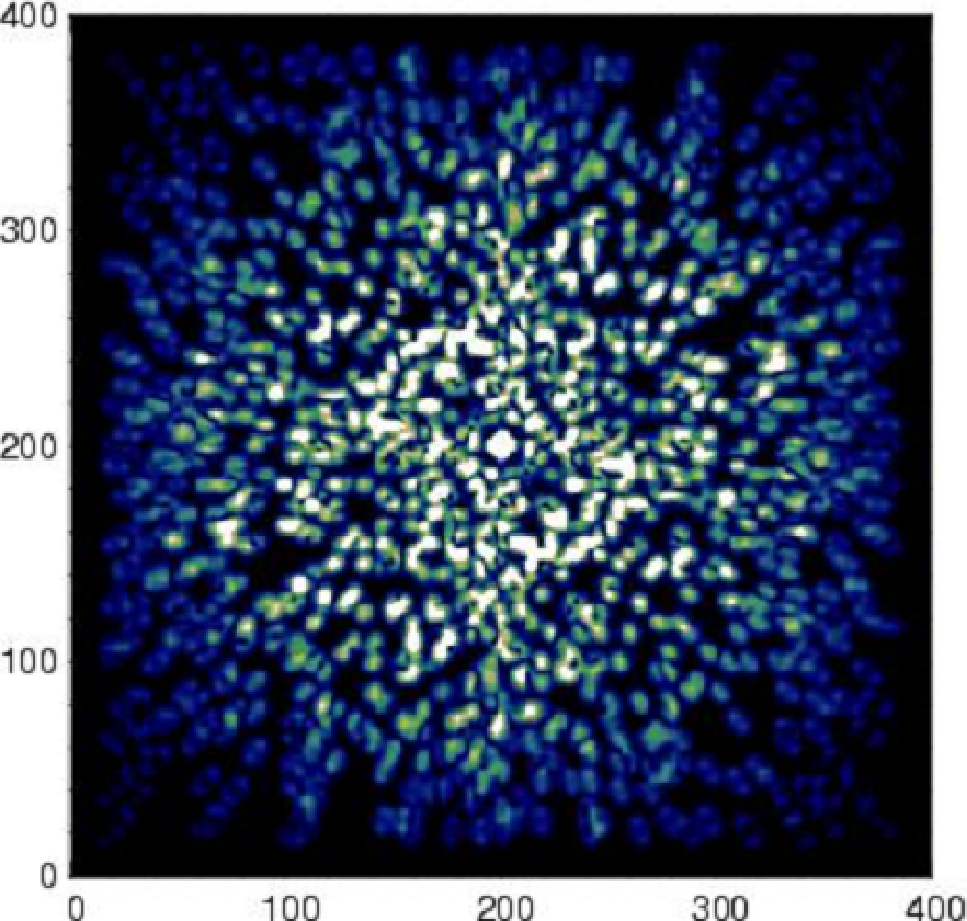}
   \caption[Effet du chromatisme sur le plan des fréquences
   spatiales]{ Effet du chromatisme sur le plan des fréquences
   spatiales. A gauche, les pics franges sont nettement définis sur
   l'image achromatique. A droite~: le chromatisme (620 nm < $\lambda$
   < 640 nm) disperse les pics franges sur plusieurs fréquences
   spatiales.}
   \label{fig:chr_fr}
\end{figure}

Afin d'optimiser
la largeur de la bande passante, nous avons choisi celle-ci telle que
$\Delta\lambda/\lambda_0 = 2 d'/\max(D')$. Or, l'influence de la bande
passante va avoir pour effet d'élargir radialement le pic frange dans
le domaine de Fourier. Un exemple est donné figure~\ref{fig:chr_fr}
où le plan de Fourier est composé de 630 pics franges. L'image de
gauche correspond à un faisceau achromatique de longueur d'onde 630
nm et celle de droite à une bande passante entre 620 et 640
nm. Une confusion des pics franges est visible.

En l'absence de déphasage, la fonction de modulation obtenue a une
 fréquence proportionnelle à $1/D'\lambda$. Lorsque l'on prend en
 compte l'effet de la bande passante, ces fréquences sont alors
 comprises entre $1/D'(\lambda_0+\Delta\lambda/2)$ et
 $1/D'(\lambda+\Delta\lambda/2)$. Ceci doit être intégré au calcul de
 la configuration non-redondate de manière à ne pas polluer
 l'information contenue aux différentes fréquences. Une modification
 de l'algorithme à une dimension peut être faite en remplaçant la
 ligne 6 de l'algorithme par: \\ \texttt{ for
 (i=positions(0)+1+max([0,$\frac{\lambda_0}{\Delta\lambda}$(dimension-i),$\frac{\lambda_0}{\Delta\lambda}$(i-1)])
 ;i<=dimension;i++) \\ }

\clearpage
\section{Le temps d'intégration}

\begin{figure}[h]
   \centering \includegraphics[width=7cm]{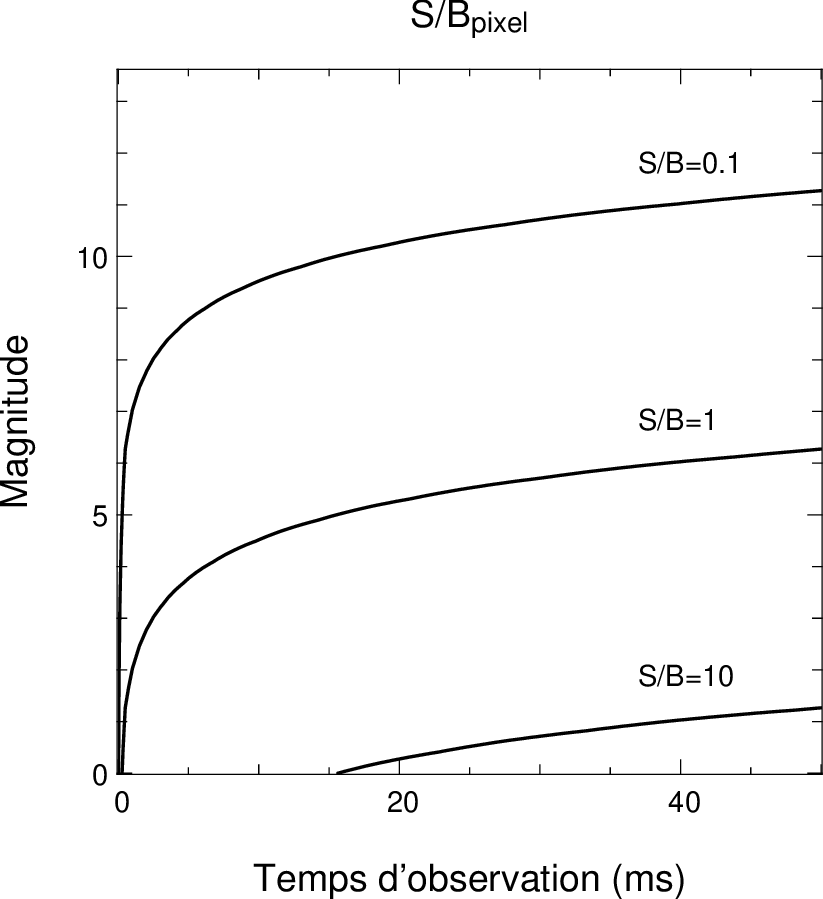}\hfill
   \includegraphics[width=7cm]{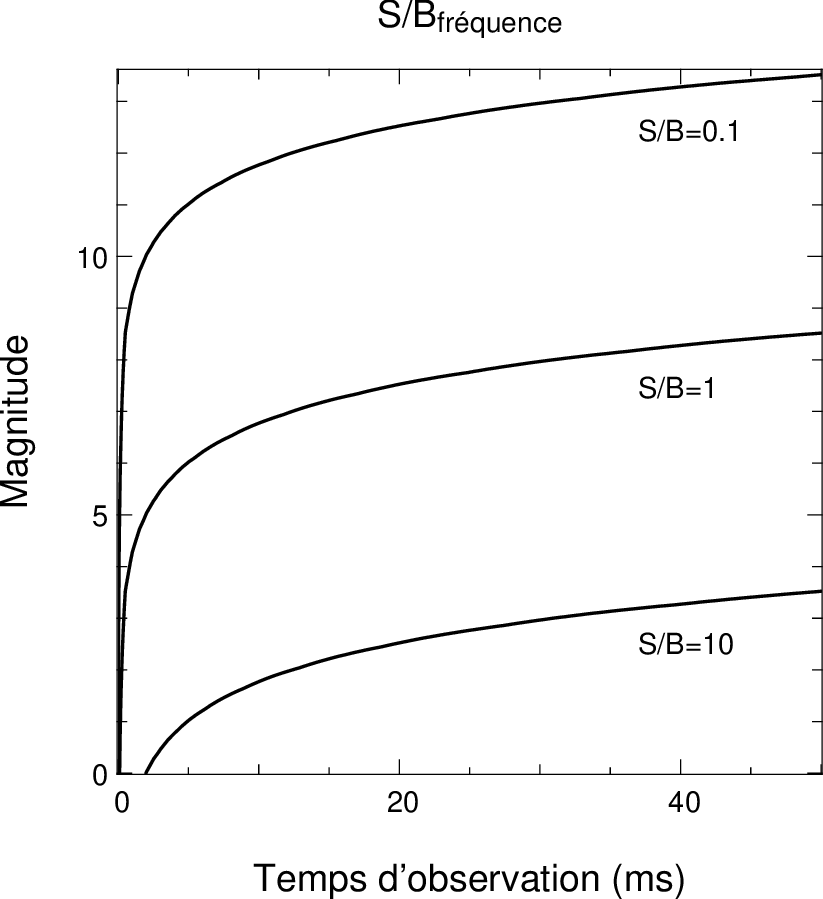} \caption[Signal sur
   bruit obtenu en fonction du temps d'observation et de la magnitude
   de l'objet observé]{Signal sur bruit obtenu en fonction du temps
   d'observation et de la magnitude de l'objet observé, dans le cas
   d'un système à réarrangement de pupille composé de 36 fibres et
   fonctionnant dans le visible. A gauche~: signal sur bruit par
   pixel. A droite~: signal sur bruit par fréquence spatiale.}
   \label{fig:sb}
\end{figure}

Les trois paragraphes précédents nous ont permis de caractériser
l'instrument. Nous avons optimisé l'injection, la bande passante et
la géométrie du réarrangement. Lors d'une observation, il va être
nécessaire d'optimiser un dernier paramètre, celui du temps
d'intégration.

Comme pour toutes les techniques nécessitant une déconvolution, le
réarrangement par fibres optiques suppose une turbulence fixe pendant
l'acquisition. Cette condition joue un rôle prépondérant sur la
qualité finale des données. Cependant, puisque la technique de codage
fréquentielle nécessite une dilution du flux de l'objet sur de
nombreux pixels, il faut optimiser le temps d'acquisition afin obtenir
un compromis entre le gel de la turbulence et le nombre de photons
reçus.

Nous avons fait ces calculs dans le cadre de l'instrument en projet à
l'observatoire. Il s'agit d'un système fibré à réarrangement de
pupille fonctionnant dans le domaine du visible.  Le
paragraphe~\ref{sc:inject} fixe une taille de sous-pupilles de 3$r_0$
et un rapport d'ouverture entre la fibre et les lentilles $\eta
=1,5$. L'efficacité de couplage est alors en moyenne $\rho=10\%$. La
pupille de sortie choisie est une configuration non-redondante en deux
dimensions de rapport $\max(D'/d') = 40$. Le nombre de pixels utilisé
sera par conséquent d'approximativement $N_{\rm pixels} =
5\,000$. Enfin, nous avons choisi d'optimiser la bande passante en
fonction de l'équation~(\ref{eq:delta_lambda}), ce qui nous donne
$\Delta\lambda/\lambda_0=0$,05 pour une efficacité interférométrique
moyenne $\mu = 0,75$. En ne considérant que le bruit de photons, nous
pouvons en déduire un rapport signal sur bruit par pixel~:
\begin{equation}
S/B_{\rm pixel}=\mu \sqrt{\frac{N_{\rm photons}}{N_{\rm pixels}} \rho
  \Delta\lambda Q t} 
\end{equation}
ainsi qu'un rapport signal sur bruit par fréquence spatiale~:
\begin{equation}
S/B_{\rm frequence}=\mu \sqrt{\frac{N_{\rm photons}}{N_{\rm Fibres} (N_{\rm Fibres}-1)} \rho
  \Delta\lambda Q t} 
\label{eq:pose}
\end{equation}
où $N_{\rm photons}$ est le nombre de photons par sous-pupille et par
unité de longueur d'onde, $Q$ l'efficacité quantique du détecteur et
$t$ le temps d'intégration par pose. En considérant une caméra à
comptage de photons, fonctionnant dans le visible ($Q=0,2$), la courbe
de gauche de la figure~\ref{fig:sb} nous donne le signal sur bruit
obtenu par pixel en fonction du temps d'observation. Si le détecteur
présente un bruit de lecture, la sensibilité de l'instrument sera plus
faible. Dans le cadre d'un instrument infrarouge par exemple, il
serait nécessaire de restreindre le nombre de fibres à recombiner
simultanément. En l'absence de bruit de lecture, le signal sur bruit
par fréquence spatiale fixe la sensibilité de l'instrument. Nous avons
montré au chapitre~\ref{sec:concept} que notre algorithme permet le
traitement de fréquences spatiales ayant un signal sur bruit aux
environs de 0,1. Il est par conséquent nécessaire de choisir le temps
de pose pour que le signal sur bruit soit, au minimum, égal à cette
valeur.

\clearpage
\section{Récapitulatif}

\subsubsection{Le choix de la taille des sous-pupilles: $d$}

Ce choix a une influence sur trois paramètres~: 
\begin{enumerate}
\item Le champ reconstructible. Celui-ci
est représenté figure~\ref{fig:Champ} en l'absence de perturbations
atmosphériques. Au premier ordre, il peut être estimé par $\lambda/d$.
\item Le nombre de sous-pupilles. En effet, plus les sous-pupilles seront
grandes, moins elles pourront être nombreuses dans la pupille principale~:
$M \lessapprox \sqrt{D/d}$.  
\item Le flux injecté dans chaque fibre. Ce
flux, proportionel à $\rho d^2$, est un des paramètres qui caractérisent
la sensibilité de l'instrument. Ainsi, pour un maximum de sensibilité,
il peut être utile de choisir une taille de sous pupille supérieure à
$r_0$. Le tableau~\ref{tb:recap_param} récapitule les différents taux
d'injection en fonction de la taille de la sous-pupille et du rapport
d'ouverture numérique $\eta$~:
\end{enumerate}
\begin{table}[h]
\caption{Taux de couplage}
\label{tb:recap_param}
\centering
\begin{tabular}{lcccccc}
\hline
\hline
d & \multicolumn{2}{c}{$1\,r_0$} & \multicolumn{2}{c}{$2\,r_0$} & 
\multicolumn{2}{c}{$3\,r_0$} \\
\hline
$\eta$     & 1,03 & 1,14 & 1,15 & 1,26 & 1,26 & 1,39\\
$\rho$     & 43 \% & 43 \% & 39,5 \%& 39 \%& 31,3 \% & 31,5 \%\\
${\rho d^2}$ & 0,43$r_0^2$  & 0,42$r_0^2$  & 0,79$r_0^2$ & 0,78$r_0^2$ & 
0,94$r_0^2$  & 0,93$r_0^2$  \\
\hline
\end{tabular}
\end{table}

Dans le cas où une optique adaptative serait utilisée, une
approximation peut être effectuée en remplaçant $r_0$ par le paramètre de
Fried généralisé $\rho_0$. Il peut être obtenu à
partir du nombre de polynomes de Zernik corrigés par l'optique
adaptative \citep[Equation (7) et (27) dans][]{2000JOSAA..17..903C}.

\subsubsection{Le choix du nombre de sous-pupilles: $M$}

Plus précisément, il y a deux termes à choisir~: $M$ le nombre de
sous-pupilles et $M_{\rm R}$ le nombre de sous-pupilles que l'on va
recombiner simultanément~:
\begin{enumerate}
\item $M$ est uniquement limité par la taille de la pupille principale
et des sous-pupilles~: $M \lessapprox \sqrt{D/d}$.  
\item $M_{\rm R}$ va conditionner la sensibilité de l'instrument. Plus
le nombre de fréquences spatiales mesurées est grand, moins
l'instrument est sensible. Il est nécessaire de déterminer le signal
sur bruit par fréquence spatiale (équation~(\ref{eq:pose})), ainsi que
par pixel (dans le cas du bruit de détecteur).
\item Il faut que le nombre de sous-pupilles $M_{\rm R}$ soit suffisant
pour permettre de déterminer les phases, les amplitudes et la
distribution spatiale d'intensité de la source. En pratique, il faut
établir les matrices~(\ref{eq:matrix_A}) et~(\ref{eq:matrix_P}) pour
vérifier que le système est inversible.
\end{enumerate}

\subsubsection{Le choix du rapport d'ouverture: $\eta$}

L'ouverture numérique de la fibre peut être mesurée expérimentalement à
partir du seuil à 5\% du cône de diffraction de l'énergie lumineuse
sortant de la fibre. Le rapport d'ouverture $\eta=ON_{\rm
lentille}/ON_{\rm fibre}$ permet alors de calculer le taux de couplage
optique. En fonction du diamètre de la sous-pupille, le maximum de
l'énergie couplée est obtenu pour~:
\begin{equation}
\eta=0,92+0,115\,d/r_0 \,.
\end{equation}
Cependant, pour gagner en champ et minimiser le bruit de confusion,
nous recommandons d'augmenter la valeur de $\eta$ d'environ 10\%. Cela
se fait au prix d'une très faible perte d'énergie couplée 
(tableau~\ref{tb:recap_param}).

\subsubsection{Le choix de la bande passante: $\Delta \nu/\nu_0$}

Le choix de la bande passante se fait selon deux contraintes~:
\begin{enumerate}
\item La première porte sur l'effet du piston atmosphérique. Suivant
le niveau de précision souhaité sur les visibilités, on consultera la
figure~\ref{fig:piston} pour déterminer une limite maximale à la bande 
passante.
\item Pour avoir une sensibilité maximale il faut que le produit
$\mu\sqrt{\Delta \nu/\nu_0}$ soit maximal (figure~\ref{fig:coher}). On
choisira par conséquent la bande passante optimale en fonction du
réarrangement, et notamment de la valeur $\max(D'/d')$ par~:
\begin{equation}
\frac{\Delta\nu}{\nu_0} \approx 2 \frac{d'}{\max(D')}
\label{eq:delta_lambda2}
\end{equation}
\end{enumerate}
Dans le cas d'un sytème avec dispersion spectrale, la limitation en
bande passante correspond à une limitation par canal spectral. 

\subsubsection{Le choix du réarrangement}

Trois critères viennent contraindre le réarrangement choisi~:
\begin{enumerate}
\item Si l'on souhaite ou non une information spectrale, on choisira
une configuration à une ou à deux dimensions.
\item Pour maximiser le facteur de cohérence, il faut
chercher à minimiser le rapport $D'/d'$, c'est à dire avoir la pupille
de sortie la plus compacte possible (figure~\ref{fig:coher}).
\item Parce que l'on souhaite que le champ de l'interférométre soit supérieur 
à celui de
la fibre, il faut que~:
\begin{equation}
\frac{D'}{d'} \geq  2 \frac{D}{d}\,.
\end{equation}
\end{enumerate}

\subsubsection{Le choix du temps d'intégration}

Le temps de pose doit satisfaire deux conditions~:
\begin{enumerate}
\item Il doit permettre une détermination correcte
de chaque fréquence spatiale pour une pose. Nos simulations,
section~\ref{sec:simu_c4}, nous ont montré que nous pouvions
reconstruire une image avec un signal sur bruit d'environ 0,1 par
fréquence spatiale instantanée. L'estimation du signal sur bruit peut
être obtenue par la relation~(\ref{eq:pose}).
\item Il doit être inférieur au temps de cohérence de
l'atmosphère. Plus le temps de pose sera faible, plus la qualité de la
mesure des visibilités sera bonne.
\end{enumerate}
Il est conseillé de pouvoir adapter le temps d'intégration à la
magnitude de l'objet observé de façon à maintenir un rapport signal-sur-bruit
par fréquence spatiale d'au moins 1.

\chapter{L'élaboration d'un démonstrateur}
\begin{center}
\end{center}
\minitoc \label{sec:demonstrateur} \vskip1cm

\clearpage

\section{Chronologie}

   \begin{figure}[h!]
   \centering \resizebox{\hsize}{!}{ 
\includegraphics{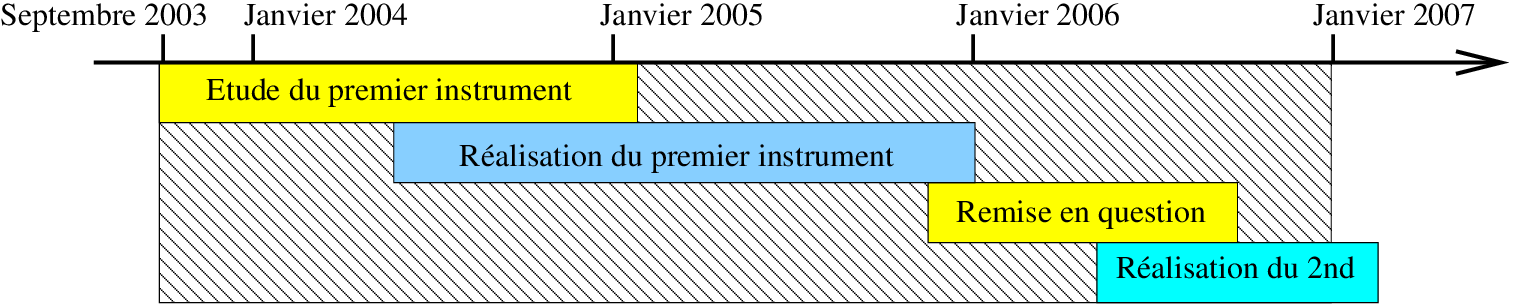} }
   \caption{ Chronologie de mon activité instrumentale au cours de ma thèse}
         \label{fig:chron}
   \end{figure}

Une grande partie de ma thèse a été consacrée à la conception d'un
démonstrateur, validant de façon expérimentale le concept du
réarrangement de pupille. La chronologie de ce travail est présentée
figure~\ref{fig:chron}.

La première année a été dévolue à la simulation et à l'étude de l'
instrument, afin d'aboutir à un concept raisonnable, pouvant être
adapté à un télescope et capable de fournir des résultats
astrophysiques. Ces critères nous ont menés au projet d'un premier
instrument, doté de 36 fibres, et suffisamment miniaturisé pour être
adaptable au foyer optique de n'importe quel télescope (section
~\ref{sec:manip_1}).

La réalisation de celui-ci a duré près deux ans, au cours desquels
nous avons pris conscience de la difficulté du positionnement des
fibres (voir section~\ref{sec:contraintes}). Fin 2005, devant
l'impossibilité à respecter le positionnement nécessaire à une
injection homogène dans l'ensemble des fibres, nous avons remis en
question la définition du premier instrument.

Nous nous sommes alors tournés vers un système de plus grande
dimension, doté de seulement 6 fibres. Chacune d'entre elles est
associée à un micro-positionneur 2 axes, permettant un ajustement
actif. Nous présentons cet instrument section~\ref{sec:manip_2}.

D'autres voies restent cependant à explorer, comme l'utilisation de
piezo-positionneurs, ou bien encore d'un miroir adaptatif segmenté.

\clearpage
\section{Les contraintes mécaniques}
\label{sec:contraintes}

\subsection{Le positionnement des fibres dans la pupille d'entrée}

Le positionnement des fibres optiques dans la pupille d'entrée est une
étape cruciale, cela pour deux raisons distinctes. La première,
directement apparente lors des premiers tests de l'instrument, porte
sur le couplage dans la fibre. En effet, plus la fibre sera loin de
l'image de l'étoile, moins de flux sera injecté dans la fibre. Le
deuxième effet est plus discret, et ne se révélera que lors de la
déconvolution des données. Il tient à la précision de l'emplacement
d'échantillonage du plan $u$-$v$ ainsi qu'au champ observé.

\subsubsection{Les contraintes de couplage}

   \begin{figure}[h!]
   \centering \includegraphics[width=8cm]{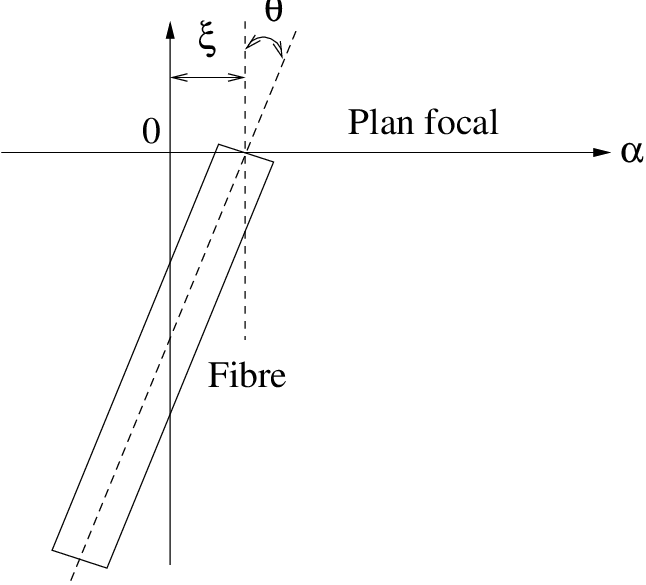}
   \caption[Positionnement de la fibre dans le plan focal de la
     lentille]{ Position de la fibre dans le plan focal de la
     lentille. Deux erreurs de positionnement sont explicitées par cette
     représentation~: $\xi$, pour le décalage, et $\theta$ pour
     l'inclinaison.  }
         \label{fig:inc_dec_leg}
   \end{figure}

   \begin{figure}[h]
   \centering \resizebox{\hsize}{!}{
\includegraphics{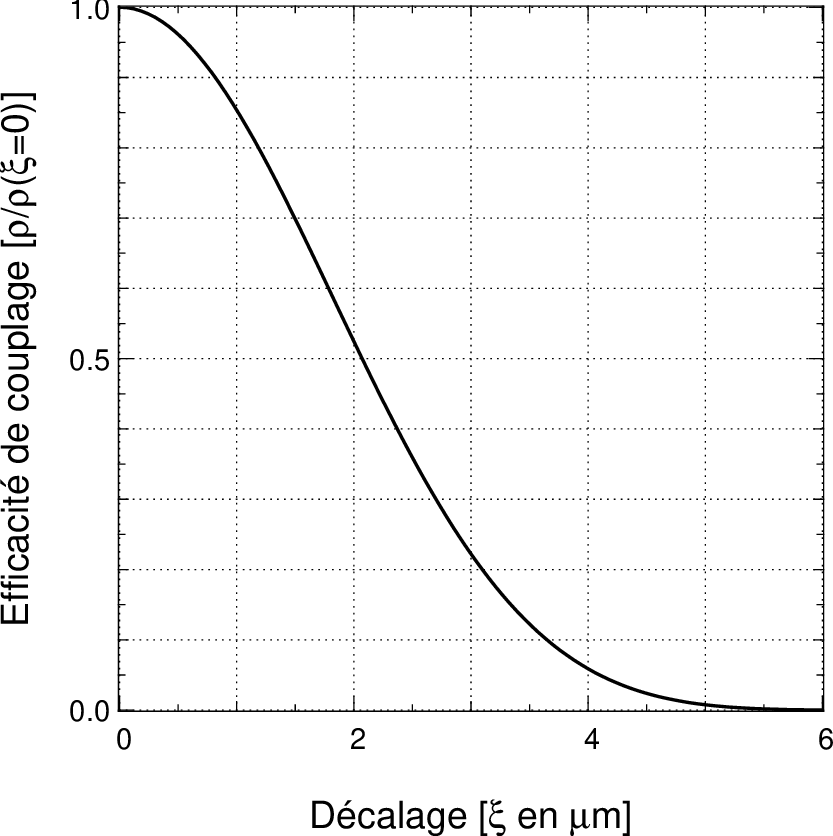}
\includegraphics{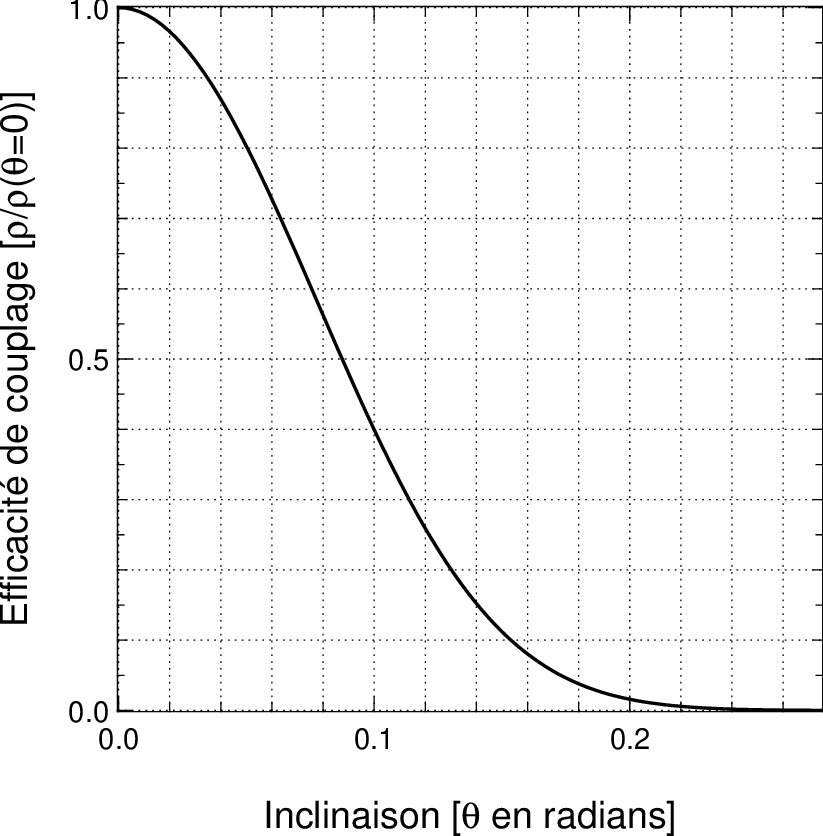}}
   \caption[Pertes en efficacité de couplage dues à l'imprécision du
   placement des fibres dans le plan focal des lentilles]{ Pertes en
   efficacité de couplage dues à l'imprécision du placement des fibres
   dans le plan focal des lentilles. A droite, effet d'un déplacement
   de la fibre. A gauche, effet d'une inclinaison (voir
   figure~\ref{fig:inc_dec_leg}).  } \label{fig:inc_dec} \end{figure}

L'efficacité de couplage dans une fibre monomode, avec ou sans
turbulences atmosphériques, a été établie section~\ref{sc:inject}. On
peut effectuer de nouveau ce travail en tennant compte de la présence
d'une inclinaison ou d'un décalage de la fibre. Le champ de la fibre
s'écrit alors~:
\begin{equation}
E_\bullet(\alpha,\beta) \propto
\exp\left(-\frac{4((\alpha-\xi/f)^2+\beta^2)f^2}{w_0^2}-\I\frac{ 2 \pi  f
\theta}{\lambda} \alpha\right)\,.
\end{equation}
où $\xi$ est le décalage de la fibre (en unités de longueur) et
$\theta$ son inclinaison (en radians). Ces deux valeurs sont
présentées figure~\ref{fig:inc_dec_leg}.  De façon similaire à
l'équation~(\ref{eq:Eo}), le champ dans le plan pupille s'écrit, après
normalisalisation~:
\begin{equation}
E_\circ(u,v) = \frac{ 2 \eta \lambda}{ d} \sqrt{ \frac{3}{\pi}} 
\exp\left(-6 \frac{ ((u-\theta f/\lambda)^2+v^2)\lambda^2}{d^2} \eta^2
+2\I\pi\xi u/f \right) \,.
\label{eq:Eo_b}
\end{equation}
Or, le champ dans la pupille, en l'absence de perturbations, est connu~:
\begin{equation}
U_\circ(u,v)= \left\{ \begin{array}{cl}
\displaystyle \frac{2\lambda}{\sqrt{\pi} d}  &\mbox{si $ \sqrt{u^2+v^2} \leq d/2\lambda $} \\
  0 &\mbox{sinon}
       \end{array} \right.
\end{equation}
L'efficacité de couplage peut être ainsi établie par le biais de
l'intégrale de recouvrement des fibres, (équations~(\ref{eq:int_rec})
et~(\ref{eq:eff_coup}))~:
\begin{equation}
\rho = \left| \iint_{\sqrt{u^2+v^2} \leq d/2\lambda}  \frac{4\sqrt{3} \eta \lambda^2}{\pi
  d^2} 
\exp\left(-6 \frac{ ((u-\theta f/\lambda)^2+v^2)\lambda^2}{d^2} \eta^2-2\I\pi\xi u/f
  \right)       dudv \right|^2\,,
\end{equation}
que l'on peut aussi écrire, en faisant intervenir l'ouverture optique
de la fibre (équation~(\ref{eq:eta})) et les changements de variables
$u'=2\lambda u/d$ et $v'=2\lambda v/d$~:
\begin{eqnarray}
\rho  &=& \left| \iint_{\sqrt{u'^2+v'^2} \leq 1}  \frac{4\sqrt{3} \eta \lambda^2}{\pi
  d^2} 
\exp\left(-\frac{3}{2} ((u'-2\theta f/ d)^2+v'^2)\lambda^2 \eta^2-\I\pi\xi u'd/f\lambda
  \right)       du'dv' \right|^2 \nonumber\\
&=& \left| \iint_{\sqrt{u'^2+v'^2} \leq 1}  \frac{ \sqrt{3} \eta}{\pi}
\exp\left(- \I  \frac{\pi \eta ON u'}{\lambda}\xi\right)
\exp\left(- \frac32  \left((u'-\frac{\theta}{ ON\eta })^2+v'^2\right) \eta^2
  \right)       du'dv' \right|^2\,,\nonumber\\
\label{eq:rho_6}
\end{eqnarray}
Pour des fibres monomodes à une longueur d'onde de
630\,nm, une valeur typique de l'ouverture numérique est de
$ON=0$,12. Ainsi, en utilisant $\eta=1$,25, l'équation~(\ref{eq:rho_6})
permet d'établir la perte en efficacité de couplage due au décalage
($\xi$) et à l'inclinaison ($\theta$) de la fibre. Ce résultat est
représenté figure~\ref{fig:inc_dec}. 

\textbf{ Ainsi, pour obtenir au moins
90\% du taux d'injection maximum, il faut~:
\begin{itemize}
\item Un placement de la fibre $\xi \leq 0$,8\,$\mu$m.
\item Une inclinaison maximale de la fibre de $\theta \leq 0$,035\,radians.
\end{itemize}
}

\subsubsection{Les contraintes observationnelles}

L'obtention de suffisamment de flux dans chacune des fibres est un
critère de positionnement nécessaire, mais non suffisant. Deux autres
paramètres viennent renforcer les contraintes.
Tout d'abord, la précision de positionnement de la fibre va avoir une
influence sur le champ. Ceci se voit clairement en notant l'analogie entre
l'angle d'incidence $\varepsilon$ de l'équation~(\ref{eq:conf}) et $\xi$
dans l'équation~(\ref{eq:rho_6}). Les deux sont reliées par la relation~:
\begin{equation}
\varepsilon=\xi/f\,.
\end{equation}

La deuxième contrainte porte sur l'échantillonnage du plan
$u$-$v$. Ainsi, si la fibre est inclinée, le flux injecté dans la fibre
proviendra d'une partie légèrement décalée de la pupille. Cela
signifie que l'information spatiale sur l'objet sera obtenue pour une
base différente. La relation donnant le déplacement dans le plan
$u$-$v$ en fonction de l'inclinaison de la fibre $\theta$ peut être
établie à partir de l'équation~(\ref{eq:Eo_b}) et est la suivante~:
\begin{equation}
\Delta u = \frac{f}\lambda \theta\,.
\end{equation}

\subsection{Le positionnement des fibres dans la pupille
  de sortie}

  \begin{figure}[h] \centering
   \includegraphics[width=8cm]{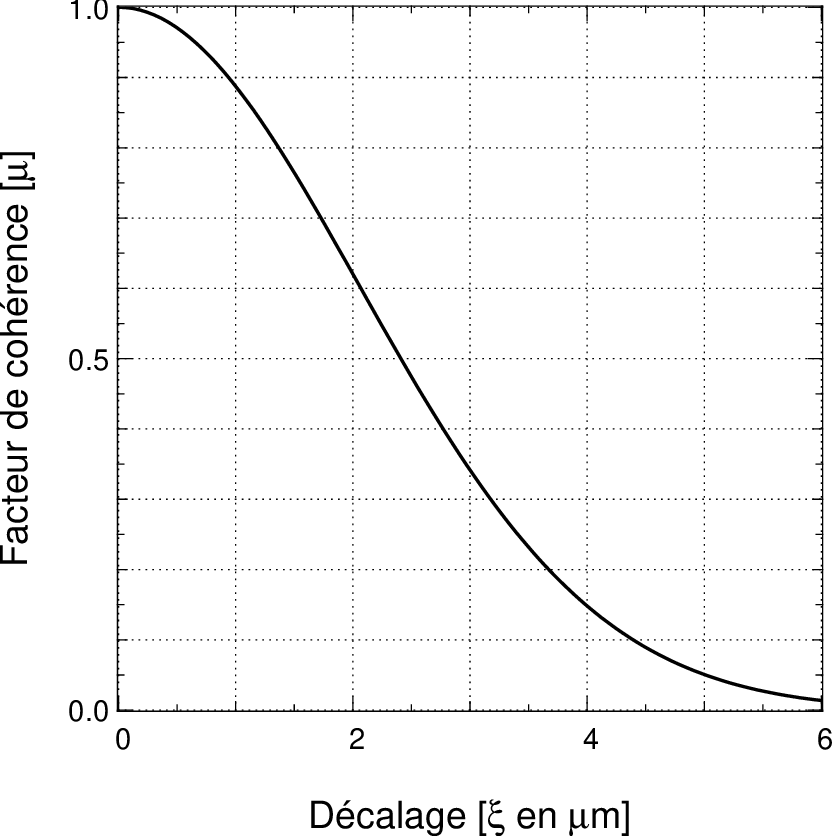} \caption{
   Diminution du facteur de couplage en fonction de l'emplacement de
   la fibre dans le plan focal de la lentille de la pupille de
   sortie.}  \label{fig:mu_sortie} \end{figure}

Contrairement à ce que l'on pourrait peut-être croire, les contraintes
de positionnement des fibres dans la pupille de sortie sont également
très strictes. Cette contrainte peut être établie à partir du facteur
de cohérence. Nous l'avons déja calculé analytiquement
section~\ref{sec:recom_spec} à partir du champ électrique dans le plan
du détecteur. Celui-ci peut être défini compte tenu d'une erreur
de positionnement $\xi$ par la relation~:
\begin{equation}
E_{1\bullet}(\alpha,\beta) = g_1 
\exp\left(-\frac{\pi^2 d'^2((\alpha-\xi/f)^2+\beta^2)}{6\lambda^2}\right) \exp(\I\pi
D'/\lambda\alpha )\,.
\end{equation}
Nous supposerons le champ de la deuxième fibre inchangé, tel qu'établi
équation~(\ref{eq:E2o}). L'image que l'on obtient sur le détecteur peut,
de la même façon, être écrite sous la forme de trois termes~:
\begin{eqnarray}
I(\alpha,\beta)&=&\underbrace{g_1^2\exp\left(-\frac{\pi^2
  d'^2((\alpha-\xi/f)^2+\beta^2)}{3\lambda^2}\right)}_{\mbox{=
  $I_1(\alpha,\beta)$}} \nonumber\\
&+&\underbrace{g_2^2\exp\left(-\frac{\pi^2
  d'^2(\alpha^2+\beta^2)}{3\lambda^2}\right)}_{\mbox{=
  $I_2(\alpha,\beta)$}}\nonumber\\
&+&2\underbrace{g_1g_2\exp\left(-\frac{\pi^2
  d'^2(\alpha^2+(\alpha-\xi/f)^2+2\beta^2)}{6\lambda^2}\right)}_{\mbox{=
  $A_{(1,2)}(\alpha,\beta)$}}\cos(2\pi D'/\lambda\alpha-\Delta\phi)\nonumber\\
\end{eqnarray}
On retrouve ici une formulation déjà utilisé
équation~(\ref{eq:Iab_dec}). Deux termes correspondant aux contributions
respectives de la première et de la deuxième fibre, et un terme
correspond à l'amplitude de modulation. On peut alors, de manière
similaire à ce que nous avons fait section~\ref{sec:fact_cohe},
établir le facteur de cohérence~:
\begin{eqnarray}
\mu&=&\frac{\displaystyle\iint_{-\infty}^{+\infty}
A_{(1,2)}(\alpha,\beta)d\alpha d\beta}{\displaystyle\sqrt{\iint_{-\infty}^{+\infty}
I_1(\alpha,\beta)d\alpha d\beta \cdot \iint_{-\infty}^{+\infty}
I_2(\alpha,\beta)d\alpha d\beta }}\nonumber\\
&=&\frac{\displaystyle\iint_{-\infty}^{+\infty}g_1g_2
\exp\left(-\frac{\pi^2
d'^2 (\alpha^2+(\alpha-\xi/f)^2+\beta^2)}{6\lambda^2}\right)  d\alpha d\beta}{ \displaystyle\sqrt{
\iint_{-\infty}^{+\infty} g_1^2  \exp\left(-\frac{\pi^2
d'^2 (\alpha^2+\beta^2)}{3\lambda^2}\right)d\alpha d\beta
\cdot
\iint_{-\infty}^{+\infty} g_2^2  \exp\left(-\frac{\pi^2
d'^2 (\alpha^2+\beta^2)}{3\lambda^2}\right)d\alpha d\beta
}}\nonumber\\
&=&\frac{\displaystyle\int_{-\infty}^{+\infty} 
\exp\left(-\frac{\pi^2
d'^2 (\alpha^2+(\alpha-\xi/f)^2)}{6\lambda^2}\right)  d\alpha}{\displaystyle
\int_{-\infty}^{+\infty}   \exp\left(-\frac{\pi^2
d'^2 \alpha^2}{3\lambda^2}\right)d\alpha}\nonumber\\
&=&\frac{\sqrt{\pi}d'\nu_0}{\sqrt{3}}\displaystyle\int_{-\infty}^{+\infty} 
\exp\left(-\frac{\pi^2
d'^2 (\alpha^2+(\xi/2f)^2)}{3\lambda^2}\right)  d\alpha\nonumber\\
&=&\exp\left(-\frac{\pi^2
d'^2 \xi^2}{12f^2\lambda^2}\right) 
\ =\ \exp(-\pi^2
ON^2 \xi^2/3\lambda^2)
\ .
\end{eqnarray}
Ainsi, nous pouvons tracer l'évolution du facteur de cohérence en
fonction de l'écart entre la position nominale des fibres dans la
pupille de sortie. C'est ce que nous avons fait
figure~\ref{fig:mu_sortie} pour les mêmes paramètres que dans la
section précédente, c'est à dire $\lambda=630$\,nm, et $ON=0$,12.  La
diminution du facteur de cohérence suit une courbe sensiblement
identique à celle du couplage obtenu pour la pupille d'entrée. La
présision demandée est donc similaire, avec un facteur de cohérence de
90\% pour une erreur de positionnement de 1\,$\mu$m.

\subsection{L'influence de la longueur des fibres optiques}

L'influence de la longueur des fibres, à la fois en terme de longueur
totale et relative, est aussi un point important lors de l'élaboration
de l'instrument.  Cependant, ce problème est extrêmement difficile à
traiter autrement qu'expérimentalement. Bien que nous n'ayons pu
effectuer ce travail, nous pouvons faire quelques remarques d'ordre
qualitatif.

Chaque fibre doit être suffisamment longue pour permettre un filtrage
efficace, tout en restant suffisamment faible. La longueur necessaire
au filtrage dépend de beaucoup de paramètres. Certains sont dus à la
conceptions de la fibre elle-même, d'autre sont propres à sont
utilisation (comme par exemple la courbure que l'on applique à
celle-ci). Des recherches en ce domaine ont été effectué à l'IRCOM
(Institut de Recherche en Communications Optiqes et Microondes), et
semblent confirmer qu'une dizaine de centimètres serait suffisant pour
obtenir un filtrage suffissant \citep{2005OptCo.244..209H}. Cependant
lorsque l'on augmente la longueur des fibres on multiplie aussi les
effets de biréfringences ou de polarisations. Si l'on souhaite
utiliser des fibres sur une longueur de l'ordre du mètre, il semble
important d'utiliser des fibres à maintient de polarisation. Dans le
but de restreindre la complexité de l'instrument, nous avons décidé
d'utiliser des longueurs de l'ordre de 20 centimètres.

Il est enfin fondamental d'égaliser en longueur les différentes fibres
optiques. Sans cela, on introduit des différences chromatiques de la
longueur des trajets optiques. Dans le cas où l'on utilise une source
chromatique chromatique, et cela particulièrement aux faibles
longueurs d'onde, il est nécessaire d'effectuer une telle
égalisation. La précisions nécessaire d'une telle égalisation n'a pas
pu être établi au cours de cette thèse.

\clearpage
\section{Les expérimentations millimétriques}
\label{sec:manip_1}

   \begin{figure}[h]
   \centering 
\resizebox{\hsize}{!}{
\includegraphics{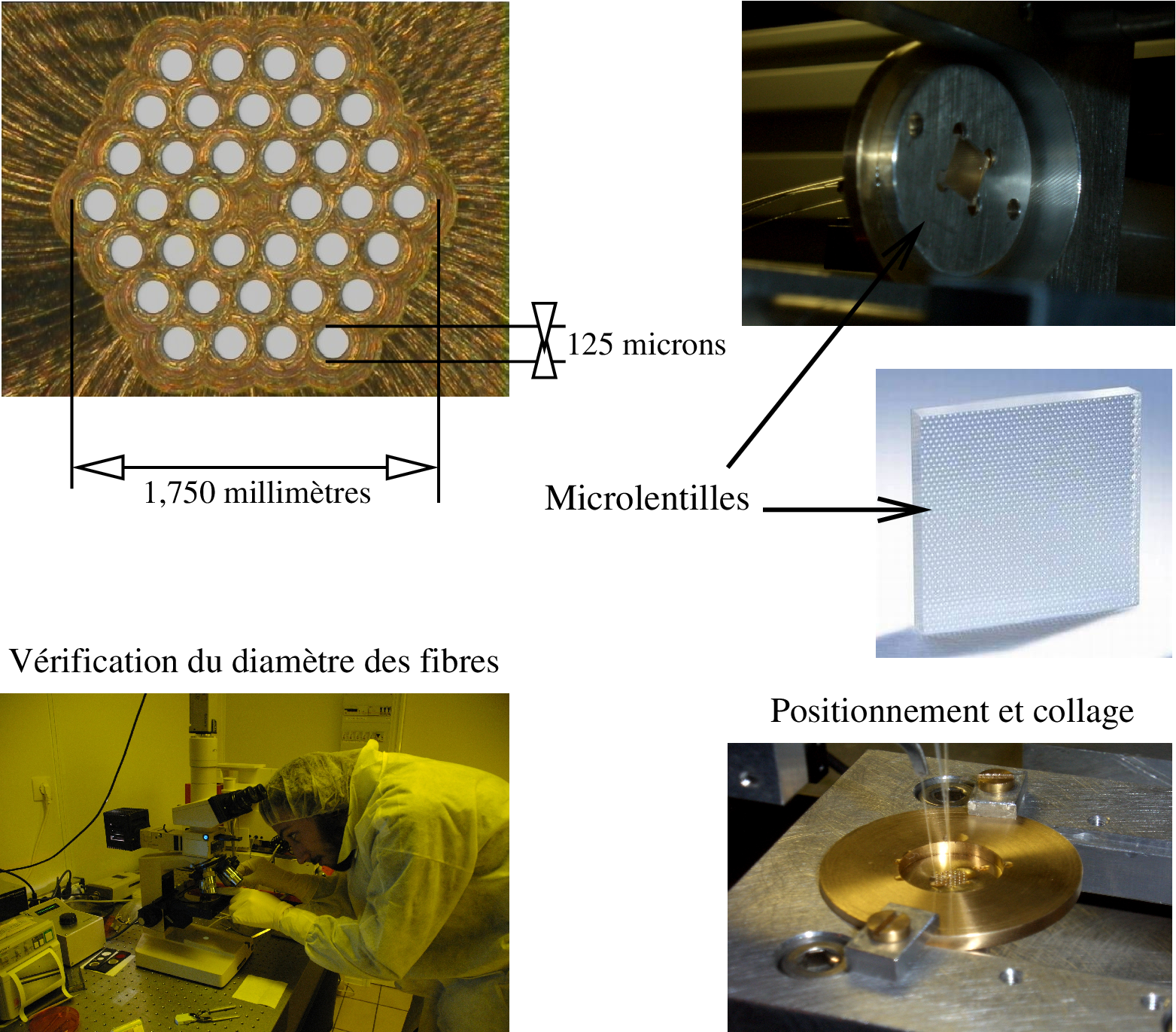}}
   \caption[Images de la conception des instruments de première
         génération]{En haut à gauche, pièce en laiton destinée à
         accueillir les 36 fibres optiques monomodes. En haut à
         droite, la même pièce, mais sur laquelle est adaptée la
         matrice de microlentilles. En bas à gauche, Keyan Bennaceur,
         stagiaire de l'observatoire, est en train de vérifier et de
         trier les fibres optiques par la taille de leur gaine
         optique. En bas à droite, positionnement par bras
         piézo-électrique des fibres optiques. } \label{fig:manip1}
         \end{figure}

  \begin{figure}[h] \centering
   \includegraphics[width=8cm]{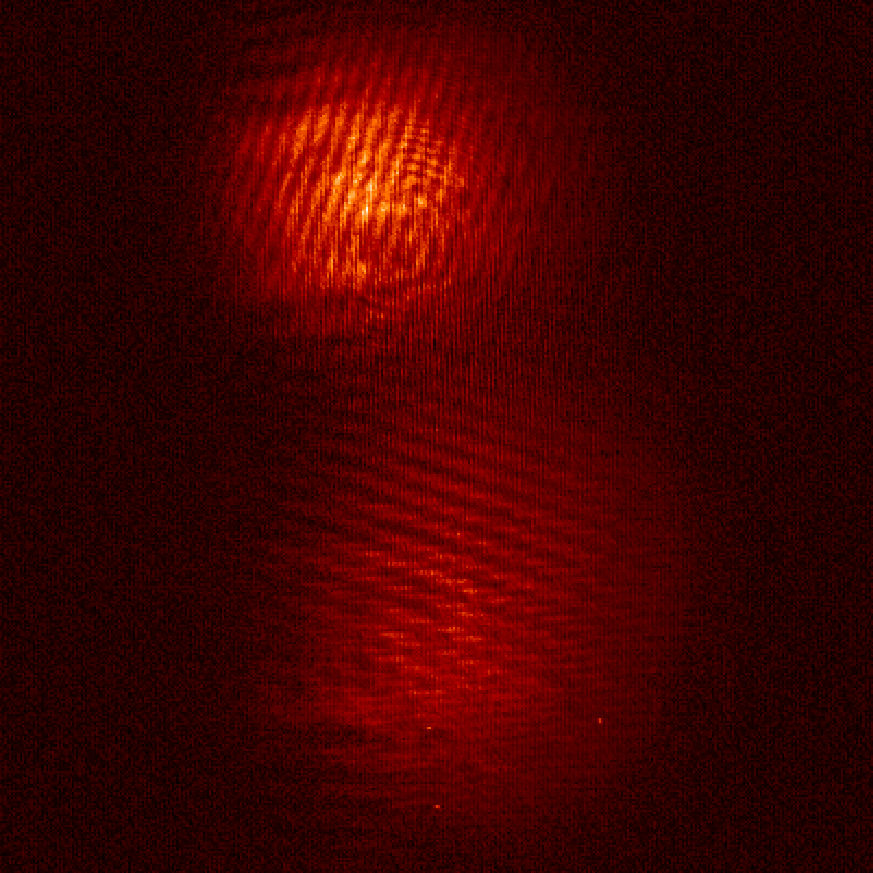}
   \caption[Figure de diffraction obtenue par focalisation de la
   pupille de sortie sur un détecteur]{ Figure de diffraction obtenue
   par focalisation de la pupille de sortie sur un détecteur. Ici, le
   collage à été arrêté à 7 fibres, car un décalage manifeste des
   fibres à été détecté, séparant les différents faisceaux sur deux
   taches Gaussienne distinctes. On peut cependant noter la très bonne
   cohérence de la lumière. } \label{fig:manip1_echec} \end{figure}

Au cours de la première partie de ma thèse, il a été question
d'appliquer les techniques de positionnement et de collage utilisées
pour les fibres multimodes (par exemple, celles utilisées à la
spectro-imagerie). Ces techniques, déjà utilisées à l'observatoire,
nécessitent l'usinage micrométrique d'une pièce de taille
millimétrique (photo en haut à gauche de la
figure~\ref{fig:manip1}). Les fibres sont ensuite individuellement
ajustées et collées. La pièce est alors ajusté à une matrice de
microlentille prévue à cet effet. Ainsi, si la précision du montage le
permet, l'ajustement se fait mécaniquement, et un grand nombre de
fibres peuvent être ajustées dans un système extrêmement compact. Dans
notre cas, nous avons travaillé sur une configuration à 36
sous-pupilles.

Si la technique est éprouvée pour l'injection dans des fibres
multimodes, il a cependant fallut l'adapter à l'injection dans des
fibres monomodes. La principale différence réside dans la taille des
c\oe urs optiques. Dans une fibre multimode, le c\oe ur est de
quelques dizaines de microns. Le diamètre d'une fibre monomode est,
lui, d'une valeur comprise entre 4 à 5 microns. La contrainte est donc
bien plus grande sur le placement des fibres. Nous avons vu
section~\ref{sec:contraintes} que les précisions demandées étaient
inférieures au micron. Ce saut quantitatif en terme d'exigence de
positionnement nous a contraint à prendre des précautions
particulières, en termes d'environnement comme de contrôle.

Nous avons en conséquence fait usiner par le laboratoire GEPI (Galaxies,
Etoiles, Physique et Instrumentation) des pièces à trous calibrés, de
tailles $125\pm3 \mu$m, équivalents à la taille de la gaine optique de
nos fibres optiques. Les fibres ont ensuite été individuellement vérifiées
et ajustées par un bras piézoélectrique dans leur emplacement
correspondant. Enfin, le collage s'est fait sous lampe UV, avec contrôle
du déplacement en temps réel par ordinateur. Au cours de ma thèse,
trois stagiaires ont participé à ce travail d'ajustement~: Keyan
Bennaceur, Eric Bughin et Kamel Houairi.

Le résultat n'a pas été à la hauteur de nos espérances. La
figure~\ref{fig:manip1_echec} représente, par exemple, un cas typique
d'échec du collage. On peut voir au moins 2 traces de diffractions
distinctes, striées de franges. La netteté des franges observées a été
un résultat des plus intéressants, car il a validé la pertinence de
poursuivre l'expérimentation à une telle longueur d'onde (ici à 730
nm). Cependant, le collage, de part le nombre élevé de fibres, la
petitesse du système, et les incertitudes mécaniques associées, s'est
avéré extrêmement difficile. Après deux ans de recherche dans cette
voie, nous avons décidé de changer d'optique, et de nous tourner vers
un système de plus grande taille, permettant une meilleure
qualification du principe.

\clearpage
\section{Une version décimétrique}
\label{sec:manip_2}

   \begin{figure}[h]
   \centering \resizebox{\hsize}{!}{
\includegraphics{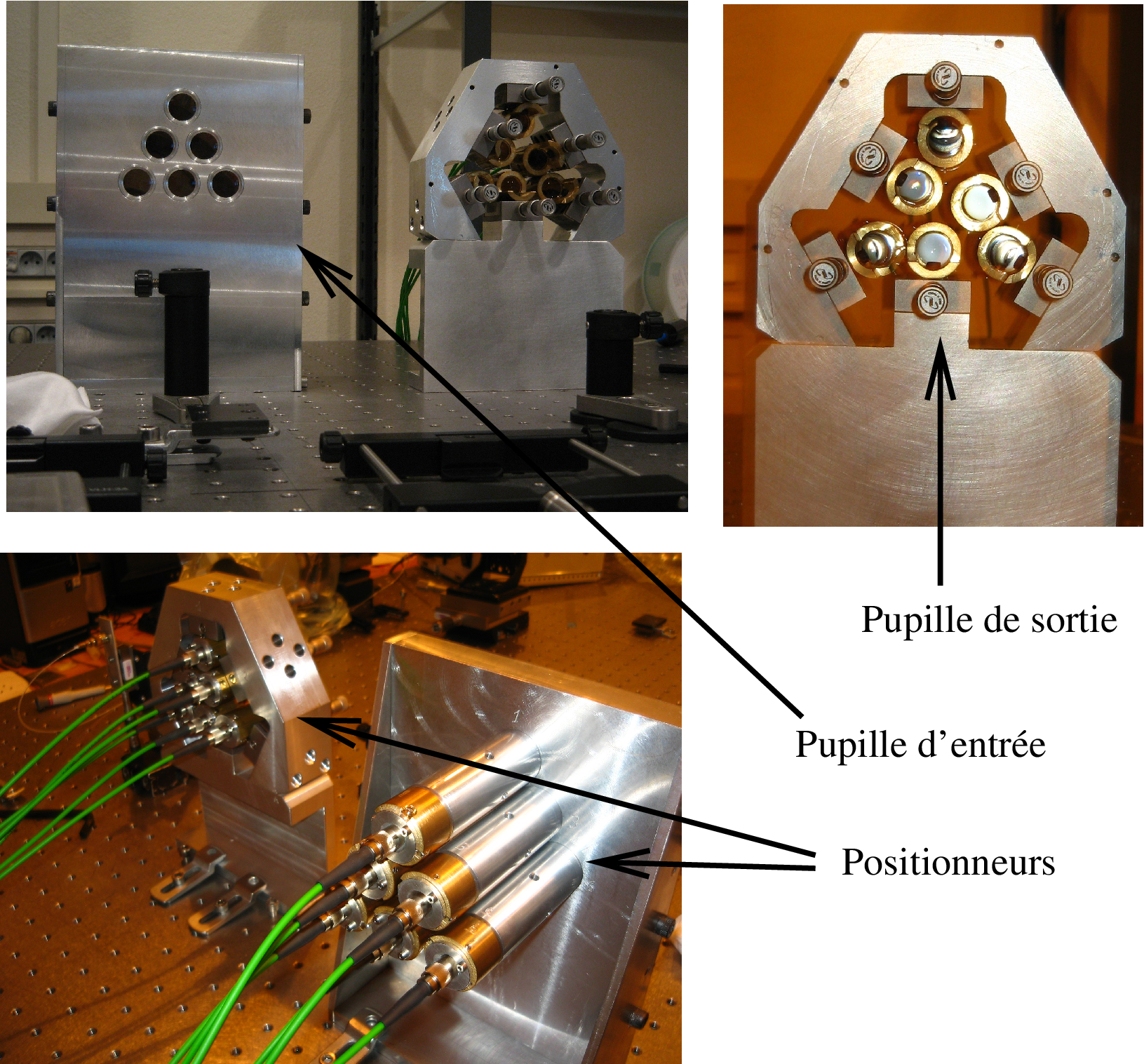}}
   \caption[Dernière génération de l'instrument]{ Dernière génération
         de l'instrument, construit dans le but simple de valider le
         concept. A gauche, on peut voir la pupille d'entrée, composée
         de 6 sous-pupilles disposées selon un triangle équilatéral. A
         droite, les 6 sous-pupilles sont réarrangées pour former une
         configuration non-redondante. Les fibres sont ici
         connectorisées, assurant une souplesse d'utilisation
         nécessaire pour tester différents types de fibres. }
         \label{fig:manip2} \end{figure}

Nous avons construit, au cours de cette année 2006, une version
de démonstration de taille décimétrique, permettant plus de souplesse
pour la caractérisation de la technique. Contrairement aux premières versions,
l'objectif n'est plus d'obtenir un système testable sur le ciel, mais
plus simplement d'établir une démonstration de la validité du
concept. L'objectif second est de concevoir un instrument suffisamment
modulable pour tester différents types de fibres, à différentes
longueurs d'onde. 

Ce système est actuellement en cours de montage. Il est présenté par
des photos figure~\ref{fig:manip2}, où l'on peut voir qu'il est
composé de seulement 6 sous-pupilles. Chaque fibre est montée sur un
micropositionneur 2 axes développés au LESIA, et réalisé au GEPI. Les
fibres ont été égalisées au LAOG (Laboratoire d'Astrophysique de
l'Observatoire de Grenoble) avec une précision inférieure au centième
de micron. Un banc de connectorisation est actuellement en
construction, pour permettre le test de différents types de
fibre. Enfin, le système de connecteur est conçu pour, éventuellement,
tester des fibres optiques à maintien de polarisation.

Les premiers tests de cet instrument sont en cours.

\cleardoublepage
\begin{figure}
  \centering  
\resizebox{\hsize}{!}{\includegraphics{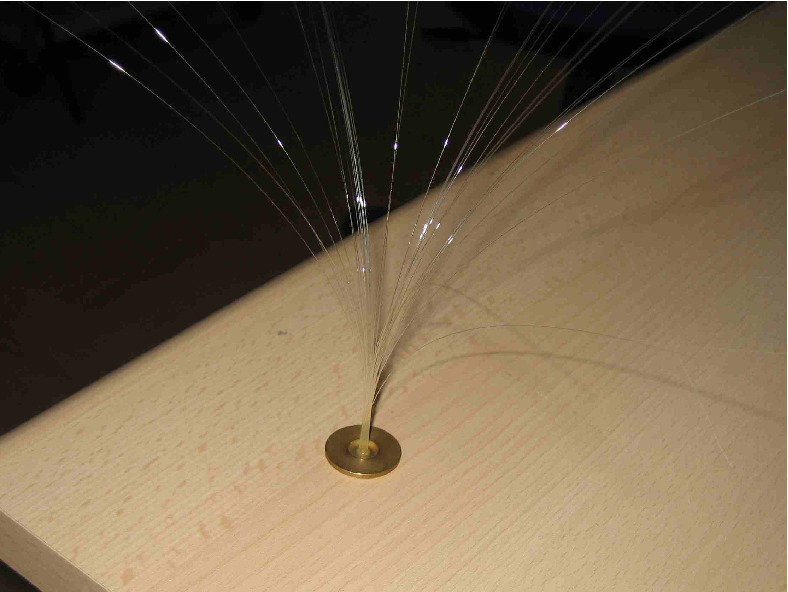}}
\end{figure}

\chapter*{Conclusion générale}
\markboth{Conclusion}{Conclusion}
\addcontentsline{toc}{part}{{\large\sf\bfseries Conclusion}}
\label{conclu}
 Mon travail de thèse a consisté à combiner deux concepts, le masquage
de pupille et le filtrage par fibres optiques monomodes, afin
d'aboutir à l'élaboration d'un instrument alliant haute résolution
spatiale et haute dynamique. Le développement technique de cet
instrument a été accompagné de recherches approfondies afin, dans un
premier temps, d'être convaincu de l'intérêt d'un tel système, puis,
de persuader mon entourage scientifique de son utilité. Il a fallu
comparer les spécificités de cette combinaison technique avec des
systèmes déjà existants comme l'optique adaptative ou
l'interférométrie des tavelures permettant d'obtenir de la haute
résolution angulaire. Ainsi, alors que ma thèse n'aurait pu être
qu'instrumentale, elle a également été consacrée à la justification,
au sens large, de l'instrument. Ceci s'est traduit par trois domaines
de recherches menés en parallèle :
\begin{itemize}
\item La construction d'un prototype (chapitre 5 et 6)
\item Le développement d'un algorithme dédié (chapitre 4)
\item L'utilisation d'un observatoire interférométrique afin de positionner cet 
instrument dans un contexte astrophysique spécifique (chapitre 2 et 3)
\end{itemize}

Reprenons brièvement chacun de ces domaines.  La construction du
prototype a nécessité une première année de travaux qui a permis de
définir un instrument optimisé pour fonctionner sur un télescope de 8
mètres, observant dans le visible, en présence de turbulences
atmosphériques moyennes. Nous avons tenté de mettre en \oe uvre une
configuration composée de 36 sous-pupilles, selon une géométrie proche
de celle du miroir primaire du télescope Keck, à Hawaï. La
réalisation de l'instrument a débuté pendant la deuxième année. Le
principal point d'achoppement a été de positionner l'ensemble des
fibres de façon mécanique à des précisions inférieures au micron. Nous
nous sommes alors aperçus que cela nécessitait un développement
spécifique, extrêmement exigeant au point de vue technologique. C'est
pourquoi il nous est apparu nécessaire, le temps nous étant compté, de
nous diriger vers un système plus simple, composé uniquement de 6
fibres, mais permettant de tester le principe. Nous travaillons encore
sur cet instrument, qui est sur le point de fournir ses premiers
résultats.

En parallèle à ce développement technologique, nous avons entrepris la
publication du concept d'un tel instrument. Or, les performances
dépendent énormément de l'algorithme de réduction des données. Afin de
pouvoir publier des simulations crédibles, il fallait présenter une
technique de réduction appropriée, permettant d'obtenir le maximum des
capacités de l'instrument. Nous nous sommes alors aperçus que les
techniques existantes, telles celle du bispectre, ne permettait pas
une reconstruction optimale. C'est pourquoi, en collaboration avec
l'Observatoire de Lyon, nous avons développé un algorithme permettant
d'obtenir le maximum de vraisemblance au sens du moindre carré de
manière simultanée sur plusieurs milliers d'acquisitions. Ce travail a
permis de démontrer que l'on pouvait éliminer le bruit de turbulence
(bruit de ``speckle''). La limitation par les seuls bruits de photon
et de detecteur est la preuve de capacités potentielles importantes
pour l'imagerie à très haute dynamique à la limite de diffraction des
télescopes.

Il est, par ailleurs, important, lors de la conception d'un instrument
interférométrique, de bénéficier de l'expérience des instruments déjà
existants. Les travaux préexistants sont d'une aide précieuse lors de
la concrétisation d'une idée qui, jusqu'alors, n'existait que sur le
papier.  Nous avons pris une part active dans l'acquisition, le
traitement, et l'analyse de données interférométriques issues de
l'interféromètre IOTA. Plus qu'une simple initiation à
l'interférométrie, cette partie de ma thèse s'est révélée des plus
intéressantes d'un point de vue intellectuel comme scientifique. Notre
approche de l'interférométrie, passant par l'imagerie en aveugle des
surfaces stellaires, n'avait alors jamais été entreprise
auparavant. \`A la différence des autres observateurs, qui préféraient
multiplier les observations sur différents objets, nous nous sommes
attachés à observer les mêmes étoiles de façon fréquente, en utilisant
les multiples bases de l'interféromètre. Nous avons ainsi pu obtenir
les images d'une série de sept étoiles évoluées révélant des
structures extrêmement complexes. Nos résultats sur Chi Cyg sont
particulièrement intéressants et novateurs. Les images ont été
obtenues à plusieures époques, et ont permis la mesure du déplacement
de la couche moléculaire. Nous avons ainsi calculé, et ce pour la
première fois, la masse de l'étoile à partir du déplacement de la
matière dans l'atmosphère étendue de celle-ci.

Le travail effectué au cours de cette thèse a ainsi une finalité qui
lui est propre. Cependant, l'objectif est à plus long terme, avec la
réalisation d'un instrument à réarrangement de pupille permettant
l'expérimentation sur le ciel. Cela va nécessiter toujours plus
d'investissements, auxquels je compte bien participer.

\appendix
\addtocontents{toc} {\protect\newpage\protect\begin{center}
{\protect\Large\protect\sf Annexes} \protect\end{center}
\protect\vskip -0.3cm }
\chapter{Articles sur le concept du réarrangement de pupille}
\vskip1cm \label{sec:artic1}

\cleardoublepage

\begin{figure}  \centering
  \resizebox{\hsize}{!}{\includegraphics{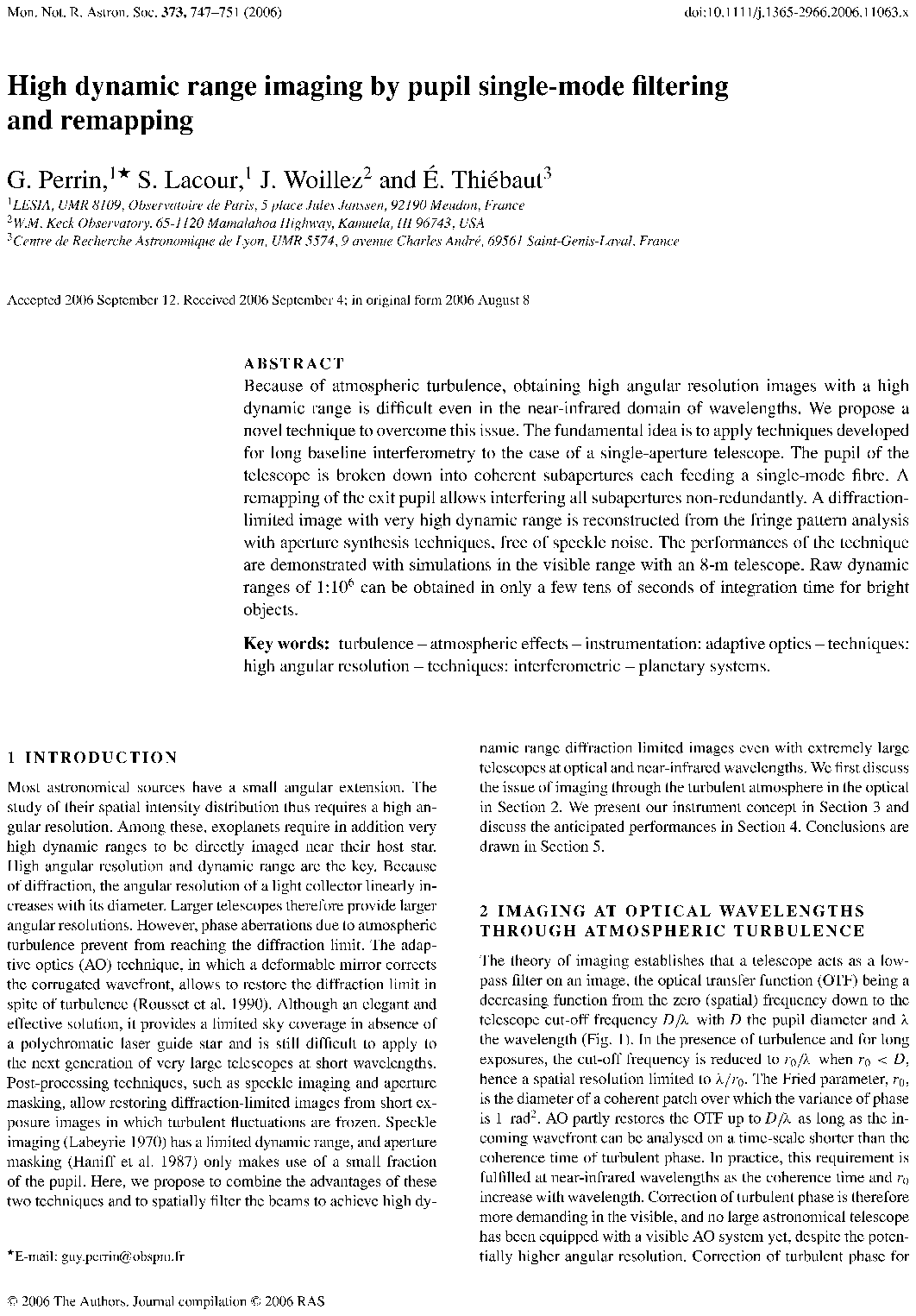}}
\end{figure}
\begin{figure}  \centering
  \resizebox{\hsize}{!}{\includegraphics{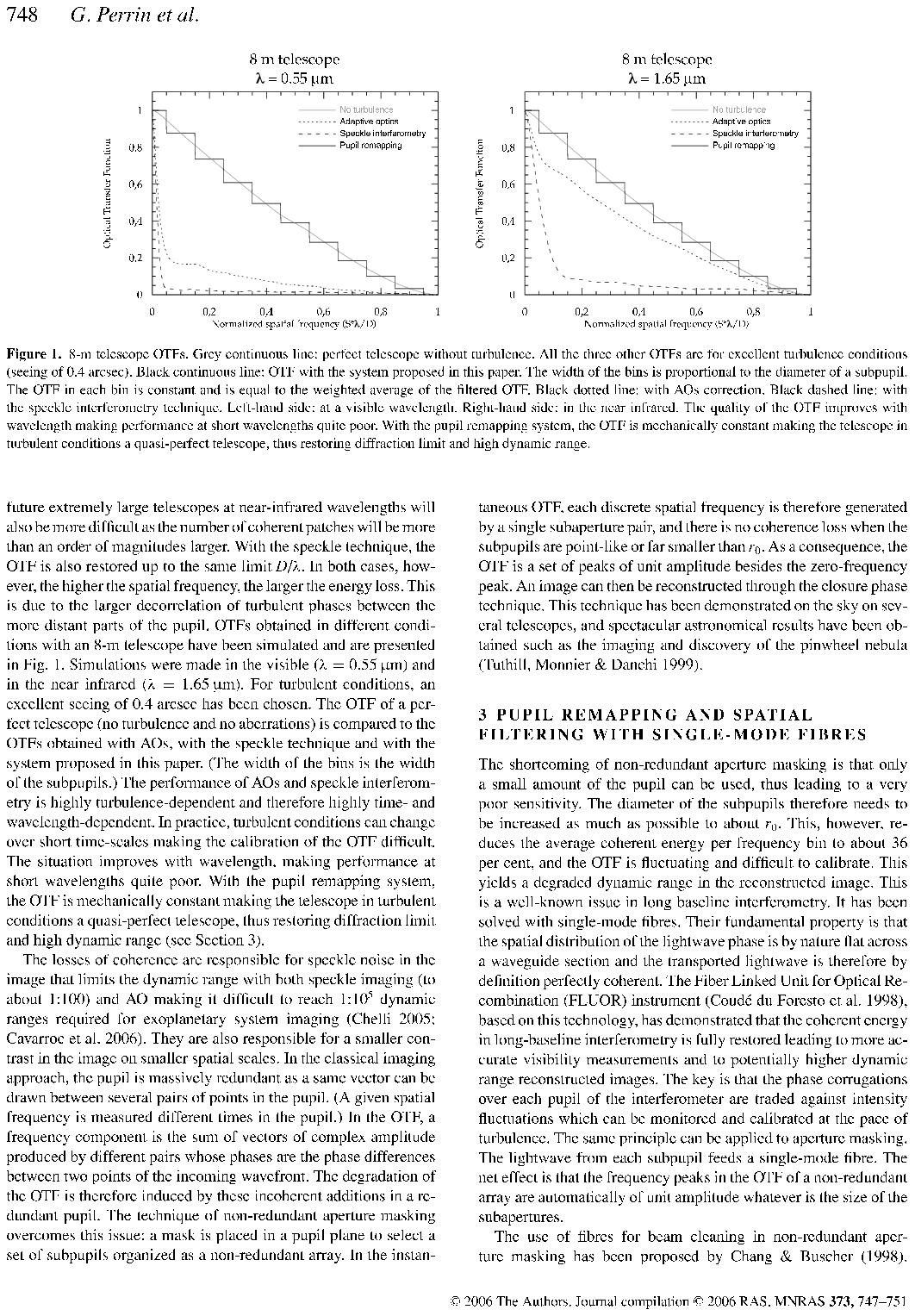}}
\end{figure}
\begin{figure}  \centering
  \resizebox{\hsize}{!}{\includegraphics{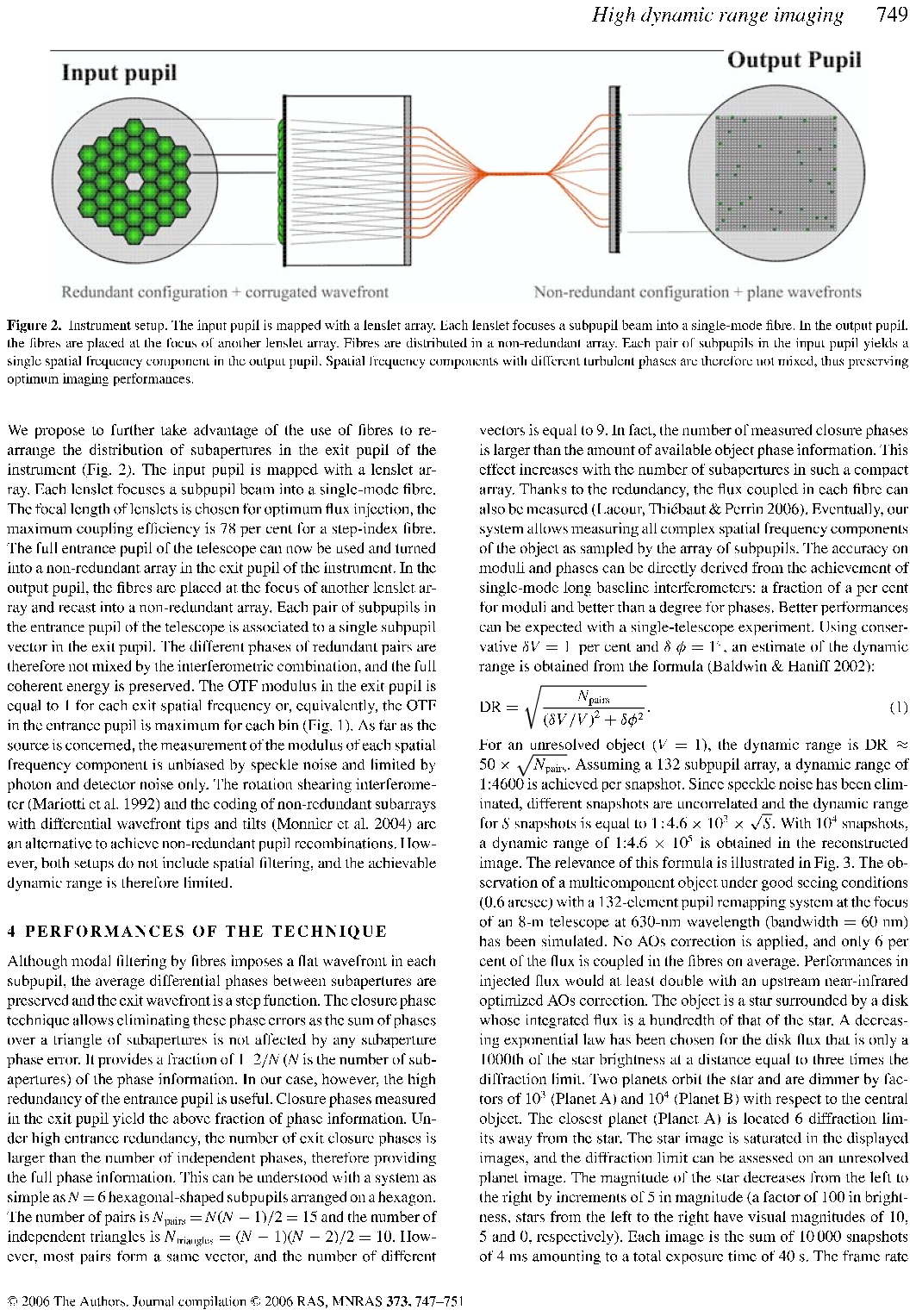}}
\end{figure}
\begin{figure}  \centering
  \resizebox{\hsize}{!}{\includegraphics{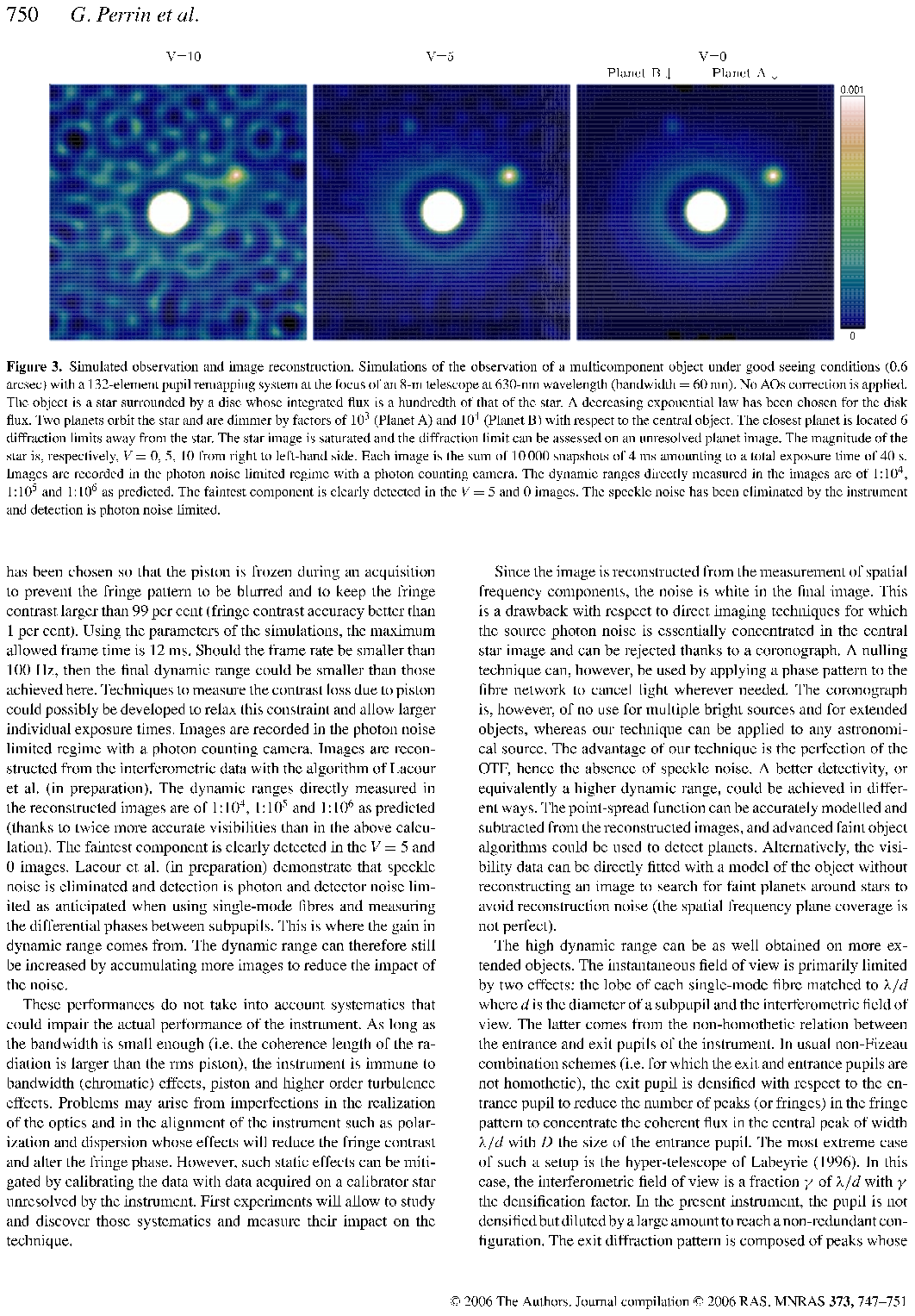}}
\end{figure}
\begin{figure}  \centering
  \resizebox{\hsize}{!}{\includegraphics{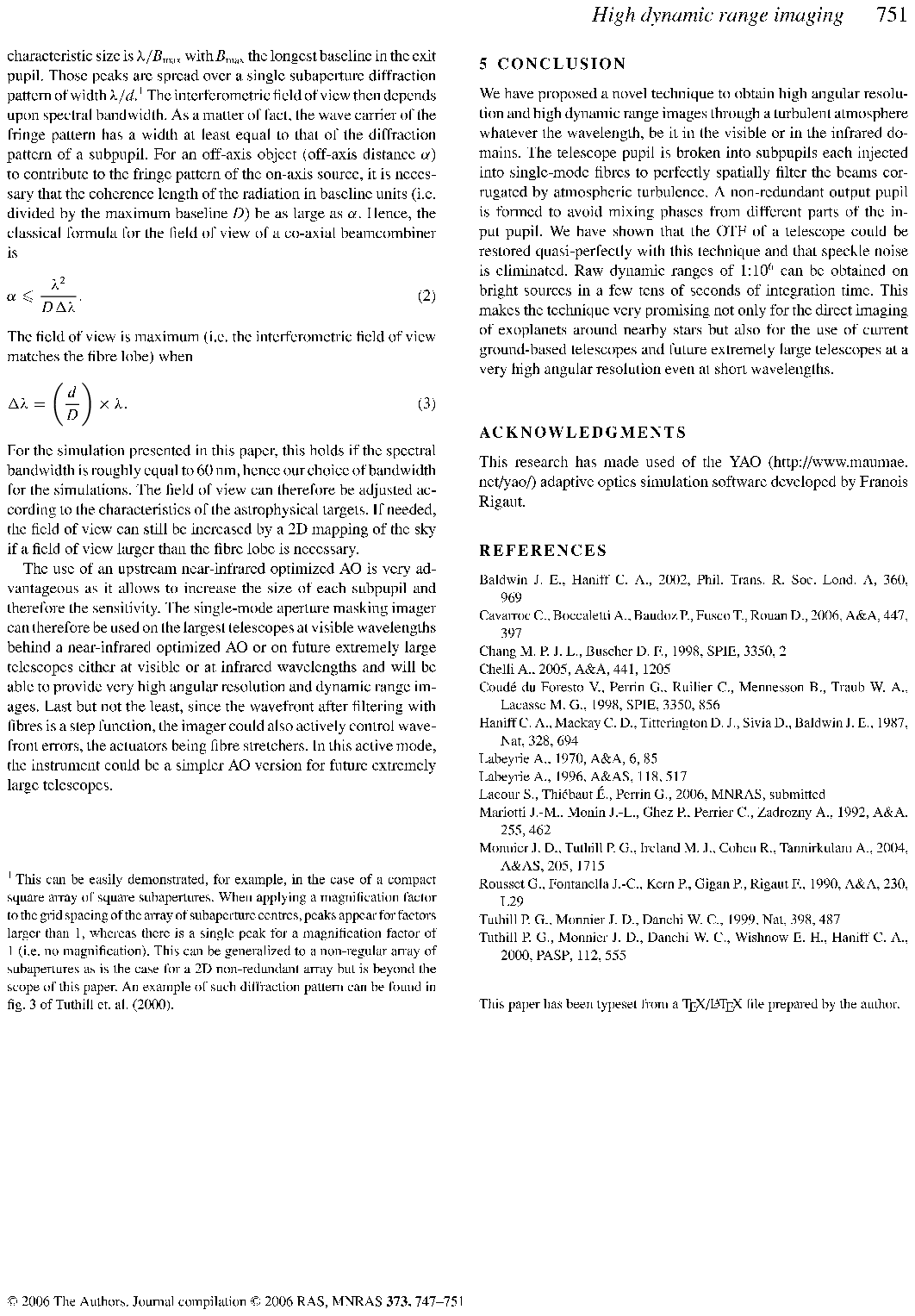}}
\end{figure}
\cleardoublepage

\begin{figure}  \centering
  \resizebox{\hsize}{!}{\includegraphics{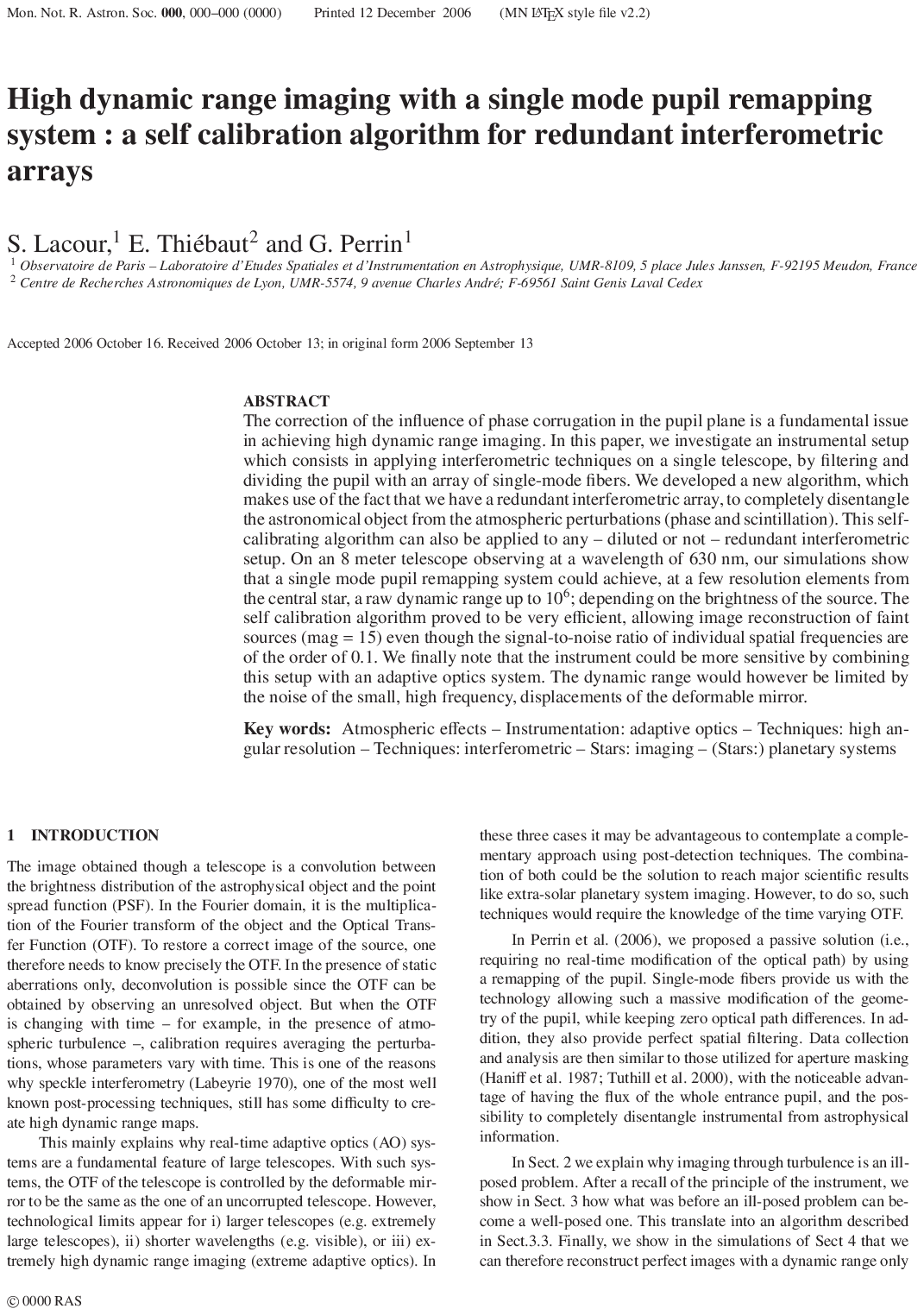}}
\end{figure}
\begin{figure}  \centering
  \resizebox{\hsize}{!}{\includegraphics{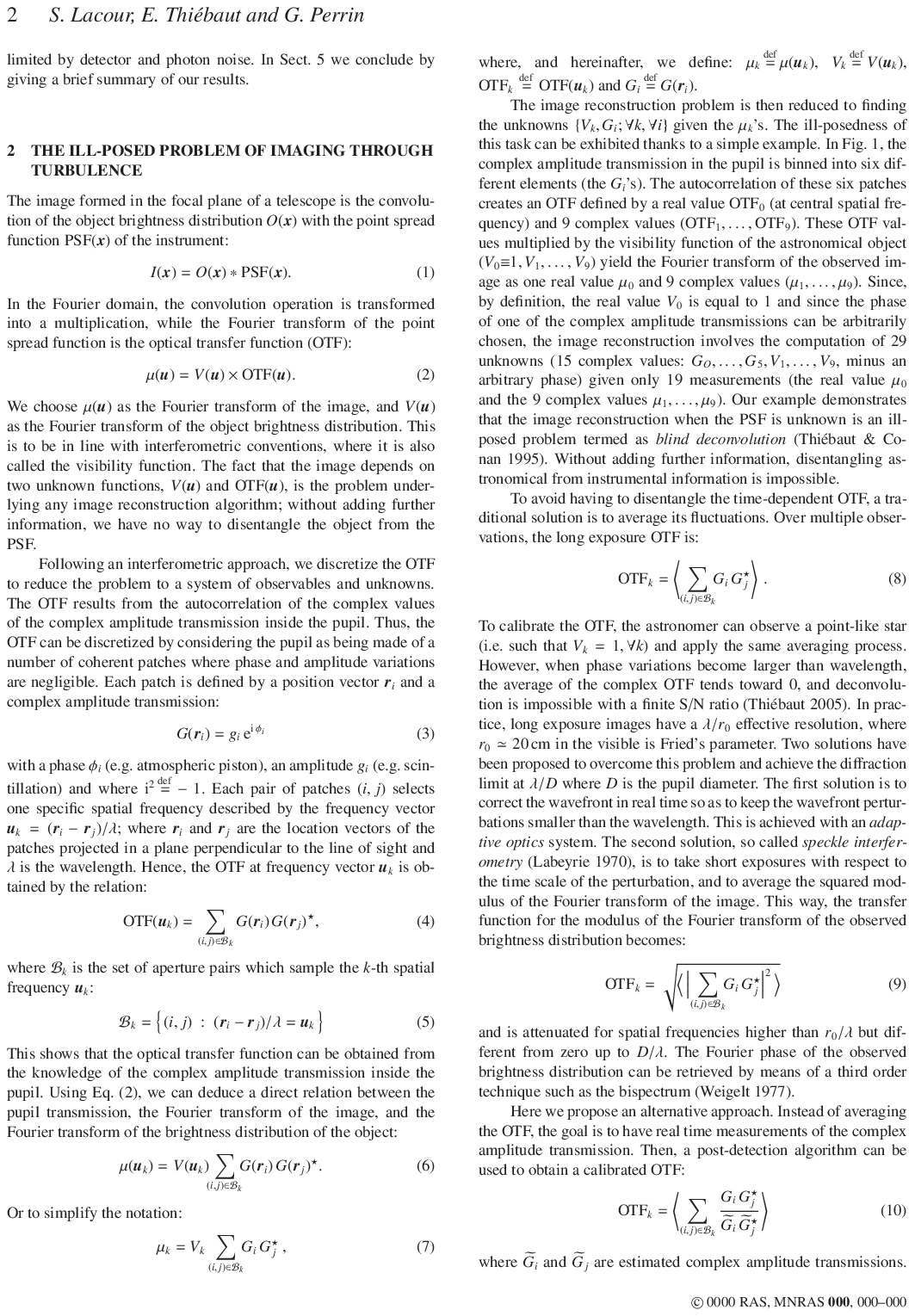}}
\end{figure}
\begin{figure}  \centering
  \resizebox{\hsize}{!}{\includegraphics{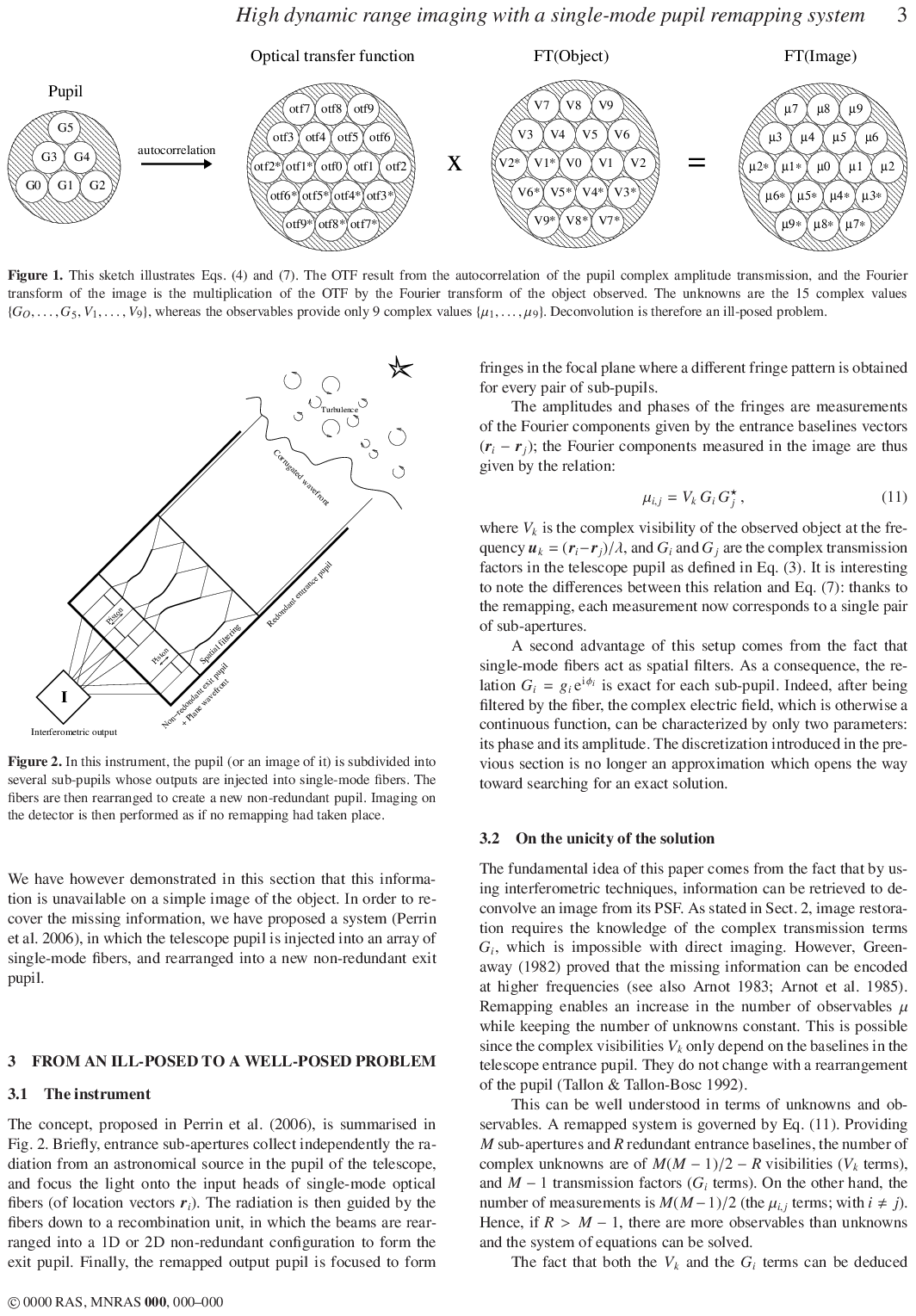}}
\end{figure}
\begin{figure}  \centering
  \resizebox{\hsize}{!}{\includegraphics{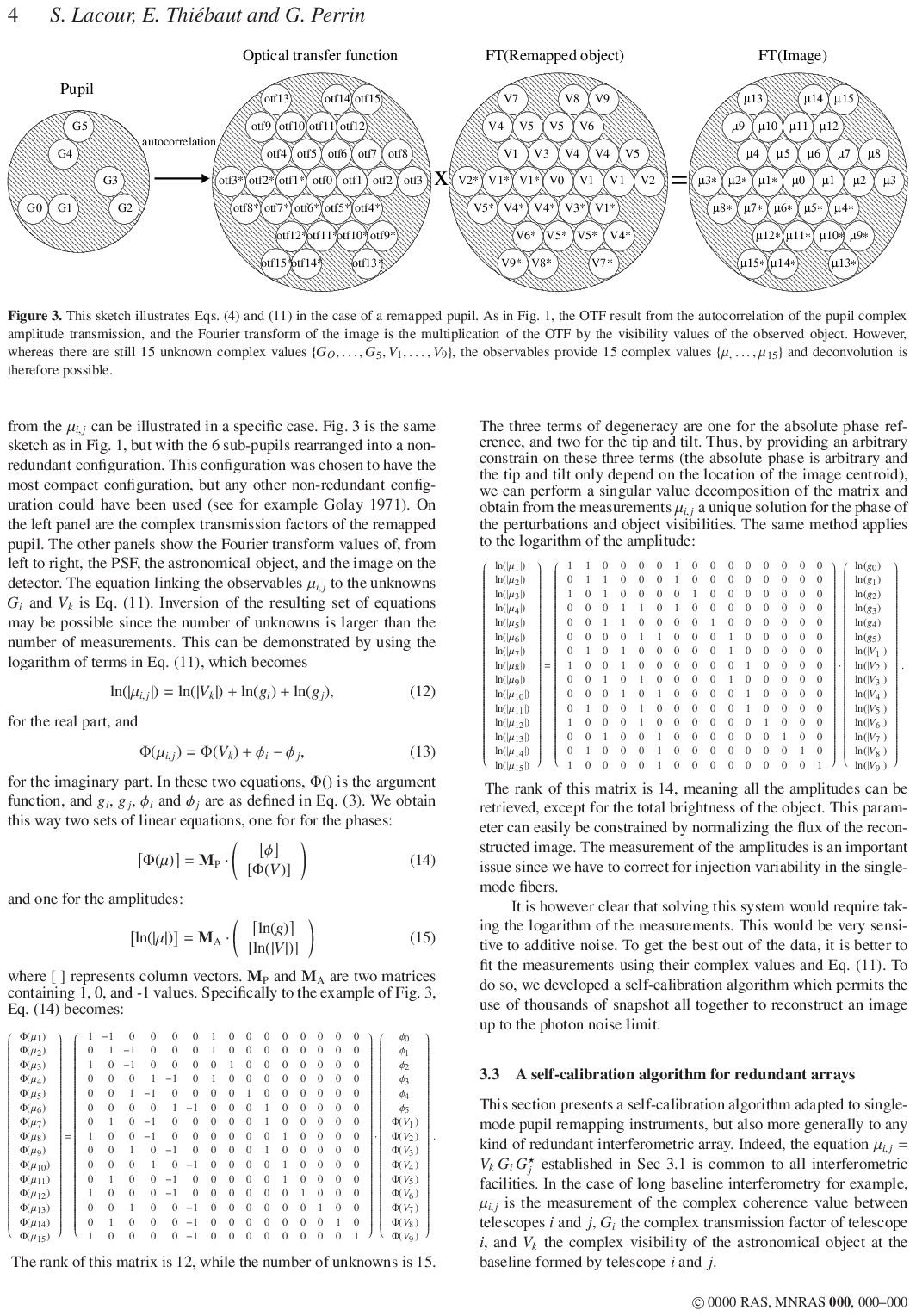}}
\end{figure}
\begin{figure}  \centering
  \resizebox{\hsize}{!}{\includegraphics{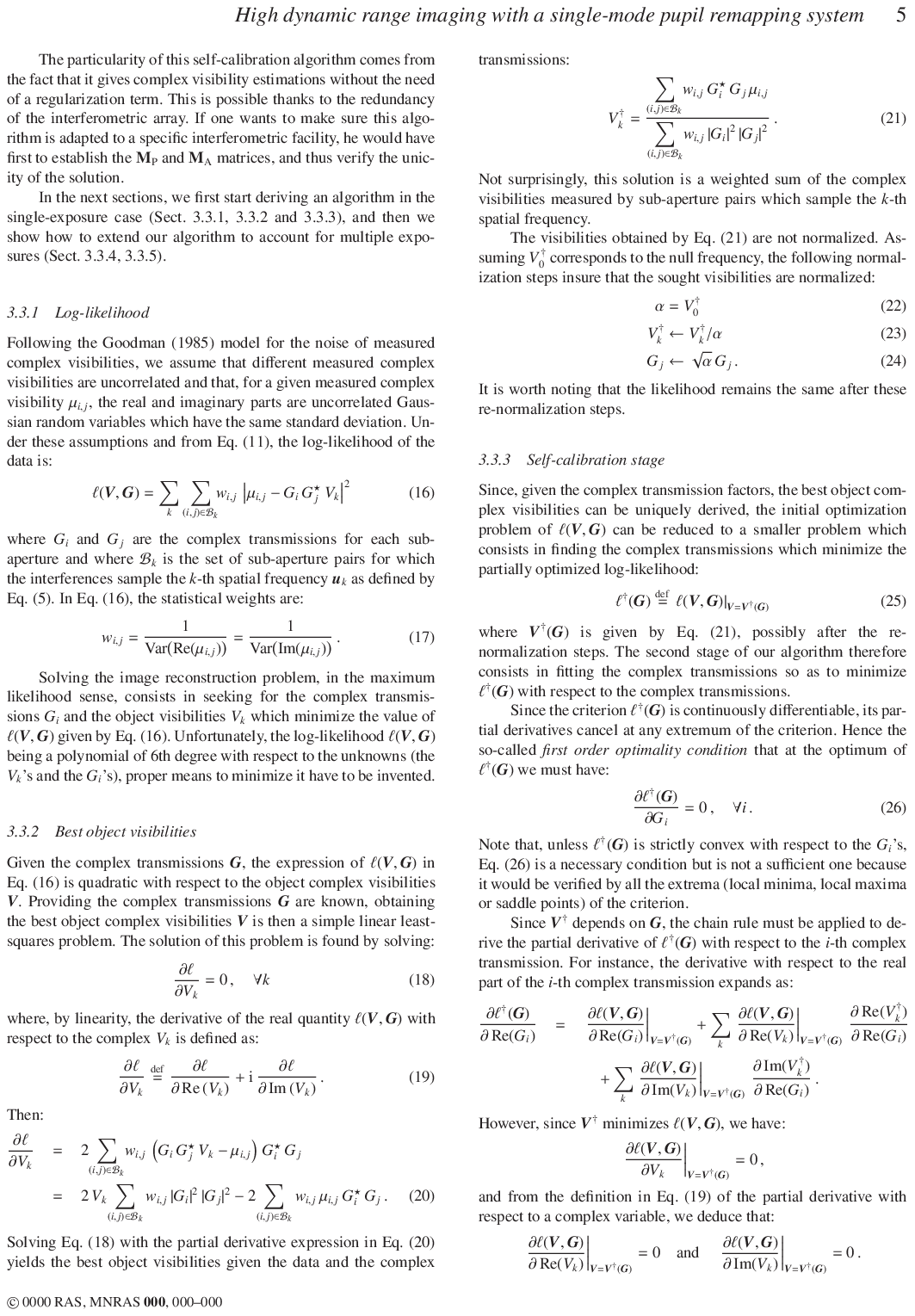}}
\end{figure}
\begin{figure}  \centering
  \resizebox{\hsize}{!}{\includegraphics{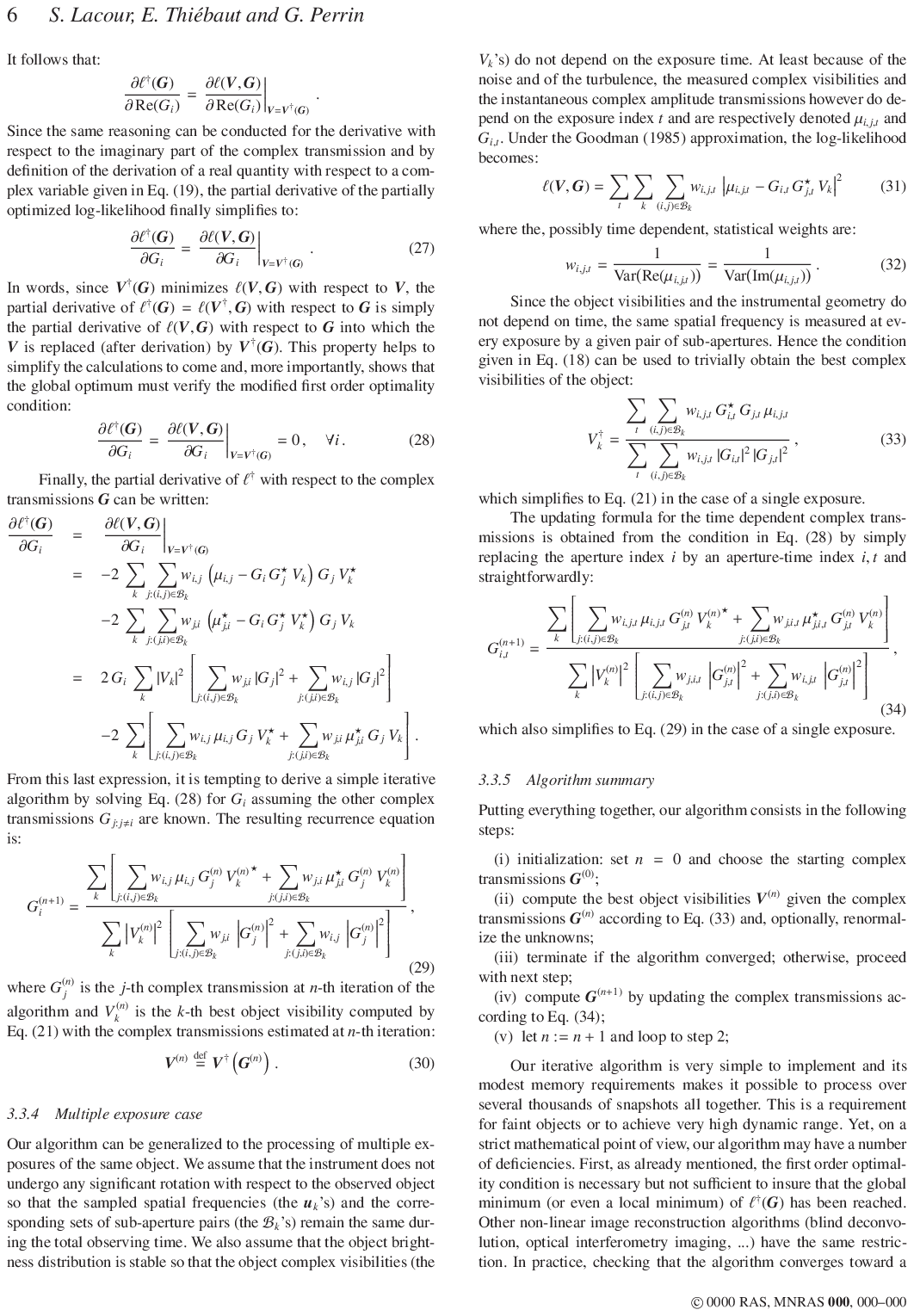}}
\end{figure}
\begin{figure}  \centering
  \resizebox{\hsize}{!}{\includegraphics{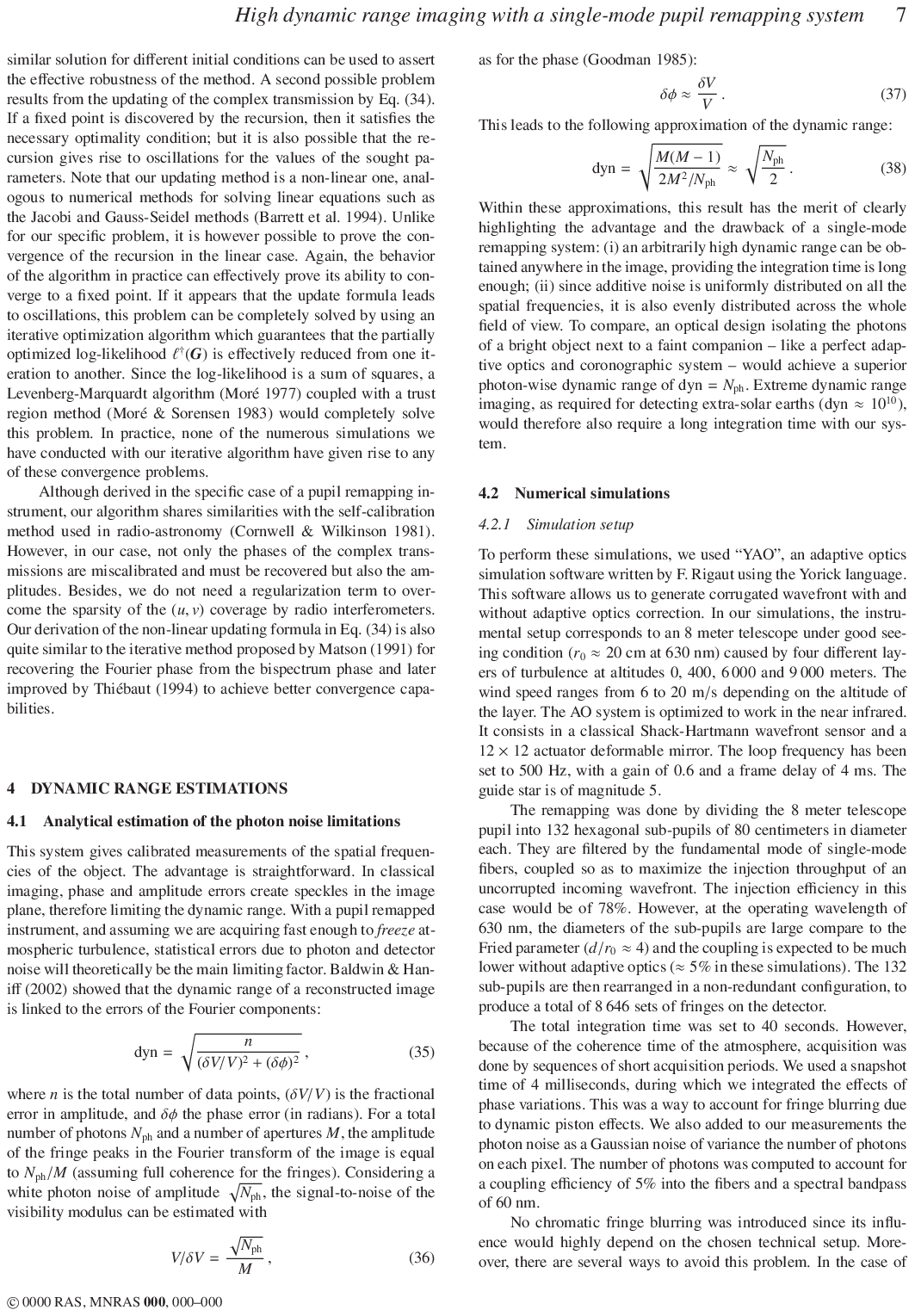}}
\end{figure}
\begin{figure}  \centering
  \resizebox{\hsize}{!}{\includegraphics{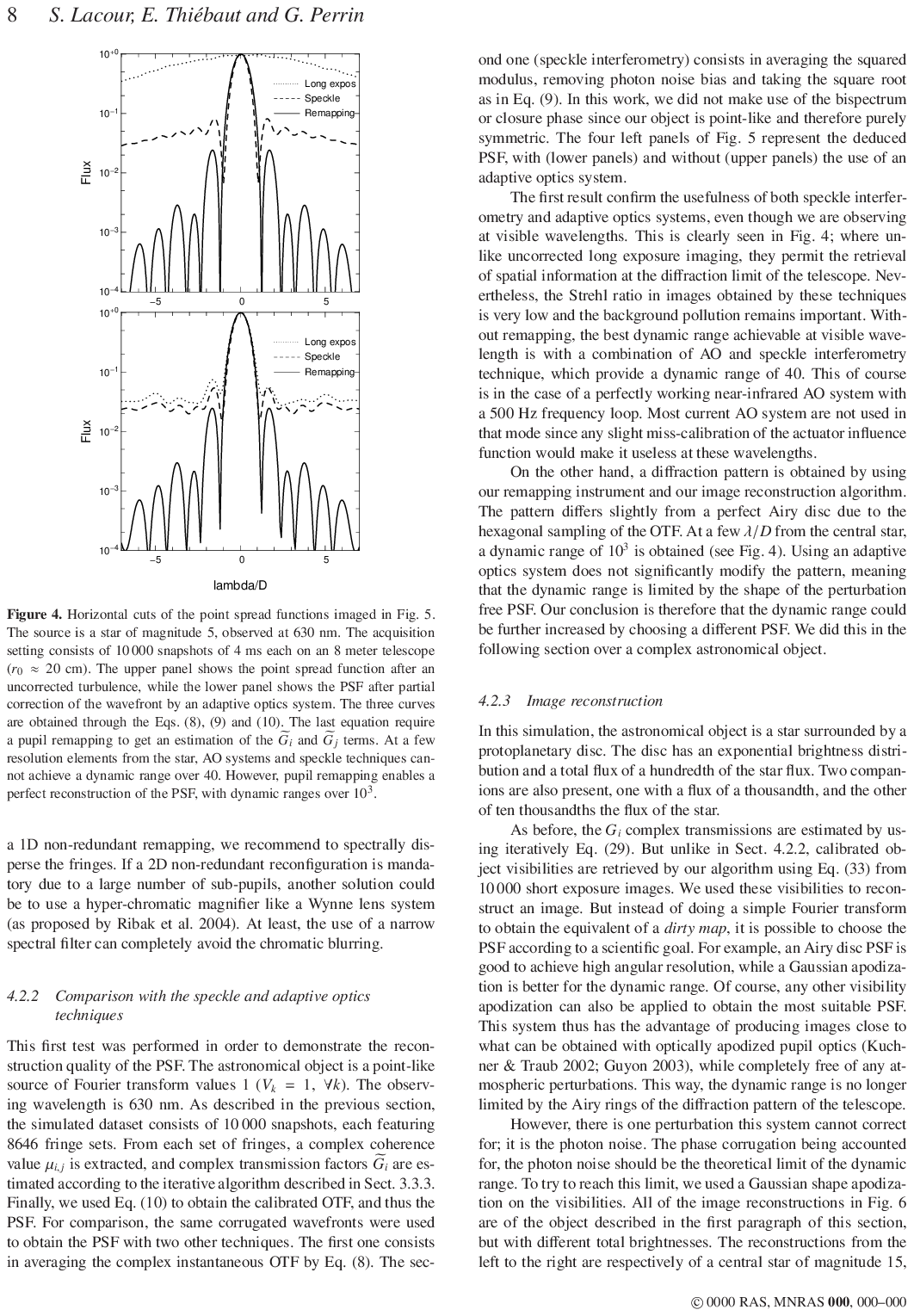}}
\end{figure}
\begin{figure}  \centering
  \resizebox{\hsize}{!}{\includegraphics{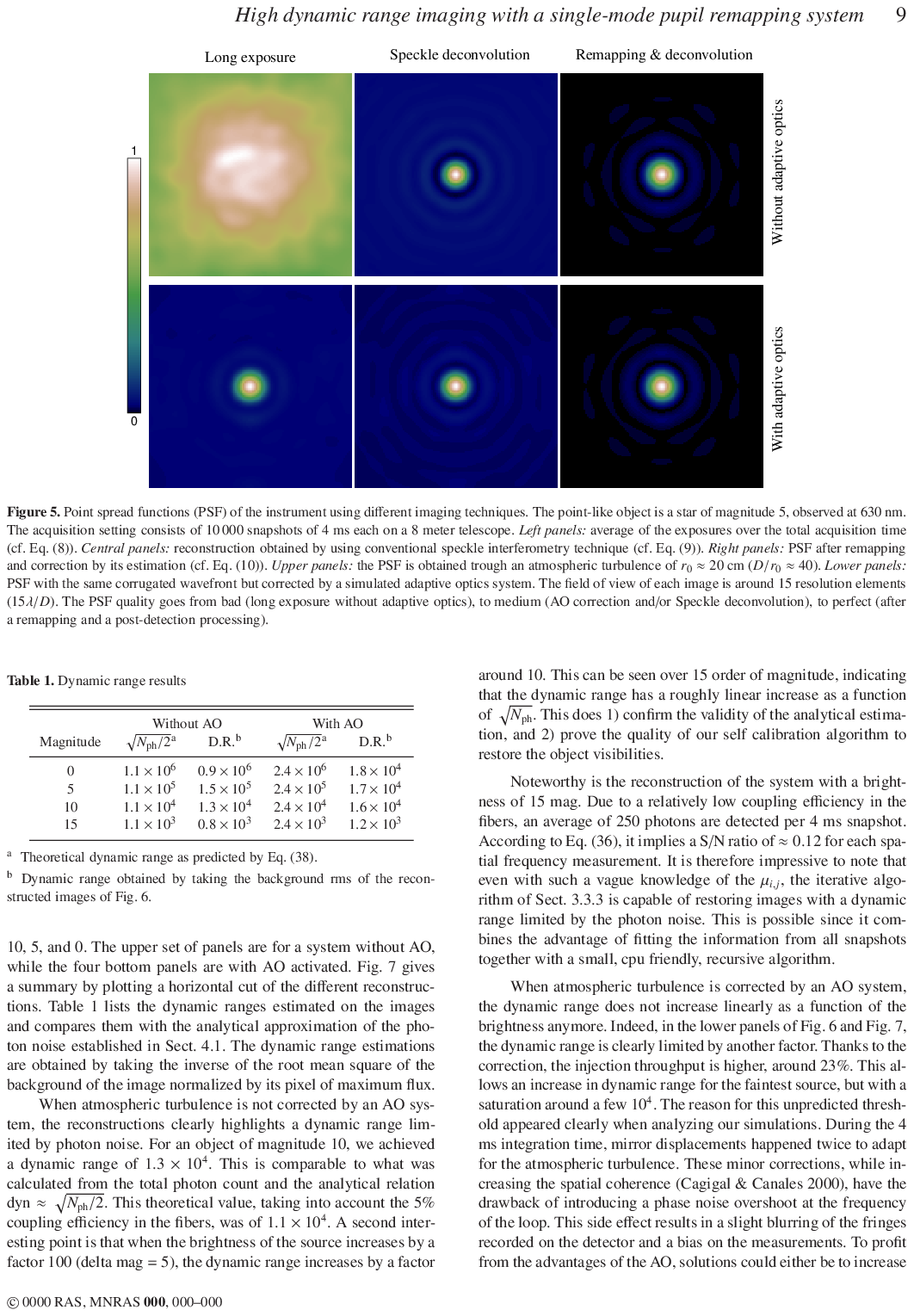}}
\end{figure}
\begin{figure}  \centering
  \resizebox{\hsize}{!}{\includegraphics{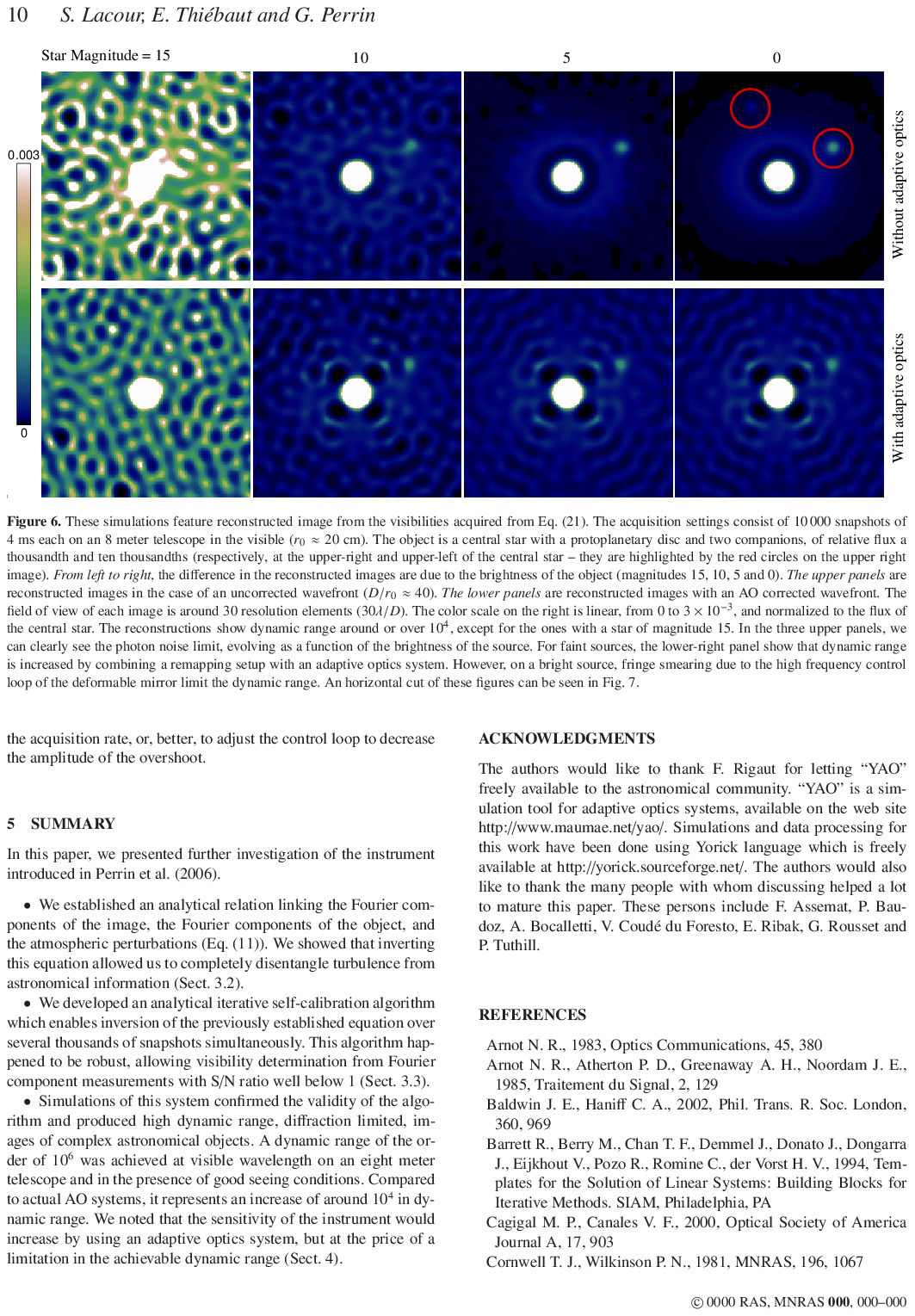}}
\end{figure}
\begin{figure}  \centering
  \resizebox{\hsize}{!}{\includegraphics{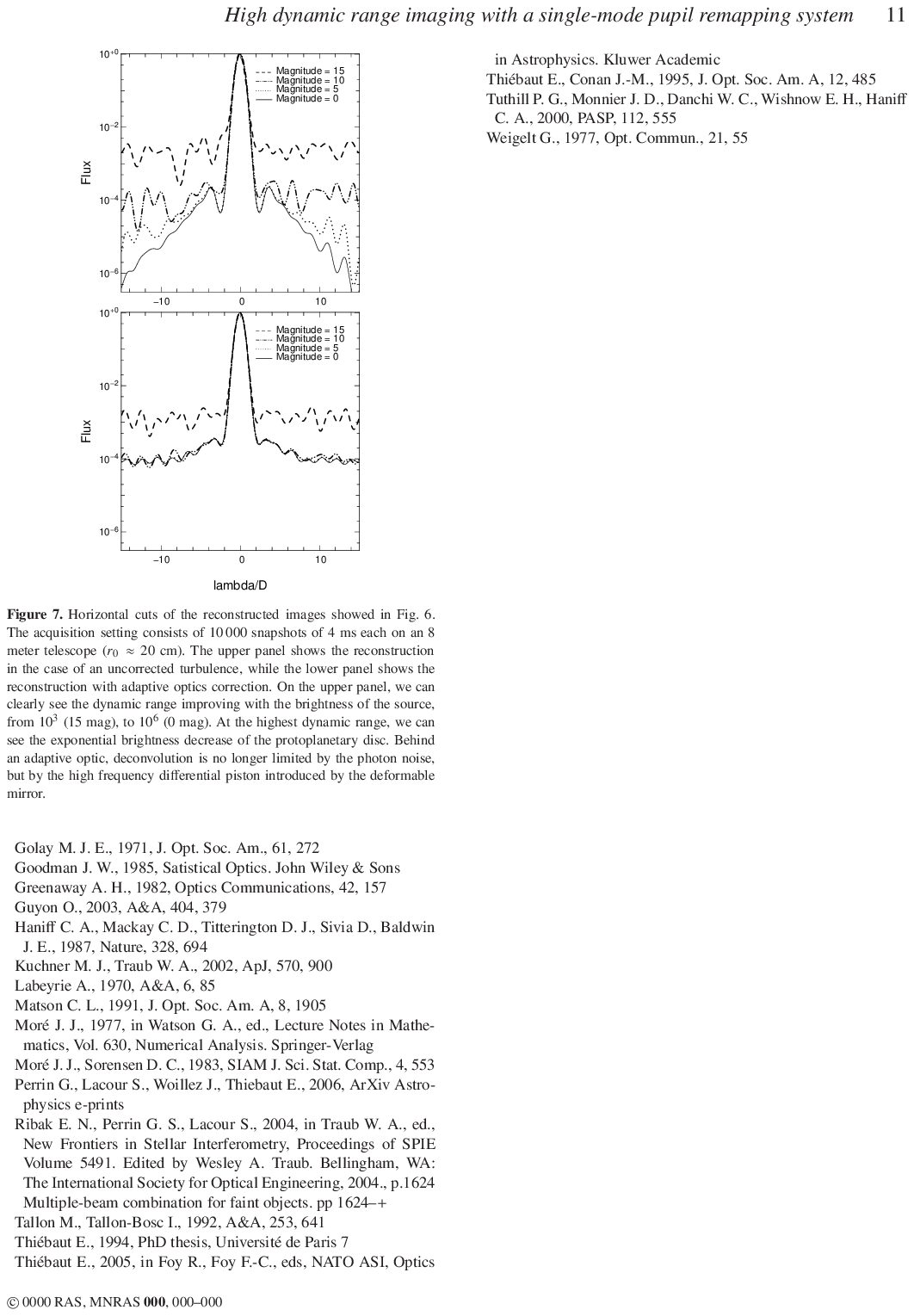}}
\end{figure}

\chapter{Autres publications}
\vskip1cm \label{sec:artic2}

Les deux articles qui suivent ne sont pas en rapport avec le sujet
central de mon travail de thèse. Cependant, ils correspondent à des
recherches qui, débutées avant le commencement de ma thèse, ont abouti
au cours de celle-ci.

\cleardoublepage
\begin{figure}  \centering
  \resizebox{\hsize}{!}{\includegraphics{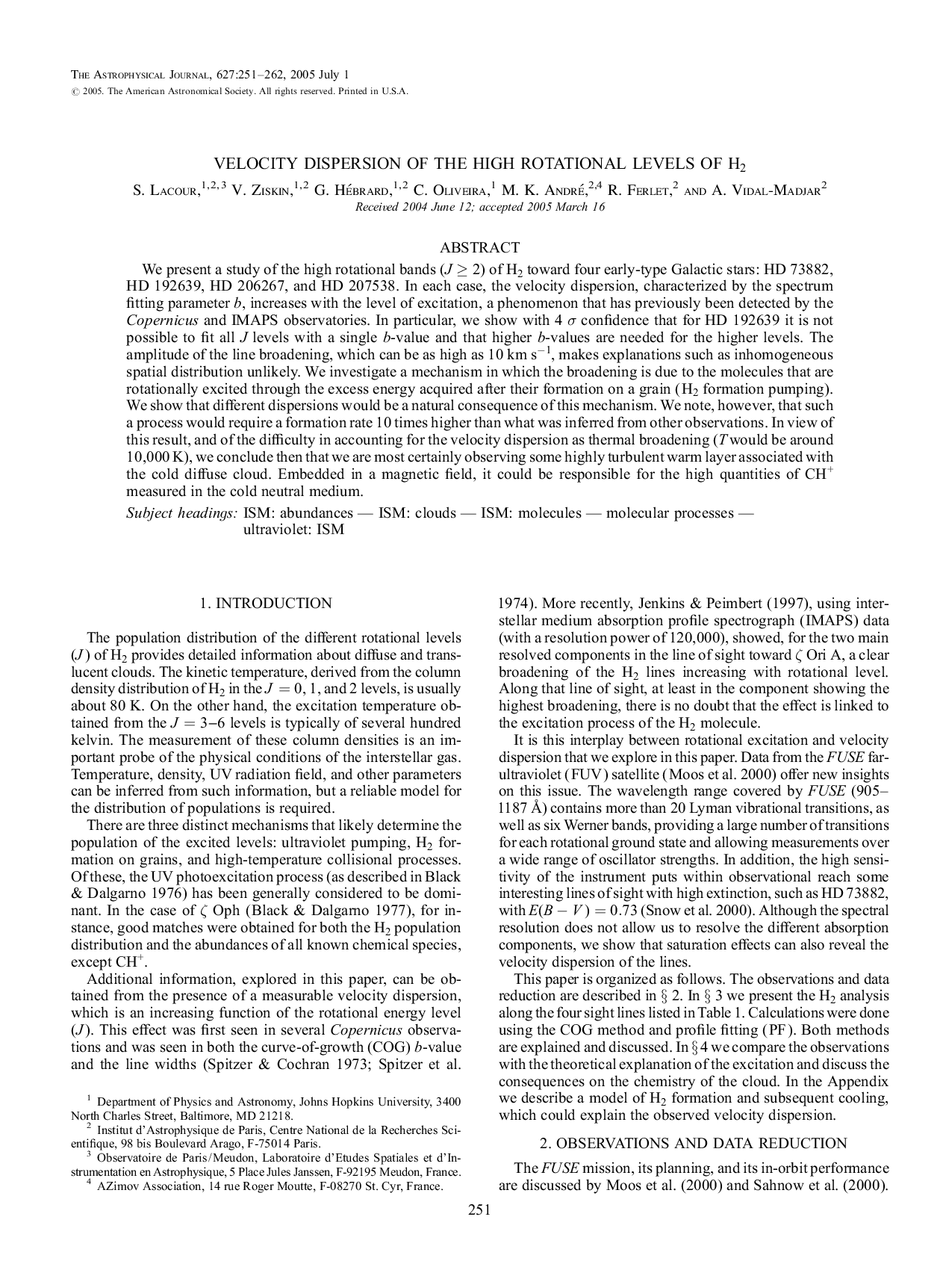}}
\end{figure}
\begin{figure}  \centering
  \resizebox{\hsize}{!}{\includegraphics{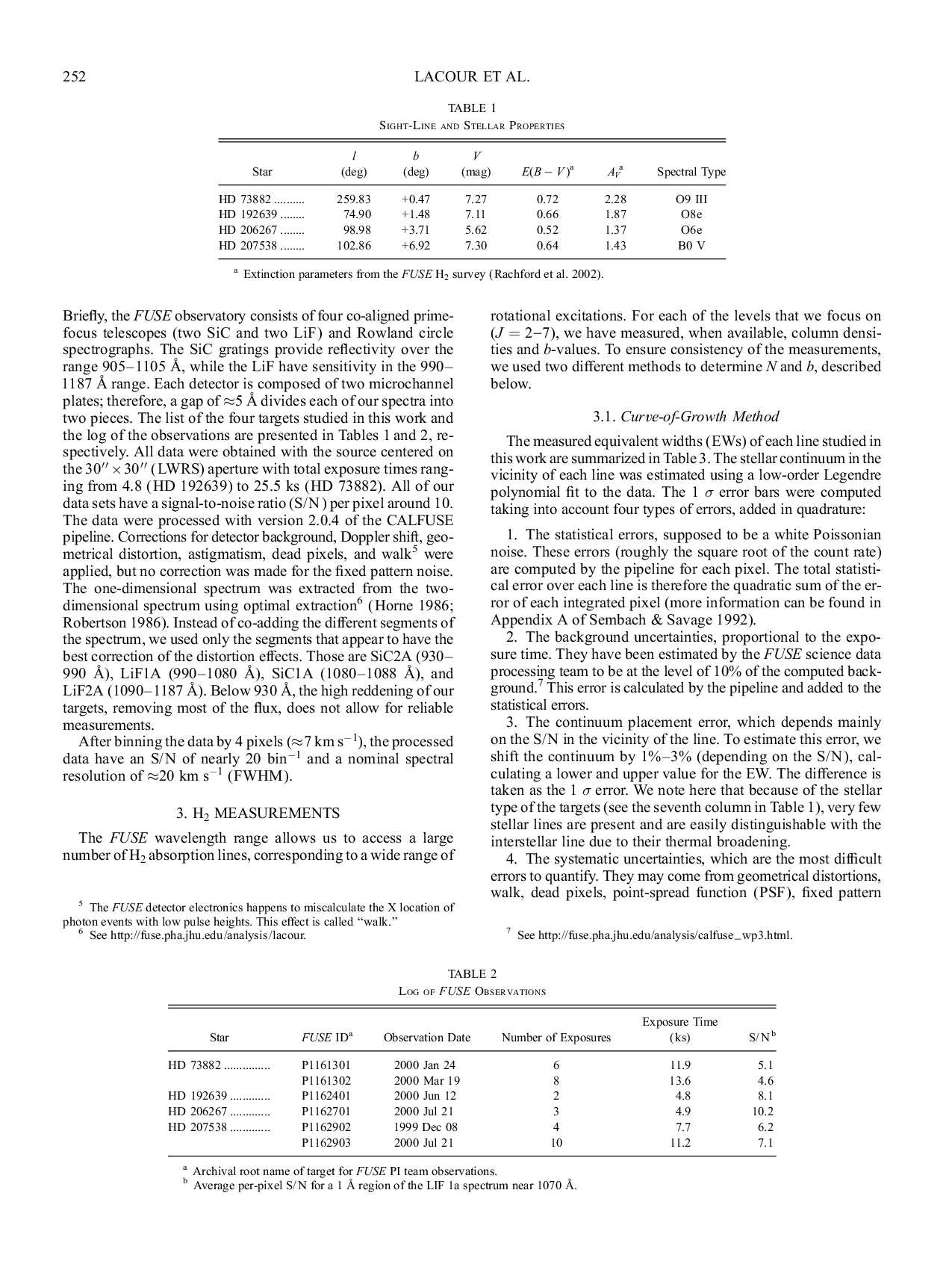}}
\end{figure}
\begin{figure}  \centering
  \resizebox{\hsize}{!}{\includegraphics{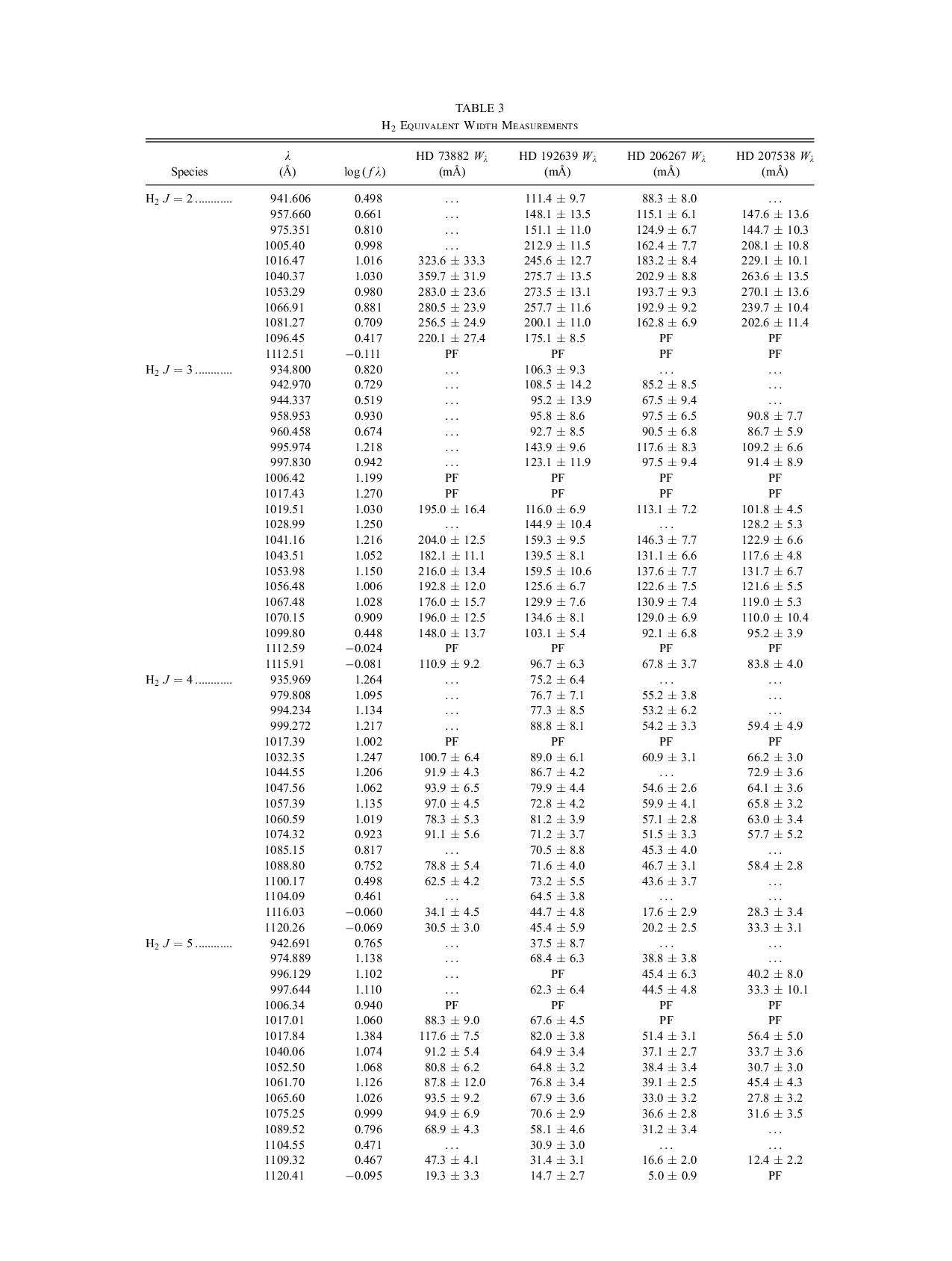}}
\end{figure}
\begin{figure}  \centering
  \resizebox{\hsize}{!}{\includegraphics{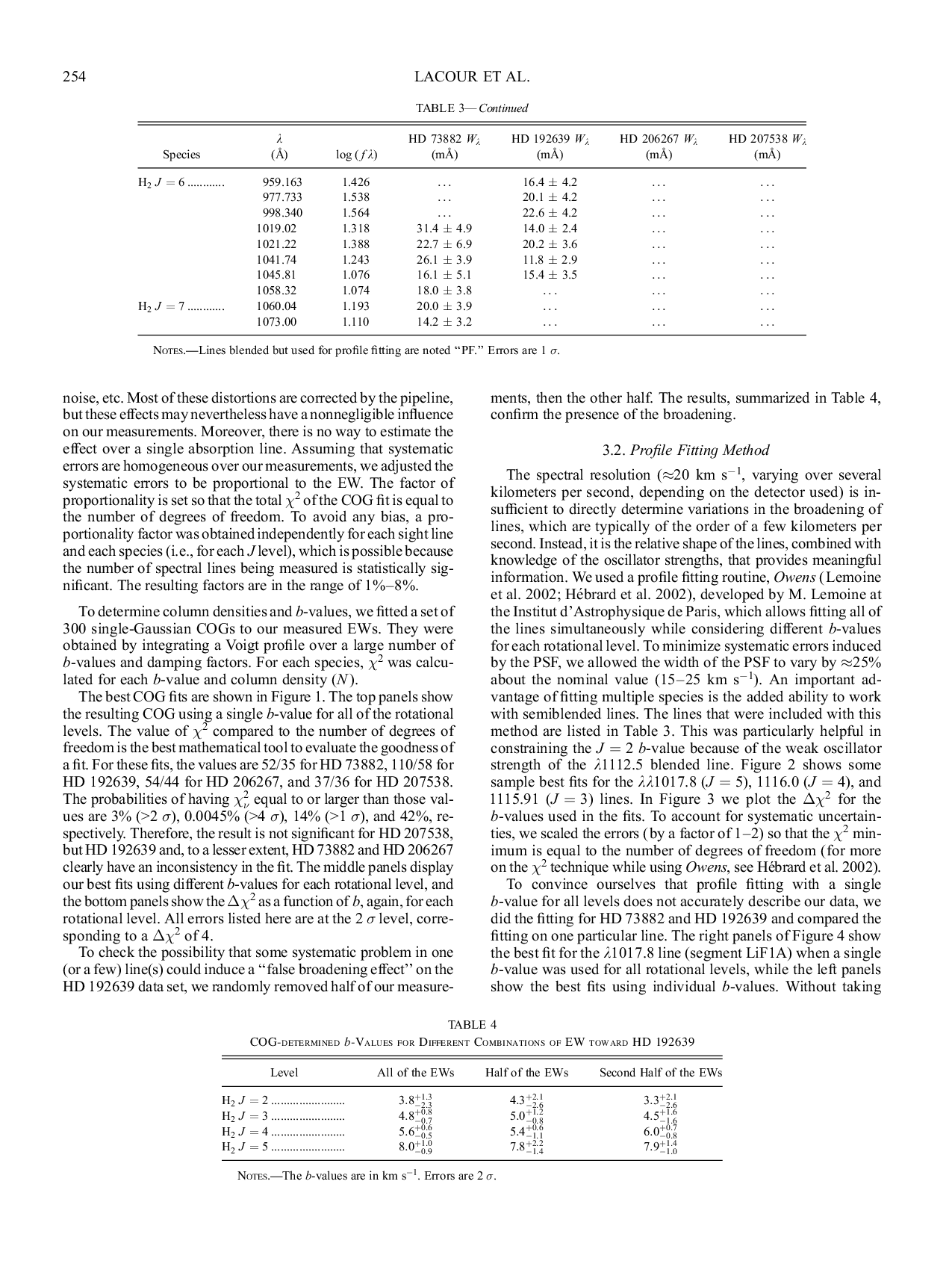}}
\end{figure}
\begin{figure}  \centering
  \resizebox{\hsize}{!}{\includegraphics{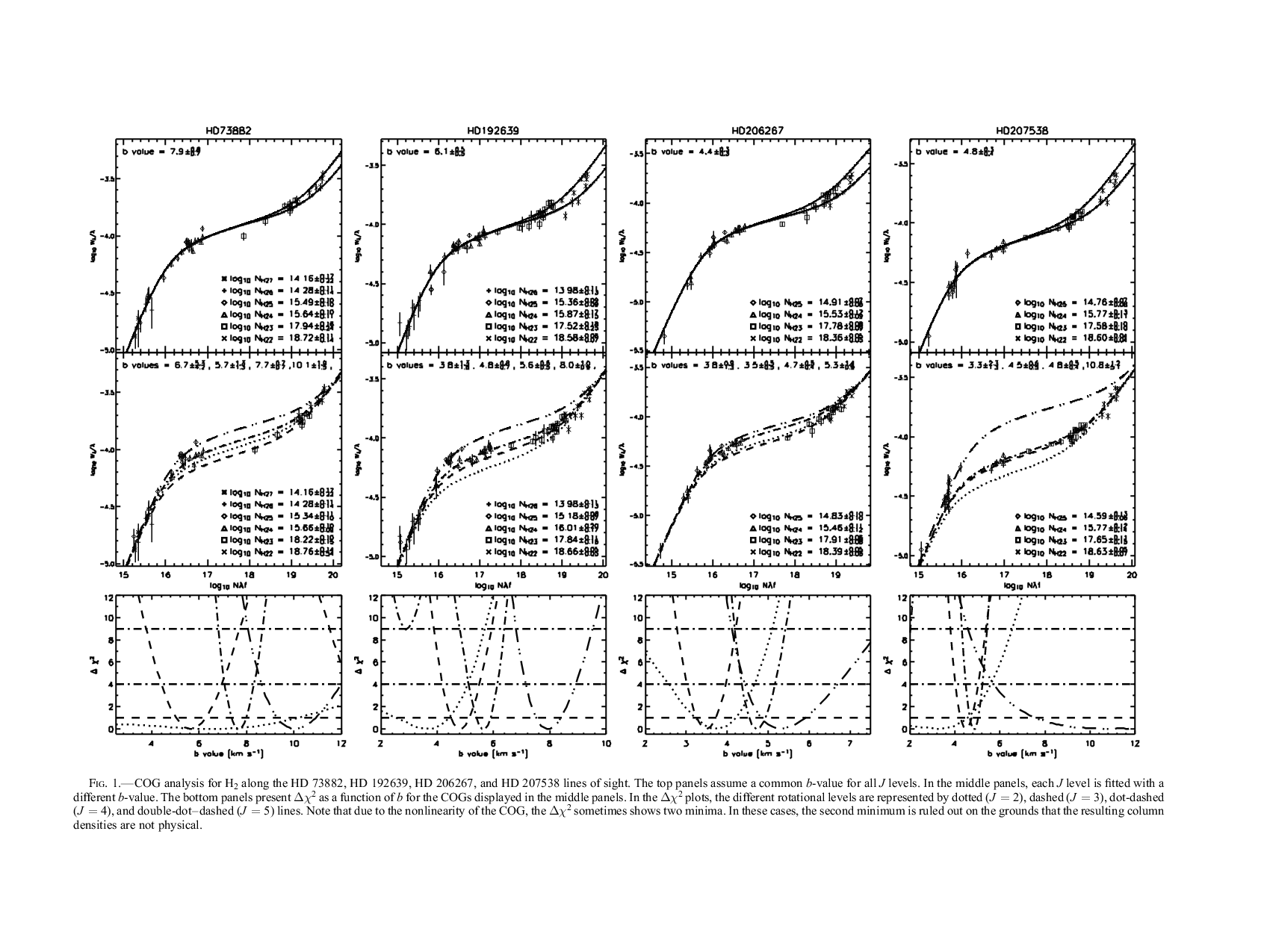}}
\end{figure}
\begin{figure}  \centering
  \resizebox{\hsize}{!}{\includegraphics{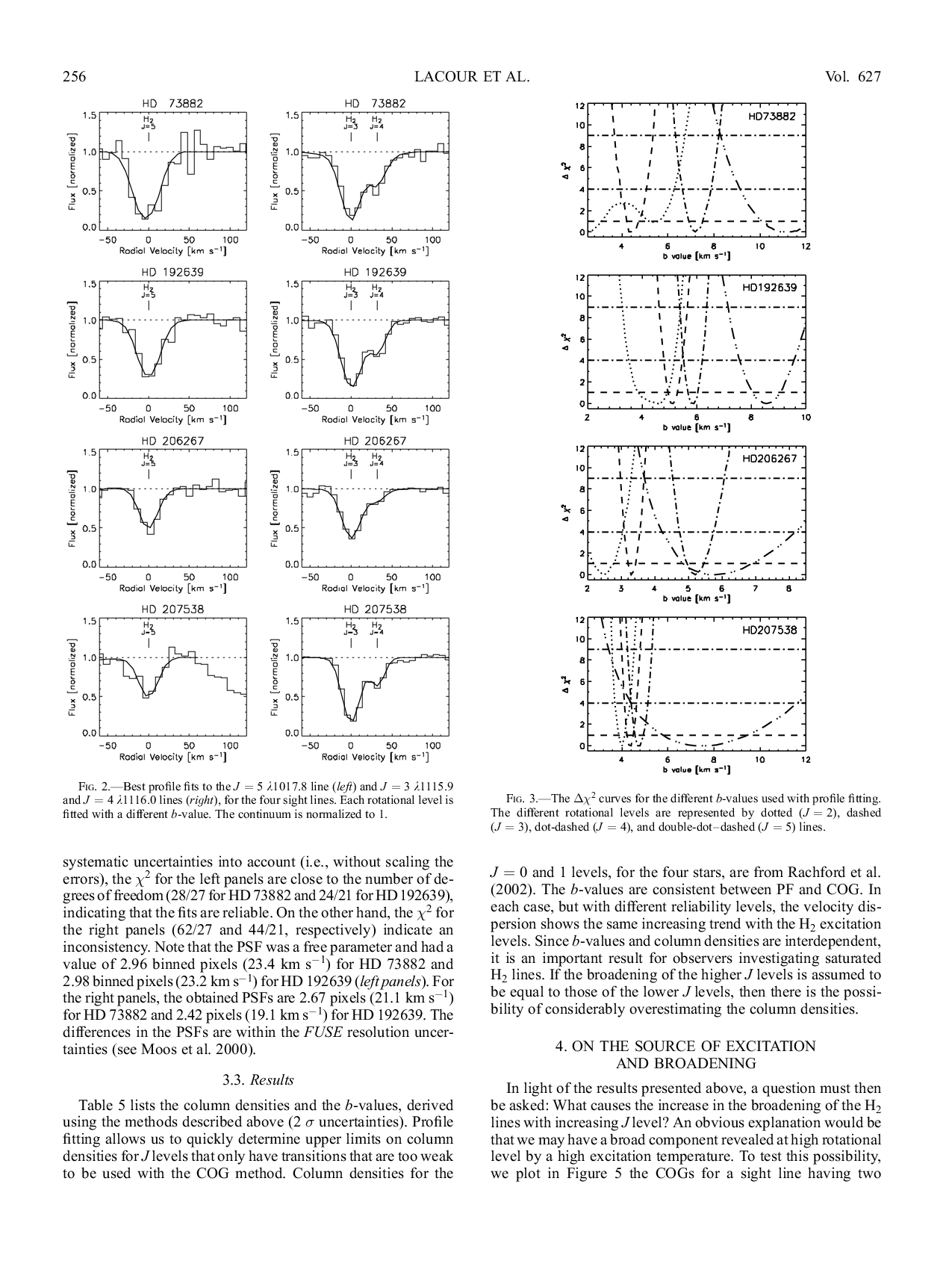}}
\end{figure}
\begin{figure}  \centering
  \resizebox{\hsize}{!}{\includegraphics{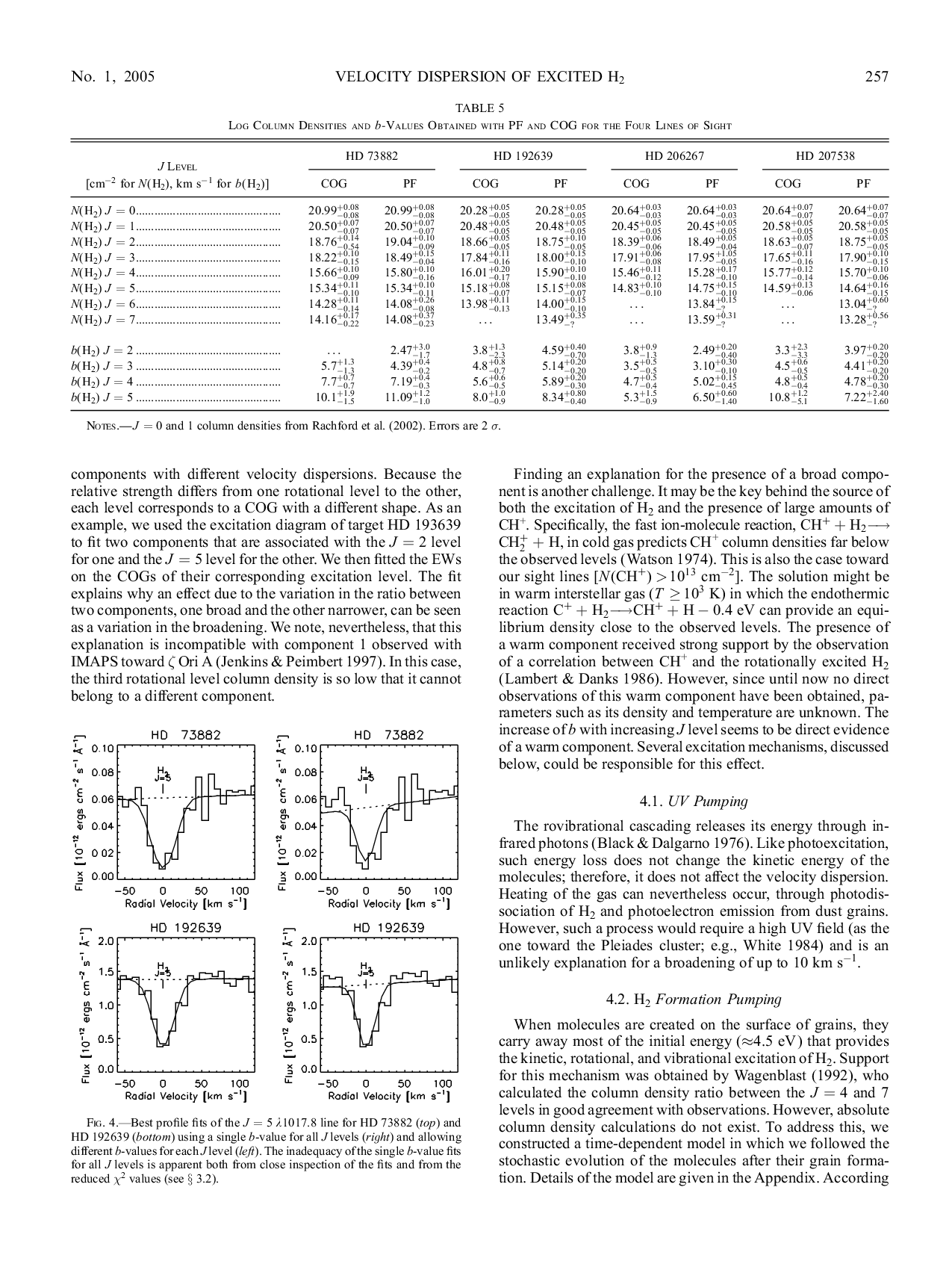}}
\end{figure}
\begin{figure}  \centering
  \resizebox{\hsize}{!}{\includegraphics{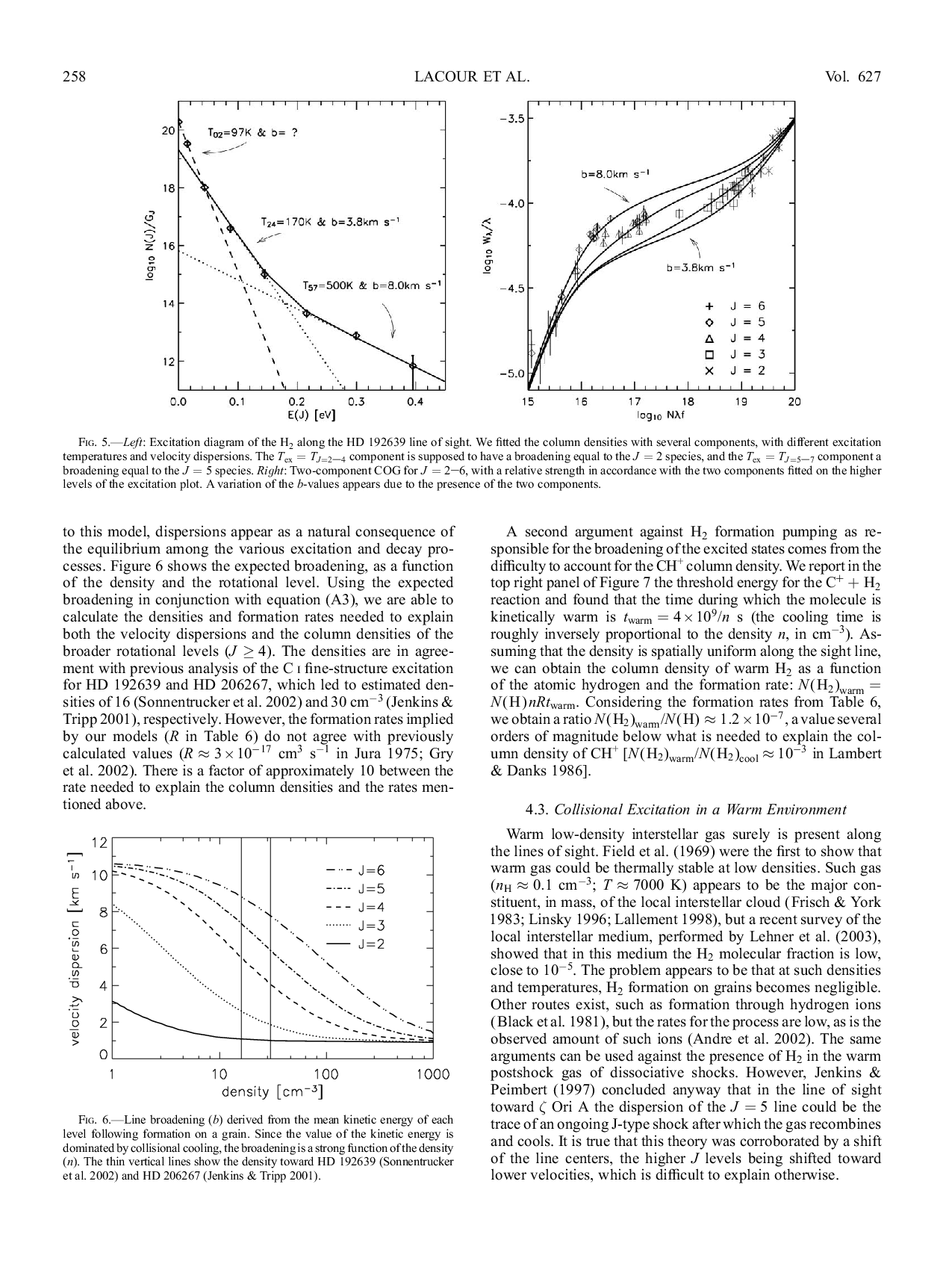}}
\end{figure}
\begin{figure}  \centering
  \resizebox{\hsize}{!}{\includegraphics{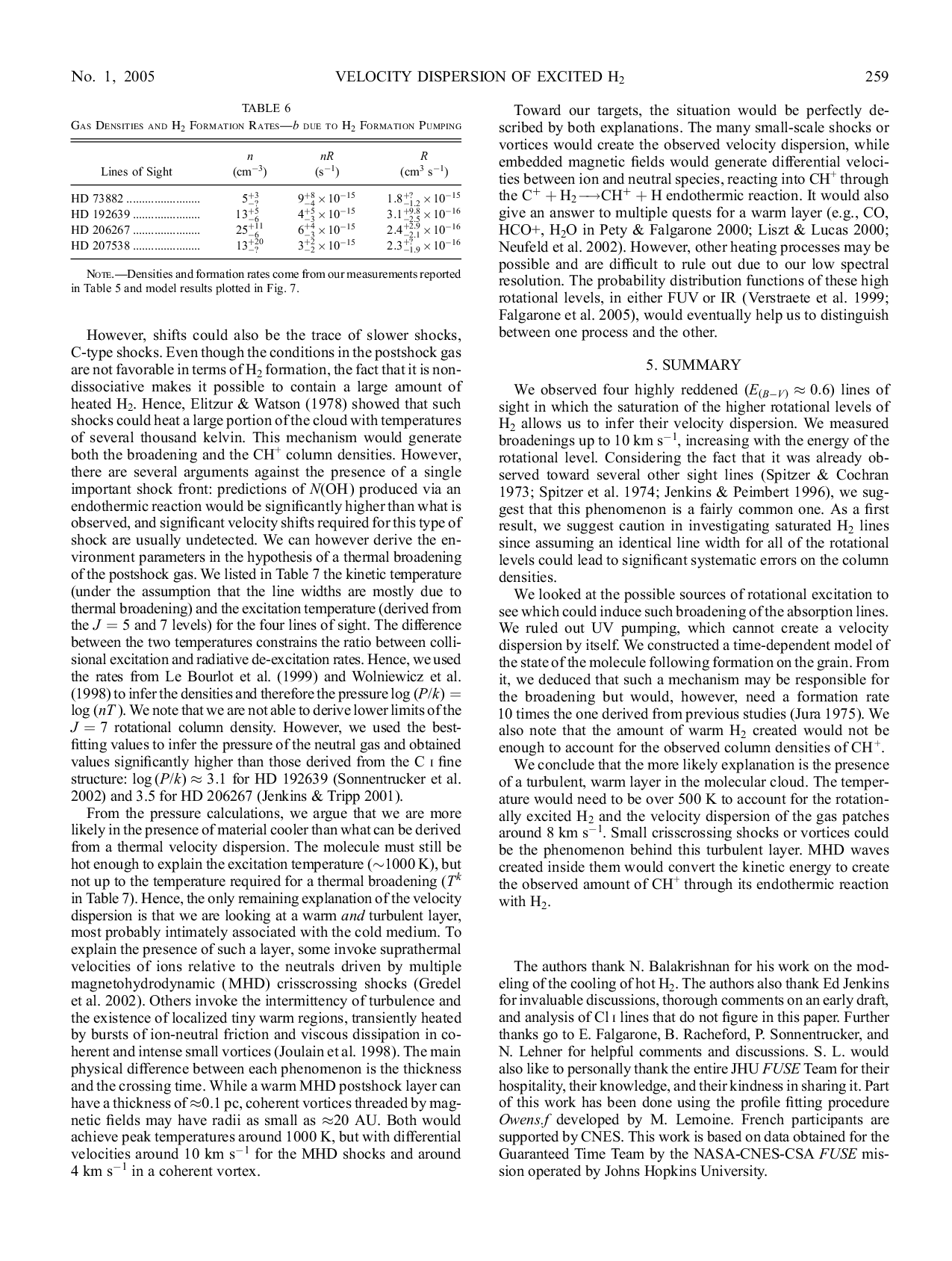}}
\end{figure}
\begin{figure}  \centering
  \resizebox{\hsize}{!}{\includegraphics{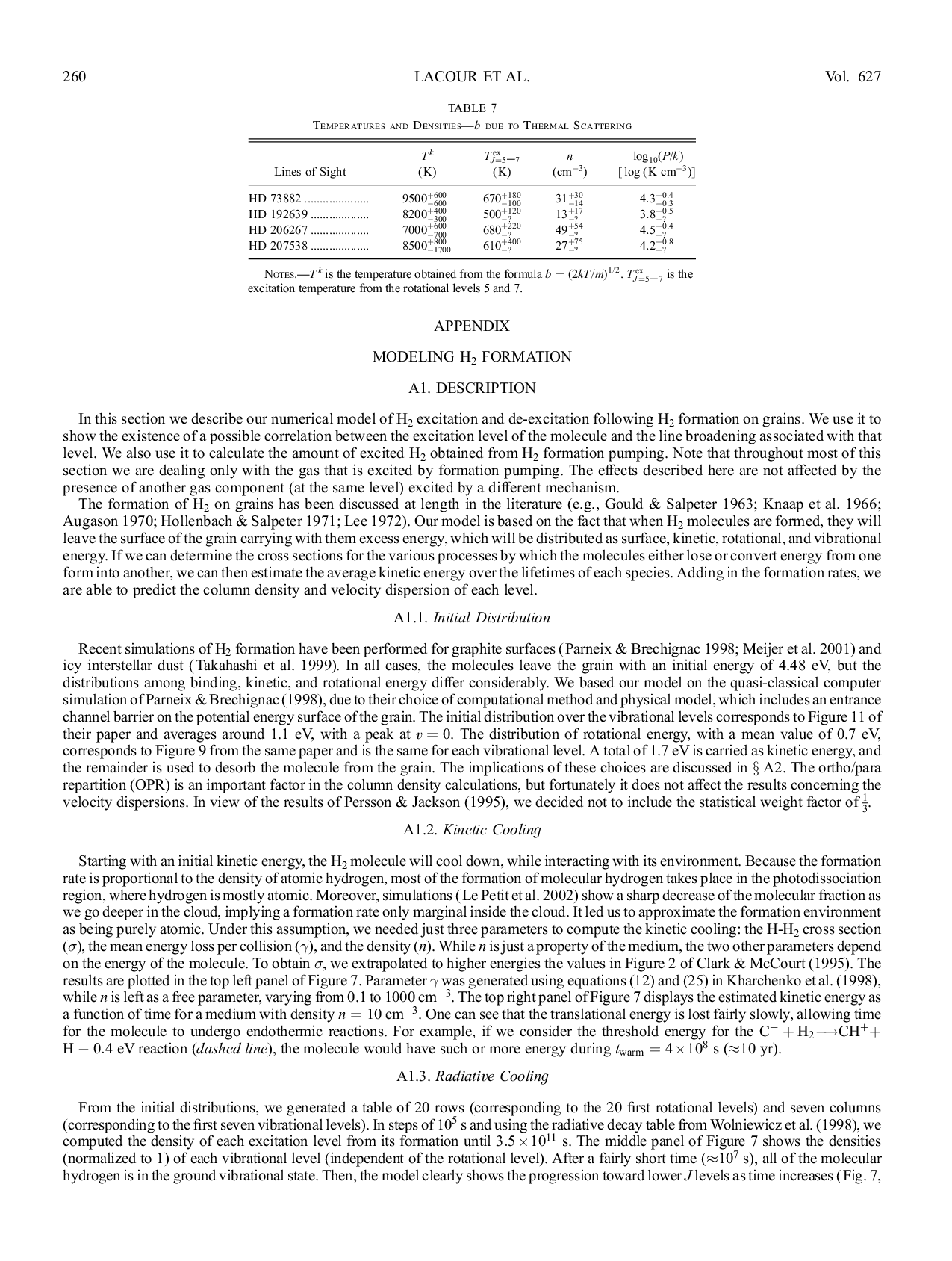}}
\end{figure}
\begin{figure}  \centering
  \resizebox{\hsize}{!}{\includegraphics{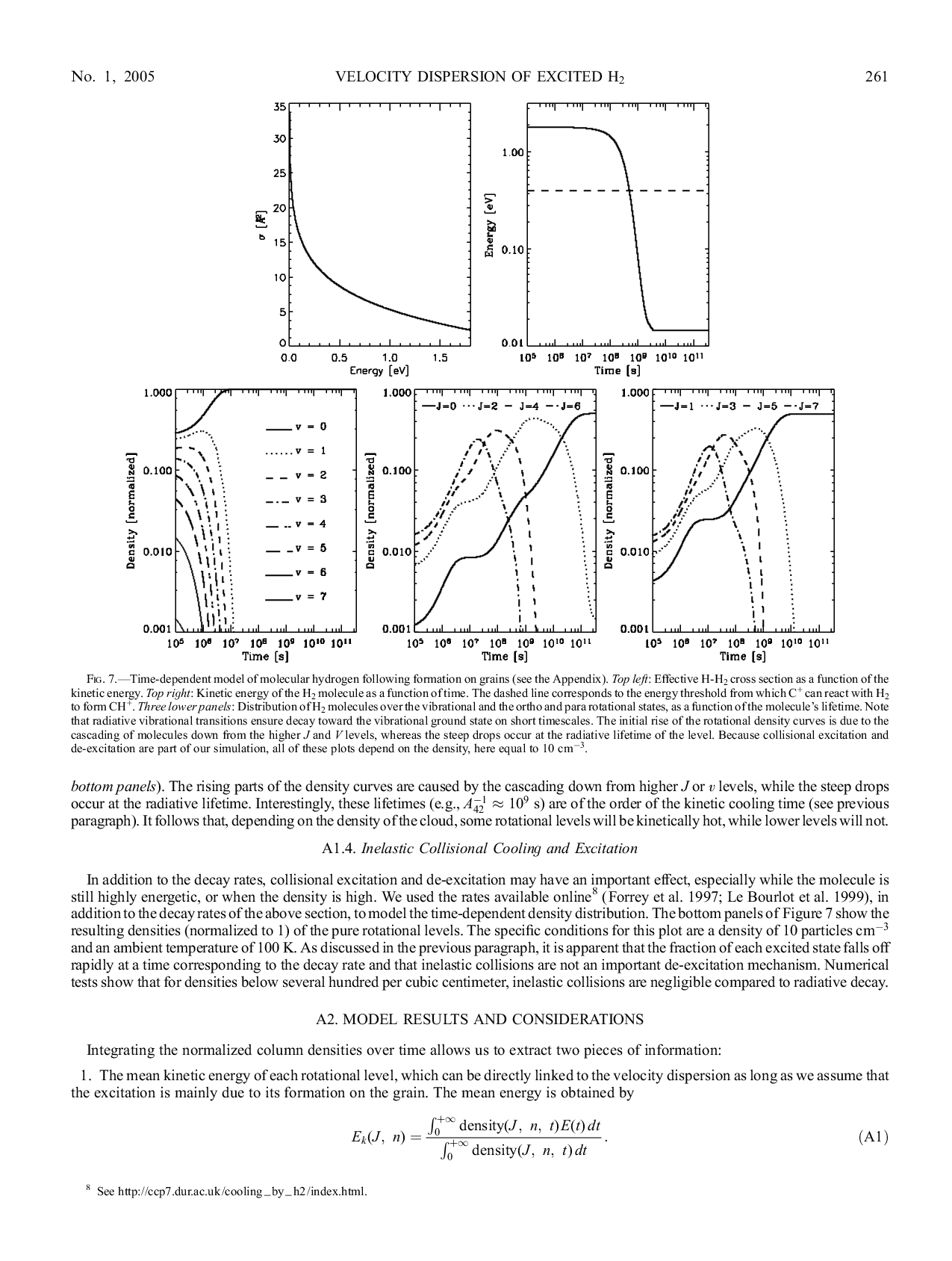}}
\end{figure}
\begin{figure}  \centering
  \resizebox{\hsize}{!}{\includegraphics{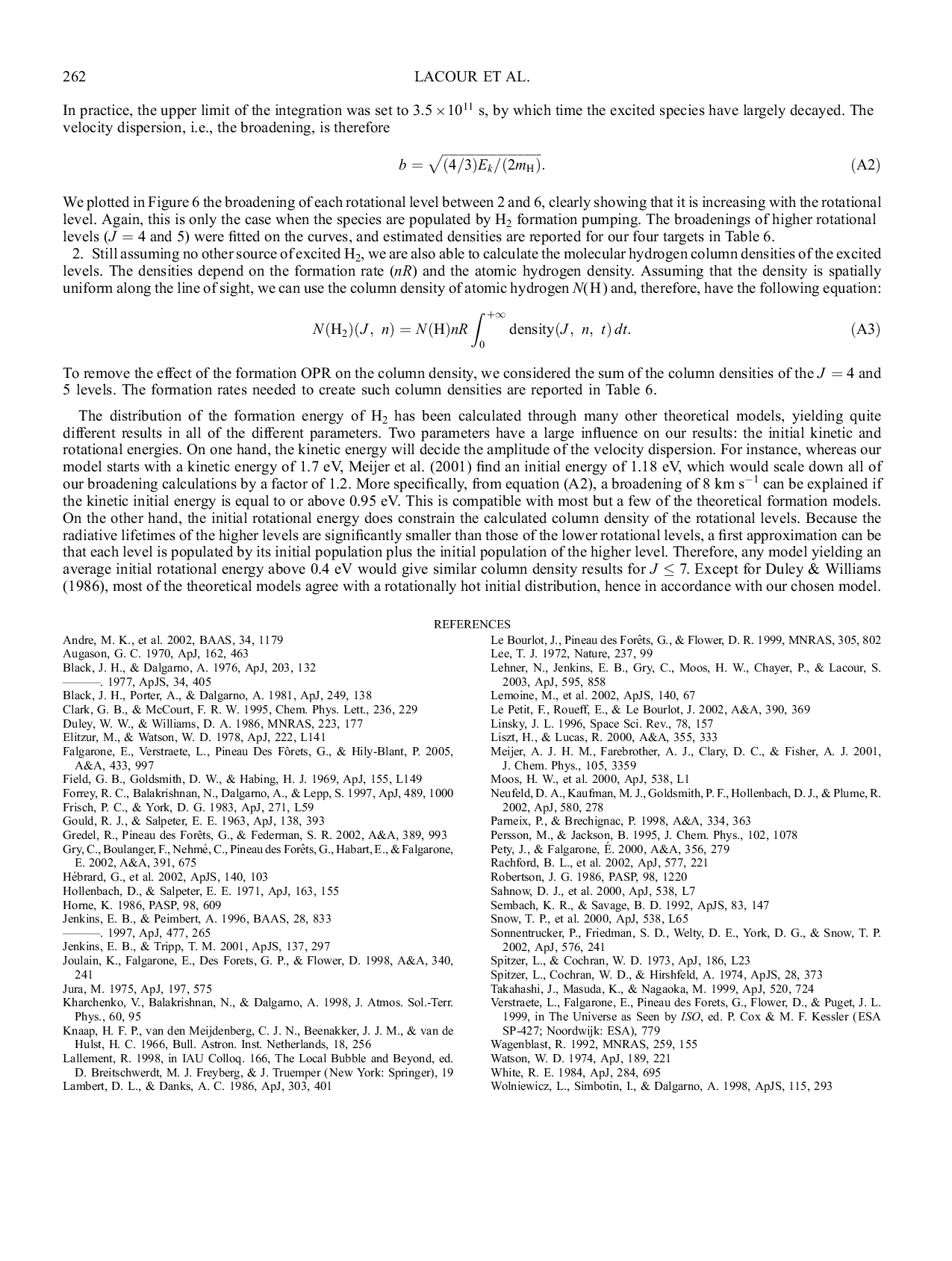}}
\end{figure}
\cleardoublepage
\begin{figure}  \centering
  \resizebox{\hsize}{!}{\includegraphics{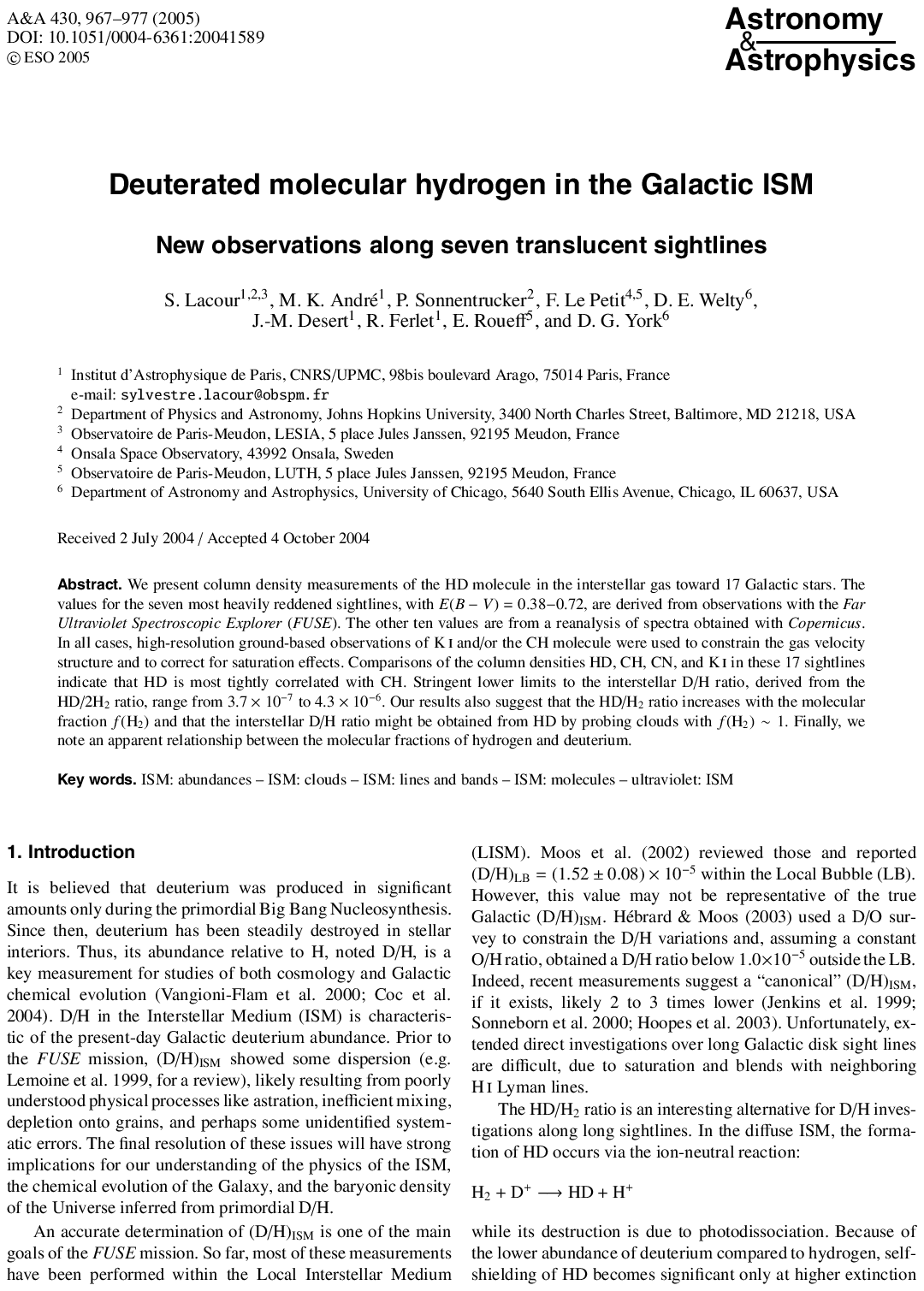}}
\end{figure}
\begin{figure}  \centering
  \resizebox{\hsize}{!}{\includegraphics{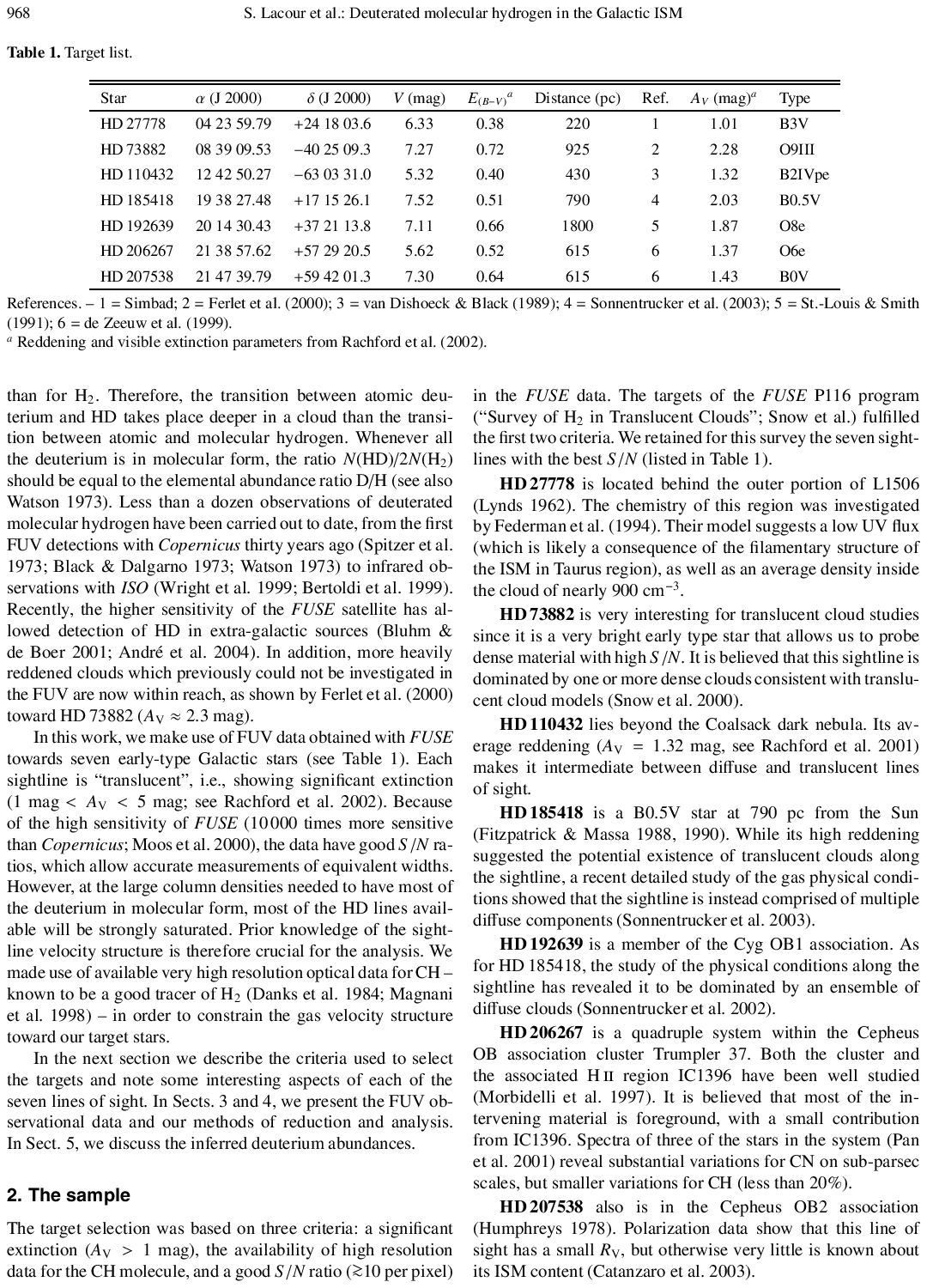}}
\end{figure}
\begin{figure}  \centering
  \resizebox{\hsize}{!}{\includegraphics{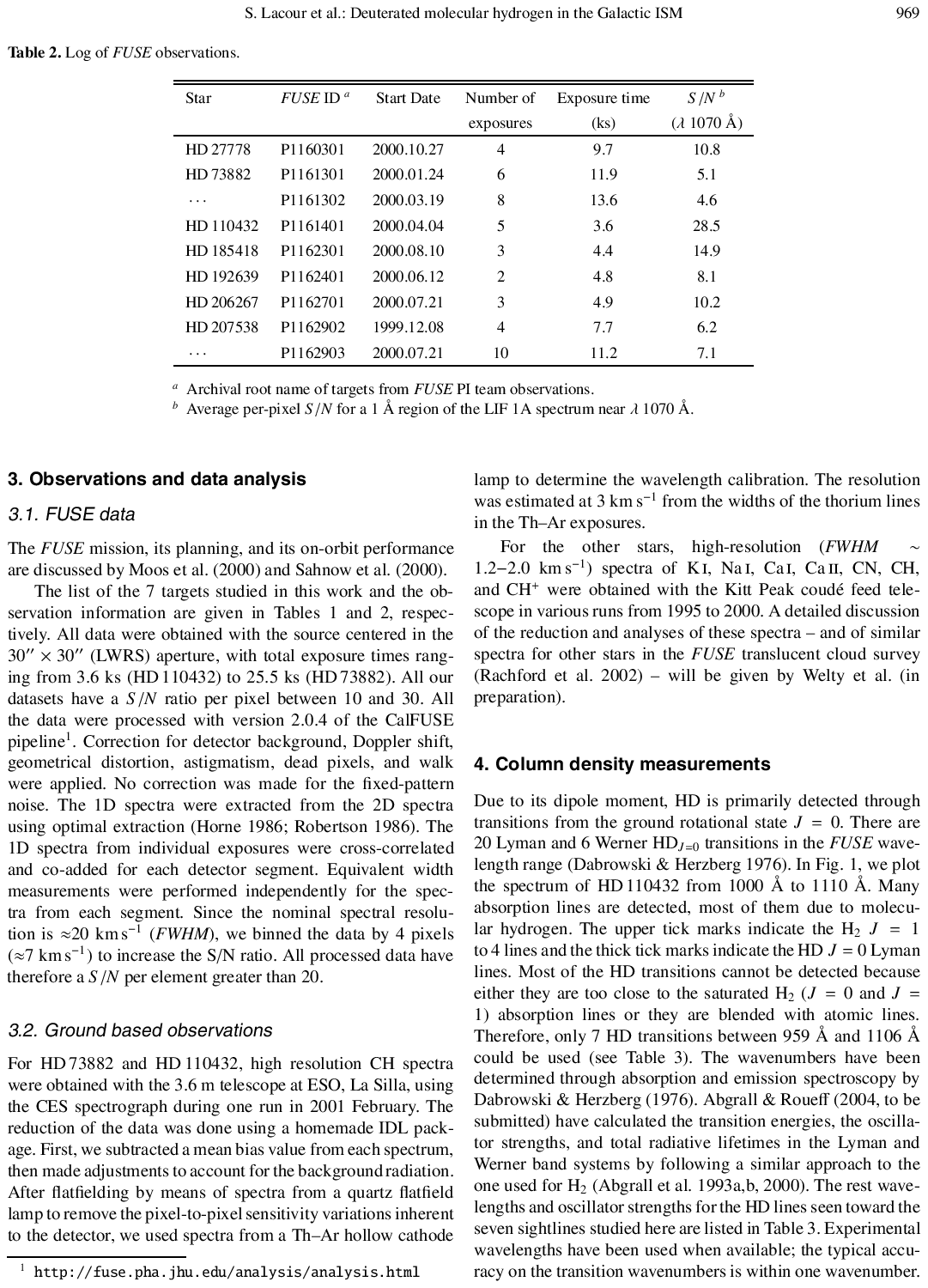}}
\end{figure}
\begin{figure}  \centering
  \resizebox{\hsize}{!}{\includegraphics{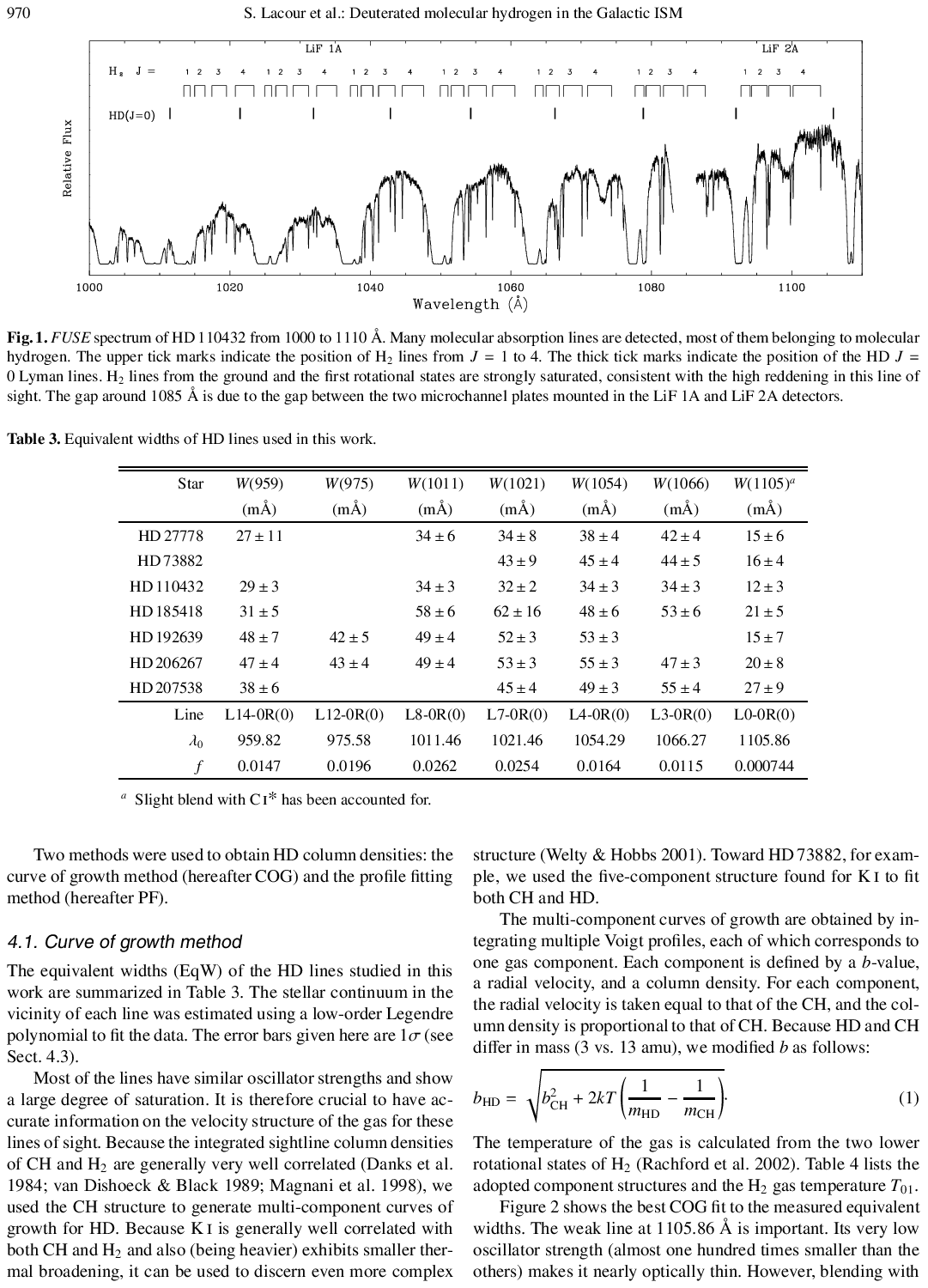}}
\end{figure}
\begin{figure}  \centering
  \resizebox{\hsize}{!}{\includegraphics{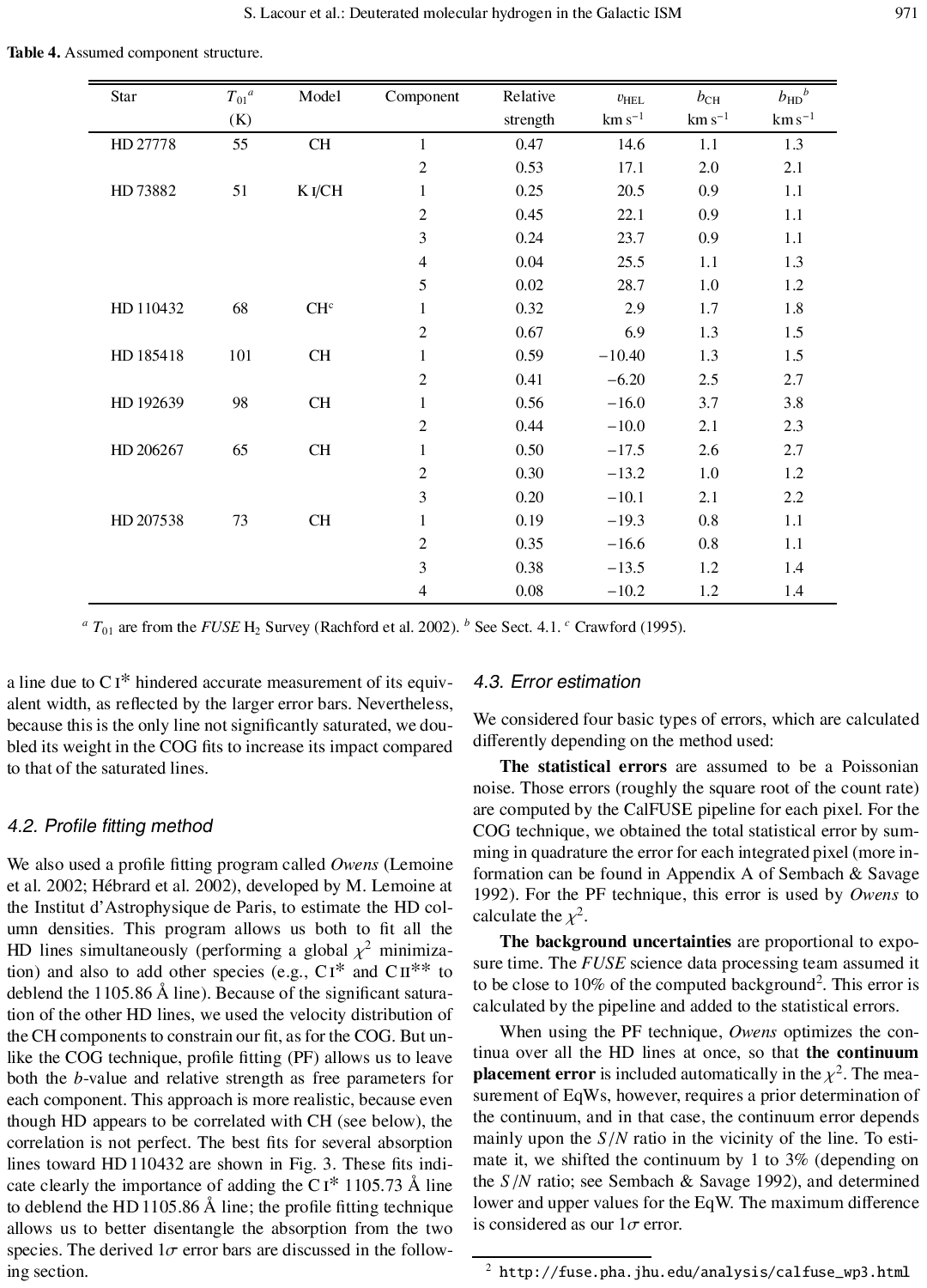}}
\end{figure}
\begin{figure}  \centering
  \resizebox{\hsize}{!}{\includegraphics{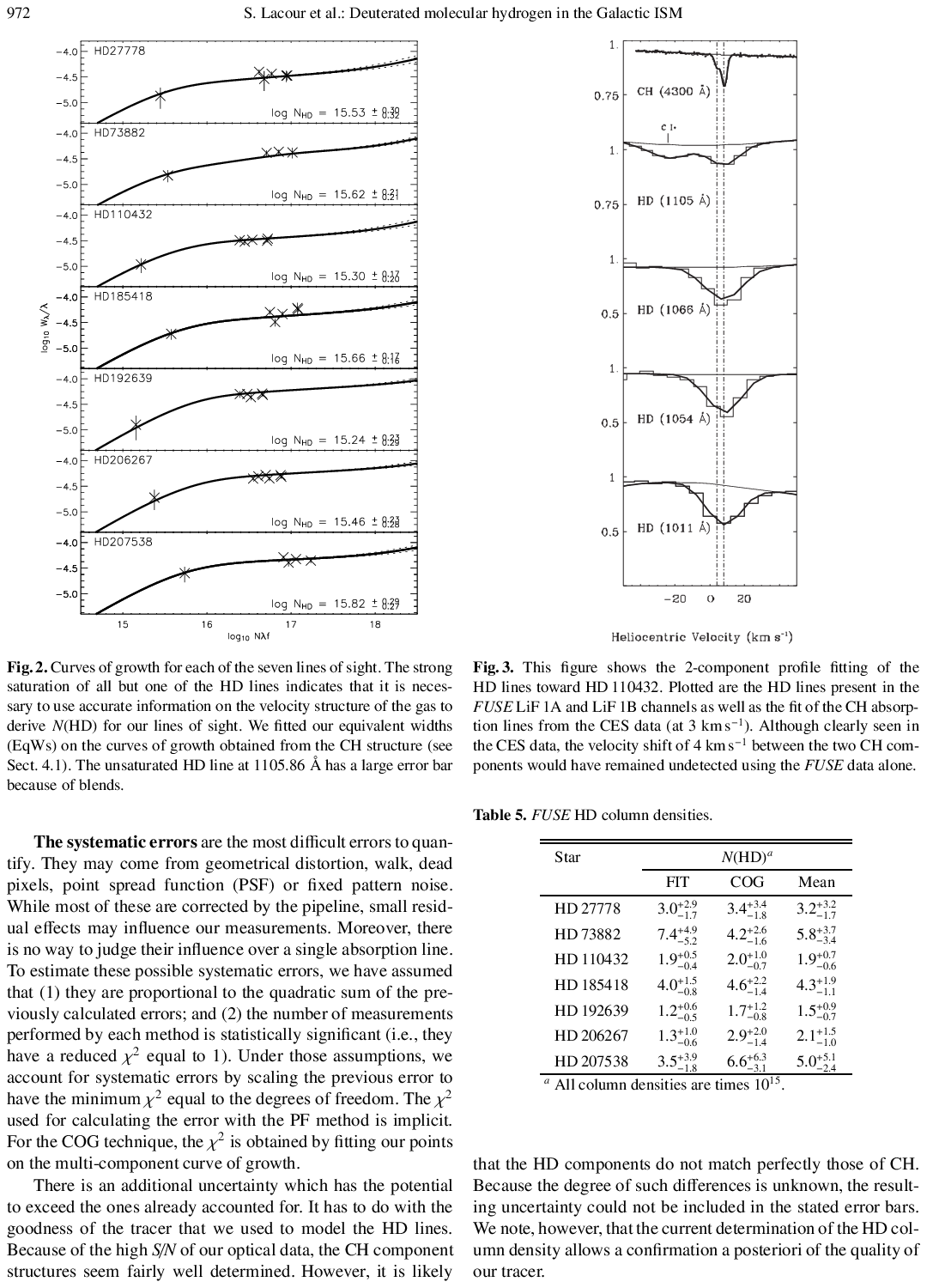}}
\end{figure}
\begin{figure}  \centering
  \resizebox{\hsize}{!}{\includegraphics{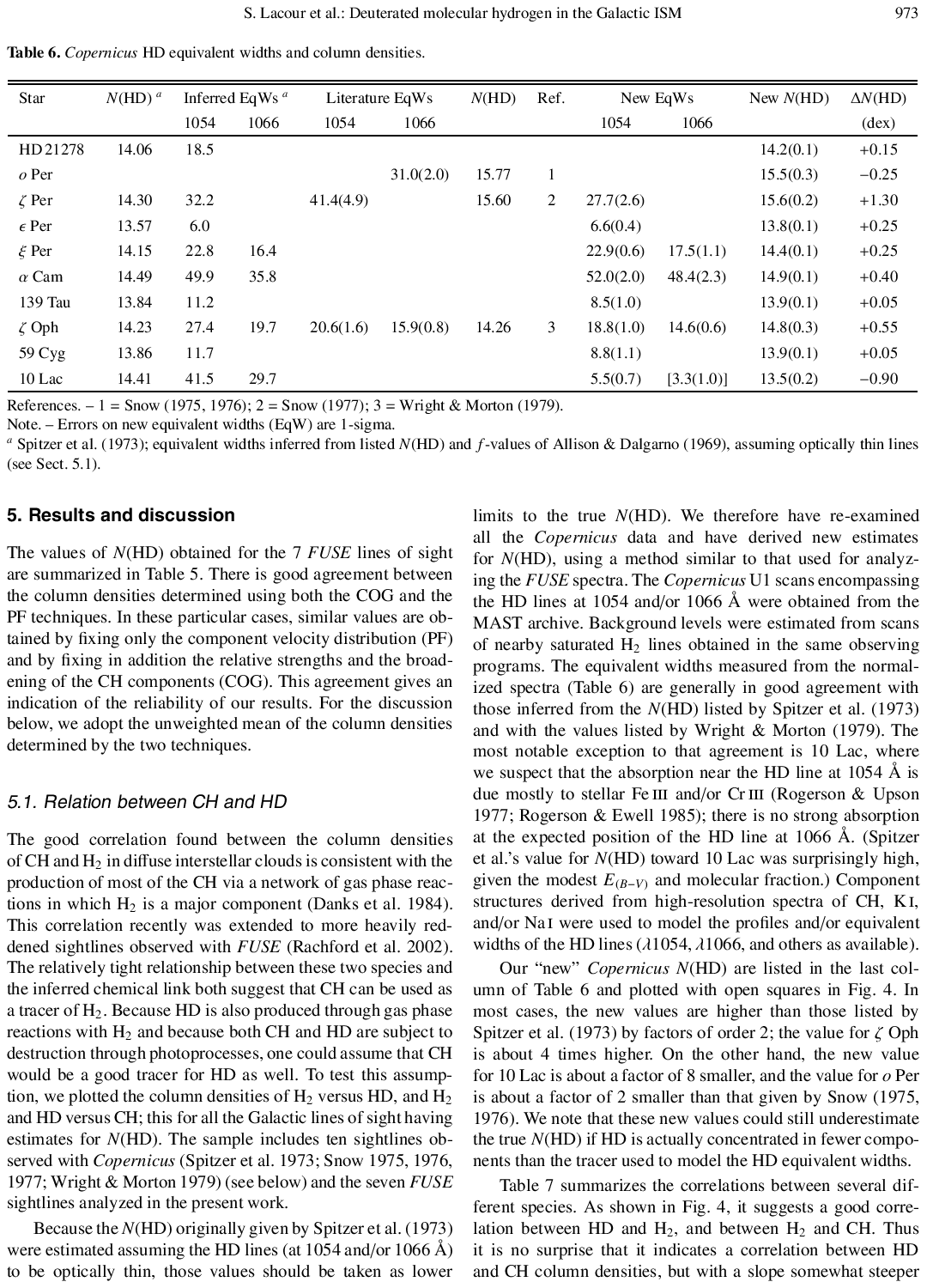}}
\end{figure}
\begin{figure}  \centering
  \resizebox{\hsize}{!}{\includegraphics{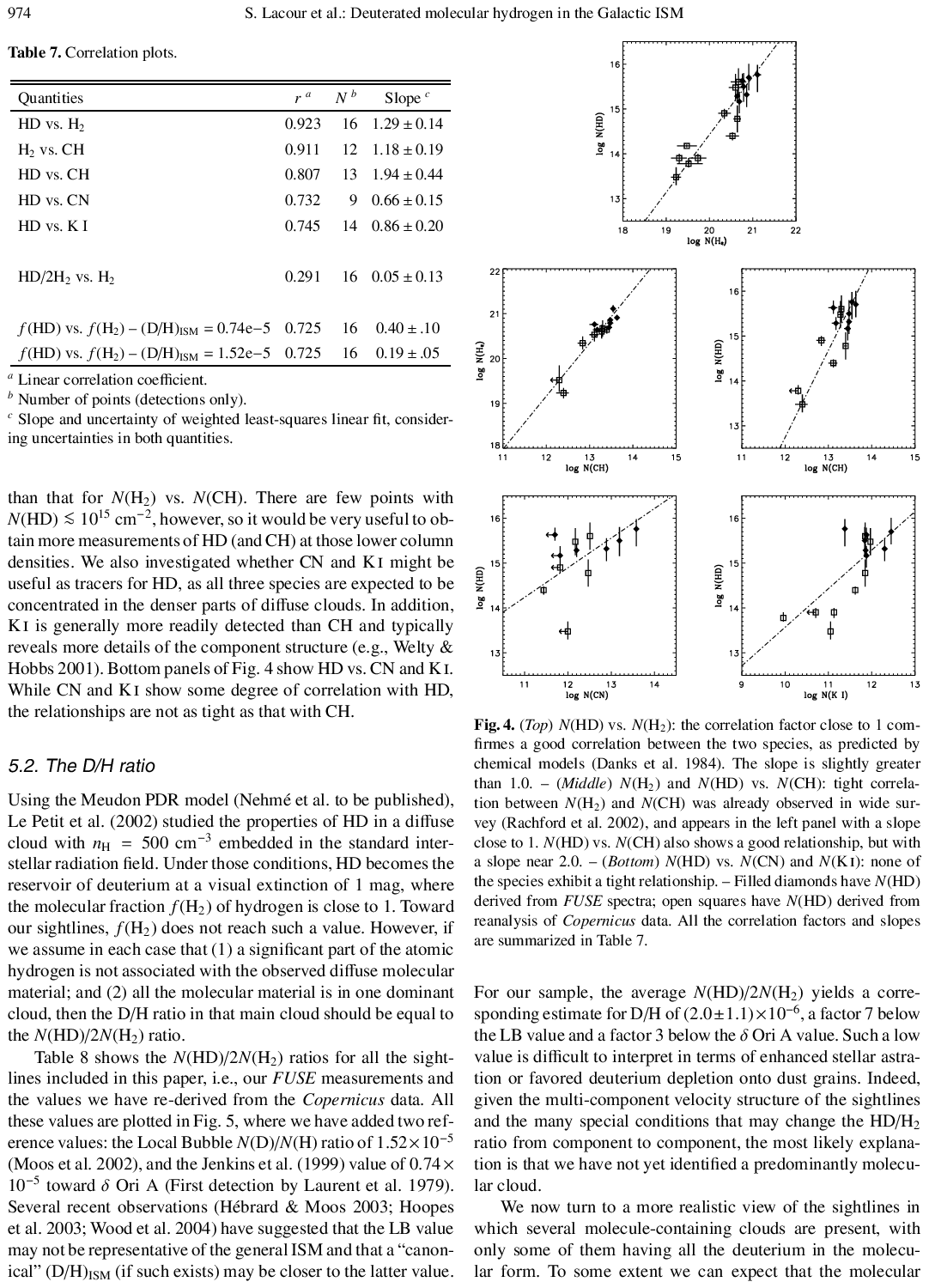}}
\end{figure}
\begin{figure}  \centering
  \resizebox{\hsize}{!}{\includegraphics{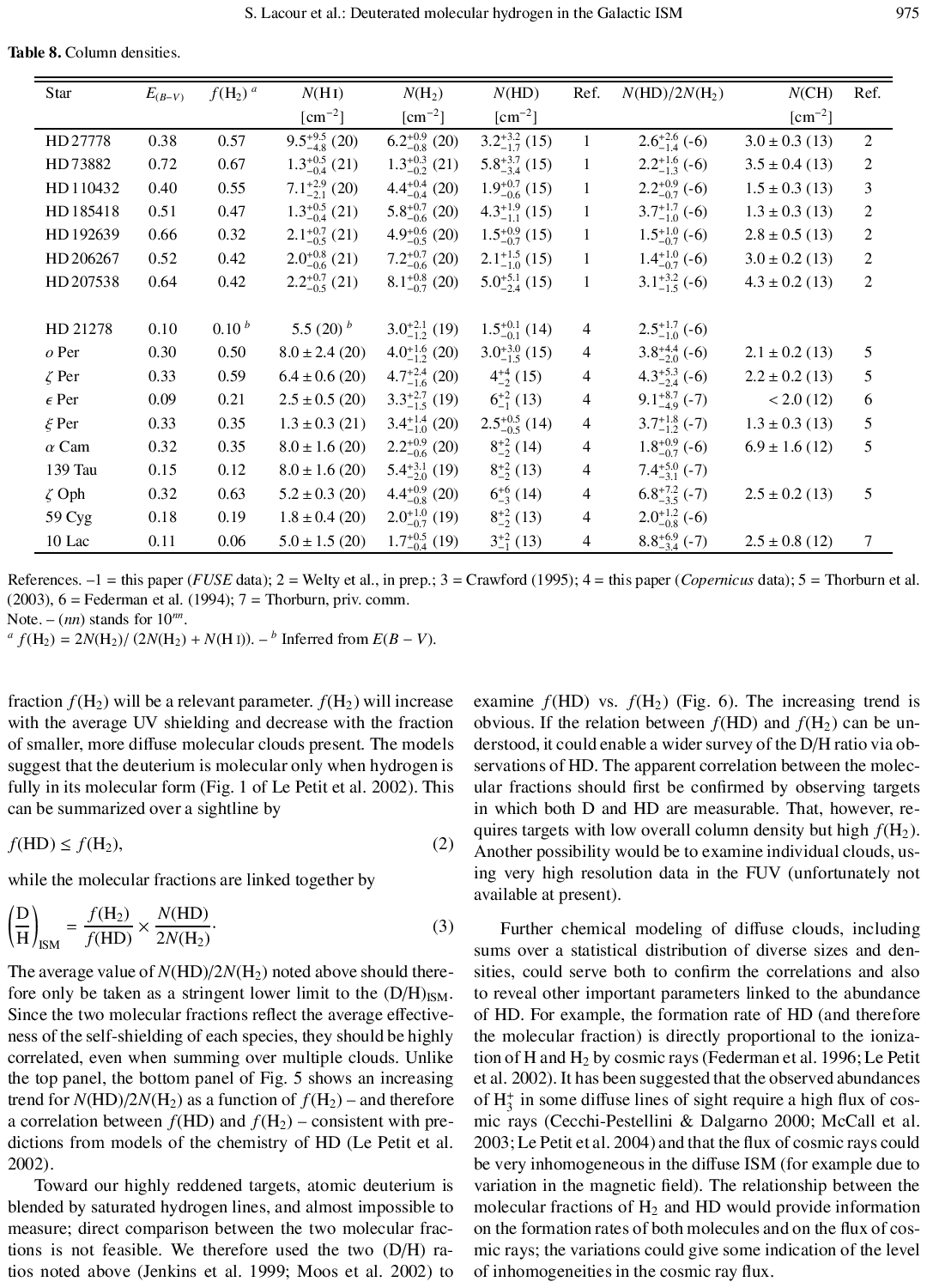}}
\end{figure}
\begin{figure}  \centering
  \resizebox{\hsize}{!}{\includegraphics{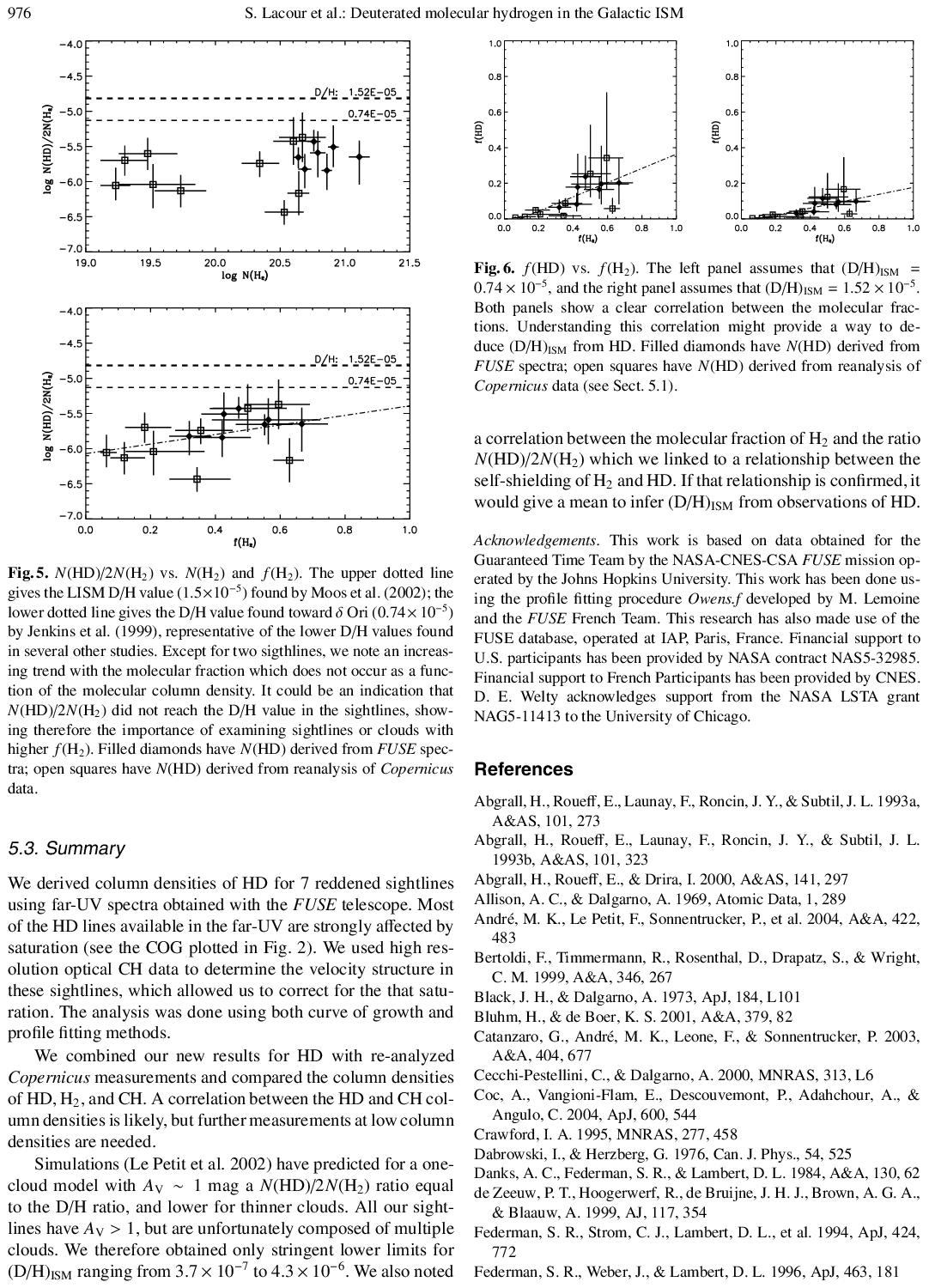}}
\end{figure}
\begin{figure}  \centering
  \resizebox{\hsize}{!}{\includegraphics{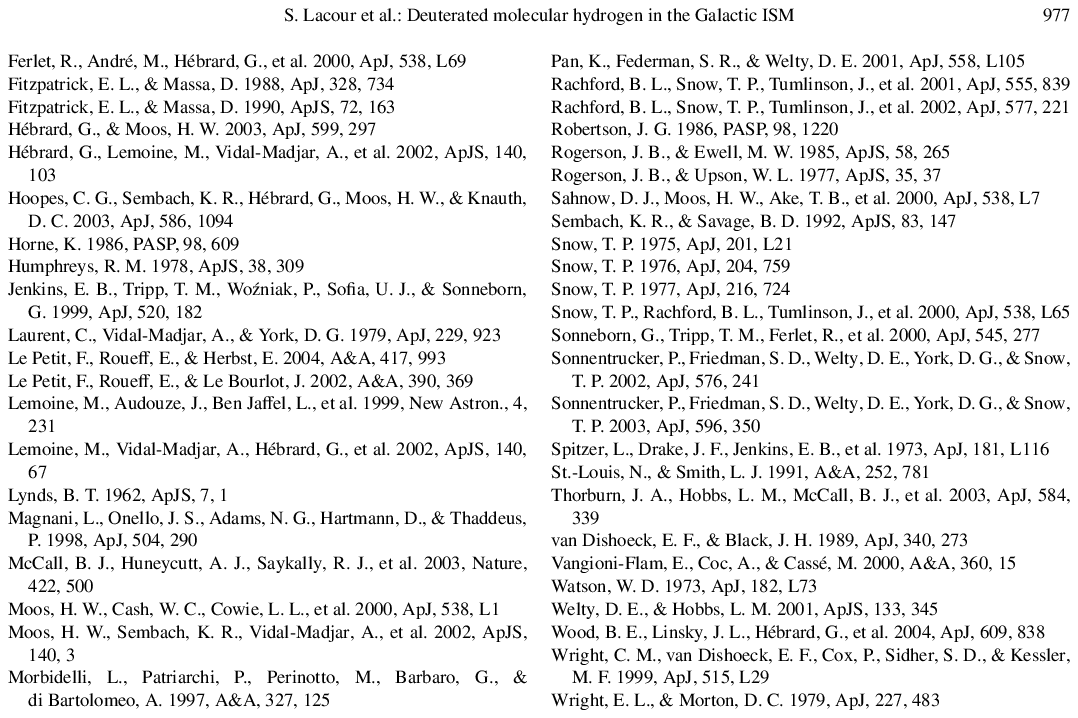}}
\end{figure}

%

\def\bibname{Bibliographie g\'en\'erale}
\addcontentsline{toc}{part}{{\large\sf\bfseries Bibliographie}}
\bibliographystyle{plainnat_fr}
\bibliography{Thesebib,Thesebib2}

\end{document}